\documentclass[12pt,draft,twoside]{article}
\usepackage{fullpage}
\usepackage{amsfonts}

\usepackage{latexsym}

\newcommand{\tensor}{\ensuremath{T^{b_1 \ldots b_s}_{a_1 \ldots a_r}}}

\newtheorem{theorem}{Theorem}

\newenvironment{mylist}[1]
   {\begin{list}{}%
      {%
      \settowidth{\labelwidth}{#1}%
      \setlength{\leftmargin}{\labelwidth}%
      \addtolength{\leftmargin}{\labelsep}}}%
   {\end{list}}

\newtheorem{hunch}{Hunch}
\newcommand{\cir}[1]{\ensuremath{(\mathit{#1})}}
\newcommand{\seg}[1]{\ensuremath{[\mathit{#1}]}}
\newcommand{\ang}[1]{\ensuremath{\langle #1 \rangle}}
\newcommand{\z}[1]{\ensuremath{{\mathbb Z}_{#1}}}
\newcommand{\x}[1]{\ensuremath{{\mathcal X}_{#1}}}
\newcommand{\Z}{\ensuremath{{\mathbb Z}}}
\newcommand{\R}{\ensuremath{{\mathbb R}}}
\newcommand{\N}{\ensuremath{{\mathbb N}}}
\newcommand{\Q}{\ensuremath{{\mathbb Q}}}
\newcommand{\X}{\ensuremath{{\mathcal X}}}
\newcommand{\F}{\ensuremath{{\mathcal F}}}
\newcommand{\M}{\ensuremath{{\mathcal M}}}
\newcommand{\A}{\ensuremath{{\mathcal A}}}
\newcommand{\U}{\ensuremath{{\mathcal U}}}
\newcommand{\V}{\ensuremath{{\mathcal V}}}
\newcommand{\W}{\ensuremath{{\mathcal W}}}
\newcommand{\Y}{\ensuremath{{\mathcal Y}}}
\newcommand{\I}{\ensuremath{{\mathcal I}}}

\newcommand{\Chi}{\raisebox{2pt}{$\chi$}}
\newcommand{\Seg}[1]{\ensuremath{{\mathcal S}_{#1}}}
\newcommand{\zf}{\ensuremath{{\mathcal Z}}}
\newcommand{\G}{\ensuremath{\Gamma}}
\newcommand{\g}[1]{\ensuremath{\Gamma\hspace{-2pt}_{#1}}}

\newlength{\wod}
\newlength{\woda}
\newcommand{\hw}{@{\hspace{\wod}}}
\newcommand{\proofend}{\hfill\mbox{\hspace{24pt}$\Box$}}

\begin{document}
\begin{titlepage}
\pagenumbering{roman}
\begin{center}\vspace*{\fill}
{\LARGE\textbf{Combinatorial Spacetimes}\\}
{\normalsize\vspace{72pt}
by\\
David Hillman\\
Ph.D., University of Pittsburgh, 1995\\
dhi@olympus.net\\
\vspace{81pt}
Submitted to the Graduate Faculty of\\
Arts and Sciences in partial fulfillment\\
of the requirements for the degree of\\
Doctor of Philosophy\\
Department of Mathematics and Statistics\\
University of Pittsburgh\\
\vspace{81pt}
defended: November 30, 1995\\
most recent correction: \today\\}
\end{center}
\vspace*{\fill}

\newpage
\thispagestyle{plain}
\setcounter{page}{2}
\mbox{}
\newpage
\thispagestyle{plain}
\begin{center}
\textbf{\large Combinatorial Spacetimes}\\
David Hillman, Ph.D.\\
University of Pittsburgh, 1995\bigskip\\
\end{center}
Combinatorial spacetimes are a class of dynamical systems in which finite
pieces of spacetime contain finite amounts of information. Most of the guiding
principles for designing these systems are drawn from general relativity: the
systems are deterministic; spacetime may be foliated into Cauchy surfaces; the
law of evolution is local (there is a light-cone structure); and the geometry
evolves locally (curvature may be present; big bangs are possible). However,
the systems differ from general relativity in that spacetime is a combinatorial
object, constructed by piecing together copies of finitely many types of
allowed neighborhoods in a prescribed manner. Hence at least initially there is
no metric, no concept of continuity or diffeomorphism. The role of
diffeomorphism, however, is played by something called a ``local equivalence
map.''

Here I attempt to begin to lay the mathematical foundations for the study of
these systems. (Examples of such systems already exist in the literature. The
most obvious is reversible cellular automata, which are flat combinatorial
spacetimes. Other related systems are structurally dynamic cellular automata, L
systems and parallel graph grammars.) In the 1+1-dimensional oriented case,
sets of spaces may be described equivalently by matrices of nonnegative
integers, directed graphs, or symmetric tensors; local equivalences between
space sets are generated by simple matrix transformations. These equivalence
maps turn out to be closely related to the flow equivalence maps between
subshifts of finite type studied in symbolic dynamics. Also, the symmetric
tensor algebra generated by equivalence transformations turns out to be
isomorphic to the abstract tensor algebra generated by commutative
cocommutative bialgebras.

In higher dimensions I attempt to follow the same basic model, which is to
define the class of $n$-dimensional space set descriptions and then generate
local equivalences between these descriptions using elementary
transformations. Here I study the case where space is a special type of colored
graph (discovered by Pezzana) which may be interpreted as an $n$-dimensional
pseudomanifold. Finally, I show how one may study the behavior of combinatorial
spacetimes by searching for constants of motion, which typically are associated
with local flows and often may be interpreted in terms of particles.

\newpage
\thispagestyle{plain}
\vspace*{74pt}\noindent {\LARGE \textbf{Preface}}\bigskip\vspace{12pt}

\noindent Though I no doubt have not entirely succeeded, my goal in writing
this document was to produce a work that even I, had I not already done this
research, could understand. To this end I have sacrificed brevity, but
hopefully not to the point of excess. Many concepts are presented in more than
one way. There are plenty of examples and pictures. An understanding of several
basic ideas in algebra and topology is assumed on occasion.

Thanks to John Baez, Tim Boykett, Andr\'{e} de Carvalho, Bob Heath and Danrun
Huang for helpful discussions. Thanks to Doug Lind for sending me a preprint
of his book with Brian Marcus on symbolic dynamics. Thanks to Jack Wagoner for
pointing out the connection of my work with John Franks' paper. Thanks to Ken
Manders, John Porter and Frank Slaughter for being on my committee. Thanks to
Erica Jen for bringing me to Los Alamos and working with me a few summers
ago. And many, many thanks to George Sparling, my advisor. He treated me as an
equal even though mathematically we are not even remotely close to being equal;
this made being his graduate student a pleasure. He was enthusiastic and
patient, and I learned many things from him. And it was he who noticed that my
blackboard scrawlings resembled the axioms of a bialgebra; I knew nothing of
bialgebras, so I do not know when or if I would have discovered this on my own.

Finally, special thanks go to me, who has had to put up with myself during over
six years of graduate school. That certainly could not have been easy.

\newpage
\thispagestyle{plain}
\vspace*{40pt}\tableofcontents

\end{titlepage}

\newpage
\pagenumbering{arabic}
\vspace*{40pt}\section[Introduction]{\LARGE Introduction}\bigskip
\label{introduction}

Pure mathematicians often say that their interest in an area of mathematics is
completely intrinsic, and quite apart from questions of relevance to the
physical world. One mathematician I know, for example, says she studies general
relativity because it is ``nice geometry.'' For whatever reason, I do not seem
to be made this way. I get my mathematical thrills when I feel that the
mathematics I am doing may be related to fundamental laws of nature. In
particular, I have a fantasy that I can make a positive contribution towards
the search for principles that will enable us to unify general relativity and
quantum mechanics in a single consistent physical \mbox{theory}.

The empirical evidence available to me indicates that my fantasies rarely come
true. I only mention this one here because it influences, in a general way, the
choices I make in the work that follows. More specific influences arise from
the following hunches of mine concerning the probable nature of a consistent
unified physical \mbox{theory}.

\begin{hunch}The universe is simple.\end{hunch}
It is not clear how one should make the notion of simplicity precise. For
heuristic purposes I will use the idea of algorithmic information theory, which
says (roughly) that the simplicity of an object can be measured by the length
of its shortest description.

There is a natural candidate for a type of universe satisfying this
criterion. Such a universe can be completely described in a finite number of
bits by specifying an initial state and a deterministic algorithm by which that
state evolves.

In the case where the law of evolution is invertible, the ``initial'' state
need not be initial in the usual sense; it must, however, be a Cauchy surface,
and the universe must be completely describable by evolving this surface
backwards and forwards in time. I will sometimes call such a state an initial
state anyway.

Quantum field theory is not appealing from this point of view, because it
(supposedly) contains randomness as an essential component. Thus it is not at
all algorithmic, and hence the associated universe is not at all simple,
according to my adopted heuristic definition of simplicity.

On the other hand, general relativity avoids this difficulty by being
deterministic. It also allows scenarios such as the big bang, which is itself
appealing from this point of view because in such a scenario there is an
initial state which is very small, hence perhaps easily described.

But smallness of the initial state in itself does not insure finite
describability. From the point of view taken here, it is a weakness of general
relativity that it makes no restrictions on the complexity of the initial
state.

If the initial state is finitely describable and it evolves algorithmically,
then subsequent states must also be finitely describable. A second hunch is
that this finite describability is a \emph{local} property of spacetime.

\begin{hunch}The universe is locally simple.\end{hunch}
The idea here is that any finite chunk of spacetime is finitely
describable. According to this view, the chunk of spacetime located at the
present moment inside this letter ``o'', while no doubt quite complicated, is
probably not infinitely complicated. Again, general relativity does not insure
this.

A natural way to insure that one's mathematical models of universes are locally
simple is to let spacetimes be combinatorial objects that are made from a
finite set of allowed neighborhoods by piecing copies of those neighborhoods
together in a prescribed manner. This insures that there are only finitely many
allowed pieces of spacetime of any given finite size.

It is certainly to be expected that some continuous models also may possess
this property of local simplicity. If so, then it seems likely that in some
sense such models are isomorphic to combinatorial models of the sort described
above. Since it is not immediately obvious which continuous models are locally
simple and which are not, it seems safer to concentrate on the combinatorial
models, which are guaranteed to be locally simple, and later to examine the
question of the local simplicity of continuous models.

There are two more hunches, somewhat different from the previous two.

\begin{hunch}The law of evolution is invertible.\end{hunch}
The idea here is that the universe is foliated by Cauchy surfaces. This is
true in general relativity, and it would also be true of quantum field theory
were it not for wave function collapse. The alternative to invertibility is
dissipation: information is lost as the universe evolves. Certainly the
universe cannot be too dissipative, since from present conditions we have very
substantial ability to reconstruct the past, and what information we do lose
about the past may easily be attributed to the spreading out of that
information rather than to the absence of invertibility. My guess is that,
rather than the universe being just a little bit dissipative, it isn't
dissipative at all.

\begin{hunch}The law of evolution is local, and so is its inverse.\end{hunch}
The idea here is to preserve the notion of maximum velocity of causal
propagation (light-cone structure) from general relativity. Quantum field
theory is consistent with this point of view in an important sense, since in
that theory probabilities are propagated in a local fashion. A particle in
quantum field theory may travel faster than light; this, however, is not
evidence against local causation, since it is naturally explained by local
actions of virtual particles and antiparticles. Later I will present an example
of a local, deterministic model in which this same phenomenon occurs.

As a natural result of possessing this point of view, I chose to study
mathematical models of spacetimes that have the following properties.\vspace{3.5pt}

\begin{itemize}\linespread{1}\normalsize
\item Spacetimes are combinatorial objects which can be constructed by piecing
\mbox{together} copies of a finite number of allowed neighborhoods in a
prescribed manner.
\item A spacetime may be described by specifying a Cauchy surface and an
invertible, local, deterministic law of evolution.
\item There is no fixed background geometry; the geometry evolves in accordance
with the local law of evolution, just as it does in general relativity;
``curvature'' happens, and big bangs are possible.
\end{itemize}

It will be noted that, apart from the first property listed above, these
systems are modeled after general relativity. What, then, of quantum field
theory? Since the models are deterministic, they would qualify as hidden
variable theories if they exhibited quantum phenomena. But the models are also
local, and it is commonly believed (see~\cite{cushing/mcmullin}) that no local,
deterministic hidden variable theory can exist, since no such theory can
violate the Bell inequalities (as quantum field theories do). So how can I hope
that these systems might be relevant to the quantum aspects of our universe? I
do hope so, for several reasons.

Firstly, the combinatorial nature of these models implies an essential
discreteness, which also seems to be present in quantum phenomena (and which in
fact was partly suggested by these phenomena). I have already found that in
many of these models particles and antiparticles play an important role. It is
natural to wonder how far towards quantum phenomena one can go simply by
inserting this sort of discreteness into models which in other respects
resemble general relativity.

Second, models such as the ones I have described here have hardly been
studied. I think it is dangerous to make far-reaching conclusions about what
properties can or cannot be present in models without having studied them. It
is far too easy in such cases to make unwarranted assumptions or to use an
inappropriate conceptual scheme. I believe that this concern may be applicable
in the case of the standard arguments against deterministic, local hidden
variable theories. What I would prefer to these arguments is a direct
verification that the models I am interested in do or do not violate the Bell
inequalities. But how is this to be done? What constitutes a measurement in
these systems, or an observable? The answer to these questions is by no means
obvious, and without one I do not see how one can judge whether or not these
systems contain quantum phenomena. (For other arguments in favor of the
possibility of local, deterministic hidden variable theories, see 't
Hooft~\cite{hooft}.)

Even if it turns out that this class of models does not include one that
closely resembles our universe, I believe that there may be much of value to be
learned from them. The natural approaches to studying combinatorial dynamical
systems are somewhat different from those used to study standard dynamical
systems. My hope is that insights may be gained into the nature of dynamical
systems in general through the interplay of these different approaches.

The best way to begin learning about combinatorial spacetimes is to study
specific examples of these systems. It turns out that it is easy to construct
such examples by hand. Chapter~\ref{examples} describes how this may be done,
and carries out several such constructions in detail.

A number of types of combinatorial dynamical systems appear in the literature
that are related to combinatorial spacetimes. The most obvious of these is
reversible cellular automata; these are a special case of combinatorial
spacetime in which spacetime is flat. L systems and their generalizations are
another example of a related class of systems. Though space tends not to be
closed in these systems, and though L systems with invertible dynamics are
rarely studied, these systems come closer to combinatorial spacetimes than do
cellular automata in the sense that they allow for a more flexible spacetime
geometry (expanding universes; curvature). These various relationships are
described in Chapter~\ref{relatedsystems}.

In combinatorial spacetimes there is, at least initially, no concept of
continuity, hence no such thing as a diffeomorphism. However, the role of
diffeomorphism in differential geometry is played here by something called a
``local invertible map.'' These maps play two roles, which in turn correspond
to two ways of looking at dynamical systems: the space+time picture and the
spacetime picture. In the first case, one views local invertible maps as laws
of evolution; in the second, one views them as equivalences between
spacetimes. This is described in Chapter~\ref{spacetime}.

Local invertible maps are different from diffeomorphisms in the sense that they
are not just maps of one space or spacetime to another, but maps of one set of
spaces or spacetimes to another. It is natural to wonder whether one might be
able to put a topology on each set of spaces such that each local map from one
space set to another is a homeomorphism. If so, then combinatorial dynamics
would be a species of topological dynamics. In 1+1 dimensions this indeed turns
out to be the case; details are in Section~\ref{flow}.

My attempts to arrive at a cohesive mathematical picture of the set of all
1+1-dimensional oriented combinatorial spacetimes are described in
Chapter~\ref{orient}. This is the largest chapter in the dissertation, for it
is here that I have achieved my highest degree of success (which is not
surprising, since these systems constitute the simplest case). The solution
that I arrived at seems sufficiently elegant as to provide a model for the sort
of setup one may hope to achieve in higher-dimensional cases. It involves
descriptions of sets of spaces and local transformations of those
descriptions. The space set descriptions may be provided in three equivalent
ways: by square matrices of nonnegative integers, by directed graphs, or by
scalars in the algebra of symmetric tensors. The transformations of these
descriptions are generated by a single type of basic matrix operation (and by
an analogous operation on graphs and on tensors). A law of evolution is
obtained by composing a sequence of these operations together to obtain a map
from the original space set description to itself.

This formulation allowed me to discover several relationships between my work
and other areas of mathematics. For example, I solved an important special case
of the problem of finding the equivalence classes of space sets under these
transformations, only to discover that this solution had been arrived at ten
years earlier by Franks~\cite{franks}, who was working in the field of symbolic
dynamics. It turns out that flow equivalence maps, a not-terribly-well-known
sort of map which arises in symbolic dynamics, are exactly the transformations
between sets of spaces that I had discovered. On another front, the algebra
obtained by adding the above-mentioned operation to the algebra of symmetric
tensors turns out to be isomorphic to the abstract tensor algebra generated by
commutative cocommutative bialgebras. This means that there is a relationship
between 1+1-dimensional combinatorial spacetimes and Hopf algebras, which,
given the relationship of Hopf algebras to so many other parts of mathematics,
is intriguing.

My investigations of combinatorial spacetimes involving other sorts of spaces
are in more of an embryonic state than those just described. Nevertheless, I
believe that I have taken some steps in the right direction. Results in the
general 1+1-dimensional case (where orientedness is no longer required) are
described in Chapter~\ref{unorient}; higher-dimensional cases are discussed in
Chapter~\ref{higherD}.

Finally, one hopes in studying these systems not just to understand the
structure of the set of all such systems, but to understand the individual
systems themselves. What happens when one of these systems evolves? How can one
know whether the evolution of a particular space under some rule will be
periodic, or whether that space will expand forever? How does one find the true
dynamical properties of one of these systems---that is, the intrinsic
properties of the system, and not those properties which depend on the fact
that the system is described in a certain manner? What levels of complexity can
one expect to arise in such systems? I have barely begun to study such
questions, and only in the case where space is one-dimensional.

Several techniques have proven useful. In Chapter~\ref{invar} I describe how to
find constants of motion, which I call ``invariants.'' These correspond to a
local flow, which in turn can often be interpreted in terms of
particles. Sometimes the particles are solitons. Some particles have associated
antiparticles. In some cases systems can be described completely in terms of
their constituent particles and their interactions. In Chapter~\ref{cauchy} I
take a different approach: in the flat case (reversible cellular automata), I
study the properties of straight-line surfaces. I was able to prove that such
systems possess a light-cone structure, and that if one performs an analogue of
a Lorentz transformation on these systems the result is another cellular
automaton. However, the new description of spacetime obtained in this way
generally is not isomorphic to the original description. The question of
whether there exists a relativistic combinatorial spacetime remains an
interesting and unanswered one.

I have made a beginning, but much remains to be done. My sense is that many
deep and beautiful mathematical results remain to be discovered in this field.

\cleardoublepage\vspace*{40pt}\section[Examples]{\LARGE Examples}\bigskip
\label{examples}

In later chapters I will place emphasis on understanding the structure of
combinatorial spacetimes in as much generality as possible. Here, however, my
goal is to give the reader a hands-on feel for the nature of these systems. To
that end, I will present a simple, practical method for generating examples of
combinatorial spacetimes, and then use this method to generate two particular
examples: one in which space is one-dimensional, and one in which it is
two-dimensional.

The method proceeds in three steps. One begins by defining a set \X\ of
spaces. Next, one drums up (by whatever means) several very simple local
invertible operators on \X\ such that these operators do not all commute with
one another. Finally, one composes these simple operators with one another in
various ways to generate new operators. Though the original operators may be
trivial, the fact that they do not all commute with one another means that one
may obtain infinitely many operators in this way, and that some of these may be
quite interesting.

Consider the case where space is one-dimensional. In the continuous case there
are two basic sorts of closed one-dimensional spaces: those that are
homeomorphic to a circle, and those that are homeomorphic to a line. Here,
however, we need to consider combinatorial structures. The natural ones to
consider are graphs in which each vertex has degree two. (A graph in which
every vertex has degree $k$ is called a \emph{k-regular graph}; hence these are
2-regular graphs.) If such a graph contains a finite number of vertices, then
it must be a circuit (hence resembling a circle); otherwise it resembles a
line.

In most of what follows I will restrict my attention to the oriented case. One
may give 2-regular graphs an orientation in a natural way by requiring them to
be directed graphs where each vertex has one incoming edge and one outgoing
edge.

We may also attach properties to these graphs. This amounts to coloring them in
some way. We may color vertices, edges, or both. Here, let us color the
vertices only. (Later, we will color edges instead of vertices. I will
sometimes use the term \emph{cell} to mean a piece of graph---either vertex or
edge---that contains a single color.)

Now we are ready to define a set of allowed one-dimensional spaces. To do this,
we must pay attention to the local simplicity requirement described in
Chapter~\ref{introduction}. To make this precise, we need to be able to measure
the size of a neighborhood of a ``point'' in space. Let us consider the
vertices to be the points. The natural measure of distance between two vertices
in a graph is the minimum number of edges on the paths that join them. Let us
use this as our measure here, and define the size of the neighborhood of a
vertex to be the maximum distance from that vertex to the other vertices in the
neighborhood. The local simplicity requirement may be interpreted to mean that
there must be only finitely many allowed vertex neighborhoods of any given
finite radius. This means, for instance, that the neighborhood of a vertex with
radius zero---that is, the vertex itself---must have only a finite number of
possible states. Hence the number of allowed colors must be finite. If the
number of allowed colors is finite, then clearly, for finite $r$, any
radius-$r$ neighborhood of a vertex will also have a finite number of possible
states. Hence this restriction on the number of allowed colors is the only one
we need.

It is also possible to impose further local restrictions on which arrangements
of colors are allowed. Indeed, as will become clear later, it is important to
consider such restrictions. However, at present it is not necessary to do
this. Let the number of allowed colors be $k$. Any choice of $k$ determines a
set \x{k} of allowed spaces: namely, the set of all oriented nonempty
$k$-colored 2-regular graphs. These are the sets of spaces which will be
considered in this chapter.

I will use parentheses to denote a connected vertex-colored oriented 2-regular
graph with a finite number of vertices (i.e., an oriented circuit). For
example, \cir{abacc} denotes an oriented circuit whose vertices are colored
consecutively by the colors $a$, $b$, $a$, $c$ and $c$, where the left-to-right
direction in the notation corresponds to the direction of the edges in the
graph. Thus \cir{abacc} means the same thing as \cir{bacca}.

It is also useful to define the concept of a \emph{segment} of an oriented
vertex-colored 2-regular graph. A segment consists of colored vertices and
directed edges. Every vertex in the segment has degree two, with one edge
pointing in and one pointing out. However, a segment is not, strictly speaking,
a colored directed graph, because not every edge in the segment is connected to
a vertex at each end. More precisely, one edge in the segment does not point
towards a vertex, and one does not point away from a vertex; otherwise, every
edge in the segment connects two distinct vertices. The edge which does not
point towards a vertex and the edge which does not point away from a vertex may
be the same edge; in this case the segment consists of this edge alone, and has
no vertices. This is the \emph{empty} segment. The length $|S|$ of a segment
$S$ is equal to the number of vertices it contains; hence the number of edges
in $S$ is equal to $|S|+1$. Square brackets will usually be used to denote
segments, though occasionally they will be omitted. For example, \seg{abc}
contains three vertices, colored $a$, $b$ and $c$, and four edges: one points
towards the $a$-colored vertex; one goes from the $a$-colored vertex to the
$b$-colored vertex; one goes from the $b$-colored vertex to the $c$-colored
vertex; and one points away from the $c$-colored vertex. The symbol \seg{\,}
represents the empty segment.

Segments may be \emph{embedded} in other segments or in spaces; this notion
will be used often in what follows. (Whenever I say later on that a segment is
``in'' or ``contained in'' a segment or space, I mean that it is embedded.) To
describe embeddings rigorously, one needs coordinates. A particular description
of a space implicitly gives coordinates to its vertices. Consider, for
instance, the space \cir{abacc}. In this case, there are five vertices; we
assign each successive vertex a coordinate in \z{5}, in ascending order. The
colors of the vertices are read off from the colors in the description in the
natural way; hence here the 0th vertex is colored $a$, the 1st vertex is
colored $b$, and so on. In addition, the edge which points towards the $i$th
vertex will be called the $i$th edge. Coordinates are assigned to segments in
the same way (though there is only one way to assign coordinates to a segment,
while a space containing $w$ vertices may be assigned coordinates in $w$
ways). If $x$ is a space and we have a coordinate system for that space in
mind, then the notation $x_{i,j}$ will refer to the segment embedded in $x$
that begins at the $i$th edge in $x$ and has length $j$. For example, there are
two possible ways to embed the segment \seg{a} in \cir{abacc}, and these are
given by $\cir{abacc}_{0,1}$ and $\cir{abacc}_{2,1}$. An embedded segment may
wrap around a space any number of times; hence $\cir{abacc}_{1,13}$ refers to
an embedding of the segment \seg{baccabaccabac} in \cir{abacc}. Similarly one
may speak of the segment $S_{i,j}$ embedded in $S$; however, here there can be
no wrapping around.

Let $S_i$ be a segment for each $i\in\z{n}$. Then we may define
$S=\seg{S_0\ldots S_{n-1}}$ to be the segment formed by overlapping the
rightmost edge of $S_i$ with the leftmost edge of $S_{i+1}$ for each
$i<n-1$. Similarly, we may define $x=\cir{S_0\ldots S_{n-1}}$ to be the space
formed by overlapping the rightmost edge of $S_i$ with the leftmost edge of
$S_{i+1}$ for each $i\in\z{n}$. In this context, it is useful to think of each
element of $K$ as a segment of length one. Then our previous definitions of
\seg{c_0\ldots c_{n-1}} and of \cir{c_0\ldots c_{n-1}} agree with the above
definitions; also, we may mix segments and colors, and define things like
\seg{Sc}. Note that sometimes I shall specify a different rule for overlapping
the segments in \seg{S_0\ldots S_{n-1}} or in \cir{S_0\ldots S_{n-1}}; however,
single-edge overlap will be the default.

The next task is to find a law of evolution. In other words, we need an
invertible operator $T:\x{k}\mapsto\x{k}$ where $T$ and $T^{-1}$ are
``local.'' For now, I shall only use the idea of locality as a heuristic
concept, rather than attempting to give it a precise definition. Heuristically,
local laws are those in which the maximum velocity at which causal effects may
be propagated is finite.

Here is an example of the sort of thing we are \emph{not} looking for. Let
$k=3$. Then the allowed colors are $a$, $b$ and $c$. Let $\bar{a}=b$ and
$\bar{b}=a$. Let $x=\cir{x_0\ldots x_{n-1}}$ be an arbitrary space, where the
subscripts of $x$ are considered modulo $n$ ($n$ may be infinite). If $x_i\neq
c$, let $L(i)$ be the smallest positive number such that $x_{i+L(i)}=c$ (if no
such cell exists, $L(i)$ is infinite). Similarly, if $x_i\neq c$, let $R(i)$ be
the smallest positive number such that $x_{i-R(i)}=c$. Let
$F(i)=L(i)+R(i)+1$. Then $F(i)$ is the size of the largest segment in $x$
containing the $i$th cell of $x$ and having the property that no cell in the
segment is colored $c$. Suppose that $T$ maps \cir{x_0\ldots x_{n-1}} to
\cir{y_0\ldots y_{n-1}} where $y_i=x_i$ when $x_i=c$ or when $F(i)$ is odd, and
$y_i=\bar{x_i}$ otherwise. Then $T$ is invertible, but not local: in order to
determine $y_i$ one needs to know the colors of cells $i$ through $i+L(i)$,
where $L(i)$ may be arbitrarily large. Since the number of cells at time $t$
which need to be examined in order to determine the color of a cell at time
$t+1$ is unbounded, there is no maximum finite ``velocity'' at which causal
effects may be propagated (where velocity is measured by considering each
application of $T$ to represent a unit of time, and each edge on a graph at
time $t$ to represent a unit of distance).

It turns out that it is not difficult to construct simple examples of local,
invertible maps $T$ such that $T$ is its own inverse. One way to do this is to
choose segments $s_0$ and $s_1$ and specify that $T$ acts on $x$ by replacing
each occurrence of $s_i$ in $x$ with $s_{1-i}$ and leaving the rest of $x$
invariant. One must choose one's segments carefully and be precise about what
``replacing'' means (sometimes overlap may be involved) so that the map is
well defined and so that $T=T^{-1}$. But this may be done; and one may easily
verify that any such $T$ (and hence its inverse, which is also $T$) is local in
the heuristic sense defined earlier.

The simplest examples of this sort involve segments of length one. For
instance, consider a map which replaces \seg{a} with \seg{b} and \seg{b} with
\seg{a}. This map is clearly well defined, and it is its own inverse. It leaves
the vertices and edges of the graph fixed, but permutes its colors. For
example, this map sends \cir{abbababb} to \cir{baababaa}.

A more complicated example is the map that replaces the segment \seg{aa} by the
segment \seg{aba} and vice versa, and leaves everything else fixed. The idea is
to operate on a graph as follows: wherever you see an \seg{aa}, insert a $b$
between the two $a$'s; and wherever you see an \seg{aba}, delete the $b$. Note
that these segments can overlap: for example, \seg{aaa} contains two
consecutive overlapping segments \seg{aa}, which means that two $b$'s are
inserted. Also, note that segments may wrap around and hence contain the same
cell more than once; hence \cir{a} contains one copy of \seg{aa}, and \cir{ab}
contains one copy of \seg{aba}. Again, it is easy to see that this map is well
defined and is its own inverse. It sends \cir{aaabbabaabb} to
\cir{abababbaababb}. Clearly this map can cause the number of vertices in the
graph to change.

Another example is the map that exchanges \seg{a} with \seg{bc}, leaving
everything else fixed. Another example is the map that exchanges \seg{ab} with
\seg{ac}, leaving everything else fixed. And so on; it is easy to write down
many such examples.

These examples provide trivial evolutions (since they have period two). But any
composition of these maps is again a local, invertible map. In addition, many
of these maps do not commute with one another. (For example, let $T$ be the map
which permutes the colors $a$ and $b$, let $U$ be the map which exchanges
\seg{aa} with \seg{aba}, and let $x=\cir{a}$. Then $UT(x)=\cir{b}$, but
$TU(x)=\cir{ba}$. Hence $T$ and $U$ do not commute.) Because of this
noncommutativity, the local, invertible maps that result from composing several
of these period-two maps together often turn out to be quite interesting.

Let us examine one such interesting case. Let $k=3$, let the colors be $a$, $b$
and $c$, and consider the map which operates on a graph in four steps, as
follows:\vspace{3.5pt}
\begin{enumerate}\linespread{1}\normalsize
\item Exchange \seg{ab} with \seg{ac}.
\item Exchange \seg{ab} with \seg{c}.
\item Exchange \seg{a} with \seg{c}.
\item Exchange \seg{a} with \seg{b}.\footnote{\linespread{1}\footnotesize Note
that the last two steps together perform the permutation $(acb)$ on the colors;
I list them separately so that each step is its own inverse. This insures that
the inverse of the entire map may be obtained by performing the same steps in
the opposite order.}
\end{enumerate}
For example, this map sends \cir{acbaabcb} to \cir{abbaaccb} via step~1, to
\cir{cbaaababb} via step~2, to \cir{abcccbcbb} via step~3, and finally to
\cir{bacccacaa} via step~4.

Suppose that the initial state is \cir{ab}. Figure~\ref{orbit183}
\begin{figure}
\linespread{1}
\fbox{\begin{picture}(425,576)\footnotesize\put(0,12){\makebox(425,552)[bl]{\hfill \begin{tabular}{rl}
0 & \cir{ab}\\
1 & \cir{cca}\\
2 & \cir{cab}\\
3 & \cir{cacca}\\
4 & \cir{bcab}\\
5 & \cir{acacca}\\
6 & \cir{bbcac}\\
7 & \cir{aacab}\\
8 & \cir{cbcca}\\
9 & \cir{acacab}\\
10 & \cir{bbcca}\\
11 & \cir{caacacac}\\
12 & \cir{cacbbb}\\
13 & \cir{cabaaa}\\
14 & \cir{ccaccb}\\
15 & \cir{cacabcaa}\\
16 & \cir{bccacacb}\\
17 & \cir{acacabba}\\
18 & \cir{bbccaac}\\
19 & \cir{aacacacb}\\
20 & \cir{cbbba}\\
21 & \cir{aaab}\\
22 & \cir{cccca}\\
23 & \cir{cacacab}\\
24 & \cir{cabbcca}\\
25 & \cir{ccaacacab}\\
26 & \cir{cacacbbcca}\\
27 & \cir{bbaacacab}\\
28 & \cir{aacbbcca}\\
29 & \cir{cbaacacac}\\
30 & \cir{caacbbb}\\
31 & \cir{cacbaaa}\\
32 & \cir{baccb}\\
33 & \cir{abcaa}\\
34 & \cir{ccacacc}\\
35 & \cir{cacabbca}\\
36 & \cir{bccaacab}\\
37 & \cir{acacacbcca}\\
38 & \cir{bbbacacac}\\
39 & \cir{aaabbb}\\
40 & \cir{ccccaaa}\\
41 & \cir{cacacaccb}\\
42 & \cir{cabbbcaa}\\
43 & \cir{ccaaacacb}\\
44 & \cir{cacaccbba}\\
45 & \cir{bbcaaab}\\
\end{tabular} \hfill
\begin{tabular}{rl}
46 & \cir{aacacccca}\\
47 & \cir{cbbcacacac}\\
48 & \cir{caaacabbb}\\
49 & \cir{caccbccaaa}\\
50 & \cir{bcaacacaccb}\\
51 & \cir{acacbbbcaa}\\
52 & \cir{bbaaacacc}\\
53 & \cir{aaccbbca}\\
54 & \cir{cbcaaacac}\\
55 & \cir{caacaccbb}\\
56 & \cir{cacbbcaaa}\\
57 & \cir{baacaccb}\\
58 & \cir{acbbcaa}\\
59 & \cir{baacacc}\\
60 & \cir{acbbca}\\
61 & \cir{baacac}\\
62 & \cir{acbb}\\
63 & \cir{baa}\\
64 & \cir{cacc}\\
65 & \cir{cabca}\\
66 & \cir{ccacab}\\
67 & \cir{cacabcca}\\
68 & \cir{bccacacab}\\
69 & \cir{acacabbcca}\\
70 & \cir{bbccaacacac}\\
71 & \cir{aacacacbbb}\\
72 & \cir{cbbbaaa}\\
73 & \cir{aaaccb}\\
74 & \cir{ccbcaa}\\
75 & \cir{caacacb}\\
76 & \cir{cacbba}\\
77 & \cir{baab}\\
78 & \cir{accca}\\
79 & \cir{bcacac}\\
80 & \cir{acabb}\\
81 & \cir{bccaa}\\
82 & \cir{cacacacc}\\
83 & \cir{cabbbca}\\
84 & \cir{ccaaacab}\\
85 & \cir{cacaccbcca}\\
86 & \cir{bbcaacacab}\\
87 & \cir{aacacbbcca}\\
88 & \cir{cbbaacacac}\\
89 & \cir{caaacbbb}\\
90 & \cir{caccbaaa}\\
91 & \cir{bcaaccb}\\
\end{tabular} \hfill
\begin{tabular}{rl}
92 & \cir{acacbcaa}\\
93 & \cir{bbacacc}\\
94 & \cir{aabbca}\\
95 & \cir{cccaacac}\\
96 & \cir{cacacacbb}\\
97 & \cir{cabbbaa}\\
98 & \cir{ccaaacb}\\
99 & \cir{cacaccba}\\
100 & \cir{bbcaab}\\
101 & \cir{aacaccca}\\
102 & \cir{cbbcacac}\\
103 & \cir{caaacabb}\\
104 & \cir{caccbccaa}\\
105 & \cir{bcaacacacb}\\
106 & \cir{acacbbba}\\
107 & \cir{bbaaac}\\
108 & \cir{aaccb}\\
109 & \cir{cbcaa}\\
110 & \cir{acacb}\\
111 & \cir{bba}\\
112 & \cir{caac}\\
113 & \cir{cacb}\\
114 & \cir{caba}\\
115 & \cir{ccab}\\
116 & \cir{cacacca}\\
117 & \cir{bbcab}\\
118 & \cir{aacacca}\\
119 & \cir{cbbcac}\\
120 & \cir{caaacab}\\
121 & \cir{caccbcca}\\
122 & \cir{bcaacacab}\\
123 & \cir{acacbbcca}\\
124 & \cir{bbaacacac}\\
125 & \cir{aacbbb}\\
126 & \cir{cbaaa}\\
127 & \cir{accb}\\
128 & \cir{bcaa}\\
129 & \cir{cacacc}\\
130 & \cir{cabbca}\\
131 & \cir{ccaacab}\\
132 & \cir{cacacbcca}\\
133 & \cir{bbacacab}\\
134 & \cir{aabbcca}\\
135 & \cir{cccaacacac}\\
136 & \cir{cacacacbbb}\\
137 & \cir{cabbbaaa}\\
\end{tabular} \hfill
\begin{tabular}{rl}
138 & \cir{ccaaaccb}\\
139 & \cir{cacaccbcaa}\\
140 & \cir{bbcaacacb}\\
141 & \cir{aacacbba}\\
142 & \cir{cbbaac}\\
143 & \cir{caaacb}\\
144 & \cir{caccba}\\
145 & \cir{bcaab}\\
146 & \cir{acaccca}\\
147 & \cir{bbcacac}\\
148 & \cir{aacabb}\\
149 & \cir{cbccaa}\\
150 & \cir{acacacb}\\
151 & \cir{bbba}\\
152 & \cir{caaac}\\
153 & \cir{caccb}\\
154 & \cir{cabcaa}\\
155 & \cir{ccacacb}\\
156 & \cir{cacabba}\\
157 & \cir{bccaab}\\
158 & \cir{acacaccca}\\
159 & \cir{bbbcacac}\\
160 & \cir{aaacabb}\\
161 & \cir{ccbccaa}\\
162 & \cir{caacacacb}\\
163 & \cir{cacbbba}\\
164 & \cir{baaab}\\
165 & \cir{acccca}\\
166 & \cir{bcacacac}\\
167 & \cir{acabbb}\\
168 & \cir{bccaaa}\\
169 & \cir{cacacaccc}\\
170 & \cir{cabbbcaca}\\
171 & \cir{ccaaacabb}\\
172 & \cir{cacaccbccaa}\\
173 & \cir{bbcaacacacb}\\
174 & \cir{aacacbbba}\\
175 & \cir{cbbaaac}\\
176 & \cir{caaaccb}\\
177 & \cir{caccbcaa}\\
178 & \cir{bcaacacb}\\
179 & \cir{acacbba}\\
180 & \cir{bbaac}\\
181 & \cir{aacb}\\
182 & \cir{cba}\\
183 & \cir{ab}\\
\end{tabular} \hfill}}\end{picture}}\linespread{1}
\caption[An evolution of a one-dimensional oriented space]{An evolution of a one-dimensional oriented space under a local,
invertible operator (period 183).}
\label{orbit183}
\end{figure}
displays the orbit of this state under the above map. (The intermediate results
of steps 1, 2 and 3 are omitted.) This evolution has period 183 and attains a
maximum width of 11. By studying the orbits of this map empirically, one
quickly arrives at the conjecture that every orbit has finite period. These
periods tend to be quite large. For instance, the orbit of \cir{abcb} has
period 898 and attains a maximum width of 18; the orbit of \cir{aabab} has
period 6,360 and attains a maximum width of 22; and the orbit of \cir{ababcb}
has period 343,904 and attains a maximum width of 29. A proof that every orbit
of this map has finite period will be given in Chapter~\ref{invar}.

Much more will be said about the case of one-dimensional spaces later on, but
now I will turn my attention to other sorts of spaces. Since our spaces are
combinatorial, we may consider spaces whose dimension is not immediately
obvious (if it is definable at all). For example, if our spaces are graphs with
colored vertices or edges, we might define a set of spaces to be all graphs
such that the radius-$r$ neighborhood of any vertex of the graph lies in some
finite set of allowed radius-$r$ neighborhoods. Perhaps there is a known way to
assign some sort of fractal dimension to spaces of this sort; I have not
pursued the question. Here I will focus instead on sets of combinatorial spaces
which are related in clearly defined ways to $n$-dimensional manifolds or
pseudomanifolds.

The usual way to construct combinatorial $n$-dimensional manifolds (or
pseudomanifolds) is to use simplicial complexes. However, one may also
represent such objects by graphs that have certain simple properties. It seems
to me that this is easier than using complexes, so this is the procedure that I
will follow here.

Let \G\ be a graph containing three types of edges (called 0-edges, 1-edges and
2-edges) such that each vertex meets exactly one edge of each type. (This
implies that each vertex has degree three, and that each edge meets two
distinct vertices.)  One may consider any such graph as a two-dimensional
manifold in the following way. If one begins at a vertex and travels along
edges of type 0 and 1 in an alternating manner (first type 0, then type 1, then
type 0, and so on), then there is only one way to proceed at any stage, and
eventually (if the graph is finite) one must return to the original vertex. The
path traced out in this manner is a circuit (which necessarily contains an even
number of edges). We will think of the edges of this circuit as the sides of a
polygonal face of a two-dimensional manifold. Any polygonal face of this sort
will be called an 01-face. Similarly, there are 02-faces and 12-faces. Every
edge in the graph is the side of exactly two polygons. For example, a type-0
edge is the side of an 01-face and an 02-face. We may construct a
two-dimensional space by gluing together all such polygonal faces along their
common sides.

It turns out that the manifold associated with \G\ is orientable if and only if
the vertices of \G\ may be divided into two groups such that each edge joins a
vertex in one group to a vertex in the other. Such a graph is said to be
\emph{bipartite}. Suppose that we are given a division of the vertices of \G\
into two such groups. Color the vertices in the first group black and those in
the second group white. Then we may assign an orientation to the manifold
associated with \G\ as follows. The orientation of each face will be specified
in terms of arrows whose tails are at black vertices on the boundary of the
face. For an 01-face, the arrows point in the direction of the 0-edges emerging
from those vertices. For an 02-face, they point in the direction of the 2-edges
emerging from the vertices. For a 12-face, they point in the direction of the
1-edges emerging from the vertices. One may verify that, due to the properties
of the coloring, the orientations of these faces must be
compatible. Conversely, given any orientation of the manifold there is exactly
one way to color the vertices of \G\ black or white so that no edge joins two
vertices having the same color and so that the arrows just mentioned point in
the appropriate directions.

\begin{figure}
\fbox{\begin{picture}(425,229)\footnotesize
\put(50,137.5){\begin{picture}(30,70)\put(15,5){\circle{10}}
\put(15,65){\circle{10}}
\qbezier(15,10)(15,35)(15,60)\put(13,35){\line(1,0){4}}
\qbezier(15,10)(0,25)(0,35)\qbezier(0,35)(0,45)(15,60)
\qbezier(15,10)(30,20)(30,35)\qbezier(30,35)(30,45)(15,60)
\put(28,34){\line(1,0){4}}\put(28,36){\line(1,0){4}}
\end{picture}}
\put(142.5,137.5){\begin{picture}(80,70)
\multiput(0,0)(50,0){2}{\put(15,5){\circle{10}}
        \put(15,65){\circle{10}}
        \qbezier(15,10)(5,25)(5,35)\qbezier(5,35)(5,45)(15,60)
        \qbezier(15,10)(25,20)(25,35)\qbezier(25,35)(25,45)(15,60)}
\multiput(0,0)(0,60){2}{\qbezier(20,5)(40,5)(60,5)\put(39,3){\line(0,1){4}}\put(41,3){\line(0,1){4}}}
\put(23,35){\line(1,0){4}}\put(53,35){\line(1,0){4}}
\end{picture}}
\put(285,133){\begin{picture}(90,79)\put(5,5){\circle{10}}
\put(85,5){\circle{10}}\put(45,28){\circle{10}}
\put(45,74){\circle{10}}\qbezier(10,5)(45,5)(80,5)
\qbezier(45,33)(45,51)(45,69)\qbezier(7.5,9.33)(25,39.5)(42.5,69.67)
\qbezier(82.5,9.33)(65,39.5)(47.5,69.67)\qbezier(9.33,7.5)(25,16.5)(40.67,25.5)
\qbezier(80.67,7.5)(65,16.5)(49.33,25.5)\put(44,3){\line(0,1){4}}
\put(46,3){\line(0,1){4}}\put(43,50){\line(1,0){4}}\put(43,52){\line(1,0){4}}
\qbezier(24,18.23)(25,16.5)(26,14.77)\qbezier(63.27,38.5)(65,39.5)(66.73,40.5)
\end{picture}}
\put(70,17){\begin{picture}(110,90)\multiput(45,5)(-40,20){2}{%
        \multiput(0,0)(0,60){2}{%
                \multiput(0,0)(60,0){2}{\circle{10}}
                \qbezier(5,0)(30,0)(55,0)\put(30,-2){\line(0,1){4}}}
        \multiput(0,0)(60,0){2}{%
                \qbezier(0,5)(0,30)(0,55)\multiput(-2,29)(0,2){2}{\line(1,0){4}}}}
\multiput(0,0)(60,0){2}{\multiput(0,0)(0,60){2}{%
        \qbezier(40.52,7.24)(25,15)(9.48,22.76)}}
\end{picture}}
\put(245,17){\begin{picture}(110,90)\multiput(45,5)(-40,20){2}{%
        \multiput(0,0)(0,60){2}{%
                \multiput(0,0)(60,0){2}{\circle{10}}
                \qbezier(5,0)(30,0)(55,0)}
        \multiput(0,0)(60,0){2}{%
                \qbezier(0,5)(0,30)(0,55)}}
\multiput(0,0)(60,0){2}{\multiput(0,0)(0,60){2}{%
        \qbezier(40.52,7.24)(25,15)(9.48,22.76)}}
\multiput(45,5)(0,60){2}{\multiput(29,-2)(2,0){2}{\line(0,1){4}}}
\multiput(45,5)(60,0){2}{\put(-2,30){\line(1,0){4}}}
\multiput(5,25)(0,60){2}{\put(30,-2){\line(0,1){4}}}
\multiput(5,25)(60,0){2}{\multiput(-2,29)(0,2){2}{\line(1,0){4}}}
\end{picture}}
\end{picture}}\linespread{1}\caption[Representation of two-dimensional manifolds
by graphs]{Representation of two-dimensional manifolds by graphs. Here the
three types of edges are indicated by zero, one or two hatch marks on the
edge. The three possible graphs with four or fewer vertices are displayed in
the upper row. In the leftmost of these, there are three faces (an 01-face, an
02-face and a 12-face), each with two sides; the 01-face is glued to the
12-face along the 1-edge, the 12-face is glued to the 02-face along the 2-edge,
and then the 02-face is glued to the 01-face along the 0-edge; the result is a
sphere. In the upper central graph, there are four faces; the 12- and 02-faces
each have 4 sides, and are glued together to form a cylinder; the two 01-faces
cap the ends of the cylinder to form a sphere. One may also verify that these
are spheres by using the Euler characteristic $\Chi=F-V/2$; in the first case
$F\!=3$ and $V\!=2$, and in the second case $F\!=4$ and $V\!=4$, so in both
cases $\Chi=2$. In the upper right-hand graph, $F\!=3$ and $V\!=4$; hence
$\Chi=1$, so this is a projective plane. The lower left-hand graph is a sphere:
the vertices have been arranged in the picture to suggest a cube, and this is
exactly what the graph represents since the faces of the suggested cube are
exactly the 01-, 02- and 12-faces of the graph. On the other hand, the lower
right-hand graph has its edge types switched on the ``front'' edges. Now
$F\!=4$ and $V\!=8$, so $\Chi=0$. Since one may color the vertices of this
graph with two colors in such a way that no edge connects two vertices of the
same color, this is orientable; hence it is a torus.}
\label{twographs}
\end{figure}
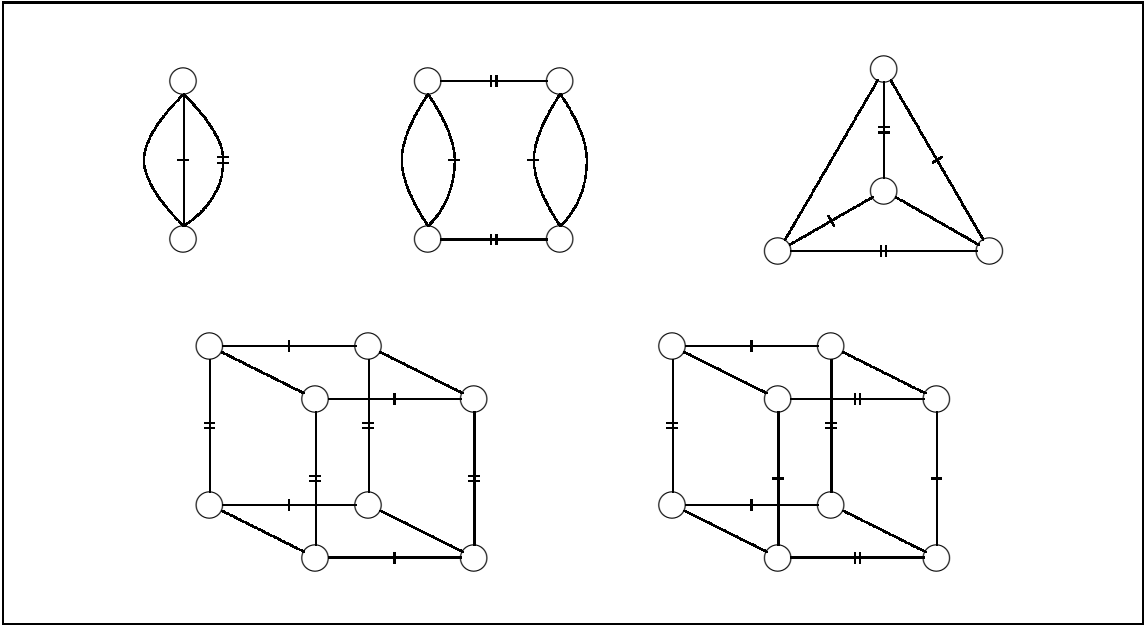

Several examples of graphs which represent two-dimensional manifolds are shown
in Figure~\ref{twographs}.
In the manifold associated with such a graph, let $V$ be the number of
vertices, $E$ be the number of edges and $F$ be the number of faces. Then the
Euler characteristic $\Chi$ equals $V-E+F$. Since each vertex in the graph has
degree 3, it follows that $3V=2E$. (Note that this means that the number of
vertices must be even, and that the number of edges must be divisible by
three.) So $\Chi=F-V/2$. By using this formula and (if necessary) determining
whether or not the graph is bipartite, one may easily determine the topology of
these manifolds.

What is especially nice about this graphical representation of two-manifolds is
that it may be generalized in a surprisingly simple way. Let \G\ be a graph
containing $n+1$ types of edges (labeled 0 through $n$) such that each vertex
meets exactly one edge of each type. Graphs of this sort have different names
in the literature; I will call them \emph{$n$-graphs}. It turns out that, for
$n>0$, every $n$-graph can be associated with an $n$-pseudomanifold.

If $T$ is a topological space, then $C(T)$, the \emph{cone} over $T$, is
defined to be the identification space $T\times [0,1]/\sim$ where $\sim$ is the
relation identifying all points in the set $\{(t,1)\,|\,t\in T\}$. The
construction of an $n$-pseudomanifold from an $n$-graph \G\ proceeds
inductively using cones.

The \emph{k-faces} of an $n$-pseudomanifold represented by an $n$-graph
correspond to the components of each subgraph that is obtained by deleting all
edges whose types are not in $K$, for any $K\subset\{0,\ldots,n\}$ containing
$k$ integers. If $K$ is empty then these components are the vertices of the
$n$-graph; the faces corresponding to these vertices are distinct
points. If $|K|=1$ then the components are pairs of vertices joined by a single
edge; the face corresponding to such a component is a line segment joining the
two $0$-faces corresponding to the vertices. If $|K|>1$ then each component
corresponds to a $(k-1)$-pseudomanifold. In this case the associated $k$-face
is the cone over this pseudomanifold.

To construct an $n$-pseudomanifold from its associated $n$-graph, one
begins by constructing each of its $n$-faces. Each $(n-1)$-face of the
pseudomanifold is contained in exactly two $n$-faces. (This corresponds to the
fact that any subset of $\{0,\ldots,n\}$ containing $n-1$ elements is in
exactly two subsets containing $n$ elements.) If two $n$-faces contain the
same $(n-1)$-face, these points are identified in the obvious way; the result
after all of these identifications are made is the desired $n$-pseudomanifold.

One standard definition of pseudomanifold is given in terms of simplicial
complexes (see~\cite{seifert/threlfall}). Let $P$ be the set of simplicial
complexes which satisfy the rules for $n$-pseudomanifolds, and let $G$ be the
set of manifolds constructed from finite connected $n$-graphs as described
above. It turns out (I will omit proofs) that there exist local 1-1
topology-preserving maps $f:G\mapsto P$ and $g:P\mapsto G$. Hence not only are
the elements of $G$ pseudomanifolds from a topological point of view, but every
topological pseudomanifold is represented in $G$. In our case, we are not
concerned just with topology, but also with geometry: the particular
combinatorial structures involved in constructing these spaces are
important. The fact that $f$ and $g$ are local and 1-1 indicates that $G$ and
$P$ are in some sense equally rich in their descriptions of pseudomanifolds in
terms of combinatorial structures.

Note also that the space constructed from an $n$-graph turns out to be
orientable if and only if the graph is bipartite. Again, I will omit the proof.

Apparently this connection between $n$-graphs and $n$-pseudomanifolds was
first discovered in the 1970s by Pezzana~\cite{pezzana}, then rediscovered by
Vince~\cite{vince}, then partially rediscovered by Lins and
Mandel~\cite{lins/mandel}. (The latter reference contains further pointers to
the literature on this subject.) Later I too rediscovered this connection in a
slightly altered form. (I was concentrating on oriented pseudomanifolds and did
not realize that, with a slight modification, my result applies to unoriented
pseudomanifolds also; and in my version there is an additional complication
which, while unnecessary, does result in the nice property that the
Poincar\'{e} dual of a pseudomanifold can be obtained simply by performing a
certain permutation on the colors of the edges of the associated graph. See
Section~\ref{hinge} for more details.) Perhaps the connection would not have
been rediscovered so often if it were better known. In my opinion it deserves
to be better known, since the definition of an $n$-graph is much simpler than
the definition of a simplicial complex.

My suggestion, then, is that $n$-graphs provide a suitable basis for a
definition of a set of $n$-dimensional spaces. We may add properties to these
graphs by coloring them in some way. Since the edges already are colored (by
their type numbers), the easiest thing is to color the vertices. Again, the
local simplicity requirement implies that the number of colors must be
finite. We may also restrict which configurations of vertices and edges are
allowed locally by choosing a radius $r$, making a list of vertex
neighborhoods of radius $r$, and defining the set of spaces to be all colored
$n$-graphs such that each neighborhood of a vertex is in this list.

A set of spaces of the sort just described automatically satisfies the
requirement that, for finite $k$, the set of allowed $k$-neighborhoods of a
vertex is finite. However, it is not clear that they satisfy the requirement
that, given a finite radius $r$, the set of allowed radius-$r$ neighborhoods of
a \emph{point in space} is finite. What should we think of as being the
``points'' in space? There are several possibilities to consider here.

First, we might decide to think of the points in space as being the vertices of
the graph. If we do this, then every set of spaces of the sort just described
satisfies the local simplicity requirement.

On the other hand, it is natural to think of the $n$-dimensional faces of a
pseudomanifold as \emph{containing} points. For example, in the two-dimensional
case we might wish to think that the points in space do not merely lie on the
boundary of a polygonal face, but inside that face. If we think of each such
face as being homeomorphic to a disk, then there are infinitely many points in
the interior of a face. But this presents a problem. Is there any way to adopt
this point of view and also hold to the view that the set of radius-$r$
neighborhoods of a point ought to be finite? We have not adopted a metric for
these points, so it is not clear what a radius-$r$ neighborhood would be. But
if, for instance, the metric has the property that the distance of a point in
the interior of a face from the boundary of that face takes on infinitely many
possible values, then this means that infinitely many points have distinct
$r$-neighborhoods if $r$ is large enough; hence local simplicity is violated.

A third possible point of view, in some sense intermediate between the first
two, is that, if there are to be finitely many ``points'' in a finite piece of
$n$-dimensional space, then those points should be thought of not as
zero-dimensional objects, but as little $n$-dimensional objects. In this case,
the natural choice for a point is an $n$-face. For example, in the
two-dimensional case we would think of the polygonal faces as themselves being
the points. Of course, it may well be that ``point'' is not the correct word to
use in this case. In any event, if our local simplicity requirement says that
the set of allowed radius-$r$ neighborhoods of \emph{n-faces} must be finite,
this places a restriction on which sets of spaces are allowed, for it means
that the number of allowed types of $n$-faces must itself be
finite. Restricting the allowed neighborhoods of a vertex does not necessarily
accomplish this. For instance, the set of all 2-graphs in which the vertices
are colored $a$ or $b$ may have polygonal faces containing as many vertices as
you like. If we wish to have finitely many $n$-faces, then there must be a
finite number of allowed vertex colors \emph{and} a maximum finite number of
vertices contained in any $n$-face.

The above discussion about ``points'' and ``radius-$r$ neighborhoods'' is, of
course, somewhat vague. This comes with the territory when one is trying to
apply heuristic principles which have not been given precise mathematical
definitions. Practically speaking, the question in this case is: do we concern
ourselves with sets of spaces in which there are arbitrarily many types of
$n$-faces, or not? Probably the best approach would be to be as general as
possible, to allow arbitrarily many types of $n$-faces; then a set of spaces
with finitely many types of $n$-faces can be studied as a special
case. However, as I am somewhat partial to the third point of view described
above, my tendency thus far has been to focus on those cases where there are
only finitely many types of $n$-faces.

Thus, my promised two-dimensional example is defined on a set of spaces of this
sort. The set of spaces is the set of 2-graphs such that there are three types
of vertices (called $a$-vertices, $b$-vertices and $c$-vertices), and such that
no face has more than 8 sides.

Let $F_s$ be the number of faces with $s$ sides in one of our spaces. Then
$\Chi=F-V/2=F_2+F_4+F_6+F_8-V/2$. But $V=(2F_2+4F_4+6F_6+8F_8)/3$. Hence
$\Chi=\frac{2}{3}F_2+\frac{1}{3}F_4+0F_6-\frac{1}{3}F_8$. We may think of each
face as carrying ``gaussian curvature,'' where the amount of curvature of $F_s$
is equal to its coefficient in this last equation, and where the total
curvature of a manifold is taken to be equal to its Euler characteristic. The
fact that faces with six sides carry zero curvature corresponds to the fact
that one may tile the plane with hexagons.  Note that curvature in this setting
only comes in multiples of one-third. Since I am allowing eight-sided polygons
in my spaces, this means that I am allowing negative curvature.

The next step is to come up with some local invertible operators on this set of
spaces. Again, I will look for maps $T$ such that $T=T^{-1}$. As was the case
in one dimension, some simple maps of this sort are obtained by exchanging a
pair of colors. Here we may exchange colors on the vertices (e.g., one such map
turns $a$-vertices into $b$-vertices, turns $b$-vertices into $a$-vertices, and
leaves the rest of the graph alone), or on the edges (one such map turns
0-edges into 1-edges, turns 1-edges into 0-edges, and leaves the rest of the
graph alone).

Other period-2 maps may be designed which do not affect the underlying graph,
but only affect its colors. A useful example is the following (for fixed $i$,
$w$, $x$, $y$, $z$): if an $i$-edge connects a $w$-vertex to an $x$-vertex,
change $w$ to $y$ and $x$ to $z$; if it connects a $y$-vertex to a $z$-vertex,
change $y$ to $w$ and $z$ to $x$. This map is only well defined if either
$w\neq x$ and $y\neq z$ or $w=x$ and $y=z$.

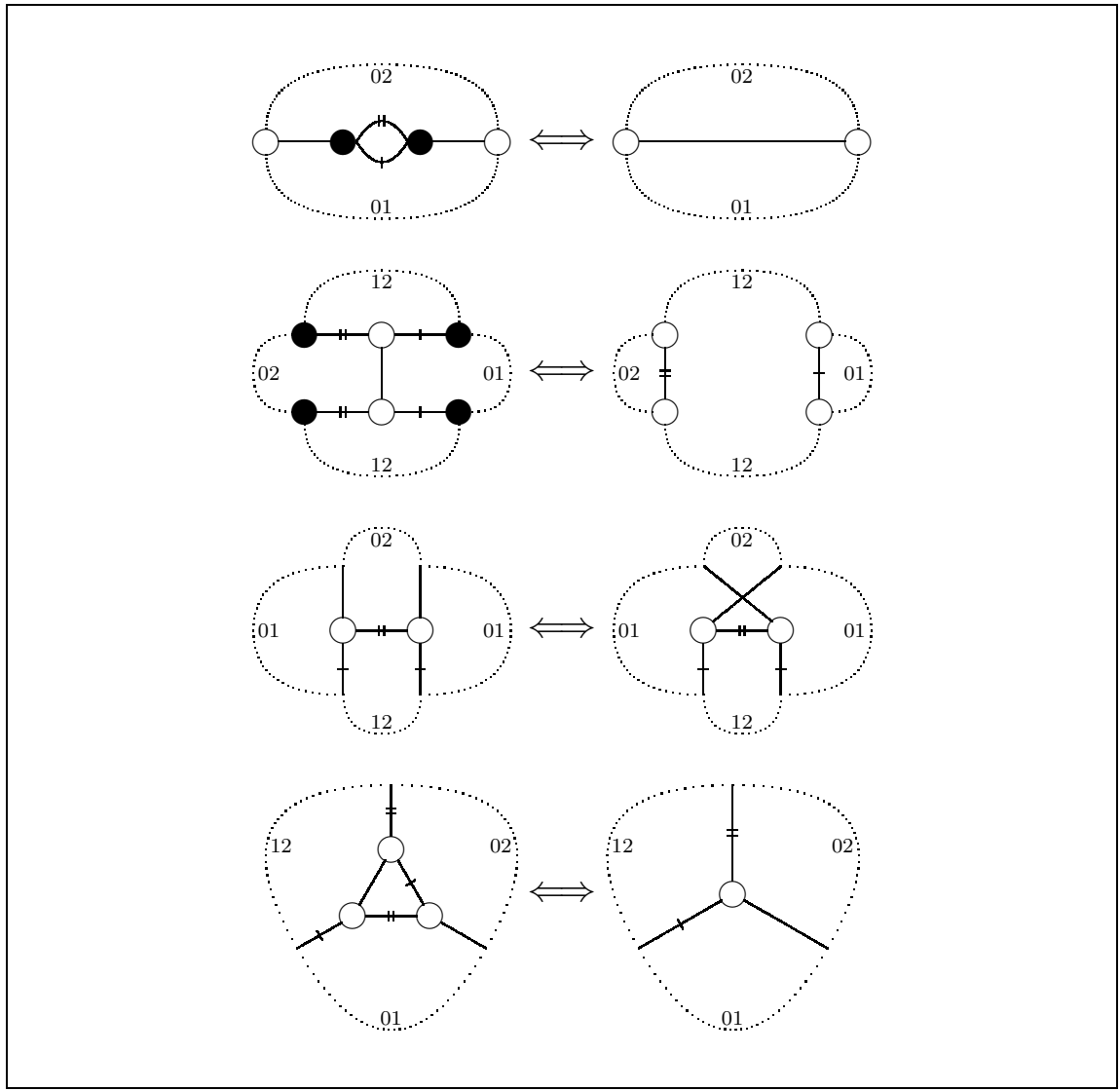
\begin{figure}
\fbox{\begin{picture}(425,414.64)\scriptsize

\put(92.5,334.64){\begin{picture}(240,60)\put(5,30){\circle{10}}
\put(35,30){\circle*{10}}\put(65,30){\circle*{10}}
\put(95,30){\circle{10}}
\put(10,30){\line(1,0){20}}\put(70,30){\line(1,0){20}}
\qbezier(40,30)(45,22)(50,22)\qbezier(50,22)(55,22)(60,30)
\qbezier(40,30)(45,38)(50,38)\qbezier(50,38)(55,38)(60,30)
\put(50,20){\line(0,1){4}}\multiput(49,36)(2,0){2}{\line(0,1){4}}
\qbezier[30](5,25)(5,0)(50,0)\qbezier[30](50,0)(95,0)(95,25)
\qbezier[30](5,35)(5,60)(50,60)\qbezier[30](50,60)(95,60)(95,35)
\put(50,2){\makebox(0,0)[b]{01}}\put(50,58){\makebox(0,0)[t]{02}}
\put(120,30){\makebox(0,0){\large $\Longleftrightarrow$}}
\put(140,0){\put(5,30){\circle{10}}
\put(95,30){\circle{10}}
\put(10,30){\line(1,0){80}}
\qbezier[30](5,25)(5,0)(50,0)\qbezier[30](50,0)(95,0)(95,25)
\qbezier[30](5,35)(5,60)(50,60)\qbezier[30](50,60)(95,60)(95,35)
\put(50,2){\makebox(0,0)[b]{01}}\put(50,58){\makebox(0,0)[t]{02}}}
\end{picture}}

\put(92.5,234.64){\begin{picture}(240,80)
\put(0,10){\multiput(0,15)(0,30){2}{\put(50,0){\circle{10}}
        \multiput(20,0)(60,0){2}{\circle*{10}}
        \multiput(25,0)(30,0){2}{\line(1,0){20}}
        \multiput(34,-2)(2,0){2}{\line(0,1){4}}
        \put(65,-2){\line(0,1){4}}}
\put(50,20){\line(0,1){20}}
\qbezier[20](20,10)(20,-10)(50,-10)\qbezier[20](50,-10)(80,-10)(80,10)
\qbezier[20](20,50)(20,70)(50,70)\qbezier[20](50,70)(80,70)(80,50)
\qbezier[11](85,15)(100,15)(100,30)\qbezier[11](100,30)(100,45)(85,45)
\qbezier[11](15,15)(0,15)(0,30)\qbezier[11](0,30)(0,45)(15,45)
\put(50,-8){\makebox(0,0)[b]{12}}\put(50,68){\makebox(0,0)[t]{12}}
\put(98,30){\makebox(0,0)[r]{01}}\put(2,30){\makebox(0,0)[l]{02}}}
\put(120,40){\makebox(0,0){\large $\Longleftrightarrow$}}
\put(140,10){\multiput(0,15)(0,30){2}{%
        \multiput(20,0)(60,0){2}{\circle{10}}}
\multiput(20,20)(60,0){2}{\line(0,1){20}}\put(78,30){\line(1,0){4}}
\multiput(18,29)(0,2){2}{\line(1,0){4}}
\qbezier[20](20,10)(20,-10)(50,-10)\qbezier[20](50,-10)(80,-10)(80,10)
\qbezier[20](20,50)(20,70)(50,70)\qbezier[20](50,70)(80,70)(80,50)
\qbezier[11](85,15)(100,15)(100,30)\qbezier[11](100,30)(100,45)(85,45)
\qbezier[11](15,15)(0,15)(0,30)\qbezier[11](0,30)(0,45)(15,45)
\put(50,-8){\makebox(0,0)[b]{12}}\put(50,68){\makebox(0,0)[t]{12}}
\put(98,30){\makebox(0,0)[r]{01}}\put(2,30){\makebox(0,0)[l]{02}}}
\end{picture}}

\put(92.5,134.64){\begin{picture}(240,80)
\put(0,10){\put(35,30){\circle{10}}\put(65,30){\circle{10}}
\put(40,30){\line(1,0){20}}\multiput(49,28)(2,0){2}{\line(0,1){4}}
\multiput(35,25)(30,0){2}{\put(0,0){\line(0,-1){20}}\put(-2,-10){\line(1,0){4}}
        \put(0,10){\line(0,1){20}}}
\qbezier[11](35,5)(35,-10)(50,-10)\qbezier[11](50,-10)(65,-10)(65,5)
\qbezier[11](35,55)(35,70)(50,70)\qbezier[11](50,70)(65,70)(65,55)
\qbezier[21](35,5)(0,5)(0,30)\qbezier[21](0,30)(0,55)(35,55)
\qbezier[21](65,5)(100,5)(100,30)\qbezier[21](100,30)(100,55)(65,55)
\put(50,-8){\makebox(0,0)[b]{12}}\put(50,68){\makebox(0,0)[t]{02}}
\put(2,30){\makebox(0,0)[l]{01}}\put(98,30){\makebox(0,0)[r]{01}}}
\put(120,40){\makebox(0,0){\large $\Longleftrightarrow$}}
\put(140,10){\put(35,30){\circle{10}}\put(65,30){\circle{10}}
\put(40,30){\line(1,0){20}}\multiput(49,28)(2,0){2}{\line(0,1){4}}
\multiput(35,25)(30,0){2}{\put(0,0){\line(0,-1){20}}\put(-2,-10){\line(1,0){4}}}
\qbezier(38.54,33.54)(65,55)(65,55)\qbezier(61.46,33.54)(35,55)(35,55)
\qbezier[11](35,5)(35,-10)(50,-10)\qbezier[11](50,-10)(65,-10)(65,5)
\qbezier[11](35,55)(35,70)(50,70)\qbezier[11](50,70)(65,70)(65,55)
\qbezier[21](35,5)(0,5)(0,30)\qbezier[21](0,30)(0,55)(35,55)
\qbezier[21](65,5)(100,5)(100,30)\qbezier[21](100,30)(100,55)(65,55)
\put(50,-8){\makebox(0,0)[b]{12}}\put(50,68){\makebox(0,0)[t]{02}}
\put(2,30){\makebox(0,0)[l]{01}}\put(98,30){\makebox(0,0)[r]{01}}}
\end{picture}}

\put(107.86,30){\begin{picture}(209.28,84.64)
\put(38.32,33.66){\put(-15,0){\circle{10}}\put(15,0){\circle{10}}
\put(0,25.98){\circle{10}}\put(-10,0){\line(1,0){20}}
\multiput(-1,-2)(2,0){2}{\line(0,1){4}}
\qbezier(-12.5,4.33)(-12.5,4.33)(-2.5,21.65)\qbezier(12.5,4.33)(12.5,4.33)(2.5,21.65)
\qbezier(-19.33,-2.5)(-19.33,-2.5)(-36.65,-12.5)
\qbezier(5.77,11.99)(7.5,12.99)(9.23,13.99)
\qbezier(-28.99,-5.77)(-27.99,-7.5)(-26.99,-9.23)
\qbezier(19.33,-2.5)(19.33,-2.5)(36.65,-12.5)
\put(0,30.98){\line(0,1){20}}\multiput(-2,39.98)(0,2){2}{\line(1,0){4}}
\qbezier[35](-36.65,-12.5)(0,-75.98)(36.65,-12.5)
\qbezier[35](-36.65,-12.5)(-73.3,50.98)(0,50.98)
\qbezier[35](36.65,-12.5)(73.3,50.98)(0,50.98)
\put(0,-42){\makebox(0,0)[b]{01}}\put(47,30){\makebox(0,0)[tr]{02}}
\put(-47,30){\makebox(0,0)[tl]{12}}}
\put(104.64,42.32){\makebox(0,0){\large $\Longleftrightarrow$}}
\put(170.96,33.66){\put(0,8.66){\circle{10}}
\qbezier(-4.33,6.16)(-4.33,6.16)(-36.65,-12.5)
\qbezier(-21.49,-1.44)(-20.49,-3.17)(-19.49,-4.9)
\qbezier(4.33,6.16)(4.33,6.16)(36.65,-12.5)
\put(0,13.66){\line(0,1){37.32}}\multiput(-2,31.32)(0,2){2}{\line(1,0){4}}
\qbezier[35](-36.65,-12.5)(0,-75.98)(36.65,-12.5)
\qbezier[35](-36.65,-12.5)(-73.3,50.98)(0,50.98)
\qbezier[35](36.65,-12.5)(73.3,50.98)(0,50.98)
\put(0,-42){\makebox(0,0)[b]{01}}\put(47,30){\makebox(0,0)[tr]{02}}
\put(-47,30){\makebox(0,0)[tl]{12}}}
\end{picture}}

\end{picture}}\linespread{1}\caption[Local invertible period-2 transformations
in two dimensions]{Local invertible period-2 transformations in two
dimensions. Here a vertex with color $a$ is represented by an open circle, and
a vertex with color $b$ is represented by a filled-in circle. A sequence of an
indeterminate number of edges which alternate between type 0 and type 1 is
represented by a dotted line with an ``01'' next to it, and so on. In each
case, the double arrow is meant to indicate that the left-to-right move and the
right-to-left move are performed simultaneously, at all locations in the
manifold containing the left or right configurations. Exact details of these
moves are given in the text.}
\label{twodtransforms}
\end{figure}

We also need period-2 maps which affect the underlying graph. Four such maps
are sketched in Figure~\ref{twodtransforms}.
The rules for these maps are somewhat complex, because they must preserve the
property that no polygonal face has more than eight sides. These rules are not
shown in the figure, but will be described below.

In the first map a 0-edge joining two $a$-vertices is transformed by inserting
a two-sided 12-face containing two $b$-vertices in the middle of the edge, and
vice versa: a two-sided 12-face containing two $b$-vertices each of which is
connected along a 0-edge to an $a$-vertex is transformed by removing the
12-face and joining the two $a$-vertices by a 0-edge. These transformations
only take place if certain conditions are satisfied. The original 0-edge is in
a 01-face and a 02-face. One must make sure that these faces have six or fewer
sides before performing the right-to-left move, since this move increases their
number of sides by two. In addition, in order to perform the move going in
either direction it is required that the dotted-line portion of the 01-face
does not contain either of the sequences $a0a$ or $a0b1b0a$ (where ``$a0a$''
refers to a connected portion of the graph containing an $a$-vertex, then a
0-edge, then an $a$-vertex). Similarly, the dotted-line portion of the 02-face
is required not to contain either of the sequences $a0a$ or $a0b2b0a$. This
insures that the 01-face and the 02-face are stable except for possibly this
single operation.

In the second map a 0-edge joins two $a$-vertices which are not in the same
12-face and which are joined by 1- and 2-edges to (not necessarily distinct)
$b$-vertices. The $a$-vertices are removed, the $b$-vertices that were joined
to the $a$-vertices by $i$-edges are joined to one another by an $i$-edge
($i\in\{1,2\}$), and each $b$-vertex is changed into an $a$-vertex. And vice
versa: if a 12-face contains exactly one occurrence of each of the sequences
$a1a$ and $a2a$, the color of each of these $a$-vertices is changed to $b$, the
1- and 2-edges joining them are removed, and two new $a$-vertices are added as
shown in the figure. Neither of these operations takes place unless two
additional conditions are satisfied. Firstly, the dotted-line sections of
12-faces (including the vertices which terminate these sections) cannot contain
the sequences $a1a$, $a2a$ or $b1a2b$ either before or after the proposed
transformation (one must check both cases since the terminal vertices change
color during the transformation). Secondly, if a 1-edge in the dotted-line
section of a 01-face or a 2-edge in the dotted-line section of a 02-face is
not also in one of the 12-faces shown in the figure, then the 12-face that it
is in must not contain exactly one occurrence of each of the sequences $a1a$
and $a2a$, nor may it contain the sequence $b1a2b$. These conditions are
somewhat complicated, but something like them is needed to insure that nothing
is going on in the neighborhood of the location of the proposed move that would
make the map ill-defined or not invertible.

The first two maps are somewhat related, respectively, to the ``two-dipole''
and ``one-dipole'' moves described by Ferri and
Gagliardi~\cite{ferri/gagliardi}. These are moves on an $n$-graph which
preserve the topology of the associated pseudomanifold. Ferri and Gagliardi
define a \emph{k-dipole} to be a pair of vertices connected to one another by
$k$ edges such that, if one removes all edges from the graph which have the
same color as those $k$ edges, the two vertices are not in the same component
of the resulting subgraph. A $k$-dipole move is the move which removes the pair
of vertices and joins the remaining edges to one another, or its
inverse. (Note that the requirement about vertices being in a separate
component is checked in our second map, but not in our first map.) Ferri and
Gagliardi show that, if one restricts attention to graphs which are associated
with manifolds (rather than pseudomanifolds), then any two graphs whose
associated manifolds have the same topology can be transformed into one another
using $k$-dipole moves. (Note that these moves are intended to act at one spot
on an $n$-graph at a time, while the maps I have been describing act all over
the graph at once, with the moves adding $k$-dipoles and the moves removing
$k$-dipoles happening in parallel.)

One may easily verify that if a graph is bipartite before applying either of
the first two maps, then it is bipartite afterwards; furthermore, these maps
preserve the Euler characteristic. Hence each of the first two maps does not
change the topology. But it is also possible to come up with maps which operate
locally on an $n$-graph and which \emph{do} change the topology. The last two
moves in Figure~\ref{twodtransforms} are of this type.

In the first of these, if a 2-edge joins two $a$-vertices $v_0$ and $v_1$ and
two 0-edges join $v_i$ to $w_i$, $i\in\{0,1\}$, then these 0-edges are removed
and new ones are inserted that join $v_i$ to $w_{1-i}$, $i\in\{0,1\}$. If the
two 0-edges are in the same face before this move, then they will be in
different faces after the move, and vice versa. The move is not performed
unless the face containing both 0-edges (either before the move or after it)
has eight or fewer sides. It also is not performed unless the 02-face and
12-face containing the 2-edge do not contain any other 2-edges joining two
$a$-vertices. The number of vertices is preserved by this operation, but the
number of faces (locally) is increased or decreased by one. Therefore, locally
this map can change $\Chi$ by one, so globally it can change the topology.

In the second of these maps, which is the last one in the figure, an $a$-vertex
is replaced by three $a$-vertices as shown (and vice versa). The move is only
allowed when there are no other $a$-vertices in the 01-face, 02-face or 12-face
given by dotted lines, and when these faces have eight or fewer sides both
before and after the move is performed. Here the number of faces is preserved,
but the number of vertices (locally) is increased or decreased by
two. Therefore again this map locally can change $\Chi$ by one, so globally it
can change the topology.

Should maps which change the topology be studied? Should we be interested in
laws of evolution in which handles and crosscaps appear and disappear? I do not
know. As has just been shown, maps can easily be constructed that act locally
on the graph, are invertible, and change the topology. (In fact, I have found
examples of maps of this sort which exhibit very complex behavior.) On the
other hand, perhaps the fact that we have a defined topology loses its
significance if we consider such maps. (It would seem that the natural analog
of a ``local'' map in the topological setting is a \emph{continuous} map. These
preserve the topology; so we seem to have two competing versions of locality
here.) Probably the best approach is to be inclusive by studying all invertible
maps that act locally on the graph; then the topology-preserving maps would be
a special case. Nevertheless, I have focused most of my efforts in the
two-dimensional case on studying topology-preserving maps.

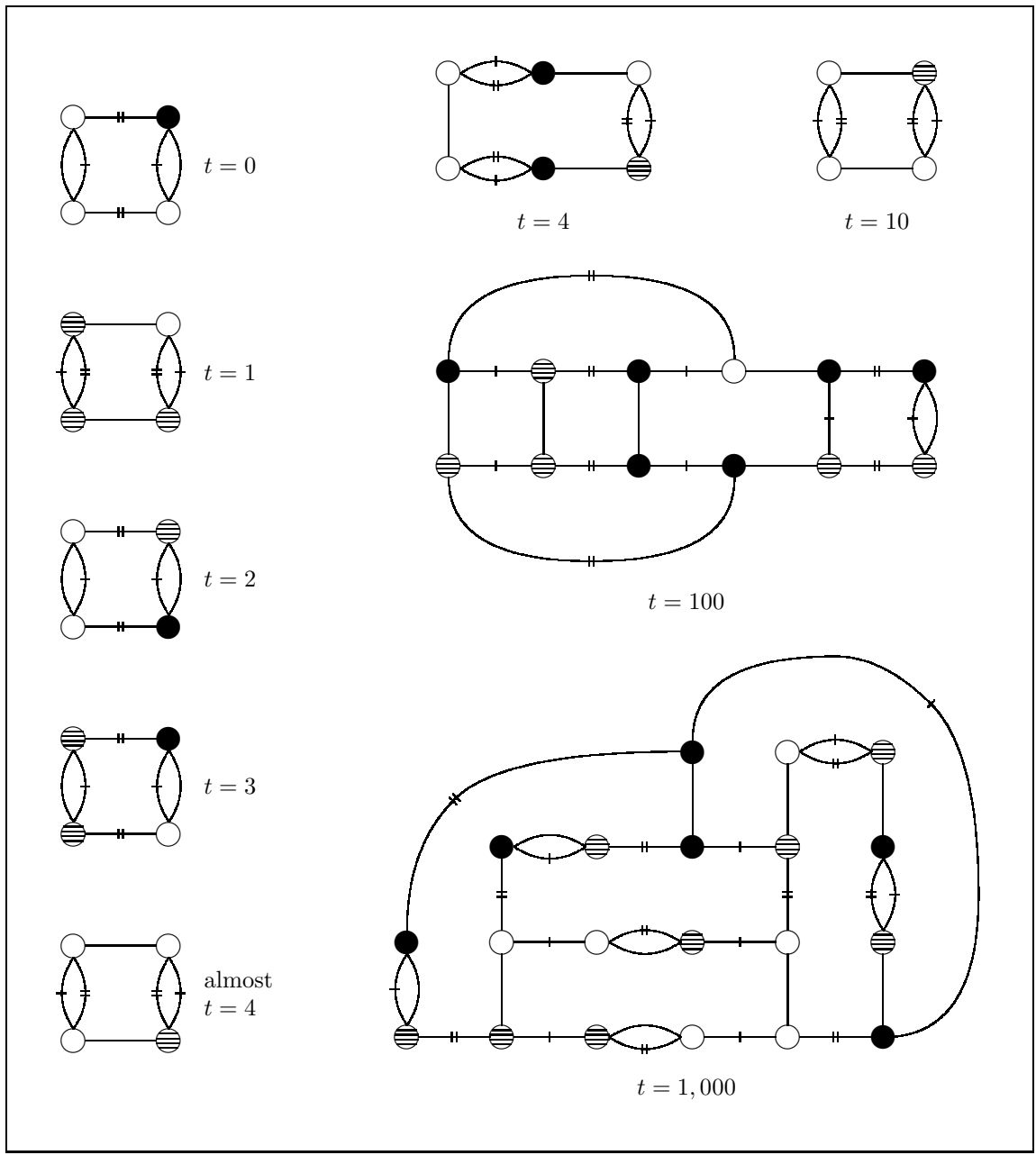
\begin{figure}
\fbox{\begin{picture}(425,475)\footnotesize

\newcommand{\vertexc}{\put(0,0){\circle{10}}
         \put(-4,-3){\line(1,0){8}}\put(-4.9,-1){\line(1,0){9.8}}
         \put(-4.9,1){\line(1,0){9.8}}\put(-4,3){\line(1,0){8}}}

\put(20,386.5){\begin{picture}(50,50)
\put(5,5){\circle{10}}
\put(45,5){\circle{10}}
\put(5,45){\circle{10}}
\put(45,45){\circle*{10}}
\put(60,25){\makebox(0,0)[l]{$t=0$}}
\multiput(10,5)(0,40){2}{\line(1,0){30}}
\multiput(5,10)(40,0){2}{
        \qbezier(0,0)(-5,7.5)(-5,15)\qbezier(-5,15)(-5,22.5)(0,30)
        \qbezier(0,0)(5,7.5)(5,15)\qbezier(5,15)(5,22.5)(0,30)}
\multiput(0,0)(0,40){2}{\multiput(24,3)(2,0){2}{\line(0,1){4}}}
\multiput(8,25)(30,0){2}{\line(1,0){4}}
\end{picture}}

\put(20,299.5){\begin{picture}(50,50)
\put(5,5){\vertexc}
\put(45,5){\vertexc}
\put(5,45){\vertexc}
\put(45,45){\circle{10}}
\put(60,25){\makebox(0,0)[l]{$t=1$}}
\multiput(10,5)(0,40){2}{\line(1,0){30}}
\multiput(5,10)(40,0){2}{
        \qbezier(0,0)(-5,7.5)(-5,15)\qbezier(-5,15)(-5,22.5)(0,30)
        \qbezier(0,0)(5,7.5)(5,15)\qbezier(5,15)(5,22.5)(0,30)}
\multiput(-2,25)(50,0){2}{\line(1,0){4}}
\multiput(8,24)(0,2){2}{\multiput(0,0)(30,0){2}{\line(1,0){4}}}
\end{picture}}

\put(20,212.5){\begin{picture}(50,50)
\put(5,5){\circle{10}}
\put(45,5){\circle*{10}}
\put(5,45){\circle{10}}
\put(45,45){\vertexc}
\put(60,25){\makebox(0,0)[l]{$t=2$}}
\multiput(10,5)(0,40){2}{\line(1,0){30}}
\multiput(5,10)(40,0){2}{
        \qbezier(0,0)(-5,7.5)(-5,15)\qbezier(-5,15)(-5,22.5)(0,30)
        \qbezier(0,0)(5,7.5)(5,15)\qbezier(5,15)(5,22.5)(0,30)}
\multiput(8,25)(30,0){2}{\line(1,0){4}}
\multiput(24,3)(2,0){2}{\multiput(0,0)(0,40){2}{\line(0,1){4}}}
\end{picture}}

\put(20,125.5){\begin{picture}(50,50)
\put(5,5){\vertexc}
\put(45,5){\circle{10}}
\put(5,45){\vertexc}
\put(45,45){\circle*{10}}
\put(60,25){\makebox(0,0)[l]{$t=3$}}
\multiput(10,5)(0,40){2}{\line(1,0){30}}
\multiput(5,10)(40,0){2}{
        \qbezier(0,0)(-5,7.5)(-5,15)\qbezier(-5,15)(-5,22.5)(0,30)
        \qbezier(0,0)(5,7.5)(5,15)\qbezier(5,15)(5,22.5)(0,30)}
\multiput(0,0)(0,40){2}{\multiput(24,3)(2,0){2}{\line(0,1){4}}}
\multiput(8,25)(30,0){2}{\line(1,0){4}}
\end{picture}}

\put(20,38.5){\begin{picture}(50,50)
\put(5,5){\circle{10}}
\put(45,5){\vertexc}
\put(5,45){\circle{10}}
\put(45,45){\circle{10}}
\put(60,25){\makebox(0,0)[l]{\parbox{40pt}{almost\\[-4pt]$t=4$}}}
\multiput(10,5)(0,40){2}{\line(1,0){30}}
\multiput(5,10)(40,0){2}{
        \qbezier(0,0)(-5,7.5)(-5,15)\qbezier(-5,15)(-5,22.5)(0,30)
        \qbezier(0,0)(5,7.5)(5,15)\qbezier(5,15)(5,22.5)(0,30)}
\multiput(-2,25)(50,0){2}{\line(1,0){4}}
\multiput(8,24)(0,2){2}{\multiput(0,0)(30,0){2}{\line(1,0){4}}}
\end{picture}}

\put(177.5,385){\begin{picture}(90,70)(0,-20)
\put(5,5){\circle{10}}
\put(45,5){\circle*{10}}
\put(85,5){\vertexc}
\put(5,45){\circle{10}}
\put(45,45){\circle*{10}}
\put(85,45){\circle{10}}
\multiput(10,5)(0,40){2}{
        \qbezier(0,0)(7.5,-5)(15,-5)\qbezier(15,-5)(22.5,-5)(30,0)
        \qbezier(0,0)(7.5,5)(15,5)\qbezier(15,5)(22.5,5)(30,0)}
\put(85,10){
        \qbezier(0,0)(-5,7.5)(-5,15)\qbezier(-5,15)(-5,22.5)(0,30)
        \qbezier(0,0)(5,7.5)(5,15)\qbezier(5,15)(5,22.5)(0,30)}
\multiput(50,5)(0,40){2}{\line(1,0){30}}
\put(5,10){\line(0,1){30}}
\multiput(25,-2)(0,50){2}{\line(0,1){4}}
\multiput(24,8)(2,0){2}{\multiput(0,0)(0,30){2}{\line(0,1){4}}}
\multiput(78,24)(0,2){2}{\line(1,0){4}}
\put(88,25){\line(1,0){4}}
\put(45,-17){\makebox(0,0){$t=4$}}
\end{picture}}

\put(337.5,385){\begin{picture}(50,70)(0,-20)
\put(5,5){\circle{10}}
\put(45,5){\circle{10}}
\put(5,45){\circle{10}}
\put(45,45){\vertexc}
\put(25,-17){\makebox(0,0){$t=10$}}
\multiput(10,5)(0,40){2}{\line(1,0){30}}
\multiput(5,10)(40,0){2}{
        \qbezier(0,0)(-5,7.5)(-5,15)\qbezier(-5,15)(-5,22.5)(0,30)
        \qbezier(0,0)(5,7.5)(5,15)\qbezier(5,15)(5,22.5)(0,30)}
\multiput(-2,25)(50,0){2}{\line(1,0){4}}
\multiput(8,24)(0,2){2}{\multiput(0,0)(30,0){2}{\line(1,0){4}}}
\end{picture}}

\put(177.5,225){\begin{picture}(210,140)(0,-20)
\put(5,40){\vertexc}
\put(45,40){\vertexc}
\put(85,40){\circle*{10}}
\put(125,40){\circle*{10}}
\put(165,40){\vertexc}
\put(205,40){\vertexc}
\put(5,80){\circle*{10}}
\put(45,80){\vertexc}
\put(85,80){\circle*{10}}
\put(125,80){\circle{10}}
\put(165,80){\circle*{10}}
\put(205,80){\circle*{10}}
\multiput(0,0)(0,40){2}{\multiput(10,40)(40,0){5}{\line(1,0){30}}}
\multiput(5,45)(40,0){3}{\line(0,1){30}}
\put(165,45){\line(0,1){30}}
\put(205,45){%
        \qbezier(0,0)(-5,7.5)(-5,15)\qbezier(-5,15)(-5,22.5)(0,30)
        \qbezier(0,0)(5,7.5)(5,15)\qbezier(5,15)(5,22.5)(0,30)}
\put(198,60){\line(1,0){4}}
\qbezier(5,35)(5,0)(65,0)\qbezier(65,0)(125,0)(125,35)
\qbezier(5,85)(5,120)(65,120)\qbezier(65,120)(125,120)(125,85)
\multiput(0,0)(0,120){2}{\multiput(64,-2)(2,0){2}{\line(0,1){4}}}
\multiput(50,40)(0,40){2}{%
        \multiput(0,0)(120,0){2}{\multiput(14,-2)(2,0){2}{\line(0,1){4}}}}
\multiput(25,38)(0,40){2}{\multiput(0,0)(80,0){2}{\line(0,1){4}}}
\put(163,60){\line(1,0){4}}
\put(105,-17){\makebox(0,0){$t=100$}}
\end{picture}}

\put(160,20){\begin{picture}(245,185)(0,-20)
\put(5,5){\vertexc}
\put(45,5){\vertexc}
\put(85,5){\vertexc}
\put(125,5){\circle{10}}
\put(165,5){\circle{10}}
\put(205,5){\circle*{10}}
\put(5,45){\circle*{10}}
\put(45,45){\circle{10}}
\put(85,45){\circle{10}}
\put(125,45){\vertexc}
\put(165,45){\circle{10}}
\put(205,45){\vertexc}
\put(45,85){\circle*{10}}
\put(85,85){\vertexc}
\put(125,85){\circle*{10}}
\put(165,85){\vertexc}
\put(205,85){\circle*{10}}
\put(125,125){\circle*{10}}
\put(165,125){\circle{10}}
\put(205,125){\vertexc}
\multiput(10,5)(40,0){2}{\line(1,0){30}}
\multiput(130,5)(40,0){2}{\line(1,0){30}}
\multiput(90,85)(40,0){2}{\line(1,0){30}}
\multiput(50,45)(80,0){2}{\line(1,0){30}}
\multiput(45,10)(0,40){2}{\line(0,1){30}}
\multiput(165,10)(0,40){3}{\line(0,1){30}}
\multiput(205,10)(0,80){2}{\line(0,1){30}}
\put(125,90){\line(0,1){30}}
\multiput(90,5)(0,40){2}{%
        \qbezier(0,0)(7.5,-5)(15,-5)\qbezier(15,-5)(22.5,-5)(30,0)
        \qbezier(0,0)(7.5,5)(15,5)\qbezier(15,5)(22.5,5)(30,0)}
\multiput(50,85)(120,40){2}{%
        \qbezier(0,0)(7.5,-5)(15,-5)\qbezier(15,-5)(22.5,-5)(30,0)
        \qbezier(0,0)(7.5,5)(15,5)\qbezier(15,5)(22.5,5)(30,0)}
\multiput(5,10)(200,40){2}{%
        \qbezier(0,0)(-5,7.5)(-5,15)\qbezier(-5,15)(-5,22.5)(0,30)
        \qbezier(0,0)(5,7.5)(5,15)\qbezier(5,15)(5,22.5)(0,30)}
\qbezier(5,50)(5,85)(25,105)\qbezier(25,105)(45,125)(120,125)
\qbezier(125,130)(125,165)(185,165)\qbezier(185,165)(205,165)(225,145)
\qbezier(225,145)(245,125)(245,65)\qbezier(245,65)(245,5)(205,5)
\multiput(5,5)(160,0){2}{\multiput(19,-2)(2,0){2}{\line(0,1){4}}}
\multiput(85,50)(0,35){2}{\multiput(19,-2)(2,0){2}{\line(0,1){4}}}
\multiput(85,0)(80,120){2}{\multiput(19,-2)(2,0){2}{\line(0,1){4}}}
\multiput(65,3)(0,40){2}{\line(0,1){4}}
\multiput(145,3)(0,40){3}{\line(0,1){4}}
\multiput(65,78)(120,50){2}{\line(0,1){4}}
\multiput(-2,25)(210,40){2}{\line(1,0){4}}
\multiput(198,64)(0,2){2}{\line(1,0){4}}
\multiput(0,0)(120,0){2}{\multiput(43,64)(0,2){2}{\line(1,0){4}}}
\qbezier(223.59,143.59)(225,145)(226.41,146.41)
\qbezier(24.3,107.12)(25.71,105.71)(27.12,104.3)
\qbezier(22.88,105.7)(24.29,104.29)(25.7,102.88)
\put(122.5,-17){\makebox(0,0){$t=1,000$}}
\end{picture}}

\end{picture}}\linespread{1}\caption[An evolution of a two-dimensional space
under a local, invertible, topology-preserving operator]{An evolution of a
two-dimensional space under a local, invertible, topology-preserving
operator. Here $a$-vertices are represented by open circles, $b$-vertices by
filled-in circles, and $c$-vertices by shaded circles.}
\label{twodevolution}
\end{figure}

Now that we have some simple operators, we may compose them with one another to
produce new operators. I found that it was necessary to compose approximately
five or more of them together before anything interesting happened. Otherwise,
the resulting maps tended to produce small, finite orbits. But I did manage to
find numerous maps which give complex orbits. My example map proceeds in five
steps, as follows:\vspace{3.5pt}
\begin{enumerate}\linespread{1}\normalsize
\item Change $a$-vertices into $c$-vertices and $c$-vertices into $a$-vertices.
\item Change 0-edges into 1-edges and 1-edges into 0-edges.
\item Change 0-edges into 2-edges and 2-edges into
0-edges.\footnote{\linespread{1}\footnotesize Again, the last two steps
together perform a (012) permutation on edge types.}
\item If a 2-edge joins a $c$-vertex to an $a$-vertex, change the $a$ to $b$;
and if it joins a $c$-vertex to a $b$-vertex, change the $b$ to $a$.
\item Perform the second operation sketched in Figure~\ref{twodtransforms}.
\end{enumerate}
This operator does not change the topology. The results of applying it to a
simple initial state are illustrated in Figure~\ref{twodevolution}.  The only
step which changes anything but the colors of edges or vertices is step 5,
which has no effect on anything until one applies the operator to the $t=3$
state. The graph marked ``almost $t=4$'' in the figure represents the state
after applying steps 1 through 4 to the $t=3$ state. This produces a 12-face
with a 1-edge and 2-edge each joining two $a$-vertices; all other requirements
for performing the right-to-left step-5 operation are also satisfied, so the
result is that two new vertices are added to the graph in the manner shown. The
manifolds in this example are spheres; this has enabled me to draw the graphs
so that each area of the page separated by lines of the graph corresponds to a
face of the manifold. (Note that this means that the portion of the page
outside of the graph corresponds to a face of the manifold.)

\clearpage

The vertex count of the manifold in this example tends to increase with
time. It does not always do this; for instance, at $t=5$ there are six
vertices, but at $t=10$ there are four. The vertex count at $t=100$ is 12; at
$t=1,000$ it is 20; at $t=10,000$ it is 120; at $t=40,000$ it is $3,916$. It
appears that the number of vertices (at least in this portion of the evolution)
grows at roughly an exponential rate, doubling every 5,000 to 7,000 time
steps. Approximately the same growth rate seems to hold if one runs the law in
reverse, though it begins more slowly: at $t=-10,000$ there are only 48
vertices, and at $t=-40,000$ there are 1,040. I have also run the same law
using a four-vertex projective plane (as shown in Figure~\ref{twographs}) with
three $a$-colored vertices and one $b$-colored vertex as my initial state. The
same growth rate seems to apply, though again the growth seems to take off more
slowly: at $t=40,000$ the vertex count is 1,280, and at $t=-40,000$ it is 976.

It is easy to count vertices; and the fact that the number of vertices
increases without exhibiting an apparent regular pattern indicates that
something complex is happening here. But what, exactly, is happening? Is it
simply a random swirling of colors and polygons, like some sort of gas? Or does
this universe eventually contain a phenomenon that might accurately be
described as a two-dimensional equivalent of Madonna? At present, there is no
way of knowing.

\cleardoublepage\vspace*{40pt}\section[Related systems]{\LARGE Related systems}\bigskip
\label{relatedsystems}

Several areas in the literature involve dynamical systems related to the sort
that I wish to study here. These include cellular automata, structurally
dynamic cellular automata, L systems and parallel graph grammars, and symbolic
dynamics.

In cellular automata, the geometry of space is usually chosen to be a
rectangular $n$-dimensional lattice (occasionally other sorts of spaces are
studied, such as hexagonal lattices, or even arbitrary graphs) whose vertices
are colored by $k$ colors; a local homogeneous map specifies how to evolve the
colors of the state at time $t$ to produce the state at time $t+1$. If we add
the requirement that the map be invertible, then these systems become
combinatorial spacetimes. However, they are a restricted form of combinatorial
spacetime, since the geometry in cellular automata does not
evolve. One-dimensional cellular automata will be discussed further in
Chapters~\ref{orient} and~\ref{cauchy}.

Ilachinski and Halpern~\cite{ilachinski/halpern} discuss the idea of allowing
the geometry in cellular automata to evolve locally. They call such systems
``structurally dynamic cellular automata.'' Further discussion appears
in~\cite{halpern,halpern/caltagirone}. Because they do not focus on the case
where the law of evolution is invertible, and because their examples (from my
point of view) are somewhat cumbersome geometrically, there is not much
concrete information to be gained from these papers that is relevant to the
discussion here. However, apart from the concept of invertibility, the basic
ideas of combinatorial spacetimes are indeed present in these papers.

Another class of related systems has been studied primarily from the point of
view of biology, linguistics, and computer science. In the one-dimensional case
these are called L systems. An L system consists of an initial state and a
local rule by which this state evolves. The state is a string of symbols. The
local rule allows the number of symbols in the string to increase or
decrease. Typically the string of symbols does not wrap around to form a
circle, and the law of evolution is typically not invertible. But otherwise,
these systems closely resemble one-dimensional combinatorial
spacetimes. Numerous books and papers have been written about them (see for
example \cite{rozenberg/salomaa,rozenberg/salomaa/eds}).

My impression has been that the point of view of those engaged in studying L
systems is generally very different from mine. For example, L systems are often
used as a way to model various forms of growth. Hence the rules of evolution
are often restricted to be such that the string only grows, and never
shrinks. Such rules cannot be invertible. An additional restriction often
imposed is that the rules are ``context-free.'' What this means is that there
is no interaction in the system. From a physical point of view, these are the
most uninteresting systems possible. Finally, the fact that closed spaces are
not considered is a difference which, from my point of view, introduces an
unnecessary complication. I suspect that it would have been quite difficult for
me to achieve the sorts of results described here had I not restricted myself
to closed spaces.

Because of this differing point of view, I have not invested much time in
studying L systems. However, this is not to say that L systems are irrelevant
to my work. Certainly there is some overlap; in fact, we will see one example
of this in Section~\ref{firstresults}.

Generalizations of L systems to higher dimensions are provided by parallel
graph grammars (see
\cite{rozenberg/salomaa/eds,ehrig/kreowski/rozenberg}). Instead of a state
being a string of symbols, here it is a colored graph; laws of evolution
involve local transformations of these graphs which are applied in parallel at
all points on the graph. Various methods have been suggested for defining
systems of this sort. Again there is often a focus on context-free systems and
growth problems, but there are schemes which do not restrict themselves in
these ways; these seem interesting from my point of view. As we have seen,
combinatorial $n$-dimensional manifolds and pseudomanifolds may be represented
by graphs, so it is possible that $n+1$-dimensional combinatorial spacetimes
may be seen as a type of parallel graph grammar. I have not investigated this
connection in detail, however.

Unlike the above approaches to combinatorial dynamical systems, my approach
combines a geometric flexibility with a focus on invertible local maps. I
believe that this offers several mathematical advantages. For example, the
class of cellular automaton models is a difficult one to study. If, however,
one \emph{expands} that class by allowing for geometry change, one obtains a
class of models which, it seems to me, is easier to comprehend as a whole. For
now many simple transformations between models are possible which were not
possible before. These can simplify the picture (examples of this will be seen
later on). In addition, if one focuses on \emph{invertible} local maps, the
gains are twofold. Firstly, it seems that nondissipative systems are in
important ways mathematically more tractable than dissipative ones. And
secondly, by studying such maps one gains information concerning
\emph{equivalences} between systems. This subject of equivalences, while of
central importance in most mathematical fields, has, I think, not been
sufficiently taken into account in the study of cellular automata.

The area of dynamical systems which perhaps has the most important relationship
with combinatorial spacetimes is symbolic dynamics
(see~\cite{lind/marcus}). The nature of this relationship is, to me, somewhat
surprising and obscure. Nevertheless, it exists.

In the study of dynamical systems, one may examine the trajectories of points
on a manifold $M$ under iteration of some diffeomorphism $f:M\mapsto M$. In
certain cases, one may identify a finite number $m$ of regions $R_i$ in $M$
whose union is $R$ such that $x\in R$ if and only if $f(x)\in R$. Given any
point $x\in R$ at time 0, one may write down a history vector $x_i$, $i\in\Z$,
where $f^i(x)$ is contained in region $R_{x_i}$ for each $i$. In certain cases,
it turns out that distinct points in $R$ always have distinct history
vectors. In addition, the set of history vectors of points in $R$ sometimes
form what is known as a \emph{subshift of finite type}. This means that there
exists some finite $r$ such that, if you write down a list of all length-$r$
sequences contained in history vectors, the set of history vectors is precisely
all vectors containing the symbols $1$ through $m$ which contain only those
allowed length-$r$ sequences.

In this case, one may put a topology on the set of allowed history vectors
$H$. A basis for this topology is given by $\{x_i\in H\,|\,x_{i_j}=a_j,\,1\leq
j\leq s\}$ for each choice of finite $s$, sequence of coordinates $i_j$, and
sequence of values $a_j$. (In other words, each set of vectors that is
obtainable by placing restrictions on finitely many coordinate values is a
basis element.)  The resulting topology is a Cantor set. One also defines a
\emph{shift map} $\sigma:H\mapsto H$ where $\sigma(x)=y$ if and only if
$y_i=x_{i+1}$. Then $\sigma$ is continuous under the Cantor set topology.

Next, one investigates \emph{topological conjugacy}. If $H_1$ and $H_2$ are
sets of history vectors with the Cantor set topology and $\sigma_1$ and
$\sigma_2$ are the associated shift maps on $H_1$ and $H_2$, then $g:H_1\mapsto
H_2$ is a conjugacy of $\sigma_1$ and $\sigma_2$ if $g$ is a homeomorphism and
$\sigma_2g=g\sigma_1$.

It turns out that topological conjugacy maps $g:H\mapsto H$ are exactly the
same thing as reversible one-dimensional cellular automaton maps (provided that
one extends the usual definition of cellular automata so that they are defined
on subshifts of finite type). This is well known. To symbolic dynamicists,
conjugacy maps are interesting as equivalence maps; to those interested in
cellular automata, the same maps (if they map a space to itself) are of
interest for their dynamical properties when iterated.

This, I thought, was as far as the connection went. But it turns out that there
is another sort of equivalence map between subshifts of finite type which
arises in symbolic dynamics. This is called a \emph{flow equivalence map}. It
is less well known. The relationship of conjugacy maps to reversible
one-dimensional cellular automata is precisely analogous to the relationship of
flow equivalence maps to 1+1-dimensional combinatorial spacetimes. That is: the
set of flow equivalence maps $g:H\mapsto H$ is identical to the set of laws of
evolution which I discovered in my 1+1-dimensional case. All of this will be
discussed in Section~\ref{flow}.

\newpage\vspace*{40pt}\section[Space+time picture and spacetime picture]{\LARGE
Space+time picture and \protect\\spacetime picture}\bigskip
\label{spacetime}

The examples in Chapter~\ref{examples} were developed from the space+time point
of view. The general setup from this perspective is as follows. There is a set
of sets of allowed spaces, and a set of local, invertible maps from one set of
allowed spaces to another. Such a map $T:\X\mapsto\X'$ is called an
\emph{equivalence} between space sets. If $\X=\X'$, then $T$ is also a
\emph{law of evolution}: it is an operator on \X, and the orbits of \X\ under
this operator are considered as spacetimes. In other words, given any $x\in\X$,
one considers the set $\ang{T,x}=\{T^t(x)\,|\,t\in\Z\}$ as a foliation of
spacetime by Cauchy surfaces contained in \X, where these surfaces are indexed
by the time parameter $t$. Here one is not concerned with the particular value
of the time parameter, but only with relative values; hence \ang{T,x} is
equivalent to $\ang{T,T^t(x)}$ for all $t\in\Z$. Also, the choice of whether
$t$ increases or decreases in a particular direction perpendicular to the
Cauchy surfaces is simply a matter of convention; hence \ang{T,x} is equivalent
to \ang{T^{-1},x}.

At first glance there would appear to be no specified local connection between
spaces from this perspective. There is a space $x$ at time $t$, and a space
$T(x)$ at time $t+1$, and these are separate objects. However, the fact that
$T$ is a local law provides an implicit local causal structure linking these
objects. So the result of evolving a space using $T$ really is a spacetime.

From this point of view, two space sets \X\ and $\X'$ are equivalent if and
only if there exists a local invertible map $U:\X\mapsto\X'$. If two space sets
are equivalent, then there is a local invertible map $T\mapsto UTU^{-1}$
between laws of evolution on \X\ and laws of evolution on $\X'$. These are the
only equivalences between laws that arise naturally in the space+time picture.

Note that an equivalence between laws induces a local equivalence between
spacetimes: if $S=UTU^{-1}$ then the map $U$ sends $T^t(x)$ to $S^t(U(x))$ for
each $t$; hence $U$ induces a local invertible map which sends \ang{T,x} to
\ang{S,U(x)}. We may think of \ang{T,x} and \ang{UTU^{-1},U(x)} as being
descriptions of the same set of spacetimes using different coordinate systems;
then maps of the form $\ang{T,x}\mapsto\ang{UTU^{-1},U(x)}$ are the allowed
coordinate transformations. From the space+time point of view, the dynamical
properties of a set of spacetimes are precisely those properties which are
invariant under these coordinate transformations. (There is obvious analogy
here with linear operators: the operator itself is independent of the basis;
the properties of the operator $T$ are those which are invariant under change
of basis, which again takes the form $UTU^{-1}$ for some invertible $U$. One
difference is that here I am only considering operators $T$ which are
themselves invertible.)

The space+time point of view is a straightforward one, but it is not the best
one. In general relativity the natural point of view is the spacetime point of
view, and this is also the case here. In the spacetime picture, two
\emph{spacetime} sets \X\ and $\X'$ are equivalent if and only if there exists
a local invertible map $U:\X \mapsto \X'$. Certain spacetime sets will contain
laws of evolution, and others will not. Those that do will admit foliations
into Cauchy surfaces, any one of which locally determines the rest of the
spacetime; those that don't will not admit such foliations.

By ``set of spacetimes'' I here simply mean some sort of set of combinatorial
objects that is defined by specifying what the allowed neighborhoods are and
how they are allowed to be pieced together. Since the word ``spacetime''
normally refers to situations in which there is some sort of dynamics, I
suppose that it would be better to restrict the use of that word to refer to
those sets of objects that admit a foliation into Cauchy surfaces as described
above. But for now I will not do this, since I have no precise general
definition which describes when laws of evolution exist and when they do not,
let alone a general method of how to detect their existence. The above idea
about foliation into Cauchy surfaces is at the moment only a heuristic one.

The main point that I wish to make about these two points of view has to do
with the equivalence of laws and with dynamical properties. In the spacetime
point of view, the law of evolution really in a sense \emph{is} the set of
spacetimes. The set of spacetimes is a set of rules concerning which
neighborhoods are allowed and how they may be glued together. If these rules
imply that there is a foliation into Cauchy surfaces, then the rules are in
effect the law of evolution, since they determine how one constructs the rest
of the spacetime from one of the Cauchy surfaces. From the spacetime point of
view, then, two laws are equivalent if and only if the associated sets of
spacetimes are equivalent. It is natural to expect that this is a broader
notion of equivalence that the space+time one: every space+time equivalence
transformation no doubt corresponds to a spacetime equivalence transformation,
but not vice versa.

This has implications regarding the nature of dynamical properties. If from the
space+time point of view we find that $P$ is a dynamical property, we cannot be
certain that $P$ will transform invariantly under the larger class of
equivalence transformations provided by the spacetime picture. In the spacetime
picture, it is only those properties which are invariant under spacetime
equivalence transformations which should be regarded as dynamical
properties. Thus the spacetime picture should allow us to get a better idea of
which properties are important from a dynamical point of view.

Unfortunately, it is difficult to adopt the spacetime point of view at the
start. In order to do so, one needs a higher-dimensional theory. At the moment
that theory is not well developed. The geometry of one-dimensional spaces is
simple; for this reason, most of the work to be discussed here is approached
from the 1+1-dimensional perspective. A two-dimensional theory is required in
order to understand these systems from a spacetime point of view. My attempts
to formulate such a theory will be discussed in Chapter~\ref{higherD}.

Despite the difficulty of adopting the spacetime point of view, it turns out to
be easy to find examples of a certain sort of equivalence between laws which is
a natural one in the spacetime picture but does not arise naturally in the
space+time picture. These will be discussed in Chapter~\ref{cauchy}.

\cleardoublepage\vspace*{40pt}\section[The 1+1-dimensional oriented case]{\LARGE The
1+1-dimensional oriented case}\bigskip
\label{orient}

In this chapter I trace the development of my ideas about 1+1-dimensional
combinatorial spacetimes in a somewhat historical manner. The path which is
traced begins with reversible cellular automata, passes through a
straightforward attempt to extend these systems, and eventually arrives at a
new point of view which is related to symbolic dynamics and to bialgebras. One
purpose of presenting the subject in this manner is that it provides an easy
introduction to combinatorial spacetimes for those familiar with cellular
automata, and demonstrates why the transition from reversible cellular automata
to the new point of view is inevitable. A second reason for proceeding in this
way is that my story is by no means completed, and contains a number of
holes. For example, though I believe that the various points of view presented
here describe the same class of systems, I have not proven this. In addition,
though the new point of view is more elegant than the older ones, it is
possible that the older ones may yet serve a purpose. I have attempted to
provide the reader with every type of tool known to me that might be useful in
further study of these systems.

\subsection{Reversible one-dimensional cellular automata}
\label{casection}

When I first became interested in exploring the subjects discussed in this
document, the only systems known to me that fit the criteria I had in mind were
reversible cellular automata (i.e., cellular automata where the laws of
evolution are invertible). These systems had been brought to my attention by
papers of Fredkin and Toffoli~\cite{fredkin/toffoli},
Toffoli~\cite{toffoli1,toffoli2}, Margolus~\cite{margolus} and
Vichniac~\cite{vichniac}. I decided to begin by exploring the simplest of these
systems: i.e., one-dimensional systems. It soon became apparent that
one-dimensional cellular automata (without the restriction of reversibility)
had been studied most extensively by
Wolfram~\cite{wolfram1,wolfram2,wolframbook}. My early research on reversible
one-dimensional cellular automata focused on the question of how to generate
all such systems; the results (and further references) are in \cite{me}. (See
also a paper by Boykett~\cite{boykett}, who independently obtained similar
results in a different and interesting way. Also see Williams~\cite{williams};
it turns out that the above results are closely related to his main theorem
concerning topological conjugacy maps between subshifts of finite type.)

The set of states in a traditional one-dimensional cellular automaton is the
set of functions
$\{f:\z{w} \mapsto K\,|\,w \in \N\,\cup\,\{\infty\}\}$ (here
$\z{\infty} \equiv \Z$) where $K$ is a fixed, finite set of allowed
``colors.'' In other words, it is a closed one-dimensional lattice of cells
where the cells are indexed by \z{w} and where each cell contains a color in
$K$. If $w=\infty$ then the lattice has the form of a line; otherwise it has
the form of a circle. The number $w$ is called the \emph{width} of the
automaton.

Each state in a cellular automaton is associated with a variable $t$ (time),
with $t \in \Z$ if the automaton is reversible and $t \in \Z^+$ otherwise. The
law of evolution $T$ associated with the automaton acts homogeneously on the
spatial lattice by examining the state of a neighborhood of a cell at time $t$
and thereby deciding what color that cell should be at time $t+1$. Typically
the neighborhood of cell $i$ is chosen to be the $2r+1$ consecutive cells
beginning at cell $i-r$ and ending at cell $i+r$ for some fixed positive
integer $r$. But one may choose any neighborhood one wishes, as long as it is
finite. For example, in what follows I will sometimes think of the neighborhood
of cell $i$ as being the $d$ consecutive cells beginning at cell $i$ and ending
at cell $i+d-1$, for some fixed positive integer $d$.

The law $T$ can therefore be described by a function mapping each possible
state of a cell's neighborhood to the color that a cell becomes at time $t+1$
if its neighborhood is in that state at time $t$. This may be represented
nicely by a table with two columns; the left-hand column lists the possible
states of the neighborhood, and the right-hand column lists the associated
color.

Several such tables for laws which will be referred to later in the text are
presented in Figure~\ref{calaws}.
\begin{figure}
\fbox{\begin{picture}(425,300)\footnotesize\put(0,0){\makebox(425,300){%
$\begin{array}{ccc}\begin{array}{c||cc|cc|cc|cc|cc}
\,&\multicolumn{2}{c|}{A}&\multicolumn{2}{c|}{B}&\multicolumn{2}{c|}{C}&\multicolumn{2}{c|}{D}&\multicolumn{2}{c}{E}\\\hline
\mathit{aa}&\hspace*{3pt}a\hspace*{3pt}&\hspace*{3pt}a\hspace*{3pt}&\hspace*{3pt}a\hspace*{3pt}&\hspace*{3pt}a\hspace*{3pt}&\hspace*{3pt}a\hspace*{3pt}&\hspace*{3pt}a\hspace*{3pt}&\hspace*{3pt}b\hspace*{3pt}&\hspace*{3pt}c\hspace*{3pt}&\hspace*{3pt}a\hspace*{3pt}&\hspace*{3pt}a\hspace*{3pt}\\
\mathit{ab}&c&c&c&c&b&a&d&a&d&c\\
\mathit{ac}&a&a&a&a&a&c&b&c&a&c\\
\mathit{ad}&c&c&c&c&b&c&d&a&d&a\\
\mathit{ba}&a&a&a&b&a&b&b&c&a&b\\
\mathit{bb}&c&c&c&d&b&b&d&a&d&c\\
\mathit{bc}&a&a&a&b&a&d&b&c&a&c\\
\mathit{bd}&c&c&c&d&b&d&d&a&d&b\\
\mathit{ca}&b&b&d&b&c&a&a&d&c&a\\
\mathit{cb}&d&d&b&d&d&a&c&b&b&d\\
\mathit{cc}&b&b&d&b&c&c&a&d&b&d\\
\mathit{cd}&d&d&b&d&d&c&c&b&c&a\\
\mathit{da}&b&b&d&a&c&b&a&d&c&b\\
\mathit{db}&d&d&b&c&d&b&c&b&b&d\\
\mathit{dc}&b&b&d&a&c&d&a&d&b&d\\
\mathit{dd}&d&d&b&c&d&d&c&b&c&b
\end{array}&&\begin{array}{c}\begin{array}{c|cccccc}
F&a&b&c&d&e&f\\ \hline
a&a&f&a&f&a&f\\
b&a&f&a&f&a&f\\
c&d&b&d&b&d&b\\
d&d&b&d&b&d&b\\
e&c&e&c&e&c&e\\
f&c&e&c&e&c&e
\end{array}\\[72pt]
\begin{array}{c|cccccc}
F^{-1}&a&b&c&d&e&f\\ \hline
a&a&c&e&c&e&a\\
b&b&d&f&d&f&b\\
c&a&c&e&c&e&a\\
d&a&c&e&c&e&a\\
e&b&d&f&d&f&b\\
f&b&d&f&d&f&b
\end{array}\end{array}\end{array}$}}
\end{picture}}\linespread{1}\caption[Some reversible one-dimensional cellular
automata]{Some invertible one-dimensional cellular automata. The first table
displays five laws ($A$ through $E$) defined on \x{4}. Each of these laws and
their inverses have the same left-hand column containing all length-2 segments
in \x{4}; this is displayed at the far left of the table. Each law name is
associated with two columns: the first is the right-hand column for the law;
the second is the right-hand column for its inverse. Laws $F$ and $F^{-1}$ are
also $d=2$ laws, but they are defined on \x{6}. Here I have used a more compact
``product'' notation which is particularly suited for $d=2$ laws: each
left-hand segment is of the form \seg{LR}, where $L$ and $R$ are colors; the
right-hand segments are placed in a matrix, where the rows are indexed by $L$
and the columns by $R$. All laws given here are ``center-reversible'' laws; for
further discussion (and another type of compact notation) see~\cite{me}.}
\label{calaws}
\end{figure}
These laws happen to be invertible; the tables for their inverses are also
shown in the figure.

\subsection{A modification}

My next step was to modify this scheme to allow for the possibility of geometry
change. Consider the case where space is a one-dimensional lattice containing a
finite number of cells and forming a circle. The only thing that geometrically
differentiates one such space from another is the number of cells in the
lattice. This number is an invariant in traditional cellular automata. The
obvious idea is to modify the setup so that the number of cells is now allowed
to change. This change should be accomplished by the action of the local law
alone: we want it to be possible that certain neighborhoods at time $t$ cause
new cells to appear at time $t+1$, while others cause cells to disappear.

If one is to insert cells or delete them at certain spots in the lattice, then
what happens to the cell coordinates? It is impossible to have them be updated
in a local way and still maintain the property that the coordinate numbers of
adjacent cells differ by one. There is an easy solution, however: dispense with
coordinates. Formally, this can be expressed as follows. Let $X$ be the
original set of states $\{f:\z{w} \mapsto K\,|\,w \in
\N\,\cup\,\{\infty\}\}$. Define a relation on $X$: $f \sim g$ if and only if
there exists an $m$ in \z{w} such that $f(i)=g(i+m)$ for all $i \in \z{w}$. It
is easy to see that this is an equivalence relation on $X$. Let \X\ be the set
of equivalence classes $\{[x]\,|\,x \in X\}$. Then the elements of \X\ are the
coordinate-free states.

Now examine again the tables in Figure~\ref{calaws}. I will now present a
different interpretation of these tables, which will be helpful in the process
of generalizing one-dimensional cellular automata to allow for geometry change.

The first step is to view the entries in the table as \emph{segments} (see
Chapter~\ref{examples}). A table consists of two length-$n$ columns of
segments, where $n=|K|^d$ ($d$ is the neighborhood size). The set of segments
in the left column will be denoted by $L$, and the set of segments in the right
column by $R$; the $i$th segment on the left side will be denoted $L_i$, and so
on. Each left-hand segment has length $d$, and each right-hand segment has
length 1.

Secondly, the left-hand side of the table may be viewed as containing not just
a list of segments, but (implicitly) a set of rules for how these segments may
be glued together to form a state. The gluing rules are as follows: if the last
$d-1$ cells of $L_i$ have the same state as the first $d-1$ cells of $L_j$,
then $L_i$ may be glued to $L_j$ by overlapping those matching $d-1$ cells.

Note that these gluing rules have two parts. The first is the matter of
\emph{adjacency}: which segments may be glued to which? When two segments are
glued together, the right end of one is glued to the left end of the other
(when both segments are aligned so that the directed edges point to the
right). Hence adjacency has a direction. In the above example the right end of
$L_i$ is glued to the left end of $L_j$; we say that $L_j$ \emph{follows}
$L_i$. The set of adjacency rules may be nicely encoded in a $n \times n$
matrix $A_{ij}$ of 0's and 1's, where $A_{ij}=1$ if and only if $L_i$ may be
followed by $L_j$. Some examples of left-hand sides of tables and their
associated adjacency matrices are given in Figure~\ref{adjacency}.
\begin{figure}
\fbox{\begin{picture}(425,146)\footnotesize\put(0,12){\makebox(425,146)[bl]{\hfill
\begin{tabular}{r|cc}
\seg{a}&1&1\\
\seg{b}&1&1\\
\end{tabular}\hfill
\begin{tabular}{r|cccc}
\seg{aa}&1&1&0&0\\
\seg{ab}&0&0&1&1\\
\seg{ba}&1&1&0&0\\
\seg{bb}&0&0&1&1\\
\end{tabular}\hfill
\begin{tabular}{r|cccccccc}
\seg{aaa}&1&1&0&0&0&0&0&0\\
\seg{aab}&0&0&1&1&0&0&0&0\\
\seg{aba}&0&0&0&0&1&1&0&0\\
\seg{abb}&0&0&0&0&0&0&1&1\\
\seg{baa}&1&1&0&0&0&0&0&0\\
\seg{bab}&0&0&1&1&0&0&0&0\\
\seg{bba}&0&0&0&0&1&1&0&0\\
\seg{bbb}&0&0&0&0&0&0&1&1\\
\end{tabular}\hfill
}}\end{picture}} \linespread{1}\caption[Some left-hand segments and their
implicit adjacency matrices]{Left-hand segments and their implicit adjacency
matrices. Shown are the cases $d=1$, $d=2$ and $d=3$ for two colors (no
neighborhood restrictions).}
\label{adjacency}
\end{figure}

The second part of the gluing rules answers this kind of question: if $L_i$ is
followed by $L_j$, then how are they to be glued together? For each $i,j$ such
that $A_{ij}=1$, there must exist a rule $g^L_{ij}$ which says how to glue
$L_i$ to $L_j$. In our example, $g^L_{ij}$ says that the gluing is to be
accomplished by overlapping the right-hand $d-1$ cells of $L_i$ with the
(identical) left-hand $d-1$ cells of~$L_j$.

Now consider the segments on the right-hand side of the table. The table also
contains implicit gluing rules for these segments. The rules are simply this:
the adjacency matrix for the right-hand segments is the same as the one for the
left-hand segments; and the right-hand segments are glued together by placing
them next to one another with no overlap.

Thus the overall picture so far is as follows. Our table consists of $n$ rows
and two columns. The entries in the table are segments. There is an $n \times
n$ adjacency matrix $A$ of 0's and 1's which specifies whether $L_j$ may follow
$L_i$ and whether $R_j$ may follow $R_i$ (they may if and only if
$A_{ij}=1$). And for each $i,j$ such that $A_{ij}=1$, there is a rule
$g^L_{ij}$ which says how to glue $L_i$ to $L_j$, and a rule $g^R_{ij}$ which
says how to glue $R_i$ to~$R_j$.

Suppose that we make a number of copies of the left-hand segments and then glue
them together using the gluing rules in such a way that each segment is
preceded by one segment and followed by another. This gives us a state of the
cellular automaton. For instance: if the left-hand segments in our table are
\seg{aa}, \seg{ab}, \seg{ba}, and \seg{bb}, we may make two copies of \seg{aa},
one copy of \seg{ab} and one copy of \seg{ba}, arrange them in a circle, and
then glue them together to form \cir{aaab}.

This situation has two important features. Firstly, it is easy to see that,
given any state of the automaton, there is one and only one way to obtain it by
gluing left-hand segments together in accordance with the rules. Secondly, if
we start with a state of the automaton and disassemble it into an ordered
collection of left-hand segments, this process is a \emph{local} one. What I
mean by this is that there is a fixed finite radius $r$ (in this case, $r=d-1$)
such that, if $S$ is a connected piece of the state, then the portion of the
segment decomposition of the state that overlaps $S$ may be determined by
looking at the state of $S$ plus its $r$ neighbors to either side. For
instance, if \cir{aababb} is the state, $d=2$, and the piece of state in
question is \seg{abb}, we need only look at \seg{abb} plus its neighbor on
either side (i.e, at \seg{babba}) in order to see that the portion of the
segment decomposition that overlaps the piece \seg{abb} is the sequence
\seg{ba}, \seg{ab}, \seg{bb}, \seg{ba}.

Now the cellular automaton law may be seen to work as follows. Suppose that
$[x]$ is the state of the cellular automaton at time $t$. One begins by
decomposing $[x]$ into an arrangement of left-hand segments and left-hand
gluing rules. This decomposition is unique. The state of the automaton at time
$t+1$ is then constructed by replacing each left-hand segment with the
associated right-hand segment, replacing each left-hand gluing rule with the
associated right-hand gluing rule, and then gluing the resulting segments
together using the new gluing rules. Since the decomposition into left-hand
segments is a local one, the resulting map is local. One additional property is
required in order for the map to be a local law of evolution: it must be
possible to iterate the map. Hence in all cellular automata tables it is the
case that every state that can be assembled from right-hand segments can also
be assembled from left-hand segments.

This may be formalized in the following way. Let $F$ be the set of functions
$f:\z{w}\mapsto\z{n}$ such that $A_{f(i),f(i+1)}=1$ for each
$i\in\z{w}$. Define a relation on $F$ as follows: $f_1\sim f_2$ if and only if
there exists a $j$ in \z{w} such that $f_1(i)=f_2(i+j)$ for all
$i\in\z{w}$. This is an equivalence relation. Let $\mathcal{F}$ be the set of
equivalence classes, and denote the equivalence class containing $f$ by
$[f]$. Recall our earlier definitions of $X$ and of \X. One may construct an
element of $X$ from $f$ by defining $l_i$ to be a copy of segment $L_{f(i)}$
for each $i\in\z{w}$, defining gluing rule $g_{i,i+1}$ to be a copy of
$g^L_{f(i),f(i+1)}$ for each $i\in\z{w}$, and then gluing the resulting
segments together using these gluing rules. This defines a map $h_L:F\mapsto
X$. In addition, since $h_L(f)=x$ implies that $h_L[f]=[x]$, we may define a
map $H_L:\mathcal{F}\mapsto\X$ by the rule $H_L([f])=[h_L(f)]$. If, on the
other hand, we use right-hand segments and right-hand gluing rules in the above
construction process, this gives us a map $H_R:\mathcal{F}\mapsto\Y$. We
require that $\Y\subset\X$. If $H_L$ is invertible and $H_L^{-1}$ is local,
then $H_L^{-1}H_R$ is a cellular automaton law.

One may also view this in terms of graphs. If $A$ is an $n\times n$ matrix of
nonnegative integers, then it corresponds to a directed graph \g{A} with $n$
vertices, where there are $A_{ij}$ edges directed from vertex $i$ to vertex
$j$. Suppose that $A$ is an adjacency matrix for a cellular automaton
table. Then $\mathcal{F}$ is simply the set of closed directed paths in
\g{A}. Thus there is an invertible map between the set of closed directed paths
in \g{A} and the set of states of the cellular automaton. One may then view the
segments as being associated with the vertices, and the closed directed paths
as recipes for assembling copies of these segments into states.

It is now easy to generalize one-dimensional cellular automata to produce
systems in which there is geometry change. We may drop the requirement that the
left-hand segments are all of length $d$ and are glued together by a
$(d-1)$-cell overlap; and we may drop the requirement that the right-hand
segments are all of length one and are glued together by placing them next to
one another. All we need is that the gluing rules produce well-defined maps
$H_L$ and $H_R$ as defined above, that $H_L$ is invertible, that $H_L^{-1}$
is local, and that $H_R(\mathcal{F})\subset H_L(\mathcal{F})$.

A further generalization will prove useful later on: instead of requiring $A$
to be a matrix of 0's and 1's, we may allow it to be a matrix of nonnegative
integers. Now $A_{ij}$ represents the number of ways in which the $j$th segment
may directly follow the $i$th segment. In other words, instead of there being
at most one gluing rule $g_{ij}$ for each ordered pair of segments, we now
allow there to be a finite set $\{g_{ij}^k\,|\,1\leq k\leq A_{ij}\}$ of such
rules. The associated graph \g{A} has $A_{ij}$ edges going from vertex $i$ to
vertex $j$. This requires us to modify the definition of $F$: if $G$ is the set
of gluing rules, then $F$ is the set of functions $f:\z{w}\mapsto G$ with the
property that if $f(m)=g_{ij}^k$ and $f(m+1)=g_{qr}^s$ then $j=q$.

Let us set up some notation to deal with these systems. A rule table will be
denoted by the tuple \ang{L,R,G^L\!,G^R\!,A} where $L$ and $R$ are sets of
segments, $G^L$ and $G^R$ are sets of gluing rules, and $A$ is an adjacency
matrix. It is also useful to consider tuples \ang{L,G^L\!,A}. Such a tuple will
be called a \emph{description} of a set of spaces. The set of spaces it
describes will be denoted by \x{\ang{L,G^L\!,A}}. It is the set of 2-regular
graphs obtained by gluing the segments in $L$ together using the adjacency
rules in $A$ and the gluing rules in $G^L$; in other words, it is the set
$H_L(\mathcal{F})$. If $H_L$ is invertible and its inverse is local, we will
say that \ang{L,G^L\!,A} is a \emph{faithful description} of
\x{\ang{L,G^L\!,A}}.

One may obtain a compact notation for a description by encoding the gluing-rule
information in the adjacency matrix. A gluing rule $g_{ij}^k$ may be thought of
as a segment which has the property that $L_i$ is embedded at the left end of
the segment and $L_j$ is embedded at the right end of the segment. For example,
if $L_i=\seg{abcb}$ and $L_j=\seg{cb}$, then $g_{ij}^k=\seg{abcb}$ means that
$L_i$ is glued to $L_j$ with a two-cell overlap, while $g_{ij}^k=\seg{abcbcb}$
means they are to be glued together with no overlap, and
$g_{ij}^k=\seg{abcbacb}$ means they are to be glued together by inserting an
$a$ between them. Consider the matrix $A$ such that $A_{ij}$ is written as a
formal sum of the segments corresponding to each gluing rule $g_{ij}^k$, with
the order of the sum corresponding to the order of the index $k$. It is useful
in this case to omit square brackets and think of the colors as free
variables. The empty segment corresponds to a product of zero variables; hence
it is denoted by 1. Now $A$ contains all gluing-rule information. The original
adjacency matrix may be obtained by setting each color variable to 1.

Here are two examples, both of which are descriptions of \x{3}:\bigskip\\
\footnotesize\hspace*{\fill}\begin{tabular}{c|ccc}
\seg{a}&\textit{aa}&\textit{ab}&\textit{ac}\\
\seg{b}&\textit{ba}&\textit{bb}&\textit{bc}\\
\seg{c}&\textit{ca}&\textit{cb}&\textit{cc}\\
\end{tabular}\hfill\begin{tabular}{c|c}
\seg{\,}&$a+b+c$\\
\end{tabular}\hspace*{\fill}\normalsize\bigskip\\
The segments are written in the left-hand columns. The left-hand description is
the standard one with length-2 segments and one-cell-overlap gluing rules. The
right-hand description involves a single empty segment (recall that this is
simply an edge) and three ways to glue two such segments together by putting a
vertex between them (one for each color).

In terms of descriptions, a rule table \ang{L,R,G^L\!,G^R\!,A} may be denoted
as \ang{D,E}, where $D=\ang{L,G^L\!,A}$, $E=\ang{R,G^R\!,A}$,
$\x{E}\subset\x{D}$, and $D$ is a faithful description.

Since I am primarily interested in the case where the law of evolution
\ang{D,E} is local and has a local inverse, this is the case that will be
treated here. Such systems will be called \emph{1+1-dimensional oriented
combinatorial spacetimes}. In this case both $H_L$ and $H_R$ must be
invertible, and their inverses must both be local. Hence $D$ and $E$ must both
be faithful descriptions. Furthermore, it must be the case that
$\x{D}=\x{E}$. Such a law, then, is simply a pair of faithful descriptions of a
set of spaces \X\ that share the same adjacency matrix. It follows that, in
order to study the set of all such invertible maps on \X, it is important to
obtain an understanding of the set of all faithful descriptions of \X. More
will be said about this topic in Section~\ref{descriptrans}.

It should be noted that I have been focusing on laws of evolution, but that this
formalism is perfectly suited to deal with local maps $T:\X\mapsto\Y$ where \X\
is not necessarily equal to \Y. Such a map is given by any pair of descriptions
\ang{D,E} such that $D$ and $E$ share the same adjacency matrix and $D$ is a
faithful description. The important case here will be when $T$ has a local
inverse. In that case it is also required that $E$ is a faithful
description. Maps of this sort will be discussed in Section~\ref{element}.

Finally, consider the orientation-reversing map on \X. In many contexts
orientation reversal is not thought of as a local map. It is local in this
context, however, because it operates on a space simply by switching the
direction of the arrows in the directed graph. This operation is completely
analogous to operations which act locally on colors. In fact, the oriented case
is a subset of the general 1+1-dimensional case (see Chapter~\ref{unorient}),
in which both oriented and unoriented space sets are allowed. In that context
one may represent oriented spaces by singling out two colors for special roles;
reversal of orientation amounts to permuting those two colors.

This map cannot be given in the form \ang{D,E}. Neither can other maps which
involve orientation reversal in an essential way. To generalize our formalism
to include these maps we must drop the assumption that the orientation of
segments always goes from left to right as written on the page. Instead, the
orientation of each segment must be given explicitly in the rule table. Now
$A_{ij}=1$ will be taken to mean that the right-hand end of $L_i$ may be glued
to the left-hand end of $L_j$. In order for this gluing to make sense, it must
be the case that if $A_{ij}=1$ then the orientations of $L_i$ and $L_j$ are the
same: they both point to the right or they both point to the left. Here is an
example of a simple rule table of this sort:\bigskip\\
\footnotesize\hspace*{\fill}$\begin{array}{c|c}\rightarrow
a\rightarrow&\leftarrow b\leftarrow\\\rightarrow b\rightarrow&\leftarrow
a\leftarrow\end{array}$\hspace*{\fill}\normalsize\bigskip\\
I have drawn the directed edges in these length-1 segments to indicate
orientation. This rule acts on \x{2} by permuting the colors and then reversing
orientation; e.g., it sends \cir{aababbb} to \cir{aaababb}.

Define a \emph{simple orientation-reversing map} to be one in which the
left-hand and right-hand segments are the same (as read from left to right)
except perhaps for directions of edges. In general, while one may compose maps
of the form \ang{D,E} with simple orientation-reversing maps in whatever way
one pleases, any such map is equal to a composition of two maps $PT$ where $T$
preserves orientation and $P$ is a simple orientation-reversing map. Hence one
may easily derive the set of all maps involving orientation reversal from the
set of orientation-preserving maps. Since accounting for orientation reversal
makes life a bit complicated, I will largely ignore it in what follows, and
focus on maps that preserve orientation.

\subsection{The class of one-dimensional space sets}

Up to now I have only discussed sets of spaces in which there are $k$ colors
and no neighborhood restrictions; i.e., any color may be
adjacent to any other. These space sets will appear frequently in the
sequel. As before, I shall refer to such a set as \x{k}, and label the $k$
colors with lower-case letters of the alphabet, beginning with $a$.

The requirement that finite pieces of spacetime contain finite amounts of
information does not restrict us to these space sets alone, however. It is
satisfied simply by allowing only a finite number of colors. We may impose
further restrictions on which patterns of colors are allowed.

Here is a natural way to do this. Choose a finite positive integer $d$. Then
list all allowed neighborhoods of diameter $d$. Let the set of spaces be that
subset of \x{k} in which each $d$-neighborhood is in the list of allowed
neighborhoods.

Such a space set is easily described by a set of segments and gluing rules. For
example, the set of segments can simply be the set of allowed
$d$-neighborhoods, and the gluing rules can be the same as those implied by the
left-hand side of a traditional cellular-automaton table (one may glue two
segments together by forming a $(d-1)$-cell overlap whenever the overlapping
cells match).

In symbolic dynamics these space sets are called \emph{subshifts of finite
type}. They arise naturally when considering the reinterpretation of the
cellular automaton rule table described above; and, as shall be seen, they are
essential in the work that follows.

One may also consider sets of spaces with other kinds of restrictions. For
example, one might wish to consider the subset of \x{2} in which the length of
a sequence of $a$'s is always even. This particular set of spaces may also be
generated by segments and gluing rules. For example: one may do it using three
segments $x=\seg{a}$, $y=\seg{a}$, and $z=\seg{b}$, where the gluing rules are
that there are no adjacency restrictions except that $x$ must be followed by
$y$, and that there is no overlap. It is easy to see that any space in this
subset of \x{2} may be decomposed into these segments in one and only one
way. However, this decomposition is not local in the sense described
earlier. For given a cell in some state whose color is $a$, in order to tell
whether the segment overlapping this cell in the decomposition is $x$ or $y$
one would need to scan to the left or right to find the first cell whose color
is $b$. This scan might extend arbitrarily far, since a sequence of $a$'s in
this space set may be as long as you like. Hence there is no finite $r$, and
the definition for locality is not satisfied. In fact, there does not exist an
\ang{L,G^L\!,A} that is a faithful description of this set of spaces.

Consider now the set of space sets that may be obtained by requiring the
segments to be of length one, letting the adjacency matrix be arbitrary, and
requiring that the gluing be done without overlap. In symbolic dynamics these
space sets are called \emph{sofic shifts}. The above example is a sofic
shift. In addition, it is easy to show that \emph{any} set of spaces generated
by segments and gluing rules is a sofic shift. (Hence all subshifts of finite
type are sofic shifts.)

It is possible that dynamical systems with laws of evolution that are in some
sense local can be designed for sofic shifts which are not subshifts of finite
type. But these laws would not be local in the sense defined above. In the next
section I will show that a set of spaces that has a faithful description must
be a subshift of finite type. We already know that every subshift of finite
type has a faithful description. Hence the subshifts of finite type are exactly
those sets of oriented one-dimensional spaces that have faithful
descriptions. For this reason, they are the sets of spaces which I will
consider in what follows.

\subsection[Initial attempts to generate 1+1-dimensional
combinatorial\protect\\spacetimes]{Initial attempts to generate
1+1-dimensional\protect\\combinatorial spacetimes}

I did not immediately arrive at the general point of view concerning segments
and gluing rules described above. Initially, I only realized that one could
view the right-hand side of the table as a set of segments that were glued
together by certain rules, and that one could modify the existing setup by
allowing these segments to be of varying lengths (including length zero)
instead of always being length one. I retained the rule that the right-hand
segments were to be glued together by placing them next to one another without
overlap. It was clear that this new setup allowed for change in the width of
the state as the system evolved. So I tried to find an algorithm which, if
allowed to run forever, would generate all invertible rules of this type. I
will call these rules the \emph{nonoverlap rules}.

In the remainder of this section, then, we will consider only rule tables of
the form \ang{L,R,G^L\!,N,A}, where $L$ is a list of all $n$ allowed segments
of width $d$ (each such distinct segment listed once, in lexographical order)
and $A$ and $G^L$ correspond to the $(d-1)$-overlap gluing rules. The symbol
$N$ denotes the nonoverlap gluing rules. The set of spaces \x{\ang{L,G^L\!,A}}
is denoted by \X, and $K$ denotes the set of allowed colors. Given
\ang{L,G^L\!,A}, our task is to find all \ang{R,N,A}'s such that
\ang{L,R,G^L\!,N,A} is a law of evolution. Thus we want to find $R$ such that
\ang{R,N,A} is a faithful description of~\X.

\subsubsection{Existence of a rule-generating algorithm}
\label{revalg}

If one does not concern oneself with invertibility, then there are infinitely
many such $R$'s for each choice of \ang{L,G^L\!,A}. For if \X\ is nonempty then
it is easy to show that it must contain a finite space $(a_1 \ldots
a_m)$. Let $S_j=[a_1 \ldots a_m]$ for every $j$. Then each space
$(S_1\ldots S_r)$ must also be in \X\ for any $r>0$. We may therefore
choose segment $R_i$ to be $[S_1\ldots S_{r_i}]$, where $r_i>0$ for each
$i$. There are infinitely many ways to choose the $R_i$'s in this manner, and
each choice results in a set of segments that generates spaces in \X.

It turns out, however, that this is not the case once one imposes two
conditions. The first is that the directed graph \g{A} associated with the
adjacency matrix $A$ is the union of its strongly connected components. (I will
also say in this case that the matrix $A$ is the union of its strongly
connected components; in general, I will often speak of a graph and of its
adjacency matrix interchangeably.) A \emph{strongly connected component} of a
directed graph is a maximal subgraph such that, for any vertices $i$ and $j$ in
the subgraph, there is a directed path contained in the subgraph that connects
$i$ to $j$. (A directed graph is \emph{strongly connected} when it has one
strongly connected component and it is equal to that component; this property
will be useful later on. The term \emph{irreducible} is also used for this
property in the literature.) The graph \g{A} is the union of its strongly
connected components if and only if every vertex in that graph is contained in
a closed finite path. Thus if this property holds it follows that each $L_i$
(each $R_i$) appears in a decomposition into segments of at least one
finite-width space in \x{\ang{L,G^L\!,A}} (\x{\ang{R,G^R\!,A}}).

The second condition is that the rule is invertible.

\begin{theorem}
For each $L$ whose adjacency matrix $A$ is the union of its strongly
connected components, there are only finitely many invertible nonoverlap rules.
\end{theorem}
\textsc{Proof.} Consider the process of constructing $R$ by adding one
right-hand segment at a time. At any moment in this process we may consider the
spaces generated by this partial set of right-hand segments. Suppose that this
assignment process is not yet complete. Consider an unassigned segment
$r_i$. Since $A$ is the union of its strongly connected components, $r_i$ must
be present in a segment decomposition of some finite-width space. This space
cannot also be generated by the assigned segments, for then the rule would not
be invertible. Hence the set of finite-width spaces not generated by the
assigned segments is nonempty. It follows that there must exist a space of
finite width $w$ that is not generated by the assigned segments, while every
space with width less than $w$ is generated by those segments. Since the rule
is invertible, this space must be decomposable into right-hand segments, each
of which cannot have width greater than $w$ (due to the nonoverlap rule). At
least one of these segments must be unassigned. Thus there must exist at least
one not-yet-assigned segment whose width is less than or equal to $w$. Require
that this segment be the next one assigned. Then there are only finitely many
ways to make this assignment. Since this is true at each stage in the
segment-assigning process, the proof is complete.\proofend\bigskip

This assignment process does not constitute an algorithm for generating
invertible nonoverlap rules unless we add a routine that checks whether
\ang{R,N,A} is a faithful description of \X. Here is such an
algorithm.

Let $R_{i,j}$ indicate the $j$th cell in $R_i$. We may construct an associated
description \ang{R',N,A'} in which each of the cells $R_{i,j}$ now
appears as a length-1 segment in $R'$. The new adjacency matrix $A'$ is
constructed as follows. If $j$ is less than the length of segment $R_i$, then
$R_{i,j}$ is succeeded only by $R_{i,j+1}$. If not, then one uses $A$ to find
all possible chains of segments that succeed $R_i$ such that each segment in
the chain is of length zero except for the terminal segment in the chain; then
$R_{i,j}$ is succeeded by the first cell in each of these terminal segments. An
example of this construction is shown in Figure~\ref{lengthonesegs}.
\begin{figure}
\fbox{\begin{picture}(425,196)\footnotesize\put(0,12){\makebox(425,196)[bl]{\settowidth{\wod}{00}
\hfill \begin{tabular}{c\hw | c\hw c\hw c\hw c\hw c\hw c\hw c\hw c\hw c}
\seg{c}   &1&1&1&0&0&0&0&0&0\\
\seg{cc}  &0&0&0&1&1&1&0&0&0\\
\seg{b}   &0&0&0&0&0&0&1&1&1\\
\seg{a}   &1&1&1&0&0&0&0&0&0\\
\seg{a}   &0&0&0&1&1&1&0&0&0\\
\seg{aca} &0&0&0&0&0&0&1&1&1\\
\seg{\,}  &1&1&1&0&0&0&0&0&0\\
\seg{\,}  &0&0&0&1&1&1&0&0&0\\
\seg{ca}  &0&0&0&0&0&0&1&1&1\\
\end{tabular}
\hfill \begin{tabular}{c\hw |c\hw c\hw c\hw c\hw c\hw c\hw c\hw c\hw c\hw c\hw c}
\seg{c} &1&1&0&1&0&0&0&0&0&0&0\\
\seg{c} &0&0&1&0&0&0&0&0&0&0&0\\
\seg{c} &0&0&0&0&1&1&1&0&0&0&0\\
\seg{b} &1&1&0&1&1&1&1&0&0&1&0\\
\seg{a} &1&1&0&1&0&0&0&0&0&0&0\\
\seg{a} &0&0&0&0&1&1&1&0&0&0&0\\
\seg{a} &0&0&0&0&0&0&0&1&0&0&0\\
\seg{c} &0&0&0&0&0&0&0&0&1&0&0\\
\seg{a} &1&1&0&1&1&1&1&0&0&1&0\\
\seg{c} &0&0&0&0&0&0&0&0&0&0&1\\
\seg{a} &1&1&0&1&1&1&1&0&0&1&0\\
\end{tabular}
\hfill}}\end{picture}}\linespread{1}
\caption[Transforming a set of variable-length segments into a set of
length-one segments]{The set $R$ of nine variable-length segments at the left,
with associated adjacency matrix $A$, is transformed into the set $R'$ of
eleven segments of length one (and new adjacency matrix $A'$) shown at
right. (None of these segments overlap.) The segments on the right are the
cells in the segments at the left, taken in order of appearance. The two sets
of segments generate the same set of spaces. Note that if $L$ is the set of all
width-2 segments containing 3 colors, taken in alphabetical order, and $G^L$ is
the 1-cell-overlap gluing rule, then the adjacency matrix for $L$ is also $A$;
it turns out that \ang{L,R,G^L\!,N,A} is the 1+1-dimensional rule described in
Chapter~\ref{examples}.}
\label{lengthonesegs}
\end{figure}

The description \ang{R',N,A'} is useful not only because it is in fact a
description of \x{\ang{R,N,A}} (which indeed it is) but also because there is a
local invertible map between the closed paths in \g{A} and the closed paths in
\g{A'}. It follows that \ang{R,N,A} is a faithful description of
\x{\ang{R,N,A}} if and only if \ang{R',N,A'} is a faithful description of
\x{\ang{R,N,A}}.

We now focus our attention entirely on \ang{R',N,A'} and attempt to
generate a certain set $P$ of sets of ordered pairs of elements of $R'$, as
follows. (In the remainder of this section I will assume that \g{A}, and
hence \g{A'}, is the union of its strongly connected components.)

Let $S$ be an arbitrary nonempty segment. Then $S$ may be constructed by
assembling the elements of $R'$ using the gluing rules in a finite number of
ways. (This number may be zero.) Let each way of constructing $S$ be
associated with an ordered pair $(i,j)$ such that $R'_i$ is the first
segment used in this particular construction of $S$ and $R'_j$ is the last
segment used in this construction. Let $P_S$ be the set of all such ordered
pairs. We say that the segment $S$ is ``associated'' with the set $P_S$. (If
$P_{S'}=P_S$ then $S'$ is also associated with $P_S$; in other words, more
than one segment can be associated with one of these sets of ordered pairs.)
If $\X\subset\x{k}$ then $P$ is the set of all elements $P_S$ for each
nonempty segment $S$ that occurs in a space in \x{k}.

There are two requirements we have about these sets $P_S$. The first
requirement is that, given any ordered pair $(i,j)$ in $P_S$, there must be
exactly one way to construct $S$ (or any other segment associated with
$P_S$) whose beginning segment is $R'_i$ and whose ending segment is $R'_j$.
For if there were two such ways, then, since \g{A'} is the union of its
strongly connected components, the segment $S$ may be completed to form a
finite space; there will then be at least two distinct ways to construct
that space using \ang{R',N,A'}, which therefore will not be a faithful
description of \x{\ang{R',N,A'}}. The second requirement is that if $|S|>1$
and the first color of $S$ is equal to the last color of $S$, then there
must be at most one element of $P_S$ which is of the form $(i,i)$. For any
such $S$ may be closed up to form a finite space in $\x{k}$ by identifying
its first and last elements. This space may or may not be in $\X$. If
\ang{R',N,A'} is to be a faithful description then there must be at most one
way to construct this space using elements of $R'$, and this will be true if
and only if at most one element of $P_S$ is of the form $(i,i)$. (This
follows since any ordered pair of this form is associated with exactly one
method of constructing $S$. To construct the associated space, one simply
identifies the identical beginning and ending segments in this construction
of $S$.)

Given $P_S$, it is a straightforward process to create $P_{\seg{Sc}}$ for each
$c \in K$. (I will call this process ``extending one cell to the right.'') The
process does not depend on a choice of $S$ (in other words, if $P_S=P_{S'}$
then $P_{\seg{Sc}}=P_{\seg{S'c}}$). All information needed to carry out the
construction is present in $P_S$ itself and in the gluing rules. Furthermore,
it is possible at each stage to insure that the two requirements described
above are met. The second requirement is easy to verify: one simply examines
the resulting $P_{\seg{Sc}}$ to see whether or not it contains two ordered
pairs of the form $(i,i)$. The first requirement cannot be checked by examining
the resulting $P_{\seg{Sc}}$; it must be checked during the process of creating
it. It fails if and only if there are elements $(w,x)$ and $(w,y)$ in $P_S$
with $x\neq y$ such that, according to the adjacency matrix $A'$, both $R'_x$
and $R'_y$ may be followed by some segment $R'_z$. If either condition fails,
we say that an error has occurred.

The process of constructing the set $P$ of all such elements $P_S$ is
initiated by creating an element $P_{\seg{c}}$ for each $c \in K$. One then
extends to the right each element $P_S$ which has been constructed but not
yet extended until an error occurs or until one cannot do this anymore. The
process is bound to terminate since there are only finitely many possible
sets of ordered pairs of elements of $R'$. It is easy to see that, in the
case of termination without error, the set of elements $P_S$ that we have
constructed must be complete: every nonempty segment $S$ will be associated
with one of the constructed sets of ordered pairs. An example of a
successfully constructed set $P$ is given in Figure~\ref{p}.
\begin{figure}
\fbox{\begin{picture}(425,202)\footnotesize\settowidth{\wod}{$\{$}
\put(0,12){\makebox(425,202)[bl]{
\hfill\begin{tabular}{rl}
\seg{a}&$\{(5,5),(6,6),(7,7),(9,9),(11,11)\}$\\
\seg{c}&$\{(1,1),(2,2),(3,3),(8,8),(10,10)\}$\\
\seg{ac}&$\{(5,1),(5,2),(7,8),(9,1),(9,2),(9,10),(11,1),(11,2),(11,10)\}$\\
\seg{aca}&$\{(7,9),(9,11),(11,11)\}$\\
\seg{caca}&$\{(3,9),(8,11),(10,11)\}$\\
\seg{ca}&$\{(3,5),(3,6),(3,7),(8,9),(10,11)\}$\\
\seg{cc}&$\{(1,1),(1,2),(2,3)\}$\\
\seg{cac}&$\{(3,1),(3,2),(3,8),(8,1),(8,2),(8,10),(10,1),(10,2),(10,10)\}$\bigskip\\
$B$&$\{\{5,9,11\},\{6,9,11\},\{7,9,11\},\{4\},\{1\},\{2\},\{3,8,10\}\}\bigskip$\\
$E$&$\{\{5,6,7\},\{9\},\{11\},\{4\},\{1,2,3\},\{1,2,8\},\{1,2,10\}\}$\\
\end{tabular}
\hfill}}\end{picture}}\linespread{1}
\caption[The set $P$ for the example in Figure~\ref{lengthonesegs}]{The set $P$
for the example in Figure~\ref{lengthonesegs}. The elements of $P$ which are
not products are each associated with a finite number of segments; these
segments are listed. Those that are products are all elements of the form $b
\times e$, $b \in B$, $e \in E$. Each of these is associated with infinitely
many segments. The fact that \emph{every} product $b \times e$ is an element of
$P$ results from the fact that \g{A'} in this case is strongly connected; this
need not be the case if \g{A'} is merely the union of its strongly connected
components. Note that if $b \in B$ and $e \in E$ then $b \cap e$ contains at
most one element.}
\label{p}
\end{figure}

When the process terminates due to an error, we know that \ang{R',N,A'} is not
a faithful description. It turns out that the converse is also true: if the
process terminates without error, then \ang{R',N,A'} must indeed be a faithful
description. This, however, is by no means obvious. What is obvious is that, if
the process terminates without error, then no \emph{finite} space is
constructible in more than one way; this is guaranteed by the first
condition. But a further argument is required in order to prove that the same
holds for infinite spaces, and that the decomposition of spaces into segments
is a local process.

To proceed, it is necessary to consider two directed graphs which are
associated with $P$; I will call these the \emph{right graph} and the
\emph{left graph}. In each case the vertices of the graph are the elements
of $P$. The edges are labeled with the elements of $K$. One edge of each
color points outward from each vertex. In the right graph, the edges
correspond to the operation of extending to the right: for any nonempty $S$
and for each $c \in K$, there is a $c$-colored edge pointing from $P_S$ to
$P_{\seg{Sc}}$. In the left graph, the edges correspond to the operation of
extending to the left.

Let $S$ be an arbitrary nonempty segment, and $c$ be an arbitrary color. Let
$B_{P_S}$ be the set of all first elements in the ordered pairs in $P_S$,
and let $E_{P_S}$ be the set of all second elements in those ordered pairs.
Then it is easy to see that $B_{P_{[\hspace{-.5pt}S\hspace{-.5pt}c]}}
\subset B_{P_S}$, and that $E_{P_{[cS\hspace{.5pt}]}} \subset E_{P_S}$. In
other words, as you travel in the direction of the edges along the right
graph, the sets $B_{P_S}$ form a nested, nonincreasing chain, and the same
is true of the sets $E_{P_S}$ as you travel in the direction of the edges
along the left graph. It follows that, as you travel along any circuit in
the right graph, $B_{P_S}$ must remain constant; the same must hold for
$E_{P_S}$ in the left graph.

I will say that $P_S$ is a \emph{product} whenever $P_S=B_{P_S} \times
E_{P_S}$.

\begin{theorem}
  Suppose that the algorithm to generate $P$ terminates without error. Let
  $S$ be a segment that begins and ends with the same color, with $|S|>1$.
  Let $S^m$ denote the segment formed by joining $m$ consecutive copies of
  $S$ together with a one-cell overlap. If there exists an $m>1$ such that
  $P_{S^m}=P_S$, then $P_S$ is a product.
\end{theorem}
\textsc{Proof.} First note that $(p,q)$ is in $P_{S^{k+1}}$ if and only if
there exists an $i$ such that $(p,i)$ is in $P_{S^k}$ and $(i,q)$ is in
$P_S$. It follows inductively that $(p,q)$ is in $P_{S^k}$ if and only if there
exist $i_0 \ldots i_k$ such that $p=i_0$, $q=i_k$, and $(i_j,i_{j+1}) \in P_S$
for each $j$, $0 \leq j<k$.

Now consider any $(p,q) \in P_S$. Since $P_S=P_{S^m}$, it follows that
$P_S=P_{S^{1+r(m-1)}}$ for each $r\in\Z^+$. Hence for each such $r$ there is a
sequence $Y_r=(i^r_0, \ldots, i^r_{1+r(m-1)})$ such that $p=i^r_0$,
$q=i^r_{1+r(m-1)}$, and $(i^r_j,i^r_{j+1}) \in P_S$ for each $j$, $0 \leq
j<1+r(m-1)$. (The superscripts of $i$ denote indices, not exponents.)

Since the number of segments in $R'$ is finite, there must exist
$r_0\in\Z^+$, $s\in\Z^+$ and $t\in\N$ such that $i_s^{r_0}=i_{s+t}^{r_0}$.
Since $(i^r_j,i^r_{j+1})$ is in $P_S$ for any $j$, it follows that
$(i^r_j,i^r_{j+k})$ is in $P_{S^k}$ for any $j$ and $k$. Hence
$(i^{r_0}_s,i^{r_0}_s)$ is in $P_{S^t}$. Let $h=i^{r_0}_s$. Then there must
be a sequence $(m_0,\ldots,m_t)$ such that $m_0=m_t=h$ and $(m_j,m_{j+1})$
is in $P_S$ for each $j$. The sequence
$(n_0,\ldots,n_t)=(m_1,\ldots,m_t,m_0,m_1)$ also has the property that
$(n_j,n_{j+1})$ is in $P_S$ for each $j$. So $(m_1,m_1)$ is in $P_{S^t}$.
Since $|S|>1$ and therefore $|S^t|>1$, there can be at most one element in
$P_{S^t}$ of the form $(i,i)$. It follows that $m_1=h$. So $(m_0,m_1)=(h,h)$
is \mbox{in $P_S$}.

From the above it follows that $(p,h)$ is in $P_{S^s}$. But since
$(h,h)$ is in $P_S$, it must also be true that $(p,h)$ is in $P_{S^{s+j}}$ for
any nonnegative $j$. Hence $(p,h)$ must be in $P_{S^{1+r(m-1)}}$ for large
enough $r$. But this is the same as $P_S$. So $(p,h)$ is in $P_S$. Similarly,
$(h,q)$ must be in $P_S$.

Next consider any $(p,q)$ and $(s,t) \in P_S$. Since $(s,h) \in P_S$ and $(h,h)
\in P_S$ and $(h,q) \in P_S$, it is easy to verify that $(s,q) \in P_{S^k}$ for
any $k>1$. Hence $(s,q) \in P_{S^m}$. But $P_{S^m}=P_S$. So $(s,q) \in
P_S$. Similarly, $(p,t) \in P_S$. Hence $P_S$ is a product.\proofend

\begin{theorem}
  Suppose that the algorithm to generate $P$ terminates without error. If
  $S$ is a segment whose first and last cells have the same color and
  $|S|>1$, then there exists a $k>0$ such that $P_{S^k}$ is a product.
\end{theorem}
\textsc{Proof.} Consider $P_{S^i}$ for all $i$. Since there are only finitely
many elements in $P$, there must exist a positive integer $k$ such that
$P_{S^k}=P_{S^{k+m}}$ for some $m>0$. Hence $P_{S^k}=P_{S^{k+qm}}$ for any
nonnegative $q$. So $P_{S^k}=P_{S^{k+km}}=P_{S^{k(1+m)}}=P_{(S^k)^{1+m}}$, and
the result follows because $S^k$ meets the conditions of the previous
theorem.\proofend

\begin{theorem}
  Suppose that the algorithm to generate $P$ terminates without error. Then
  $P_S$ is in a circuit in the right graph (or left graph) if and only if it
  is a product.
\end{theorem}
\textsc{Proof.} In what follows I will use the rule that \seg{ST} means that
$S$ and $T$ overlap by one cell, rather than the usual rule that they do not
overlap.

Suppose that $P_S$ is a product and that $(p,q)$ is in $P_S$. Then $(i,q)$ and
$(p,j)$ are also in $P_S$ for each $i\in B_{P_S}$ and each $j\in
E_{P_S}$. Since \g{A'} is the union of its strongly connected components, it
contains a directed path from the $q$th vertex to the $p$th vertex. Hence there
exists a segment $T$ which begins with the color of $R'_q$ and ends with the
color of $R'_p$ such that $P_{\seg{ST}}$ contains $(i,p)$ for each $i\in
B_{P_S}$. It follows that $P_{\seg{STS}}$ contains $(i,j)$ for each $i\in
B_{P_S}$ and for each $j\in E_{P_S}$; i.e., it contains $B_{P_S} \times
E_{P_S}$. Furthermore, $P_{\seg{STS}}$ is obtained from $P_S$ by extending to
the right, and it is also obtained from $P_S$ by extending to the left. Hence
$B_{P_{[\hspace{-.5pt}ST\!\hspace{.3pt}S\hspace{.5pt}]}} \subset B_{P_S}$ and
$E_{P_{[\hspace{-.5pt}ST\!\hspace{.3pt}S\hspace{.5pt}]}} \subset E_{P_S}$. It
follows that $P_{\seg{STS}}$ is contained in $B_{P_S} \times E_{P_S}$. Hence
$P_S=P_{\seg{STS}}$. Thus $P_S$ is in a circuit in the right graph and in the
left graph.

Now suppose that $P_S$ is in a circuit in the right graph. Then
$P_S=P_{\seg{ST}}$ for some segment $T$ that begins and ends with the same
color, with $|T|>1$. It follows also that $P_S=P_{[ST^m]}$ for any $m>0$. By
the previous theorem, there exists a $k>0$ such that $P_{T^k}$ is a product.
Consider any element $(p,q)$ in $P_S$ such that $q$ is in $B_{P_{T^k}}$.
Then $(p,i)$ is in $P_{[ST^k]}$ (and hence in $P_S$) for each $i\in
E_{P_{T^k}}$. Since this is true for each such $(p,q)\in P_S$, and since
there are no elements in $P_{[ST^k]}$ (and hence in $P_S$) except for those
that are generated in this way, it follows that $P_S$ must be a
product.\proofend\bigskip

Let $B$ be the set of all $B_{P_S}$ such that $P_S$ is a product. Let $E$ be
the set of all $E_{P_S}$ such that $P_S$ is a product.

\begin{theorem} Suppose that the algorithm to generate $P$ terminates without
error. Then the intersection of an element of $B$ with an element of $E$ must
contain less than two elements.
\end{theorem}
\textsc{Proof.} Suppose that $P_S$ and $P_T$ are products, and that $i$ and $j$
are distinct elements of $B_{P_S} \cap E_{P_T}$. Then $P_S$ and $P_T$ are
nonempty, so there exists $p\in B_{P_T}$ and $q\in E_{P_S}$. Let \seg{TS} be
the segment where $T$ and $S$ overlap by one cell (this makes sense since the
cell in common must have the same color). Then there are at least two distinct
ways to construct \seg{TS} using elements of $R'$ that begin with $R'_p$ and
end with $R'_q$ (one way passes through $R'_i$ and the other through
$R'_j$). This cannot be, since it violates the first requirement for
$P_{\seg{TS}}$.\proofend

\begin{theorem} Suppose that the algorithm to generate $P$ terminates without
error. Then \ang{R',N,A'} is a faithful description.
\end{theorem}
\textsc{Proof.} Let $S$ be a length-1 segment. Consider the set of segments
$T$ such that $B_{P_{\seg{ST}}}\in B$ but $B_{P_{\seg{ST'}}}\notin B$ if
$T=\seg{T'c}$ for some $c$. This set is finite since for long enough $T$
$P_{\seg{ST}}$ is a product. Let $t$ be the maximum number of cells in any
such segment $T$, where the maximum is taken over all possible $S$. Similarly,
consider the set of segments $U$ such that $E_{P_{\seg{US}}}\in E$ but
$E_{P_{\seg{U'S}}}\notin E$ if $U=\seg{cU'}$ for some $c$, and let $u$ be the
maximum number of cells in any such segment $U$.

Now suppose we are examining a particular cell in a space, and considering the
decomposition of this space into elements of $R'$. This cell must be the
leftmost cell in one of the segments \seg{ST} described above; hence, in the
decomposition of the space into segments, the cell must be represented by an
element of $B_{P_{[\hspace{-.5pt}ST]}}$. The cell must also be the rightmost
cell in one of the segments \seg{US} described above; hence, in the
decomposition of the space into segments, the cell must be represented by an
element of $E_{P_{[U\!S\hspace{.5pt}]}}$. But $B_{P_{[\hspace{-.5pt}ST]}}
\in B$ and $E_{P_{[U\!S\hspace{.5pt}]}} \in E$; hence they intersect in at most
one element. If they do not intersect, then the space in question is not in
\x{\ang{R',N,A'}}. If they intersect in exactly one element $i$, then $R_i$ is
the segment which must be assigned to the cell in the decomposition. Hence
there is only one way to decompose the space into segments, and this
decomposition may be carried out in a local manner where the neighborhood
examined has size $t+u+1$.\proofend\bigskip

The segments \seg{UST}, as described in the above proof, determine which
element of $R'$ is assigned to the length-1 segment $S$ in the
decomposition. I call the set of these segments (for all such $U$, $S$ and
$T$), with the segment $S$ identified in each and assigned a segment $R'_i$,
a set of \emph{forcing strings} for \ang{R',N,A'}. They contain all the
information needed to compute the decomposition of a space. An example of a
set of forcing strings is given in Figure~\ref{force}.
\begin{figure}
\fbox{\begin{picture}(425,301)\footnotesize
\put(0,12){\makebox(425,301)[bl]{
\hfill\begin{tabular}[t]{rl}
\multicolumn{2}{c}{\seg{ST}}\\
\hline
$\{R'_6,R'_9,R'_{11}\}$&\textit{aa}\\
$\{R'_5,R'_9,R'_{11}\}$&\textit{ab}\\
$\{R'_7,R'_9,R'_{11}\}$&\textit{aca}\\
$\{R'_5,R'_9,R'_{11}\}$&\textit{acb}\\
$\{R'_5,R'_9,R'_{11}\}$&\textit{acc}\\
$\{R'_4\}$&\textit{b}\\
$\{R'_3,R'_8,R'_{10}\}$&\textit{ca}\\
$\{R'_1\}$&\textit{cb}\\
$\{R'_2\}$&\textit{cca}\\
$\{R'_1\}$&\textit{ccb}\\
$\{R'_1\}$&\textit{ccc}\\
\end{tabular}
\hfill\begin{tabular}[t]{rl}
\multicolumn{2}{c}{\seg{US}}\\
\hline
\textit{aa}&$\{R'_5,R'_6,R'_7\}$\\
\textit{ba}&$\{R'_5,R'_6,R'_7\}$\\
\textit{aaca}&$\{R'_9\}$\\
\textit{baca}&$\{R'_9\}$\\
\textit{acaca}&$\{R'_{11}\}$\\
\textit{bcaca}&$\{R'_{11}\}$\\
\textit{ccaca}&$\{R'_9\}$\\
\textit{bca}&$\{R'_{11}\}$\\
\textit{cca}&$\{R'_5,R'_6,R'_7\}$\\
\textit{b}&$\{R'_4\}$\\
\textit{aac}&$\{R'_1,R'_2,R'_8\}$\\
\textit{bac}&$\{R'_1,R'_2,R'_8\}$\\
\textit{acac}&$\{R'_1,R'_2,R'_{10}\}$\\
\textit{bcac}&$\{R'_1,R'_2,R'_{10}\}$\\
\textit{ccac}&$\{R'_1,R'_2,R'_8\}$\\
\textit{bc}&$\{R'_1,R'_2,R'_{10}\}$\\
\textit{cc}&$\{R'_1,R'_2,R'_3\}$\\
\end{tabular}
\hfill\textit{\begin{tabular}[t]{r@{\hspace{0 pt}}c@{\hspace{0 pt}}l}
\multicolumn{3}{c}{$R'_1$}\\
\hline
&\underline{c}&b\\
&\underline{c}&cb\\
&\underline{c}&cc\bigskip\\
\multicolumn{3}{c}{$R'_2$}\\
\hline
&\underline{c}&ca\bigskip\\
\multicolumn{3}{c}{$R'_5$}\\
\hline
a&\underline{a}&b\\
a&\underline{a}&cb\\
a&\underline{a}&cc\\
b&\underline{a}&b\\
b&\underline{a}&cb\\
b&\underline{a}&cc\\
cc&\underline{a}&b\\
cc&\underline{a}&cb\\
cc&\underline{a}&cc\\
\end{tabular}\hfill\begin{tabular}[t]{r@{\hspace{0 pt}}c@{\hspace{0 pt}}l}
\multicolumn{3}{c}{$R'_3$}\\
\hline
c&\underline{c}&a\bigskip\\
\multicolumn{3}{c}{$R'_6$}\\
\hline
a&\underline{a}&a\\
b&\underline{a}&a\\
cc&\underline{a}&a\bigskip\\
\multicolumn{3}{c}{$R'_7$}\\
\hline
a&\underline{a}&ca\\
b&\underline{a}&ca\\
cc&\underline{a}&ca\bigskip\\
\multicolumn{3}{c}{$R'_8$}\\
\hline
aa&\underline{c}&a\\
ba&\underline{c}&a\\
cca&\underline{c}&a\\
\end{tabular}\hfill\begin{tabular}[t]{r@{\hspace{0 pt}}c@{\hspace{0 pt}}l}
\multicolumn{3}{c}{$R'_4$}\\
\hline
&\underline{b}&\bigskip\\
\multicolumn{3}{c}{$R'_9$}\\
\hline
aac&\underline{a}&\\
bac&\underline{a}&\\
ccac&\underline{a}&\bigskip\\
\multicolumn{3}{c}{$R'_{10}$}\\
\hline
aca&\underline{c}&a\\
bca&\underline{c}&a\\
b&\underline{c}&a\bigskip\\
\multicolumn{3}{c}{$R'_{11}$}\\
\hline
acac&\underline{a}&\\
bcac&\underline{a}&\\
bc&\underline{a}&\\
\end{tabular}}\hfill}}\end{picture}}\linespread{1}
\caption[Forcing strings for the example in Figure~\ref{lengthonesegs}]{Forcing
  strings for the example in Figure~\ref{lengthonesegs}. Here the segments
  \seg{ST} and \seg{US} are given, along with a list of the ways in which
  they restrict the element of $R'$ assigned to $S$. Then a set of forcing
  strings is listed. There is one set of forcing strings for each element of
  $R'$. The cell which is forced to be that element of $R'$ is underlined.
  Instead of listing the set of forcing strings \seg{UST}, I have chosen
  here to list a smaller set of forcing strings. Smaller sets such as this
  can often be constructed. (In general a set of forcing strings can be
  defined to be a set of segments, each having an identified cell assigned
  some $R'_i$, with the property that, for any cell in an allowed space, a
  segment in the set can be embedded in the space so that the identified
  cell of the segment maps to the chosen cell in the space.) For example:
  since here $R'_2$ is associated only with the single \seg{ST} segment
  \seg{cca}, all you need to force \seg{c} to be $R'_2$ is a $ca$ after it;
  you don't have to list all seven strings \seg{Ucca}. This is the unique
  smallest set of forcing strings for this example; in the general case,
  however, it is not always possible to find such a unique smallest set.}
\label{force}
\end{figure}

\begin{theorem}
  If \hspace{2 pt}\g{A'} is the union of its strongly connected components,
  then \ang{R',N,A'} is a faithful description if and only if there is a
  unique way to decompose each finite space in \x{\ang{R',N,A'}} into
  segments.
\end{theorem}
\textsc{Proof.} The forward implication is trivial. Suppose that there is a
unique way to decompose each finite space in \x{\ang{R',N,A'}} into
segments, and we attempt to generate $P$. We have already shown that
violation of the first requirement means there are two ways to construct a
finite space, so this cannot happen. If the second requirement is violated,
then there are two ways to construct some segment $S$ both of which begin
with the same element $R'_i$ and end with the same element $R'_j$. Since
\g{A'} is the union of its strongly connected components, there exists a
segment $T$ which may be constructed by beginning with $R'_j$ and ending
with $R'_i$. Let \cir{ST} denotes the space where $S$ and $T$ overlap by one
cell at each end. This space clearly may be constructed in two different
ways. Hence this requirement may not be violated either; so the algorithm to
compute $P$ must terminate without error, which means that \x{\ang{R',N,A'}}
must indeed be a faithful description.\proofend

\begin{theorem}
  If \ang{R',N,A'} is a faithful description, then \x{\ang{R',N,A'}} is a
  subshift of finite type.
\end{theorem}
\textsc{Proof.} Consider the set of forcing strings described above, where
the maximal length of $T$ is $t$ and the maximal length of $U$ is $u$. It is
a straightforward procedure to construct all allowed segments of length
$t+u+2$ that are generated by \ang{R',N,A'}. My claim is that
\x{\ang{R',N,A'}} is exactly that subset $\Y$ of \x{k} obtained by requiring
each $(t+u+2)$-neighborhood to be one of these segments. Surely
$\x{\ang{R',N,A'}}\subseteq\Y$. Let $x$ be any space in \Y. Any
length-$(t+u+1)$ neighborhood in $x$ contains a forcing string which
determines that, in any decomposition of $x$ into segments, the $(u+1)$st
cell of this neighborhood must be represented by some particular segment
$R'_i$. Since any cell in $x$ is the $(u+1)$st cell in an
$(t+u+1)$-neighborhood, its representation in the decomposition is
determined. Thus there is at most one way to decompose $x$ into segments.
The question remaining is whether the decomposition determined by the
$(t+u+1)$-neighborhoods is a valid one: if $R'_i$ and $R'_j$ are assigned to
neighboring cells, is it allowed for $R'_j$ to follow $R'_i$?  The answer is
obtained by examining the $(t+u+2)$-neighborhood containing the cell
assigned to $R'_i$ as its $(u+1)$st cell and the cell assigned to $R'_j$ as
its $(u+2)$nd cell. This neighborhood is constructible using the description
\ang{R',N,A'}. But it contains the two $(t+u+1)$-neighborhoods which fix the
representations of its $(u+1)$st cell and $(u+2)$nd cell to be $R'_i$ and
$R'_j$, respectively. Thus it must be allowed for $R'_j$ to follow $R'_i$.
So the decomposition is a valid one; hence $x\in\x{\ang{R',N,A'}}$. So
$\Y\subseteq\x{\ang{R',N,A'}}$. So $\x{\ang{R',N,A'}}=\Y$, which is a
subshift of finite~type.\proofend\bigskip

The proof of the above theorem provides the final tool needed to complete our
algorithm for generating all rules \ang{L,R,G^L\!,N,A} given a faithful
description \ang{L,G^L\!,A}. The algorithm proceeds by generating each
candidate for $R$ in turn. For each candidate, we run the algorithm to generate
$P$. If it terminates without error, then \ang{R,N,A} is a faithful
description. Furthermore, we know that the set of spaces \x{\ang{R,N,A}} is
given by a list of its allowed $(t+u+2)$-neighborhoods, as described above;
hence it is determined by a list of its allowed $m$-neighborhoods for any
$m\geq t+u+2$. Similarly, the set of spaces \x{\ang{L,G^L\!,A}} is determined
by a list of its allowed $m$-neighborhoods for any $m\geq d$. We therefore
compute the set of allowed $m$-neighborhoods for \x{\ang{L,G^L\!,A}} and for
\x{\ang{R,N,A}}, where $m=\mathrm{max}\hspace{1pt}\{t+u+2,d\}$. If these sets
are the same, then $\x{\ang{L,G^L\!,A}}=\x{\ang{R,N,A}}$; in this case
\ang{L,R,G^L\!,N,A} is a valid rule.

\subsubsection{Standard forms and the shift}
\label{shift}

Consider a no-overlap rule described by a table whose left-hand segments have
width $d$. We may easily describe the same rule using a table whose left-hand
segments have width $e$ for any $e>d$. This can be done as follows: if $L_i$
and $R_i$ refer to the segments in the original table, $L'_j$ is a left-hand
segment in the new table, and the subsegment of $L'_j$ consisting of its first
$d$ cells is $L_i$, let $R'_j=R_i$.

This is one example of the fact that there are many ways to describe the same
rule. In this case one may easily find the smallest possible $d$, and express
the rule using that value of $d$. This is sufficient to give a canonical form
in the case of cellular automata. However, once one dispenses with coordinates,
it is no longer sufficient; certain tables which previously described different
rules now describe the same rule. When in addition one allows right-hand
segments to have lengths different from one, one obtains a greater freedom of
description which results in the fact that some rules may be described in a
great many ways for a fixed value of $d$.

For example, consider the following four tables.\bigskip\\
{\textit{\footnotesize\hspace*{\fill}\begin{tabular}{r|l}
aa&a\\
ab&a\\
ba&b\\
bb&b\\
\end{tabular}\hfill\begin{tabular}{r|l}
aa&a\\
ab&\textup{-}\\
ba&ba\\
bb&b\\
\end{tabular}\hfill\begin{tabular}{r|l}
aa&a\\
ab&ab\\
ba&\textup{-}\\
bb&b\\
\end{tabular}\hfill\begin{tabular}{r|l}
aa&a\\
ab&b\\
ba&a\\
bb&b\\
\end{tabular}\hspace*{\fill}}}\bigskip\\
The first and last of these tables are traditional cellular automaton tables
which represent (slightly) different laws. Given the convention that the
neighborhood of a cell begins at that cell and extends to the right, the first
table represents the identity law, and the last represents the law in which the
colors of the cells are shifted one unit to the left. Once we dispense with
coordinates, however, the difference between these two laws disappears.
Clearly, this is a good thing. Our choice of convention regarding the position
of the neighborhood is arbitrary, and merely determines where we write down the
state at time $t+1$ relative to the state at time $t$. The choice is a
representational one; it has no effect on the intrinsic behavior of the
cellular automaton. This would not be the case if the laws of evolution being
studied depended on the coordinates. But they do not; they are translation
invariant. By removing the coordinates, we have removed the need for this
convention.

The right-hand columns of the second and third tables contain segments of
length zero, one and two. (A dash stands for the empty segment.) Hence these
are not traditional cellular-automaton tables. Nevertheless, it is easy to
verify that these two tables also represent the identity law. Additional ways
of representing this same law are given in Figure~\ref{identity}, which
displays all 41 ways to describe the identity law on \x{2} when $d=3$. For a
given value of $d$ the identity law seems to have more representations than
other laws, but other laws do often have multiple representations.
\begin{figure}
\fbox{\begin{picture}(425,423)\footnotesize\put(0,12){\makebox(425,423)[bl]{%
\hfill\itshape\begin{tabular}{r||c|c|c|c|c|c|c|c|c|c|c|c|c|c|}
aaa&a&a &a &a &a &a &a &a &a &a &a &a  &a &a \\
aab&a&\textup{-} &aa&a &a &a &\textup{-} &\textup{-} &aa&aa&a &\textup{-}  &a &a \\
aba&a&a &\textup{-} &ab&a &\textup{-} &ab&a &b &\textup{-} &ab&\textup{-}  &b &\textup{-} \\
abb&a&a &\textup{-} &a &ab&\textup{-} &a &ab&\textup{-} &b &ab&\textup{-}  &\textup{-} &b \\
baa&b&ba&b &\textup{-} &b &ba&a &ba&\textup{-} &b &\textup{-} &baa&a &ba\\
bab&b&b &ba&\textup{-} &b &ba&\textup{-} &b &a &ba&\textup{-} &ba &a &ba\\
bba&b&b &b &bb&\textup{-} &b &bb&\textup{-} &bb&\textup{-} &b &b  &bb&\textup{-} \\
bbb&b&b &b &b &b &b &b &b &b &b &b &b  &b &b \bigskip\\

aaa&a &a &a  &a &a  &a  &a&a  &a  &a &a  &a &a &a \\
aab&\textup{-} &aa&a  &\textup{-} &\textup{-}  &a  &a&\textup{-}  &aab&aa&\textup{-}  &\textup{-} &a &ab\\
aba&ab&b &ab &b &\textup{-}  &ba &b&ab &\textup{-}  &b &ba &b &ba&\textup{-} \\
abb&ab&b &abb&\textup{-} &b  &\textup{-}  &b&abb&\textup{-}  &bb&\textup{-}  &b &b &\textup{-} \\
baa&a &\textup{-} &\textup{-}  &aa&baa&\textup{-}  &a&a  &\textup{-}  &\textup{-} &a  &aa&\textup{-} &a \\
bab&\textup{-} &a &\textup{-}  &a &ba &\textup{-}  &a&\textup{-}  &ab &a &\textup{-}  &a &\textup{-} &ab\\
bba&b &b &\textup{-}  &bb&\textup{-}  &bba&b&\textup{-}  &b  &\textup{-} &bba&b &ba&b \\
bbb&b &b &b  &b &b  &b  &b&b  &b  &b &b  &b &b &b \bigskip\\

aaa&a &a  &a &a &a &a &a &a &a &a &a &a &a\\
aab&a &aab&\textup{-} &b &\textup{-} &ab&a &ab&b &\textup{-} &b &ab&b\\
aba&b &\textup{-}  &ba&\textup{-} &b &a &ba&\textup{-} &a &ba&\textup{-} &a &a\\
abb&bb&b  &b &\textup{-} &bb&\textup{-} &bb&b &\textup{-} &bb&b &b &b\\
baa&a &\textup{-}  &a &aa&aa&\textup{-} &\textup{-} &a &a &a &aa&\textup{-} &a\\
bab&a &ab &\textup{-} &ab&a &b &\textup{-} &ab&b &\textup{-} &ab&b &b\\
bba&\textup{-} &\textup{-}  &ba&b &\textup{-} &ba&a &\textup{-} &ba&a &\textup{-} &a &a\\
bbb&b &b  &b &b &b &b &b &b &b &b &b &b &b\\
\end{tabular}\hfill}}\end{picture}}\linespread{1}
\caption[The 41 nonoverlap descriptions of the identity law on \x{2} when the
width of the left-hand segments is three]{The 41 descriptions of the identity
law on \x{2} when the width of the left-hand segments is three. The left-hand
segments are shown to the left of the double line; the remaining columns
represent right-hand segments. These descriptions are obtainable from one
another by shift operations.}
\label{identity}
\end{figure}

It turns out that all of these equivalent descriptions are related by a simple
transformation and its inverse. Suppose that we are given a law table and we
choose a length-$(d-1)$ segment $S$. Let $R^1$ be the set of right-hand segments
whose associated left-hand segments end with $S$, and let $R^2$ be the set of
right-hand segments whose associated left-hand segments begin with $S$. Then
$R^1$ and $R^2$ are two subsets of the set of right-hand segments such that
every segment in $R^1$ must be followed by some segment in $R^2$ and every
segment in $R^2$ must be preceded by some segment in $R^1$. If every segment in
$R^1$ ends in a cell with color $j$, then we may delete these cells from the
segments in $R^1$ and then add a cell with color $j$ to the beginning of each
segment in $R^2$. The resulting table represents the same law as the original
one. Similarly, if every segment in $R^2$ begins with a cell with color $j$,
then we may delete these cells from the segments in $R^2$ and then add a cell
with color $j$ to the end of each segment in $R^1$; the law remains
unchanged. I call these two operations the \emph{left shift} and \emph{right
shift} operations.

Consider the first of the four tables listed above. In this table, $R_1$ is
\seg{a}. The segments that can come after $R_1$ are $R_1$ and $R_2$; these are
both \seg{a}. Hence the presence of $R_1$ in the decomposition of a space
forces the presence of \seg{aa} in that space, where the first $a$ is
associated with $R_1$ itself and the second $a$ is associated with the segment
that follows $R_1$. Note that this is the largest contiguous segment which is
determined in this way; if we look at the second segment past $R_1$, this can
be anything, hence the corresponding cell may be $a$ or $b$; and the segment
before $R_1$ may be $R_1$ or $R_3$, which again corresponds to a cell which is
either $a$ or $b$. I will say that \seg{aa} is the segment that is
\emph{forced} by $R_1$.

This definition is problematical if an infinite number of cells are forced by
$R_i$, since segments are by definition finite. If \g{A} is the union of its
strongly connected components, then such a situation can only arise if the
strongly connected component containing the $i$th vertex consists of a single
circuit. I will call such a component \emph{trivial}. In this section, I will
assume that \g{A} is the union of its strongly connected components, and that
none of these components is trivial.

Given a nonoverlap representation \ang{L,R,G^L\!,N,A} of a rule, we
may now construct another representation \ang{L,R',G^L\!,G^{R'}\!\!,A} of
the same rule by letting $R'_i$ be the segment forced by $R_i$ and letting the
gluing rules include the proper amount of overlap. For example, in the table we
were just discussing each new segment $R'_i$ turns out to be exactly the same
as $L_i$; the gluing rule becomes a 1-cell overlap in each case. I will call
the representation obtained in this way the \emph{forcing segment
representation}.

Notice that, if we apply the same procedure to each of the remaining three
tables in our list, the same new representation results. This is an example of
the following easily verified fact: the left shift and right shift operations
do not alter the associated forcing segment representation. Thus by using the
forcing segment representation we may remove the problem of multiplicity of
representation caused by left and right shifts. (An additional advantage of
this representation in the case of the identity law is that both sides of the
table are identical, which makes it obvious that the law represented is the
identity law.)

It is also possible to verify the following facts, though I will not do so
here:
\begin{itemize}
\item Suppose that $R'_i$ is glued to $R'_j$ with an $m$-cell overlap. Then if
$R'_i$ is followed by $R'_k$ they will be glued with an $m$-cell overlap; and
if $R'_j$ is preceded by $R'_h$ they will also be glued with an $m$-cell
overlap. If $L_i$ ends with a $(d-1)$-cell segment $S$, then $m$ is the largest
number such that, for all $j$ such that $L_j$ begins with $S$, the first $m$
cells of $R'_j$ are the same. Similarly, if $L_j$ begins with a $(d-1)$-cell
segment $S$, then $m$ is the largest number such that, for all $i$ such that
$L_i$ ends with $S$, the last $m$ cells of $R'_i$ are the same.
\item If one begins with any nonoverlap representation \ang{L,R,G^L\!,N,A} and
performs left shifts repeatedly until this can no longer be done, one ends up
with a representation which may be obtained from the associated forcing segment
representation by chopping off the right-hand portions of each segment $R_i$
that overlap with segments that follow it. An analogous result holds if one
performs right shifts repeatedly until this can no longer be done. The process
of performing repeated left shifts (or right shifts) is guaranteed to terminate
so long as our assumption about \g{A} holds.
\end{itemize}
It follows that another way to remove the problem of multiplicity of
representation caused by left and right shifts is to use the (unique)
representation obtained by shifting left (or right) repeatedly until this can
no longer be done. I will call these the \emph{left representation} and the
\emph{right representation}.

Illustrations of these various canonical forms and of their relation to one another
are provided in Figure~\ref{shiftcanon}.
\begin{figure}
\fbox{\begin{picture}(425,270)\footnotesize\put(0,0){\makebox(425,270){%
\hfill\textit{\begin{tabular}{c||c|c|c}
aa&c&c&c\\
ab&cc\underline{a}&cc&cca\\
ac&b&b&b\\
ba&$\overline{\mathit{a}}$&a&\textup{-}\\
bb&$\overline{\mathit{a}}$\underline{a}&a&a\\
bc&$\overline{\mathit{a}}$ca&aca&ca\\
ca&\textup{-}&\textup{-}&\textup{-}\\
cb&\underline{a}&\textup{-}&a\\
cc&ca&ca&ca\\
\end{tabular}\hfill\begin{tabular}{c||c|c|c}
aaaa&$\overline{\mathit{aaa}}$a\settowidth{\wod}{aaa}\hspace{-\wod}\underline{\mbox{\hspace{\wod}}}&a&a\\
aaab&$\overline{\mathit{aaa}}$\underline{b}&aaa&b\\
aaba&\underline{$\overline{\mathit{b}}$aab}&\textup{-}&aab\\
aabb&\underline{$\overline{\mathit{b}}$}&\textup{-}&\textup{-}\\
abaa&$\overline{\mathit{baa\underline{b}}}$&baa&\textup{-}\\
abab&$\overline{\mathit{baab}}$ab\settowidth{\wod}{bab}\hspace{-\wod}\underline{\mbox{\hspace{\wod}}}&baa&ab\\
abba&\underline{$\overline{\mathit{b}}$}&\textup{-}&\textup{-}\\
abbb&\underline{$\overline{\mathit{b}}$aabaab}&\textup{-}&aabaab\\
baaa&$\overline{\mathit{b}}$\underline{aaa}&b&aaa\\
baab&$\overline{\mathit{b}}$\underline{b}&b&b\\
baba&$\overline{\mathit{bab}}$aab\settowidth{\wod}{baab}\hspace{-\wod}\underline{\mbox{\hspace{\wod}}}&ba&aab\\
babb&$\overline{\mathit{ba\underline{b}}}$&ba&\textup{-}\\
bbaa&\underline{$\overline{\mathit{b}}$}&\textup{-}&\textup{-}\\
bbab&\underline{$\overline{\mathit{b}}$ab}&\textup{-}&ab\\
bbba&$\overline{\mathit{baabaa\underline{b}}}$&baabaa&\textup{-}\\
bbbb&$\overline{\mathit{baabaab}}$aab\settowidth{\wod}{baabaab}\hspace{-\wod}\underline{\mbox{\hspace{\wod}}}&baa&aab\\
\end{tabular}}\hfill}}\end{picture}}\linespread{1}\caption[Three canonical
forms]{Canonical forms for two rules. The first rule is the 1+1-dimensional
rule described in Chapter~\ref{examples}. The second rule operates by first
exchanging segments \seg{aba} with \seg{abba} and then exchanging segments
\seg{bb} with \seg{baab}. Three canonical forms are given for each rule. The
first is the forcing segment representation. In each segment, the section that
overlaps any segment that precedes it is overlined, and the section that
overlaps any segment that follows it is underlined. The next two forms are the
left and right representations. Note that the left representation is obtained
by removing the underlined portion from each segment in the forcing segment
representation, and the right representation is obtained by removing the
overlined portion. It takes just one right shift to move from the left
representation to the right representation of the first rule, but a large
number of right shifts are needed to do the same thing for the second
rule. (Note that the second rule is symmetric under orientation reversal; this
is associated with a symmetry in its forcing segment representation and in a
symmetric relationship between its left and right representations.)}
\label{shiftcanon}
\end{figure}

These canonical forms allow us to describe a rule $T:\X\mapsto\X$ in a unique
way. However, the objects of primary interest here are not the maps themselves,
but equivalence classes of rules. As described in Chapter~\ref{spacetime}, if
$T:\X \mapsto \X$ is a rule, then $T$ is equivalent to $UTU^{-1}$ where $U:\X
\mapsto \X'$ is a local invertible map. In particular, when $\X=\X'$ the map $T
\mapsto UTU^{-1}$ is an equivalence map between rules on the space set \X.

It would be nice to know when two rules on \X\ were equivalent. Then when we
generated rules on \X\ we could store one representative for each equivalence
class. Unfortunately, the problem of determining such equivalence is completely
unsolved at the moment. I can sometimes tell that two rules are not
equivalent. For example, sometimes it is possible to show that the number of
orbits of size $n$ in one law is different from the number of orbits of size
$n$ in a second law. In this case those two laws cannot be equivalent. (An
equivalence map, as shown in Chapter~\ref{spacetime}, is also an equivalence on
orbits, and it sends orbits of size $n$ to orbits of size $n$.) And, of course,
given a rule I can generate lots of equivalent rules by operating on the
original rule using different $U$'s. But at present it is not at all clear to
me that there exists a general method for proving the equivalence or
nonequivalence of rules; if there is one, I have no idea how it might work.

Nevertheless, in our representation of rules it is useful to take into account
certain simple equivalences. For example, let $U$ be a map which acts on a
space by changing the color of each cell according to some permutation. This
map is local and invertible. If it maps \X\ to itself, then the map $T \mapsto
UTU^{-1}$ is an equivalence. We may obtain a description of $UTU^{-1}$ from a
description \ang{L,R,G^L\!,N,A} of $T$ by permuting all colors in $L$
and $R$ and then rearranging the rows in each (and the rows and columns in $A$,
simultaneously) so that the new $L$ has its segments listed in lexicographical
order.

The map $P$ that reverses orientation is another simple example of a local,
invertible map. Again, if $P$ maps \X\ to itself then it induces an equivalence
map $T \mapsto PT\!P$ (here $P=P^{-1}$), and it is straightforward to obtain a
description of $PT\!P$ from a description \ang{L,R,G^L\!,N,A} of $T$ by
switching the order of cells in all segments, transposing the matrix $A$, and
then rearranging the rows and columns as in the previous example.

Finally, one may take a table \ang{L,R,G^L\!,N,A} to stand for more than one
rule, as follows. Let $U:\X\mapsto\X$ be a permutation of colors. If one
applied the permutation specified by $U$ to the colors in the $R$ segments
only, the resulting table describes the rule that results if you first apply
the original rule and then apply $U$. A rule table may be taken to stand for
all rules that result from permuting the colors of the $R$ segments in this
way.

All of this comes in handy when one is generating rules, because it makes it
possible to reduce storage requirements. The idea is that each stored table
represents all equivalence classes under color permutation and orientation
reversal of all rules obtained by permuting the colors in the $R$ segments of
the table. Given a generated rule table, one computes all of the rule tables
that it represents; then one stores only that table among them that comes first
lexographically.

\subsubsection{Zeta functions and the matching condition}

If one is given \ang{L,G^L\!,A} and wishes to generate rules
\ang{L,R,G^L\!,N,A}, the above tools are inadequate. There are simply too many
candidates for $R$. Very few of them turn out to be good candidates. A
successful rule-generating program of this sort needs additional tools. One
such tool is provided by the the following theorem.

\begin{theorem} If $A$ and $B$ are square matrices over a commutative ring with
identity, then $\mathrm{tr}\,A^n=\mathrm{tr}\,B^n$ for all $n>0$ if and only if
$\det(I-xA)=\det(I-xB)$.
\label{zeta}
\end{theorem}
\textsc{Proof.} We first define a formal power series $\displaystyle\F^M \equiv
\sum_{n=0}^\infty M^{n+1}x^n$ for any square matrix $M$ over a commutative
ring. What we wish to prove is that $\mathrm{tr}\,\F^A=\mathrm{tr}\,\F^B$ if and only if $\det
(I-xA)=\det (I-xB)$. Since \[\F^M = \sum_{n=0}^\infty M^{n+1}x^n =
M+\sum_{n=1}^\infty M^{n+1}x^n = M+xM(\sum_{n=0}^\infty M^{n+1}x^n) =
M+xM\F^M,\] it follows that $(I-xM)\F^M=M$. Write $\F^M=[\F^M_1 \ldots \F^M_r]$
and $M=[M_1 \ldots M_r]$, where the $\F^M_i$'s and $M_i$'s are column
vectors. Then $(I-xM)\F^M_i=M_i$ for each $i$. Since the constant term of
$\det (I-xM)$ is 1, it is a unit in the ring of formal power series; hence we
may apply Cramer's rule:
\[\F^M_{ij}=\frac{\det C(i,j)}{\det (I-xM)},\] where $C(i,j)$ denotes the matrix
obtained from $I-xM$ by replacing the $j$th column with $M_i$. Hence\vspace{-8pt} \[\mathrm{tr}\,\F^M=\frac{\displaystyle{\sum_{i=1}^r}\det C(i,i)}{\det
(I-xM)}=\frac{\rule[-6pt]{0pt}{6pt}-\frac{d}{dx}\det(I-xM)}{\det(I-xM)},\]
where $\frac{d}{dx}$ means the formal derivative operator. Let
$\displaystyle\det(I-xA)=\sum_{i=0}^ma_ix^i$ and\vspace{-8pt}
$\displaystyle\det(I-xB)=\sum_{i=0}^mb_ix^i$. Then\newpage\vspace*{-25pt}
\begin{eqnarray*}
\mathrm{tr}\,\F^A & = & \mathrm{tr}\,\F^B\quad\Longleftrightarrow\\[10pt]
\frac{\displaystyle-\sum_{i=1}^mia_ix^{i-1}}{\displaystyle\sum_{i=0}^ma_ix^i} & =
&\frac{\displaystyle-\sum_{i=1}^mib_ix^{i-1}}{\displaystyle\sum_{i=0}^mb_ix^i}\quad\Longleftrightarrow\\[8pt]
(\sum_{i=1}^mia_ix^{i-1})(\sum_{i=0}^mb_ix^i) & = &
(\sum_{i=1}^mib_ix^{i-1})(\sum_{i=0}^ma_ix^i).
\end{eqnarray*}
By using the fact that $a_0=b_0=1$ and equating coefficients of $x^i$ on both
sides, one may easily show that $a_i=b_i$ for all $i$.\proofend\bigskip

Note that $\det(I-xA)=\det(I-xB)$ if and only if the characteristic polynomials
$\det(Ix-A)$ and $\det(Ix-B)$ differ by a factor of $x^k$ for some $k \in \Z$.

Consider a faithful description \ang{L,N,A} where each segment in $L$ has
length one. Let \X=\x{\ang{L,N,A}}, and let $K$ be the set of colors used in
\X. Let $G$ be the free group whose generators are the elements of $L$, and let
$R(G)$ be the group ring of $G$ over \Z. We may create a new matrix $\tilde{A}$
over $R(G)$ by setting $\tilde{A}_{ij}=L_i$ whenever $A_{ij}=1$, and setting
$\tilde{A}_{ij}=0$ otherwise. Then $(\tilde{A}^n)_{ij}$ contains a formal sum
of a set of length-$n$ words in the alphabet $L$. There is a natural one-to-one
correspondence between this set of words and the set of allowed recipes for
gluing together $n+1$ elements of $L$, beginning with $L_i$ and ending with
$L_j$, to form a segment. (The word contains the first $n$ of these elements of
$L$, in order; it omits the final $L_j$.) If $i=j$ then the first and last
elements of $L$ in this construction process may be identified. Hence each word
$s$ in the formal sum in a diagonal element of $\tilde{A}^n$ may be considered
as a means of constructing an element $x \in \X$ that contains $n$ cells. I
will say in this case that $x$ is \emph{associated} with $s$.

A given element $x \in \X$ may be associated with more than one word in the
diagonal of $\tilde{A}$. If $S$ is any segment, let
$\mathrm{rot}\hspace{1pt}(\seg{Sc})=\seg{cS}$ for each $c$ such that the segment
\seg{Sc} is allowed. If S is a segment and $x=\cir{S}$ then, using the fact
that \X\ is a faithful description, it is easy to show that there is a
one-to-one correspondence between the words associated with $x$ in the diagonal
of $\tilde{A}$ and the elements of the set $\{\mathrm{rot}^z(S)\,|\,z \in
\Z^+\}$. Thus the number of words associated with $x$ in the diagonal of
$\tilde{A}$ depends only on $x$ itself, and not on the particular description
\ang{L,N,A}.

Now consider the free group $\hat{G}$ generated by the elements of $K$, the
group ring $R(\hat{G})$ of $\hat{G}$ over \Z, and the ring map $h:L \mapsto K$
which sends a segment $L_i$ to the color of its single cell. Let
$\hat{A}_{ij}=h(\tilde{A}_{ij})$. From the discussion in the above paragraph,
it follows that if $\x{\ang{L,N,A}}=\x{\ang{M,N,B}}$, where each segment in $M$
also has length one, we must have $\mathrm{tr}\,\hat{A}^n=\mathrm{tr}\,\hat{B}^n$ for each $n>0$.

This is where Theorem~\ref{zeta} comes in. It does not apply directly to
matrices over $R(\hat{G})$ since this is a noncommutative ring. But suppose we
have a ring map $f:R(\hat{G})\mapsto \bar{R}$ where $\bar{R}$ is a commutative
ring with identity, and let $\bar{A}_{ij}=f(\hat{A}_{ij})$ and
$\bar{B}_{ij}=f(\hat{B}_{ij})$. Then if $\x{\ang{L,N,A}}=\x{\ang{M,N,B}}$ as
above, we must also have $\mathrm{tr}\,\bar{A}^n=$\linebreak$\mathrm{tr}\,\bar{B}^n$ for each
$n>0$. Now we may apply the theorem and conclude that\linebreak
$\det(I-x\bar{A})=\det(I-x\bar{B})$.

Two such ring maps are relevant here, which I will call $f_c$ and $f_w$ (the
$c$ stands for ``colors'' and the $w$ for ``widths'').

Let $r(\hat{G})$ be the abelianized version of $R(\hat{G})$, and let
$f_c:R(\hat{G})\mapsto r(\hat{G})$ be the natural map. If
$\bar{A}_{ij}=f_c(\hat{A}_{ij})$, then $\bar{A}$ is the same as $\hat{A}$
except that its entries are taken to be free commuting variables. In this case,
the ordering of colors in a word $s$ in a matrix element $(\hat{A}^n)_{ij}$ is
removed in the corresponding word $s'$ in $(\bar{A}^n)_{ij}$; what remains in
$s'$ is a count of how many cells of each color are present in $s$. Let
$K=\{c_1,\ldots,c_k\}$, $\bar{A}_{ij}=f_c(\hat{A}_{ij})$, and
$\bar{B}_{ij}=f_c(\hat{B}_{ij})$. Then $\det(I-x\bar{A})=\det(I-x\bar{B})$ if
and only if \x{\ang{L,N,A}} and \x{\ang{M,N,B}} contain the same number of
spaces that have exactly $n_i$ cells with the color $c_i$, $1\leq i\leq k$, for
every set $\{n_i\,|\,1\leq i\leq k\}$ of nonnegative integers.

Let $f_w:R(\hat{G})\mapsto \Z$ be the ring map that is induced by sending $zg$
to $z$ for each $z \in Z$ and each $g \in \hat{G}$. If
$\bar{A}_{ij}=f_c(\hat{A}_{ij})$, then $\bar{A}$ is just $A$. In this case, if
a matrix element $(\hat{A}^n)_{ij}$ is a sum of $m$ words in $K$ for some
nonnegative integer $m$, then the corresponding matrix element
$(\bar{A}^n)_{ij}$ is $m$. Let $\bar{A}_{ij}=f_c(\hat{A}_{ij})$ and
$\bar{B}_{ij}=f_c(\hat{B}_{ij})$. Then $\det(I-x\bar{A})=\det(I-x\bar{B})$ if
and only if \x{\ang{L,N,A}} and \x{\ang{M,N,B}} contain the same number of
spaces that have exactly $m$ cells for every positive integer $m$.

The above discussion is closely related to the topic of \emph{dynamical zeta
functions} (see~\cite{lind/marcus,ruelle}). Let $T:X\mapsto X$ be an invertible
map on any set $X$. Let $p_n(T)$ denote the number of elements $x\in X$ such
that $T^n(x)=x$. One defines the zeta function $\zeta_T(t)$ as follows:
\[\zeta_T(t)=\exp\left(\sum_{n=1}^{\infty}\frac{p_n(T)}{n}t^n\right).\] In
dynamical systems one is concerned with subshifts of finite type represented by
matrices $A$ of nonnegative integers. Let $E$ be the set of edges $g_{ij}^k$ in
\g{A}. Let $X$ be the set of functions $f:\Z\mapsto E$ such that, for each
$z\in\Z$, if $f(z)=g_{ij}^m$ then $f(z+1)=g_{jk}^n$ for some $k$ and $n$. Let
$\sigma_A$ be the shift operation on $X$: if $f\in X$ then
$(\sigma_A(f))(z)=f(z+1)$. In this case it was shown by Bowen and
Lanford~\cite{bowen/lanford} that
\[\zeta_{\sigma_A}(t)=[\hspace{1pt}\det(I-tA)]^{-1}.\] From this one may easily
deduce the results obtained above in the case of $f_w$.

\begin{theorem}
\label{matrixthm}
Let $A$ be an $n \times n$ matrix over a commutative ring. Let \g{A} be the
directed graph with $n$ vertices such that there is an edge going from vertex
$i$ to vertex $j$ if and only if $A_{ij}\neq 0$. Denote any such edge by
$(i,j)$. Let \Y\ be the set of all sets of disjoint circuits in \g{A}. If
$Y\in\Y$, let $V_Y$ and $E_Y$ be the sets of vertices and edges appearing in
$Y$. Let $\Y'$ be the subset of \Y\ such that, if $Y\!\in\Y'$ and $i,j\in V_Y$,
then the $i$th and $j$th rows (or columns) of $A$ are not equal. If $Y\!\in\Y$,
let
\raisebox{0pt}[0pt][0pt]{$\displaystyle\overline{Y}=\hspace{-8pt}\prod_{(i,j)\in
E_Y}\hspace{-8pt}A_{ij}$}. Then
$\displaystyle\det(I-A)=\sum_{Y\in\,\Y}(-1)^{|Y|}\overline{Y}=\sum_{Y\in\,\Y'}(-1)^{|Y|}\overline{Y}$.\vspace{-8pt}
\end{theorem}
\textsc{Proof.} Let $\pi_n$ be the set of permutations of
$N=\{1,\ldots,n\}$. Then
$\displaystyle\det(I-A)=\sum_{\sigma\in\pi_n}\mathrm{sgn}\,\sigma\prod_{i=1}^n(I-A)_{i\sigma(i)}$
(by a standard formula for determinants). So
$\displaystyle\det(I-A)=\sum_{k=0}^n\sum_S\sum_{\tau}(-1)^k\mathrm{sgn}\,\tau\prod_{i\in
S}A_{i\tau(i)}$, where $S$ is a subset of $N$ containing $k$ elements and
$\tau$ is a permutation of $S$. In the cyclic decomposition of $\tau$ let
$|\tau|$ be the number of cycles, $o$ be the number of odd cycles and $e$ be
the number of even cycles. Then $\mathrm{sgn}\,\tau=(-1)^e$. Since also
$(-1)^k=(-1)^o$, it follows that
$(-1)^k\mathrm{sgn}\,\tau=(-1)^{e+o}=(-1)^{|\tau|}$. So
$\displaystyle\det(I-A)=\sum_{k=0}^n\sum_S\sum_{\tau}(-1)^{|\tau|}\prod_{i\in
S}A_{i\tau(i)}$.

If $\tau$ is such that $A_{i\tau(i)}=0$ for some $i$, then the associated term
$\displaystyle\prod_{i\in S}A_{i\tau(i)}$ in the above formula is
zero. Otherwise, each cycle in the cyclic decomposition of $\tau$ corresponds
to a circuit in \g{A}, and the set of all such cycles corresponds to a set of
disjoint circuits in \g{A}. Thus each such $\tau$ is associated with an element
$Y\!\in\Y$, and in this case $|\tau|=|Y|$ and $\displaystyle\prod_{i\in
S}A_{i\tau(i)}=\overline{Y}$. It follows that
$\displaystyle\det(I-A)=\sum_{Y\in\,\Y}(-1)^{|Y|}\overline{Y}\!\!$.

Let $A_S$ be the matrix obtained from $A$ by deleting those rows and columns
whose indices are not in $S$. Then
$\displaystyle\sum_{\tau}(-1)^{|\tau|}\prod_{i\in
S}A_{i\tau(i)}=(-1)^k\det(A_S)$. Therefore, in our formula for $\det(I-A)$ we
need not sum over those values of $S$ which have the property that
$\det(A_S)=0$. This will clearly be the case when there exist distinct indices
$i,j\in S$ such that rows (or columns) $i$ and $j$ of $A$ are equal. When seen
in terms of \g{A}, this means that we may compute $\det(I-A)$ by summing over
$\Y'$ instead of \Y.\proofend\bigskip

Theorems resembling the one above appear in~\cite{brualdi/ryser}.

\begin{theorem}
\label{detsegthm}
Let $\X=\ang{L,N,A}$, where the segments in $L$ have arbitrary length and $K$
is the set of colors appearing in \X. Let \Y\ be the set of all sets of ways to
assemble spaces using elements of $L$ such that no element of $L$ is used
twice. Let $\Y'$ be the subset of \Y\ such that, if $Y\!\in\!\Y'$ and $L_i$ and
$L_j$ are used in $Y\!\!$, then the $i$th and $j$th rows (or columns) of $A$
are not equal. If there are $n_{ik}$ cells with color $c_k$ in segment $L_i$
for each $i$ and $k$, let $\bar{A}$ be the matrix over $r(\hat{G})$ obtained
from $A$ by setting $\bar{A}_{ij}$ to $\displaystyle\sum_kn_{ik}f_c(c_k)$ if
$A_{ij}=1$, and setting $\bar{A}_{ij}=0$ otherwise. If $Y\!\in\!\Y$ and, for
each $c_i\in K$, $n_i$ denotes the number of cells in the spaces associated
with $Y$ whose color is $c_i$, then let
\raisebox{0pt}[0pt][0pt]{$\displaystyle\overline{Y}=\sum_in_if_c(c_i)$}. Then
$\displaystyle\det(I-\bar{A})=\sum_{Y\in\Y}(-1)^{|Y|}\overline{Y}=\sum_{Y\in\Y'}(-1)^{|Y|}\overline{Y}$.
\end{theorem}
\textsc{Proof.} This is a straightforward application of
Theorem~\ref{matrixthm}.\proofend

\begin{theorem}
\label{matchthm}
Let $\X=\x{\ang{L,G^L\!,A}}$ where the elements of $L$ are width-$d$ segments,
no segment occurs twice in $L$, and the gluing rules $G^L$ are the
$(d-1)$-overlap rules. Let \ang{R,N,A} be any nonoverlap description. Let $\Y$
be the set of subsets of \X\ in which no length-$(d-1)$ segment appears
twice. Let $T:\X\mapsto\x{\ang{R,N,A}}$ be the natural map given by the rule
table \ang{L,R,G^L\!,N,A}. If $Y\!\in\Y$, let $\overline{Y}$ be the
abelianization of the spaces in $Y\!\!$, as before (it is the product of the
colors in $Y\!\!$, where each color is considered to be a free commuting
variable). Similarly, let $\overline{T(Y)}$ be the abelianization of
$T(Y)$. Then the number of spaces having $n_i$ cells with color $c_i$ is the
same in \X\ and \x{\ang{R,N,A}} for each sequence $n_i$ of nonnegative numbers
if and only if
$\displaystyle\sum_{Y\in\Y}(-1)^{|Y|}(\overline{Y}-\overline{T(Y)})=0$.
\end{theorem}
\textsc{Proof.} If we delete every cell except the leftmost one in each segment
in $L$, we obtain a set of segments $L'$ such that $\X=\x{\ang{L',N,A}}$. Using
the method illustrated in Figure~\ref{lengthonesegs}, we may create
\ang{R',N,A'} such that each segment in $R'$ has length one and
$\x{\ang{R,N,A}}=\x{\ang{R',N,A'}}$. Consider the colors in $K$ as free commuting
variables. Let $\bar{A}$ be the matrix obtained from $A$ by letting
$\bar{A}_{ij}=c$ when $A_{ij}=1$ and the cell in $L'_i$ is colored $c$, and
letting $\bar{A}_{ij}=0$ otherwise. Let $\bar{A'}$ be defined similarly (this
time using $R'$ instead of $L'$). Then $\det(I-x\bar{A})=\det(I-x\bar{A'})$ if
and only if \x{\ang{L',N,A}} and \x{\ang{R',N,A'}} contain the same number of
spaces that have exactly $n_i$ cells colored $c_i$, $1\leq i\leq |K|$,
for every set $\{n_i\,|\,1\leq i\leq |K|\}$ of nonnegative integers. Since the
colors are free commuting variables, and the power of $x$ in one of these
determinants is equal to the number of colors that are present in that term,
the $x$'s carry no information. So $\det(I-\bar{A})=\det(I-\bar{A'})$ if
and only if \x{\ang{L',N,A}} and \x{\ang{R',N,A'}} contain the same number of
spaces that have exactly $n_i$ cells colored $c_i$, $1\leq i\leq |K|$,
for every set $\{n_i\,|\,1\leq i\leq |K|\}$ of nonnegative integers.

Let \U\ be the set of all sets of ways to assemble spaces using \ang{L',N,A}
such that no element of $L'$ is used twice. Let $\U'$ be the subset of \U\ such
that, if $U\!\in\U'$ and $L'_i$ and $L'_j$ are used in $U\!$, then the $i$th and
$j$th rows (or columns) of $A$ are not equal. Let \V\ and $\V'$ be defined
similarly with respect to \ang{R,N,A}. Let \W\ and $\W'$ be defined similarly
with respect to \ang{R',N,A'}.

Let $B$ be the matrix obtained from $A$ by letting $B_{ij}$ equal the product
of colors in $R_i$ if $A_{ij}=1$, and letting $B_{ij}=0$ otherwise. (Again the
colors are considered as commuting variables.) By Theorem~\ref{detsegthm},
$\displaystyle\det(I-\bar{A'})=\sum_{Y\in\,\W}(-1)^{|Y|}\overline{Y}$ and
$\displaystyle\det(I-B)=\sum_{Y\in\,\V}(-1)^{|Y|}\overline{Y}\!$,
where $\overline{Y}$ is as defined in that theorem. There is a natural
one-to-one correspondence between the ways of generating a space using
\ang{R,N,A} and the ways of generating a space using \ang{R',N,A'}. This
correspondence induces an invertible map $h:\V\mapsto\W$ such that
$\overline{Y}=\overline{h(Y)}$ for each $Y\!\in\V$. Hence
$\det(I-\bar{A'})=\det(I-B)$. So $\det(I-\bar{A})=\det(I-B)$ if and only if
\x{\ang{L',N,A}} and \x{\ang{R',N,A'}} contain the same number of spaces that
have exactly $n_i$ cells colored $c_i$, $1\leq i\leq |K|$, for every set
$\{n_i\,|\,1\leq i\leq |K|\}$ of nonnegative integers.

By Theorem~\ref{detsegthm},
$\displaystyle\det(I-\bar{A})=\sum_{Y\in\,\,\U'}(-1)^{|Y|}\overline{Y}$ and
$\displaystyle\det(I-B)=\sum_{Y\in\,\V'}(-1)^{|Y|}\overline{Y}$. The theorem
follows trivially once one notices that the sets of spaces associated with the
elements of $\U'$ are just the elements of \Y\, and that the sets of spaces
associated with the elements of $\V'$ are just the elements of
$T(\Y)$.\proofend\bigskip

Theorem~\ref{matchthm} says that a matching condition holds between the
left-hand and right-hand sides of a valid rule table \ang{L,R,G^L\!,N,A}. Write
the sets of spaces in \Y\ with $|Y|$ odd in one column and with $|Y|$ even in
another column, and consider the colors in these spaces as free commuting
variables. What we would like is to be able to pair up equal terms, one from
the odd column and one from the even column, so that each term is in a pair. To
do so, we add terms as needed to either column. These terms, with appropriate
sign, constitute $\det(I-x\bar{A})$. Now substitute $T(Y)$ for each element
$Y\!\in\Y$ in these two columns and consider the colors of $T(Y)$ as free
commuting variables; do not alter the previously added terms from the
determinant. The theorem says that if you again pair up equal terms, one from
each column, then none will be left over. The situation is illustrated in
Figure~\ref{matching}.
\begin{figure}
\fbox{\begin{picture}(425,192)\footnotesize\put(0,0){\makebox(425,192){%
\hfill\begin{tabular}{c|c}
1&\\
\cir{a}&$a$\\
\cir{b}&$b$\\
\cir{c}&$c$\\
\cir{ab}&\cir{a}\cir{b}\\
\cir{ac}&\cir{a}\cir{c}\\
\cir{bc}&\cir{b}\cir{c}\\
\cir{abc}&\cir{ab}\cir{c}\\
\cir{acb}&\cir{ac}\cir{b}\\
\cir{a}\cir{b}\cir{c}&\cir{bc}\cir{a}\\
\end{tabular}\hfill\begin{tabular}{c|c}
1&\\
\cir{c}&$a$\\
\cir{a}&$b$\\
\cir{ca}&$c$\\
\cir{cca}&\cir{c}\cir{a}\\
\cir{b}&\cir{c}\cir{ca}\\
\cir{aca}&\cir{a}\cir{ca}\\
\cir{ccaca}&\cir{cca}\cir{ca}\\
\cir{ba}&\cir{b}\cir{a}\\
\cir{c}\cir{a}\cir{ca}&\cir{aca}\cir{c}\\
\end{tabular}\hfill\begin{tabular}{c|c}
1&1\\
$c$&$a$\\
$a$&$b$\\
$ac$&$c$\\
$ac^2$&$ac$\\
$b$&$ac^2$\\
$a^2c$&$a^2c$\\
$a^2c^3$&$a^2c^3$\\
$ab$&$ab$\\
$a^2c^2$&$a^2c^2$\\
\end{tabular}\hfill}}\end{picture}}\linespread{1}\caption[The matching
condition for colors]{The matching condition for colors. This is condition on
\ang{L,G^L\!,A} and \ang{R,N,A} which holds whenever the two descriptions
generate the same number of spaces with $n_i$ cells colors $c_i$ for any set of
nonnegative integers $n_i$. Here we consider \ang{L,R,G^L\!,N,A} as shown on
the left-hand side of Figure~\ref{shiftcanon}; hence $d=2$ and $\X=\x{3}$. In
the left-hand table we enter the sets of spaces in which no $(d-1)$-length
segment appears twice; this means simply that no color appears twice. The sets
of spaces containing an odd number of spaces are placed on the left; the sets
with an even number of spaces are placed on the right. Here I have lined them
up so that sets that are next to one another in the table contain the same
colors. The first four rows contain sets of spaces which are matched instead by
products of colors written without parentheses; these correspond to the terms
in $\det(I-\bar{A})=1-a-b-c$, written with positive terms on the left and
negative terms on the right. (Notice that the first entry on the top of the
right-hand column is the empty set of spaces; there are no cells in these
spaces, so the product of their colors is 1.) Each space in the left-hand table
is replaced with the image of that space under the rule \ang{L,R,G^L\!,N,A} in
the middle table, with the terms corresponding to the determinant left
intact. The colors in sets of spaces in the middle table (considered as free
commuting variables) are multiplied together to produce the right-hand
table. If this rule is invertible then there must be a one-to-one
correspondence between entries in the left-hand column of this table and those
in the right-hand column in which corresponding entries are equal (there is
such a correspondence in this case).}
\label{matching}
\end{figure}

If one is given \ang{L,G^L\!,A} and seeks to find all laws $T$ given by rule
tables of the form \ang{L,R,G^L\!,N,A}, the matching condition may be used to
construct an algorithm to generate partial information about the description
\ang{R,N,A}. Let $v^i_j\in\Z^+$ denote the number of cells with color $c_j$
that are in segment $R_i$. Then if a space $x$ generated by \ang{L,N,A}
decomposes into $n^x_i$ copies of $L_i$ for each $i$, the number of colors
$c_j$ in $T(x)$ is given by $\displaystyle\sum_{i=1}^n n^x_iv^i_j$. The
goal of the algorithm will be to find every $v$ that satisfies the matching
condition and one additional condition. The additional condition is that the
number of cells in each space $x$ must be positive; i.e., for any $x$ we must
have $\displaystyle\sum_{j=1}^{|K|}\sum_{i=1}^n n^x_iv^i_j>0$.

Given any solution $v$, we may construct another solution in the following
way. Choose a length-$(d-1)$ segment $S$, and a vector $z_j$, $1\leq j\leq
|K|$. If we add $z$ to each $v^i$ such that $L_i$ ends with $S$, and then
subtract $z$ from each $v^i$ such that $L_i$ begins with $S$, then the numbers
of colors in $T(x)$, as computed by the above formula, is left invariant. This
is the commutative analogue of the left shift and right shift operations
described in Section~\ref{shift}. The operation may make $v^i_j$ negative for
some $i$ and $j$. For the moment, let us allow this. We will seek all solutions
$v$ such that $v^i_j\in\Z$. However, we will continue to require that the
number of cells in each space must be positive.

I will say that $v$ is equivalent to $v'$ whenever $v'$ may be obtained from
$v$ by a sequence of commutative shift operations (i.e., adding and subtracting
$z$ as described above). Instead of looking for all solutions $v$, then, we may
look for representative of each of the equivalence classes whose members
satisfy the two conditions.

This makes the computation simpler, because now we can set some of the $v^i$ to
zero. To do this, it is useful to consider a certain directed graph \G. The
vertices of \G\ correspond to the allowed length-$(d-1)$ segments of \X; the
edges correspond to the allowed length-$d$ segments of \X, with the $i$th edge
going from vertex $S$ to vertex $S'$ if and only if $L_i$ begins with $S$ and
ends with $S'$. In order to set variables to zero, we construct a spanning tree
of \G. To do this, we first choose as the root of this tree any vertex $S$ in
\G. Next we repeatedly choose edges to put in the tree. The edge is chosen so
that it points from a vertex in the tree to a vertex not in the tree. (I am
assuming that \g{A}, and hence \G, is strongly connected; if so, then such an
edge can always be found if there is a vertex not yet in the tree.) Call this
latter vertex $S'$. The edge just added corresponds to some segment $L_i$ which
ends with $S'$. Subtract $v^i$ from each $v^j$ that ends with $S'$, and add it
to each $v^j$ that begins with $S'$. This sets $v^i$ to zero (since it cannot
both begin and end with $S'$, since it is in a tree), and does not affect $v^j$
if the edge associated with $L_j$ is already in the tree (hence any variable
that has been set to zero remains zero). We may thus continue adding edges and
setting variables to zero until every vertex is in the tree (and hence it is a
spanning tree). If \G\ contains $V$ vertices, then the completed spanning tree
contains $V-1$ edges. Hence each equivalence class contains a representative
$v$ such that $v^i=0$ for any $V-1$ segments $L_i$ whose associated edges
constitute a spanning tree of \G.

For example, if \ang{L,G^L\!,A} is given by the left-hand column on the left
side of Figure~\ref{shiftcanon} (i.e., if it is the standard description of
\x{3} for $d=2$) then the tree \G\ has three vertices (corresponding to the
length-1 segments \seg{a}, \seg{b} and \seg{c}) and nine edges (corresponding
to the nine length-2 segments). One spanning tree of \G\ is given by the edges
$L_7$ and $L_8$ (which correspond to the length-2 segments \seg{ca} and
\seg{cb}). Hence we may set $v^7_j$ and $v^8_j$ to zero for each $j$, $1\leq
j\leq |K|$.

Now we may replace each set of spaces in our matching table with a sum of
variables which tells how many cells of each color are in the image of these
sets of spaces under the mapping $T$. In Figure~\ref{matchvar} this is shown
for the matching table on the left side of Figure~\ref{matching},
\begin{figure}
\fbox{\begin{picture}(425,192)\footnotesize\put(0,0){\makebox(425,192){%
\hfill\begin{tabular}{c|c}
1&\\
\cir{a}&$a$\\
\cir{b}&$b$\\
\cir{c}&$c$\\
\cir{ab}&\cir{a}\cir{b}\\
\cir{ac}&\cir{a}\cir{c}\\
\cir{bc}&\cir{b}\cir{c}\\
\cir{abc}&\cir{ab}\cir{c}\\
\cir{acb}&\cir{ac}\cir{b}\\
\cir{a}\cir{b}\cir{c}&\cir{bc}\cir{a}\\
\end{tabular}\hfill\begin{tabular}{c|c}
$(0,0,0)$&$(0,0,0)$\\
$v^1$&$(1,0,0)$\\
$v^5$&$(0,1,0)$\\
$v^9$&$(0,0,1)$\\
$v^2+v^4$&$v^1+v^5$\\
$v^3$&$v^1+v^9$\\
$v^6$&$v^5+v^9$\\
$v^2+v^6$&$v^2+v^4+v^9$\\
$v^3+v^4$&$v^3+v^5$\\
$v^1+v^5+v^9$&$v^6+v^1$\\
\end{tabular}\hfill}}\end{picture}}\linespread{1}\caption[The matching
table in terms of variables]{The matching table in terms of variables. The
left-hand table is the same as the left-hand table in Figure~\ref{matching}. In
the right-hand table, sets of spaces have been replaced by variables. Each
variable $v^i$ is a vector, so the table entries are vectors. The $j$th
component of the vector in an entry associated with a set of spaces is the
number of cells colored by the $j$th color that are present in the image of
that set of spaces under the map $T$. (The first color is $a$, the second is
$b$ and the third is $c$.) Here we have set $v^7$ and $v^8$ to zero (see text);
hence these variables have been omitted. Since the columns must match pairwise,
it follows that the sums of the columns must be equal. This means that
$v^2+v^3+v^4+v^6=(1,1,1)+v^1+v^5+v^9$. So
$v^2=(1,1,1)+v^1+v^5+v^9-v^3-v^4-v^6$; this may be substituted into the
right-hand table to reduce the number of variables by one. (Note that here I am
using summation rather than product to denote the commutative group operation;
this differs from the notation used in Figure~\ref{matching}.)}
\label{matchvar}
\end{figure}
which is the table corresponding to the example of \ang{L,G^L\!,A} given
above. Of course, the variables which we have set to zero may be omitted.

One relationship between the variables may be read off immediately. Clearly the
sum of all entries in the odd column must equal the sum of all entries in the
even column (since the elements in the two columns must match pairwise). This
gives us a linear relation among the variables which, if it is nontrivial,
enables us to eliminate one variable. (See Figure~\ref{matchvar} for an
example.)

Now we begin the matching process. At any stage before the end of the process,
some of the entries in the two columns of the table still contain variables,
and some contain only a constant term. One begins by striking out matching
pairs of constant-term entries, one from each column.

Define the width of a constant term $c$ to be
\raisebox{0pt}[0pt]{$\displaystyle\sum_{j=1}^{|K|}c_j$}. Consider the
non-constant entries that correspond to a single space. These are in the odd
column. In the completed table, each such entry will have a positive width. Let
$m$ be the minimum such width. Then every non-constant entry in the even column
will have width greater than $m$ (since it corresponds to a set of spaces
containing one of the spaces whose width is at least $m$ plus at least one
other space). Hence there must be an unmatched constant entry of width $m$ in
the even column. Furthermore, there must be no unmatched constant entry of
width less than $m+1$ in the odd column, because this cannot be matched.

Hence, the next step is to find those entries with smallest unmatched width
$m$. These must exist (it cannot be that every constant entry is matched), and
they must all be in the even column, for otherwise the table cannot be
successfully completed. Every such entry must be matched with a non-constant
entry in the odd column that corresponds to a single space. So next we choose
one of the constant entries of width $m$ and one of the non-constant entries
that correspond to a single space, and we set these to be equal. This gives us
a nontrivial linear equation which allows us to eliminate one of the
variables. By repeating this process, eventually all variables are eliminated;
if the sides of the table match, then we have a solution. If we perform our
selections in all possible ways, we get all solutions.

Given such a solution $v$, which is a representative of an equivalence class of
solutions (and which may contain negative entries), it is a straightforward
process to generate all members of this equivalence class which contain only
nonnegative entries. Each such member gives us a possible configuration of the
$R$ segments in terms of numbers of cells of each color. By ordering these
cells in all possible ways and applying the algorithm described in
Section~\ref{revalg} to detect whether the resulting \ang{R,N,A} is a faithful
description of \x{\ang{L,G^L\!,A}}, one may obtain all $R$ such that
\x{\ang{L,R,G^L\!,N,A}} is a law of evolution.

It is a strange and unexplained fact that, in over one hundred cases in which I
have generated representatives $v$ of equivalence classes of solutions as
described above, there has always turned out to be exactly one corresponding
$R$ (up to equivalence by shift transformation) such that
\x{\ang{L,R,G^L\!N,A}} is a law of evolution.

\subsubsection{First results}
\label{firstresults}

At a somewhat early stage in my investigations I made an extended attempt to
generate nonoverlapping rule tables by computer. At this point I had not
considered subshifts of finite type in general, so my attempt only involved the
space sets \x{k}. In fact, due to difficulty of computation, I only obtained
useful results for \x{2}, \x{3} and~\x{4}.

This effort involved at least a thousand hours of programming and many
overnight runs on an 80386D-20 CPU. Numerous programming insights and
algorithmic improvements were required. The final best algorithm made heavy use
of the matching condition described in the previous section.

The algorithm begins by letting each variable $v^i$ represent the number of
cells in segment $R_i$ mod 2. The matching condition must hold for these $v^i$,
but one cannot use the matching algorithm described at the end of the previous
section because the widths of constant-term table entries (i.e., the number of
cells in the images under $T$ of the space sets corresponding to these entries)
are unknown. Hence all mod 2 solutions are computed by brute force.

Next these solutions are used as templates for finding all mod 4 solutions to
the matching condition by brute force. Next the mod 4 solutions are used as
templates for finding all solutions to the matching condition for the case
where $v^i\in\Z$; here $v^i$ represents the number of cells in $R_i$ (except
that it can be negative, due to shift operations). Now the widths of
constant-term table entries are known, so the matching algorithm can be
employed (one simply pretends that all cells are the same color). Finally, the
solutions for widths are used as templates for finding the solutions where
$v^i_j$ is the number of cells in $R_i$ colored by the $j$th color; again the
matching algorithm is used, and then the desired set of $R$'s is computed (as
described at the end of the previous section).

The efficiency of the algorithm depends also on its use of information about
standard forms and equivalences between laws described in
Section~\ref{shift}. I will not go into the details here.

The results, which took approximately eight hours of CPU time to obtain in the
final version, are given in Figures~\ref{bigruletable1},~\ref{bigruletable2}
and~\ref{bigruletable3}.  In a sense these results are not very satisfying,
since most (if not all) of the laws found here may be found much more easily by
hand, via the method described in Chapter~\ref{examples}. But they constitute
the only large set of examples that I have generated by computer; and they do
have the virtue of being complete in the sense that they constitute all
\ang{L,R,G^L\!,N,A} for the given \ang{L,G^L\!,A}.

In the remainder of this section I will present a brief phenomenological tour
of the 1+1-dimensional systems that I have encountered so far. These include
the systems
\begin{figure}
\linespread{.98}\fbox{\begin{picture}(425,412)\footnotesize\settowidth{\wod}{\ }
\put(0,212){\makebox(425,80)[bl]{\hfill\textit{\begin{tabular}{c\hw||\hw c\hw|}
aa&a\\
ab&b\\
ba&a\\
bb&b\\
\end{tabular}\hfill\begin{tabular}{c\hw||\hw c\hw|\hw c\hw|}
aaa&a &a\\
aab&b &b\\
aba&\textup{-} &a\\
abb&ab&b\\
baa&aa&a\\
bab&b &b\\
bba&\textup{-} &a\\
bbb&ab&b\\
\end{tabular}\hfill\begin{tabular}{c\hw||\hw c\hw|\hw c\hw|\hw c\hw|\hw c\hw|}
aa&a &a  &a  &a\\
ab&\textup{-} &\textup{-}  &\textup{-}  &b\\
ac&\textup{-} &b  &ba &c\\
ba&ba&c  &c  &a\\
bb&c &ba &b  &b\\
bc&c &bab&bba&c\\
ca&ca&aa &a  &a\\
cb&b &\textup{-}  &\textup{-}  &b\\
cc&b &b  &ba &c\\
\end{tabular}\hfill\begin{tabular}{c\hw||\hw c\hw|\hw
c\hw|\hw c\hw|\hw c\hw|\hw c\hw|}
aaaa&a  &a&a &a  \\ 
aaab&\textup{-}  &b&b &\textup{-}  \\ 
aaba&bba&a&a &ba \\ 
aabb&a  &b&b &abb\\ 
abaa&\textup{-}  &a&\textup{-} &\textup{-}  \\ 
abab&\textup{-}  &b&ab&bb \\ 
abba&aba&a&a &a  \\ 
abbb&bbb&b&b &b  \\
baaa&a  &a&aa&a  \\
baab&\textup{-}  &b&b &\textup{-}  \\
baba&ba &a&a &a  \\
babb&\textup{-}  &b&b &b  \\
bbaa&\textup{-}  &a&\textup{-} &\textup{-}  \\
bbab&\textup{-}  &b&ab&bb \\
bbba&a  &a&a &a  \\
bbbb&b  &b&b &b  \\
\end{tabular}}\hfill}}

\put(0,12){\makebox(425,120)[bl]{\hfill\textit{\begin{tabular}{c\hw||\hw c\hw|\hw
c\hw|\hw c\hw|\hw c\hw|\hw c\hw|\hw c\hw|\hw c\hw|\hw c\hw|\hw c\hw|\hw
c\hw|\hw c\hw|\hw c\hw|\hw c\hw|\hw c\hw|\hw c\hw|\hw c\hw|\hw c\hw|\hw
c\hw|\hw c\hw|}
aaaa&a   &a &a &a &a  &a  &a   &a  &a   &a  &a   &a  &a   &a     &a     \\
aaab&\textup{-}   &b &b &b &\textup{-}  &b  &b   &b  &b   &b  &b   &b  &b   &b     &b     \\
aaba&ba  &\textup{-} &\textup{-} &ba&ab &\textup{-}  &ab  &\textup{-}  &\textup{-}   &\textup{-}  &ab  &\textup{-}  &\textup{-}   &aab   &aab   \\
aabb&a   &a &ab&\textup{-} &bab&ab &\textup{-}   &ab &a   &aab&\textup{-}   &aab&aab &\textup{-}     &\textup{-}     \\
abaa&\textup{-}   &a &aa&\textup{-} &aa &\textup{-}  &aa  &aa &aa  &a  &\textup{-}   &\textup{-}  &a   &a     &\textup{-}     \\
abab&\textup{-}   &b &b &ab&\textup{-}  &aab&b   &\textup{-}  &b   &b  &aab &ab &\textup{-}   &b     &ab    \\
abba&abba&\textup{-} &\textup{-} &a &\textup{-}  &\textup{-}  &\textup{-}   &\textup{-}  &\textup{-}   &\textup{-}  &\textup{-}   &\textup{-}  &\textup{-}   &\textup{-}     &\textup{-}     \\
abbb&bbb &ba&ab&bb&ab &ab &abab&ab &aba &aab&abab&aab&aab &aabaab&aabaab\\
baaa&a   &aa&a &aa&a  &aaa&a   &a  &a   &aa &aaa &aaa&aa  &aa    &aaa   \\
baab&\textup{-}   &bb&b &b &\textup{-}  &b  &b   &bab&bb  &b  &b   &b  &baab&b     &b     \\
baba&bba &\textup{-} &\textup{-} &ba&aab&\textup{-}  &ab  &aab&\textup{-}   &\textup{-}  &ab  &\textup{-}  &ab  &aab   &aab   \\
babb&\textup{-}   &a &ab&\textup{-} &b  &ab &\textup{-}   &b  &a   &aab&\textup{-}   &aab&b   &\textup{-}     &\textup{-}     \\
bbaa&\textup{-}   &ba&aa&\textup{-} &aa &\textup{-}  &aa  &aa &abaa&a  &\textup{-}   &\textup{-}  &a   &a     &\textup{-}     \\
bbab&\textup{-}   &ab&b &ab&\textup{-}  &aab&b   &\textup{-}  &b   &b  &aab &ab &\textup{-}   &b     &ab    \\
bbba&a   &\textup{-} &\textup{-} &a &\textup{-}  &\textup{-}  &\textup{-}   &\textup{-}  &\textup{-}   &\textup{-}  &\textup{-}   &\textup{-}  &\textup{-}   &\textup{-}     &\textup{-}     \\
bbbb&b   &ba&ab&b &ab &ab &ab  &ab &aba &aab&ab  &aab&aab &aab   &aab   \\
\end{tabular}}\hfill}}\end{picture}}\linespread{1}
\caption[Computer-generated nonoverlap rules]{Computer-generated nonoverlap
rules. All such rules were found in five cases: $k=2$, $d=2$; $k=2$, $d=3$;
$k=3$, $d=2$; $k=2$, $d=4$; $k=4$, $d=2$ (here $k$ is the number of colors and
$d$ is the width of the left-hand segments). Results for the first four cases
are presented in this figure; results for the last case are presented in the
next two figures. Each right-hand column represents the class of right-hand
columns obtained by permuting the colors of the column, and is given in its
maximally-shifted-to-the-right form. (Note that the results for $k=2$, $d=2$
are a subset of the results for $k=2$, $d=3$, which are in turn a subset of the
results for $k=2$, $d=4$.)}
\label{bigruletable1}
\end{figure}\clearpage\begin{figure}\linespread{.98}\fbox{\begin{picture}(425,612)\footnotesize\settowidth{\wod}{\ }
\put(0,12){\makebox(425,200)[bl]{\hfill\itshape\begin{tabular}{c\hw||\hw c\hw|\hw
c\hw|\hw c\hw|\hw c\hw|\hw c\hw|\hw c\hw|\hw c\hw|\hw c\hw|\hw c\hw|\hw
c\hw|\hw c\hw|\hw c\hw|\hw c\hw|\hw c\hw|}
aa &a  &a  &a  &a  &a  &a  &a  &a  &a  &a  &a  &a  &a  &a  \\
ab &\textup{-}  &\textup{-}  &\textup{-}  &\textup{-}  &\textup{-}  &\textup{-}  &\textup{-}  &\textup{-}  &\textup{-}  &\textup{-}  &\textup{-}  &\textup{-}  &\textup{-}  &\textup{-}  \\
ac &\textup{-}  &\textup{-}  &\textup{-}  &\textup{-}  &\textup{-}  &\textup{-}  &\textup{-}  &\textup{-}  &\textup{-}  &\textup{-}  &\textup{-}  &\textup{-}  &\textup{-}  &\textup{-}  \\
ad &\textup{-}  &\textup{-}  &\textup{-}  &\textup{-}  &\textup{-}  &\textup{-}  &\textup{-}  &\textup{-}  &\textup{-}  &\textup{-}  &\textup{-}  &\textup{-}  &\textup{-}  &\textup{-}  \\
ba &b  &b  &b  &b  &b  &b  &b  &b  &b  &b  &b  &b  &ba &ba \\
bb &c  &c  &c  &c  &c  &c  &c  &ca &ca &ca &ca &ca &b  &c  \\
bc &c  &c  &c  &d  &d  &c  &c  &d  &c  &ca &ca &ca &b  &b  \\
bd &c  &c  &c  &c  &c  &ca &da &ca &ca &ca &ca &ca &c  &c  \\
ca &da &da &da &ca &da &da &daa&da &caa&da &da &da &da &ca \\
cb &d  &ca &ca &da &ca &ca &da &c  &d  &c  &c  &d  &c  &d  \\
cc &ca &d  &ca &da &ca &ca &da &c  &d  &c  &d  &c  &c  &d  \\
cd &ca &d  &ca &da &ca &c  &c  &c  &d  &c  &d  &c  &b  &d  \\
da &caa&caa&caa&daa&caa&caa&ca &caa&da &caa&caa&caa&ca &da \\
db &ca &d  &d  &d  &d  &d  &d  &d  &c  &d  &d  &c  &d  &b  \\
dc &d  &ca &d  &c  &c  &d  &d  &ca &ca &d  &c  &d  &d  &c  \\
dd &d  &ca &d  &d  &d  &d  &d  &d  &c  &d  &c  &d  &d  &b  \bigskip\\

aa &a  &a  &a  &a  &a  &a   &a  &a   &a  &a   &a   &a   &a &a \\
ab &\textup{-}  &\textup{-}  &\textup{-}  &\textup{-}  &\textup{-}  &\textup{-}   &\textup{-}  &\textup{-}   &\textup{-}  &\textup{-}   &\textup{-}   &\textup{-}   &\textup{-} &\textup{-} \\
ac &\textup{-}  &\textup{-}  &\textup{-}  &\textup{-}  &\textup{-}  &\textup{-}   &\textup{-}  &\textup{-}   &\textup{-}  &\textup{-}   &\textup{-}   &\textup{-}   &\textup{-} &\textup{-} \\
ad &\textup{-}  &b  &b  &b  &b  &b   &b  &b   &b  &b   &b   &b   &b &b \\
ba &ba &c  &c  &c  &c  &c   &c  &c   &c  &c   &c   &c   &ca&ca\\
bb &c  &d  &d  &d  &d  &b   &b  &ba  &ba &ba  &ba  &ba  &c &c \\
bc &c  &d  &d  &d  &d  &ba  &baa&b   &ba &ba  &ba  &ba  &c &c \\
bd &c  &db &db &db &bab&bb  &bb &bab &db &bab &bab &baab&da&da\\
ca &ca &da &da &da &da &d   &d  &d   &da &d   &d   &d   &cb&db\\
cb &d  &b  &ba &ba &ba &baa &ba &baa &d  &b   &baa &baa &d &d \\
cc &d  &ba &b  &ba &ba &baa &ba &baa &d  &baa &baa &baa &d &d \\
cd &d  &bab&bb &bab&db &baab&bab&baab&bab&baab&baab&bab &db&cb\\
da &da &aa &aa &aa &aa &aaa &aaa&aaa &aa &aaa &aaa &aaa &a &a \\
db &b  &a  &\textup{-}  &\textup{-}  &\textup{-}  &a   &aa &\textup{-}   &\textup{-}  &aa  &\textup{-}   &\textup{-}   &\textup{-} &\textup{-} \\
dc &b  &\textup{-}  &a  &\textup{-}  &\textup{-}  &\textup{-}   &\textup{-}  &a   &\textup{-}  &\textup{-}   &\textup{-}   &\textup{-}   &\textup{-} &\textup{-} \\
dd &b  &b  &ab &b  &b  &ab  &aab&b   &b  &b   &b   &b   &b &b \bigskip\\

aa &a &a &a &a &a &a  &a  &a  &a  &a    &a   &a    &a  &a  \\
ab &\textup{-} &\textup{-} &\textup{-} &\textup{-} &\textup{-} &\textup{-}  &\textup{-}  &\textup{-}  &\textup{-}  &\textup{-}    &\textup{-}   &\textup{-}    &\textup{-}  &\textup{-}  \\
ac &\textup{-} &\textup{-} &\textup{-} &\textup{-} &\textup{-} &\textup{-}  &\textup{-}  &\textup{-}  &\textup{-}  &\textup{-}    &\textup{-}   &\textup{-}    &\textup{-}  &\textup{-}  \\
ad &b &b &b &b &b &ba &ba &ba &ba &ba   &ba  &ba   &ba &ba \\
ba &ca&ca&ca&ca&cb&c  &c  &c  &c  &c    &c   &c    &c  &c  \\
bb &c &d &d &d &d &b  &b  &d  &d  &b    &b   &b    &b  &d  \\
bc &c &d &d &d &d &b  &b  &d  &d  &b    &b   &b    &b  &d  \\
bd &db&cb&cb&da&ca&da &da &da &da &bba  &bba &bba  &dba&bba\\
ca &da&da&db&cb&db&bba&dba&bba&dba&d    &d   &d    &da &da \\
cb &d &c &c &c &c &d  &d  &b  &b  &ba   &baa &baa  &d  &b  \\
cc &d &c &c &c &c &d  &d  &b  &b  &baa  &ba  &baa  &d  &b  \\
cd &cb&db&da&db&da&dba&bba&dba&bba&baaba&baba&baaba&bba&dba\\
da &a &a &a &a &a &a  &a  &a  &a  &aa   &aa  &aa   &a  &a  \\
db &\textup{-} &\textup{-} &\textup{-} &\textup{-} &\textup{-} &\textup{-}  &\textup{-}  &\textup{-}  &\textup{-}  &a    &\textup{-}   &\textup{-}    &\textup{-}  &\textup{-}  \\
dc &\textup{-} &\textup{-} &\textup{-} &\textup{-} &\textup{-} &\textup{-}  &\textup{-}  &\textup{-}  &\textup{-}  &\textup{-}    &a   &\textup{-}    &\textup{-}  &\textup{-}  \\
dd &b &b &b &b &b &ba &ba &ba &ba &ba   &aba &ba   &ba &ba \\
\end{tabular}\hfill}}\end{picture}}\linespread{1}
\caption[Some of the nonoverlap rules for $k=4$, $d=2$]{Some of the nonoverlap
rules for $k=4$, $d=2$.}
\label{bigruletable2}
\end{figure}\clearpage\begin{figure}\linespread{.98}\fbox{\begin{picture}(425,612)\footnotesize\settowidth{\wod}{\ }
\put(0,12){\makebox(425,200)[bl]{\hfill\itshape\begin{tabular}{c\hw||\hw c\hw|\hw
c\hw|\hw c\hw|\hw c\hw|\hw c\hw|\hw c\hw|\hw c\hw|\hw c\hw|\hw c\hw|\hw
c\hw|\hw c\hw|\hw c\hw|\hw c\hw|\hw c\hw|}
aa &a  &a  &a  &a  &a    &a    &a    &a    &a    &a    &a    &a  &a  &a  \\
ab &\textup{-}  &\textup{-}  &\textup{-}  &\textup{-}  &\textup{-}    &\textup{-}    &\textup{-}    &\textup{-}    &\textup{-}    &\textup{-}    &\textup{-}    &\textup{-}  &\textup{-}  &\textup{-}  \\
ac &\textup{-}  &\textup{-}  &\textup{-}  &\textup{-}  &b    &\textup{-}    &\textup{-}    &\textup{-}    &\textup{-}    &\textup{-}    &\textup{-}    &\textup{-}  &\textup{-}  &\textup{-}  \\
ad &ba &ba &ba &ba &ab   &ba   &ba   &ba   &ba   &ba   &ba   &ba &ba &ba \\
ba &c  &c  &c  &c  &c    &c    &c    &c    &c    &c    &c    &ca &ca &ca \\
bb &d  &b  &d  &d  &cb   &b    &b    &b    &b    &bba  &bba  &b  &b  &c  \\
bc &d  &b  &d  &d  &d    &b    &b    &b    &b    &bba  &bba  &b  &b  &c  \\
bd &dba&dba&bba&dba&d    &bbaa &baaba&bbaba&bbaba&bbaa &bbaba&d  &d  &d  \\
ca &da &bba&dba&bba&a    &bbaba&d    &d    &bbaa &bbaba&bbaa &bba&cba&bba\\
cb &b  &d  &b  &b  &b    &bba  &baa  &bba  &bba  &b    &b    &c  &c  &b  \\
cc &b  &d  &b  &b  &b    &bba  &baa  &bba  &bba  &b    &b    &c  &c  &b  \\
cd &bba&da &da &da &ab   &d    &bba  &bbaa &d    &d    &d    &cba&bba&cba\\
da &a  &a  &a  &a  &c    &a    &aa   &a    &a    &a    &a    &a  &a  &a  \\
db &\textup{-}  &\textup{-}  &\textup{-}  &\textup{-}  &cb   &\textup{-}    &\textup{-}    &\textup{-}    &\textup{-}    &\textup{-}    &\textup{-}    &\textup{-}  &\textup{-}  &\textup{-}  \\
dc &\textup{-}  &\textup{-}  &\textup{-}  &\textup{-}  &d    &\textup{-}    &\textup{-}    &\textup{-}    &\textup{-}    &\textup{-}    &\textup{-}    &\textup{-}  &\textup{-}  &\textup{-}  \\
dd &ba &ba &ba &ba &d    &ba   &ba   &ba   &ba   &ba   &ba   &ba &ba &ba \bigskip\\

aa &a  &a  &a  &a  &a  &a    &a    &a    &a    &a    &a    &a  &a  &a  \\
ab &\textup{-}  &\textup{-}  &\textup{-}  &\textup{-}  &\textup{-}  &\textup{-}    &\textup{-}    &\textup{-}    &\textup{-}    &\textup{-}    &\textup{-}    &\textup{-}  &\textup{-}  &\textup{-}  \\
ac &\textup{-}  &\textup{-}  &\textup{-}  &\textup{-}  &\textup{-}  &\textup{-}    &\textup{-}    &\textup{-}    &\textup{-}    &\textup{-}    &\textup{-}    &b  &\textup{-}  &\textup{-}  \\
ad &ba &ba &ba &ba &ba &ba   &ba   &baa  &baa  &baa  &baa  &c  &ba &ba \\
ba &ca &ca &ca &ca &bba&bbaa &bbaa &c    &c    &bbaa &bbaa &d  &cba&cba\\
bb &c  &b  &b  &c  &c  &b    &bba  &b    &b    &b    &ba   &b  &b  &c  \\
bc &c  &b  &b  &c  &c  &b    &bba  &b    &b    &b    &ba   &bb &b  &c  \\
bd &d  &bba&cba&bba&d  &c    &c    &bbaa &babaa&c    &c    &bac&d  &d  \\
ca &cba&cba&bba&cba&cba&bbaba&bbaba&d    &d    &babaa&babaa&aa &bba&bba\\
cb &b  &c  &c  &b  &b  &bba  &b    &ba   &ba   &ba   &b    &a  &c  &b  \\
cc &b  &c  &c  &b  &b  &bba  &b    &ba   &ba   &ba   &b    &ab &c  &b  \\
cd &bba&d  &d  &d  &ca &d    &d    &babaa&bbaa &d    &d    &c  &ca &ca \\
da &a  &a  &a  &a  &a  &a    &a    &a    &a    &a    &a    &a  &a  &a  \\
db &\textup{-}  &\textup{-}  &\textup{-}  &\textup{-}  &\textup{-}  &\textup{-}    &\textup{-}    &\textup{-}    &\textup{-}    &\textup{-}    &\textup{-}    &\textup{-}  &\textup{-}  &\textup{-}  \\
dc &\textup{-}  &\textup{-}  &\textup{-}  &\textup{-}  &\textup{-}  &\textup{-}    &\textup{-}    &\textup{-}    &\textup{-}    &\textup{-}    &\textup{-}    &b  &\textup{-}  &\textup{-}  \\
dd &ba &ba &ba &ba &ba &ba   &ba   &baa  &baa  &baa  &baa  &c  &ba &ba \bigskip\\

aa &a  &a  &a &a &a  &a    &a &a&a&a&a&a&a\\
ab &\textup{-}  &\textup{-}  &\textup{-} &\textup{-} &\textup{-}  &\textup{-}    &\textup{-} &a&a&a&a&a&b\\
ac &b  &b  &b &b &b  &\textup{-}    &b &a&b&b&b&b&c\\
ad &c  &c  &ab&ab&ca &ba   &ab&b&b&b&b&b&d\\
ba &d  &d  &c &c &d  &c    &c &b&b&c&c&c&a\\
bb &ca &ca &d &d &c  &b    &cb&b&b&c&c&c&b\\
bc &cb &cab&d &d &cb &b    &d &b&a&d&d&d&c\\
bd &cac&cac&cb&cb&cca&bbaa &d &a&a&d&d&d&d\\
ca &a  &a  &a &a &a  &d    &a &c&c&a&a&a&a\\
cb &\textup{-}  &\textup{-}  &b &ab&\textup{-}  &bba  &ab&c&c&a&a&a&b\\
cc &b  &b  &b &b &b  &bba  &b &d&d&b&b&d&c\\
cd &c  &c  &ab&b &ca &bbaba&b &d&d&b&b&d&d\\
da &aa &aa &c &c &a  &a    &c &d&d&c&d&c&a\\
db &\textup{-}  &\textup{-}  &d &d &\textup{-}  &\textup{-}    &cb&d&d&c&c&c&b\\
dc &ab &b  &d &d &b  &\textup{-}    &d &c&c&d&c&b&c\\
dd &c  &c  &cb&cb&ca &ba   &d &c&c&d&d&b&d\\
\end{tabular}\hfill}}\end{picture}}\linespread{1}
\caption[The remaining nonoverlap rules for $k=4$, $d=2$]{The remaining
nonoverlap rules for $k=4$, $d=2$.}
\label{bigruletable3}
\end{figure}\clearpage
\noindent described in Figures~\ref{bigruletable1},~\ref{bigruletable2}
and~\ref{bigruletable3}, as well as the reversible cellular automata
which I generated in~\cite{me}.

In the case of reversible cellular automata, the period of each evolutionary
orbit is necessarily finite. This is because there are only finitely many
spaces having a given width $w$, and this number $w$ is conserved. Hence there
must be a $p_x$ such that $x=T^{p_x}(x)$.

In the very simplest cases, there is some smallest positive constant $p$ such
that $p_x$ divides $p$ for all $x$. For example, if $T$ is the identity law on
any set \X\, then $p=1$. If $T$ is rule $A$ from Figure~\ref{calaws},
then $p=2$. If $T$ is rule $D$ from Figure~\ref{calaws}, then $p=4$.

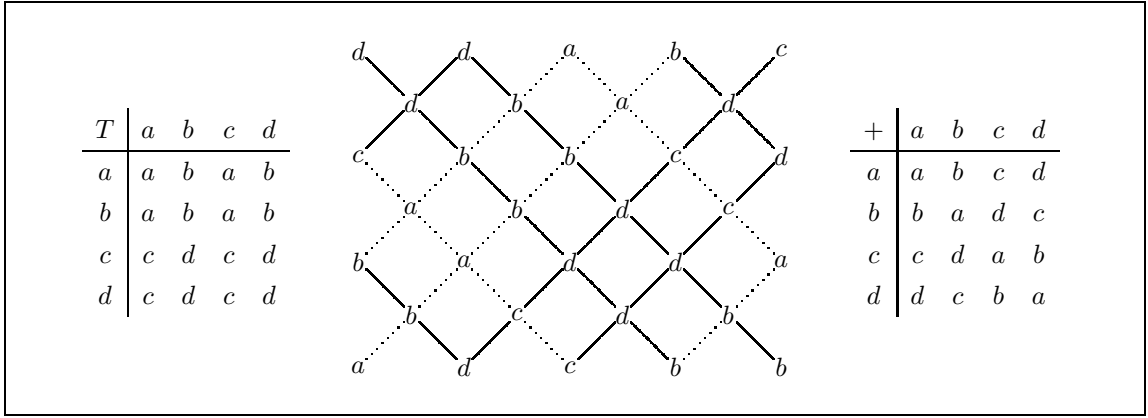
\begin{figure}
\fbox{\begin{picture}(425,150)\footnotesize\put(0,0){\makebox(425,150){%
\hspace*{\fill}\raisebox{-4pt}{$\begin{array}{c|cccc}
T&a&b&c&d\\\hline
a&a&b&a&b\\
b&a&b&a&b\\
c&c&d&c&d\\
d&c&d&c&d
\end{array}$}\hfill\raisebox{-60pt}[60pt][60pt]{\begin{picture}(160,120)
        \put(0,0){\makebox(0,0){$a$}}
        \put(40,0){\makebox(0,0){$d$}}
        \put(80,0){\makebox(0,0){$c$}}
        \put(120,0){\makebox(0,0){$b$}}
        \put(160,0){\makebox(0,0){$b$}}
        \put(20,20){\makebox(0,0){$b$}}
        \put(60,20){\makebox(0,0){$c$}}
        \put(100,20){\makebox(0,0){$d$}}
        \put(140,20){\makebox(0,0){$b$}}
        \put(0,40){\makebox(0,0){$b$}}
        \put(40,40){\makebox(0,0){$a$}}
        \put(80,40){\makebox(0,0){$d$}}
        \put(120,40){\makebox(0,0){$d$}}
        \put(160,40){\makebox(0,0){$a$}}
        \put(20,60){\makebox(0,0){$a$}}
        \put(60,60){\makebox(0,0){$b$}}
        \put(100,60){\makebox(0,0){$d$}}
        \put(140,60){\makebox(0,0){$c$}}
        \put(0,80){\makebox(0,0){$c$}}
        \put(40,80){\makebox(0,0){$b$}}
        \put(80,80){\makebox(0,0){$b$}}
        \put(120,80){\makebox(0,0){$c$}}
        \put(160,80){\makebox(0,0){$d$}}
        \put(20,100){\makebox(0,0){$d$}}
        \put(60,100){\makebox(0,0){$b$}}
        \put(100,100){\makebox(0,0){$a$}}
        \put(140,100){\makebox(0,0){$d$}}
        \put(0,120){\makebox(0,0){$d$}}
        \put(40,120){\makebox(0,0){$d$}}
        \put(80,120){\makebox(0,0){$a$}}
        \put(120,120){\makebox(0,0){$b$}}
        \put(160,120){\makebox(0,0){$c$}}
        \multiput(3,83)(20,20){2}{\qbezier(0,0)(7,7)(14,14)}
        \multiput(3,43)(20,20){4}{\qbezier[7](0,0)(7,7)(14,14)}
        \multiput(3,3)(20,20){6}{\qbezier[7](0,0)(7,7)(14,14)}
        \multiput(43,3)(20,20){6}{\qbezier(0,0)(7,7)(14,14)}
        \multiput(83,3)(20,20){4}{\qbezier(0,0)(7,7)(14,14)}
        \multiput(123,3)(20,20){2}{\qbezier[7](0,0)(7,7)(14,14)}
        \multiput(37,3)(-20,20){2}{\qbezier(0,0)(-7,7)(-14,14)}
        \multiput(77,3)(-20,20){4}{\qbezier[7](0,0)(-7,7)(-14,14)}
        \multiput(117,3)(-20,20){6}{\qbezier(0,0)(-7,7)(-14,14)}
        \multiput(157,3)(-20,20){6}{\qbezier(0,0)(-7,7)(-14,14)}
        \multiput(157,43)(-20,20){4}{\qbezier[7](0,0)(-7,7)(-14,14)}
        \multiput(157,83)(-20,20){2}{\qbezier(0,0)(-7,7)(-14,14)}
\end{picture}}\hfill\raisebox{-4pt}{$\begin{array}{c|cccc}
+&a&b&c&d\\\hline
a&a&b&c&d\\
b&b&a&d&c\\
c&c&d&a&b\\
d&d&c&b&a
\end{array}$}\hspace*{\fill}
}}\end{picture}}\linespread{1}\caption[A rule with four noninteracting types of
particles]{A cellular automaton rule $T$ with four noninteracting types of
particles that move diagonally and do not change direction. The rule table is
given at left in product form (see Figure~\ref{calaws}). At center, a portion
of an evolution is displayed (time moves up). The two types of particles that
move up the diagonal to the right are displayed by solid and dotted lines, and
similarly for the two that move up the diagonal to the left. The vertex color
specifies which types of particles cross at that point; for instance, $b$ means
that a dotted line going up and to the right crosses a solid line going up and
to the left. The right-hand table shows one way to interpret the four colors as
a four-group. Each space in \x{4} having $w$ cells can be interpreted as being
the direct sum of $w$ copies of this four-group. Since $T(x+y)=T(x)+T(y)$ for
any space $x$ and $y$ of width $w$, $T$ is said to be ``linear.'' It is natural
in this setting to interpret the dotted-line particles as background (since $a$
is the identity element for this choice of group). One associates a
two-dimensional field with each cell at time $t$ by setting the colors of all
other cells at time $t$ to the identity element and then evolving the
resulting space using $T$. Here the field associated with a cell is given by
the solid-line particles (if any) that pass through the cell. The entire
evolution of the cellular automaton may be obtained by summing all
fields of cells at time $t$.}
\label{rule2}
\end{figure}

Another common occurrence is that there exists a constant $c$ such that $p_x$
divides $cw_x$ for each $x$ (here $w_x$ is width of $x$). A simple example of
this (with $c=1$) is given by rule $C$ from Figure~\ref{calaws}. This rule may
be interpreted as containing four kinds of particles (see Figure~\ref{rule2}).
Each diagonal line in the lattice contains a particle. Two particles always
travel up and to the right, and the other two travel up and to the left. The
particles do not interact. The state of a cell in the automaton records which
pair of particles are crossing at that point. Since there is no interaction,
after $w_x$ time steps the system returns to its original state. Clearly one
may generalize this example so that there are $m$ particles that travel up and
to the right and $n$ that travel up and to the left; each such system is a
reversible cellular automaton.

In another class of cellular automata, there is a constant $c$ such that, if
one examines $p_x$ for many random initial states $x$, the values of $p_x$
cluster around $cw_x$. In several cases of this sort I found that these systems
contain solitons, and that the solitons completely describe the
system. Typically there are several types of solitons, with some types going to
the left and others to the right. When these particles cross there is a slight
jog in their trajectories and then they continue onward in their original
directions. The value of $p_x$ in such cases tends to be a linear function of
the number of solitons of the various types. For random initial conditions,
these numbers tend to cluster around a certain value which is proportional to
the width. I believe that this situation is probably characteristic of those
systems whose periods display this property. An example is given by rule $E$
from Figure~\ref{calaws}. The behavior of this system is analyzed in detail in
Chapter~\ref{invar}, where I examine the existence of constants of motion and
particles in combinatorial spacetimes.

In other cases, there is typically no obvious pattern to the values of $p_x$. I
have found one case, however, in which, for a certain restricted set of initial
conditions, the period is an exponential function of the width. This case is
rule $F$ from Figure~\ref{calaws}. If the initial state $x$ contains a
single $b$ against a background of $a$'s, then empirically I have found that
$p_x=(5)(2^{w_x-2})-1$. For large enough widths, the appearance of this system
is that of a river against a fixed background. The river very gradually
widens. Periodically a region of $a$'s appears in the river, widening to a
maximum size and then shrinking to zero. My guess based on observations is that
this system is counting in base two: the size of the river is related to the
number of nonzero bits needed to represent the number; the number is
incremented every five time steps; the region of $a$'s are the carry bits being
transmitted locally throughout the system (the larger the region, the farther
the carry bits have to be transmitted). But I have not worked out the details.

A property which is useful in analyzing the behavior of some laws is that of
\emph{linearity} (this is also referred to as \emph{additivity} in the
literature). For some discussion of this property (and other references),
see~\cite{me}. The idea is that, if $\X(w)=\{x\in\X\,|$ x has width~$w\}$, then
it may be possible to define each $\X(w)$ to be an abelian group in such a way
that the law of evolution $T$ acts linearly on these groups; i.e., if
$x\in\X(w)$ and $y\in\X(w)$ then $T(x+y)=T(x)+T(y)$. (Note that this definition
only works for cellular automata, since otherwise there is no guarantee that
$T(x)$ and $T(y)$ have the same width.) If the set of spaces is \x{k} then the
easiest way to obtain a group structure is to define an abelian group operation
on $K$ and then considering the set of spaces of width $w$ to be the direct sum
$K^w$. In linear rules there is no interaction; spacetime may be considered to
be a superposition of fields, where each field is associated with a cell in a
preselected Cauchy surface. One example has already been presented in
Figure~\ref{rule2}; two more are given in Figure~\ref{linear}.
\begin{figure}
\fbox{\begin{picture}(425,222)\footnotesize\put(0,0){\makebox(425,222){%
$\begin{array}{cccccccccc}\begin{array}{c|cccc}
&a&b&c&d\\\hline
a&a&c&a&c\\
b&a&c&a&c\\
c&d&b&d&b\\
d&d&b&d&b\\
\multicolumn{5}{c}{\,}\\
\multicolumn{5}{c}{\,}\\
&a&b&c&d\\\hline
a&a&b&c&d\\
b&b&a&d&c\\
c&c&d&a&b\\
d&d&c&b&a
\end{array}&&&&&\setlength{\wod}{3.5pt}\begin{array}{c\hw c\hw c\hw c\hw c\hw c\hw c}
c& &b& &c& &d\\
 &d& &b& &d\\
c& &c& &d\\
 &b& &d\\
c& &d\\
 &d\\
c&\\
 &b\\
c& &b\\
 &d& &b\\
c& &c& &b\\
 &b& &d& &b\\
c& &d& &c& &b
\end{array}&&&&\setlength{\wod}{0pt}\begin{array}{c\hw c\hw c\hw c\hw c\hw c\hw c\hw c\hw c\hw c\hw
c\hw c\hw c\hw c\hw c\hw c\hw c\hw c\hw c\hw c\hw c\hw c\hw c\hw c\hw c}
1& &0& &0& &0& &1& &0& &0& &0& &1& &0& &0& &0& &1\\
 &1& &1& &1& &1& &0& &0& &0& &0& &1& &1& &1& &1\\
 & &1& &0& &1& &0& &0& &0& &0& &0& &1& &0& &1\\
 & & &1& &1& &0& &0& &0& &0& &0& &0& &1& &1\\
 & & & &1& &0& &0& &0& &0& &0& &0& &0& &1\\
 & & & & &1& &1& &1& &1& &1& &1& &1& &1\\
 & & & & & &1& &0& &1& &0& &1& &0& &1\\
 & & & & & & &1& &1& &0& &0& &1& &1\\
 & & & & & & & &1& &0& &0& &0& &1\\
 & & & & & & & & &1& &1& &1& &1\\
 & & & & & & & & & &1& &0& &1\\
 & & & & & & & & & & &1& &1\\
 & & & & & & & & & & & &1
\end{array}\end{array}$}}
\end{picture}}\linespread{1}\caption[Two linear systems]{Two
linear systems. The upper left-hand table is the product form of law $B$ from
Figure~\ref{calaws}. This rule is linear if \x{4} is given the structure of a
four-group as shown in the lower left-hand table. Here $a$ is the identity. A
portion of the fields associated with $b$, $c$ and $d$ are displayed in the
center diagram (time moves up). One should view this diagram as extending to
the left and right indefinitely; all cells not shown contain $a$'s. (Since for
each $k\in\{b,c,d\}$ this diagram contains a row containing one $k$ and the
rest $a$'s, it describes all three fields.) The rightmost diagram describes a
portion of the field associated with 1 for the $d=2$ cellular automaton on
\z{2} where the law of evolution is addition mod 2. The law is linear on
\z{2}. All entries not displayed are 0. This rule is not invertible (it is a
2-1 map); hence the field only extends in the forward-time direction. Note that
the numbers displayed are the binomial coefficients mod 2. It turns out that
these two systems are closely related (see Chapter~\ref{cauchy}).}
\label{linear}
\end{figure}

The definition of linearity given above is one which, in my view, needs to be
generalized. If $T$ is linear then we ought to be able to define linearity so
that $UTU^{-1}$ is also linear for any $U$. At present this is not the case,
since $UTU^{-1}$ may not be a cellular automaton. We could specify that any law
$T$ is to be called linear if and only if there exists a $U$ such that
$UTU^{-1}$ is a linear cellular automaton according to the above
definition. However, this will only work if whenever $T$ and $UTU^{-1}$ are
both cellular automata then $T$ is linear if and only if $UTU^{-1}$ is. I doubt
very much that this is the case. In addition, I have found rules which have no
interaction and which have the superposition property but which do not seem to
be linear according to the present definition. It would seem that this
superposition property is the important one dynamically. For this property to
hold, there is no reason why $\X(w)$ needs to be a group; for instance, it
might be a subset of a group, or perhaps just a commutative semigroup. More
work needs to be done here.

In the case of cellular automata, it is easy to create a two-dimensional
display of the evolving spacetime on a computer screen. There is a limit to how
much such a display can reveal. For instance, one typically cannot see
explicitly the particles in a system on the screen unless one makes a special
effort to reveal them. A particle is typically encoded in any one of a number
of color patterns which may stretch across several cells; this is difficult to
pick out visually. However, when I am able to detect mathematically that a
system contains particles, the visual display usually has a certain sort of
look: it has the appearance of lines streaming along at some angle, and other
lines streaming along at other angles, with the lines crossing and interacting.

In addition to these particle-containing systems, there are many systems which
have this same sort of look but for which I have not been able to find
particles. My guess is that there is a way to analyze these systems in terms of
particles. At present my technology for finding particles in systems is quite
limited, so it is easily possible that particles exist in these systems whose
detection is beyond this technology. Also, it may be that my working
mathematical definition of particle is not broad enough to adequately capture
the phenomenon.

Besides the ``particle look'' described above, there is another sort of visual
appearance that reversible cellular automata can have. Here there are no
streaming lines, or if there are then they are interrupted by large gaps. These
gaps seem characteristic of such systems. They are regions of a particular
color, or having a particular pattern, which appear and disappear and which can
be arbitrarily large. In these systems I have rarely found particles, and I
have never been able to analyze such a system entirely in terms of
particles. In fact, up to this point I have not been able to analyze the
behavior of these systems at all.

Though it seems unwise to place too much reliance on visual clues, I do believe
the unadjusted visual appearance of these systems can contain useful hints as
to the type of system involved. In the case of those 1+1-dimensional
combinatorial spacetimes which are not cellular automata, however, such clues
are at present available to me only in a much cruder form. These systems are
not (at least not self-evidently) flat, so there is no easy way to display
their two-dimensional evolution on a computer screen. It would be useful to
develop a method to do this. For now I have settled for a one-dimensional
display (the state of the system at time $t$), as well as a readout of the
width of the state, the number of cells of each color, and other statistical
information.

Let $T:\X\mapsto\X$ be a cellular automaton law. Let $U:\X\mapsto\X'$ be an
equivalence map. Then the map $UTU^{-1}:\X'\mapsto\X'$ is a 1+1-dimensional
combinatorial spacetime, but it is quite often no longer a cellular automaton
(according to the usual definition of cellular automaton). That is: the width
of a space $x\in\X'$ might be different from the width of $UTU^{-1}(x)$. Thus
cellular-automatonness (in the traditional sense) is not an invariant property
of 1+1-dimensional combinatorial spacetimes. It is tempting to think of
cellular automata as ``flat,'' and as analogous to Minkowski space in general
relativity. The situation there is similar: there are metrics which are not the
Minkowski metric but which are equivalent to the Minkowski metric under a
change of variables. In order to make flatness an invariant property, one
defines all such metrics to be flat. Similarly, it is natural to define all
combinatorial spacetime rules $T:\X\mapsto\X$ to be flat whenever there exists
a $U:X\mapsto X'$ such that $UTU^{-1}$ is a reversible cellular automaton law.

In the case where $T$ is flat, every orbit of $T$ must be finite, since
$T\mapsto UTU^{-1}$ maps orbits of period $p$ into orbits of period $p$. In
many cases, it appears empirically that laws in which the width is not a
constant of motion do in fact have this property. We have already seen this in
the 1+1-dimensional example in Chapter~\ref{examples}. This example is extreme,
compared to many other examples, in that the orbits are very large and in that
there is great variation in size of spaces within an orbit. In every other case
I have seen so far, whenever the period is always finite then the orbits are
smaller and the widths do not change very much within an orbit. My guess is
that the example in Chapter~\ref{examples} cannot be transformed into a
cellular automaton, but that these other examples can. I succeeded in doing so
for one such example, which is illustrated in Figure~\ref{changetoca}.
\begin{figure}
\fbox{\begin{picture}(425,336)\footnotesize\put(0,0){\makebox(425,336){%
\hspace*{\fill}\begin{tabular}{c|c}
\textit{aa}&\textit{a}\\
\textit{ab}&\textit{-}\\
\textit{ac}&\textit{-}\\
\textit{ad}&\textit{-}\\
\textit{ba}&\textit{c}\\
\textit{bb}&\textit{b}\\
\textit{bc}&\textit{b}\\
\textit{bd}&\textit{b}\\
\textit{ca}&\textit{da}\\
\textit{cb}&\textit{d}\\
\textit{cc}&\textit{ba}\\
\textit{cd}&\textit{ba}\\
\textit{da}&\textit{baa}\\
\textit{db}&\textit{ba}\\
\textit{dc}&\textit{d}\\
\textit{dd}&\textit{d}\\
\end{tabular}\hfill\begin{tabular}{c|cccccccc|c}
\textit{a}&1&1&1&1&1&1&1&1&\textit{e}\\
\textit{ba}&0&1&1&1&1&1&1&1&\textit{f}\\
\textit{baa}&1&1&1&1&1&1&1&1&\textit{g}\\
\textit{b}&0&1&1&1&1&1&1&1&\textit{h}\\
\textit{ca}&1&1&1&1&1&1&1&1&\textit{i}\\
\textit{c}&0&1&1&1&1&1&1&1&\textit{j}\\
\textit{da}&1&1&1&1&1&1&1&1&\textit{k}\\
\textit{d}&0&1&1&1&1&1&1&1&\textit{l}\\
\end{tabular}\hfill\begin{tabular}{c|c}
\textit{e*}&$e$\\
\textit{f*}&$j$\\
\textit{g*}&$i$\\
\textit{h*}&$h$\\
\textit{i*}&$k$\\
\textit{jf}&$l$\\
\textit{jg}&$l$\\
\textit{jh}&$l$\\
\textit{ji}&$f$\\
\textit{jj}&$f$\\
\textit{jk}&$f$\\
\textit{jl}&$f$\\
\textit{k*}&$g$\\
\textit{lf}&$f$\\
\textit{lg}&$f$\\
\textit{lh}&$f$\\
\textit{li}&$l$\\
\textit{lj}&$l$\\
\textit{lk}&$l$\\
\textit{ll}&$l$\\
\end{tabular}%
\hspace*{\fill}}}\end{picture}}\linespread{1}\caption[Transforming an expanding
law into a cellular automaton]{Transforming an expanding law into a cellular
automaton. The left-hand table is a representation of the original law
$T_1:\x{4}\mapsto\x{4}$. The central table contains two nonoverlap descriptions
$D$ and $E$ associated with the same adjacency matrix $A$, given in the central
8 columns of this table. The left-hand and right-hand columns contain the
segments for descriptions $D$ and $E$, respectively. Let $U=\ang{D,E}$. The
right-hand table depicts a reversible cellular automaton law
$T_2:\X_E\mapsto\X_E$. (Note that each asterisk in this table stands for any
color that is allowed by the adjacency matrix to be in that position.) One may
verify that $\X_D=\x{4}$, and that $T_2=UT_1U^{-1}$.}
\label{changetoca}
\end{figure}

Just as it is not a simple matter to determine whether there exists a change of
variables that transforms a given metric into the Minkowski metric, I do not
know an algorithm to determine whether, for a given $T$, there exists a $U$
such that $UTU^{-1}$ is a reversible cellular automaton law. I suspect,
however, that the solution to the latter problem may be simpler than the
solution to the former one. The solution would seem to be related to the
problem of analyzing systems in terms of constants of motion (see
Chapter~\ref{invar}). The width is a constant of motion of a cellular automaton
$T$. The map $T\mapsto UTU^{-1}$ must send this constant of motion into another
constant of motion (which may no longer be the width). If we knew which
properties a constant of motion needed to have if it were to be a
transformation of the width constant of motion, and if we could detect whether
a system possessed such a constant of motion, this would seem to go a long way
towards determining whether the desired $U$ exists.

If it is not true that every orbit of a rule $T$ is periodic, then it seems
usually to be the case that almost all orbits of $T$ are infinite. (Sometimes
it seems that all of them are infinite.) In this case the easiest property of
the system to study is the function $w_x(t)$, that is, the width of the system
at time $t$ when the initial state is $x$.

A commonly occurring phenomenon is that, for large enough $|t|$, there exists a
$c_x$ such that $w_x(t+1)=w_x(t)+c_x$. A simple example is the first rule
pictured in Figure~\ref{rules11and15}.
\begin{figure}
\fbox{\begin{picture}(425,214)\footnotesize\put(0,0){\makebox(425,214){%
$\begin{array}{cccccccccc}\begin{array}{c|c}
\mathit{aa}&a\\
\mathit{ab}&\mathrm{-}\\
\mathit{ac}&c\\
\mathit{ba}&b\\
\mathit{bb}&\mathit{ca}\\
\mathit{bc}&\mathit{cac}\\
\mathit{ca}&\mathit{aa}\\
\mathit{cb}&\mathrm{-}\\
\mathit{cc}&c
\end{array}&&\begin{array}{c}
\cir{caaaacaaaaaaa}\\
\cir{caaacaaaaaa}\\
\cir{caacaaaaa}\\
\cir{cacaaaa}\\
\cir{bcaaa}\\
\cir{bacaa}\\
\cir{baaca}\\
\cir{baaab}\\
\cir{baaaaba}\\
\cir{baaaaabaa}\\
\cir{baaaaaabaaa}\\
\cir{baaaaaaabaaaa}\\
\cir{baaaaaaaabaaaaa}
\end{array}&&&&&\begin{array}{c|c}
\mathit{aa}&c\\
\mathit{ab}&\mathrm{-}\\
\mathit{ac}&b\\
\mathit{ba}&a\\
\mathit{bb}&\mathit{bc}\\
\mathit{bc}&\mathit{bcb}\\
\mathit{ca}&\mathit{cc}\\
\mathit{cb}&\mathrm{-}\\
\mathit{cc}&b
\end{array}&&\begin{array}{c}
\cir{bcbbcbcbbcbbc}\\
\cir{bcbbcbcb}\\
\cir{bcbbc}\\
\cir{bcb}\\
\cir{bc}\\
\cir{b}\\
\cir{c}\\
\cir{a}\\
\cir{ba}\\
\cir{cba}\\
\cir{acba}\\
\cir{baacba}\\
\cir{cbabaacba}
\end{array}\end{array}$}}
\end{picture}}\linespread{1}\caption[Two expanding 1+1-dimensional systems]{Two
expanding 1+1-dimensional systems. A $d=2$ nonoverlap description is given for
each rule, followed by a portion of an evolution. Time moves up.}
\label{rules11and15}
\end{figure}
This is the rule which first exchanges segments \seg{ba} and \seg{ca}, and then
exchanges segments \seg{b} and \seg{ca}. The action of this rule is
simple. Wherever there is a $b$ followed by one or more $a$'s, one of the $a$'s
is removed. Wherever there is a $b$ which is not followed by an $a$, it is
changed into \textit{ca}. Wherever there is a $c$ followed by one or more
$a$'s, another $a$ is inserted after the $c$. Hence there are finitely many
locations in a space where the law has an effect, and the number of locations
is preserved. The number $n$ of locations is equal to the number of $b$'s plus
the number of \textit{ca}'s. Each $b$ is eventually turned into a \textit{ca},
but the reverse does not occur. When all $b$'s have become \textit{ca}'s, from
then on the law adds an $a$ at each location; so from that point forward the
width increases by $n$ each time the law is applied. Similarly, if one goes
backwards in time eventually every location is associated with a $b$. From that
point backward, the width increases by $n$ each time the inverse of the law is
applied.

Another common occurrence is that, for large enough $|t|$ and certain initial
states $x$, $w_x(t)$ satisfies a linear recursion. An example is given by the
second rule pictured in Figure~\ref{rules11and15}. This rule (call it $T$) may
be obtained by first exchanging \seg{ba} and \seg{ca}, then exchanging \seg{b}
and \seg{ca}, and then performing an (\textit{acb}) permutation on colors. If
there exists a $t_0$ such that $T^{t_0}(x)$ has no $a$'s, then from then on the
rule acts by replacing \seg{b} with \seg{bc} and \seg{c} with \seg{b}. Let
$k_t$ be the number of $k$-colored cells in $T^t(x)$. Then if $t>t_0+1$ it
follows that $b_t=b_{t-1}+c_{t-1}=w_x(t-1)$ and
$c_t=b_{t-1}=b_{t-2}+c_{t-2}=w_x(t-2)$. Hence $w_t=b_t+c_t=w_x(t-1)+w_x(t-2)$;
so the widths satisfy the Fibonacci recursion. Similarly, if there is a $t_1$
such that $T^{t_1}(x)$ has no \seg{bc}'s and you proceed backwards in time,
from then on $T_{-1}$ acts by replacing \seg{c} with \seg{a}, \seg{b} with
\seg{c} and \seg{a} with \seg{ba}. If $t<t_1-2$ it follows that
$a_t=c_{t-1}+a_{t-1}$, $b_t=a_{t-1}$ and $c_t=b_{t-1}$, so
$b_t=c_{t-2}+a_{t-2}=b_{t-3}+c_{t-3}+a_{t-3}=w_x(t-3)$, so
$w_x(t)=a_t+b_t+c_t=c_{t-1}+a_{t-1}+w_x(t-3)+b_{t-1}=w_x(t-1)+w_x(t-3)$. So the
widths follow a different linear recursion in the reverse direction.

While these recursions may be present for many initial states $x$, there may be
exceptions. In this example, for instance, there exist many $x$ such that no
$t_0$ exists. If \seg{bcba} is present in $x$, then it must be present in
$T^t(x)$ for all $t$. Suppose there is a $t_2$ such that $T^{t_2}(x)$ has only
$b$'s and $c$'s except possibly for some $a$'s that are in \seg{bcba}
segments. From then on, $T$ acts as it does when $t>t_0$, with one exception:
if \seg{b} is followed by $a$ then, instead of replacing \seg{b} with \seg{bc},
\seg{ba} is replaced with \seg{a}. It follows that for $t>t_2+1$ we have
$a_t=a_{t-1}$, $b_t=b_{t-1}+c_{t-1}-a_t=w_x(t-1)-2a_{t_2}$, and
$c_t=b_{t-1}-a_t=b_{t-2}+c_{t-2}-2a_{t-2}=w_x(t-2)-3a_{t_2}$. Hence
$w_x(t)=w_x(t-1)+w_x(t-2)-4a_{t_2}$; so the widths follow a nonhomogeneous
linear recursion relation. I have not completed my analysis of $T$, but as far
as I know it is possible that there does exist such a $t_2$ for every $x$.

I have found many examples of linear recursions of this type; some are quite
complex. Though my focus here has been on recursions involving $w_x$, in doing
so I have also mentioned recursions involving numbers of colors. It may well be
that other types of recursion are often relevant, possibly involving larger
segments. At any rate, the role of linear recursion in these systems seems
worth exploring. Perhaps it already has been. For example, the map on strings
of $b$'s and $c$'s which sends \seg{b} to \seg{bc} and seg{c} to \seg{b} is
well known; it is called the \emph{Fibonacci morphism}. It has been studied in
the context of L systems (see \cite{berstel}). As mentioned earlier, most work
in L systems does not involve invertibility (and note that our rule $T$ is
necessarily more complex than the Fibonacci morphism because that morphism is
not invertible). However, here is an instance where the analysis of
noninvertible rules can shed light on at least some aspects of the behavior of
invertible rules.

In both rules pictured in Figure~\ref{rules11and15}, there is little or no
interaction. (Certainly there is no interaction in the first of these rules.)
Such systems resemble particle collisions: noninteracting particles approach
one another, collide, and the resulting particles disperse. (This resemblance
is clearly illusory in the first example, since there is in fact no
interaction; in other instances I suspect the description is more apt.)
Typically the sorts of spaces one encounters in the noninteracting past portion
of the evolution and in the noninteracting future portion of the evolution are
distinct in an obvious sort of way.

And then there are the rest of the rules, which are not flat and whose widths
display no obvious pattern. Here no doubt there is often very complex
behavior. I have not made a serious attempt to understand these rules, and have
been content simply to watch them and to observe their typical rates of
growth. It seems in general that rules either expand at a linear rate (e.g.,
the first rule in Figure~\ref{rules11and15}) or at an exponential rate (e.g.,
the second rule in Figure~\ref{rules11and15}). In the exponential case, if one
is able to determine that the expansion is based on a linear recursion then one
may calculate the rate precisely; otherwise I must depend on statistics, and it
is difficult to collect many of these since the systems quickly exceed the
bounds of my computer. It seems visually that the most complex systems are
those that expand exponentially at the slowest rates.

It seems likely that more progress in understanding expanding 1+1-dimensional
systems can be made by analyzing them in terms of constants of motion and their
associated particles. Algorithms for detecting invariants of this type will be
described in Chapter~\ref{invar}. Earlier I mentioned their usefulness in
studying flat systems; it turns out that such invariants are present in
expanding systems as well. Thus far I have detected the presence of these
invariants in many cases, but for the most part have not attempted to make use
of them.

Here ends my tour. Clearly these systems provide a very large (perhaps
inexhaustible) supply of research problems. Of course, if one's goal is to gain
physical insights about the real world then not all solutions to these problems
are likely to be important. Still, some experience with the analysis of the
behavior of individual 1+1-dimensional systems is likely to yield methodology
that is applicable to higher-dimensional cases.

\subsection{Transformations of space set descriptions}
\label{descriptrans}

The foregoing methodology for generating 1+1-dimensional laws of evolution
suffers from several limitations. While designing it, I had not yet realized
that allowing flexibility in gluing rules was useful (though this had become
more evident after analyzing the properties of the shift). In addition, it took
some time for me to see that allowing my sets of spaces to be subshifts of
finite type, rather than just \x{k}, was a natural thing to do and required
virtually no alteration of my formalism. Also, I did not consider maps
$T:\X\mapsto\X'$ where \X\ and $\X'$ were not the same, which turn out to be
important. Finally, I had not completely understood that the primary objects of
study ought to be the faithful descriptions themselves, rather than rule
tables.

My main breakthrough came after I began to consider whether there were any
natural transformations that changed a faithful description of \X\ into another
faithful description of \X. My hope was that, from a single faithful
description of \X\ and a few simple transformations, one could generate all
faithful descriptions of \X.

One benefit of this would be that my current procedures for generating rules
involved much trial and error, but this new one, if it existed, would not.
Every description generated from a faithful description of \X\ would be
guaranteed to be a faithful description of \X. The problem of finding rules
would be reduced to the problem of detecting which of these descriptions had
the same adjacency matrix.

The indices assigned to segments in a faithful description of \X\ are
arbitrary; hence one rather trivial invertible transformation is simply to
permute these indices. In terms of the adjacency matrix, this means performing
a permutation $\sigma$ on both the row indices and column indices
simultaneously. Similarly, if we allow the possibility that $A_{ij}>1$ (as I
shall do from now on), then the order of the gluing rules $g_{ij}^k$, $1\leq
k\leq A_{ij}$, is arbitrary; so another rather trivial invertible
transformation is to permute these gluing rules.

In addition, I soon came up with two nontrivial invertible transformations. I
call them \emph{splitting} and \emph{doubling}.

Let \ang{L,G,A} be a faithful description of \X, where the description is in
standard form with respect to the shift (each segment is extended to the left
and right as much as possible). Choose $i$. Let $F_i=\{g^k_{ij}\,|\,1\leq j\leq
|L|,\ 1\leq k\leq A_{ij}\}$. Then $F_i$ is the set of gluing rules in which
$L_i$ is the leftmost segment. Suppose that $H_1 \cup H_2=F_i$ and $H_1 \cap
H_2=\emptyset$. Then we may construct a new faithful description of \X\ as
follows. First, remove segment $L_i$ and replace it with two copies of itself,
$L_{i_1}$ and $L_{i_2}$. Next, if $g^k_{ii}$ is in $H_j$ ($j\in\{1,2\}$),
replace it with two copies of itself, $g^k_{i_ji_1}$ and $g^k_{i_ji_2}$. Next,
if $g^k_{im}$ is in $H_j$ ($j\in\{1,2\}$), replace it with one copy of itself,
$g^k_{i_jm}$. Next, replace each $g^k_{mi}$ with two copies of itself,
$g^k_{mi_1}$ and $g^k_{mi_2}$. Finally, extend $L_{i_1}$ and $L_{i_2}$ to the
left if possible to make sure the result is in standard form, and extend all
gluing rules in columns $i_1$ and $i_2$ to the left as needed so that they end
with $L_{i_1}$ and $L_{i_2}$. I call this a \emph{splitting}
transformation. The idea is simple: in the old decomposition of spaces in \X,
some copies of $L_i$ are glued to the segments that follow them using a rule in
$H_1$, and others are glued to the segments that follow them using a rule in
$H_2$; the transformation simply gives new names to these two groups of $L_i$
segments. In this case I say that the new transformation is obtained from the
old one by \emph{splitting to the right}; there is an analogous transformation,
which I call \emph{splitting to the left}, in which one focuses on gluing rules
in which $L_i$ is the rightmost segment, rather than the leftmost segment.

Here is an example:\bigskip\\
\footnotesize\begin{tabular}{c|ccc}
\multicolumn{4}{c}{\normalsize (A)}\\
\hline\seg{a}&\textit{aa}&\textit{ab}&\textit{ac}\\
\seg{b}&\textit{ba}&\textit{bb}&\textit{bc}\\
\seg{c}&\textit{ca}&\textit{cb}&\textit{cc}\\
\multicolumn{4}{c}\,\\
\multicolumn{4}{c}\,\\
\end{tabular}\hfill$\Longleftrightarrow$\hfill\begin{tabular}{c|cccc}
\multicolumn{5}{c}{\normalsize (B)}\\
\hline\seg{a}&\textit{aa}&\textit{aab}&0&\textit{ac}\\
\seg{ab}&0&0&\textit{ab}&0\\
\seg{b}&\textit{ba}&\textit{bab}&\textit{bb}&\textit{bc}\\
\seg{c}&\textit{ca}&\textit{cab}&\textit{cb}&\textit{cc}\\
\multicolumn{5}{c}\,\\
\end{tabular}\hfill$\Longleftrightarrow$\hfill\begin{tabular}{c|ccccc}
\multicolumn{6}{c}{\normalsize (C)}\\
\hline\seg{a}&\textit{aa}&\textit{aab}&0&0&\textit{ac}\\
\seg{ab}&0&0&\textit{ab}&0&0\\
\seg{ab}&\textit{aba}&\textit{abab}&0&\textit{abb}&\textit{abc}\\
\seg{b}&\textit{ba}&\textit{bab}&0&\textit{bb}&\textit{bc}\\
\seg{c}&\textit{ca}&\textit{cab}&0&\textit{cb}&\textit{cc}\\
\end{tabular}\normalsize\bigskip\\
(A) is the standard $d=1$ description of \x{3}. This is transformed by
splitting \seg{a} to the right, where $H_1=\{g_{11}^1,g_{13}^1\}$ and
$H_2=\{g_{12}^1\}$. This means that two copies of \seg{a} are made; however,
the second of these is always followed by a $b$, so it becomes \seg{ab}. The
fact that segment $L_{i_2}$ has been extended to the right means that each
gluing-rule segment in column $i_2$ of the matrix must also be extended to the
right so that it ends in \textit{ab} instead of in $a$. The resulting
description (B) is then transformed by splitting \seg{b} to the left, where
$H_1=\{g_{23}^1\}$ and $H_2=\{g_{33}^1,g_{43}^1\}$. Two copies of \seg{b} are
made, but the first of these is always preceded by an $a$, so it becomes
\seg{ab}; the gluing-rule segments in row $i_1$ are also extended to the left
so that they begin with \textit{ab} instead of with $b$. Notice that, in terms
of an adjacency matrix, a transformation which splits $L_i$ to the right is
obtained as follows: first obtain an $(n+1)\times n$ matrix by replacing row
$i$ with two adjacent rows $i_1$ and $i_2$ such that these rows sum to row $i$;
then obtain an $(n+1)\times (n+1)$ matrix by replacing column $i$ with two
adjacent columns $i_1$ and $i_2$ which are copies of column $i$.

Now again consider \ang{L,G,A}. Choose $i$. Then we may obtain a new faithful
description of \X\ as follows. First replace $L_i$ with two copies of itself,
$L_{i_1}$ and $L_{i_2}$. Next, replace each $g^k_{ji}$ with a copy of itself,
$g^k_{ji_1}$; and replace each $g^k_{ij}$ with a copy of itself,
$g^k_{i_2j}$. Finally, add a single gluing rule $g^1_{i_1i_2}=L_i$; in other
words, $L_{i_1}$ is followed only by $L_{i_2}$, and they are glued by
overlapping them completely. I call this a \emph{doubling} transformation. The
new description is the same as the old one, except that each occurrence of
$L_i$ in a decomposition of a space using the old description is replaced in
the new description by two successive copies of $L_i$ (named $L_{i_1}$ and
$L_{i_2}$), where these two copies of $L_i$ overlap completely.

Here is an example:\bigskip\\
\footnotesize\hspace*{\fill}\hspace*{\fill}\hspace*{\fill}\begin{tabular}{c|ccccc}
\multicolumn{6}{c}{\normalsize (C)}\\
\hline\seg{a}&\textit{aa}&\textit{aab}&0&0&\textit{ac}\\
\seg{ab}&0&0&\textit{ab}&0&0\\
\seg{ab}&\textit{aba}&\textit{abab}&0&\textit{abb}&\textit{abc}\\
\seg{b}&\textit{ba}&\textit{bab}&0&\textit{bb}&\textit{bc}\\
\seg{c}&\textit{ca}&\textit{cab}&0&\textit{cb}&\textit{cc}\\
\end{tabular}\hfill$\Longleftrightarrow$\hfill\begin{tabular}{c|cccc}
\multicolumn{5}{c}{\normalsize (D)}\\
\hline\seg{a}&\textit{aa}&\textit{aab}&0&\textit{ac}\\
\seg{ab}&\textit{aba}&\textit{abab}&\textit{abb}&\textit{abc}\\
\seg{b}&\textit{ba}&\textit{bab}&\textit{bb}&\textit{bc}\\
\seg{c}&\textit{ca}&\textit{cab}&\textit{cb}&\textit{cc}\\
\multicolumn{5}{c}\,\\
\end{tabular}\hspace*{\fill}\hspace*{\fill}\hspace*{\fill}\normalsize\bigskip\\
The segment \seg{ab} in (D) is doubled to obtain (C). In terms of adjacency
matrices, a doubling transformation involves inserting a row of zeros before
row $i$ and a column of zeros after column $i$, and then setting the entry at
the intersection of the new row and column to 1.

Splits and doublings are invertible (their inverses are called \emph{unsplits}
and \emph{un\-doublings}). The types of splitting and doubling operations that
can be performed on a faithful description are completely determined by its
adjacency matrix $A$. (Choose an $i$. Then for each way to divide the integers
from 1 to $A_{ij}$ into two groups for each $j$, there is a corresponding split
to the right; and for each way to divide the integers from 1 to $A_{ji}$ into
two groups for each $j$, there is a corresponding split to the left. Also, for
each $i$ there is a corresponding doubling.) However, it seems at first glance
that the same cannot be said of the inverse operations, at least in the case of
unsplits. To see this, consider the unsplit-to-the-left operation which goes
from (C) to (B). Here segments $L_3$ and $L_4$ are combined to form a single
segment. The new segment is determined by aligning $L_3$ and $L_4$ on the right
and seeing what they have in common as you scan to the left. In this case
$L_3=\seg{ab}$ and $L_4=\seg{b}$, so they have \seg{b} in common when they are
aligned on the right. The condition on the adjacency matrix which makes the
unsplit possible is that the third and fourth rows of this matrix be
identical. However, we also need an additional condition on the gluing
rules. To transform $L_3$ into the new segment \seg{b}, one removes $a$ from
the left. Similarly, one must remove $a$ from the left of each gluing rule in
row 3. ($L_4$ is the same as the new segment, so nothing needs to be done to
row 4.) In order for the unsplit operation to be performed, corresponding
gluing rules in rows 3 and 4 must now be identical. These become the gluing
rules that begin with \seg{b} in (B).

The adjacency matrix in (D) contains a (24) symmetry, by which I mean that one
may exchange indices 2 and 4 and leave the matrix invariant (since the second
and fourth rows are the same, and the second and fourth columns are the
same). Above I have shown that one may perform the following sequence of moves
on (D): first double the second segment to obtain (C), then unsplit segments 3
and 4 to the left to obtain (B), then unsplit segments 1 and 2 to the right to
obtain (A). Can we exchange indices 2 and 4 in (D) to produce a description
(D$'$), and then perform these same moves on (D$'$)? If so, then we would
arrive at a description whose adjacency matrix was the same as that of (A), and
these two descriptions of \x{3} would thus provide us with a law of
evolution. The first move is no problem; we simply double \seg{c} instead of
\seg{ab}:\bigskip\\
\footnotesize\hspace*{\fill}\hspace*{\fill}\hspace*{\fill}\begin{tabular}{c|cccc}
\multicolumn{5}{c}{\normalsize (D$'$)}\\
\hline\seg{a}&\textit{aa}&\textit{ac}&0&\textit{aab}\\
\seg{c}&\textit{ca}&\textit{cc}&\textit{cb}&\textit{cab}\\
\seg{b}&\textit{ba}&\textit{bc}&\textit{bb}&\textit{bab}\\
\seg{ab}&\textit{aba}&\textit{abc}&\textit{abb}&\textit{abab}\\
\multicolumn{5}{c}\,\\
\end{tabular}\hfill$\Longleftrightarrow$\hfill\begin{tabular}{c|ccccc}
\multicolumn{6}{c}{\normalsize (C$'$)}\\
\hline\seg{a}&\textit{aa}&\textit{ac}&0&0&\textit{aab}\\
\seg{c}&0&0&\textit{c}&0&0\\
\seg{c}&\textit{ca}&\textit{cc}&0&\textit{cb}&\textit{cab}\\
\seg{b}&\textit{ba}&\textit{bc}&0&\textit{bb}&\textit{bab}\\
\seg{ab}&\textit{aba}&\textit{abc}&0&\textit{abb}&\textit{abab}\\
\end{tabular}\hspace*{\fill}\hspace*{\fill}\hspace*{\fill}\normalsize\bigskip\\
Next we need to do an unsplit-to-the-left operation which combines segments 3
and 4. If we align these segments (\seg{c} and \seg{b}) on the right, we see
that they have nothing in common; hence the new segment must be \seg{\,}. This
means that we must shorten each gluing rule in row 3 by removing $c$ on the
left, and in row 4 by removing $b$ on the left. The resulting rows are the
same, so the unsplit may be indeed be performed:\bigskip\\
\footnotesize\hspace*{\fill}\hspace*{\fill}\hspace*{\fill}\begin{tabular}{c|ccccc}
\multicolumn{6}{c}{\normalsize (C$'$)}\\
\hline\seg{a}&\textit{aa}&\textit{ac}&0&0&\textit{aab}\\
\seg{c}&0&0&\textit{c}&0&0\\
\seg{c}&\textit{ca}&\textit{cc}&0&\textit{cb}&\textit{cab}\\
\seg{b}&\textit{ba}&\textit{bc}&0&\textit{bb}&\textit{bab}\\
\seg{ab}&\textit{aba}&\textit{abc}&0&\textit{abb}&\textit{abab}\\
\end{tabular}\hfill$\Longleftrightarrow$\hfill\begin{tabular}{c|cccc}
\multicolumn{5}{c}{\normalsize (B$'$)}\\
\hline\seg{a}&\textit{aa}&\textit{ac}&0&\textit{aab}\\
\seg{c}&0&0&$c$&0\\
\seg{\,}&\textit{a}&\textit{c}&\textit{b}&\textit{ab}\\
\seg{ab}&\textit{aba}&\textit{abc}&\textit{abb}&\textit{abab}\\
\multicolumn{5}{c}\,\\
\end{tabular}\hspace*{\fill}\hspace*{\fill}\hspace*{\fill}\normalsize\bigskip\\
Next we need to do an unsplit-to-the-right operation which combines segments 1
and 2. If we align these segments (\seg{a} and \seg{c}) on the left, again we
see that they have nothing in common; hence the new segment must be
\seg{\,}. When each gluing rule is shortened in column 1 by removing $a$ on the
right, and in column 2 by removing $c$ on the right, the resulting columns are
the same, so again the unsplit may be performed:\bigskip\\
\footnotesize\hspace*{\fill}\hspace*{\fill}\hspace*{\fill}\begin{tabular}{c|cccc}
\multicolumn{5}{c}{\normalsize (B$'$)}\\
\hline\seg{a}&\textit{aa}&\textit{ac}&0&\textit{aab}\\ \seg{c}&0&0&$c$&0\\
\seg{\,}&\textit{a}&\textit{c}&\textit{b}&\textit{ab}\\
\seg{ab}&\textit{aba}&\textit{abc}&\textit{abb}&\textit{abab}\\
\end{tabular}\hfill$\Longleftrightarrow$\hfill\begin{tabular}{c|ccc}
\multicolumn{4}{c}{\normalsize (A$'$)}\\
\hline\seg{\,}&\textit{a}&\textit{c}&\textit{aab}\\
\seg{\,}&1&\textit{b}&\textit{ab}\\
\seg{ab}&\textit{ab}&\textit{abb}&\textit{abab}\\
\multicolumn{4}{c}\,\\
\end{tabular}\hspace*{\fill}\hspace*{\fill}\hspace*{\fill}\normalsize\bigskip\\
(Recall that the entry 1 in the above table refers to the empty segment.) The
two descriptions (A) and (A$'$) have the same adjacency matrix. Hence they
determine an evolutionary law \ang{\mathrm{(A),(A}'\mathrm{)}}. One may verify
that this law is the one in which \seg{ab} and \seg{c} are exchanged. For
example, the space \cir{abbc} is constructed using (A) by arranging copies of
segments 1, 2, 2, 3 in a circle and then gluing them together (since each entry
in the adjacency matrix is 1, there is only one way to do this). If we do the
same thing using (A$'$), gluing segments 1 to 2 gives \seg{c}, gluing another
copy of segment 2 to the end of this gives \seg{cb}, gluing segment 3 to the
end of this gives \seg{cbab}, and gluing the beginning and ending segments
gives \cir{cbab}, which indeed is what one obtains from \cir{abbc} by replacing
\seg{ab} with \seg{c} and \seg{c} with \seg{ab}.

The law that exchanges \seg{ab} with \seg{ac} may be constructed in a similar
fashion. Beginning again with (A), we split the second segment to the left with
$H_1=\{g_{12}^1\}$, then split the fourth segment to the left with
$H_1=\{g_{14}^1\}$. The adjacency matrix of the resulting description has a
(24) symmetry. So we permute indices 2 and 4 and then perform the operations in
reverse: unsplit segments 4 and 5 to the left, then unsplit segments 1 and 2 to
the left. The resulting description (A$''$) has the same adjacency matrix as
(A), and the law \ang{\mathrm{(A),(A}''\mathrm{)}} is the one which exchanges
\seg{ab} with \seg{ac}. These transformations are shown in
Figure~\ref{descriptransforms}.
\begin{figure}\linespread{1.19}
\fbox{\begin{picture}(425,576)\footnotesize\put(222.5,0){\line(0,1){576}}
\put(4,12){\makebox(425,564)[bl]{%
\begin{tabular}{lc|ccccc}
&\seg{a}&\textit{aa}&\textit{ab}&\textit{ac}\\
(A)&\seg{b}&\textit{ba}&\textit{bb}&\textit{bc}\\
&\seg{c}&\textit{ca}&\textit{cb}&\textit{cc}\\
\,\\
&\seg{a}&\textit{aa}&\textit{aab}&0&\textit{ac}\\
&\seg{ab}&0&0&\textit{ab}&0\\
\raisebox{6pt}[0pt]{(B)}&\seg{b}&\textit{ba}&\textit{bab}&\textit{bb}&\textit{bc}\\
&\seg{c}&\textit{ca}&\textit{cab}&\textit{cb}&\textit{cc}\\
\,\\
&\seg{a}&\textit{aa}&\textit{aab}&0&0&\textit{ac}\\
&\seg{ab}&0&0&\textit{ab}&0&0\\
(C)&\seg{ab}&\textit{aba}&\textit{abab}&0&\textit{abb}&\textit{abc}\\
&\seg{b}&\textit{ba}&\textit{bab}&0&\textit{bb}&\textit{bc}\\
&\seg{c}&\textit{ca}&\textit{cab}&0&\textit{cb}&\textit{cc}\\
\,\\
&\seg{a}&\textit{aa}&\textit{aab}&0&\textit{ac}\\
&\seg{ab}&\textit{aba}&\textit{abab}&\textit{abb}&\textit{abc}\\
\raisebox{6pt}[0pt]{(D)}&\seg{b}&\textit{ba}&\textit{bab}&\textit{bb}&\textit{bc}\\
&\seg{c}&\textit{ca}&\textit{cab}&\textit{cb}&\textit{cc}\\
\,\\
&\seg{a}&\textit{aa}&\textit{ac}&0&\textit{aab}\\
&\seg{c}&\textit{ca}&\textit{cc}&\textit{cb}&\textit{cab}\\
\raisebox{6pt}[0pt]{(D$'$)}&\seg{b}&\textit{ba}&\textit{bc}&\textit{bb}&\textit{bab}\\
&\seg{ab}&\textit{aba}&\textit{abc}&\textit{abb}&\textit{abab}\\
\,\\
&\seg{a}&\textit{aa}&\textit{ac}&0&0&\textit{aab}\\
&\seg{c}&0&0&\textit{c}&0&0\\
(C$'$)&\seg{c}&\textit{ca}&\textit{cc}&0&\textit{cb}&\textit{cab}\\
&\seg{b}&\textit{ba}&\textit{bc}&0&\textit{bb}&\textit{bab}\\
&\seg{ab}&\textit{aba}&\textit{abc}&0&\textit{abb}&\textit{abab}\\
\,\\
&\seg{a}&\textit{aa}&\textit{ac}&0&\textit{aab}\\
&\seg{c}&0&0&$c$&0\\
\raisebox{6pt}[0pt]{(B$'$)}&\seg{\,}&\textit{a}&\textit{c}&\textit{b}&\textit{ab}\\
&\seg{ab}&\textit{aba}&\textit{abc}&\textit{abb}&\textit{abab}\\
\,\\
&\seg{\,}&\textit{a}&\textit{c}&\textit{aab}\\
(A$'$)&\seg{\,}&1&\textit{b}&\textit{ab}\\
&\seg{ab}&\textit{ab}&\textit{abb}&\textit{abab}\\
\end{tabular}}}
\put(230.5,12){\makebox(211.5,564)[bl]{%
\begin{tabular}{lc|ccccc}
&\seg{a}&\textit{aa}&\textit{ab}&\textit{ac}\\
(A)&\seg{b}&\textit{ba}&\textit{bb}&\textit{bc}\\
&\seg{c}&\textit{ca}&\textit{cb}&\textit{cc}\\
\,\\
&\seg{a}&\textit{aa}&\textit{ab}&0&\textit{ac}\\
&\seg{ab}&\textit{aba}&0&\textit{abb}&\textit{abc}\\
\raisebox{6pt}[0pt]{(E)}&\seg{b}&\textit{ba}&0&\textit{bb}&\textit{bc}\\
&\seg{c}&\textit{ca}&0&\textit{cb}&\textit{cc}\\
\,\\
&\seg{a}&\textit{aa}&\textit{ab}&0&\textit{ac}&0\\
&\seg{ab}&\textit{aba}&0&\textit{abb}&0&\textit{abc}\\
(F)&\seg{b}&\textit{ba}&0&\textit{bb}&0&\textit{bc}\\
&\seg{ac}&\textit{aca}&0&\textit{acb}&0&\textit{acc}\\
&\seg{c}&\textit{ca}&0&\textit{cb}&0&\textit{cc}\\
\,\\
&\seg{a}&\textit{aa}&\textit{ac}&0&\textit{ab}&0\\
&\seg{ac}&\textit{aca}&0&\textit{acb}&0&\textit{acc}\\
(F$''$)&\seg{b}&\textit{ba}&0&\textit{bb}&0&\textit{bc}\\
&\seg{ab}&\textit{aba}&0&\textit{abb}&0&\textit{abc}\\
&\seg{c}&\textit{ca}&0&\textit{cb}&0&\textit{cc}\\
\,\\
&\seg{a}&\textit{aa}&\textit{ac}&0&\textit{ab}\\
&\seg{ac}&\textit{aca}&0&\textit{acb}&\textit{acc}\\
\raisebox{6pt}[0pt]{(E$''$)}&\seg{b}&\textit{ba}&0&\textit{bb}&\textit{bc}\\
&\seg{\,}&\textit{a}&0&\textit{b}&\textit{c}\\
\,\\
&\seg{a}&\textit{aa}&\textit{ac}&\textit{ab}\\
(A$''$)&\seg{\,}&$a$&$b$&$c$\\
&\seg{\,}&$a$&$b$&$c$\\
\,\\
\,\\
\,\\
\,\\
\,\\
\,\\
\,\\
\,\\
\,\\
\,\\
\end{tabular}}}
\end{picture}}\linespread{1}\caption[The $\seg{ab}\Leftrightarrow\seg{c}$ and
$\seg{ab}\Leftrightarrow\seg{ac}$ transformations]{The
$\seg{ab}\Leftrightarrow\seg{c}$ and $\seg{ab}\Leftrightarrow\seg{ac}$
transformations.}
\label{descriptransforms}
\end{figure}

Note also that, if (G) is the description obtained from (A) by permuting
indices 1 and 2, then the law \ang{\mathrm{(A),(G)}} is the (\emph{ab}) color
permutation; and if (H) is the description obtained from (A) by permuting
indices 1 and 3, then the law \ang{\mathrm{(A),(H)}} is the (\emph{ac}) color
permutation.

Given the left-hand description (A), I have now described how to obtain, via a
series of transformations, the right-hand description of each of the four laws
which are composed to produce the example 1+1-dimensional law in
Chapter~\ref{examples}. Given these transformations, it turns out that we may
construct the right-hand description of the example law, given that its
left-hand description is (A), in a straightforward way. To do this, we need the
following theorems.

\begin{theorem}
\label{conjugates}
Let $T=\ang{D,E}$ be a law on \X, and let $f$ be a transformation of space
sets. Then $T=\ang{fD,fE}$ whenever $fD$ and $fE$ are well defined.
\end{theorem}
\textsc{Proof.} Let $L_i$ and $g_{ij}^k$ denote the segments and gluing rules
in $D$, and let $R_i$ and $h_{ij}^k$ denote the segments and gluing rules in
$E$. Let $U=\ang{fD,fE}$. Let $L'_i$ and $G_{ij}^k$ denote the segments and
gluing rules in $fD$, and let $R'_i$ and $H_{ij}^k$ denote the segments and
gluing rules in $fE$.

Suppose that $T(x)=y$. Then there exist finite sequences of indices $i_k$ and
$j_k$, $1\leq k\leq m$, such that
$(L_{i_1}g_{i_1i_2}^{j_1}L_{i_2}\dots L_{i_m}g_{i_mi_1}^{j_m})$ is the
decomposition of $x$ into segments and gluing rules under $D$, and
$(R_{i_1}h_{i_1i_2}^{j_1}R_{i_2}\dots R_{i_m}h_{i_mi_1}^{j_m})$ is the
decomposition of $y$ into segments and gluing rules under $E$.

Suppose that $f$ is given by a permutation $\sigma$ of segment indices. This
means that $L'_{\sigma(i)}=L_i$ and $G_{\sigma(i)\sigma(j)}^k=g_{ij}^k$, and
that $R'_{\sigma(i)}=R_i$ and $H_{\sigma(i)\sigma(j)}^k=h_{ij}^k$. So
$(L'_{\sigma(i_1)}G_{\sigma(i_1)\sigma(i_2)}^{j_1}L'_{\sigma(i_2)}\dots
L'_{\sigma(i_m)}G_{\sigma(i_m)\sigma(i_1)}^{j_m})$ is the decomposition of $x$
into segments and gluing rules under $fD$, and
$(R'_{\sigma(i_1)}H_{\sigma(i_1)\sigma(i_2)}^{j_1}R'_{\sigma(i_2)}\dots
R'_{\sigma(i_m)}H_{\sigma(i_m)\sigma(i_1)}^{j_m})$ is the decomposition of $y$
into segments and gluing rules under $fE$. Since the indices involved in these
two decompositions are identical, it follows that $U(x)=y$. A similar argument
shows that $U(x)=y$ if $f$ is a permutation of gluing rules $g_{ij}^k$ and
$h_{ij}^k$ for fixed $i$ and~$j$.

I have written the argument in detail for permutations to show how this sort of
argument works; now I will sketch the remainder of the proof, since the details
are cumbersome. Suppose that $f$ is a splitting to the right. This means that a
row index $q$ is chosen, along with sets $H_1$ and $H_2$. To produce the
decomposition of $x$ under $fD$ from the decomposition of $x$ under $D$ one
proceeds as follows: replace $g$ by $G$ and $L$ by $L'$; then replace
$L'_qG_{qj}^k$ by $L'_{q_r}G_{q_rj}^k$ if $g_{qj}^k\in H_r$
($r\in\{1,2\}$). The identical procedure is carried out to produce the
decomposition of $y$ under $fE$. Hence the indices in both decompositions are
the same, so $U(x)=y$. A similar argument holds if $f$ is a splitting to the
left. If $f$ is a doubling, this means that an index $q$ is chosen. To produce
the decomposition of $x$ under $fD$ from the decomposition of $x$ under $D$ one
proceeds as follows: replace $g$ by $G$ and $L$ by $L'$; replace $G_{iq}^k$ by
$G_{iq_1}^k$; replace $G_{qj}^k$ by $G_{q_2j}^k$; replace $L'_q$ by
$L'_{q_1}G_{q_1q_2}^1L'_{q_2}$. Again the identical procedure is carried out to
produce the decomposition of $y$ under $fE$; hence the indices in both
decompositions are the same, so $U(x)=y$.

Since each $f$ is invertible, it follows that $U(x)=y$ if $f$ is an unsplitting
or undoubling transformation. Since $T=U$ for each of the basic
transformations, it must also hold if $f$ is any composition of these
transformations.\proofend\bigskip\\ It follows from this that if
\ang{D,f^{-1}gfD} represents a law of evolution then \ang{fD,gfD} represents
the same law. As an application of this, note that
\ang{\mathrm{(D),(D}'\mathrm{)}}, \ang{\mathrm{(C),(C}'\mathrm{)}} and
\ang{\mathrm{(B),(B}'\mathrm{)}} are also representations of the law that
exchanges \seg{ab} with \seg{c}, and that \ang{\mathrm{(F),(F}''\mathrm{)}} and
\ang{\mathrm{(E),(E}''\mathrm{)}} are representations of the law that exchanges
\seg{ab} with~\seg{ac}.

\begin{theorem}
\label{composition}
Suppose that $f$ and $g$ are transformations of space set descriptions, that
$fD$ and $fgD$ are defined, and that $T=\ang{fD,D}$ and $U=\ang{gD,D}$ are laws
of evolution. Then $TU=\ang{fgD,D}$.
\end{theorem}
\textsc{Proof.} Since $fD$ and $fgD$ are defined, it follows from the previous
theorem that $U=\ang{fgD,fD}$. The result follows from the fact that if
$U=\ang{D,E}$ and $T=\ang{E,F}$ are laws of evolution then
$TU=\ang{D,F}$.\proofend\bigskip\\
As a corollary, note that if $T=\ang{D,fD}$ and $U=\ang{D,gD}$ then
$T^{-1}=\ang{fD,D}$ and $U^{-1}=\ang{gD,D}$, so $U^{-1}T^{-1}=\ang{gfD,D}$, so
$TU=\ang{D,gfD}$. That is our situation here. Suppose that the example law is
given by \ang{\mathrm{(A),(U)}}. To obtain (U), we begin with (A) and apply
transformations in reverse order. Hence we begin with the transformation
associated with the (\emph{ab}) color permutation; in other words, we exchange
indices 1 and 2. Next we perform the transformation associated with the
(\emph{ac}) color permutation; i.e., we exchange indices 1 and 3. Next we
perform the transformations associated with the law that exchanges \seg{ab}
with \seg{c}, and finally we perform the transformations associated with the
law that exchanges \seg{ab} with \seg{ac}. These transformations are shown in
Figure~\ref{exampledescrip}.
\begin{figure}\linespread{1.19}
\fbox{\begin{picture}(425,564)\footnotesize\put(207.5,0){\line(0,1){564}}
\put(-1,12){\makebox(425,552)[bl]{%
\begin{tabular}{lc|ccccc}
&\seg{a}&\textit{aa}&\textit{ab}&\textit{ac}\\
(A)&\seg{b}&\textit{ba}&\textit{bb}&\textit{bc}\\
&\seg{c}&\textit{ca}&\textit{cb}&\textit{cc}\\
\,\\
&\seg{b}&\textit{bb}&\textit{ba}&\textit{bc}\\
(G)&\seg{a}&\textit{ab}&\textit{aa}&\textit{ac}\\
&\seg{c}&\textit{cb}&\textit{ca}&\textit{cc}\\
\,\\
&\seg{c}&\textit{cc}&\textit{ca}&\textit{cb}\\
(I)&\seg{a}&\textit{ac}&\textit{aa}&\textit{ab}\\
&\seg{b}&\textit{bc}&\textit{ba}&\textit{bb}\\
\,\\
&\seg{c}&\textit{cc}&\textit{cca}&0&\textit{cb}\\
&\seg{ca}&0&0&\textit{ca}&0\\
\raisebox{6pt}[0pt]{(J)}&\seg{a}&\textit{ac}&\textit{aca}&\textit{aa}&\textit{ab}\\
&\seg{b}&\textit{bc}&\textit{bca}&\textit{ba}&\textit{bb}\\
\,\\
&\seg{c}&\textit{cc}&\textit{cca}&0&0&\textit{cb}\\
&\seg{ca}&0&0&\textit{ca}&0&0\\
(K)&\seg{ca}&\textit{cac}&\textit{caca}&0&\textit{caa}&\textit{cab}\\
&\seg{a}&\textit{ac}&\textit{aca}&0&\textit{aa}&\textit{ab}\\
&\seg{b}&\textit{bc}&\textit{bca}&0&\textit{ba}&\textit{bb}\\
\,\\
&\seg{c}&\textit{cc}&\textit{cca}&0&\textit{cb}\\
&\seg{ca}&\textit{cac}&\textit{caca}&\textit{caa}&\textit{cab}\\
\raisebox{6pt}[0pt]{(L)}&\seg{a}&\textit{ac}&\textit{aca}&\textit{aa}&\textit{ab}\\
&\seg{b}&\textit{bc}&\textit{bca}&\textit{ba}&\textit{bb}\\
\,\\
&\seg{c}&\textit{cc}&\textit{cb}&0&\textit{cca}\\
&\seg{b}&\textit{bc}&\textit{bb}&\textit{ba}&\textit{bca}\\
\raisebox{6pt}[0pt]{(M)}&\seg{a}&\textit{ac}&\textit{ab}&\textit{aa}&\textit{aca}\\
&\seg{ca}&\textit{cac}&\textit{cab}&\textit{caa}&\textit{caca}\\
\,\\
&\seg{c}&\textit{cc}&\textit{cb}&0&0&\textit{cca}\\
&\seg{b}&0&0&\textit{b}&0&0\\
(N)&\seg{b}&\textit{bc}&\textit{bb}&0&\textit{ba}&\textit{bca}\\
&\seg{a}&\textit{ac}&\textit{ab}&0&\textit{aa}&\textit{aca}\\
&\seg{ca}&\textit{cac}&\textit{cab}&0&\textit{caa}&\textit{caca}\\
\end{tabular}}}
\put(210.5,12){\makebox(211.5,552)[bl]{%
\begin{tabular}{lc|ccccc}
&\seg{c}&\textit{cc}&\textit{cb}&0&\textit{cca}\\
&\seg{b}&0&0&$b$&0\\
\raisebox{6pt}[0pt]{(O)}&\seg{\,}&\textit{c}&\textit{b}&\textit{a}&\textit{ca}\\
&\seg{ca}&\textit{cac}&\textit{cab}&\textit{caa}&\textit{caca}\\
\,\\
&\seg{\,}&\textit{c}&\textit{b}&\textit{cca}\\
(P)&\seg{\,}&1&\textit{a}&\textit{ca}\\
&\seg{ca}&\textit{ca}&\textit{caa}&\textit{caca}\\
\,\\
&\seg{\,}&\textit{c}&\textit{b}&0&\textit{cca}\\
&\seg{b}&\textit{b}&0&\textit{ba}&\textit{bca}\\
\raisebox{6pt}[0pt]{(Q)}&\seg{a}&\textit{a}&0&\textit{aa}&\textit{aca}\\
&\seg{ca}&\textit{ca}&0&\textit{caa}&\textit{caca}\\
\,\\
&\seg{\,}&\textit{c}&\textit{b}&0&\textit{cca}&0\\
&\seg{b}&\textit{b}&0&\textit{ba}&0&\textit{bca}\\
(R)&\seg{a}&\textit{a}&0&\textit{aa}&0&\textit{aca}\\
&\seg{cca}&\textit{cca}&0&\textit{ccaa}&0&\textit{ccaca}\\
&\seg{ca}&\textit{ca}&0&\textit{caa}&0&\textit{caca}\\
\,\\
&\seg{\,}&\textit{c}&\textit{cca}&0&\textit{b}&0\\
&\seg{cca}&\textit{cca}&0&\textit{ccaa}&0&\textit{ccaca}\\
(S)&\seg{a}&\textit{a}&0&\textit{aa}&0&\textit{aca}\\
&\seg{b}&\textit{b}&0&\textit{ba}&0&\textit{bca}\\
&\seg{ca}&\textit{ca}&0&\textit{caa}&0&\textit{caca}\\
\,\\
&\seg{\,}&\textit{c}&\textit{cca}&0&\textit{b}\\
&\seg{cca}&\textit{cca}&0&\textit{ccaa}&\textit{ccaca}\\
\raisebox{6pt}[0pt]{(T)}&\seg{a}&\textit{a}&0&\textit{aa}&\textit{aca}\\
&\seg{\,}&1&0&\textit{a}&\textit{ca}\\
\,\\
&\seg{\,}&\textit{c}&\textit{cca}&\textit{b}\\
(U)&\seg{a}&$a$&\textit{aa}&\textit{aca}\\
&\seg{\,}&1&$a$&\textit{ca}\\
\,\\
\,\\
\,\\
\,\\
\end{tabular}}}
\end{picture}}\linespread{1}\caption[Obtaining the 1+1-dimensional law in
Chapter~\ref{examples} via space set transformations]{Obtaining the
1+1-dimensional law in Chapter~\ref{examples} via space set
transformations. The law is represented by \ang{\mathrm{(A),(U)}}.}
\label{exampledescrip}
\end{figure}

So far, we have been lucky: every time we have attempted to do an unsplit
transformation (given that the necessary condition on the adjacency matrix was
satisfied), we have been successful. Now I will give an example where this does
not happen. Consider the following description of \x{3}:\bigskip\\
\footnotesize\hspace*{\fill}\begin{tabular}{c|c}
\multicolumn{2}{c}{\normalsize (V)}\\
\hline\seg{\,}&$a+b+c$\\
\end{tabular}\hspace*{\fill}\normalsize\bigskip\\
Suppose that we wish to express the law that exchanges \seg{ab} with \seg{ac}
using (V) as the left-hand description. Theorem~\ref{conjugates} gives us a
means to attempt to do this. Suppose we can find a transformation $f$ that maps
(V) to (A). We already have a transformation $g$ which maps (A) to (A$''$). The
theorem says that \ang{\mathrm{(A)},\mathrm{(A}''\mathrm{)}}, which represents
the desired law, is equal to
\ang{f^{-1}\mathrm{(A)},f^{-1}\mathrm{(A}''\mathrm{)}} if
$f^{-1}\mathrm{(A}''\mathrm{)}$ is defined. The latter representation is just
\ang{\mathrm{(V)},f^{-1}gf\mathrm{(V)}}, which is what we are after.

We may indeed obtain (A) from (V) by first splitting segment 1 to the left with
$H_1=\{g_{11}^1\}$, and then splitting segment 2 to the left with
$H_1=\{g_{12}^1,g_{22}^1\}$:\bigskip\\
\footnotesize\hspace*{\fill}\hspace*{\fill}\begin{tabular}{c|c}
\multicolumn{2}{c}{\normalsize (V)}\\
\hline\seg{\,}&$a+b+c$\\
\multicolumn{2}{c}\,\\
\multicolumn{2}{c}\,\\
\end{tabular}\hfill$\Longleftrightarrow$\hfill\begin{tabular}{c|cc}
\multicolumn{3}{c}{\normalsize (W)}\\
\hline\seg{a}&\textit{aa}&\textit{ab}$+$\textit{ac}\\
\seg{\,}&$a$&$b+c$\\
\multicolumn{3}{c}\,\\
\end{tabular}\hfill$\Longleftrightarrow$\hfill\begin{tabular}{c|ccc}
\multicolumn{4}{c}{\normalsize (A)}\\
\hline\seg{a}&\textit{aa}&\textit{ab}&\textit{ac}\\
\seg{b}&\textit{ba}&\textit{bb}&\textit{bc}\\
\seg{c}&\textit{ca}&\textit{cb}&\textit{cc}\\
\end{tabular}\hspace*{\fill}\hspace*{\fill}\normalsize\bigskip\\
Thus we have our $f$. Now we wish to see if it is possible to apply $f^{-1}$ to
(A$''$); if so, then we have our desired representation. To begin to attempt
this transformation, first we perform an unsplit-to-the-left operation that
combines segments 2 and 3. This can be done:\bigskip\\
\footnotesize\hspace*{\fill}\hspace*{\fill}\hspace*{\fill}\begin{tabular}{c|ccc}
\multicolumn{4}{c}{\normalsize (A$''$)}\\
\hline\seg{a}&\textit{aa}&\textit{ac}&\textit{ab}\\ \seg{\,}&$a$&$b$&$c$\\
\seg{\,}&$a$&$b$&$c$\\
\end{tabular}\hfill$\Longleftrightarrow$\hfill\begin{tabular}{c|cc}
\multicolumn{3}{c}{\normalsize (X)}\\
\hline\seg{a}&\textit{aa}&\textit{ac}$+$\textit{ab}\\
\seg{\,}&$a$&$b+c$\\
\multicolumn{3}{c}\,\\
\end{tabular}\hspace*{\fill}\hspace*{\fill}\hspace*{\fill}\normalsize\bigskip\\
Next we attempt to perform an unsplit-to-the-left operation that combines
segments 1 and 2. But now we are stuck. If we align segments 1 and 2 on the
right, we see that the new segment must be \seg{\,}, and that we must remove
$a$ from the left of each gluing-rule segment in row 1. But now rows 1 and 2
are supposed to be the same, and they are not: the second entry in row 1 is
$c+b$, and the second entry in row 2 is $b+c$. The order here is
important. Somehow we are supposed to merge the $c$ in the first row with the
$b$ in the second row, and vice versa. Our formalism does not allow this, so
the operation fails.

It is important to notice, however, that this operation fails only because the
formalism used here, though more general than the original no-overlap
formalism, is still too restrictive. There \emph{is} a sense in which this
operation can be carried out. The resulting adjacency matrix, as in (V), is a
$1\times 1$ matrix whose single entry is 3. This corresponds to a graph with
one vertex and three edges. The segment corresponding to the single vertex may
be taken to be the empty segment. The first edge corresponds to the color
$a$. The second edge corresponds to $c$ if the previous color is $a$, and to
$b$ otherwise. The third edge corresponds to $b$ if the previous color is $a$,
and to $c$ otherwise. One may easily verify that, if (Y) is interpreted in this
way, then it is indeed a faithful description of \x{3}, and
\ang{\mathrm{(V),(Y)}} is a representation of the law that exchanges \seg{ab}
with \seg{ac}.

In fact, the gluing-rule condition on unshift transformations can be
ignored. If the adjacency matrix satisfies the rules for an unshift
transformation, then one may make sense out of that transformation. One way to
do this is to further generalize space set descriptions, which means obtaining
an expanded view of what a vertex and edge may represent. While obtaining such
an expanded view is a worthwhile goal, there is also an easier way to
proceed. We may dispense with complicated descriptions of space sets, and work
only with adjacency matrices. This will be shown in the next section.

\subsection{Elementary maps and matrix moves}
\label{element}

Consider an $n\times n$ matrix $A$ of nonnegative integers. We may associate
any such matrix with a faithful description of a set of spaces \x{A} as
follows. Let each vertex in \g{A} be assigned the empty segment. Let each edge
$a_{ij}^k$ ($1\leq i,j\leq n$, $1\leq k\leq A_{ij}$) in \g{A} be assigned a
length-1 segment where the color of the cell in this segment is given simply by
the edge label $a_{ij}^k$, considered as an abstract symbol. Clearly any such
description is faithful, since every edge is colored and since there is only
one edge containing any given color. I will call descriptions of this sort
\emph{matrix descriptions}.

In order to avoid the subscripts and superscripts associated with the names
$a_{ij}^k$ for the colors in \x{A}, it will often be useful to give alternative
letter names to these colors. This may be done by writing down a gluing matrix
associated with $A$ containing these alternate names. In these gluing matrices,
since each gluing relation consists of a length-one segment, it is simpler to
omit plus signs. Hence, for instance, I might say that $A$ is the $1\times 1$
matrix $(abc)$; this simply means that $a$ stands for $a_{11}^1$, $b$ stands
for $a_{11}^2$, and $c$ stands for $a_{11}^3$. It will sometimes be convenient
to assign the same names to colors in different matrix descriptions. This
causes no problems; however, it is important that every edge in a given matrix
description be assigned a distinct color. To simplify notation, I will use the
same name $A$ to refer to an adjacency matrix and to its associated matrix
description; the intended usage should be clear from the context.

In this section, rather than focusing on the many ways in which one may
describe a space set, I will focus on equivalence maps between space sets given
by matrix descriptions. If \ang{L,G,A} is a description of a space set \X, and
if one considers the space set \x{A} given by the matrix description $A$, then
these two descriptions have the same adjacency matrix; hence \X\ and \x{A} are
locally equivalent. So we are not losing any equivalence classes of space sets
by restricting ourselves to matrix descriptions.

Let $f$ be one of the elementary transformations of space set descriptions
defined in the previous section: that is, it is a permutation of matrix
indices, or a permutation of gluing rules $g_{ij}^k$ for fixed $i$ and $j$, or
it is a splitting, doubling, unsplitting or undoubling transformation. For each
such transformation $f$ that maps a space set description $D$ with adjacency
matrix $A$ to a description $fD$ with adjacency matrix $B$, we will associate
an equivalence map $T:\x{A}\mapsto\x{B}$ where $T=\ang{fA,B}$. Note
that $T$ is always well defined, since if $A$ is a matrix description and $f$
is one of these elementary transformations then $fA$ is always well
defined.

As an example, let $A$ be the $1\times 1$ adjacency matrix (3), and let the
matrix description $A$ be given by the gluing matrix $(\mathit{abc})$. This is
exactly the description (V) given in the previous section. There we produced
(W) from (V) by performing a transformation $f$ which split segment 1 to the
left with $H_1=\{g_{11}^1\}$. The adjacency matrix $B$ for (W) is
{\footnotesize $\left(\begin{array}{cc}1&2\\1&2\end{array}\right)$}. Let
{\footnotesize $\left(\begin{array}{cc}d&\mathit{ef}\\a&\mathit{bc}\end{array}\right)$} be the
gluing matrix for the matrix description $B$. The map $T$ associated with $f$
is given by\bigskip\\
\hspace*{\fill}\settowidth{\wod}{\,\,\,}\begin{tabular}{c\hw c\hw c\hw c\hw
c\hw c\hw c}
$T$&$=$&\ang{f\!A,B}&$=$&\ang{\mathrm{(W)},B}&$=$&\footnotesize$\displaystyle\left<\begin{array}{c|cc}\seg{a}&\mathit{aa}&\mathit{ab}+\mathit{ac}\\\seg{\,}&a&b+c\end{array}\right.$\raisebox{-9pt}{,\hspace*{6pt}}$\displaystyle\left.\begin{array}{c|cc}\seg{\,}&d&e+f\\\seg{\,}&a&b+c\end{array}\right>$\normalsize\raisebox{-9pt}{.}\\
\end{tabular}\hspace*{\fill}\bigskip\\
$T$ acts by changing the color of a cell if it is preceded by an
$a$-colored cell; in this case it changes $a$ to $d$, $b$ to $e$ and $c$ to
$f$. Note that the fact that some of the colors in $A$ and $B$ are the same
simplifies the description of $T$, because it makes it possible for $T$ to
leave certain colors alone.

Now consider a sequence of elementary transformations $f_i$ mapping
\ang{L_i,G_i,A_i} to \ang{L_{i-1},G_{i-1},A_{i-1}}, $1\leq i\leq n$, where
$A_0=A_n=A$, $D=\ang{L_n,G_n,A}$ and \X=\x{D}. Then there is a law of evolution
$T=\ang{L_0,L_n,G_0,G_n,A}=\ang{f_1\dots f_nD,D}$ on \X. For each $f_i$ there
is an associated equivalence map $T_i:\x{A_i}\mapsto\x{A_{i-1}}$. Let
$U=\ang{D,A}$. Then $UTU^{-1}$ is a law of evolution on \x{A} that is
equivalent to $T$. The composition $U=T_1T_2\dots T_n$ is a map from \x{A} to
\x{A}; hence it is also a law of evolution on \x{A}.

\begin{theorem} If $T=\ang{f_1\dots f_nA,A}$ and $A_i$, $f_i$, $T_i$ and $U$
are as defined above for each $i$, then $UTU^{-1}=T_1\dots T_n$.
\end{theorem}
\textsc{Proof.} By definition, $T_i=\ang{f_iA_i,A_{i-1}}$, $1\leq i\leq
n$. Then \[T_i=\ang{f_if_{i+1}\dots f_nD,A_{i-1}}\ang{f_iA_i,f_if_{i+1}\dots
f_nD}\] by the rule of composition, since all descriptions involved are well
defined and since the descriptions paired in angle brackets share the same
adjacency matrix. By \mbox{Theorem}~\ref{conjugates} it follows that
$T_i=\ang{f_i\dots f_nD,A_{i-1}}\ang{A_i,f_{i+1}\dots f_nD}$. Therefore
\[\begin{array}{rcl}T_1\dots T_n&=&(\ang{f_1\dots f_nD,A_0}\ang{A_1,f_2\dots
f_nD})(\ang{f_2\dots f_nD,A_1}\ang{A_2,f_3\dots
f_nD})\\&&\dots(\ang{f_{n-1}f_nD,A_{n-2}}\ang{A_{n-1},f_nD})(\ang{f_nD,A_{n-1}}\ang{A_n,D}).\end{array}\]
Cancelling adjacent pairs of inverse maps gives $T_1\dots T_n=\ang{f_1\dots
f_nD,A_0}\ang{A_n,D}=\ang{D,A}\ang{f_1\dots f_nD,D}\ang{A,D}=UTU^{-1}$ (using
the rules for composition and the fact that $A_0=A_n=A$).\proofend\bigskip

The above theorem says that if $T$ is any law generated by elementary
transformations of space set descriptions, then $T$ is equivalent to a law
which is the composition of ``elementary maps'' of the form \ang{fA,A} (where
$A$ is a matrix description and $f$ is an elementary transformation of space
set descriptions). On the other hand, let $D$ be any description with adjacency
matrix $A=A_0=A_n$, let each $f_i$ ($1\leq i\leq n$) be an elementary
transformation which maps descriptions with adjacency matrix $A_i$ to
descriptions with adjacency matrix $A_{i-1}$ as above, and suppose that
\ang{f_1\dots f_nD,D} is not defined. In this case each rule
$T_i=\ang{f_iA_i,A_{i-1}}$ is still defined, so $T=T_1\dots T_n$ is a law of
evolution on \x{A}. This is why our inability noted in the previous section to
perform certain transformations of descriptions can be diagnosed as being the
result of a too restrictive definition of a description; the desired map is
indeed well defined in all such cases.

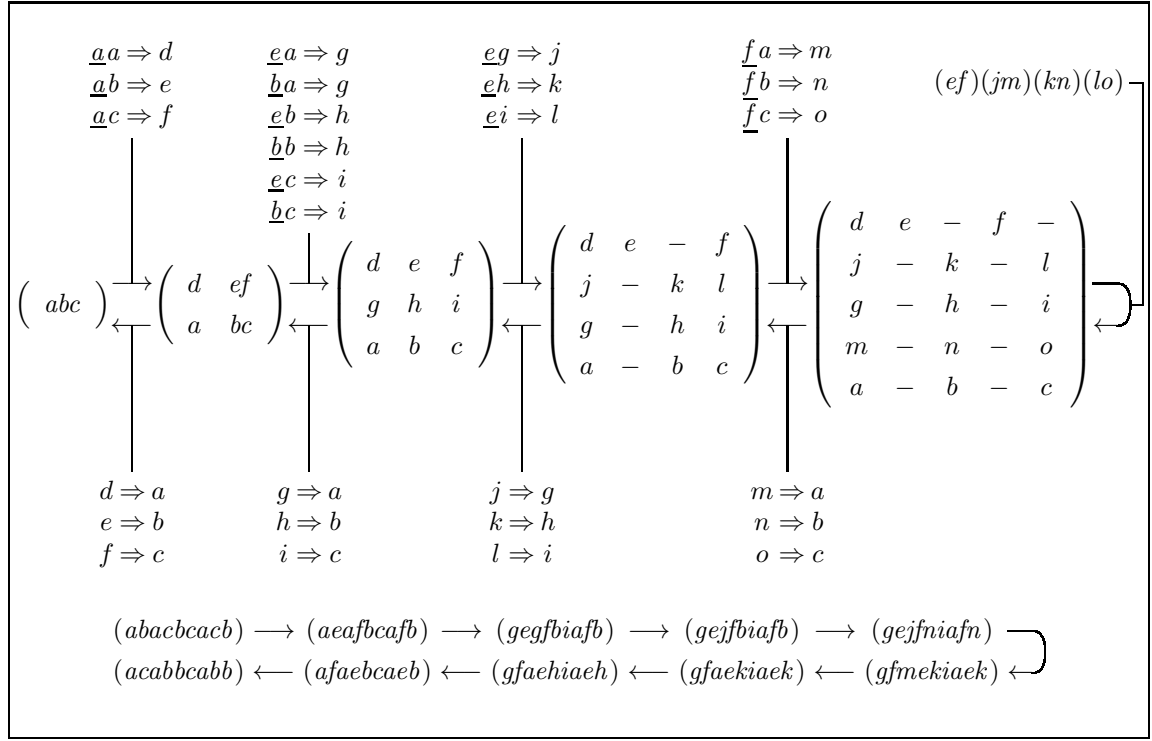
\begin{figure}
\fbox{\begin{picture}(425,272)\footnotesize\put(0,50){\makebox(425,222){%
\settowidth{\woda}{$\longrightarrow$}\setlength{\wod}{2pt}$\left(\begin{array}{c}
\mathit{abc}
\end{array}\right)$\raisebox{8pt}{$\longrightarrow$\hspace{-.5\woda}%
\begin{picture}(0,0)
\put(0,2.3){\line(0,1){55}}
\put(0,59.3){\linespread{1}\footnotesize\makebox(0,0)[b]{%
$\begin{array}{c\hw c\hw c}
\mathit{\underline{a}a}&\Rightarrow&d\\
\mathit{\underline{a}b}&\Rightarrow&e\\
\mathit{\underline{a}c}&\Rightarrow&f
\end{array}$}}\end{picture}\hspace{-.5\woda}}\raisebox{-8pt}{$\longleftarrow$\hspace{-.5\woda}%
\begin{picture}(0,0)
\put(0,2.3){\line(0,-1){55}}
\put(0,-54.7){\linespread{1}\footnotesize\makebox(0,0)[t]{%
$\begin{array}{c\hw c\hw c}
d&\Rightarrow&a\\
e&\Rightarrow&b\\
f&\Rightarrow&c
\end{array}$}}\end{picture}\hspace{.5\woda}}$\left(\begin{array}{cc}
d&\mathit{ef}\\
a&\mathit{bc}
\end{array}\right)$\raisebox{8pt}{$\longrightarrow$\hspace{-.5\woda}%
\begin{picture}(0,0)
\put(0,2.3){\line(0,1){19}}
\put(0,23.3){\linespread{1}\footnotesize\makebox(0,0)[b]{%
$\begin{array}{c\hw c\hw c}
\mathit{\underline{e}a}&\Rightarrow&g\\
\mathit{\underline{b}a}&\Rightarrow&g\\
\mathit{\underline{e}b}&\Rightarrow&h\\
\mathit{\underline{b}b}&\Rightarrow&h\\
\mathit{\underline{e}c}&\Rightarrow&i\\
\mathit{\underline{b}c}&\Rightarrow&i
\end{array}$}}\end{picture}\hspace{-.5\woda}}\raisebox{-8pt}{$\longleftarrow$\hspace{-.5\woda}%
\begin{picture}(0,0)
\put(0,2.3){\line(0,-1){55}}
\put(0,-54.7){\linespread{1}\footnotesize\makebox(0,0)[t]{%
$\begin{array}{c\hw c\hw c}
g&\Rightarrow&a\\
h&\Rightarrow&b\\
i&\Rightarrow&c
\end{array}$}}\end{picture}\hspace{.5\woda}}$\left(\begin{array}{ccc}
d&e&f\\
g&h&i\\
a&b&c
\end{array}\right)$\raisebox{8pt}{$\longrightarrow$\hspace{-.5\woda}%
\begin{picture}(0,0)
\put(0,2.3){\line(0,1){55}}
\put(0,59.3){\linespread{1}\footnotesize\makebox(0,0)[b]{%
$\begin{array}{c\hw c\hw c}
\mathit{\underline{e}g}&\Rightarrow&j\\
\mathit{\underline{e}h}&\Rightarrow&k\\
\mathit{\underline{e}i}&\Rightarrow&l
\end{array}$}}\end{picture}\hspace{-.5\woda}}\raisebox{-8pt}{$\longleftarrow$\hspace{-.5\woda}%
\begin{picture}(0,0)
\put(0,2.3){\line(0,-1){55}}
\put(0,-54.7){\linespread{1}\footnotesize\makebox(0,0)[t]{%
$\begin{array}{c\hw c\hw c}
j&\Rightarrow&g\\
k&\Rightarrow&h\\
l&\Rightarrow&i
\end{array}$}}\end{picture}\hspace{.5\woda}}$\left(\begin{array}{cccc}
d&e&\mathrm{-}&f\\
j&\mathrm{-}&k&l\\
g&\mathrm{-}&h&i\\
a&\mathrm{-}&b&c
\end{array}\right)$\raisebox{8pt}{$\longrightarrow$\hspace{-.5\woda}%
\begin{picture}(0,0)
\put(0,2.3){\line(0,1){55}}
\put(0,59.3){\linespread{1}\footnotesize\makebox(0,0)[b]{%
$\begin{array}{c\hw c\hw c}
\mathit{\underline{f}a}&\Rightarrow&m\\
\mathit{\underline{f}b}&\Rightarrow&n\\
\mathit{\underline{f}c}&\Rightarrow&o
\end{array}$}}\end{picture}\hspace{-.5\woda}}\raisebox{-8pt}{$\longleftarrow$\hspace{-.5\woda}%
\begin{picture}(0,0)
\put(0,2.3){\line(0,-1){55}}
\put(0,-54.7){\linespread{1}\footnotesize\makebox(0,0)[t]{%
$\begin{array}{c\hw c\hw c}
m&\Rightarrow&a\\
n&\Rightarrow&b\\
o&\Rightarrow&c
\end{array}$}}\end{picture}\hspace{.5\woda}}$\left(\begin{array}{ccccc}
d&e&\mathrm{-}&f&\mathrm{-}\\
j&\mathrm{-}&k&\mathrm{-}&l\\
g&\mathrm{-}&h&\mathrm{-}&i\\
m&\mathrm{-}&n&\mathrm{-}&o\\
a&\mathrm{-}&b&\mathrm{-}&c
\end{array}\right)$\begin{picture}(19,0)
        \put(0,10.3){\line(1,0){9}}\qbezier(9,10.3)(14,10.3)(14,2.3)
        \qbezier(14,2.3)(14,-5.7)(9,-5.7)\put(0,-8){$\leftarrow$}
        \put(14,2.3){\line(1,0){5}}\put(19,2.3){\line(0,1){84}}
        \put(19,86.3){\line(-1,0){5}}
        \put(12,86.3){\makebox(0,0)[r]{$(\mathit{ef})(\mathit{jm})(\mathit{kn})(\mathit{lo})$}}
\end{picture}}}
\put(0,5){\makebox(425,50){\setlength{\wod}{3pt}$\begin{array}{c\hw c\hw c\hw c\hw c\hw c\hw c\hw
c\hw c}
\cir{abacbcacb}&\longrightarrow&\cir{aeafbcafb}&\longrightarrow&\cir{gegfbiafb}&\longrightarrow&\cir{gejfbiafb}&\longrightarrow&\cir{gejfniafn}\\
\cir{acabbcabb}&\longleftarrow&\cir{afaebcaeb}&\longleftarrow&\cir{gfaehiaeh}&\longleftarrow&\cir{gfaekiaek}&\longleftarrow&\cir{gfmekiaek}
\end{array}$\begin{picture}(17,0)(2,.5)
        \put(0,10.3){\line(1,0){9}}\qbezier(9,10.3)(14,10.3)(14,2.3)
        \qbezier(14,2.3)(14,-5.7)(9,-5.7)\put(0,-8){$\leftarrow$}
\end{picture}}}
\end{picture}}\linespread{1}\caption[Generating a law by composing elementary
maps]{Generating a law by composing elementary maps. Here the law $T$ is the
one that exchanges \seg{ab} with \seg{ac}. Let $f_i$, $1\leq i\leq 5$, be the
transformations given earlier such that $f_1\mathrm{(V)}=\mathrm{(W)}$,
$f_2\mathrm{(W)}=\mathrm{(A)}$, $f_3\mathrm{(A)}=\mathrm{(E)}$,
$f_4\mathrm{(E)}=\mathrm{(F)}$ and $f_5\mathrm{(F)}=\mathrm{(F}'\mathrm{)}$. We
attempted to obtain $T$ in the form
\ang{\mathrm{(V)},(f_1)^{-1}(f_2)^{-1}(f_3)^{-1}(f_4)^{-1}f_5f_4f_3f_2f_1\mathrm{(V)}},
but were not able to successfully apply $(f_1)^{-1}$.  Here we associate an
elementary map $T_i$ with each $f_i$; the law is then
$(T_1)^{-1}(T_2)^{-1}(T_3)^{-1}(T_4)^{-1}(T_5)^{-1}T_4T_3T_2T_1$ (since the
$f$'s are on the right, we reverse their order and take inverses; since
$f_5=(f_5)^{-1}$ this is the same as taking the $f$'s in their original
order). I have chosen gluing matrices in such a way that the maps $T_i$ are as
simple as possible. Each arrow between matrices represents such a map; a
description of each map is attached. The notation
``$\mathit{\underline{a}b}\Rightarrow e$'' means that if a cell $b$ is preceded by
an $a$, its color is changed to $e$. Pairs of maps that are above and below one
another in the diagram are inverses of one another. At bottom an example is
given: the $T_i$'s are applied in order to an original space $x$; the result is
that the map which exchanges \seg{ab} and \seg{ac} has been performed.}
\label{abac}
\end{figure}

For example, at the end of the previous section we attempted to generate a
description of the law that exchanges \seg{ab} with \seg{ac} by transforming
(V) into (Y), but we had trouble writing this down. The above theorem provides
us with an alternative way to describe this law in terms of composition of
elementary maps. Details are provided in Figure~\ref{abac}.

It is possible, then, to adopt the following point of view. Consider the set of
matrices of nonnegative integers. To each such matrix, associate a matrix
description. For any matrix description $A$ in this set of descriptions, there
are only finitely many elementary maps \ang{f\!A,B} where $B$ is another
description in this set. The entire setup may be represented by a directed
graph \G\ in which there is a one-to-one correspondence between vertices and
matrix descriptions and between edges and elementary maps (the edge
corresponding to \ang{f\!A,B} points from $A$ to $B$). There are infinitely
many vertices, and each vertex has finite degree.

Choose a vertex $A$ in \G. The set of vertices in the component of \G\
containing $A$ correspond to the set of matrices $B$ of nonnegative integers
such that \x{A} and \x{B} are related by the equivalence relation generated by
elementary maps. Consider any finite closed path beginning and ending at
$A$. This corresponds to a law of evolution on \x{A} (obtained by composing the
maps associated with the path's edges in the order given). Let $F_A$ be the set
of all such paths. These paths form a group under composition; it is isomorphic
to the fundamental group of \G. Some of the paths correspond to the identity
law on \x{A}. (For example, if $A$ is the matrix description $(\mathit{abc})$
and $\alpha$ is the path corresponding to the law that exchanges \seg{ab} with
\seg{ac}, then $\alpha^2$ is the identity on $A$.)  The set of all such paths
is a subgroup $I_A$ of $F_A$. It is a normal subgroup, since if $\alpha\in F_A$
and $\beta\in I_A$ then $\alpha\beta\alpha^{-1}$ must correspond to the
identity law on \x{A}, so it is in $I_A$. The quotient group $F_A/I_A$
corresponds to the set of laws on \x{A} (for $\alpha$ and $\beta$ correspond to
the same law on \x{A} if and only if $\alpha\beta^{-1}$ corresponds to the
identity law on \x{A}, and hence is in $I_A$).

Let $B$ be another vertex in the same component as $A$. Then there is a path
$\gamma$ going from $A$ to $B$. This path corresponds to an equivalence map
$U:\x{A}\mapsto\x{B}$. The map from $F_A$ to $F_B$ given by
$\alpha\mapsto\gamma\alpha\gamma^{-1}$ is an isomorphism of groups. It is also
an isomorphism of the subgroups $I_A$ and $I_B$. Hence it is an isomorphism of
$F_A/I_A$ and $F_B/I_B$. Said another way: every law $T$ on \x{A} that
corresponds to some path $\alpha$ is equivalent to a law $UTU^{-1}$ on \x{B}
which corresponds to the path $\gamma\alpha\gamma^{-1}$, and vice versa (switch
roles of $A$ and $B$). To find all non-equivalent laws generated by elementary
maps, then, it suffices to choose one vertex $A_{\sigma}$ in each component
$\sigma$ of \G, and to generate the group $F_{A_{\sigma}}/I_{A_{\sigma}}$ for
each such vertex $A_{\sigma}$.

This is a nice point of view, and one might hope that the set of maps generated
in this manner might contain all the maps one needs. More precisely, one might,
via wishful thinking, come to believe the following:\bigskip\\
\textbf{Conjecture}\hspace*{3pt} \textit{The elementary maps generate all
orientation-preserving local maps with local inverses on oriented
1+1-dimensional space sets given by the set of matrix descriptions $A$ such
that every vertex of \,\g{A} is in a circuit.}\bigskip\\
I do in fact believe this, but I have not proven it. Recall that all I did
earlier was to come up with a few transformations of space set descriptions. I
never showed that every description of a space set \x{D} could be generated by
applying these transformations to $D$. In fact, it is not completely clear what
it would mean to generate ``every description,'' since it is clear that my
definition of space set descriptions is inadequate. However, at minimum it
would be nice to know that all descriptions given by the current definition can
be so generated. This would imply that the laws in $F_A/I_A$ include all laws
on \x{A} that are of the form \ang{D,E} where $D$ and $E$ satisfy the current
definition of description. I have found no reason to believe that this is not
the case, but this is no substitute for a proof.

Note that if the conjecture holds then the situation is similar to that of
invertible linear operators in finite-dimensional linear algebra. There we have
a theorem that the set of all such operators is generated by the elementary
linear operators. Here the situation is a bit more complicated because our
elementary maps may not be operators; they may go from \x{A} to \x{B} where
$A\neq B$. But otherwise it is the same.

I have not yet attempted to write a computer program to generate laws on \X\ by
generating the group $F_A/I_A$. To do so, it would be nice to have some
understanding of the structure of $I_A$. I do not have any such understanding,
nor do I know whether it is possible to gain one; perhaps it is extremely
complicated. Nevertheless, without knowing anything about $I_A$ one could still
easily write an algorithm which finds all closed paths of length $n$ that begin
and end at $A$. Each such path would correspond to a law. Many laws found in
this way would be the same as one another, so it would be useful to put the
laws in some sort of standard form in order to detect this (some new techniques
would need to be developed here, but I do not think this would be
difficult). By doing this for each $n$, in increasing order, one would in the
limit generate all such laws. I expect that the resulting algorithm would be
far more efficient than the one described earlier in this chapter.

The condition on matrix descriptions mentioned in the conjecture (that every
vertex of \g{A} be in a circuit) may easily be removed by allowing another
rather trivial elementary transformation, which I will call
\emph{reducing}. The move is this: if a vertex contains only incoming edges and
no outgoing edges, or only outgoing edges and no incoming edges, then that
vertex and all edges connecting to that vertex may be removed. The inverse
move, \emph{unreducing}, is also allowed.

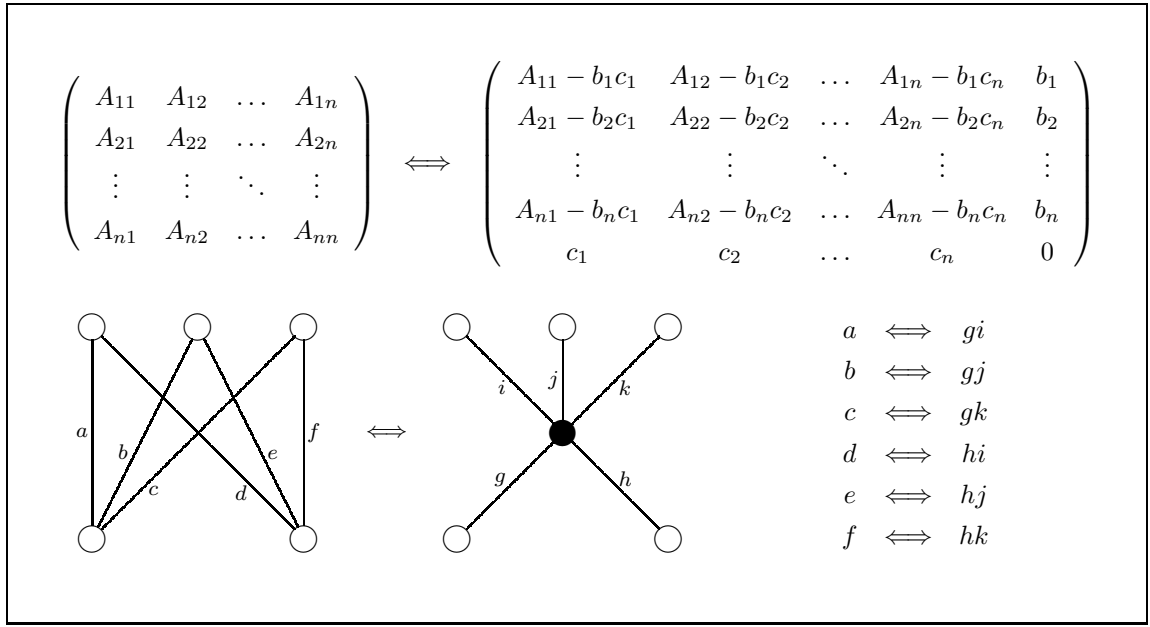
\begin{figure}
\fbox{\begin{picture}(425,228)\footnotesize
\put(0,114){\makebox(425,114){$\begin{array}{ccc}\left(\begin{array}{cccc}
A_{11}&A_{12}&\ldots&A_{1n}\\
A_{21}&A_{22}&\ldots&A_{2n}\\
\vdots&\vdots&\ddots&\vdots\\
A_{n1}&A_{n2}&\ldots&A_{nn}
\end{array}\right)&\Longleftrightarrow&\left(\begin{array}{ccccc}
A_{11}-b_1c_1&A_{12}-b_1c_2&\ldots&A_{1n}-b_1c_n&b_1\\
A_{21}-b_2c_1&A_{22}-b_2c_2&\ldots&A_{2n}-b_2c_n&b_2\\
\vdots&\vdots&\ddots&\vdots&\vdots\\
A_{n1}-b_nc_1&A_{n2}-b_nc_2&\ldots&A_{nn}-b_nc_n&b_n\\
c_1&c_2&\ldots&c_n&0
\end{array}\right)\end{array}$}}
\put(12,12){\begin{picture}(425,114)\scriptsize
\put(195,57){
        \put(0,0){\circle*{10}}\put(-40,-40){\circle{10}}
        \put(-40,40){\circle{10}}\put(0,40){\circle{10}}
        \put(40,40){\circle{10}}\put(40,-40){\circle{10}}
        \put(-20,-20){\makebox(0,0)[br]{$g$\,}}
        \put(20,-20){\makebox(0,0)[bl]{\,$h$}}
        \put(-20,20){\makebox(0,0)[tr]{$i$\,}}
        \put(0,20){\makebox(0,0)[r]{$j$\,}}
        \put(20,20){\makebox(0,0)[tl]{\,$k$}}
        \qbezier(-36.46,-36.46)(0,0)(36.46,36.46)
        \qbezier(36.46,-36.46)(0,0)(-36.46,36.46)
        \qbezier(0,0)(0,17.5)(0,35)}
\put(128.5,57){\makebox(0,0){$\Longleftrightarrow$}}
\put(57,57){
        \put(-40,-40){\circle{10}}\put(-40,40){\circle{10}}
        \put(0,40){\circle{10}}\put(40,40){\circle{10}}
        \put(40,-40){\circle{10}}
        \put(-40,0){\makebox(0,0)[r]{$a$\,}}
        \put(40,0){\makebox(0,0)[l]{\,$f$}}
        \put(-25,-10){\makebox(0,0)[br]{$b$\,}}
        \put(25,-10){\makebox(0,0)[bl]{\,$e$}}
        \put(-20,-20){\makebox(0,0)[tl]{\,$c$}}
        \put(20,-20){\makebox(0,0)[tr]{$d$\,}}
        \qbezier(-36.46,-36.46)(0,0)(36.46,36.46)
        \qbezier(36.46,-36.46)(0,0)(-36.46,36.46)
        \qbezier(-40,-35)(-40,0)(-40,35)
        \qbezier(40,-35)(40,0)(40,35)
        \qbezier(-37.76,-35.52)(-20,0)(-2.24,35.52)
        \qbezier(37.76,-35.52)(20,0)(2.24,35.52)}
\put(252,10){\footnotesize\makebox(153,94){$\begin{array}{ccc}
a&\Longleftrightarrow&gi\\
b&\Longleftrightarrow&gj\\
c&\Longleftrightarrow&gk\\
d&\Longleftrightarrow&hi\\
e&\Longleftrightarrow&hj\\
f&\Longleftrightarrow&hk\end{array}$}}
\end{picture}}
\end{picture}}\linespread{1}\caption[The elementary space set
transformation]{The elementary space set transformation. The transformation is
shown here in three ways. In terms of matrices, the $n\times n$ matrix $A$ is
transformed into an $(n+1)\times (n+1)$ matrix $A'$ by writing down an
$(n+1)$st row $c_i$, $1\leq i\leq n+1$, and an $(n+1)$st column $b_j$, $1\leq
j\leq n+1$, with $b_{n+1}=c_{n+1}=0$, and then subtracting $b_ic_j$ from entry
$(i,j)$ for each $i$ and $j$. The transformation is allowed if $A'$ is a matrix
of nonnegative integers. An example of how this transformation affects a graph
is given. Instead of using arrows, I will use the convention that edges
emerging from the top of a vertex point away from that vertex, and that edges
emerging from the bottom of a vertex point toward that vertex. Here an open
circle denotes a ``free'' vertex: that is, it refers to some unknown vertex (so
different free vertices may refer to the same vertex). A filled-in circle
denotes a ``fixed'' vertex: all edges connecting to it are shown (hence no free
vertex refers to a fixed vertex). Hence the right-to-left transformation is
simply the following: remove a vertex which has no edge going from that vertex
to itself, and replace each length-2 path through that vertex with a length-1
path. The path replacement in terms of segments is listed on the right.}
\label{elem}
\end{figure}

This picture may be simplified somewhat by revising the definition of
elementary map. It turns out that we may replace the maps associated with
splitting, doubling and reducing with a single type of map, described in
Figure~\ref{elem}.
Henceforth when I say ``elementary transformation'' I will be referring to this
move and its inverse. The move will also be referred to as ``adding a vertex''
or ``removing a vertex.''

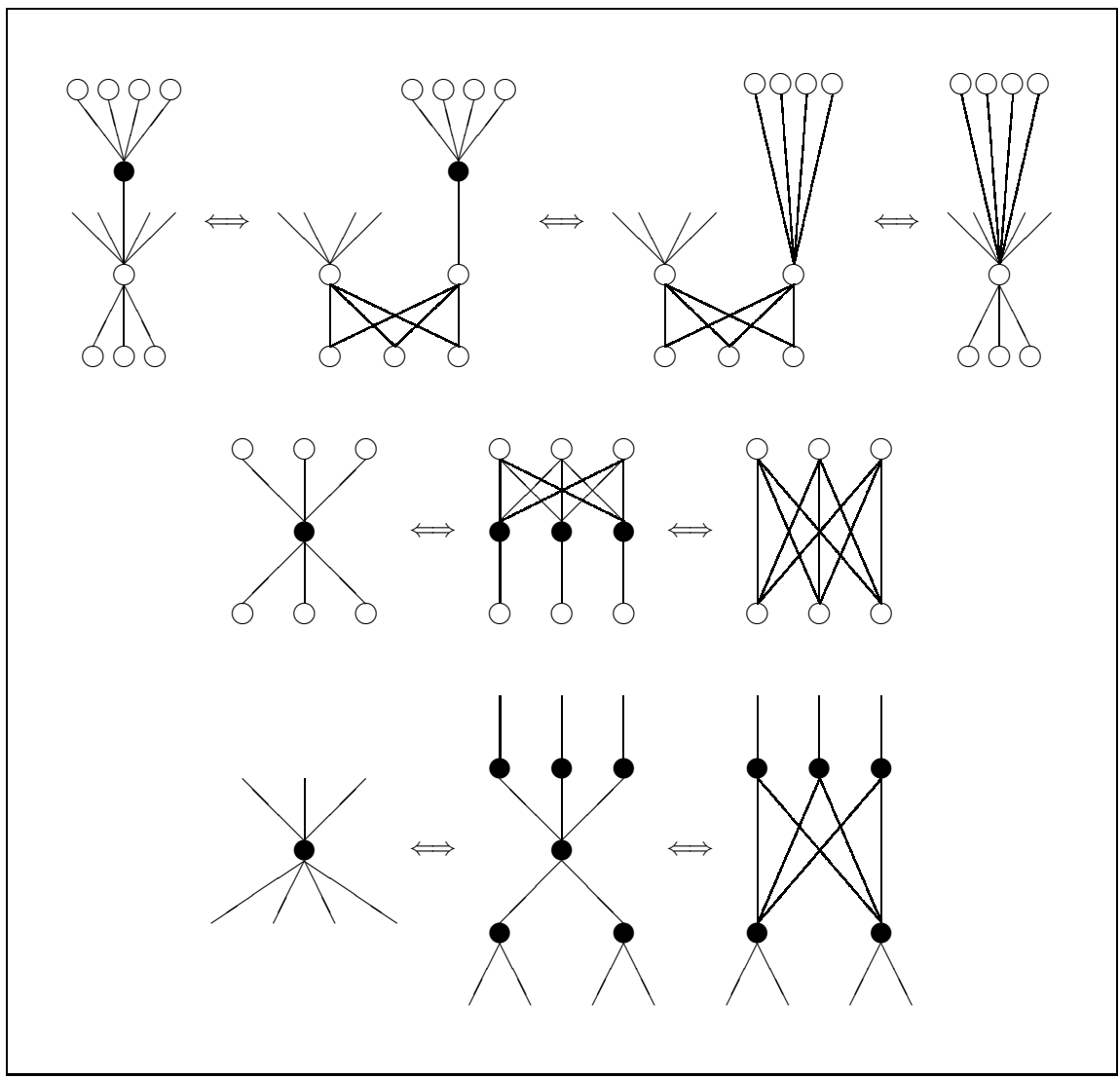
\begin{figure}
\fbox{\begin{picture}(425,408)\footnotesize
\put(0,88){
\put(42.5,220){
\put(0,0){\circle{8}}\put(0,40){\circle*{8}}
\put(0,44){\line(-1,4){6}}\put(0,44){\line(-3,4){18}}
\put(0,44){\line(1,4){6}}\put(0,44){\line(3,4){18}}
\multiput(-18,72)(12,0){4}{\circle{8}}\put(0,4){\line(0,1){32}}
\put(0,4){\line(-1,2){10}}\put(0,4){\line(-1,1){20}}
\put(0,4){\line(1,2){10}}\put(0,4){\line(1,1){20}}
\put(0,-4){\line(-1,-2){12}}\put(-12,-32){\circle{8}}
\put(0,-4){\line(1,-2){12}}\put(0,-32){\circle{8}}
\put(0,-4){\line(0,-1){24}}\put(12,-32){\circle{8}}
}\put(82.5,240){\makebox(0,0){$\Longleftrightarrow$}}\put(122.5,220){
\put(0,0){\circle{8}}
\put(0,4){\line(-1,2){10}}\put(0,4){\line(-1,1){20}}
\put(0,4){\line(1,2){10}}\put(0,4){\line(1,1){20}}
\multiput(0,-32)(25,0){3}{\circle{8}}\multiput(0,-4)(50,0){2}{\line(0,-1){24}}
\qbezier(0,-4)(12.5,-16)(25,-28)\qbezier(0,-4)(25,-16)(50,-28)
\qbezier(50,-4)(37.5,-16)(25,-28)\qbezier(50,-4)(25,-16)(0,-28)
}\put(172.5,220){
\put(0,0){\circle{8}}\put(0,40){\circle*{8}}
\put(0,44){\line(-1,4){6}}\put(0,44){\line(-3,4){18}}
\put(0,44){\line(1,4){6}}\put(0,44){\line(3,4){18}}
\multiput(-18,72)(12,0){4}{\circle{8}}\put(0,4){\line(0,1){32}}
}\put(212.5,240){\makebox(0,0){$\Longleftrightarrow$}}\put(252.5,220){
\put(0,0){\circle{8}}
\put(0,4){\line(-1,2){10}}\put(0,4){\line(-1,1){20}}
\put(0,4){\line(1,2){10}}\put(0,4){\line(1,1){20}}
\multiput(0,-32)(25,0){3}{\circle{8}}\multiput(0,-4)(50,0){2}{\line(0,-1){24}}
\qbezier(0,-4)(12.5,-16)(25,-28)\qbezier(0,-4)(25,-16)(50,-28)
\qbezier(50,-4)(37.5,-16)(25,-28)\qbezier(50,-4)(25,-16)(0,-28)
}\put(302.5,220){
\put(0,0){\circle{8}}
\qbezier(0,4)(-7.5,37)(-15,70)\qbezier(0,4)(-2.5,37)(-5,70)
\qbezier(0,4)(2.5,37)(5,70)\qbezier(0,4)(7.5,37)(15,70)
\multiput(-15,74)(10,0){4}{\circle{8}}
}\put(342.5,240){\makebox(0,0){$\Longleftrightarrow$}}\put(382.5,220){
\put(0,0){\circle{8}}
\qbezier(0,4)(-7.5,37)(-15,70)\qbezier(0,4)(-2.5,37)(-5,70)
\qbezier(0,4)(2.5,37)(5,70)\qbezier(0,4)(7.5,37)(15,70)
\multiput(-15,74)(10,0){4}{\circle{8}}
\put(0,4){\line(-1,2){10}}\put(0,4){\line(-1,1){20}}
\put(0,4){\line(1,2){10}}\put(0,4){\line(1,1){20}}
\put(0,-4){\line(-1,-2){12}}\put(-12,-32){\circle{8}}
\put(0,-4){\line(1,-2){12}}\put(0,-32){\circle{8}}
\put(0,-4){\line(0,-1){24}}\put(12,-32){\circle{8}}
}\put(112.5,120){
\put(0,0){\circle*{8}}
\put(0,4){\line(0,1){24}}\put(0,4){\line(-1,1){24}}
\put(0,4){\line(1,1){24}}\multiput(-24,32)(24,0){3}{\circle{8}}
\put(0,-4){\line(0,-1){24}}\put(0,-4){\line(-1,-1){24}}
\put(0,-4){\line(1,-1){24}}\multiput(-24,-32)(24,0){3}{\circle{8}}
}\put(162.5,120){\makebox(0,0){$\Longleftrightarrow$}}\put(212.5,120){
\multiput(-24,0)(24,0){3}{\circle*{8}}\multiput(-24,-4)(24,0){3}{\line(0,-1){24}}
\multiput(-24,4)(24,0){3}{\line(0,1){24}}\multiput(0,4)(24,0){2}{\line(-1,1){24}}
\multiput(0,4)(-24,0){2}{\line(1,1){24}}\multiput(-24,32)(24,0){3}{\circle{8}}
\qbezier(-24,4)(0,16)(24,28)\qbezier(24,4)(0,16)(-24,28)
\multiput(-24,-32)(24,0){3}{\circle{8}}
}\put(262.5,120){\makebox(0,0){$\Longleftrightarrow$}}\put(312.5,120){
\multiput(-24,32)(24,0){3}{\circle{8}}
\multiput(-24,-28)(24,0){3}{\line(0,1){56}}
\multiput(0,0)(24,0){2}{\qbezier(-24,-28)(-12,0)(0,28)}
\multiput(0,0)(-24,0){2}{\qbezier(24,-28)(12,0)(0,28)}
\qbezier(-24,-28)(0,0)(24,28)\qbezier(24,-28)(0,0)(-24,28)
\multiput(-24,-32)(24,0){3}{\circle{8}}
}\put(112.5,-4){
\put(0,0){\circle*{8}}\put(0,4){\line(-1,1){24}}\put(0,4){\line(0,1){24}}
\put(0,4){\line(1,1){24}}\put(0,-4){\line(-1,-2){12}}\put(0,-4){\line(-3,-2){36}}
\put(0,-4){\line(1,-2){12}}\put(0,-4){\line(3,-2){36}}
}\put(162.5,-4){\makebox(0,0){$\Longleftrightarrow$}}\put(212.5,-4){
\put(0,0){\circle*{8}}
\multiput(-24,32)(24,0){3}{\circle*{8}}\multiput(-24,36)(24,0){3}{\line(0,1){24}}
\put(0,4){\line(-1,1){24}}\put(0,4){\line(0,1){24}}
\put(0,4){\line(1,1){24}}\put(0,-4){\line(-1,-1){24}}
\put(0,-4){\line(1,-1){24}}\put(-24,-32){\circle*{8}}\put(24,-32){\circle*{8}}
\multiput(-24,-36)(48,0){2}{\put(0,0){\line(-1,-2){12}}\put(0,0){\line(1,-2){12}}}
}\put(262.5,-4){\makebox(0,0){$\Longleftrightarrow$}}\put(312.5,-4){
\multiput(-24,32)(24,0){3}{\circle*{8}}\multiput(-24,36)(24,0){3}{\line(0,1){24}}
\put(-24,-32){\circle*{8}}\put(24,-32){\circle*{8}}
\multiput(-24,-36)(48,0){2}{\put(0,0){\line(-1,-2){12}}\put(0,0){\line(1,-2){12}}}
\multiput(-24,-28)(48,0){2}{\line(0,1){56}}\qbezier(-24,-28)(0,0)(24,28)
\qbezier(-24,28)(0,0)(24,-28)\qbezier(-24,-28)(-12,0)(0,28)
\qbezier(24,-28)(12,0)(0,28)}
}
\end{picture}}\linespread{1}\caption{Proof of the equivalence of splitting,
doubling and reducing moves to vertex removal moves (see text for further
details). In the first row the vertex below the black vertex is split right
(convention: arrows point upwards in the diagram) to produce the second
diagram; then the edge pointing towards the black vertex is undoubled to
produce the third diagram; then the two vertices that were produced by the
split are unsplit to produce the fourth diagram. In the second row the black
vertex is split left twice to produce the second diagram; then the move from
the first row is applied three times to produce the third diagram. In the third
row a vertex is added to each upper edge and two vertices are added at bottom
to produce the second diagram (now the edges to be split are at bottom); the
original vertex is removed to produce the third diagram; and the final step
(not shown in the picture) is to remove the top three vertices.}
\label{elemproof}
\end{figure}

\begin{theorem} The splitting, doubling and reducing moves and their inverses
are equivalent to the elementary transformation.
\end{theorem}
\textsc{Proof.} Let $k$ be the in-degree of the black vertex in
Figure~\ref{elem}. If $k=0$ then removing the vertex is the same move as the
reducing move. If $k=1$ then removing the vertex can be accomplished by a
split, then an undoubling, then an unsplit, as shown in the first row of
Figure~\ref{elemproof}.
If $k>1$ then removing the vertex can be accomplished
by $k-1$ splits followed by $k$ applications of the removing-the-vertex operation
when $k=1$, as shown in the second row of the figure.

Conversely, the reducing move and undoubling moves are special cases of
removing a vertex. Let $k$ be the in-degree and $j$ be the out-degree of a
vertex being split to the left. To accomplish the split, we begin by adding $j$
vertices so as to replace each of the $j$ outgoing edges with two consecutive
edges.. Then two vertices are added so that the two chosen subsets of the $k$
incoming edges are separated. Now the original vertex is removed. Then the $j$
vertices that had been added are removed. This is shown in the third row of the
figure.\proofend

\subsection{Equivalence classes of space sets}
\label{equiv}

Given results in previous sections, it should now be apparent that the problem
of finding the equivalence classes of space sets under local equivalence maps
reduces to the problem of finding the equivalence classes of matrices of
nonnegative integers (or, equivalently, of directed graphs) under elementary
transformations.

I was able to solve this problem for an important special case: namely, the
case where the graph contains exactly one nontrivial connected component, and
no other connected component. In this case, it turns out that a certain set of
invariants completely determines the equivalence classes.

\subsubsection[Some useful space set transformations]{A new matrix representation, and some useful space set\protect\\transformations}

Consider the $n\times n$ nonnegative integer matrix $A$ and the pair of
consecutive elementary transformations described in Figure~\ref{addrow}.
\begin{figure}
\fbox{\begin{picture}(425,184)\footnotesize\put(0,0){\makebox(425,184){%
$\begin{array}{cccc}
\left(\begin{array}{ccc}
A_{11}&\cdots&A_{1n}\\
\vdots&\ddots&\vdots\\
A_{n1}&\cdots&A_{nn}
\end{array}\right)&\Longrightarrow&\left(\begin{array}{ccccc}
A_{11}&\cdots&A_{1(n-1)}&1&A_{1n}-1\\
A_{21}&\cdots&A_{2(n-1)}&0&A_{2n}\\
\vdots&\ddots&\vdots&\vdots&\vdots\\
A_{n1}&\cdots&A_{n(n-1)}&0&A_{nn}\\
0&\cdots&0&1&0
\end{array}\right)&\Longrightarrow\\[45pt]
\multicolumn{4}{c}{\left(\begin{array}{cccc}
A_{11}+A_{n1}&\cdots&A_{1(n-1)}+A_{n(n-1)}&A_{1n}+A_{nn}-1\\
A_{21}&\cdots&A_{2(n-1)}&A_{2n}\\
\vdots&\ddots&\vdots&\vdots\\
A_{n1}&\cdots&A_{n(n-1)}&A_{nn}
\end{array}\right)}
\end{array}$}}\end{picture}}\linespread{1}\caption[Adding columns]{An
equivalence transformation involving row addition. Here first an elementary
equivalence transformation is performed in which an $(n+1)$st vertex is added,
as shown. Then the $n$th vertex is removed. The net result is that row $n$ has
been added to row 1 and then 1 has been subtracted from $A_{1n}$. The operation
may only be performed if $A_{1n}>0$ (so that each entry in these three matrices
is nonnegative). (In terms of \g{A}, this operation first removes an edge $e$
going from vertex 1 to vertex $n$. Then, for each formerly existing two-edge
path that went along $e$ from vertex 1 to vertex $n$ and then from vertex $n$
to vertex $k$, a new edge is added going from 1 to $k$.) By composing this
operation with index permutations, one may perform the operation in which row
$j$ is added to row $i$ and then 1 is subtracted from $A_{ij}$ for any $i\neq
j$.}
\label{addrow}
\end{figure}
The net result is that $A$ has been transformed to another $n\times n$ matrix
by adding row $j$ to row $i$ and subtracting 1 from $A_{ij}$. By an analogous
pair of transformations, one may add column $i$ to column $j$ and subtract 1
from $A_{ij}$. Since the transformations are invertible, one may also subtract
row $j$ from row $i$ and add 1 to $A_{ij}$, or subtract column $i$ from column
$j$ and add 1 to $A_{ij}$. In all of these cases, the transformations are only
allowed if all entries in the intermediate and final matrices are
nonnegative. This means two things. Firstly, $A_{ij}$ must be at least 1
\emph{before} the moves involving row or column addition, and it must be at
least 1 \emph{after} the moves involving row or column subtraction. And
secondly, after a subtraction the resulting matrix must not contain negative
integers.

A simplification can be obtained as follows. Instead of representing \x{A} by
the matrix $A$, let us represent it by the matrix $B=A-I$, where $I$ is the
identity matrix. I call $B$ a \emph{reduced} representation of \x{A}. The set
of allowed reduced matrices is the set of all matrices $B$ such that
$B_{ij}\geq 0$ if $i\neq j$, and $B_{ii}\geq -1$. In terms of the reduced
matrix, the pair of transformations given in Figure~\ref{addrow} becomes
simpler: while before we added row $j$ to row $i$ and then subtracted 1 from
$A_{ij}$, now all we do is add row $j$ to row $i$. (This follows since 1 has
already been subtracted from $A_{jj}$.) A similar result holds for
columns. Hence the advantage of using the reduced matrix is that the
above-described transformations now become standard row and column operations
from linear algebra: one may add or subtract one row or column from another row
or column. As before, there are restrictions. Firstly, one must check that
$B_{ij}$ is at least 1 before adding row $j$ to row $i$ or column $i$ to column
$j$, and that it is at least 1 after subtracting row $j$ from row $i$ or column
$i$ from column $j$. And secondly, after a subtraction the resulting matrix
must be a valid reduced matrix.

\subsubsection[Invariants of integer matrices]{Invariants of integer matrices
under a subset of standard row and\protect\\column operations}
\label{integermatrices}

What happens if we consider only $n\times n$ matrices for fixed $n$, allow the
entries of these matrices to be \emph{arbitrary} integers, and try to find the
equivalence classes generated by the operations of adding and subtracting one
row or column from another? This is the problem that I will address in this
section. It turns out that it may be solved by making a slight adaptation to a
standard algorithm from linear algebra. The solution involves a set of matrix
invariants which completely determine the equivalence classes. As it turns out,
these invariants are also invariants of reduced matrices under equivalence
transformations.

The algorithm in question is one for reducing a matrix of polynomials to Smith
normal form, as described, for example, in Hoffman and Kunze
(\cite{hoffman/kunze}, pp.\ 253-261). There are some slight differences,
because the operation of multiplying a row or column by a constant is allowed
for polynomial matrices, but is not allowed here. However, the basic methods of
the algorithm still go through. I describe it below in detail, in part because
the ideas in this algorithm will be used to prove the main invariants theorem
later on.

\begin{theorem}
\label{integerthm}
Let $B$ be an $n\times n$ integer matrix. Let the allowed operations on $B$
consist of adding an integer multiple of row $i$ to row $j$ or of column $i$ to
column $j$, where $i\neq j$. Then $B$ may be transformed by a finite sequence
of these allowed operations into a unique $n\times n$ diagonal matrix $D$,
where $D_{ii}$ divides $D_{jj}$ if $i<j$, and where all entries are nonnegative
except possibly for $D_{nn}$.
\end{theorem}
\textsc{Proof.} The main tool used here is a variant of the Euclidean algorithm
for finding a greatest common divisor of two positive integers. In this case
the integers may be positive, negative or zero. Here we will consider
``greatest common divisor'' to mean greatest in \emph{absolute value}. Hence a
set of integers either has a greatest common divisor of zero (in which case the
integers are all zero) or it has two greatest common divisors ($d$ and $-d$ for
some positive integer $d$).

We begin by considering the first row. Either it contains a greatest common
divisor of its entries or it does not. If it does not, then it must contain two
nonzero entries $B_{1i}$ and $B_{1j}$ with $|B_{1i}|>|B_{1j}|$. If $B_{1i}$ and
$B_{1j}$ have the same sign, then subtract column $j$ from column $i$;
otherwise, add column $j$ to column $i$. In either case, the sum of the
absolute values of the entries in row 1 decreases. Now repeat the process. If
row 1 does not yet contain a greatest common divisor of its entries, perform
another column operation which decreases the sum of the absolute values of its
entries. This sum cannot decrease past zero; therefore, row 1 will eventually
contain a greatest common divisor of its entries in some entry $B_{1k}=g_1$. By
adding multiples of column $k$ to the other columns, each entry in row 1 may be
set to $g_1$.

Either $g_1$ divides all of the entries in $B$ or it does not. Suppose it does
not divide some entry $B_{ij}$. Now we do the same thing to column $j$ as we
did above to row 1: add or subtract rows in order to decrease the sum of the
absolute values of the entries in column $j$ until column $j$ contains a
greatest common divisor of its entries in some entry $B_{mj}=g_2$; then set
each entry in column $j$ to $g_2$. Note that $g_2$ is a greatest common divisor
of the entries which were in column $j$ before these operations were carried
out; hence it divides $g_1$. Since in addition $g_2$ is not divisible by $g_1$,
it must be true that $|g_2|<|g_1|$.

Now either $g_2$ divides all of the entries in $B$ or it does not. If not, we
repeat the process to produce $g_3$, $g_4$, and so on. Since $|g_{i+1}|$ is
always strictly less than $|g_i|$, this process must terminate. We end up with
an entry $B_{ij}=g_r$ which divides all of the entries in $B$. If $i$ does not
equal 1, then by adding a multiple of row $i$ to row 1 we may set
$B_{1j}=g_r$. Now $g_r$ is in some entry $B_{1j}$ in row 1. If $j=1$ then by
adding a multiple of column 1 to column 2 we may set $B_{12}=g_r$. Now
$B_{1k}=g_r$ for some $k>1$. By subtracting a multiple of column $k$ from
column 1, we may set $B_{11}=|g_r|$. Now by adding multiples of row 1 to the
other rows, we may set $B_{i1}=0$ for $i>1$; and by adding multiples of column
1 to the other columns, we may set $B_{1j}=0$ for $j>1$.

Now the situation is as illustrated in Figure~\ref{hoffkunzalg}.
\begin{figure}
\fbox{\begin{picture}(425,204)\footnotesize
\put(0,100){\makebox(425,104){$\begin{array}{cccc}\left(\,\,\begin{array}{|ccccccc|}
\hline
\,&\,&\,&\,&\,&\,&\,\\
\,&\,&\,&\,&\,&\,&\,\\
\multicolumn{7}{|c|}{\raisebox{0pt}[0pt][0pt]{\large $S_1$}}\\
\,&\,&\,&\,&\,&\,&\,\\
\,&\,&\,&\,&\,&\,&\,\\
\hline\end{array}\,\,\right)&\Longrightarrow&\left(\begin{array}{c|cccc|}
\multicolumn{1}{c}{|g_1|}&0&0&\cdots&\multicolumn{1}{c}{0}\\ \cline{2-5}
0&\,&\,&\,&\,\\
0&\,&\,&\,&\,\\
\vdots&\multicolumn{4}{|c|}{\raisebox{7pt}[0pt][0pt]{\large $S_2$}}\\
0&\,&\,&\,&\,\\
\cline{2-5}\end{array}\,\,\right)&\Longrightarrow\end{array}$}}
\put(0,2){\makebox(425,104){$\begin{array}{ccccc}\left(\begin{array}{cc|ccc|}
|g_1|&\multicolumn{1}{c}{0}&0&\cdots&\multicolumn{1}{c}{0}\\
0&\multicolumn{1}{c}{|g_2|}&0&\cdots&\multicolumn{1}{c}{0}\\ \cline{3-5}
0&0&\,&\,&\,\\
\vdots&\vdots&\multicolumn{3}{|c|}{\raisebox{0pt}[0pt][0pt]{\large $S_3$}}\\
0&0&\,&\,&\,\\
\cline{3-5}\end{array}\,\,\right)&\Longrightarrow&\cdots&\Longrightarrow&\left(\begin{array}{ccccc}
|g_1|&0&0&\cdots&0\\
0&|g_2|&0&\cdots&0\\
0&0&|g_3|&\cdots&0\\
\vdots&\vdots&\vdots&\ddots&\vdots\\
0&0&0&\cdots&g_n\end{array}\right)\end{array}$}}
\end{picture}}\linespread{1}\caption[Reduction of integer matrices to canonical
form]{Reduction of integer matrices to canonical form.}
\label{hoffkunzalg}
\end{figure}
The above procedure can now be applied to submatrix $S_2$; the row and column
operations required do not affect the first row or column. The absolute value
of the greatest common divisors of the entries of $S_2$ is placed in $B_{22}$,
the rest of row 2 and column 2 is set to zero, and the process is repeated on
$S_3$. One may continue in this manner until one reaches $S_n$, a $1\times 1$
matrix. No operations may be performed on this matrix; in particular, the sign
of its single entry $S_{nn}$ cannot be changed. The result of all of these
operations is a diagonal matrix $D$ in which $D_{ii}$ divides $D_{jj}$ for
$i<j$, and in which $D_{ii}>0$ for $1\leq i<n$, as required.

It is easy to show that, for $1\leq k\leq n$, the number
$\delta_k=\prod_{i=1}^kD_{ii}$ is equal to a greatest common divisor of the
$k\times k$ minors of $D$. If $k<n$ then $\delta_k$ is the absolute value of
the greatest common divisors of the $k\times k$ minors; and $\delta_n$ is the
(signed) determinant of $D$. It is also a standard result
(see~\cite{hoffman/kunze}, pp.\ 157, 259) that the allowed row and column
operations do not change the greatest common divisors of the $k\times k$~minors
of a matrix, nor do they change its determinant.  Therefore, the matrix $B$ may
be transformed into only one matrix $D$ satisfying the conditions in the
theorem.\proofend\bigskip

It follows that the equivalence class in which the matrix $B$ belongs may be
precisely identified by the associated matrix $D$, or, equivalently, by the
determinant of $B$ and by the absolute values of the greatest common divisors
of its $k\times k$ minors. I will refer to the list of these absolute values,
ordered from highest to lowest, as the \emph{list of invariant factors} of $B$.

\subsubsection[Invariants of reduced matrices]{Invariants of reduced matrices
under equivalence transformations}

The next task is to find out what role the determinant and the list of
invariant factors play when we switch our attention to reduced matrices and
equivalence transformations of space sets. An elementary equivalence
transformation maps an $n\times n$ matrix to an $(n+1)\times (n+1)$ matrix. It
turns out that this transformation maps the determinant of the matrix and its
invariant factors in a natural way.

\begin{theorem}
Let $T:A\mapsto A'$ be an elementary equivalence transformation, where $A$ is
an $n\times n$ matrix and $A'$ is an $(n+1)\times (n+1)$ matrix. Then
$\det(I-A)=\det(I-A')$, and the list of invariant factors of $I-A$ and of
$I-A'$ are the same once trailing 1's are removed.
\end{theorem}
\textsc{Proof.} The situation is illustrated in Figure~\ref{matrixinvars}.
\begin{figure}
\fbox{\begin{picture}(425,212)\footnotesize\settowidth{\wod}{2}\addtolength{\wod}{1pt}
\put(0,98){\makebox(425,114){$\begin{array}{cc}\left(\begin{array}{ccccc}
1-A_{11}+b_1c_1&\hspace{\wod}-\,A_{12}+b_1c_2&\ldots&\hspace{\wod}-\,A_{1n}+b_1c_n&-b_1\\
\hspace{\wod}-\,A_{21}+b_2c_1&1-A_{22}+b_2c_2&\ldots&\hspace{\wod}-\,A_{2n}+b_2c_n&-b_2\\
\vdots&\vdots&\ddots&\vdots&\vdots\\
\hspace{\wod}-\,A_{n1}+b_nc_1&\hspace{\wod}-\,A_{n2}+b_nc_2&\ldots&1-A_{nn}+b_nc_n&-b_n\\
-c_1&-c_2&\ldots&-c_n&1
\end{array}\right)&\Longrightarrow\end{array}$}}
\put(0,0){\makebox(425,114){$\begin{array}{ccc}\left(\begin{array}{ccccc}
1-A_{11}&\hspace{\wod}-\,A_{12}&\ldots&\hspace{\wod}-\,A_{1n}&0\\
\hspace{\wod}-\,A_{21}&1-A_{22}&\ldots&\hspace{\wod}-\,A_{2n}&0\\
\vdots&\vdots&\ddots&\vdots&\vdots\\
\hspace{\wod}-\,A_{n1}&\hspace{\wod}-\,A_{n2}&\ldots&1-A_{nn}&0\\
-c_1&-c_2&\ldots&-c_n&1
\end{array}\right)&\Longrightarrow&\left(\begin{array}{ccccc}
1-A_{11}&\hspace{\wod}-\,A_{12}&\ldots&\hspace{\wod}-\,A_{1n}&0\\
\hspace{\wod}-\,A_{21}&1-A_{22}&\ldots&\hspace{\wod}-\,A_{2n}&0\\
\vdots&\vdots&\ddots&\vdots&\vdots\\
\hspace{\wod}-\,A_{n1}&\hspace{\wod}-\,A_{n2}&\ldots&1-A_{nn}&0\\
0&0&\ldots&0&1
\end{array}\right)\end{array}$}}
\end{picture}}\linespread{1}\caption[Invariants of elementary matrix
transformations]{Invariants of elementary matrix transformations. The first
matrix shown is $I-A'$, where $A'$ was obtained from $A$ as described in
Figure~\ref{elem}. The second matrix is obtained from the first one by adding
$b_i$ times the last row to the $i$th row for each $i\leq n$. The third matrix
is obtained from the second one by adding $c_j$ times the last column to the
$j$th column for each $j\leq n$. From this one may easily conclude that
$\det(I-A)=\det(I-A')$ and that the list of invariant factors of $I-A$ is
the same as that of $I-A'$ once trailing 1's have been deleted.}
\label{matrixinvars}
\end{figure}
Since the bottom matrix in the figure has been obtained from $I-A'$ by adding
multiples of rows or columns to one another, it has the same determinant and
set of invariant factors as does $I-A'$. But this bottom matrix contains $I-A$
as an $n\times n$ submatrix. If we apply the standard algorithm to reduce this
to Smith normal form, only the submatrix will be affected because of the zeroes
in the (n+1)st rows and columns. Hence the resulting matrix will have the same
list of invariant factors as does $I-A$, except for an additional trailing
1. Also, its determinant is clearly equal to $\det(I-A)$.\proofend\bigskip

Henceforth when I refer to a list of invariant factors I will mean that
trailing 1's have been deleted. Given that the elementary equivalence
transformations generate the entire set of equivalence transformations, it
follows that the determinant and the list of invariant factors are invariants
of equivalence maps on reduced matrices. Since the greatest common divisor of
the $n\times n$ minors of $B$ is just $|\det B|$, it follows that all
information is contained in the list of invariant factors except when $\det B$
is nonzero, in which case the additional information provided by $\det B$ is
its sign.

\subsubsection{Positive reduced matrices}

The invariants found above do not completely determine the equivalence classes
of reduced matrices under equivalence transformations. This can easily be seen
by first noting that the number of strongly connected components of a matrix is
preserved under equivalence transformations. Given a strongly connected matrix,
it is easy to construct another matrix with the same invariants whose graph
contains several strongly connected components. For example, suppose that $A$
is a strongly connected $n\times n$ matrix such that $n$ is even and
$\det(I-A)>0$. Let $D$ be the diagonal matrix such that $D_{ii}$ is equal to
the $i$th invariant factor of $I-A$. Then $D+I$ is a matrix of nonnegative
integers whose invariants are the same as those of $A$. But $D+I$ contains
$n$ strongly connected components, and cannot be in the same equivalence class
as $A$ since $n>1$.

Even if we restrict our attention to matrices with one strongly connected
component, the invariants do not completely determine the equivalence
classes. Consider the (unreduced) matrices $A_1=(1)$ and
$A_2=\left(\begin{array}{cc}2&1\\1&2\end{array}\right)$. In both cases
$\det(I-A_i)=0$ and the list of invariant factors contains a single zero
entry. But one may easily verify that every matrix in the equivalence class of
$A_1$ has a \emph{trivial} strongly connected component (consisting of a single
circuit). This is not true of $A_2$. Hence these matrices are in different
equivalence classes.

It turns out that this is the only exception. There is a single equivalence
class consisting of all matrices containing one trivial strongly connected
component and no other strongly connected components. If we restrict our
attention to the set of matrices which have one nontrivial strongly connected
component and no other strongly connected components, then the invariants found
so far completely determine the equivalence classes. A proof of this assertion
is the subject of the next section. This proof requires the following theorem,
which asserts the existence of a relevant property which distinguishes this set
of matrices from all others.

\begin{theorem}
Let $B$ be an $n\times n$ reduced matrix. Then \g{B+I} contains one nontrivial
strongly connected component and no other strongly connected components if and
only if $B$ is equivalent via equivalence transformations to a reduced $k\times
k$ matrix in which all entries are positive for each $k\geq n$.
\end{theorem}
\textsc{Proof.} Suppose $B$ can be transformed to a matrix $M$ in which all
entries are positive. Then \g{B+I} must contain exactly one strongly connected
component, since \g{M+I} is strongly connected and the number of strongly
connected components is invariant under equivalence transformations. But it
cannot be trivially strongly connected, since no reduced matrix in that
equivalence class has all entries positive.

Conversely, suppose that \g{B+I} contains one nontrivial strongly connected
component and no other strongly connected component. If vertex $i$ of \g{B+I}
is not in a strongly connected component, then $(B+I)_{ii}$ must be zero, so
the $i$th vertex may be removed via an elementary transformation. We may
continue to remove vertices in this way until the resulting graph is strongly
connected.

Now suppose that $B$ contains an element $B_{ii}=-1$. We may remove vertex $i$
using an elementary equivalence transformation, thereby reducing the dimension
of $B$ by 1. This operation may be repeated until there are no more negative
diagonal elements (this must happen eventually since the $1\times 1$ reduced
matrix with diagonal entry $-1$ is not strongly connected).

Now suppose that all of the entries of $B$ are nonnegative and that
$B_{ij}=0$. Since $B$ is strongly connected, its associated graph contains a
directed path connecting vertex $i$ to vertex $j$. The path may be chosen so
that no two vertices on the path are the same except possibly for the first and
last vertices (if $i=j$); for if we have a path in which any pair of vertices
other than the first and last are the same, we may shorten it by removing the
segment of the path connecting that vertex pair. Furthermore, we may require
that every edge in the path connects two distinct vertices. This follows
automatically from the above if $i\neq j$. Suppose $i=j$. Then $B_{ii}=0$. But
this means that $n$ cannot be 1, since then $B$ would be trivially strongly
connected. Since $n>1$ and $B$ is strongly connected there must be an edge
connecting $i$ to some other vertex $k$, and then a path with no vertices
repeated from $k$ back to $i$; this path satisfies our requirements.

Let such a path be $m_1,m_2,\ldots,m_k$ with $m_1=i$ and $m_k=j$. Since $m_1$
connects to $m_2$ and $m_1\neq m_2$, it follows that $B_{m_1m_2}>0$ and that we
may add row $m_2$ to row $m_1$. We know that $B_{m_2m_3}>0$ since $m_2$
connects to $m_3$ and $m_2\neq m_3$; since row $m_2$ was added to row $m_1$,
this means that now $B_{m_1m_3}>0$. If $k>3$ then $m_1\neq m_3$, so we may now
add row $m_3$ to row $m_1$. Continuing in the same manner, we eventually add
row $m_{k-1}$ to row $m_1$, which causes $B_{m_1m_k}=B_{ij}$ to become greater
than zero. No entry of $B$ has been decreased by these operations, so there is
now at least one fewer zero entry in $B$. If the whole process is repeated
often enough, eventually all entries in $B$ will be positive.

Given an $m\times m$ matrix $B$ with all entries positive, one may apply an
elementary equivalence transformation followed by a row operation to produce an
$(m+1)\times(m+1)$ matrix with all entries nonnegative, as shown in
Figure~\ref{expanding}.
\begin{figure}
\fbox{\begin{picture}(425,114)\footnotesize
\put(0,0){\makebox(425,114){$\begin{array}{ccc}\left(\begin{array}{ccccc}
B_{11}&B_{12}&\ldots&B_{1m}&0\\
B_{21}&B_{22}&\ldots&B_{2m}&0\\
\vdots&\vdots&\ddots&\vdots&\vdots\\
B_{m1}&B_{m2}&\ldots&B_{mm}-1&1\\
0&0&0&1&-1
\end{array}\right)&\Longrightarrow&\left(\begin{array}{ccccc}
B_{11}&B_{12}&\ldots&B_{1m}&0\\
B_{21}&B_{22}&\ldots&B_{2m}&0\\
\vdots&\vdots&\ddots&\vdots&\vdots\\
B_{m1}&B_{m2}&\ldots&B_{mm}-1&1\\
B_{m1}&B_{m2}&\ldots&B_{mm}&0
\end{array}\right)\end{array}$}}
\end{picture}}\linespread{1}\caption[A reduced $m\times m$ matrix with all
entries positive is equivalent to a reduced $(m+1)\times(m+1)$ matrix with all
entries positive]{A reduced $m\times m$ matrix with all entries positive is
equivalent to a reduced $(m+1)\times(m+1)$ matrix with all entries
positive. The matrix $B$ is transformed to produce the matrix $B'$, shown at
left. Since $B'_{(m+1)m}=1$, we may add row $m$ to row $m+1$ to produce the
matrix shown at right. All entries of the right-hand matrix are nonnegative, so
this matrix can be transformed into one with all entries positive.}
\label{expanding}
\end{figure}
By our previous results, this may be transformed into an $(m+1)\times(m+1)$
matrix with all entries positive. Clearly one may continue to expand $B$ in
this manner. Hence $B$ is equivalent to a $k\times k$ matrix with all entries
positive for each $k\geq m$.\proofend\bigskip

If $B$ is equivalent to a reduced matrix all of whose entries are positive, I
will call it a \emph{positive reduced matrix}.

Finally, note that if $B$ is a $k\times k$ matrix with all entries positive and
$k>1$, then it is equivalent to infinitely many such $k\times k$ matrices. For
since all entries are positive one may generate equivalent matrices with all
entries positive by adding rows or columns to one another with wild abandon.

\begin{theorem}
Let $B_1$ and $B_2$ be positive reduced matrices. Suppose that
$\det(-B_1)=\det(-B_2)$ and that the lists of invariant factors of $B_1$ and
$B_2$ are identical. Then $B_1$ and $B_2$ are in the same equivalence class.
\end{theorem}
\textsc{Proof.} To begin, note that $B_1$ or $B_2$ may be expanded by
elementary equivalence transformations until they are both $n\times n$ matrices
for some $n$. All that is required, then, is that we prove the theorem for the
case that $B_1$ and $B_2$ are both $n\times n$ matrices.

The strategy here, as in the proof of Theorem~\ref{integerthm}, is to provide
an algorithm which reduces any positive reduced $n\times n$ matrix $B$ to a
canonical form, which is also an $n\times n$ matrix. There will be only one
canonical form for each distinct set of invariants; hence it will follow that
any two positive reduced $n\times n$ matrices with the same set of invariants
must reduce to the same canonical form.

In the proof of Theorem~\ref{integerthm}, this canonical form was a diagonal
matrix. This will not be satisfactory here since diagonal $n\times n$ matrices
are not connected when $n>1$; hence they are not positive reduced
matrices. Instead, the canonical form used here is built around the circuit
$(12\ldots n)$. This is illustrated in Figure~\ref{submatrices}.
\begin{figure}
\fbox{\begin{picture}(425,350)\footnotesize
\put(0,234){\makebox(425,116){$\begin{array}{cccc}\left(\,\,\begin{array}{|ccccccccc|}
\hline
\,&\,&\,&\,&\,&\,&\,&\,&\,\\
\,&\,&\,&\,&\,&\,&\,&\,&\,\\
\,&\,&\,&\,&\,&\,&\,&\,&\,\\
\multicolumn{9}{|c|}{\raisebox{7pt}[0pt][0pt]{\large $S_1$}}\\
\,&\,&\,&\,&\,&\,&\,&\,&\,\\
\,&\,&\,&\,&\,&\,&\,&\,&\,\\
\hline\end{array}\,\,\right)&\Longrightarrow&\left(\begin{array}{c|ccccc|}  \cline{2-6}
0&\,&\,&\,&\,&\,\\
0&\,&\,&\,&\,&\,\\
0&\multicolumn{5}{|c|}{\raisebox{0pt}[0pt][0pt]{\large $S_2$}}\\
\vdots&\,&\,&\,&\,&\,\\
0&\,&\,&\,&\,&\,\\ \cline{2-6}
\multicolumn{1}{c}{g_1}&0&0&\cdots&0&\multicolumn{1}{c}{0}\\
\end{array}\,\,\right)&\Longrightarrow\end{array}$}}
\put(0,118){\makebox(425,122){$\begin{array}{cccccc}\left(\begin{array}{cc|cccc|}
0&\multicolumn{1}{c}{g_2}&0&\cdots&0&\multicolumn{1}{c}{0}\\ \cline{3-6}
0&0&\,&\,&\,&\,\\
0&0&\,&\,&\,&\,\\
\vdots&\vdots&\multicolumn{4}{|c|}{\raisebox{7pt}[0pt][0pt]{\large $S_3$}}\\
0&0&\,&\,&\,&\,\\ \cline{3-6}
g_1&\multicolumn{1}{c}{0}&0&\cdots&0&\multicolumn{1}{c}{0}\\
\end{array}\,\,\right)&\Longrightarrow\left(\begin{array}{ccc|ccc|}
0&g_2&\multicolumn{1}{c}{0}&\cdots&0&\multicolumn{1}{c}{0}\\
0&0&\multicolumn{1}{c}{g_3}&\cdots&0&\multicolumn{1}{c}{0}\\ \cline{4-6}
0&0&0&\,&\,&\,\\
\vdots&\vdots&\vdots&\multicolumn{3}{|c|}{\raisebox{0pt}[0pt][0pt]{\large $S_4$}}\\
0&0&0&\,&\,&\,\\ \cline{4-6}
g_1&0&\multicolumn{1}{c}{0}&\cdots&0&\multicolumn{1}{c}{0}\\
\end{array}\,\,\right)&\Longrightarrow&\cdots&\Longrightarrow\end{array}$}}
\put(0,2){\makebox(425,122){$\begin{array}{ccc}\left(\begin{array}{cccccc}
0&g_2&0&\cdots&0&0\\
0&0&g_3&\cdots&0&0\\
0&0&0&\ddots&\vdots&\vdots\\
\vdots&\vdots&\vdots&\cdots&g_{n-1}&g_{n-1}\\
0&0&0&\cdots&g_{n-1}&g_{n-1}+g_n\\
g_1&0&0&\cdots&0&0\end{array}\right)&\mathrm{or}&\left(\begin{array}{cccccc}
0&g_2&0&\cdots&0&0\\
0&0&g_3&\cdots&0&0\\
0&0&0&\ddots&\vdots&\vdots\\
\vdots&\vdots&\vdots&\cdots&g_{n-1}&g_{n-1}+g_n\\
0&0&0&\cdots&g_{n-1}&g_{n-1}\\
g_1&0&0&\cdots&0&0\end{array}\right)\end{array}$}}
\end{picture}}\linespread{1}\caption[Reduction of reduced matrices to canonical
form]{Reduction of reduced matrices to canonical form. Note the similarities to
Figure~\ref{hoffkunzalg}. Again, $g_i$ divides $g_{i+1}$ for each $i$. Here,
however, one does not always proceed to the final stage. If one arrives at a
state where every element of $S_i$ is equal to some number $g_i$, then this is
the canonical form. In this case $\det(I-A)$ (which is $\det(-B)$) is zero; the
list of invariant factors begins with $n-i$ zeros and continues with the
numbers $g_i$ through $g_1$ (omitting trailing 1's). Otherwise one keeps on
with the process. If one reaches $S_{n-1}$ and all elements of this $2\times 2$
matrix are not equal, then the determinant is nonzero and the reduction process
terminates in one of two possible ways (in one $\det(I-A)$ is positive, and in
the other it is negative); in this case the numbers $g_n$ through $g_1$ (with
trailing 1's omitted) constitute the list of invariant factors.}
\label{submatrices}
\end{figure}

There will be two variables associated with the algorithm. The variable $c$
refers to the submatrix $S_c$ being worked on; it also refers to a designated
column. The variable $r$ refers to a designated row. To begin the algorithm,
$c$ is set to 1 and $r$ is set to $n$. In addition, using the previous theorem
we initially transform $B$ so that all of its entries are positive.

The algorithm proceeds by performing a sequence of row and column operations on
$B$ in order to transform $S_c$ in a certain way. Unlike the case in
Theorem~\ref{integerthm}, here we must check each such operation to make sure
that it is allowed. One portion of this checking is to insure that the result
of any subtraction is an allowed reduced matrix. This will be checked as we go;
in fact, the algorithm will insure that $B$ always consists of nonnegative
integers. The other portion of the checking is to insure that $B_{ij}>0$ before
adding row $j$ to row $i$ or column $i$ to column $j$ and after subtracting row
$j$ from row $i$ or column $i$ from column $j$. If $B_{ij}$ lies in $S_c$, then
this too will be checked as we go. If it does not, then there is a trick which
insures that the desired operation will always be allowed. The trick depends on
the following fact: the algorithm guarantees that all entries in the $r$th row
and the $n$th column of $S_c$ are positive whenever an application of the trick
is required.

Suppose that we wish to add column $i$ to column $j$, or to subtract column $i$
from column $j$, where both columns contain elements of $S_c$. Instead of
attempting to perform the desired operation directly, we may do the
following:\vspace{3.5pt}
\begin{enumerate}\linespread{1}\normalsize
\item Add columns $c-1, c-2, \ldots, 1$ to column $j$, in that order.
\item Add column $i$ to column $j$, or subtract column $i$ from column $j$ (if
allowed).
\item Subtract columns $1, 2, \ldots, c-1$ from column $j$, in that order.
\end{enumerate}
We may add column $c-1$ to column $j$ because $B_{(c-1)j}$ is in row $r$ of
$S_c$, and hence is guaranteed by the algorithm to be positive. This sets
$B_{(c-2)j}=g_{c-1}>0$, so we may add column $c-2$ to column $j$, which sets
$B_{(c-3)j}=g_{c-2}>0$, so we may add column $c-3$ to column $j$, and so
on. Finally we add column 1 to column $j$, which sets $B_{nj}=g_1$. The result
is that each element of column $j$ that is not in $S_c$ is now positive. After
the desired operation is performed, we may subtract column 1 from column $j$
because $B_{1j}=g_2>0$; this zeroes $B_{nj}$. Then we may subtract column 2
from column $j$ because $B_{2j}=g_3>0$; this zeroes $B_{1j}$. And so
on. Finally, we may subtract column $c-1$ from column $j$ because $B_{(c-1)j}$
is in row $r$ of $S_c$ and hence is guaranteed by the algorithm to be
positive. Step 3 reverses the effects of step 1, and the net result is that the
desired column operation (if allowed) has been performed.

A similar trick works if $c>1$ and one wishes to add row $j$ to row $i$, or to
subtract row $j$ from row $i$, where rows $j$ and $i$ contain elements of
$S_c$:\vspace{3.5pt}
\begin{enumerate}\linespread{1}\normalsize
\item Add rows $n, 1, 2, \ldots, c-2$ to row $i$, in that order.
\item Add row $j$ to row $i$, or subtract row $j$ from row $i$.
\item Subtract rows $c-2, c-1, \ldots, 1, n$ from row $j$, in that order.
\end{enumerate}
We may add row $n$ to row $i$ because $B_{in}$ is guaranteed by the algorithm
to be positive. This sets $B_{i1}=g_1$, so we may add row 1 to row $i$, which
sets $B_{i2}=g_2$, so we may add row 2 to row $i$, and so on. The result is
that each element in row $j$ that is not in $S_c$ is made positive. After the
desired operation is performed, we may subtract row $c-2$ from row $i$ because
$B_{i(c-2)}=g_{c-2}>0$; this zeroes $B_{i(c-1)}$. Then we may subtract row
$c-3$ from row $i$ because $B_{i(c-3)}=g_{c-3}>0$; this zeroes
$B_{i(c-2)}$. And so on. Finally we may subtract row $n$ from row $i$ since
$B_{in}$ is guaranteed by the algorithm to be positive; this zeroes
$B_{i1}$. Step 3 reverses the effects of step 1, and the net result is that the
desired row operation (if allowed) has been performed.

The trick for columns is illustrated in Figure~\ref{trick}.
\begin{figure}
\fbox{\begin{picture}(425,166)\footnotesize\put(0,0){\makebox(425,166){%
$\begin{array}{cccccc}
\left(\begin{array}{cccc}0&2&0&0\\0&0&12&4\\0&0&8&4\\1&0&0&0\end{array}\right)&
\Longrightarrow&
\left(\begin{array}{cccc}0&2&2&0\\0&0&12&4\\0&0&8&4\\1&0&0&0\end{array}\right)&
\Longrightarrow&
\left(\begin{array}{cccc}0&2&2&0\\0&0&12&4\\0&0&8&4\\1&0&1&0\end{array}\right)&
\Longrightarrow\\[36pt]
\left(\begin{array}{cccc}0&2&2&0\\0&0&8&4\\0&0&4&4\\1&0&1&0\end{array}\right)&
\Longrightarrow&
\left(\begin{array}{cccc}0&2&2&0\\0&0&8&4\\0&0&4&4\\1&0&0&0\end{array}\right)&
\Longrightarrow&
\left(\begin{array}{cccc}0&2&0&0\\0&0&8&4\\0&0&4&4\\1&0&0&0\end{array}\right)
\end{array}$}}\end{picture}}\linespread{1}\caption[The trick for columns]{The
trick for performing column operations on $S_c$. Here $S_c$ is the $2\times 2$
submatrix obtained from $B$ by deleting columns 1 and 2 and rows 1 and
4. We wish to subtract column 4 from column 3 in order to modify $S_c$. To do
this, it is necessary that $B_{43}$ is positive after the operation is
performed. So we add column 2 to column 3 (which we can do since $B_{23}>0$)
and then add column 1 to column 3 (which we can do since now $B_{13}>0$). Now
all elements of column 3 are positive, so we may do the desired subtraction
(since $B_{43}>0$ after the subtraction). Next we subtract column 1 from column
3 (which we can do since $B_{13}>0$ after the subtraction), and then subtract
column 2 from column 3 (which we can do since $B_{23}>0$ after the
subtraction). The net effect is that we have subtracted column 4 from column
3.}
\label{trick}
\end{figure}
In the remainder of this section, when I say, for instance, that we may add
column $i$ to column $j$ since $B_{ij}$ is not in $S_c$, I mean that we can add
it by using the trick. Note that if all entries of $S_c$ are positive both
before and after a row or column operation on $S_c$, then that operation
(perhaps via the trick) is allowed.

Now for the algorithm proper. To start, all entries of $S_c$ are
positive. Consider some row $m$ of $S_c$. (Whenever I refer to a row or column,
I will refer to it by its index in $B$, but I mean to imply that this is a row
or column containing entries in $S_c$.) Suppose that there exist columns $i$
and $j$ such that $B_{mi}>B_{mj}$. Let $h_1$ be the greatest common divisor of
the entries in this row. We would like to reduce the sum of these entries by
subtracting column $j$ from column $i$. At the same time, we wish to preserve
the condition that all entries of $S_c$ are positive. This is not possible if
there is a row $k$ such that $B_{ki}\leq B_{kj}$. However, if this is the case
then we may add row $m$ repeatedly to row $k$; if we do it often enough, then,
since $B_{mi}>B_{mj}$, eventually we will have $B_{ki}>B_{kj}$. Thus we can
create the situation where $B_{ki}>B_{kj}$ for each $k$. Now we may subtract
column $j$ from column $i$, thereby reducing the sum of the entries in row
$m$. If there still exist columns $i$ and $j$ such that $B_{mi}>B_{mj}$, we
repeat the procedure. Eventually, the process must terminate: now the entries
of row $m$ will all have the same value. Since our operations on row $m$ did
not change the greatest common divisor of the row, this value must be $h_1$.

Let $g_c$ be the greatest common divisor of the elements of $S_c$. If $h_1\neq
g_c$, then there exists a column such that the greatest common divisor of its
elements is $h_2$, with $h_2<h_1$. By repeating the procedure described in the
above paragraph, we may set all of the elements of that column to $h_2$. If
$h_2\neq g_c$, then there exists a row such that the greatest common divisor of
its elements is $h_3$, with $h_3<h_2$, and we may again transform $S_c$ so that
all elements in this row are equal to $h_3$. And so on. Since the sequence
$h_1,h_2,h_3,\ldots$ is a decreasing sequence of positive integers, this
process must terminate. We end up with a row or column whose entries are all
equal to $g_c$. By repeating the procedure described in the preceding paragraph
one or two more times, we may set all of the elements of column c to $g_c$.

The next step is to see if there is a column $j$ in which all entries are not
equal. If not, then all columns are multiples of column $c$. By subtracting
column $c$ from each of the other columns as many times as needed, we may
transform $S_c$ to a matrix in which all entries are equal to $g_c$. (The
subtractions are allowed since again all entries in $S_c$ remain positive.) The
resulting matrix $B$ is our canonical form, so we exit the algorithm.

If, on the other hand, there does exist such a column $j$, then there is an
element $B_{ij}$ in $S_c$ which is less than some other element in that
column.

If $c=n-1$ then we subtract column $n-1$ from column $n$ as many times
as we can while preserving the fact that each entry of $S_{n-1}$ is
positive. The resulting matrix $B$ is our canonical form. Now $S_{n-1}$
contains three entries which are equal to $g_{n-1}$ and one larger entry $k$
that is divisible by $g_{n-1}$. We set $g_n=k-g_{n-1}$ and exit the algorithm.

Suppose now that $c<n-1$. If $i\neq r$ then we choose any row $k$ with $k\neq
i$ and $k\neq r$. We now perform the sequence of operations shown in
Figure~\ref{rotaterows}.
\begin{figure}
\fbox{\begin{picture}(425,60)\footnotesize\put(0,0){\makebox(425,60){%
$\begin{array}{lllllllllllll}
\mathbf{r}&\,&\mathbf{r}&\,&\mathbf{r}&\,&\mathbf{r+i+k}&\,&\mathbf{i}&\,&\mathbf{i}&\,&\mathbf{i}\\
\mathbf{k}&\Longrightarrow&\mathbf{k}&\Longrightarrow&\mathbf{k+r}&\Longrightarrow&\mathbf{k+r}&\Longrightarrow&\mathbf{k+r}&\Longrightarrow&\mathbf{k+r}&\Longrightarrow&\mathbf{r}\\
\mathbf{i}&\,&\mathbf{i+k}&\,&\mathbf{i+k}&\,&\mathbf{i+k}&\,&\mathbf{i+k}&\,&\mathbf{k}&\,&\mathbf{k}
\end{array}$}}\end{picture}}\linespread{1}\caption[Rotation of three
rows]{Rotation of three rows using row addition and subtraction. The three rows
in the table above represent rows $r$, $k$ and $i$ in the matrix. The symbols
$\mathbf{r}$, $\mathbf{k}$ and $\mathbf{i}$ represent the row vectors
containing the initial values of these rows. The following operations are
performed: row $k$ is added to row $i$, row $r$ is added to row $k$, row $i$ is
added to row $r$, row $k$ is subtracted from row $r$, row $r$ is subtracted
from row $i$, and row $i$ is subtracted from row $k$. The result is that the
row vectors have been rotated. If all entries of the three rows are positive at
the beginning of the procedure, then they are also positive at each stage
during the course of the procedure.}
\label{rotaterows}
\end{figure}
The result is a rotation of rows: row $i$ has moved to row $r$, row $r$ to row
$k$, and row $k$ to row $i$.

So far, every entry of $S_c$ has been positive before and after each row and
column operation, so these operations are all allowed. Now, however, we want to
create some zero entries, so we need to be more careful.

Now there exists some row $k$ such that $B_{rj}<B_{kj}$. Suppose there is a
column $m>c$ such that $B_{rm}\geq B_{km}$. Then we may add column $j$ to
column $m$ as many times as needed until this is no longer the case. In this
way we obtain the situation in which $B_{rm}<B_{km}$ for each $m>c$. Now we
subtract row $r$ from row $k$. The result is that $B_{kc}=0$ and all other
entries are positive. This subtraction is allowed since $B_{kr}$ is not in
$S_c$.

Now choose row $i$ to be any row except for row $k$ and row $r$. It is allowed
to add row $k$ to row $i$ as many times as one wishes; for $B_{ik}$ is
positive, and remains so after any such operation. In this way one may insure
that $B_{im}>B_{rm}$ for each $m>c$, while $B_{ic}=B_{rc}=g_c$. Now row $r$ may
be subtracted from row $i$, since $B_{ir}$ is not in $S_c$. The result is that
$B_{ic}=0$, while the rest of row $i$ remains positive. By doing this for each
$i$, one may set each element of column $c$ to zero except for $B_{rc}$, which
still equals $g_c$, while leaving the rest of $S_c$ positive. Now let column
$j$ be any column other than column $c$. Column $c$ may now be repeatedly
subtracted from column $j$ until $B_{rj}=0$; the operations are allowed since
the value of $B_{cj}$ after each subtraction is positive (in fact, the only
value affected by the operation is $B_{rj}$). This is repeated for each
$j$. (Note that now for the first time we are setting elements of row $r$ and
column $n$ of $S_c$ to zero, so the trick may no longer be allowed; but this
does not matter, since the trick is no longer needed.) The result is that all
entries in column $c$ and in row $r$ are zero except for $B_{rc}$, which equals
$g_c$, and all other entries of $S_c$ are positive. These remaining positive
entries constitute our submatrix $S_{c+1}$. We increment $c$, let $r=c-1$, and
repeat the algorithm.\proofend\bigskip

Soon after proving this theorem I realized that the invariants I had found were
also invariants under reversible cellular automaton maps, which I knew were
equivalent to topological conjugacy maps of subshifts of finite type, and were
studied in symbolic dynamics. I wrote to Jack Wagoner, asking him if these
invariants had already been discovered in that context. He directed me to a
1984 paper by John Franks (\cite{franks}) which, to my surprise, contained the
exact theorem proven above, with an almost identical proof.

Unbeknownst to me, in symbolic dynamics there had naturally arisen a type of
map between subshifts of finite type called a \emph{flow equivalence map}. It
turns out that the set of all such maps is identical to the set of maps
generated by my elementary space set transformations. The nature of this
connection between symbolic dynamics and combinatorial spacetimes will be the
subject of the next section.

The history of the discovery of the invariants is as follows. Parry and
Sullivan~\cite{parry/sullivan} showed that $\det(I-A)$ is an invariant; they
also found the first characterization of flow equivalence in terms of simple
matrix operations. Then Bowen and Franks~\cite{bowen/franks} discovered not
only that the list of invariant factors was an invariant, but that these
factors were associated with an invariant abelian group, known as the
Bowen-Franks group. This will be discussed in Section~\ref{bowenfranks}.

\subsection{Flow equivalence of subshifts of finite type}
\label{flow}

In Chapter~\ref{relatedsystems} I briefly discussed the general setting of
symbolic dynamics within the field of dynamical systems. Here I begin by
defining some of the same concepts discussed earlier in a more formal way, and
then introduce the concept of flow \mbox{equivalence}.

Let $A$ be an $n\times n$ matrix of nonnegative integers. Let $E_A$ be the set
of edges of \g{A}. Give $E_A$ the discrete topology. Let \Z\ denote the
integers. Then the product space $E_A^{\Z}$ contains a subspace
$\Sigma_A=\{\alpha\in E_A^{\Z}\mid$ the head of edge $\alpha_i$ meets the tail
of edge $\alpha_{i+1}$ for all $i\in\Z\}$. Let
$\sigma_A:\Sigma_A\mapsto\Sigma_A$ be the mapping such that if
$\sigma_A(\alpha)=\beta$ then $\beta_i=\alpha_{i-1}$ for each $i\in\Z$. Then
$\sigma_A$ is a homeomorphism, and is called the \emph{subshift of finite type}
associated with the matrix $A$.

Given any homeomorphism $h:X\mapsto X$ one may define an equivalence relation
$\sim$ on $X\times\R$ by the rule $(x,s+1)\sim (h(x),s)$. Consider the
identification space $Y=X\times\R/\sim$, and denote elements of $Y$ by
$[(x,s)]$, $x\in X$, $s\in\R$. Then the map $\phi:Y\times\R\mapsto Y$ given by
$\phi_{t}[(x,s)]=[(x,s+t)]$ is a flow on $Y$, and is called the
\emph{suspension flow} of $h$. The space $Y$ is called the \emph{suspension} of
$h$.

Two flows $\phi_1:Y_1\times\R\mapsto Y_1$ and $\phi_{2}:Y_{2}\times\R\mapsto
Y_2$ are said to be \emph{topologically equivalent} if there exists a
homeomorphism $h:Y_1\mapsto Y_2$ which maps orbits of $\phi_1$ onto orbits of
$\phi_2$ and preserves their orientation. If the subshifts of finite type
$\sigma_A$ and $\sigma_B$ have topologically equivalent suspension flows, then
$\sigma_A$ and $\sigma_B$ (and the associated matrices $A$ and $B$) are said to
be \emph{flow equivalent}.

Consider the following symmetric relations between square matrices of
nonnegative integers:
\begin{mylist}{(PS)}
\item[\textmd{(W)}] $A\sim B$ whenever $A=RS$ and $B=SR$ for some (not
necessarily square) matrices $R$ and $S$ of nonnegative integers.  (This is due
to Williams~\cite{williams}.)
\item[\textmd{(PS)}] $A\sim B$ whenever $A_{21}=1$, the remaining entries of
row~2 and column~1 of $A$ are zero, and $B$ is obtained from $A$ by deleting
row~2 and column~1. (Also write $B\sim A$ in this case, to make it symmetric.)
\end{mylist}
Parry and Sullivan~\cite{parry/sullivan} showed that the flow equivalence
relation on matrices of nonnegative integers is generated by the relations (W)
and (PS).

Another useful relation was found by Franks~\cite{franks}:
\begin{mylist}{(PS)}
\item[\textmd{(F)}] $A\sim B$ (and $B\sim A$) if, for distinct integers $i$
and $j$ such that $A_{ij}>0$, $B$ can be obtained from $A$ by subtracting 1
from the ($i,j$)th entry and then adding row $j$ to row $i$.
\end{mylist}
Note that this is the same relation shown in Figure~\ref{addrow}. A
similar relation exists for columns. Franks showed that matrices of
nonnegative integers that satisfy these relations are flow equivalent.

\begin{theorem} The equivalence relation on matrices of nonnegative integers
generated by elementary transformations is the same as the flow equivalence
relation.
\end{theorem}
\textsc{Proof}. Let (V) denote the vertex removal operation (and its inverse). It
suffices to show the following: (V) $\Rightarrow$ (PS); (V) $\Rightarrow$ (W);
(F) and (W) $\Rightarrow$ (V).

The first implication is easy. Consider (PS). In \g{A} there is an edge from
vertex 2 to vertex 1, and there are no other edges leaving vertex 2 or arriving
at vertex 1. The move is to combine vertices 2 and 1 into a single vertex
having the predecessors of vertex 2 and the successors of vertex 1. This is the
undoubling operation. It is a special case of (V), since the move is achieved
by removing either of vertices 1 or 2.

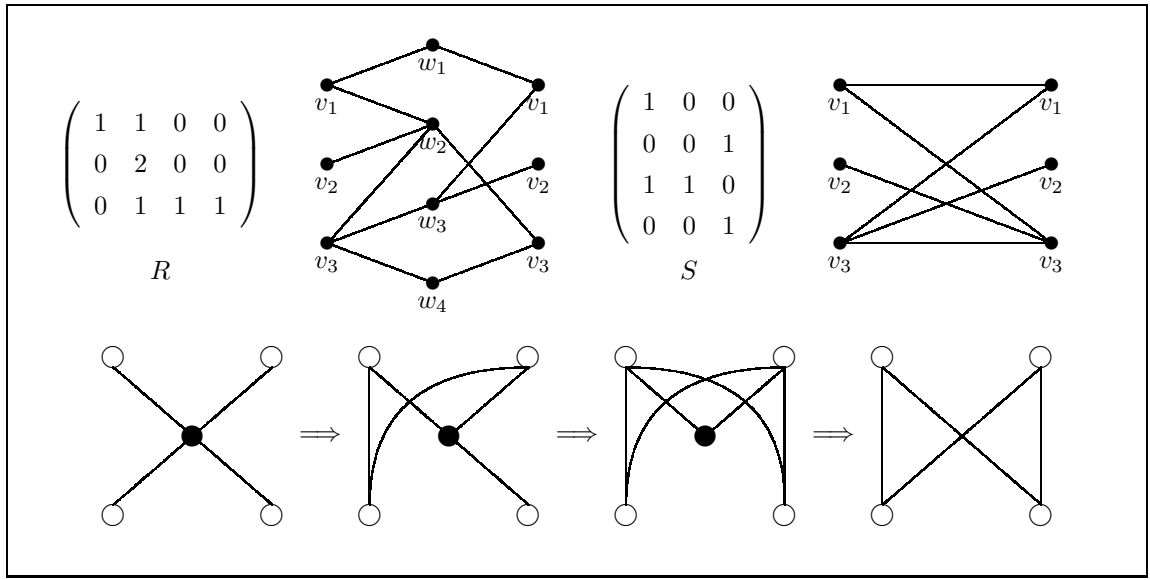
\begin{figure}[t!]
\fbox{\begin{picture}(425,210)\footnotesize

\put(0,-42){
\put(118,150){
\multiput(0,0)(80,0){2}{\multiput(0,15)(0,30){3}{\circle*{5}}}
\multiput(40,0)(0,30){4}{\circle*{5}}
\qbezier(0,75)(20,82.5)(40,90)\qbezier(0,75)(20,67.5)(40,60)
\qbezier(0,45)(20,52.5)(40,60)\qbezier(0,15)(20,37.5)(40,60)
\qbezier(0,15)(20,22.5)(40,30)\qbezier(0,15)(20,7.5)(40,0)
\qbezier(40,0)(60,7.5)(80,15)\qbezier(40,30)(60,37.5)(80,45)
\qbezier(40,30)(60,52.5)(80,75)\qbezier(40,60)(60,37.5)(80,15)
\qbezier(40,90)(60,82.5)(80,75)
\multiput(0,7)(80,0){2}{
        \put(0,0){\makebox(0,0){$v_3$}}
        \put(0,30){\makebox(0,0){$v_2$}}
        \put(0,60){\makebox(0,0){$v_1$}}}
\put(40,-8){
        \put(0,0){\makebox(0,0){$w_4$}}
        \put(0,30){\makebox(0,0){$w_3$}}
        \put(0,60){\makebox(0,0){$w_2$}}
        \put(0,90){\makebox(0,0){$w_1$}}}
}\put(312,150){
\multiput(0,0)(80,0){2}{\multiput(0,15)(0,30){3}{\circle*{5}}}
\put(0,75){\line(1,0){80}}
\qbezier(0,75)(40,45)(80,15)
\qbezier(0,45)(40,30)(80,15)
\put(0,15){\line(1,0){80}}
\qbezier(0,15)(40,30)(80,45)\qbezier(0,15)(40,45)(80,75)
\put(0,15){\line(1,0){80}}
\multiput(0,7)(80,0){2}{
        \put(0,0){\makebox(0,0){$v_3$}}
        \put(0,30){\makebox(0,0){$v_2$}}
        \put(0,60){\makebox(0,0){$v_1$}}}
}\put(55,195){\makebox(0,0){$\left(\begin{array}{cccc}1&1&0&0\\0&2&0&0\\0&1&1&1\end{array}\right)$}}
\put(255,195){\makebox(0,0){$\left(\begin{array}{ccc}1&0&0\\0&0&1\\1&1&0\\0&0&1\end{array}\right)$}}
\put(55,155){\makebox(0,0){$R$}} \put(255,155){\makebox(0,0){$S$}} }
\put(37,20){
\put(0,0){\multiput(0,0)(60,0){2}{\multiput(0,0)(0,60){2}{\circle{8}}}\put(30,30){\circle*{8}}
\qbezier(0,4)(15,17)(30,30)\qbezier(30,30)(45,17)(60,4)
\qbezier(0,56)(15,43)(30,30)\qbezier(30,30)(45,43)(60,56)
}\put(97,0){\multiput(0,0)(60,0){2}{\multiput(0,0)(0,60){2}{\circle{8}}}\put(30,30){\circle*{8}}
\put(0,4){\line(0,1){52}}\qbezier(0,4)(0,56)(60,56)\qbezier(30,30)(45,17)(60,4)
\qbezier(0,56)(15,43)(30,30)\qbezier(30,30)(45,43)(60,56)
}\put(194,0){\multiput(0,0)(60,0){2}{\multiput(0,0)(0,60){2}{\circle{8}}}\put(30,30){\circle*{8}}
\put(0,4){\line(0,1){52}}\qbezier(0,4)(0,56)(60,56)
\put(60,4){\line(0,1){52}}\qbezier(60,4)(60,56)(0,56)
\qbezier(0,56)(15,43)(30,30)\qbezier(30,30)(45,43)(60,56)
}\put(291,0){\multiput(0,0)(60,0){2}{\multiput(0,0)(0,60){2}{\circle{8}}}
\multiput(0,4)(60,0){2}{\line(0,1){52}}
\qbezier(0,4)(30,30)(60,56)\qbezier(60,4)(30,30)(0,56) }
\multiput(78.5,30)(97,0){3}{\makebox(0,0){$\Longrightarrow$}} }
\end{picture}}\linespread{1}\caption{Proof that the two equivalence relations
are the same. In the first row, matrices $R$ and $S$ are interpreted as the
graph pictured between them, where pairs of vertices labelled $v_i$ are
identified for each $i$ to obtain the graph. Arrows go from left to right. $R$
indicates which edges go between the leftmost $v$'s and the $w$'s; $S$
indicates which edges go between the $w$'s and the rightmost $v$'s. The $w$'s
are removed to obtain the graph at right (again the $v$'s are identified to
obtain the graph). This graph is described by the matrix $RS$. If the $v$'s
were removed instead, the resulting graph would be described by $SR$. In the
second row, the leftmost configuration is transformed by removing the lower
edges one by one, and then removing the central vertex (note that the latter
move is a type (W) move). The result is that the elementary vertex-removal
operation has been performed.}
\label{flowproof}
\end{figure}

Now consider (W). Let $R$ be an $n\times m$ matrix and $S$ be an $m\times n$
matrix (of nonnegative integers). Let $A=RS$ and $B=SR$. The picture to
consider here is as follows (see Figure~\ref{flowproof}).
Let $v_i$, $1\leq i\leq n$, and $w_j$, $1\leq j\leq m$, be vertices of a graph
\G. Let $R_{ij}$ be the number of edges from $v_i$ to $w_j$, and let $S_{ji}$
be the number of edges from $w_j$ to $v_i$. Then \g{A} is the graph which
results from removing each of the vertices $w_j$ from \G\ via the (V)
transformation; and \g{B} is the graph which results from removing each of the
vertices $v_i$ from \G\ via the (V) transformation. So a (W) transformation may
be accomplished by successively applying $m$ inverse (V) transformations
followed by $n$ (V) transformations.

Now consider (F). The graphical picture here is similar to the one for (V),
except that here it is an edge being removed rather than a vertex.  The edge to
be removed (call it $e$) connects vertex $i$ to vertex $j$, where $i\neq
j$. For each edge from vertex $j$ to vertex $k$ (where $k$ is any vertex in the
graph), there is a two-edged path from $i$ to $k$ whose first edge is $e$. This
path will no longer be present when $e$ is removed. The transformation (F)
replaces each such path with a new edge from $i$ to $k$.

Suppose that in some graph \g{A} there are no edges connecting vertex $i$ to
itself. Using (F) repeatedly, one may remove in turn each of the $m$ edges
which are directed towards $i$. None of the edges added in this process involve
$i$; so the only remaining edges that involve $i$ are directed away from
$i$. Those edges and the vertex $i$ itself may now be removed by a (W)
transformation. The resulting transformation (see Figure~\ref{flowproof}) is
identical to the (V) transformation that removes vertex $i$. So (V) is
accomplished by $m$ applications of (F) and a single application of
(W).\proofend\bigskip

The fact that these equivalence relations are the same is by no means a
coincidence. It turns out that space sets and equivalence maps between them are
hidden in this picture.

Consider the suspension $\F\hspace{-2pt}_A$ associated with a set of spaces
\x{A}. The spaces in \x{A} are closely related to the \emph{flow lines} of
$\F\hspace{-2pt}_A$. Topologically, these flow lines turn out to be exactly the
arcwise-connected components of $\F\hspace{-2pt}_A$. As subspaces of $\F\hspace{-2pt}_A$, the
flow lines are homeomorphic to lines or circles. They are circles when the
corresponding element of \x{A} is periodic: i.e., it is of the form
\cir{SSS\ldots S} or \cir{\ldots SSS \ldots} for some segment $S$. For a given
$S$, each such element of \x{A} corresponds to the same flow line in
$\F\hspace{-2pt}_A$. The presence of infinitely many such spaces associated
with a given segment $S$ is in a sense a redundancy in \x{A}, since locally
they look the same (and therefore, for example, if $T$ is a local equivalence
map sending \cir{S} to \cir{U} it will also send \cir{SS} to
\cir{UU}). T$\F\hspace{-2pt}_A$.hus we do not lose anything significant if we
throw all such spaces except for the single space \cir{S} out of the space set
\x{A}. If we do so, then there is a natural invertible map between the elements
of \x{A} and the flow lines of $\F\hspace{-2pt}_A$.

Since the flow lines are the arcwise-connected components, any homeomorphism
$h:\F\hspace{-2pt}_A\mapsto\F\hspace{-2pt}_B$ must map flow lines to flow
lines. Hence it maps spaces to spaces. Since it is a homeomorphism, this
induced map on space sets is invertible. It turns out that these maps are
exactly our orientation-preserving space set equivalence maps. They preserve
orientation because of the definition of flow equivalence, which specifies that
the homeomorphism must preserve the direction of flow lines. If we drop this
requirement and consider the set of all homeomorphisms between suspensions of
subshifts of finite type, we simply pick up those maps which involve some
orientation reversal; thus these constitute exactly the entire set of oriented space
set equivalence maps.

Of course, these assertions require proof. I have only recently managed to
prove them, and do not have time to include the proofs here. In fact, I have
proven that the map on space sets induced by a flow equivalence map can be
represented by a nonoverlap rule table; this makes possible at least a partial
resolution of the conjecture of Section~\ref{element}. These results will
appear in a paper shortly.

While the same maps are studied in symbolic dynamics and in 1+1-dimensional
combinatorial spacetimes, the roles of these maps in the two fields are
different. In fact, the roles of dynamical maps and equivalence maps in
combinatorial spacetimes are reversed in symbolic dynamics. In symbolic
dynamics the dynamical system being discussed is the shift map. The flow
equivalence map is a type of equivalence between shift dynamical systems. But
from the combinatorial spacetime point of view the shift map is not a dynamical
map at all. In fact, it is an equivalence map: given spaces with coordinates,
any two spaces related by a sequence of shifts are considered to be
equivalent. And, of course, the flow equivalence maps are the ones whereby
spaces evolve; hence they are the dynamical maps.

The existence of this connection between flow equivalence and combinatorial
spacetimes means that results from one field may be applied to the other. For
example, it turns out that a relationship has been discovered between flow
equivalence maps and a special type of $\mathrm{C}^*$-algebra called a
Cuntz-Krieger algebra; these algebras are therefore also related to
1+1-dimensional combinatorial spacetimes. I have not explored this
relationship. See~\cite{huang1,huang2,huang3,huang4} for further details and
references.

While the connection with $\mathrm{C}^*$-algebras is apparently restricted to a
small class of those algebras, it turns out that another connection exists
between 1+1-dimensional combinatorial spacetimes and algebra which is much more
general. This is the subject of the next section.

\subsection{Bialgebras}
\label{bialgebra}

Bialgebras are objects of much recent study, largely in the guise of Hopf
algebras (or quantum groups). The latter are bialgebras with some extra
structure (the antipode); connections have arisen between these bialgebras and
statistical mechanics, knot theory and many other areas of mathematics. It
turns out that combinatorial spacetimes is yet another area that is connected
with bialgebras. (It follows that flow equivalence, and hence symbolic
dynamics, is also connected to bialgebras; this, I believe, has not been
noticed up to now.)

I will begin by presenting the definition of a bialgebra (see
Sweedler~\cite{sweedler}, Abe~\cite{abe}) and its abstract tensor
representation. I then define tensors \tensor\ for each $r,s \geq 0$. Each of
the standard objects in a bialgebra is represented by one of these tensors: the
unit scalar is $T$; the unit is $T^a$; the counit is $T_a$; the Kronecker delta
is $T_a^b$; multiplication is $T_{ab}^c$; comultiplication is $T_a^{bc}$. It
turns out that the axioms of a bialgebra can be described in this context by
two related axioms. (This reformulation of the bialgebra axioms may itself be
of interest; I have not seen it in the literature.)

The fact that there are two axioms rather than one is really only due to
notation. The two axioms embody a single principle. This principle has to do
with directed paths through abstract tensor diagrams. In such diagrams a tensor
is considered as a vertex with some extra structure resulting from the
noncommutativity and noncocommutativity of the tensor. An upper index labels
the tail of a directed edge, and a lower index labels the head of a directed
edge; when two such indices are the same (due to contraction), they refer to
the same edge. The extra structure at each vertex means that, at a vertex, the
edges entering the vertex are ordered, as are the edges leaving the
vertex. This ordering induces an ordering on all directed paths ending at the
vertex and on all directed paths beginning at the vertex. The axiom then says
the following: if a tensor subdiagram does not contain closed directed paths,
it may be replaced by any other tensors subdiagram that does not contain closed
directed paths provided that the number of paths between each pair of terminal
vertices and the ordering of paths at those vertices are preserved.

In the case of scalars, it is possible to write down the two axioms as a single
axiom. Now suppose that the bialgebra is commutative and cocommutative. In this
case there is no ordering at vertices; the diagrams of scalars are just
directed graphs, and the bialgebra axiom for scalars turns out to be exactly
the same as the elementary space set transformation.

My approach to the problems discussed below was inspired by
Kauffman~\cite{kauffman}, and in particular by his diagrammatic use of abstract
tensors. This in turn is related to work of Penrose~\cite{penrose2}.

\subsubsection{The abstract tensor representation}

Let $A$ be a vector space over a field $K$ and let $\otimes$ represent the
tensor product. Define \(\tau : A \otimes A \mapsto A \otimes A\) (the
``twist'' map) to be the bilinear map sending $a \otimes b$ to $b \otimes a$
for each $a,b \in A$. Suppose there are multilinear maps \(\mu : A \otimes A
\mapsto A\) (``multiplication''), \(u : K \mapsto A\) (``unit''), \(\Delta : A
\mapsto A \otimes A\) (``comultiplication''), and \(\epsilon : A \mapsto K\)
(``counit'') satisfying the following axioms (for all $a,b,c \in A$):
\settowidth{\wod}{(AAA2A)}\begin{description}
\item[{\normalfont\makebox[\wod][r]{(A1a)\,}}] Associativity: \(\mu(a \otimes \mu(b \otimes c)) =
          \mu(\mu(a \otimes b) \otimes c)\)
\item[{\normalfont\makebox[\wod][r]{(A1b)\,}}] Coassociativity: \((1 \otimes \Delta)(\Delta(a)) = (\Delta
          \otimes 1)(\Delta(a))\)
\item[{\normalfont\makebox[\wod][r]{(A2a)\,}}] \(\mu(u(1) \otimes a) = \mu(a \otimes u(1)) = a\)
\item[{\normalfont\makebox[\wod][r]{(A2b)\,}}] \((\epsilon \otimes 1)(\Delta(a)) = (1 \otimes \epsilon)(\Delta(a)) = a\)
\item[{\normalfont\makebox[\wod][r]{(A3)\,}}] \(\epsilon(u(1)) = 1\)
\item[{\normalfont\makebox[\wod][r]{(A4a)\,}}] \(\Delta(u(1)) = u(1) \otimes u(1)\)
\item[{\normalfont\makebox[\wod][r]{(A4b)\,}}] \(\epsilon(\mu(a,b)) = \epsilon(a)\epsilon(b)\)
\item[{\normalfont\makebox[\wod][r]{(A5)\,}}] \(\Delta(\mu(a,b)) = (\mu \otimes \mu)((1 \otimes
          \tau \otimes 1)((\Delta \otimes \Delta)(a \otimes b)))\)
\end{description}
Then \ang{A,\mu,u,\Delta,\epsilon} is called a \emph{bialgebra}. If in
addition it satisfies
\begin{description}
\item[{\normalfont\makebox[\wod][r]{\,}}] Commutativity: \(\mu(a \otimes b) = \mu(b \otimes a)\)
\item[{\normalfont\makebox[\wod][r]{\,}}] Cocommutativity: \(\Delta(a) = \tau(\Delta(a))\)
\end{description}
then it is called a \emph{commutative cocommutative bialgebra}.

Let $a$ and $b$ be any two elements of $A$. If $A$ is finite-dimensional as a
vector space, then we may write $a = V^a$ and $b = W^b$, and there exist
tensors \(M_{ab}^c, M^{ab}_c, U_a\) and $U^a$ such that \(\mu(a,b) =
M_{ab}^cV^aW^b, \Delta(a) = M^{ab}_cV^c, u(1) = U^a,\) and
\(\epsilon(a) = U_aV^a\). These tensors satisfy the following axioms:
\begin{description}
\item[{\normalfont\makebox[\wod][r]{(B1a)\,}}] Associativity: \(M_{ax}^dM_{bc}^x = M_{ab}^xM_{xc}^d\)
\item[{\normalfont\makebox[\wod][r]{(B1b)\,}}] Coassociativity: \(M^{ax}_dM^{bc}_x = M^{ab}_xM^{xc}_d\)
\item[{\normalfont\makebox[\wod][r]{(B2a)\,}}] \(M_{xa}^bU^x = M_{ax}^bU^x = \delta_a^b\)
\item[{\normalfont\makebox[\wod][r]{(B2b)\,}}] \(M^{xa}_bU_x = M^{ax}_bU_x = \delta_b^a\)
\item[{\normalfont\makebox[\wod][r]{(B3)\,}}] \(U_aU^a = 1\)
\item[{\normalfont\makebox[\wod][r]{(B4a)\,}}] \(M^{ab}_cU^c = U^aU^b\)
\item[{\normalfont\makebox[\wod][r]{(B4b)\,}}] \(M_{ab}^cU_c = U_aU_b\)
\item[{\normalfont\makebox[\wod][r]{(B5)\,}}] \(M_{ab}^xM_x^{cd} = M_a^{wx}M_b^{yz}M_{wy}^cM_{xz}^d\)
\end{description}
In this context commutativity is the relation \(M_{ab}^c = M_{ba}^c\), and
cocommutativity is the relation \(M^{ab}_c = M^{ba}_c\).

If we now drop the assumption of finite-dimensionality and instead think of the
tensors $M_{ab}^c$, $M^{ab}_c$, $U^a$, $U_a$ and $\delta^a_b$ simply as symbols
which obey the rules of tensor algebra plus the above axioms, then we have the
\emph{abstract tensor representation} of a bialgebra.

\subsubsection{A simplification}

One may now define tensors \tensor\ for any nonnegative integers $r,s$
inductively as follows. Let \(T = 1, T_a = U_a, T^a = U^a, T_a^b =
\delta_a^b\). If $r > 1$ then let \(\tensor = T_{a_1 \ldots a_{r-2} w}^{b_1
\ldots b_s}M_{a_{r- 1}a_r}^w\). If $s > 1$ then let \(\tensor = T_{a_1 \ldots
a_r}^{b_1 \ldots b_{s-2} w}M^{b_{s-1} b_s}_w\). Note that \(T_{ab}^c =
\delta_w^cM_{ab}^w = M_{ab}^c\) and \(T^{ab}_c = \delta^w_cM^{ab}_w =
M^{ab}_c\). Hence each of our original tensors is equal to one of the tensors
\tensor; thus the abstract tensor representation of a bialgebra may be
completely expressed in terms of these new tensors.

It follows from the above definition that, if $r$ and $s$ are nonzero, then
\tensor\ acts on $r$ bialgebra elements (i.e., on $r$ vectors) by multiplying
them together in a particular order and then performing $s-1$
comultiplications.

\begin{figure}
\fbox{\begin{picture}(425,180)\footnotesize
\put(20,100){
\qbezier(0,0)(10,20)(20,40)\put(20,40){\circle*{5}}\qbezier(20,40)(30,20)(40,0)
\put(30,20){\circle*{5}}\qbezier(30,20)(25,10)(20,0)\put(20,40){\line(0,1){20}}
\put(45,30){\makebox(0,0){=}}\put(50,0){
\qbezier(0,0)(10,20)(20,40)\put(20,40){\circle*{5}}\qbezier(20,40)(30,20)(40,0)
\put(10,20){\circle*{5}}\qbezier(10,20)(15,10)(20,0)\put(20,40){\line(0,1){20}}
}}
\put(170,110){
\qbezier(0,0)(5,10)(10,20)\put(10,20){\circle*{5}}\qbezier(10,20)(15,10)(20,0)
\put(10,20){\line(0,1){20}}\put(0,0){\circle*{5}}
\put(25,20){\makebox(0,0){=}}
\put(30,0){\qbezier(0,0)(5,10)(10,20)\put(10,20){\circle*{5}}\qbezier(10,20)(15,10)(20,0)
\put(10,20){\line(0,1){20}}\put(20,0){\circle*{5}}
\put(25,20){\makebox(0,0){=}}\put(40,0){\line(0,1){40}}
\put(40,20){\circle*{5}}}
}\put(300,120){
\multiput(0,0)(0,20){2}{\circle*{5}}\put(0,0){\line(0,1){20}}
\put(15,10){\makebox(0,0){=}}\put(30,10){\circle*{5}}
\put(45,10){\makebox(0,0){=}}
\put(60,10){\raisebox{-2pt}{\parbox{1in}{\linespread{1}\footnotesize the\\empty\\\raisebox{-1pt}{diagram}}}}
}\put(20,30){
\qbezier(0,40)(5,30)(10,20)\put(10,20){\circle*{5}}\qbezier(10,20)(15,30)(20,40)
\put(10,20){\line(0,-1){20}}\put(10,0){\circle*{5}}
\put(23,20){\makebox(0,0){=}}\multiput(36,0)(13,0){2}{\put(0,40){\line(0,-1){20}}
\put(0,20){\circle*{5}}}
}
\put(111,20){
\qbezier(0,0)(5,10)(10,20)\qbezier(10,20)(15,10)(20,0)
\multiput(10,20)(0,20){2}{\circle*{5}}\put(10,20){\line(0,1){20}}
\qbezier(10,40)(5,50)(0,60)\qbezier(10,40)(15,50)(20,60)
\put(23,30){\makebox(0,0){=}}
\multiput(36,0)(20,0){2}{\put(0,0){\line(0,1){60}}
\multiput(0,20)(0,20){2}{\circle*{5}}}
\put(-4,0){\qbezier(40,20)(50,30)(60,40)\qbezier(60,20)(50,30)(40,40)}
}
\put(209,20){
\qbezier(0,0)(10,20)(20,40)\qbezier(20,0)(10,20)(0,40)\put(10,20){\circle*{5}}
\qbezier(10,20)(20,30)(30,40)\multiput(10,5)(0,30){2}{\makebox(0,0){...}}
\put(31,20){\makebox(0,0){=}}
\put(38,0){\qbezier(0,0)(10,20)(20,40)\qbezier(20,0)(10,20)(0,40)\put(10,20){\circle*{5}}
\qbezier(20,40)(15,50)(10,60)\qbezier(20,40)(25,50)(30,60)\put(20,40){\circle*{5}}
\multiput(10,5)(0,30){2}{\makebox(0,0){...}}}
}
\put(309,40){
\qbezier(0,0)(20,20)(40,40)\qbezier(40,0)(20,20)(0,40)
\put(20,20){\line(0,-1){20}}\put(20,35){\makebox(0,0){...}}
\multiput(13,5)(14,0){2}{\makebox(0,0){...}}\put(20,20){\circle*{5}}
\put(43,20){\makebox(0,0){=}}
\put(46,0){
\qbezier(0,0)(20,20)(40,40)\qbezier(40,0)(20,20)(0,40)
\put(20,20){\line(0,-1){40}}\put(20,0){\circle*{5}}
\put(20,35){\makebox(0,0){...}}\multiput(13,5)(14,0){2}{\makebox(0,0){...}}
\put(20,20){\circle*{5}}
}}
\end{picture}}\linespread{1}\caption{Diagrams of the bialgebra axioms. Each
tensor \tensor\ is represented by a vertex. Edges are ordered from left to
right at the bottom and top of the vertex; they enter from the bottom and exit
from the top. Edges adjacent to only one vertex (``free'' edges) come in two
types: ``upper'' edges point away from the vertex, and ``lower'' edges point
towards the vertex. In each axiom, pairs of upper edges from the left and right
side of the equation are identified if they occupy corresponding positions in
the diagram (and the same goes for lower edges). Shown (in order) are diagrams
for axioms (C1a), (C2a), (C3), (C4a), (C5), (C6a) and (C7a). For
$i\in\{1,2,4,6,7\}$, the diagram of axiom (C$i$b) can be obtained from the
diagram of axiom (C$i$a) by turning it upside down.  By convention the product
of zero tensors is 1; hence the empty diagram represents 1.}
\label{axioms}
\end{figure}
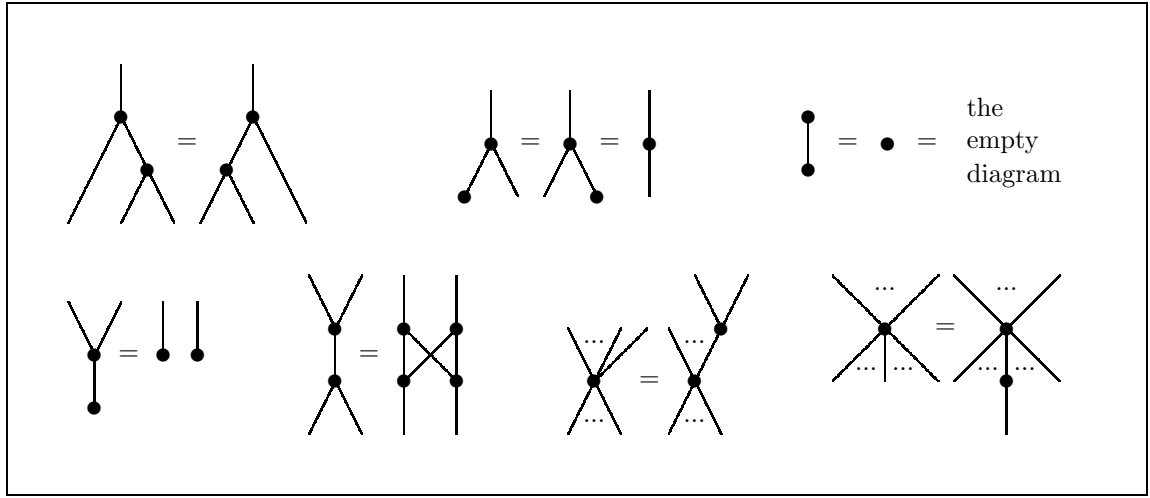

Now we may rewrite axioms (B1) through (B5) using \tensor's, add several
additional axioms to capture the defining properties of the \tensor's, and
thereby obtain an expression of the abstract tensor representation of a
bialgebra which is written entirely in terms of the new tensors.
\begin{description}
\item[{\normalfont\makebox[\wod][r]{(C1a)\,}}] Associativity: \(T_{ax}^dT_{bc}^x = T_{ab}^xT_{xc}^d\)
\item[{\normalfont\makebox[\wod][r]{(C1b)\,}}] Coassociativity: \(T^{ax}_dT^{bc}_x = T^{ab}_xT^{xc}_d\)
\item[{\normalfont\makebox[\wod][r]{(C2a)\,}}] \(T_{xa}^bT^x = T_{ax}^bT^x = T_a^b\)
\item[{\normalfont\makebox[\wod][r]{(C2b)\,}}] \(T^{xa}_bT_x = T^{ax}_bT_x = T_b^a\)
\item[{\normalfont\makebox[\wod][r]{(C3)\,}}] \(T_aT^a = T = 1\)
\item[{\normalfont\makebox[\wod][r]{(C4a)\,}}] \(T^{ab}_cT^c = T^aT^b\)
\item[{\normalfont\makebox[\wod][r]{(C4b)\,}}] \(T_{ab}^cT_c = T_aT_b\)
\item[{\normalfont\makebox[\wod][r]{(C5)\,}}] \(T_{ab}^xT_x^{cd} = T_a^{wx}T_b^{yz}T_{wy}^cT_{xz}^d\)
\item[{\normalfont\makebox[\wod][r]{(C6a)\,}}] If $r>1$ then \(\tensor = T_{a_1 \ldots a_{r-2} w}^{b_1
\ldots b_s}T^w_{a_{r-1} a_r}\)
\item[{\normalfont\makebox[\wod][r]{(C6b)\,}}] If $s>1$ then \(\tensor = T_{a_1 \ldots a_r}^{b_1 \ldots b_{s-2} w}T_w^{b_{s-1} b_s}\)
\item[{\normalfont\makebox[\wod][r]{(C7a)\,}}] If $1 \leq j \leq r$ then \(\tensor = T_{a_1
\ldots a_{j-1} w a_{j+1} \ldots a_r}^{b_1 \ldots b_s}T^w_{a_j}\)
\item[{\normalfont\makebox[\wod][r]{(C7b)\,}}] If $1 \leq j \leq s$ then \(\tensor
= T_{a_1 \ldots a_r}^{b_1 \ldots b_{j-1} w b_{j+1} \ldots b_s}T_w^{b_j}\)
\end{description}
The axioms (C1) through (C7) define a tensor algebra in the \tensor's which is
equivalent to the original tensor algebra. They are pictured diagrammatically in
Figure~\ref{axioms}.

The value of using these new tensors lies in the fact that one may now express
the original tensor algebra using many fewer axioms.
\begin{theorem}\label{c1}
If \(1 \leq j \leq r,\) \(0 \leq s\) and \(0 \leq m\), then
\begin{description}
\item[{\normalfont\makebox[\wod][r]{(D1a)\,}}] \(T_{a_1 \ldots a_r}^{b_1 \ldots b_s}T_{c_1 \ldots
    c_m}^{a_j} = T_{a_1 \ldots a_{j-1} c_1 \ldots c_m a_{j+1} \ldots a_r}^{b_1
    \ldots b_s}\)
\item[{\normalfont\makebox[\wod][r]{(D1b)\,}}] \(T_{a_1 \ldots a_s}^{b_1 \ldots b_r}T_{b_j}^{c_1
    \ldots c_m} = T_{a_1 \ldots a_s}^{b_1 \ldots b_{j-1} c_1 \ldots c_m b_{j+1}
    \ldots b_r}\)
\end{description}
\end{theorem}
\textsc{Proof sketch}. If $m>0$, then (D1a) says that a certain rearrangement
of the parentheses in a multiplication does not change anything, which follows
from (C1a); and (D1b) says that a certain rearrangement of the parentheses in a
comultiplication does not change anything, which follows from (C1b). If $m = 0$
then the result follows by applying (C2a), (C2b), (C3), (C7a), (C7b), (D1a) and
(D1b) (in the case where $m>0$), and induction on $r$ and~$s$.\proofend\bigskip

In what follows, the symbol $\prod$ will refer to the tensor product.

\begin{theorem}
\label{c2}
If $r \geq 0$ and $s \geq 0$, then
\begin{description}
\item[{\normalfont\makebox[\wod][r]{(D2)\,}}] \(\tensor = (\prod_{i=1}^{r} T_{a_i}^{x_{i,1} \ldots
x_{i,s}}) (\prod_{j=1}^{s} T_{x_{1,j} \ldots x_{r,j}}^{b_j})\)
\end{description}
\end{theorem}
\textsc{Proof}.

\emph{Case 1}. $r=0$ and $s=0$. Then (D2) says (via our convention for
defining a product of zero \tensor's) that $T = 1$, which is true by
definition.

\emph{Case 2}. $r>0$ and $s=0$. Then (D2) says that \(T_{a_1 \ldots a_r} =
T_{a_1}T_{a_2} \ldots T_{a_r}\). This is clearly true if $r = 1$. Suppose that
$r > 1$ and that the statement holds for $r-1$. Then
\begin{quote}
\begin{tabular}{lcll}
\smallskip\(T_{a_1 \ldots a_r}\)&$=$&\(T_{a_1 \ldots a_{r-2} x}T^x_{a_{r-1}a_r}\)&(by (C6a))\\
\smallskip&$=$&\(T_{a_1}T_{a_2} \ldots T_{a_{r-2}}T_xT^x_{a_{r-1}a_r}\)&(by inductive hypothesis)\\
&$=$&\(T_{a_1}T_{a_2} \ldots T_{a_{r-2}}T_{a_{r-1}}T_{a_r}\)&(by (C4b)).\\
\end{tabular}
\end{quote}

\emph{Case 3}. $r=0$ and $s>0$. Similar to case 2.

\emph{Case 4}. $r = 1$ and $s \geq 1$. Then (D2) says that \[T_{a_1}^{b_1
\ldots b_s} = T_{a_1}^{x_{1,1} \ldots
x_{1,s}}T_{x_{1,1}}^{b_1}T_{x_{1,2}}^{b_2} \ldots T_{x_{1,s}}^{b_s},\] and this
follows from (C7b).

\emph{Case 5}. $r\geq 1$ and $s=1$. Similar to case 4.

\emph{Case 6}. $r=2$ and $s=2$. Then (D2) says that \[T_{a_1 a_2}^{b_1 b_2}
= T_{a_1}^{x_{1,1} x_{1,2}}T_{a_2}^{x_{2,1} x_{2,2}}T_{x_{1,1}
x_{2,1}}^{b_1}T_{x_{1,2} x_{2,2}}^{b_2};\] but this is just (C5).

\emph{Case 7}. $r\geq 2$ and $s\geq 2$. We use induction on $r+s$.  Case 6
shows that (D2) holds if $r+s=4$. Now assume that $r+s>4$ and that $r>2$. Then
\begin{quote}
\begin{tabular}{lcl}
\smallskip\tensor&$=$&\(T_{a_1 \ldots a_{r-2} w}^{b_1 \ldots b_s}T_{a_{r-1} a_r}^w\)\\
\smallskip&$=$&\((\prod_{i=1}^{r-2}T_{a_i}^{x_{i,1} \ldots x_{i,s}})T_w^{x_{w,1} \ldots x_{w,s}}(\prod_{j=1}^{s}T_{x_{1,j} \ldots
x_{r-2,j} x_{w,j}}^{b_j})T_{a_{r-1} a_r}^w\)\\
\smallskip&$=$&\((\prod_{i=1}^{r-2}T_{a_i}^{x_{i,1} \ldots x_{i,s}})(\prod_{j=1}^{s}T_{x_{1,j} \ldots x_{r-2,j} x_{w,j}}^{b_j})T_{a_{r-1} a_r}^{x_{w,1} \ldots x_{w,s}}\)\\
\medskip&$=$&\((\prod_{i=1}^{r-2}T_{a_i}^{x_{i,1} \ldots
x_{i,s}})(\prod_{j=1}^{s}T_{x_{1,j} \ldots x_{r-2,j} x_{w,j}}^{b_j})\,\otimes\)\\
&&\(T_{a_{r-1}}^{x_{r-1,1} \ldots
x_{r-1,s}}T_{a_r}^{x_{r,1} \ldots x_{r,s}}(\prod_{k=1}^{s}T_{x_{r-1,k} x_{r,k}}^{x_{w,k}}
)\)\\
\end{tabular}
\end{quote}
(by applying (C6a), then the inductive hypothesis, then (C6a), then the
inductive hypothesis), and now the desired result (D2) follows by applying
(C6a) to remove all instances of indices involving $w$. A similar argument
holds if we assume that $s>2$ instead of assuming that $r>2$.\proofend\bigskip

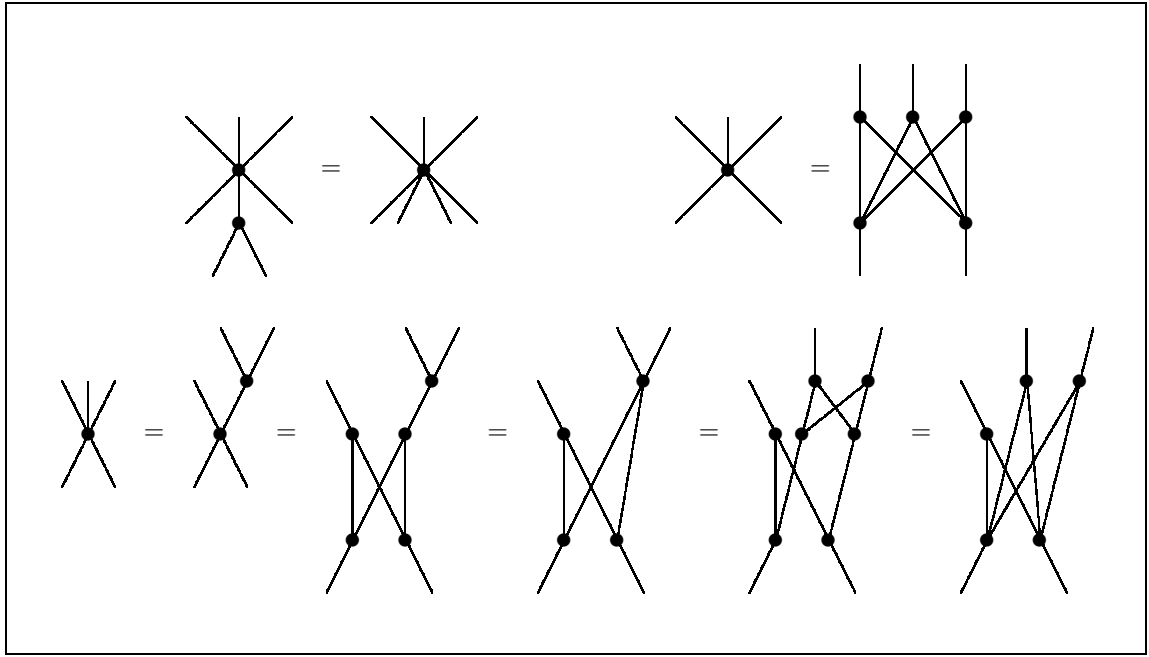
\begin{figure}
\fbox{\begin{picture}(425,240)\footnotesize
\put(18,60){
\qbezier(0,0)(10,20)(20,40)\qbezier(0,40)(10,20)(20,0)\put(10,20){\circle*{5}}
\put(10,20){\line(0,1){20}}\put(35,20){\makebox(0,0){=}}
\put(50,0){
\qbezier(0,0)(10,20)(20,40)\qbezier(0,40)(10,20)(20,0)\put(10,20){\circle*{5}}
\qbezier(20,40)(25,50)(30,60)\qbezier(20,40)(15,50)(10,60)
\put(20,40){\circle*{5}}\put(35,20){\makebox(0,0){=}}
\put(50,0){
\qbezier(0,-40)(25,10)(50,60)\qbezier(40,-40)(20,0)(0,40)
\multiput(10,-20)(20,0){2}{\multiput(0,0)(0,40){2}{\circle*{5}}\put(0,0){\line(0,1){40}}}
\put(40,40){\circle*{5}}\qbezier(40,40)(35,50)(30,60)
\put(65,20){\makebox(0,0){=}}
\put(80,0){
\qbezier(0,-40)(25,10)(50,60)\qbezier(40,-40)(20,0)(0,40)
\multiput(10,-20)(0,40){2}{\circle*{5}}\put(30,-20){\circle*{5}}
\put(40,40){\circle*{5}}\qbezier(30,-20)(35,10)(40,40)\qbezier(40,40)(35,50)(30,60)\put(10,-20){\line(0,1){40}}
\put(65,20){\makebox(0,0){=}}
\put(80,0){
\qbezier(0,-40)(5,-30)(10,-20)\qbezier(40,-40)(20,0)(0,40)
\qbezier(10,-20)(17.5,10)(25,40)\put(25,40){\line(0,1){20}}\qbezier(30,-20)(40,20)(50,60)
\multiput(20,20)(20,0){2}{\multiput(0,0)(5,20){2}{\circle*{5}}}
\qbezier(20,20)(32.5,30)(45,40)\qbezier(40,20)(32.5,30)(25,40)
\multiput(10,-20)(0,40){2}{\circle*{5}}\put(30,-20){\circle*{5}}
\put(10,-20){\line(0,1){40}}
\put(65,20){\makebox(0,0){=}}
\put(80,0){
\qbezier(0,-40)(5,-30)(10,-20)\qbezier(40,-40)(20,0)(0,40)
\qbezier(10,-20)(17.5,10)(25,40)\put(25,40){\line(0,1){20}}\qbezier(30,-20)(40,20)(50,60)
\multiput(25,40)(20,0){2}{\circle*{5}}
\qbezier(10,-20)(27.5,10)(45,40)\qbezier(30,-20)(27.5,10)(25,40)
\multiput(10,-20)(0,40){2}{\circle*{5}}\put(30,-20){\circle*{5}}
\put(10,-20){\line(0,1){40}}
}}}}}}
\put(65,160){
\qbezier(0,0)(20,20)(40,40)\put(20,0){\line(0,1){40}}\qbezier(40,0)(20,20)(0,40)\put(20,20){\circle*{5}}\put(20,0){\circle*{5}}\qbezier(20,0)(15,-10)(10,-20)\qbezier(20,0)(25,-10)(30,-20)\put(55,20){\makebox(0,0){=}}
\put(70,0){
\qbezier(0,0)(20,20)(40,40)\put(20,20){\line(0,1){20}}\qbezier(40,0)(20,20)(0,40)\put(20,20){\circle*{5}}\qbezier(20,20)(15,10)(10,0)\qbezier(20,20)(25,10)(30,0)
}}
\put(250,160){
\qbezier(0,0)(20,20)(40,40)\qbezier(40,0)(20,20)(0,40)\put(20,20){\line(0,1){20}}\put(20,20){\circle*{5}}\put(55,20){\makebox(0,0){=}}
\put(70,0){
\multiput(0,0)(40,0){2}{\circle*{5}}\multiput(0,40)(20,0){3}{\circle*{5}}
\multiput(0,-20)(40,0){2}{\line(0,1){80}}\put(20,40){\line(0,1){20}}
\qbezier(0,0)(20,20)(40,40)\qbezier(40,0)(20,20)(0,40)
\qbezier(0,0)(10,20)(20,40)\qbezier(40,0)(30,20)(20,40)
}}
\end{picture}}\linespread{1}\caption{The new bialgebra axioms. The top row
gives diagrams for examples of axioms (D1a) and (D2). The bottom row shows how
Case 7 of Theorem~\ref{c2} is proven for $r=2$ and $s=3$.}
\label{newaxioms}
\end{figure}

Examples of diagrams for these new axioms, as well as an illustration of the
proof of Case 7 above, appear in Figure~\ref{newaxioms}.

\begin{theorem}
\label{c1c2converse}
The axioms of a bialgebra are completely generated by \textup{(D1a)}, \textup{(D1b)} and \textup{(D2)}.
\end{theorem}
\textsc{Proof sketch}. This one is very easy. (C1a), (C2a), (C6a) and (C7a)
follow from one or two applications of (D1a). (C1b), (C2b), (C6b)
and (C7b) follow from one or two applications of (D1b). (C5) is a
special case of (D2). (C4a) follows by applying (D1b) and then
(D2). (C4b) follows by applying (D1a) and then (D2). (C3) follows
by applying either (D1a) or (D1b) and then (D2).\proofend\bigskip

Theorems \ref{c1}, \ref{c2} and \ref{c1c2converse} imply the following:
\begin{theorem} The axioms of the abstract tensor representation of a
bialgebra are equivalent to the axioms \textup{(D1a)}, \textup{(D1b)} and \textup{(D2)}.
\end{theorem}

This is the best we can do if our notation consists entirely of
\tensor's. By adding a few elements to our notation, we can do a
little bit better.

For instance, it would be nice to write axioms (C7a) and (C7b) simply as follows:
\begin{description}
\item[{\normalfont\makebox[\wod][r]{(E1)\,}}] $T_a^b=\delta_a^b$.
\end{description}
Graphically, $\delta_a^b$ represents an edge. The axiom then says that you can
insert a vertex in the middle of any edge. This is simpler, so let us allow
it. It means that in our diagrams we are now allowing there to be edges that
are not attached to any vertices.

I have already described the notion of free and fixed vertices in terms of
graphs (see Figure~\ref{elem}). Here I would like to adapt these notions to the
case of tensor diagrams in the \tensor's. A fixed vertex refers to a \tensor,
and is represented by a filled-in circle. A free vertex refers to a
\emph{location} on some unknown \tensor, and is represented by an open circle
(or oval). Free vertices come in two types, ``upper'' and ``lower'': all edges
adjacent to an upper vertex point towards it, and all edges adjacent to a lower
vertex point away from it. If several edges hit a free vertex, then they are
ordered from left to right just as at fixed vertices. They are thought of as
being adjacent to one another; they are an \emph{ordered clump} of edges
located someplace on an unknown \tensor. As before, free vertices may refer to
locations on the same unknown \tensor, and no free vertex refers to a location
on any of the fixed vertices.

The terms ``free edge,'' ``upper edge'' and ``lower edge'' will continue to be
used (see Figure~\ref{axioms}). The free edges are those which are connected to
at most one vertex; that vertex may either be fixed or free.

I will use the symbols $P_{\alpha}^{a_1\ldots a_n}$ and $P^{\alpha}_{a_1\ldots
a_n}$ to denote the two kinds of free vertices. Greek letters will label free
vertices; roman letters will label free edges.

\begin{figure}
\fbox{\begin{picture}(425,160)\footnotesize
\put(87.5,20){
\qbezier(0,0)(10,10)(20,20)\qbezier(20,20)(30,10)(40,0)\put(20,0){\line(0,1){60}}
\multiput(20,20)(0,20){2}{\circle*{5}}
\qbezier(40,60)(20,40)(0,20)\qbezier(0,60)(20,40)(40,20)
\put(55,30){\makebox(0,0){$\Longrightarrow$}}
\put(70,0){
\qbezier(0,0)(10,10)(20,20)\qbezier(20,20)(30,10)(40,0)\put(20,0){\line(0,1){60}}
\multiput(20,20)(0,20){2}{\circle*{5}}
\qbezier(40,60)(20,40)(0,20)\qbezier(0,60)(20,40)(40,20)
\multiput(30,10)(-10,0){3}{\circle*{5}}
\put(55,30){\makebox(0,0){$\Longrightarrow$}}
\put(70,0){
\qbezier(0,0)(5,5)(10,10)\qbezier(30,10)(35,5)(40,0)\put(20,0){\line(0,1){60}}
\put(20,40){\circle*{5}}\qbezier(10,10)(15,25)(20,40)\qbezier(30,10)(25,25)(20,40)
\qbezier(40,60)(20,40)(0,20)\qbezier(0,60)(20,40)(40,20)
\multiput(30,10)(-10,0){3}{\circle*{5}}
\put(55,30){\makebox(0,0){$\Longrightarrow$}}
\put(70,0){
\qbezier(0,0)(10,20)(20,40)\qbezier(20,40)(30,20)(40,0)\put(20,0){\line(0,1){60}}
\put(20,40){\circle*{5}}
\qbezier(40,60)(20,40)(0,20)\qbezier(0,60)(20,40)(40,20)
}}}}
\put(20,100){
\qbezier(1.77,1.77)(20,20)(38.23,38.23)\qbezier(38.23,1.77)(20,20)(1.77,38.23)
\put(20,20){\circle*{5}}
\put(20,20){\line(0,1){17.5}}
\multiput(0,0)(40,0){2}{\circle{5}}\multiput(0,40)(20,0){3}{\circle{5}}
\put(50,20){\makebox(0,0){$\Longrightarrow$}}
\put(60,0){
\qbezier(1.77,1.77)(20,20)(38.23,38.23)\qbezier(38.23,1.77)(20,20)(1.77,38.23)
\put(20,30){\line(0,1){7.5}}\multiput(10,10)(20,0){2}{\circle*{5}}
\multiput(10,30)(10,0){3}{\circle*{5}}\multiput(10,10)(20,0){2}{\line(0,1){20}}
\qbezier(10,10)(15,20)(20,30)\qbezier(30,10)(25,20)(20,30)
\multiput(0,0)(40,0){2}{\circle{5}}\multiput(0,40)(20,0){3}{\circle{5}}
\put(50,20){\makebox(0,0){$\Longrightarrow$}}
\put(67,0){
\qbezier(1.77,1.77)(20,20)(38.23,38.23)\qbezier(38.23,1.77)(20,20)(1.77,38.23)
\multiput(0,2.5)(40,0){2}{\line(0,1){35}}
\qbezier(1.12,2.24)(10,20)(18.88,37.76)\qbezier(21.12,37.76)(30,20)(38.88,2.24)
\multiput(0,0)(40,0){2}{\circle{5}}\multiput(0,40)(20,0){3}{\circle{5}}
}}}
\put(245,100){
\qbezier(0,0)(20,20)(40,40)\qbezier(40,0)(20,20)(0,40)\put(20,20){\circle*{5}}
\put(20,20){\line(0,1){20}}
\put(50,20){\makebox(0,0){$\Longrightarrow$}}
\put(60,0){
\qbezier(0,0)(20,20)(40,40)\qbezier(40,0)(20,20)(0,40)\put(20,20){\circle*{5}}
\put(20,20){\line(0,1){20}}\multiput(10,10)(20,0){2}{\circle*{5}}
\multiput(10,30)(10,0){3}{\circle*{5}}
\put(50,20){\makebox(0,0){$\Longrightarrow$}}
\put(60,0){
\qbezier(0,0)(20,20)(40,40)\qbezier(40,0)(20,20)(0,40)
\put(20,30){\line(0,1){10}}\multiput(10,10)(20,0){2}{\circle*{5}}
\multiput(10,30)(10,0){3}{\circle*{5}}\multiput(10,10)(20,0){2}{\line(0,1){20}}
\qbezier(10,10)(15,20)(20,30)\qbezier(30,10)(25,20)(20,30)
}}}
\end{picture}}\linespread{1}\caption{Proof that (E1) and (E2) constitute a
valid axiom set for bialgebras. In the upper left, (D2) is applied to the
central vertex and then (D1a) is applied three times and (D1b) two times to
eliminate the fixed vertices. The result is that (E2) has been performed. In
the upper right, (E1) is applied to each of the five edges, and then (E2) is
applied to the central vertex; the result is that (D2) has been performed. At
bottom, (E1) is applied to the bottom three edges, then (E2) is applied to the
central vertex, then the inverse of (E1) is applied three times; the result is
that (D1a) has been performed.}
\label{eproof}
\end{figure}
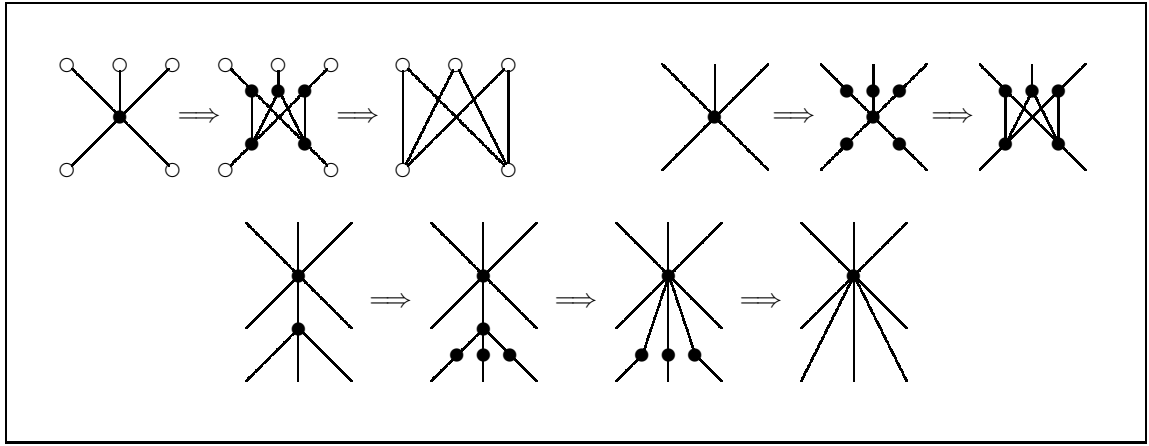

\begin{theorem} The axioms \textup{(E1)} and
\begin{description}
\item[{\normalfont\makebox[\wod][r]{(E2)\,}}] \((\prod_{i=1}^rP_{\alpha_i}^{a_i})\tensor
(\prod_{j=1}^sP^{\beta_j}_{b_j})=(\prod_{i=1}^rP_{\alpha_i}^{x_{i,1} \ldots
x_{i,s}})(\prod_{j=1}^sP^{\beta_j}_{x_{1,j} \ldots x_{r,j}})\)
\end{description}
are equivalent to the axioms \textup{(D1a)}, \textup{(D1b)} and \textup{(D2)}.
\end{theorem}
\textsc{Proof sketch.} (See Figure~\ref{eproof} for illustrations.)
We have already seen that (E1) is a consequence of (C7a) and (C7b), and hence
of (D1a), (D1b) and (D2). (E2) may be obtained by applying (D2) and then
removing all fixed vertices by applying (D1a) and (D1b). Conversely, (D1a) may
be obtained by applying (E1) to each free edge, then applying (E2), then
applying the inverse of (E1) to remove each vertex that had been added by
applying (E1). (D2b) may be obtained similarly. And (D2) may be obtained by
applying (E1) to each free edge and then applying (E2).\proofend

\subsubsection{The bialgebra principle}

By now it should be obvious that there is really only one basic sort of thing
going on with these bialgebra axioms.

The axiom (E1) is pretty trivial, and it is \emph{almost} a consequence of
(E2). If there are vertices at both ends of an edge, then applying (E1) to that
edge is indeed a special case of (E2). Hence (E1) is only needed when free
edges are involved. It seems that the existence of (E1) is mostly a matter of
notation; the main principle behind the bialgebra axioms is contained in (E2).

In (E2), one transforms a vertex with $r$ lower edges and $s$ upper
edges. There are $rs$ paths through this vertex. The vertex is removed, but the
paths are maintained. Each length-2 path that had passed through the vertex is
replaced with a length-1 path. And the ordering of paths is also maintained, in
a sense to be explained below.

A more expanded view of this principle may be obtained by considering a tensor
subdiagram $G$ that contains no directed loops and no free edges but which may
contain free vertices. Let the free lower vertices of $G$ be given by $v_1,
\ldots, v_m$ and the free upper vertices by $w_1, \ldots, w_n$.

Consider the set of directed paths \(P = \{ \alpha_1, \ldots, \alpha_k \}\)
that begin at a $v$ vertex and end at a $w$ vertex. For each $i$, $1 \leq i
\leq m$, let $\alpha_r, \alpha_s \in P$ be any two paths that begin at
$v_i$. If we start at $v_i$ and travel forwards along $\alpha_r$, there must be
some point where we diverge for the first time from $\alpha_s$. If $\alpha_r$
goes to the left at that point and $\alpha_s$ goes to the right, write
$\alpha_r >_v \alpha_s$; otherwise write $\alpha_s >_v \alpha_r$. Now for each
$j$, $1 \leq j \leq n$, let $\alpha_r, \alpha_s \in P$ be any two paths that
end at $w_j$. If we start at $w_j$ and travel backwards along $\alpha_r$, there
must be some point where we diverge for the first time from $\alpha_s$. If
$\alpha_r$ goes to the left at that point and $\alpha_s$ goes to the right,
write $\alpha_r >_w \alpha_s$; otherwise write $\alpha_s >_w \alpha_r$.

Now we may construct a new subdiagram $G'$ consisting of free vertices $v_1,
\ldots, v_m$ and $w_1, \ldots, w_n$ and no other vertices. For each directed
path in $G$ going from $v_i$ to $w_j$ there is a corresponding edge in $G'$
going from $v_i$ to $w_j$, and these are all of the edges. It remains to
determine the ordering of the edges at each vertex $v_i$ and $w_j$. This is
given by the relations $>_v$ and $>_w$: if $\alpha_r$ and $\alpha_s$ both begin
at $v_i$, then the tail of the edge corresponding to $\alpha_r$ is to the left
of the tail of the edge corresponding to $\alpha_s$ if and only if $\alpha_r
>_v \alpha_s$; and if $\alpha_r$ and $\alpha_s$ both end at $w_j$, then the
head of the edge corresponding to $\alpha_r$ is to the left of the head of the
edge corresponding to $\alpha_s$ if and only if $\alpha_r >_w \alpha_s$.

\begin{theorem} Let $G_1$ and $G_2$ be two subdiagrams containing no directed
loops and no free edges, but containing free vertices $v_1, \ldots, v_m$ and
$w_1, \ldots, w_n$ as described above ($m, n \geq 0$). Then $G_1' = G_2'$ if
and only if $G_1$ is equivalent to $G_2$ via axiom (E).
\end{theorem}
\textsc{Proof sketch}. Consider first the definition of (E). If $G_1$ is the
diagram associated with the left-hand side of this definition and $G_2$ is the
diagram associated with the right-hand side, then it is easy to see that $G_2 =
G_1'$, where the path $a_ib_j$ through $G_1$ has been replaced by the edge
$x_{ij}$ in $G_2$. This one-to-one map between the paths through $G_1$ and the
paths through $G_2$ preserves the ordering of these paths at the free vertices.

Now consider what happens when $G_1$ is arbitrary and one transforms it using
(E). This transformation consists of a one-to-one mapping of paths through a
subdiagram which preserves their ordering at the free vertices of the
subdiagram. It is easy to verify that the transformation therefore also
consists of a one-to-one mapping of paths through $G_1$ which preserves their
ordering at the free vertices of $G_1$.  If $G_2$ is obtained from $G_1$ by
repeated applications of (E), then there is a one-to-one mapping of paths
through $G_1$ with paths through $G_2$ which preserves their ordering at the
free vertices. But then it follows that $G_1' = G_2'$.

Conversely, one may transform $G_1$ using (E) to remove each fixed vertex in
turn until it no longer contains a fixed vertex. Since these transformations
preserve paths and their orderings at free vertices, the resulting diagram must
be $G_1'$. If one does the same thing to $G_2$, one must obtain $G_2'$. Thus if
$G_1' = G_2'$ then $G_1$ and $G_2$ are equivalent via (E).\proofend\bigskip

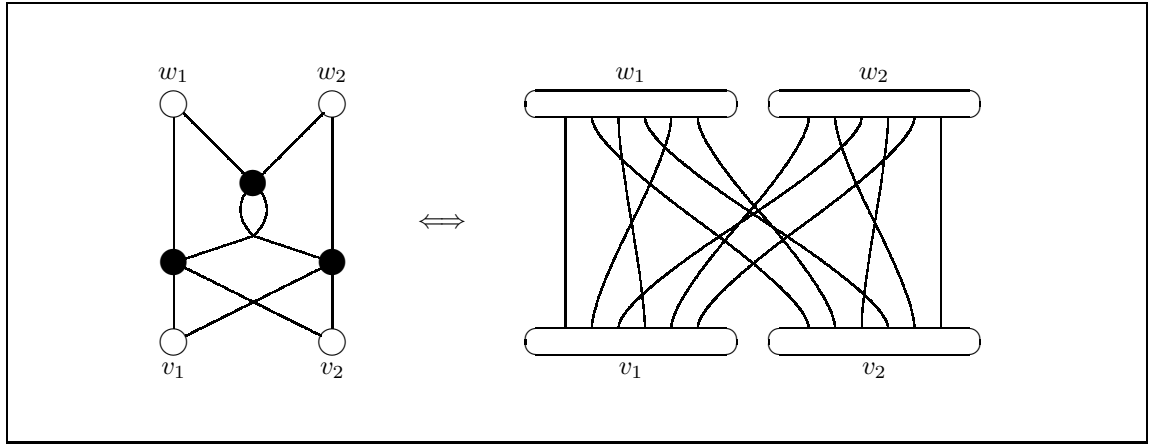
\begin{figure}
\fbox{\begin{picture}(425,160)\footnotesize
\put(193,30){
\put(40,5){\oval(80,10)}\put(132,5){\oval(80,10)}
\put(40,95){\oval(80,10)}\put(132,95){\oval(80,10)}
\put(40,-6){\makebox(0,0){$v_1$}}
\put(132,-6){\makebox(0,0){$v_2$}}
\put(40,106){\makebox(0,0){$w_1$}}
\put(132,106){\makebox(0,0){$w_2$}}
\qbezier(15,10)(15,20)(15,50)\qbezier(15,50)(15,80)(15,90)
\qbezier(25,10)(25,20)(40,50)\qbezier(40,50)(55,80)(55,90)
\qbezier(35,10)(35,20)(81,50)\qbezier(81,50)(127,80)(127,90)
\qbezier(45,10)(45,20)(40,50)\qbezier(40,50)(35,80)(35,90)
\qbezier(55,10)(55,20)(81,50)\qbezier(81,50)(107,80)(107,90)
\qbezier(65,10)(65,20)(106,50)\qbezier(106,50)(147,80)(147,90)
\qbezier(107,10)(107,20)(66,50)\qbezier(66,50)(25,80)(25,90)
\qbezier(117,10)(117,20)(92,50)\qbezier(92,50)(65,80)(65,90)
\qbezier(127,10)(127,20)(132,50)\qbezier(132,50)(137,80)(137,90)
\qbezier(137,10)(137,20)(91,50)\qbezier(91,50)(45,80)(45,90)
\qbezier(147,10)(147,20)(132,50)\qbezier(132,50)(117,80)(117,90)
\qbezier(157,10)(157,20)(157,50)\qbezier(157,50)(157,80)(157,90)
}
\put(161.5,80){\makebox(0,0){$\Longleftrightarrow$}}
\put(60,30){
\multiput(0,5)(60,0){2}{\multiput(0,0)(0,90){2}{\circle{10}}}
\multiput(0,35)(60,0){2}{\circle*{10}}\put(30,65){\circle*{10}}
\multiput(0,10)(60,0){2}{\line(0,1){80}}
\qbezier(0,35)(30,20)(55.53,7.24)\qbezier(60,35)(30,20)(4.47,7.24)
\qbezier(30,65)(15,80)(3.54,91.46)\qbezier(30,65)(45,80)(56.46,91.46)
\qbezier(0,35)(15,40)(30,45)\qbezier(30,45)(40,55)(30,65)
\qbezier(60,35)(45,40)(30,45)\qbezier(30,45)(20,55)(30,65)
\put(0,-6){\makebox(0,0){$v_1$}}
\put(60,-6){\makebox(0,0){$v_2$}}
\put(0,106){\makebox(0,0){$w_1$}}
\put(60,106){\makebox(0,0){$w_2$}}
}
\end{picture}}\linespread{1}\caption{The bialgebra principle. The diagram at
left is transformed into the diagram at right by removing vertices via axiom
(E2). Paths between free vertices, and path ordering at those vertices, is
preserved.}
\label{principle}
\end{figure}

This theorem is illustrated in Figure~\ref{principle}.

\subsubsection{Bialgebras, flow equivalence and space set transformations}

Now we shall consider the abstract tensor representation of commutative
cocommutative bialgebras. In these bialgebras multiplication and
comultiplication are commutative. This implies that our tensors \tensor\ are
symmetric. Thus in our tensor diagrams there is no longer any ordering at
vertices. The above theorem remains true with all statements about ordering
deleted; thus the bialgebra principle becomes simply the preservation of paths.

\begin{theorem} Equivalence classes of one-dimensional oriented space sets
under elementary space set transformations are scalars in the abstract tensor
representation of commutative cocommutative bialgebras.
\end{theorem}
\textsc{Proof.} A scalar diagram is a tensor diagram with no free edges. In the
commutative cocommutative case, scalar diagrams are directed graphs (and every
directed graph is a scalar diagram).\footnote{\linespread{1}\footnotesize There
is actually one exception: $\delta_a^a$ represents a scalar diagram, but it is
not a directed graph. This does not matter.} Since there are no free edges in
these diagrams, (E1) is implied by (E2). Hence (E2) generates all equivalent
diagrams. And in the commutative cocommutative case, (E2) is the same as
(V).\proofend\bigskip

An obvious corollary is that flow equivalence classes of subshifts of finite
type are scalars in the abstract tensor representation of commutative
cocommutative \mbox{bialgebras}.

We may say more than this. A one-dimensional oriented space set is represented
by a matrix of nonnegative integers or by its associated directed graph. We now
see that it is also represented by an associated scalar diagram in the abstract
tensor representation of a commutative cocommutative bialgebra. The bialgebra
axioms generate local invertible maps between space sets. If a finite sequence
of bialgebra axioms transforms a scalar diagram to itself, the resulting map
may be iterated; it is a law of evolution.

It is tempting to consider the possibility that a dynamical interpretation of
bialgebras might be obtained when those bialgebras are not necessarily
commutative or cocommutative. Each scalar diagram would again represent a set
of spaces. But I do not know what the spaces would be. Would they again be the
closed paths through the diagram? What, then, is the function of ordering at
vertices? I have not managed to make sense of this.

\subsection{The Bowen-Franks group}
\label{bowenfranks}

The invariant factors of $I-A$, which earlier were found to be invariant
under elementary transformations of $A$, resemble the structure constants of an
abelian group. In fact, they \emph{are} the structure constants of an abelian
group.

Bowen and Franks~\cite{bowen/franks} proved that if matrices $A$ and $B$ are
flow equivalent then the abelian groups ${\Z}^n/(I-A){\Z}^n$ and
${\Z}^n/(I-B){\Z}^n$ are isomorphic. The group ${\Z}^n/(I-A){\Z}^n$ is called
the \emph{Bowen-Franks group} associated with $A$, and is denoted by
$B\hspace{-1.5pt}F(A)$.

We can describe this group in terms of commuting generators and relations. Let
$s_1,\ldots,s_n$ be the standard basis for ${\Z}^n$. Then the $s_i$'s generate
the group \mbox{${\Z}^n/(I-A){\Z}^n$}. If
$x=\sum_{i=1}^{n}x_is_i\in(I-A){\Z}^n$, then $x$ is in the column space of
$I-A$. Then in the group ${\Z}^n/(I-A){\Z}^n$ we must have
$\sum_{i=1}^{n}(I-A)_{ij}s_i=0$ for each $j$ between 1 and $n$; and these
generate all of the relations on the $s_i$'s. In other words, if
$s=[s_1,\ldots,s_n]$ is a row vector, then the desired group is given by the
commuting generators $s_i$ and the relations $s(I-A)=0$.

If one operates on $I-A$ by adding a multiple of row $j$ to row $i$ or a
multiple of column $j$ to column $i$ with $i\neq j$, then
$B\hspace{-1.5pt}F(A)$ is not affected. The column operation does not change
the column space of $A$; hence it does not change the group. The row operation
has the effect of replacing generator $s_j$ with $s'_j=s_j-s_i$; again this
does not change the group. If we use these operations to reduce $(I-A)$ to the
diagonal form given in Figure~\ref{hoffkunzalg}, the relations $s(I-A)=0$ are
of the form $g_is_i=0$, where the $g_i$'s are the invariants factors. If
$g_i=1$ then the generator $s_i$ plays no role. If $g_i=0$ then there are no
constraints on $s_i$. Thus $B\hspace{-1.5pt}F(A)$ is a direct product of cyclic
groups whose orders are given by the list of invariant factors (with order 0
corresponding to the cyclic group \Z).

Now consider the directed graph \g{A}. Associate the generator $s_j$ with each
edge emerging from vertex $j$. The equation $s(I-A)=0$ is equivalent to
$sI=sA$, which is equivalent to the set of equations
$s_j=\sum_{i=1}^ns_iA_{ij}$ for each $j$. This says that the sum of the
generators on the edges entering a vertex is equal to the value of the
generator on each of the edges leaving that vertex, and that these are all of
the relations on the generators.

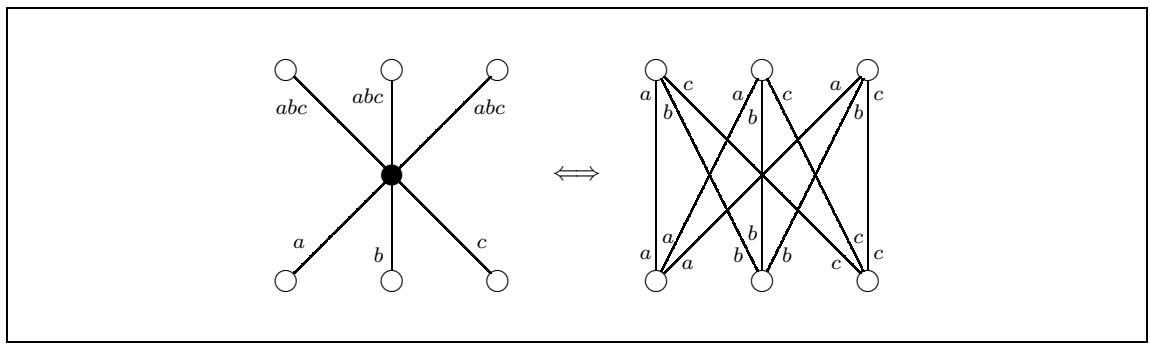
\begin{figure}
\fbox{\begin{picture}(425,120)\scriptsize
\put(102.5,20){
\multiput(0,0)(140,0){2}{
        \multiput(0,0)(0,80){2}{\multiput(0,0)(40,0){3}{\circle{8}}}
        \qbezier(2.83,2.83)(40,40)(77.17,77.17)
        \qbezier(77.17,2.83)(40,40)(2.83,77.17)
        \put(40,4){\line(0,1){72}}}
\put(110,40){\makebox(0,0){\footnotesize $\Longleftrightarrow$}}
\put(40,40){\circle*{8}}
\multiput(140,4)(80,0){2}{\line(0,1){72}}
\qbezier(141.79,3.58)(160,40)(178.21,76.42)
\qbezier(181.79,76.42)(200,40)(218.21,3.58)
\qbezier(141.79,76.42)(160,40)(178.21,3.58)
\qbezier(181.79,3.58)(200,40)(218.21,76.42)
\put(5,14){\makebox(0,0){$a$}}
\put(35,10){\makebox(0,0){$b$}}
\put(74,14){\makebox(0,0){$c$}}
\put(2,66){\makebox(0,0){$abc$}}
\put(31,70){\makebox(0,0){$abc$}}
\put(77,66){\makebox(0,0){$abc$}}
\put(140,0){\put(-4,10){\makebox(0,0){$a$}}
\put(4.5,16){\makebox(0,0){$a$}}
\put(12,6){\makebox(0,0){$a$}}
\put(31,10){\makebox(0,0){$b$}}
\put(36.5,18){\makebox(0,0){$b$}}
\put(49.5,10){\makebox(0,0){$b$}}
\put(84,10){\makebox(0,0){$c$}}
\put(76.5,16){\makebox(0,0){$c$}}
\put(68,6){\makebox(0,0){$c$}}
\put(-4,70){\makebox(0,0){$a$}}
\put(4.5,64){\makebox(0,0){$b$}}
\put(12,74){\makebox(0,0){$c$}}
\put(31,70){\makebox(0,0){$a$}}
\put(36.5,62){\makebox(0,0){$b$}}
\put(49.5,70){\makebox(0,0){$c$}}
\put(84,70){\makebox(0,0){$c$}}
\put(76.5,64){\makebox(0,0){$b$}}
\put(68,74){\makebox(0,0){$a$}}
}}
\end{picture}}\linespread{1}\caption{Invariance of groups and semigroups under
space set transformations and bialgebra axioms.}
\label{elemopsemi}
\end{figure}

In Figure~\ref{elemopsemi}
the effect of an elementary space set transformation of \g{A} on the group
$B\hspace{-1.5pt}F(A)$ is shown. Each incoming vertex carries a value equal to
the product of the generators attached to the edges arriving at that vertex. On
the left-hand side these values are multiplied together and then sent to each
outgoing vertex. On the right-hand side the values are sent individually to
each outgoing vertex. In both cases, at each outgoing vertex the generators are
then multiplied with whatever other generators are on edges pointing towards
that vertex to produce the generator on the edges leaving the vertex. Thus the
net effect is the same in both cases, so $B\hspace{-1.5pt}F(A)$ is indeed an
invariant of space set transformations (and of flow equivalence).

We can say more, however. Suppose that we drop the requirement of
commutativity. Then the diagram in Figure~\ref{elemopsemi} represents a scalar
in a bialgebra. Now the generators are no longer commutative; they generate a
group. The figure shows that this group is left invariant by the elementary
transformation, since the order of multiplication is the same on the left-hand
and right-hand sides. Since the elementary transformation is the only one
needed to generate the equivalent representations of scalars, the group is
invariant under the bialgebra axioms.

Furthermore, let us now drop the requirement that these generators have
inverses. Now they are generators of a semigroup with identity. (The identity
is needed since if no generators point towards a vertex then the generators
attached to the edges leaving the vertices must be set equal to the identity.)
Clearly the elementary transformation does not change the semigroup either. So
the semigroup is invariant under the bialgebra axioms.

The fact that these groups and semigroups are invariant under the bialgebra
axioms means that we can think of them as themselves being bialgebras.

Let $K$ be a field and $S$ a semigroup with identity. Consider the free unitary
$K$-module $KS$ generated by $S$. This is a vector space with dimension equal
to the order of $S$. We may now define multilinear maps \(\mu : KS \otimes KS
\mapsto KS\), \(u : K \mapsto KS\), \(\Delta : KS \mapsto KS \otimes KS\), and
\(\epsilon : KS \mapsto K\) by setting \(\mu(s,t)=st\), \(u(1)=1_S\),
\(\Delta(s)=s \otimes s\) and \(\epsilon(s)=1\) for all $s,t \in S$, and
extending linearly. Then \ang{KS,\mu,u,\Delta,\epsilon} is a bialgebra; it is
called the \emph{semigroup bialgebra} of $S$ over $K$ (and it is called
the \emph{group bialgebra} of $S$ over $K$ if $S$ is a group).

These types of bialgebras are well known (see~\cite{abe} for further
details). Clearly these rules for multiplication and comultiplication exactly
parallel the conditions that we assigned to the generators of our group or
semigroup. Thus we may think of this group or semigroup as being a group
bialgebra or semigroup bialgebra (defined on some arbitrary field $K$) that is
assigned to an abstract bialgebra scalar.

Note that there are actually two invariant groups or semigroups that may be
obtained from a scalar using this technique. We may reverse the roles of upper
and lower indices and then perform the same construction. The result will be
another group or semigroup. In the commutative cocommutative case, if we
restrict ourselves to groups then this reversal amounts to finding
$B\hspace{-1.5pt}F(A^T)$. Since the invariant factors of $A^T$ are the same as
those of $A$, these groups are necessarily isomorphic. This does not seem to be
the case when semigroups are used instead of groups, and it does not seem
likely to be the case when the restrictions of commutativity and
cocommutativity are removed.

Let us now focus on the commutative cocommutative case. Let $S(A)$ represent
the semigroup obtained using our original definition (as opposed to the one
obtained by switching the roles of upper and lower indices).

\begin{theorem} If $A-I$ is a positive reduced matrix, then
$S(A)=B\hspace{-1.5pt}F(A)$.
\end{theorem}
\textsc{Proof.} Let $A$ be an $n\times n$ matrix. If $A-I$ is positive then it
may be transformed so that each of its entries are positive, while leaving
$S(A)$ unchanged. Suppose that this transformation has been carried out. Then
the semigroup relations are that $s_i=s_i+\sum_{j=1}^n(A-I)_{ji}s_j$ for each
$i$, $1\leq i\leq n$. Since each $(A-I)_{ji}$ is positive, it follows that each
generator of $S(A)$ may be written as a sum of generators of $S(A)$ in which
every generator appears at least once. Since the elements of $S(A)$ are sums of
generators of $S(A)$, it follows that for any $x\in S(A)$ there exists $y\in
S(A)$ such that $x=y+s_1$. Let $e=\sum_{j=1}^n(A-I)_{j1}s_j$. Then
$s_1=s_1+e$. So $x+e=y+s_1+e=y+s_1=x$. So $e$ is the identity on $S(A)$. Since
$(A-I)_{j1}>0$ for each $j$, it follows that
$s_i+\sum_{j=1}^n((A-I)_{j1}-\delta_{ij})s_j=e$. So each $s_i$ is invertible. So
each element of $S(A)$ is invertible. So $S(A)$ is a group. So
$S(A)=B\hspace{-1.5pt}F(A)$.\proofend\bigskip

Thus semigroups are no better than groups for discriminating the equivalence
classes of positive matrices. (One might have hoped that they would pick up the
information contained in $\mathrm{sgn}\,\det(I-A)$.) On the other hand,
consider the matrices $A=(1)$ and
$B=\left(\begin{array}{cc}2&1\\1&2\end{array}\right)$. Here
$B\hspace{-1.5pt}F(A)=B\hspace{-1.5pt}F(B)=\Z$ (and
$\mathrm{sgn}\,\det(I-A)=\mathrm{sgn}\,\det(I-B)=0$). But these matrices are in
different equivalence classes, since $B$ is positive but $A$ isn't. Since $B$
is positive, the above theorem says that $S(B)=\Z$. But the single relation for
$S(A)$ is $s_1=s_1$; hence $S(A)=\Z^+$. So in this case semigroups are more
powerful than groups.

I have found other instances in which semigroups distinguish matrices that
belong to different equivalence classes, while groups do not. Clearly, though,
semigroups do not do the whole job. Are the semigroups plus
$\mathrm{sgn}\,\det(I-A)$ sufficient to distinguish the equivalence classes? I
do not know. Also, I cannot help but wonder whether there might be another
algebraic structure (again some type of bialgebra) which could be attached to
scalars and which would entirely distinguish these equivalence
classes. However, I have not found any such structure.

A relevant fact, which has only just come to my attention, is that recently
Danrun Huang~\cite{huang1,huang2,huang3,huang4}, coming from the flow
equivalence/$\mathrm{C}^*$-algebra point of view, has completely solved the
equivalence class problem. I have not had an opportunity to review his
results. Another result which I have not yet digested is the recent discovery
by Boyle and Handelman~\cite{boyle/handelman} of an ``ordered'' abelian group
which is an invariant of flow equivalence and which is a complete invariant in
the case of positive matrices. Here, then, is a single object containing
information of both the Bowen-Franks group and the sign of the determinant, at
least in the positive-matrix case. This bodes well for the existence of a
single algebraic object that is a complete invariant in all cases.

It is natural to ask: what is the set of semigroups that can be obtained from
these diagrams? Again, I have no answer.

Though the relevance to combinatorial spacetimes is not clear, we may also ask
the same sorts of questions in the noncommutative case. What are the
equivalence classes of abstract scalars in a bialgebra? How far do these groups
and semigroups go towards distinguishing them? And what is the set of groups
and semigroups that may be obtained in this way?

Finally, let us return to the commutative case. Consider a sequence $\sigma$ of
elementary equivalence transformations which transform a matrix $A$ to
itself. This sequence of transformations corresponds to a local invertible
operator $T_{\sigma}:\x{A}\mapsto\x{A}$. It also corresponds to an automorphism
$f_{\sigma}$ of $B\hspace{-1.5pt}F(A)$; for as one makes these transformations
one can track the generators which are attached to the edges and see which map
to which. Is this automorphism a property of $T$, or is it only a property of
the particular sequence $\sigma$ of transformations used to obtain $T$?

The question reduces to the following: if $T_{\sigma}$ is the identity law on
\x{A}, is $f_{\sigma}$ the identity automorphism of $B\hspace{-1.5pt}F(A)$? If
not, then the answer to the original question is no, since then
$f_{\sigma^2}\neq f_{\sigma}$ but $T_{\sigma^2}=T_{\sigma}$. And if so, then
the answer to the original question is yes, since if $T_{\sigma}=T_{\tau}$ then
$T_{\sigma^{-1}\tau}$ is the identity on \x{A}, so $f_{\sigma^{-1}\tau}$ is the
identity automorphism of $B\hspace{-1.5pt}F(A)$, so
$f_{\tau}=f_{\sigma\sigma^{-1}\tau}=f_{\sigma}f_{\sigma^{-1}\tau}=f_{\sigma}$.

It would be nice if one could canonically associate an automorphism of a
Bowen-Franks group with an operator $T$. (At present, for instance, I know no
easy way to determine whether a set of operators $T_i$ generate another
operator $U$. These automorphisms would provide a small portion of the answer:
if the automorphisms associated with the $T_i$'s did not generate the
automorphism associated with $U_i$, then the $T_i$'s could not generate
$U_i$.) Similarly, it would be interesting to know whether automorphisms of
other algebraic flow-equivalence invariants are canonically associated with
operators. One might even hope that there is some such algebraic invariant with
the property that there is an invertible map between its automorphisms and the
operators. On my more optimistic days, I do hope this.

\newpage\vspace*{40pt}\section[The general 1+1-dimensional case]{\LARGE The
general 1+1-dimensional case}\bigskip
\label{unorient}

Our one-dimensional spaces up to now have been colored directed graphs. A
natural generalization is obtained by removing the directedness from the
graphs. Now segments no longer have an orientation; \seg{ab} is the same as
\seg{ba}. Similarly, \cir{abc} is the same as \cir{cba}. Again we may define a
space set by listing all allowed neighborhoods of a given width $w$; the space
set is the set of all unoriented vertex-colored 2-regular graphs having the
property that every embedded width-$w$ segment is in the list.

Suppose that one has defined an oriented space set \X\ by listing the allowed
width-$w$ segments. Choose two colors $L$ and $R$ which do not appear in
\X. Replace each oriented segment \seg{c_1c_2\ldots c_w} with the three
unoriented width-$3w$ segments \seg{LRc_1LRc_2LR\ldots LRc_w},
\seg{Rc_1LRc_2LR\ldots LRc_wL} and \seg{c_1LRc_2LR\ldots LRc_wLR}. The
resulting list of unoriented segments determine a new space set \Y\ in which
the spaces are unoriented vertex-colored 2-regular graphs. The implied map
between \X\ and \Y\ is local and invertible, and its inverse is local; so \X\
is equivalent to \Y.

Thus in the new setup orientedness is a property of a particular set of allowed
neighborhoods, rather than a built-in property. It is not necessary to use
colors $L$ and $R$ in this way to define an oriented space set; any other space
set which is locally equivalent to one defined in this way deserves the name
``oriented.''

Note that if we define an oriented space set using $L$ and $R$ then the
orientation-reversal map is performed simply by permuting the colors $L$ and
$R$. Thus the orientation-reversal map is not different in character from other
local equivalence maps.

In the general one-dimensional case there are two kinds of segments:
symmetrical and asymmetrical. A segment \seg{c_0\ldots c_k} is symmetrical if
and only if $c_i=c_{k-i}$ for each $i$. Hence the unoriented empty segment is
symmetrical, and each unoriented length-1 segment is symmetrical. The
symmetrical segments have the property that they remain unchanged when their
ends are reversed. All oriented segments are asymmetrical.

If a space set is defined by a list of allowed unoriented length-$w$ segments,
it cannot be unoriented if one of its spaces contains a symmetrical length-$w$
segment. For suppose that $S$ is such a symmetrical segment. Choose one of its
ends. Since $S$ is in a space, it follows that we may add segments indefinitely
to that end. Since $S$ is symmetrical, we may add the exact same segments, in
the exact same order, to the other end of $S$. The result is an infinite space
which is symmetrical about the center of $S$. Since the space is symmetrical,
it cannot be part of an oriented space set. (No space made of directed edges is
symmetrical about a point.)

Hence, for example, any nonempty set of spaces defined by a list of allowed
unoriented width-1 segments is unoriented. Thus \x{2} is oriented, but the set
of spaces defined by the list of unoriented segments \seg{a} and \seg{b} is
not.

Suppose that a set of spaces is defined by a list of allowed unoriented
asymmetrical width-$w$ segments. Discard those segments which do not appear in
a space. Suppose that the edges of the remaining segments can be replaced with
directed edges in such a way that segments can only be glued together so that
all edges in the resulting segment point in the same direction. Then the space
set is oriented. And if this is not possible, the space set is not
oriented. For if it is not possible, then there must be allowed segments $S$
and $U$ such that $S=\seg{s_1\ldots s_w}$ and $V=\seg{s_1\ldots s_wUs_w\ldots
s_1}$ is a segment that appears in some space. Choose one end of $V$ and extend
$V$ indefinitely at that end. Extend $V$ indefinitely in exactly the same way
at the other end. Though the resulting infinite space $x$ may not be
symmetrical (because $U$ may not be symmetrical), it does have the property
that, for any $n$, there are segments $T$ and $R$ such that $T=\seg{t_1\ldots
t_n}$ and \seg{t_1\ldots t_nRt_n\ldots t_1} is embedded in $x$. This means that
an orientation may not be \emph{locally} assigned to $x$; hence the space set
cannot be locally equivalent to any of the space sets described in
Chapter~\ref{orient}. (This discussion shows that such space sets might better
be called ``locally oriented.'' However, I will continue to use my original
terminology.)

One may express space set descriptions using adjacency matrices in a manner
similar to that used in the oriented case. Here is an example:\bigskip\\
\hspace*{\fill}\footnotesize\begin{tabular}{c|cc}
\seg{a}&\textit{aa}&\textit{ab}\\
\seg{b}&\textit{ba}&\textit{bb}\\
\end{tabular}\normalsize\hspace*{\fill}\bigskip\\
It is the space set defined by the list of allowed unoriented width-1 segments
\seg{a} and \seg{b}. This should be interpreted in the same way as in the
oriented case, except that now there may be two closed directed paths through
the matrix that correspond to the same space. For example, the space
\cir{aababb} corresponds to the paths \cir{\mathrm{1,1,2,1,2,2}} and
\cir{\mathrm{2,2,1,2,1,1}}, where the numbers listed refer to consecutive
vertices in the path (the edges in the path are implicit in this case).

In the above example all segments attached to vertices are symmetrical. A
further wrinkle must be added if this is not the case. Here is an
example:\bigskip\\
\hspace*{\fill}\footnotesize\begin{tabular}{c|cccc}
\seg{a}&\textit{aa}&0&\textit{ab}&0\\
\seg{b}&0&\textit{bb}&0&\textit{ba}\\
\seg{ab}&0&\textit{ab}&0&0\\
\seg{ba}&\textit{ba}&0&0&0\\
\end{tabular}\normalsize\hspace*{\fill}\bigskip\\
Here \seg{ab} and \seg{ba} are taken to represent the same asymmetrical segment
\seg{ab}. One assembles spaces by travelling along a directed path in the
associated directed graph. While travelling, one may encounter either end of an
asymmetrical segment first. Thus any such segment must be present in two
different ways.

Note that both of these matrices possess a symmetry property: if one reverses
the direction of each segment, transposes the matrix, and reverses the order of
pairs of indices which correspond to the same asymmetrical segment, the result
is the same as the original matrix. This corresponds to the fact that the
successors of an asymmetrical segment aligned in one direction with respect to
the directed edges in the graph must be the same as the predecessors of the
segment aligned in the other direction. For any closed path through the
directed graph that corresponds to a space, there is another (not necessarily
distinct) closed path corresponding to the same space which traverses the
vertices in the opposite order. For instance, in our second example the paths
corresponding to the space \cir{aababb} are the distinct paths
\cir{\mathrm{1,1,3,2,4,1,3,2,2,4}} and \cir{\mathrm{3,2,2,4,1,3,2,4,1,1}}
(again the edges are implicit). Note that the second path was obtained from the
first by listing the vertices in reverse order and then replacing each vertex
corresponding to an asymmetrical segment with the vertex corresponding to the
opposite alignment of the same segment. The symmetry property insures that for
each directed path through the matrix the corresponding directed path obtained
by transforming the original path in this manner does in fact exist.

Note that the second description can be obtained from the first by performing
two successive elementary transformations. In the first transformation a vertex
is inserted in the middle of the edge connecting segment \seg{a} to segment
\seg{b}. But this operation violates our symmetry condition. So we need also to
perform the second transformation, which inserts a vertex in the middle of the
edge connecting segment \seg{b} to \seg{a}. The segments associated with the
two new vertices represent a single asymmetrical unoriented segment aligned in
the two possible ways with respect to the directed edges.

Thus it appears that the situation is as follows. General one-dimensional space
sets may be represented by descriptions possessing the above symmetry
property. Transformations of space set descriptions are the same as our
transformations of oriented space set descriptions, except that they must
preserve the symmetry condition. This means that either a vertex corresponding
to a symmetrical segment is created or removed, or two vertices corresponding
to the two alignments of an asymmetrical segment are created or removed.

So far I have only worked through a small number of examples. Presumably, if
one transforms a simple description often enough, eventually one encounters the
difficulties of notation that we came up against in the oriented case. These
difficulties may be more severe since both symmetrical and asymmetrical
segments are involved. Again, it may be possible to avoid the problem by
dealing instead with some analogy of the matrix descriptions and elementary
space set transformations used in the oriented case. Since some segments are
symmetrical and some are not, the matrix descriptions and allowed
transformations in the general case are liable to be somewhat complicated. I
have not tried to work this out. Also, I have not investigated the question of
invariants (or equivalence classes of space sets) under these transformations;
some of these invariants would appear to be related to the invariants found in
the oriented case, since similar transformations are involved.

Now consider the situation from the bialgebra point of view. The problem with
the space set descriptions given above, from this point of view, is that they
are in some sense nonlocal. That is: a space may be represented by two paths
through the graph, rather than one. If moves on space set descriptions are to
be given by local axioms, then we do not want this.

Suppose first that there are no symmetrical segments. Then the general case
differs from the oriented case only in that there is no distinction between
upper and lower ends of a vertex. In other words, while a vertex still has two
sides (it is asymmetrical), now an edge can connect any side of one with any
side of another.

This means that we need to add a \emph{metric tensor} to the bialgebra. That
is: we need tensors $g_{ab}$ and $g^{ab}$ which satisfy the following axioms:
\begin{enumerate}\linespread{1}\normalsize
\item $g_{ab}=g_{ba}$ and $g^{ab}=g^{ba}$;
\item $g_{ab}g^{ac}=\delta_b^c$;
\item $U^ag_{ab}=U_b$ and $U_ag^{ab}=U^b$;
\item $M_{ab}^cg^{ad}g^{be}g_{cf}=M^{de}_f$ and
$M^{ab}_cg_{ad}g_{be}g^{cf}=M_{de}^f$.
\end{enumerate}
Graphically, a metric tensor is simply an edge that connects an upper index to
an upper index or a lower index to a lower index. The axioms imply that, by
using metric tensors, any tensor $U$ or $M$ can be turned upside down.

The other axioms remain the same as they were before. I believe that this new
system completely describes one-dimensional space sets that contain no
symmetrical segments.

In the oriented case, spaces corresponded to closed directed paths through the
scalar diagram. Once the metric is added, the rules change slightly: now spaces
correspond to closed paths through the diagram which are constructed using the
rule that one may travel along an edge or a vertex without turning around. (The
vertices are two-sided: one arrives at a vertex at one end, proceeds through it
to the other end, and then continues along an edge.) Due to the presence of the
metric tensor, it is possible that the same path heads upwards at one point and
downwards at another. In fact, it may even head along the same edge in two
different directions during different parts of the same path.

A simple example of such a space set is given by the scalar
$M_{ab}^cM_c^{de}g^{ab}g_{de}$. We may express this space set in terms of
segments by assigning segments with two different colors to each edge in the
scalar diagram. The two different colors insure that the associated segment is
asymmetrical. In this diagram there are three edges: the edges associated with
the two metric tensors (associate with them the segments \seg{ab} and \seg{de}
in the obvious way), and the edge associated with the index $c$ that joins the
two $M$ vertices (associate with this edge the segment \seg{cf}, with the $c$
below the $f$ in the diagram). The diagram states that the length-4 segments
\seg{cfde}, \seg{cfed}, \seg{abcf} and \seg{bacf} are allowed, and that these
may be glued together with a 2-cell overlap in any desired way to form a
space. Spaces therefore include \cir{cfdefcba}, \cir{cfedfcba},
\cir{cfdefcbacfedfcba}, and so on. In short, any space has the form \seg{\ldots
cfS_1fcT_1cfS_2fcT_2\ldots cfS_nfcT_n\ldots} where $S_i$ is \seg{de} aligned in
either of the two possible ways, and $T_i$ is \seg{ab} aligned in either of the
two possible ways. This is locally equivalent to the set of spaces obtained
from this one by deleting all of the $c$'s and~$f$'s.

Now suppose that there are also symmetrical segments. Our tensor edges are
asymmetrical, so they are not useful for representing a symmetrical
segment. Besides, if the two ends of a symmetrical segment are represented by
two ends of an object in our scalar diagram, this poses a problem. For the
entire diagram must be symmetrical with respect to these ends: what one
encounters as one travels outwards from one end must be exactly the same as
what one encounters when one travels outwards from the other end. If our
scalars are composed of tensors and of local rules for hooking them together,
then this symmetry condition cannot be imposed. And local transformation rules
cannot preserve such a symmetry.

The solution is to represent a symmetrical segment as an object having only
\emph{one} end: namely, as a vector or covector. The idea, then, is to add new
tensors $V_a$ and $V^a$ to our tensor algebra. These tensors satisfy the
following axioms:
\begin{enumerate}\linespread{1}\normalsize
\item $V_ag^{ab}=V^b$ and $V^ag_{ab}=V_b$;
\item $V_aU^a=V^aU_a=1$;
\item $V_zM^z_{ab}=V_wM^{wx}_aM^{yz}_bV_zg_{xy}$ and
$V^zM_z^{ab}=V^wM_{wx}^aM_{yz}^bV^zg^{xy}$.
\end{enumerate}
The idea behind this new tensor is that it is a \emph{turn-around point}. Again
spaces are represented by closed paths through the scalar diagram. The new
wrinkle is this: if the path travels along an edge which ends at a tensor $V_a$
or $V^a$, it reverses course and heads back along that edge in the opposite
direction. Thus the $U$ tensors represent dead ends where one is not allowed to
turn around; the $V$ tensors represent mirrors. They are symmetry points in the
space set description.

One may represent the space set described by a scalar in terms of segments
by associating length-2 asymmetrical segments with each edge, as before. For
example, consider the scalar $V_aV^a$. Associate the segment \seg{ab} with the
single edge (with the $a$ below the $b$). To find the spaces, we follow closed
paths through the scalar diagram, bouncing at the mirrors. Thus this scalar
represents a space set whose spaces are \cir{abba}, \cir{abbaabba}, and so
on. The space set contains all spaces constructible from the symmetrical
segment \seg{abba} where the nonoverlap gluing rule is used. Clearly this is
locally equivalent to the set of spaces determined by the single allowed
length-1 segment \seg{a}.

Another example is given by the scalar $V^aM_a^{bc}V_bV_c$. Here we may
associate \seg{ab} with edge $a$, \seg{cd} with edge $b$, and \seg{ef} with
edge $c$ (with $a$ below $b$, $c$ below $d$ and $e$ below $f$). The resulting
space set is the set of spaces constructible from \seg{abcddcba} and
\seg{abeffeba} using the nonoverlap rule. Thus it is locally equivalent to the
set of spaces determined by the allowed length-1 segments \seg{a} and \seg{b}.

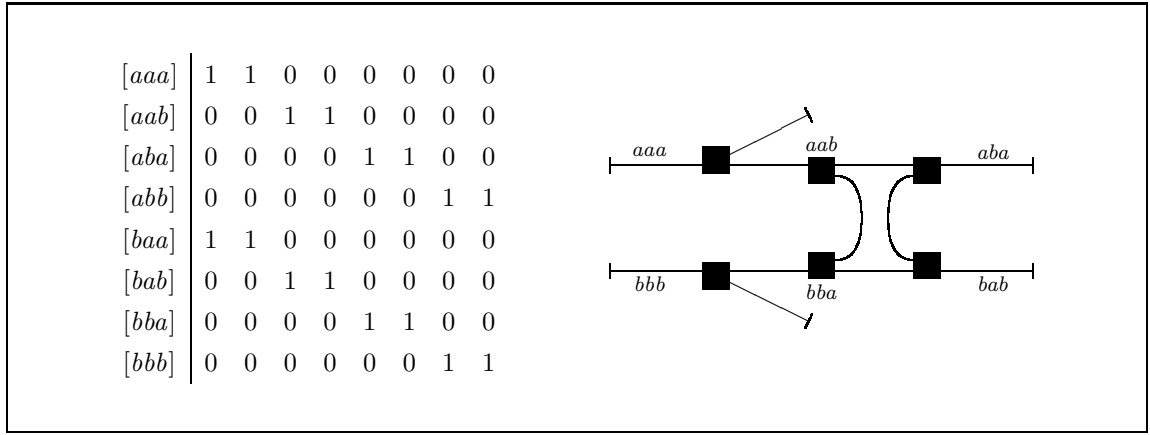
\begin{figure}
\fbox{\begin{picture}(425,156)\scriptsize
\put(35,20){\footnotesize$\begin{array}[b]{c|cccccccc}
\seg{aaa}&1&1&0&0&0&0&0&0\\
\seg{aab}&0&0&1&1&0&0&0&0\\
\seg{aba}&0&0&0&0&1&1&0&0\\
\seg{abb}&0&0&0&0&0&0&1&1\\
\seg{baa}&1&1&0&0&0&0&0&0\\
\seg{bab}&0&0&1&1&0&0&0&0\\
\seg{bba}&0&0&0&0&1&1&0&0\\
\seg{bbb}&0&0&0&0&0&0&1&1
\end{array}$}
\put(225,8){
\multiput(0,50)(0,40){2}{\line(1,0){160}}
\multiput(0,0)(0,44){2}{\put(35,43){\rule{10pt}{10pt}}}
\multiput(0,0)(0,36){2}{\multiput(75,47)(40,0){2}{\rule{10pt}{10pt}}}
\put(45,46){\line(2,-1){30}}\put(45,94){\line(2,1){30}}
\multiput(0,47)(0,40){2}{\multiput(0,0)(160,0){2}{\line(0,1){6}}}
\qbezier(85,86)(95,86)(95,70)\qbezier(95,70)(95,54)(85,54)
\qbezier(115,86)(105,86)(105,70)\qbezier(105,70)(105,54)(115,54)
\put(15,95){\makebox(0,0){\textit{aaa}}}
\put(15,45){\makebox(0,0){\textit{bbb}}}
\put(80,98){\makebox(0,0){\textit{aab}}}
\put(80,42){\makebox(0,0){\textit{bba}}}
\put(145,95){\makebox(0,0){\textit{aba}}}
\put(145,45){\makebox(0,0){\textit{bab}}}
\qbezier(73.66,28.32)(75,31)(76.34,33.68)
\qbezier(73.66,111.68)(75,109)(76.34,106.32)
}
\end{picture}}\linespread{1}\caption{Bialgebra representation of an unoriented
space set. The space set is the one defined by the unoriented width-3 segments
\seg{aaa}, \seg{aab}, \seg{aba}, \seg{abb}, \seg{bab} and \seg{bbb}. The
adjacency matrix for this space set is shown at left; gluing is done by 2-cell
overlap. Segments \seg{aab} and \seg{abb} each correspond to two matrix rows
and columns, since they are asymmetrical. On the right is the associated tensor
diagram. One may think of the upward direction (in terms of tensor indices) as
being the left-to-right direction in this diagram; since a metric is present,
such directions do not really matter. The black boxes represent the standard
bialgebra tensors; the short lines with single edges attached represent the
vectors $V_a$ and $V^a$. The diagram is equivalent to the scalar given by the
expression
$V^aT_a^{bc}V_bT_c^{de}T_{df}^gV_gV^hT_h^{ij}V_jT_i^{kl}T_{ml}^nV_ng_{ek}g^{fm}$.}
\label{general1d}
\end{figure}

Now consider a space set defined by a list of allowed unoriented width-$w$
segments. We may associate a scalar with this space set as follows. Each
symmetrical segment is represented by a tensor of the form $V_xT^x_{a_1\ldots
a_m}$ or $V^xT_x^{a_1\ldots a_m}$ (it doesn't matter which). (The tensors
\tensor\ are the bialgebra tensors defined in Section~\ref{bialgebra}.) Each
asymmetrical segment is represented by a tensor of the form \tensor; one end of
the segment is arbitrarily associated with the top end of the tensor, and the
other end is associated with the bottom. The number of free edges in each case
should be equal to the number of segment ends the given segment end is adjacent
to. It remains to connect the free edges. If a segment end is adjacent to
itself, the corresponding free edge is traced with a new vector $V_x$ or
$V^x$. This procedure is justified because an end glued to itself is indeed a
symmetry point of the space set: if $S=\seg{s_1\ldots s_w}$ and the right end
of $S$ can be adjacent to itself, then the rightmost $w-1$ cells of $S$ must
constitute a symmetrical segment $S'$, so the segment formed by gluing two
copies of $S$ together at their right-hand ends is $s_1S's_1$, which is
symmetrical. If two distinct segment ends are adjacent, the corresponding free
edges are traced or joined by $g_{ab}$ or $g^{ab}$. An example is shown in
Figure~\ref{general1d}.

The first two axioms given above for the $V$'s are straightforward, but the
third requires explanation. In the general case there are two obvious types of
matrix transformations: removal of a vertex corresponding to an symmetrical
segment, and removal of two vertices corresponding to an asymmetrical segment. It
is reasonable to hope that these two types of moves generate all desired
transformations. The second type of move involves only asymmetrical segments;
hence it is generated by the standard bialgebra axioms. But the first type of
move requires something extra. The moves of this sort look like this (see
Figure~\ref{symmetrymoves}):
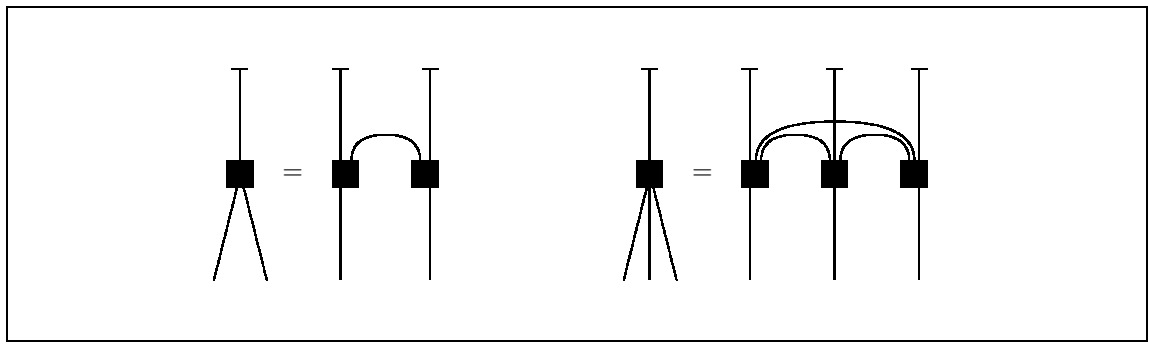
\begin{figure}
\fbox{\begin{picture}(425,120)\footnotesize
\put(85,20){
\put(0,40){\line(0,1){40}}\put(-3,80){\line(1,0){6}}
\qbezier(0,40)(-5,20)(-10,0)\qbezier(0,40)(5,20)(10,0)
\put(-5,35){\rule{10pt}{10pt}}
\put(20,40){\makebox(0,0){=}}
\put(38,0){\line(0,1){80}}\put(72,0){\line(0,1){80}}
\put(35,80){\line(1,0){6}}\put(69,80){\line(1,0){6}}
\qbezier(42,45)(42,55)(55,55)\qbezier(55,55)(68,55)(68,45)
\put(35,35){\rule{10pt}{10pt}}\put(65,35){\rule{10pt}{10pt}}
}
\put(240,20){
\put(0,40){\line(0,1){40}}\put(-3,80){\line(1,0){6}}
\qbezier(0,40)(-5,20)(-10,0)\qbezier(0,40)(5,20)(10,0)
\put(0,40){\line(0,-1){40}}
\put(-5,35){\rule{10pt}{10pt}}
\put(20,40){\makebox(0,0){=}}
\put(38,0){\line(0,1){80}}\put(70,0){\line(0,1){80}}\put(102,0){\line(0,1){80}}
\put(35,80){\line(1,0){6}}\put(67,80){\line(1,0){6}}\put(99,80){\line(1,0){6}}
\multiput(0,0)(30,0){2}{\qbezier(42,45)(42,55)(55,55)\qbezier(55,55)(68,55)(68,45)}
\qbezier(40,45)(40,60)(70,60)\qbezier(70,60)(100,60)(100,45)
\multiput(0,0)(30,0){3}{\put(35,35){\rule{10pt}{10pt}}}
}
\end{picture}}\linespread{1}\caption{Removal of a symmetry point for $n=2$ and
$n=3$. The rule for $n=2$, along with the bialgebra and metric axioms,
generates the rules for all $n$.}
\label{symmetrymoves}
\end{figure}
\[V_xT^x_{a_1\ldots a_n}=(\prod_{i=1}^nV_{c_{ii}}T_{a_i}^{a_{i1}\ldots
a_{in}})(\hspace{-10pt}\prod_{1\leq i<j\leq
n}\hspace{-10pt}g_{c_{ij}c_{ji}}).\] On the left-hand side there is one path
from each lower edge back to itself, and one path from any lower edge to any
other lower edge. When this symmetry point is removed, each of these paths must
be explicitly replaced. This is what happens on the right-hand side. (Note that
there end up being more $V$'s after the ``removal'' of the original $V$ that
there were beforehand. If a tensor has a $V$ in it, it can only be equivalent
to tensors with $V$'s in them.) The third axiom given above for the $V$'s is
exactly this move for $n=2$. It turns out that this is sufficient, in
conjunction with the standard bialgebra axioms, to generate the moves for any
$n$. For suppose inductively that the moves have been generated for all $n\leq
m$. Then\bigskip\\
$\noindent\begin{array}{lcl} \displaystyle V_xT^x_{a_1\ldots
a_{m+1}}&=&\displaystyle V_xT^x_{a_1\ldots
a_{m-1}y}T^y_{a_ma_{m+1}}\end{array}\smallskip\\\noindent\begin{array}{lcl}
\,&=&\displaystyle(\prod_{i=1}^{m-1}V_{c_{ii}}T_{a_i}^{c_{i1}\ldots
c_{im}})(\hspace{-10pt}\prod_{1\leq i<j\leq
m}\hspace{-10pt}g_{c_{ij}c_{ji}})V_{c_{mm}}T_y^{c_{m1}\ldots
c_{mm}}T_{a_ma_{m+1}}^y\smallskip\\
\,&=&\displaystyle(\prod_{i=1}^{m-1}V_{c_{ii}}T_{a_i}^{c_{i1}\ldots
c_{im}})(\hspace{-10pt}\prod_{1\leq i<j\leq
m}\hspace{-10pt}g_{c_{ij}c_{ji}})V_{c_{mm}}T_z^{c_{m1}\ldots
c_{m(m-1)}}T_y^{zc_{mm}}T_{a_ma_{m+1}}^y\smallskip\\
\,&=&\displaystyle(\prod_{i=1}^{m-1}V_{c_{ii}}T_{a_i}^{c_{i1}\ldots
c_{im}})(\hspace{-10pt}\prod_{1\leq i<j\leq
m}\hspace{-10pt}g_{c_{ij}c_{ji}})V_{c_{mm}}T_z^{c_{m1}\ldots
c_{m(m-1)}}T^z_{qr}T^{c_{mm}}_{st}T_{a_m}^{qs}T_{a_{m+1}}^{rt}\smallskip\\
\,&=&\displaystyle(\prod_{i=1}^{m-1}V_{c_{ii}}T_{a_i}^{c_{i1}\ldots
c_{im}})(\hspace{-10pt}\prod_{1\leq i<j\leq
m}\hspace{-10pt}g_{c_{ij}c_{ji}})V_uT^{uv}_sT^{wx}_tV_xg_{vw}T_{qr}^{c_{m1}\ldots
c_{m(m-1)}}T_{a_m}^{qs}T_{a_{m+1}}^{rt}\smallskip\\
\,&=&\displaystyle(\prod_{i=1}^{m-1}V_{c_{ii}}T_{a_i}^{c_{i1}\ldots
c_{im}})(\hspace{-10pt}\prod_{1\leq i<j\leq
m}\hspace{-10pt}g_{c_{ij}c_{ji}})V_uV_xg_{vw}T_q^{q_1\ldots
q_{m-1}}T_r^{r_1\ldots
r_{m-1}}(\prod_{i=1}^{m-1}T_{q_ir_i}^{c_{mi}})T_{a_m}^{quv}T_{a_{m+1}}^{rwx}\smallskip\\
\,&=&\displaystyle(\prod_{i=1}^{m-1}V_{c_{ii}}T_{a_i}^{c_{i1}\ldots
c_{im}})(\hspace{-15pt}\prod_{1\leq i<j\leq
m-1}\hspace{-15pt}g_{c_{ij}c_{ji}})V_uV_xg_{vw}T_{a_m}^{q_1\ldots
q_{m-1}uv}T_{a_{m+1}}^{r_1\ldots
r_{m-1}wx}(\prod_{i=1}^{m-1}T^{q'_ir'_i}_{c_{im}}g_{q'_iq_i}g_{r'_ir_i})
\end{array}$\bigskip\\
which equals the desired result once several index substitutions and
simplifications are carried out. (The third axiom for the $V$'s is used in
going from line 4 to line 5; otherwise only the bialgebra and metric axioms are
used.) It seems promising, then, that the axioms given above are the only ones
needed, and that scalars in the bialgebra tensors, the metric tensors and the
$V$'s, along with the bialgebra axioms and the axioms given in this section,
completely describe general one-dimensional space sets and their local
equivalence transformations.

Though the relevance to combinatorial spacetimes is unclear, it may also be
interesting to investigate bialgebras that are not commutative or cocommutative
but that have tensors $g$ and $V$ satisfying the above axioms. The axioms may
need to be adjusted slightly so that the orderings of paths is in some sense
preserved (the definition of ordering of a path which switches direction
midstream may need to be considered). It would be interesting to know whether
bialgebras of this sort have already arisen in some other context.

\cleardoublepage\vspace*{40pt}\section[Cauchy surfaces]{\LARGE Cauchy surfaces}\bigskip
\label{cauchy}

In general relativity, there is more than one way to foliate spacetime with
Cauchy surfaces. One might expect the same to hold here. When one defines a
1+1-dimensional combinatorial spacetime, one does so in terms of a chosen
foliation. Are other foliations possible? Can we find other one-dimensional
surfaces besides the chosen ones whose states causally determine the entire
spacetime? Do these systems have a ``light-cone'' structure? These are the
sorts of questions that will be addressed here.

I have so far only studied these questions in the context of one-dimensional
cellular automata. (By cellular automata I mean not the usual definition where
the set of spaces is \x{k}, but the generalization obtained by allowing the set
of spaces to be any subshift of finite type.) Despite the naturalness of the
questions, and the fact that cellular automata have been around for quite a
while, I don't know of anyone else who has done this.

Since the lattices in these cases are fixed and regular, the natural initial
approach is to examine one-dimensional surfaces that are associated with
straight lines determined by two points on the lattice. Let time proceed
upwards and space be horizontal in our original specification of the cellular
automaton. If one chooses any two lattice points, this determines a line with
slope $m$, with $m \in \Q \cup \{\infty\}$. The set of lines having this slope
foliate spacetime. We may now ask: is it possible to redescribe this system as
a combinatorial spacetime in which the spaces are associated with these lines?
In order to show that this is the case, several things must be verified. First,
one must verify that the one-dimensional surfaces associated with the slope-$m$
lines are Cauchy surfaces. Second, one must verify that the set of allowed
surfaces associated with these lines constitute a subshift of finite type. And
finally, one must determine that the law of evolution associated with these
lines is a local one.

The main result of this chapter is that every reversible cellular automaton has
a light-cone structure. That is: there are slopes $m_1<0$ and $m_2>0$ such
that, if $m_1<m<m_2$, then one may redescribe the system as a cellular
automaton whose spaces are associated with the slope-$m$ lines in the
original spacetime lattice.

I will call any redescription of this sort a \emph{rotation}. Rotations of a
cellular automaton may be thought of as coordinate transformations, and as
means of finding equivalent descriptions of the original system. Interestingly,
the subshifts of finite type obtained in this way typically are not equivalent
to one another via a local invertible map. Thus rotational equivalences are
different from the ones found using a $UTU^{-1}$ transformation, as described
in Chapter~\ref{spacetime}.

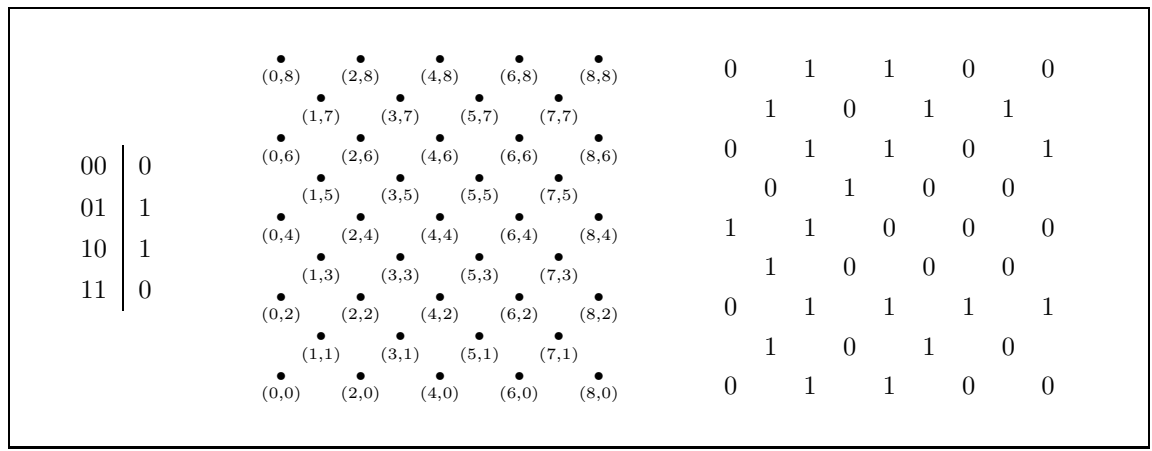
\begin{figure}
\newcounter{xcoord}
\newcounter{tcoord}
\fbox{\begin{picture}(425,160)\footnotesize\put(0,0){\makebox(75,160){%
\begin{tabular}{c|c}
00&0\\
01&1\\
10&1\\
11&0\\
\end{tabular}}}
\put(95,14){\begin{picture}(150,146)
\setcounter{tcoord}{-1}\tiny
\multiput(0,0)(0,30){4}{%
   \addtocounter{tcoord}{1}\setcounter{xcoord}{-2}%
   \multiput(0,0)(30,0){5}{%
      \addtocounter{xcoord}{2}%
      \put(5,10){\circle*{3}}%
      \put(0,0){\makebox(10,10)[bc]{(\thexcoord,\thetcoord)}}}%
   \addtocounter{tcoord}{1}\setcounter{xcoord}{-1}%
   \multiput(15,15)(30,0){4}{%
      \addtocounter{xcoord}{2}%
      \put(5,10){\circle*{3}}%
      \put(0,0){\makebox(10,10)[bc]{(\thexcoord,\thetcoord)}}}
   }
\addtocounter{tcoord}{1}\setcounter{xcoord}{-2}%
\multiput(0,120)(30,0){5}{%
   \addtocounter{xcoord}{2}%
   \put(5,10){\circle*{3}}%
   \put(0,0){\makebox(10,10)[bc]{(\thexcoord,\thetcoord)}}}%
\end{picture}}
\put(265,17){\begin{picture}(150,146)
\footnotesize
\put(0,0){\makebox(10,10)[bc]{0}}
\put(30,0){\makebox(10,10)[bc]{1}}
\put(60,0){\makebox(10,10)[bc]{1}}
\put(90,0){\makebox(10,10)[bc]{0}}
\put(120,0){\makebox(10,10)[bc]{0}}
\put(15,15){\makebox(10,10)[bc]{1}}
\put(45,15){\makebox(10,10)[bc]{0}}
\put(75,15){\makebox(10,10)[bc]{1}}
\put(105,15){\makebox(10,10)[bc]{0}}
\put(0,30){\makebox(10,10)[bc]{0}}
\put(30,30){\makebox(10,10)[bc]{1}}
\put(60,30){\makebox(10,10)[bc]{1}}
\put(90,30){\makebox(10,10)[bc]{1}}
\put(120,30){\makebox(10,10)[bc]{1}}
\put(15,45){\makebox(10,10)[bc]{1}}
\put(45,45){\makebox(10,10)[bc]{0}}
\put(75,45){\makebox(10,10)[bc]{0}}
\put(105,45){\makebox(10,10)[bc]{0}}
\put(0,60){\makebox(10,10)[bc]{1}}
\put(30,60){\makebox(10,10)[bc]{1}}
\put(60,60){\makebox(10,10)[bc]{0}}
\put(90,60){\makebox(10,10)[bc]{0}}
\put(120,60){\makebox(10,10)[bc]{0}}
\put(15,75){\makebox(10,10)[bc]{0}}
\put(45,75){\makebox(10,10)[bc]{1}}
\put(75,75){\makebox(10,10)[bc]{0}}
\put(105,75){\makebox(10,10)[bc]{0}}
\put(0,90){\makebox(10,10)[bc]{0}}
\put(30,90){\makebox(10,10)[bc]{1}}
\put(60,90){\makebox(10,10)[bc]{1}}
\put(90,90){\makebox(10,10)[bc]{0}}
\put(120,90){\makebox(10,10)[bc]{1}}
\put(15,105){\makebox(10,10)[bc]{1}}
\put(45,105){\makebox(10,10)[bc]{0}}
\put(75,105){\makebox(10,10)[bc]{1}}
\put(105,105){\makebox(10,10)[bc]{1}}
\put(0,120){\makebox(10,10)[bc]{0}}
\put(30,120){\makebox(10,10)[bc]{1}}
\put(60,120){\makebox(10,10)[bc]{1}}
\put(90,120){\makebox(10,10)[bc]{0}}
\put(120,120){\makebox(10,10)[bc]{0}}
\end{picture}}
\end{picture}}\linespread{1}\caption[The addition-mod-2 rule]{The
addition-mod-2 rule. On the left is the rule table; at center is a portion of
an empty
lattice with coordinates $(x,t)$ attached; at right is a portion of a lattice
containing cell values. The law of evolution is $c(x,t)+c(x,t+2)=c(x+1,t+1)
\bmod 2$.}
\label{mod2}
\end{figure}

Let us begin with a slightly surprising example: a $d=2$ cellular automaton on
\x{2} which is \emph{not} reversible. This system is displayed in
Figure~\ref{mod2}.
For $d=2$ cellular automata it is often convenient to stagger the lattice rows
so that two cells at time $t$ determine the value of the cell that is directly
between them at time $t+1$. Here I will let $K$, the set of allowed colors, be
the set \z{2} (rather than $\{a,b\}$). In this case, the rule $T$ of the
automaton is simply addition mod 2.

I will adopt the convention that lattice points (which I will also call cells
or vertices) are two units apart in the horizontal and vertical directions
(this way all lattice points have integral coordinates). Assign coordinates
$(x,t)$ to the lattice. Then $(x,t)$ are the coordinates of a lattice point
whenever $x+t=0 \bmod 2$. Denote the color of the lattice point with
coordinates $(x,t)$ by $c(x,t)$. Then the law $T$ says that
$c(x+1,t+1)=c(x,t)+c(x+2,t)$.

It follows that $c(x,t)+c(x+1,t+1)=c(x+2,t)$, and also that
$c(x+1,t+1)+c(x+2,t)=c(x,t)$. Suppose the colors of all cells $(2x,0)$ are
specified (call this state $x$). Assign a color to $(-1,-1)$ arbitrarily. Then
$c(1,-1)$ is determined by $c(-1,-1)$ and $c(0,0)$, $c(3,-1)$ is in turn
determined by $c(1,-1)$ and $c(2,0)$, and so on; in this way $c(-1+2n,-1)$ is
determined for each $n\geq 0$. Similarly, $c(-3,-1)$ is determined by
$c(-1,-1)$ and $c(-2,0)$, $c(-5,-1)$ is determined by $c(-3,-1)$ and $c(-4,0)$,
and so on; in this way $c(-1-2n,-1)$ is determined for each $n<0$. Each of the
two possible choices for $c(-1,-1)$ corresponds to a state $y$ such that
$T(y)=x$. Hence $T$ is two-to-one.

Now consider the lines $\{(z+2n,z)\,|\,z \in \Z\}$ for each $n \in \Z$. These
lines foliate the spacetime. The equation $c(x,t)+c(x+1,t+1)=c(x+2,t)$ implies
that the colors of the line $\{(z+2(n+1),z)\,|\,z\in\Z\}$ may be obtained from
the colors of the line $\{(z+2n,z)\,|\,z\in\Z\}$ by applying a simple rule. The
rule, in fact, is just addition mod 2. This is the same rule as $T$. Thus the
system may be broken into space+time in a different way: let space be the
direction along the lines $\{(z,z)\,|\,z\in\Z\}$, and let
$c(x,t)+c(x+1,t+1)=c(x+2,t)$ be the law of evolution. Our new description, if
abstracted from the lattice and considered as a 1+1-dimensional combinatorial
spacetime, is identical to the original one.

The same clearly applies if we consider the lines $\{(z+2n,-z)\,|\,z \in \Z\}$;
in this case the law of evolution is $c(x+2,t)+c(x+1,t+1)=c(x,t)$. Given some
choice of colors on the line $\{(z,-z)\,|\,z\in\Z\}$, this law determines the
colors of the lines $\{(z-2n,-z)\,|\,z\in\Z\}$ for each $n>0$.

Next, I wish to examine the lines which go in the direction of the vector
$(3,1)$. Here I must be more careful about my definitions. The set of lattice
points which lie directly on a line travelling in this direction through the
point $(2n,0)$ is $\{(2n+3z,z)\,|\,z \in \Z\}$. These points are too spread out
for the set of points to be a spatial surface in some space+time decomposition
of this system. In order to define a one-dimensional spatial surface in the
direction given by a vector $(a,b)$, one typically has to include more lattice
points than those which lie exactly on a line.

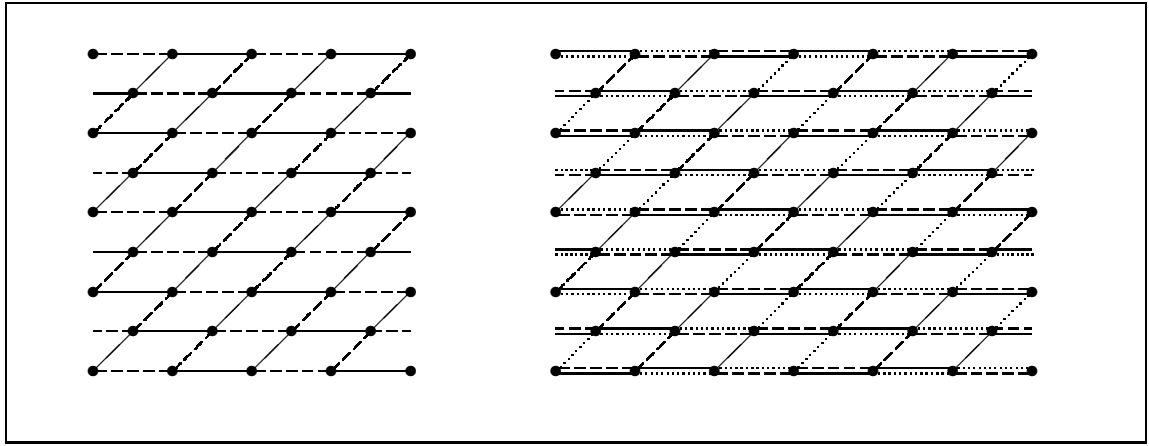
\begin{figure}
\newsavebox{\hora}
\newsavebox{\horb}
\newsavebox{\horc}
\newsavebox{\verta}
\newsavebox{\vertb}
\newsavebox{\vertc}
\newsavebox{\horshorta}
\newsavebox{\horfirstb}
\newsavebox{\horlastb}
\newsavebox{\horshortc}
\savebox{\hora}(30,0){\begin{picture}(30,0)\put(0,0){\line(1,0){30}}\end{picture}}
\savebox{\horb}(30,0){\begin{picture}(30,0)\put(0,0){\line(1,0){5}}\put(7,0){\line(1,0){4}}\put(13,0){\line(1,0){4}}\put(19,0){\line(1,0){4}}\put(25,0){\line(1,0){5}}\end{picture}}
\savebox{\horc}(30,0){\begin{picture}(30,0)\qbezier[15](0,0)(15,0)(30,0)\end{picture}}
\savebox{\verta}(15,15){\begin{picture}(15,15)\put(0,0){\line(1,1){15}}\end{picture}}
\savebox{\vertb}(15,15){\begin{picture}(15,15)\multiput(0,0)(4.1,4.1){4}{\qbezier(0,0)(1.3,1.3)(2.7,2.7)}\end{picture}}
\savebox{\vertc}(15,15){\begin{picture}(15,15)\qbezier[10](0,0)(7.5,7.5)(15,15)\end{picture}}
\savebox{\horshorta}(15,0){\begin{picture}(15,0)\put(0,0){\line(1,0){15}}\end{picture}}
\savebox{\horfirstb}(15,0){\begin{picture}(15,0)\put(0,0){\line(1,0){3}}\multiput(5,0)(6,0){2}{\line(1,0){4}}\end{picture}}
\savebox{\horlastb}(15,0){\begin{picture}(15,0)\multiput(0,0)(6,0){2}{\line(1,0){4}}\put(12,0){\line(1,0){3}}\end{picture}}
\savebox{\horshortc}(15,0){\begin{picture}(15,0)\qbezier[8](0,0)(7.5,0)(15,0)\end{picture}}
\fbox{\begin{picture}(425,160)\footnotesize
\put(25,14){\begin{picture}(150,146)
\multiput(0,0)(0,30){4}{%
   \multiput(0,0)(30,0){5}{\put(5,10){\circle*{4}}}
   \multiput(15,15)(30,0){4}{\put(5,10){\circle*{4}}}
   }
\multiput(0,120)(30,0){5}{\put(5,10){\circle*{4}}}
\multiput(0,0)(0,60){2}{%
   \put(5,10){\usebox{\horb}}
   \put(35,10){\usebox{\vertb}}
   \put(50,25){\usebox{\horb}}
   \put(80,25){\usebox{\vertb}}
   \put(95,40){\usebox{\horb}}}
\multiput(0,30)(0,60){2}{%
   \put(5,10){\usebox{\hora}}
   \put(35,10){\usebox{\verta}}
   \put(50,25){\usebox{\hora}}
   \put(80,25){\usebox{\verta}}
   \put(95,40){\usebox{\hora}}}
\multiput(0,0)(0,60){2}{%
   \put(5,10){\usebox{\verta}}
   \put(20,25){\usebox{\hora}}
   \put(50,25){\usebox{\verta}}
   \put(65,40){\usebox{\hora}}
   \put(95,40){\usebox{\verta}}
   \put(110,55){\usebox{\horshorta}}}
\put(0,30){\thinlines%
   \put(5,10){\usebox{\vertb}}
   \put(20,25){\usebox{\horb}}
   \put(50,25){\usebox{\vertb}}
   \put(65,40){\usebox{\horb}}
   \put(95,40){\usebox{\vertb}}
   \put(110,55){\usebox{\horlastb}}}
\multiput(0,15)(0,60){2}{%
   \put(5,10){\usebox{\horfirstb}}
   \put(20,10){\usebox{\vertb}}
   \put(35,25){\usebox{\horb}}
   \put(65,25){\usebox{\vertb}}
   \put(80,40){\usebox{\horb}}
   \put(110,40){\usebox{\vertb}}}
\put(0,45){%
   \put(5,10){\usebox{\horshorta}}
   \put(20,10){\usebox{\verta}}
   \put(35,25){\usebox{\hora}}
   \put(65,25){\usebox{\verta}}
   \put(80,40){\usebox{\hora}}
   \put(110,40){\usebox{\verta}}}
\put(0,0){%
   \put(35,10){\usebox{\hora}}
   \put(65,10){\usebox{\verta}}
   \put(80,25){\usebox{\hora}}
   \put(110,25){\usebox{\verta}}
   \put(95,10){\usebox{\hora}}
   \put(5,115){\usebox{\horshorta}}
   \put(20,115){\usebox{\verta}}
   \put(35,130){\usebox{\hora}}}
\put(0,0){%
   \put(65,10){\usebox{\horb}}
   \put(95,10){\usebox{\vertb}}
   \put(110,25){\usebox{\horlastb}}
   \put(5,100){\usebox{\vertb}}
   \put(20,115){\usebox{\horb}}
   \put(50,115){\usebox{\vertb}}
   \put(65,130){\usebox{\horb}}
   \put(5,130){\usebox{\horb}}}
\end{picture}}
\put(200,14){\begin{picture}(210,146)
\multiput(0,0)(0,30){4}{%
   \multiput(0,0)(30,0){7}{\put(5,10){\circle*{4}}}
   \multiput(15,15)(30,0){6}{\put(5,10){\circle*{4}}}
   }
\multiput(0,120)(30,0){7}{\put(5,10){\circle*{4}}}
\newcommand{\abc}{\put(0,-1){\usebox{\hora}}\put(0,1){\usebox{\horb}}\put(0,0){\usebox{\vertc}}}
\newcommand{\bca}{\put(0,-1){\usebox{\horb}}\put(0,1){\usebox{\horc}}\put(0,0){\usebox{\verta}}}
\newcommand{\cab}{\put(0,-1){\usebox{\horc}}\put(0,1){\usebox{\hora}}\put(0,0){\usebox{\vertb}}}
\multiput(5,10)(0,90){2}{\multiput(0,0)(90,0){2}{\put(0,0){\abc}\put(30,0){\cab}\put(60,0){\bca}}}
\multiput(5,25)(0,90){2}{\put(0,-1){\usebox{\horshorta}}\put(0,1){\usebox{\horfirstb}}\multiput(15,0)(90,0){2}{\put(0,0){\cab}\put(30,0){\bca}}\put(75,0){\abc}\put(165,-1){\usebox{\horshorta}}\put(165,1){\usebox{\horlastb}}\put(165,0){\usebox{\vertc}}}
\put(5,40){\multiput(0,0)(90,0){2}{\put(0,0){\cab}\put(30,0){\bca}\put(60,0){\abc}}}
\put(5,130){\multiput(0,0)(90,0){2}{\put(0,-1){\usebox{\horc}}\put(0,1){\usebox{\hora}}\put(30,-1){\usebox{\horb}}\put(30,1){\usebox{\horc}}\put(60,-1){\usebox{\hora}}\put(60,1){\usebox{\horb}}}}
\put(5,55){\put(0,-1){\usebox{\horshortc}}\put(0,1){\usebox{\horshorta}}\multiput(15,0)(90,0){2}{\put(0,0){\bca}\put(30,0){\abc}}\put(75,0){\cab}\put(165,-1){\usebox{\horshortc}}\put(165,1){\usebox{\horshorta}}\put(165,0){\usebox{\vertb}}}
\put(5,70){\multiput(0,0)(90,0){2}{\put(0,0){\bca}\put(30,0){\abc}\put(60,0){\cab}}}
\put(5,85){\put(0,-1){\usebox{\horfirstb}}\put(0,1){\usebox{\horshortc}}\multiput(15,0)(90,0){2}{\put(0,0){\abc}\put(30,0){\cab}}\put(75,0){\bca}\put(165,-1){\usebox{\horlastb}}\put(165,1){\usebox{\horshortc}}\put(165,0){\usebox{\verta}}}
\end{picture}}
\end{picture}}\linespread{1}\caption[Foliations of a two-dimensional
lattice]{Foliations of the two-dimensional lattice used in the addition-mod-2
rule. On the left is the $[1,1]$ foliation. Each path made by solid lines and
each path made by dashed lines represents a portion of a one-dimensional
surface. Each vertex is in two surfaces. On the right is the $[2,1]$ foliation;
here there are solid lines, dashed lines and dotted lines, and each vertex is
in three surfaces.}
\label{foliate}
\end{figure}

In this case, and in many of those which I will consider here, an appropriate
choice is to consider the path generated by moving from the point $(x,t)$ to
$(x+a,y+b)$ ($a$, $b\geq 0$) by first heading in either a positive horizontal
or positive vertical direction and then changing course by a 45-degree angle at
the moment required in order to arrive at the correct destination. Typically,
each vertex which lies on this path is included in the surface. For example, to
move from $(x,t)$ to $(x+3,t+1)$, I first move horizontally to the right,
arriving at $(x+2,t)$, and then upwards and to the right at a 45-degree angle,
arriving at $(x+3,t+1)$. Each of the vertices $(x,t)$, $(x+2,t)$ and
$(x+3,t+1)$ are included in the surface. I then repeat the procedure, adding
$(x+5,t+1)$ and $(x+6,t+2)$ to the surface, and so on. Examples of the
foliation of spacetime into surfaces in this manner are given in
Figure~\ref{foliate}.

For now, I will only consider surfaces whose slopes are between 1 and $-1$. If
a surface is described by proceeding $h$ vertices horizontally to the right and
then $k$ vertices on a 45-degree angle up and to the right, I will say that the
surface is an $[h,k]$ surface. If it is obtained by moving $h$ vertices
horizontally to the right and then $k$ vertices on a 45-degree angle down and
to the right, I will say that it is an $[h,-k]$ surface.

There are $h+k$ different types of vertices in an $[h,k]$ surface. Types 1
through $h$ are those situated on a horizontal segment of the surface,
beginning with the leftmost vertex on this segment (which is also the rightmost
vertex on a slope-1 segment) and ending with the next-to-rightmost
vertex. Types $h+1$ through $h+k$ are those situated on a slope-1 segment of
the surface, beginning with the leftmost vertex on this segment (which is also
the rightmost vertex of a horizontal segment) and ending with the
next-to-rightmost vertex. For fixed $h$ and $k$, each vertex in the lattice is
in $h+k$ different $[h,k]$ surfaces; it is a different type of vertex in each
surface. If $(x,t)$ is a vertex on an $[h,k]$ surface, then
$\{(x+n(2h+k),t+nk)\,|\,n\in\Z\}$ is the set of vertices on the surface which
have the same type as $(x,t)$. Let $(x,t)$ be a type-1 vertex in an $[h,k]$
surface. I will say that the line associated with this surface is the line in
the direction $(2h+k,k)$ which passes through $(x,t)$.

It is necessary for our purposes that each set of surfaces that foliate
spacetime be linearly ordered; for if we are to be able to redescribe spacetime
by making these surfaces be the spaces, we must be able to assign each such
surface a time coordinate in a sensible way. It is therefore necessary to
exclude the situation in which several surfaces are associated with the same
line. Consider an $[nh,nk]$ surface with $n>1$. Then the line associated with
the surface containing $(0,0)$ as a type-1 vertex is the same as the line
associated with the surface containing $(2h+k,k)$ as a type-1 vertex, even
though these surfaces are distinct. This situation is avoided if one considers
only $[h,k]$ surfaces for relatively prime $h$ and $k$. Then no two $[h,k]$
surfaces are associated with the same line, and the set of $[h,k]$ surfaces is
linearly ordered in the same way that the set of associated lines is naturally
ordered in the plane.

If our set of surfaces $[h,k]$ turns out to be a subshift of finite type, then
one must be able to describe it in the usual manner, as a set of closed
one-dimensional spaces made from sequences of colors in some finite set
$K$. Typically, one may do this by considering each group of $h+k$ consecutive
vertices on the surface (beginning with a type-1 vertex and ending with a
type-$(h+k)$ vertex) as a single cell in the one-dimensional space. Each state of
those $h+k$ vertices which is allowed by $T$ is assigned a unique color. The
adjacency rules are then determined: for example, \seg{abc} ($a,b,c \in K$) is
allowed if and only if the sections of the surface corresponding with $a$, $b$
and $c$ are allowed to follow one another consecutively. If the adjacency rules
may be completely described by listing all allowed neighborhoods of width $d$
for some finite $d$, then $[h,k]$ is a subshift of finite type.

My convention for choosing the vertices that are in a surface having direction
$(a,b)$ is an arbitrary one. A more general method of specifying a surface in
direction $(a,b)$ ($a+b=0 \bmod 2$) is as follows: fix a vertex $(x,t)$; choose
a nonnegative integer $m$; define sequences of integers $x_i$ and $t_i$ with
$x_i+t_i=0 \bmod 2$, $1\leq i\leq m$; and let the surface consist of the
vertices $(x+na+x_i,t+nb+t_i)$ for each $n \in \Z$ and for each $i$, $1\leq
i\leq m$. Let this surface be identified with the line in direction $(a,b)$
through $(x,t)$. If we now let $(x,t)$ vary, we get a set of surfaces
associated with lines having the direction $(a,b)$. Clearly, if we consider any
lattice line which has this direction, we may choose $(x,t)$ to be a vertex on
this line and thereby obtain a surface associated with this line. In addition,
if no vertex lies on the interior of a line segment joining $(x,t)$ to
$(x+a,t+b)$, then no two surfaces in this set are associated with the same
line; in this case we may assign a time parameter to the surfaces in this set
in a natural way. One may now attempt to describe the set of surfaces as a
subshift of finite type by considering the set of vertices
$\{(x+na+x_i,t+nb+t_i)\,|\,1\leq i\leq m\}$ as a single cell for each $n \in
\Z$ and then proceeding as described above to find colors and adjacency rules.

In an important case, however, the particular choice of description of a set of
surfaces in the direction $(a,b)$ does not matter. Suppose that we have two
sets of surfaces $S_1$ and $S_2$, each associated with lines having the
direction $(a,b)$. Suppose in addition that $S_1$ and $S_2$ are each
describable as a subshift of finite type, and that one may formulate a
space+time description of spacetime in such a way that either $S_1$ or $S_2$ is
the set of spaces. Then I claim that the two descriptions of spacetime obtained
in this way are equivalent via a local, invertible map. I will not prove this
rigorously here, but the argument goes roughly as follows. Let $S_3$ be the set
of surfaces in direction $(a,b)$ such that the surface in this set associated
with a particular line is equal to the union of the surfaces associated with
that line in $S_1$ and $S_2$. Then $S_3$ is a set of surfaces of the sort
described earlier, and it is a set of Cauchy surfaces. Furthermore, the colors
of the vertices that are in a surface in $S_3$ and not in the corresponding
surface in $S_1$ are locally determined by the colors of the surface in $S_1$,
because $S_1$ is a Cauchy surface. Hence there is a one-to-one correspondence
between the cells in $S_1$ and those in $S_3$, and it is straightforward to
describe a local, invertible map between them. The same relationship holds
between $S_2$ and $S_3$, and therefore between $S_1$ and $S_2$.

I now return to considering surfaces with direction $(3,1)$ in our
example. These are surfaces of type $[1,1]$. We will take the ``cells'' in
these surfaces to be pairs of consecutive horizontal vertices. Since our
original spaces were in \x{2}, each vertex may have two colors, so our cell has
four possible states, corresponding to four colors in a new color set $K'$. Let
us for the moment assume that these are all of the restrictions on states of
our surface; i.e., that each vertex in this surface may be colored 0 or 1,
and that any coloring of this sort is compatible with our original law $T$.

The surfaces are associated with a time parameter $T$ in the following way: if
$(x,0)$ and $(x+2,0)$ constitute a cell in the surface at time $T$, then
$(x-2,0)$ and $(x,0)$ constitute a cell in the surface at time $T+1$.

Suppose that we know the states of two consecutive cells in our surface at time
$T$. That means we know $c(x,t)$, $c(x+2,t)$, $c(x+3,t+1)$, $c(x+5,t+1)$. Then
$c(x+1,t+1)=c(x,t)+c(x+2,t)$, so the colors of these two cells determine the
color of the cell constituted by $(x+1,t+1)$ and $(x+3,t+1)$, which is in the
surface at time $T+1$. Clearly the colors of all cells in the $T+1$ surface may
be determined in this way. Hence all states at times greater than $T$ are
determined by the state at time $T$ in a local way. Also,
$c(x+4,t)=c(x+2,t)+c(x+3,t+1)$, so the colors of these two cells also determine
the color of the cell constituted by $(x+2,t)$ and $(x+4,t)$, which is in the
surface at time $T-1$. The colors of all cells in the $T-1$ surface may be
determined in this way also. Hence all states at times less than $T$ are also
determined by the state at time $T$ in a local way.

Thus our original law $T$ implies that the colors on any one of these surfaces
determine all of the remaining colors of the spacetime. It is easy to verify
that the resulting spacetime is consistent with $T$; this validates our
assumption that the vertices in $(3,1)$ surfaces may be colored in an arbitrary
manner. Moreover, the surfaces in this set are Cauchy surfaces. This is part of
what is surprising about this example: the original representation of the
system is in terms of a two-to-one law, but it turns out to be representable in
terms of an invertible law.

\begin{figure}
\fbox{\begin{picture}(425,100)\footnotesize
\put(20,20){\multiput(40,0)(60,20){4}{
        \qbezier(-10,0)(-10,10)(20,10)\qbezier(20,10)(50,10)(50,0)
        \qbezier(50,0)(50,-10)(20,-10)\qbezier(20,-10)(-10,-10)(-10,0)}
        \multiput(0,0)(60,20){4}{
        \qbezier[20](-10,0)(-10,10)(20,10)\qbezier[20](20,10)(50,10)(50,0)
        \qbezier[20](50,0)(50,-10)(20,-10)\qbezier[20](20,-10)(-10,-10)(-10,0)}
        \multiput(0,0)(40,0){7}{\multiput(0,0)(20,60){2}{\circle*{4}}}
        \multiput(20,20)(40,0){2}{\circle*{4}}\put(100,20){\makebox(0,0){$a$}}
        \put(140,20){\makebox(0,0){$b$}}\multiput(180,20)(40,0){3}{\circle*{4}}
        \multiput(0,40)(40,0){3}{\circle*{4}}\put(120,40){\makebox(0,0){$c$}}
        \put(160,40){\makebox(0,0){$d$}}\put(200,40){\makebox(0,0){$e$}}
        \put(240,40){\circle*{4}}}
\put(280,0){\makebox(145,100){$\begin{array}{c|cccc}
&00&11&01&10\\\hline
00&00&01&00&01\\
11&00&01&00&01\\
01&10&11&10&11\\
10&10&11&10&11
\end{array}$}}
\end{picture}}\linespread{1}\caption[Coordinate transformation in
addition-mod-2 rule]{A coordinate transformation of the addition-mod-2
rule. The cells in the selected Cauchy surface are shown by solid-line
ovals. Orientation of the new surface goes up and to the right. The colors in
the two consecutive cells \seg{ab} and \seg{de} determine the colors of cell
\seg{cd} in accordance with the right-hand table. Hence the colors of the
selected Cauchy surface determine the colors of the surface whose cells are
shown by dotted-line ovals. This rule is equivalent to rule $B$ in
Figure~\ref{calaws} (let $a=[00]$, $b=[11]$, $c=[01]$ and $d=[10]$). The
addition-mod-2 rule and rule $B$ are linear; for details, see
Figure~\ref{linear}.}
\label{rule1}
\end{figure}
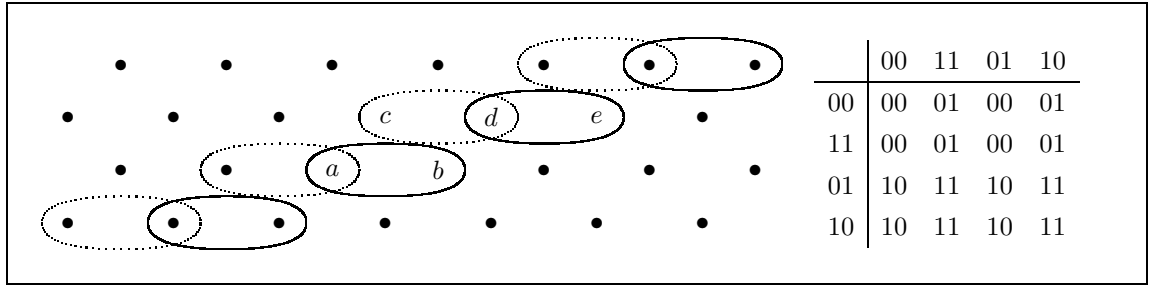

Our original law was on \x{2}; the new law $T'$ is on \x{4}. Note that \x{2}
and \x{4} have different Bowen-Franks groups, and hence are not equivalent sets
of spaces. On the other hand, it is a consequence of the above discussion that
there does exist a local equivalence between the sets of \emph{spacetimes}
generated by $T$ and $T'$. The rule $T'$ is displayed in Figure~\ref{rule1}.

In our example the $(3,1)$ surfaces are Cauchy surfaces, but they are not
alone. In fact, there are only three sets of surfaces along lattice lines that
are \emph{not} Cauchy surfaces. These are the surfaces corresponding to the
lines with slope 0, 1 and $-1$. Every other set of straight-line surfaces is a
set of Cauchy surfaces. Here I am not referring merely to surfaces of type
$[h,k]$ or $[h,-k]$. As we shall see shortly, even the set of surfaces
corresponding to \emph{vertical} lines is a set of Cauchy surfaces. This is the
second unexpected feature of the addition-mod-2 rule: it contradicts one's
usual expectations about light cones.

Consider a set of $[h,k]$ surfaces for fixed $h$ and $k$. The claim is that for
$h>0$ and $k>0$ a surface of this type is a Cauchy surface. Suppose that we fix
the colors on one such surface, which contains type-1 vertex $(x,t)$. Then $T$
means that the colors in cells $(x,t)$ through $(x+2h,t)$ determine the colors
in cells $(x+1,t+1)$ through $(x+2h-1,t+1)$. In other words, we may evolve a
horizontal segment of the surface one step forward in time using $T$. If we do
this to each horizontal segment in the same way, we end up with another $[h,k]$
surface. The process may now be repeated. Eventually the colors of all cells in
the future of our original surface are determined. Next, notice that the colors
in cells $(x+2h,t)$ through $(x+2h+k,t+k)$ determine the colors in cells
$(x+2h+2,t)$ through $(x+2h+k+1,t+k-1)$. In other words, we may evolve a
slope-1 segment of the surface one step backward in time using the relation
$c(x,t)+c(x+1,t+1)=c(x+2,t)$. If we do this to each slope-1 segment, we end up
with another $[h,k]$ surface. The process may be repeated. Eventually the
colors of all cells in the past of our original surface are determined.

This demonstrates that each $[h,k]$ surface with $h>0$ and $k>0$ is a Cauchy
surface. Again, it is easy to verify that the resulting spacetime forced by the
surface is consistent with $T$; hence we may assign colors 0 and 1
arbitrarily to the surface without causing a contradiction. Hence the set of
$[h,k]$ surfaces, for fixed $h$ and $k$, is a subshift of finite type. In fact,
this subshift of finite type must be \x{2^{h+k}}, since there are $h+k$
vertices in each cell, and each vertex has two states, and hence the cell has
$2^{h+k}$ states. It is easy to verify that the state of the surface at time
$T+1$ is caused by the state of the surface at time $T$ in a local way, and
vice versa. Hence if $h>0$ and $k>0$ (and $h$ and $k$ are relatively prime)
this system may be described as a 1+1-dimensional combinatorial spacetime where
the $[h,k]$ surfaces are the spaces. By the left-right symmetry of $T$, it
follows that the same is true if we replace $[h,k]$ with $[h,-k]$.

The proof that the addition-mod-2 system only has three non-Cauchy surfaces
(among its straight-line surfaces) now follows by noting that this system
possesses a threefold symmetry, where the symmetry transformations send
straight lines into straight lines. The spaces for law $T$ are
horizontal. Consider the lines with slope 1 and $-1$. We may rotate the
horizontal lines in either a clockwise or counterclockwise direction and stop
as soon as they have slope 1 or $-1$. As we do this, the spaces corresponding
to intervening lines are Cauchy surfaces. The spaces corresponding to the lines
with slope 1 and $-1$, on the other hand, are not Cauchy surfaces. Consider
again the definition of $T$: $c(x,t)+c(x+2,t)=c(x+1,t+1)$. The lines with slope
1 and $-1$ are associated with this definition in the following way: those with
slope 1 are the ones joining $(x,t)$ and $(x+1,t+1)$, and those with slope $-1$
are the ones joining $(x+2,t)$ and $(x+1,t+1)$.

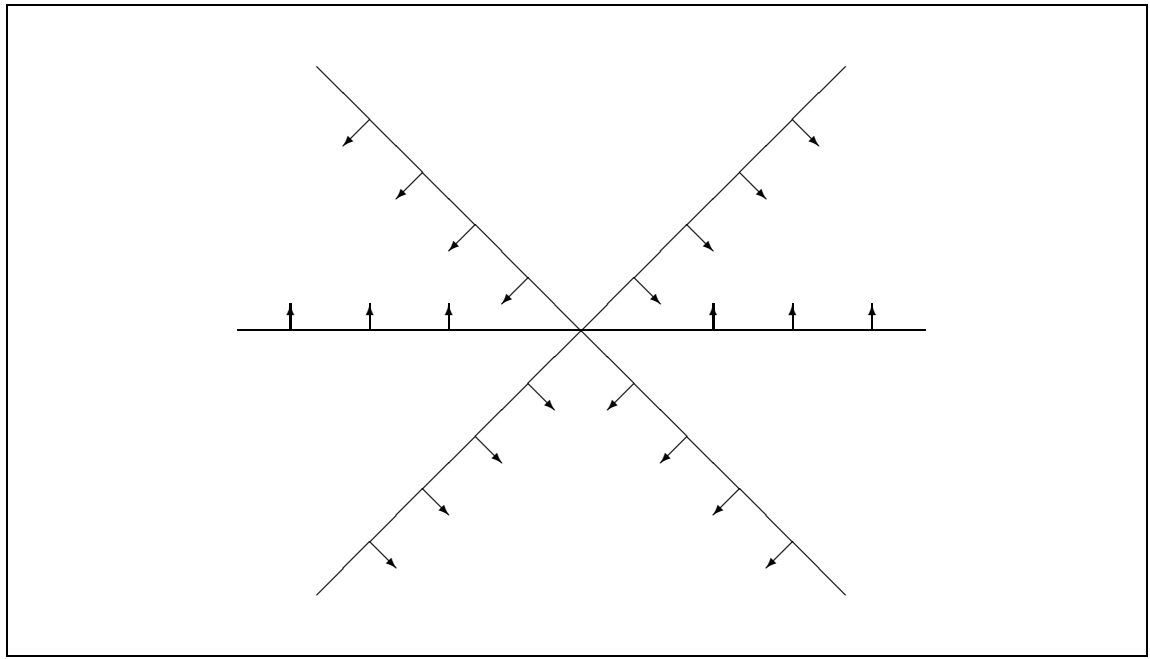
\begin{figure}
\fbox{\begin{picture}(425,240)\linespread{1}\footnotesize\put(0,0){\makebox(425,240){
\begin{picture}(200,200)
\put(-30,100){\line(1,0){260}}
\put(0,200){\line(1,-1){200}}
\put(0,0){\line(1,1){200}}
\put(20,20){\vector(1,-1){10}}
\put(40,40){\vector(1,-1){10}}
\put(60,60){\vector(1,-1){10}}
\put(80,80){\vector(1,-1){10}}
\put(120,120){\vector(1,-1){10}}
\put(140,140){\vector(1,-1){10}}
\put(160,160){\vector(1,-1){10}}
\put(180,180){\vector(1,-1){10}}
\put(180,20){\vector(-1,-1){10}}
\put(160,40){\vector(-1,-1){10}}
\put(140,60){\vector(-1,-1){10}}
\put(120,80){\vector(-1,-1){10}}
\put(80,120){\vector(-1,-1){10}}
\put(60,140){\vector(-1,-1){10}}
\put(40,160){\vector(-1,-1){10}}
\put(20,180){\vector(-1,-1){10}}
\put(-10,100){\vector(0,1){10}}
\put(20,100){\vector(0,1){10}}
\put(50,100){\vector(0,1){10}}
\put(150,100){\vector(0,1){10}}
\put(180,100){\vector(0,1){10}}
\put(210,100){\vector(0,1){10}}
\end{picture}}}
\end{picture}}\linespread{1}\caption[Threefold symmetry of the addition-mod-2
rule]{The threefold symmetry of the addition-mod-2 rule. Shown are the three
orientations of the straight-line surfaces which are not Cauchy surfaces. All
three surfaces are causal in one direction only (this direction is indicated by
the arrows); the rule for determining the next state in each case is addition
mod 2. Straight-line surfaces having all other orientations are Cauchy
surfaces.}
\label{threefold}
\end{figure}

Now consider the surfaces with slope 1 as spaces for another representation of
$T$. Here the rule $T$ is given by $c(x,t)+c(x+1,t+1)=c(x+2,t)$. Thus the lines
with slopes 1 and $-1$ in our original version correspond here to the ones
joining $(x,t)$ to $(x+2,t)$ and the ones joining $(x+1,t+1)$ to
$(x+2,t)$. These are the lines with slope 0 and $-1$ (this is reasonable, since
we expect these lines not to be Cauchy surfaces). If we rotate the lines with
slope 1 in either direction until they have slope 0 or slope $-1$, then the
surfaces corresponding to intervening lines must be Cauchy surfaces. But this
means that all lines with slope greater than 1 or less than $-1$ must
correspond to Cauchy surfaces (see Figure~\ref{threefold}).

In constructing surfaces corresponding to lines with slope greater than 1 or
less than $-1$, we must deviate slightly from our previous method for
constructing surfaces. The problem is that if we simply include in the surface
the vertices that we cross while travelling vertically, these vertices will not
be sufficient. The reason is that the vertices lying on a vertical line do not
constitute a Cauchy surface. However, the vertices lying on two adjacent
vertical lines do constitute a Cauchy surface. So every time we move
vertically, we must include the vertices lying on two adjacent vertical lines.

For example: we may construct a surface corresponding to a vertical line by
including the vertices $(x,t+n)$ and $(x+1,t+1+n)$ for each $n\in\Z$. A cell in
this surface may be taken to be constituted by the two vertices $(x,t+n)$ and
$(x+1,t+1+n)$; thus each cell has four colors and there are no adjacency
restrictions. So the vertical surfaces correspond to \x{4}. There is only one
set of surfaces corresponding to \x{4} on the lines having slope between 0 and
45 degrees; recall that these were the $(3,1)$ surfaces, and that the rule
corresponding to this set of surfaces is shown in Figure~\ref{rule1}. It
follows that the vertical surfaces must correspond under the threefold symmetry
transformations to the $(3,1)$ surfaces. It is easy to verify that this is the
case.

I have developed this example in so much detail both because it is an
interesting example and because it illustrates many of the techniques involved
in making rotational transformations. The subject of rotational transformations
is a nice application of the theory developed in the preceding portion of this
document. For example, an understanding of subshifts of finite type is required
in order to break up a surface into cells in a reasonable way. And without the
notion of equivalent space sets, it would have been difficult to make sense of
all of the different ways in which one may design Cauchy surfaces that lie
along a particular line; with the notion, all such ways are seen to be
equivalent.

Subshifts of finite type have been largely ignored by cellular automata
theorists, who typically concentrate (in the one-dimensional case) on the sets of
spaces \x{k}. The above example, however, shows that they can be quite relevant
to the study of systems defined on \x{k}. Using subshifts of finite type, we
showed that the addition-mod-2 rule describes a system which, seen from almost
any other angle, is reversible. It is natural to ask: does something similar
hold for other seemingly irreversible one-dimensional cellular automata?

To this end, I examined the most famous one-dimensional cellular automaton
rules. I call these the Wolfram rules; they are the $d=3$ rules defined on
\x{2}, and were popularized by Wolfram~\cite{wolframbook}. The accepted view is
that these only contain a few trivial reversible rules. However, I have found
that 12 different Wolfram rules are reversible after a rotation (see
Figure~\ref{wolfreverse}).
\begin{figure}
\fbox{\begin{picture}(425,146)\linespread{1}\footnotesize\put(11,0){\makebox(325,146){%
\begin{tabular}{c|cccccccccccc}
&15&30&45&51&60&90&105&106&150&154&170&204\\
\hline
\\[-10pt]
000&1&0&1&1&0&0&1&0&0&0&0&0\\
001&1&1&0&1&0&1&0&1&1&1&1&0\\
010&1&1&1&0&1&0&0&0&1&0&0&1\\
011&1&1&1&0&1&1&1&1&0&1&1&1\\
100&0&1&0&1&1&1&0&0&1&1&0&0\\
101&0&0&1&1&1&0&1&1&0&0&1&0\\
110&0&0&0&0&0&1&1&1&0&0&0&1\\
111&0&0&0&0&0&0&0&0&1&1&1&1\\
\end{tabular}}}\put(355,35){
\begin{picture}(45,75)
\put(0,0){\put(0,0){\line(1,0){45}}
\put(0,15){\line(1,0){45}}
\put(15,30){\line(1,0){15}}
\put(0,0){\line(0,1){15}}
\put(15,0){\line(0,1){30}}
\put(30,0){\line(0,1){30}}
\put(45,0){\line(0,1){15}}
\put(7.5,7.5){\circle*{3}}
\put(22.5,7.5){\circle*{3}}
\put(22.5,22.5){\circle*{3}}
\put(0,45){\put(0,0){\line(1,0){45}}
\put(0,15){\line(1,0){45}}
\put(15,30){\line(1,0){15}}
\put(0,0){\line(0,1){15}}
\put(15,0){\line(0,1){30}}
\put(30,0){\line(0,1){30}}
\put(45,0){\line(0,1){15}}
\put(22.5,7.5){\circle*{3}}
\put(22.5,22.5){\circle*{3}}
\put(37.5,7.5){\circle*{3}}}}\end{picture}}
\end{picture}}\linespread{1}\caption[Wolfram rules which are reversible
after a rotation]{The twelve Wolfram rules that are reversible after a
rotation. These are $d=3$ rules on \x{2}. The rules are listed in the left-hand
table. Each number at the top of a right-hand column in this table is equal to
that column considered as a binary number; this is the usual way to identify
Wolfram rules. On the right, the key relationships which cause most of these
rules to be reversible after a rotation are displayed. Here, the colors of the
marked cells determine the color of the unmarked cell (time moves up).}
\label{wolfreverse}
\end{figure}
The proof (in all but the trivial cases) is almost identical to our proof that
the $(3,1)$ surfaces in the addition-mod-2 rule are Cauchy surfaces. Note that
this proof depended only on the fact that the colors at $(x,t)$ and $(x+1,t+1)$
determined the color at $(x+2,t)$. Similarly, in these cases one may easily
verify that the colors at $(x,t)$, $(x+1,t)$ and $(x+1,t+1)$ determine the
color at $(x+2,t)$, or else that the colors at $(x+2,t)$, $(x+1,t)$ and
$(x+1,t+1)$ determine the color at $(x,t)$ (here I am using standard
rectangular lattice coordinates). In either case, one may use this fact to
verify that a certain surface is Cauchy.

I suspect that these 12 Wolfram rules may be all of the Wolfram rules with this
property, because they are exactly those Wolfram rules with the property that
the set of allowed infinite spaces does not change after application of the
rule. (In other words, the ``regular language'' of the states of the system is
invariant under the rule; see Wolfram~\cite{wolframbook}, pp.\ 523-526.) For
example, the addition-mod-2 rule $T$ (which is identical to Wolfram rule 60)
has this property: the set $T(\x{2})$ contains the same infinite spaces as does
\x{2} (but fewer finite spaces; $T(\x{2})$ contains half as many width-$w$
spaces as does \x{2} for any finite $w$). Perhaps this property is related to
reversibility.

I turn now to a derivation of the main result of this chapter.

Let $T$ be a one-dimensional reversible cellular automaton defined in the
standard way using a left-hand column with width-$d_0$ segments. Let us
temporarily use a rectangular lattice with standard coordinates $(x,t)$, and
specify that the colors of the cells $(x,t)$ through $(x+d_0-1,t)$ determine
the color of cell $(x,t+1)$. Since $T^{-1}$ exists and is also a cellular
automaton, it may be represented in the standard way using a left-hand column
with width-$d_1$ segments for some $d_1\in\N$. Then there exists $k\in\Z$ such
that the colors of the cells $(x,t+1)$ through $(x+d_1-1,t+1)$ determine the
color of the cell $(x+k,t)$.

Next, we widen the boundaries of the width-$d_0$ segments and the width-$d_1$
segments until a certain symmetry condition is reached. If $k<d_1-1$ then let
$e_1=d_1-1$ and $b_0=k-d_1+1$; otherwise let $e_1=k$ and $b_0=0$. If $k<d_0-1$
then let $b_1=k-d_0+1$ and $e_0=d_0-1$; otherwise let $b_1=0$ and $e_0=k$. Then
the color of $(x,t+1)$ is forced by the colors of the segment from $(x+b_0,t)$
to $(x+e_0,t)$, and the color of $(x+k,t)$ is forced by the colors of the
segment from $(x+b_1,t+1)$ to $(x+e_1,t+1)$. Let $c_0$ be the relative position
of the cell forced by the first segment, and $c_1$ be the relative position of
the cell forced by the second segment. Then $c_0=0$ and $c_1=k$. The symmetry
property may now be seen as follows: $c_1-b_0=e_1-c_0$ and
$c_0-b_1=e_0-c_1$. Hence $e_0-b_0=e_1-b_1$; i.e., the segments are of the same
length $d=e_0-b_0+1$.

\begin{figure}
\newsavebox{\blob}
\savebox{\blob}(15,15){\begin{picture}(15,15)\qbezier(0,12)(1.5,13.5)(3,15)\qbezier(0,9)(3,12)(6,15)\qbezier(0,6)(4.5,10.5)(9,15)\qbezier(0,3)(6,9)(12,15)\qbezier(0,0)(7.5,7.5)(15,15)\qbezier(3,0)(9,6)(15,12)\qbezier(6,0)(10.5,4.5)(15,9)\qbezier(9,0)(12,3)(15,6)\qbezier(12,0)(13.5,1.5)(15,3)\end{picture}}
\fbox{\begin{picture}(425,165)
\put(30,105){\put(0,0){\line(1,0){45}}\multiput(0,15)(0,15){2}{\line(1,0){75}}\multiput(0,0)(15,0){4}{\line(0,1){30}}\multiput(60,15)(15,0){2}{\line(0,1){15}}\put(0,15){\usebox{\blob}}\put(15,0){\usebox{\blob}}}
\put(140,116){$\Longrightarrow$}
\put(195,105){\put(0,0){\line(1,0){90}}\put(0,15){\line(1,0){120}}\put(30,30){\line(1,0){90}}\multiput(0,0)(15,0){2}{\line(0,1){15}}\multiput(30,0)(15,0){5}{\line(0,1){30}}\multiput(105,15)(15,0){2}{\line(0,1){15}}\put(60,0){\usebox{\blob}}\put(45,15){\usebox{\blob}}}
\put(350,116){$\Longrightarrow$}
\put(30,37.5){\put(0,0){\line(1,0){90}}\put(0,15){\line(1,0){112.5}}\put(22.5,30){\line(1,0){90}}\multiput(0,0)(15,0){7}{\line(0,1){15}}\multiput(22.5,15)(15,0){7}{\line(0,1){15}}\put(60,0){\usebox{\blob}}\put(37.5,15){\usebox{\blob}}}
\put(187.5,48.5){$\Longleftrightarrow$}
\put(256.5,30){\multiput(0,15)(0,15){2}{\line(1,0){90}}\multiput(37.5,0)(0,45){2}{\line(1,0){15}}\multiput(37.5,0)(0,30){2}{\multiput(0,0)(15,0){2}{\line(0,1){15}}\put(0,0){\usebox{\blob}}}\multiput(0,15)(15,0){7}{\line(0,1){15}}}
\end{picture}}\linespread{1}\caption[The central representation]{The central
representation of a one-dimensional reversible cellular automaton. In the example
shown here, $T$ is a $d=3$ law and $T^{-1}$ is a $d=5$ law. (Time moves up.) In
the upper left-hand figure, the lattice representation is chosen so that a
3-neighborhood at time $t$ forces the cell above the left-hand end of that
neighborhood at time $t+1$. Given such a choice, the 5-neighborhood shown at
time $t+1$ will force some cell at time $t$; in this example we assume it
forces the second cell from the left. (Forced cells are shaded.) In the first
transformation, we extend the 3-neighborhood three cells to the left and the
5-neighborhood one cell to the left so that a symmetry property holds. Next,
the row at time $t+k$ is shifted $.5k$ cells to the left, so that the cell at
time $t+1$ forced by the 6-neighborhood at time $t$ is in the center of that
neighborhood. The result is that the law may be seen as a $d=6$ law with causal
properties as shown in the last figure.}
\label{center}
\end{figure}
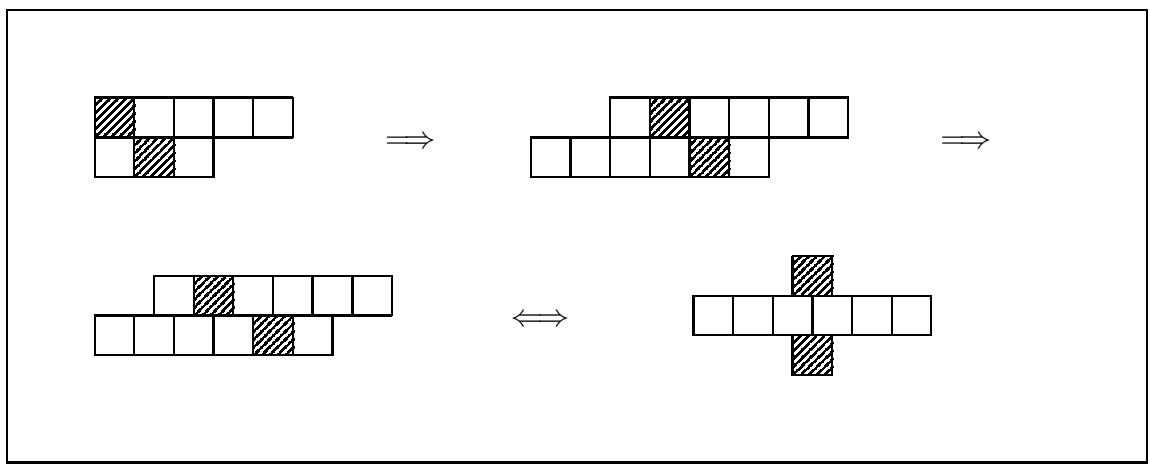

Now we may construct a representation of $T$ which will be useful in proving
the main theorem. The structure of the lattice for this representation depends
on $d$. If $d$ is odd, it is a standard rectangular lattice, but now the
coordinates of the lattice points are $(2x,t)$ for integers $x$ and $t$. If $d$
is even, the lattice is the sort that we considered in the addition-mod-2 case,
with coordinates $(x,t)$, $x+t=0 \bmod 2$, as before. We represent $T$ on one
of these lattices by specifying that the state of the segment in cells $(x,t)$
through $(x+2(d-1),t)$ forces the color of $(x+d-1,t+1)$. Because of the
symmetry condition, it must also be the case that the state of that same
segment forces the color of $(x+d-1,t-1)$. In other words, the state of a
segment of width $d$ at time $t$ forces the state of the cell whose
$x$-coordinate is exactly in the center of that segment at times $t+1$ and
$t-1$. The situation is pictured in Figure~\ref{center}.

Any representation of $T$ that is constructed in this manner will be called a
\emph{central representation} of $T$.\footnote{\linespread{1}\footnotesize The
procedure just described is adapted from one which I used in~\cite{me} to
construct ``center-reversible'' representations. These were central
representations with the additional restriction that $d=2$. Note, however, that
the idea of equivalence used in~\cite{me} is not the same as the one used
here. I claimed there that every reversible one-dimensional cellular automaton
had a center-reversible representation. Part of the proof of this involved a
claim that every one-dimensional cellular automaton was equivalent to a $d=2$
cellular automaton. The equivalence map used to verify this latter claim, while
no doubt a useful sort of map, is not a locally invertible one.

While I am on the topic of~\cite{me}, let me take the opportunity to correct
several errors that appeared there. The proof of Theorem~1 was a bit
garbled. In the last paragraph of page 286, it says ``Then each element of
$Y_i\times X_j$ can occur in at most one set $D_m$.'' Instead, this should read
``Then each set $D_m$ may contain at most one element of $Y_i\times X_j$.''
Similarly, the next paragraph begins ``Since each element of $Y_i\times X_j$
can occur in at most one set $D_m$,'' and instead this should read ``Since each
set $D_m$ may contain at most one element of $Y_i\times X_j$.'' Finally, on
page 292 the number 1200 should be changed to 1224, and in the figure on that
page where it says there are 12 distinct laws for $k=28$, the 12 should be
changed to 36.}

\begin{theorem} Let $T$ be a reversible one-dimensional cellular automaton with a
width-$d$ central representation. Let $-\frac{1}{d-1}<m<\frac{1}{d-1}$. (If
$d=1$ then let $-\infty<m<\infty$.) Then a rotation of $T$ exists in which the
spaces of the rotated system are associated with the slope-$m$ lines in the
original spacetime lattice.
\end{theorem}
\textsc{Proof.} It is only necessary to prove this for $0<m<\frac{1}{d-1}$; the
rest follows by symmetry. Let the direction of lines with slope $\frac{1}{d-1}$
be called the \emph{diagonal}. Assign new coordinates $[X,T]$ to the lattice
using the horizontal and the diagonal as coordinate axes, retaining the
original origin, and rescaling the horizontal axis so that vertices are one
unit apart. (I will use brackets to indicate that the new coordinates are being
used.) Then $x=2X+(d-1)T$ and $t=T$. If $[X,T]$ is a vertex, then so is
$[X+h,T+k]$ for any integers $h$ and $k$. In terms of these new coordinates,
the state of cells $[X,T]$ through $[X+d-1,T]$ forces the state of $[X,T+1]$
and of $[X+d-1,T-1]$ for any $X$ and $T$.

The lines with slopes between 0 and $\frac{1}{d-1}$ in the original coordinates
have slopes between 0 and $\infty$ in the new coordinates. Hence there is a
one-to-one correspondence between the allowed slopes and the set of ordered
pairs $(h,k)$ where $h$ and $k$ are positive and relatively prime. For each
such ordered pair $(h,k)$, the line $\{[X+nh,T+nk]\,|\,n\in\Z\}$ has one of the
allowed slopes, and all such lines may be obtained in this way.

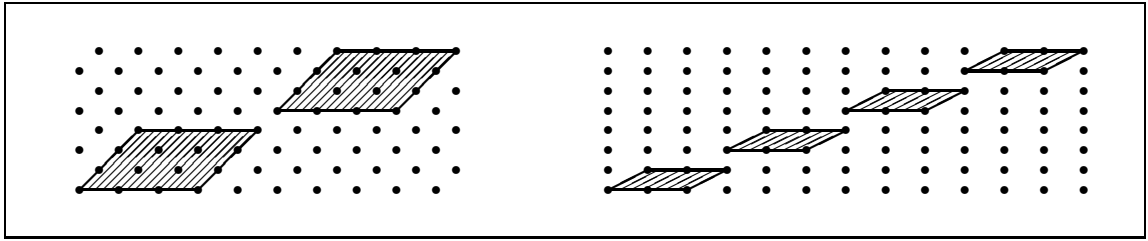
\begin{figure}
\fbox{\begin{picture}(425,82.5)\thicklines
\put(25,15){\multiput(0,0)(0,15){4}{\multiput(0,0)(15,0){10}{\put(0,0){\circle*{3}}\put(7.5,7.5){\circle*{3}}}}\multiput(0,0)(75,30){2}{\multiput(0,0)(45,0){2}{\line(1,1){22.5}}\multiput(0,0)(22.5,22.5){2}{\line(1,0){45}}\thinlines\multiput(3,0)(3,0){14}{\line(1,1){22.5}}}}
\put(225,15){\multiput(0,0)(15,0){13}{\multiput(0,0)(0,7.5){8}{\circle*{3}}}\multiput(0,0)(45,15){4}{\multiput(0,0)(30,0){2}{\line(2,1){15}}\multiput(0,0)(15,7.5){2}{\line(1,0){30}}\thinlines\multiput(4.286,0)(4.286,0){6}{\line(2,1){15}}}}
\end{picture}}\linespread{1}\caption[Surfaces in two cellular-automaton
lattices]{Surfaces in two cellular-automaton lattices. The laws here are
assumed to be given in their central representations; time moves up. The first
figure shows two cells of a surface with $[h,k]=[3,4]$ for a $d=2$ law; the
second figure shows four cells of a surface with $[h,k]=[1,2]$ for a $d=3$
law.}
\label{surfaces}
\end{figure}

Let $h$ and $k$ be positive and relatively prime. I will define the surface
associated with the line $\{[X+nh,T+nk]\,|\,n\in\Z\}$ to be the set of vertices
$\{[X+nh+i,T+nk+j]\,|\,n\in\Z,\ 0\leq i\leq h+d-2,\ 0\leq j\leq k-1\}$. Each
such surface consists of a series of rectangles (parallelograms in the original
coordinates); some examples are given in Figure~\ref{surfaces}.
I will call this surface the $[X,T,h,k]$ surface. If
$\{[X+nh+i,T+nk+j]\,|\,0\leq i\leq h+d-2,\ 0\leq j\leq k-1\}$ is a rectangle in
the $[X,T,h,k]$ surface, I will call this the $[X,T,h,k,n]$ rectangle. In
addition, the smallest rectangle that contains the $w$ consecutive rectangles
from $[X,T,h,k,n]$ through $[X,T,h,k,n+w-1]$ will be referred to as the
$[X,T,h,k,n,w]$ rectangle.

It is easy to show that any such surface is a Cauchy surface. Using the forward
law, the colors in cells $\{[X+nh+i,T+(n+1)k-1]\,|\,0\leq i\leq h+d-2\}$ force
the colors in cells $\{[X+nh+i,T+(n+1)k]\,|\,0\leq i\leq h-1\}$ for each
$n$. It follows that the state of the [X,T,h,k] surface forces the state of the
$[X,T+1,h,k]$ surface. If one now repeats this process, eventually one forces
the state of all cells in the future of the original surface. The state of all
cells in the past of the surface is forced in exactly the same way, using the
reverse law.

Let each $[X,T,h,k,n]$ rectangle be considered as a cell in the $[X,T,h,k]$
surface. The next task is to show that the surface is a subshift of finite
type. Thus we must find $w\in\N$ and a list of strings of $w$ consecutive cells
such that the only restriction on the allowed arrangements of cells in this
surface is that any string of $w$ consecutive cells must be in this list.

Consider any rectangle in the $[X,T]$ coordinates (by this I mean a set of
vertices $\{[i,j]\,|\,i_0\leq i\leq i_1,\ j_0\leq j\leq j_1\}$). I will say
that this rectangle is \emph{self-consistent} when the following holds: for any
segment containing the $d$ consecutive cells $[i,j]$ through $[i+d-1,j]$ such
that all of these cells are in the rectangle, if $[i,j+1]$ is in the rectangle
then its color must be the one forced by evolving the segment forward in time,
and if $[i+d-1,j-1]$ is in the rectangle then its color must be the one forced
by evolving the segment backward in time.

Our rule will be that a $K$-coloring of the $[X,T,h,k,n]$ rectangle in the
$[X,T,h,k]$ surface constitutes an allowed color of the associated cell if it
is self-consistent. The rule for determining the allowed $w$-neighborhoods of
these cells is similar. Consider $w$ consecutive rectangles $[X,T,h,k,n]$
through $[X,T,h,k,n+w-1]$. If a $K$-coloring of the $[X,T,h,k,n,w]$ rectangle
(which is the smallest rectangle that contains these $w$ rectangles) is
self-consistent, then so are the induced colorings of the $w$ rectangles inside
it; hence these colorings correspond to allowed colors of the associated
cells. The sequences of colorings of $w$ consecutive cells that are obtained in
this way are exactly the allowed $w$-neighborhoods.

Consider the $w+1$ consecutive cells $[X,T,h,k,n]$ through $[X,T,h,k,n+w]$, and
suppose that the colorings of the first $w$ cells and of the last $w$ cells are
allowed. Then the $[X,T,h,k,n,w]$ and $[X,T,h,k,n+1,w]$ rectangles are
self-consistent. I wish to choose $w$ so that the smallest rectangle containing
the $w+1$ consecutive cells, namely the $[X,T,h,k,n,w+1]$ rectangle, is
self-consistent.

\begin{figure}
\fbox{\begin{picture}(425,112)\thicklines
\put(15,15){\setlength{\unitlength}{1.2pt}\begin{picture}(330,70)\multiput(0,0)(0,20){4}{\multiput(0,0)(20,0){17}{\put(0,0){\circle*{3}}\put(10,10){\circle*{3}}}}\multiput(0,0)(80,20){4}{\multiput(0,0)(60,0){2}{\line(3,1){30}}\multiput(0,0)(30,10){2}{\line(1,0){60}}\thinlines\multiput(5,0)(5,0){11}{\line(3,1){30}}}\multiput(0,0)(80,20){2}{\thinlines\put(60,0){\line(1,0){40}}\put(150,50){\line(1,0){40}}\put(30,10){\line(3,1){120}}\put(100,0){\line(3,1){120}}}\multiput(157.5,52.5)(-57.5,-52.5){2}{\multiput(0,0)(7.5,2.5){8}{\qbezier[5](0,0)(10,0)(20,0)}}\end{picture}}
\end{picture}}\linespread{1}\caption[Self-consistency of the
lattice]{Self-consistency of the lattice if it is assumed that a surface is a
subshift of finite type determined by its width-$w$ neighborhoods. In the
example shown, the law is displayed in a $d=4$ central representation, and the
surface shown has $[h,k]=[1,2]$. Note that rectangles in [X,T] coordinates show
up as parallelograms on this lattice. The vertex at the lower left has
coordinates $[X+nh,T+nk]$. The four regions shaded with lines are cells in the
surface. The lines surrounding the parallelograms $[X,T,h,k,n,w]$ and
$[X,T,h,k,n+1,w]$ are drawn for $w=3$. The ``upper'' and ``lower''
parallelograms are shaded with dots.}
\label{selfconsistent}
\end{figure}
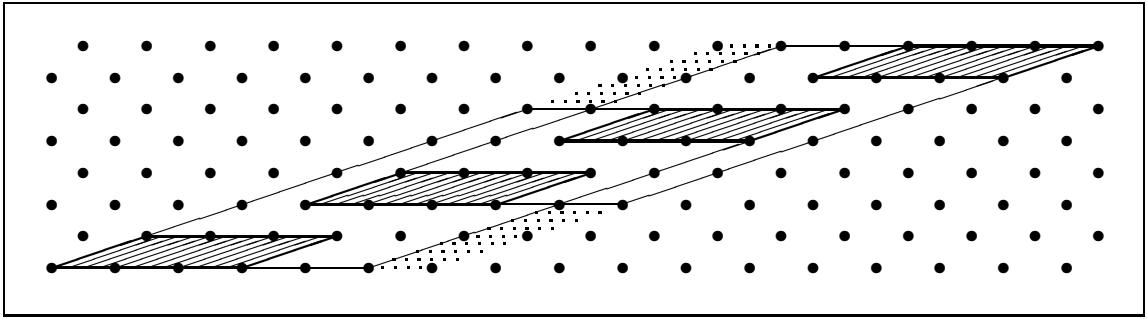

The situation is pictured in Figure~\ref{selfconsistent}.
The colors of the vertices in the $w+1$ cells are given. The remaining cells
form two subrectangles of $[X,T,h,k,n,w+1]$: one is in the bottom right-hand
corner, and the other is in the top left-hand corner. By applying $T^{-1}$, all
colors in the lower subrectangle are determined; by applying $T$, all colors in
the upper subrectangle are determined. It remains to check for
self-consistency. To do this, we must check all applications of $T$ which
involve vertices in the lower subrectangle and all applications of $T^{-1}$
which involve vertices in the upper subrectangle.

The basic method of proving self-consistency is as follows. Consider any
segment of $d$ consecutive vertices that contains at least one vertex in the
lower subrectangle. Let that segment contain the vertices $\{[i,j]\,|\,i_0\leq
i\leq i+d-1\}$. If the rows of $[X,T,h,k,n,w+1]$ are wide enough, then they
will contain $\{[i,j+1]\,|\,i_0-d+1\leq i\leq i+d-1\}$. (All that matters here
is the width, since the rows are guaranteed to go far enough to the right; the
only question is whether they go far enough to the left.) If $w>1$, the colors
in the $j$th row are guaranteed, by construction, to be the ones forced by the
colors in the $(j+1)$st row using $T^{-1}$. In this case, this means that every
color of the length-$d$ segment is forced by vertices above it. But that means
that if $c$ is the color which results from applying $T$ to the length-$d$
segment, then the color of vertex $[i_0,j+1]$ must be $c$ (for otherwise $T$
would not be the inverse of $T^{-1}$). Since there are $h$ vertices in a row of
the lower subrectangle, it turns out that, in order to insure that the proper
vertices are in row $j+1$ of $[X,T,h,k,n,w+1]$ it is sufficient to require that
the rows of $[X,T,h,k,n,w+1]$ contain $h+2d-2$ vertices. A simple calculation
shows that the rows of $[X,T,h,k,n,w+1]$ actually contain $(w+1)h+d-1$
vertices. Thus we may insure consistency for the lower subrectangle by
requiring that $w>1$ and that $(w+1)h+d-1\geq h+2d-2$; the latter relation may
be written more simply as $wh\geq d-1$.

A similar argument applies to the upper subrectangle. The final result is that,
for any $w$ such that $w>1$ and $wh\geq d-1$, $[X,T,h,k,n,w+1]$ must be
self-consistent. This means that the subrectangle $[X+1,T,h,k,w]$, which
contains $w$ consecutive cells in the $[X+1,T,h,k]$ surface, is
self-consistent. Also, the subrectangle $[X,T+1,h,k,w]$, which contains $w$
consecutive cells in the $[X,T+1,h,k]$ surface, is self-consistent. Now suppose
that the state of the surface $[X,T,h,k]$ consists entirely of allowed
sequences of $w$ colors. Then the state of the $[X+1,T,h,k]$ surface forced by
the state of $[X,T,h,k]$ must also consist of allowed strings of $w$ colors;
the same goes for the state of the $[X,T+1,h,k]$ surface. By repeating the
procedure, it follows that the state of every $[X',T',h,k]$ surface in the
spacetime forced by the state of the $[X,T,h,k]$ surface must only contain
allowed strings of $w$ colors. In addition, the entire spacetime forced by
$[X,T,h,k]$ must not contain any inconsistencies, for these have been checked
at each stage. Hence every such state $[X,T,h,k]$ is allowed; so the set of
surfaces associated with the lines $\{[X+nh,T+nk]\,|\,n\in\N\}$ is the subshift
of finite type determined by the list of allowed strings of $w$ cells.

Thus our original system may be redescribed as a new system where the spaces
are the surfaces in the direction of the lines with slope $m$, the set of
allowed spaces is a subshift of finite type, and any such space determines the
rest of spacetime via a new invertible law $T'$. It is easy to verify that $T'$
must be local. So the new description is a reversible cellular
automaton.\proofend\bigskip

Let $S^1_q$ be the subspace of the circle $S^1$ consisting of all points with
rational coordinates. If we associate the set of allowed slopes with $\S^1_q$
in the natural way, then this theorem implies that the set $\M_T$ of slopes
which correspond to allowed rotations is an open subset of $\S^1_q$. However,
the theorem does not tell us how to find $\M_T$. Given our original
description, we may find part of $\M_T$. We might then check the diagonal to
see whether a rotation in which the diagonals are the spaces is possible. If
so, we may reapply the above theorem to expand our knowledge of which slopes
are in $\M_T$. But it is not clear how one may verify that an open region of
$\S^1_q$ is \emph{not} contained in $\M_T$.

In my investigations of various diagonals, I have found that some are Cauchy
surfaces (for instance, if $T$ is the identity law on any set of spaces \X\
then every straight line corresponds to a Cauchy surface except for the
vertical). Others are causal in one direction only (for instance, in the
addition-mod-2 rule the diagonals force the past but not the future). Still
others are not causal in either direction. In every case, I have found
that there is a set of surfaces associated with the diagonal which is a
subshift of finite type. But I have not investigated all that many
diagonals, and do not have any general results on this.

We have seen that, given law $T$, there is a map $f_T:\M_T\mapsto\A$, where \A\
is the set of equivalence classes of subshifts of finite type discussed in
Section~\ref{equiv}. It would be interesting to learn more about these
maps. (Actually, the objects one would wish to study here are the equivalence
classes of such maps that are generated by rotations.) What is the set of all
possible maps of this sort? Given any two elements of \A, does there exist an
$f_T$ with both of those elements in its range? The maps describe a local
property of a point in spacetime. To what extent does this property
characterize the spacetime?

It may be that these questions can only be properly investigated once the
approach described in this chapter has been generalized. How does one define
the set of Cauchy surfaces in a general 1+1-dimensional combinatorial
spacetime? I believe that the answer to this question is closely related to the
problem of finding an appropriate definition of \emph{geodesic} in the
combinatorial-spacetime setting. I have not solved this, but do not believe
that it is a difficult problem. (The answer, I think, is related to the
subjects discussed in Chapter~\ref{invar}.) Presumably the straight-line
surfaces in cellular automata correspond to Cauchy surfaces; in general, one
would hope that each geodesic could be investigated as to whether a surface
lying along the geodesic is Cauchy, whether such a surface is a subshift of
finite type, and so on. Clearly there is much work to be done in this area.

The rotations which I have described here are reminiscent of Lorentz
transformations in Minkowski spacetime. There is a natural analogy between
Minkowski spacetime (vis-a-vis other spacetimes in general relativity) and
reversible cellular automata (vis-a-vis other combinatorial spacetimes). Both
of these are \emph{flat} spacetimes.\footnote{\linespread{1}\footnotesize Just
as in general relativity one considers any metric to represent a flat spacetime
if there is a coordinate transformation sending it to a Minkowski metric, so
here it is natural to consider any combinatorial spacetime to be flat if it may
be transformed via a local, invertible map to a cellular automaton.} In both
cases one is concerned with straight-line surfaces, and with rotations
corresponding to those surfaces. In both cases there is an open set of allowed
rotations. And, of course, in both cases locality plays an important role,
which no doubt has something to do with why in both cases similar light-cone
structures are found. However, Minkowski spacetime has one property which
cellular automata typically do not have. That property is \emph{relativity}: if
you perform a rotation in Minkowski spacetime, the laws of nature do not
change. That is, the description of the law of nature in terms of the new set
of spaces is identical to the description of that law in terms of the original
set of spaces. This leads to a natural question: does there exist a reversible
1+1-dimensional cellular automaton which also has this property?

It would be very nice to find such an automaton. I have made some attempts,
without success. In the case of reversible cellular automata the first
prerequisite for law $T:\X\mapsto \X$ to be equivalent to $T':\X'\mapsto\X'$ to
be equivalent is that \X and $\X'$ be equivalent as sets of spaces. Thus the
problem of finding a relativistic reversible 1+1-dimensional cellular automaton
is related to the question of whether there exists a $T$ such that the function
$f_T:\M_T\mapsto\A$ is constant. In all the examples I have studied, these
functions have been far from constant. However, the problem of obtaining
general results in this case seems to be a difficult one.

If such an automaton were found, one might even imagine that length contraction
and time dilation could be a natural consequence. Perhaps so, but at this point
such a thing would be difficult to detect, since it is not at all clear how an
observer would measure time or distance in cellular automata. For example:
there are many equivalent descriptions of the same cellular automaton. The
number of vertices contained in an observer can vary wildly in such
transformations (as can the number of vertices in the objects surrounding the
observer). What is the natural length scale that the observer would experience?
I suspect there is one, but do not know how to make this precise at present.

An additional puzzle is the following. One property of transformations in
cellular automata which is not present in transformations of Minkowski
spacetime is that the size of cells in the cellular automaton transformation
(at least in the particular transformation given here) depends on the
complexity of the slope as a rational number. If the slope is $\frac{50}{99}$,
the transformed cell size will be way bigger than it would have been if the
slope were $\frac{50}{100}$. Perhaps this difference may be transformed away
somehow; certainly the $UTU^{-1}$-type transformations do not respect cell
size. Nevertheless, it is natural to wonder whether this property corresponds
to an intrinsic feature of cellular automata which is very different from
Minkowski spacetime.

In summary, I am quite in the dark as to whether a relativistic
1+1-dimensional reversible cellular automaton can exist. I suspect that
advances in a number of areas will be needed before this question can be
satisfactorily answered.\footnote{\linespread{1}\footnotesize Recently I
  discovered a paper by Milnor~(\cite{milnor}) which addresses some of the
  issues described above. His investigation of information density in
  cellular automata seems relevant to understanding the question of whether
  a cellular automaton can be relativistic. Also, it appears to me now that
  the main theorem by Williams~(\cite{williams}) on topological conjugacy of
  subshifts of finite type could be profitably used to aid the analysis
  attempted here. As the dissertation must be handed in momentarily, I
  cannot comment further at this time. \emph{Note added May 1998:} There are
  further problems in interpreting these systems as relativistic which I do
  not have time to go into in detail. Briefly, the local equivalence maps I
  have considered preserve information content. Thus information is an
  intrinsic property of these systems, and hence is related to the metric
  (if there is one). For a reversible 1+1-dimensional cellular automaton to
  be relativistic, this implies that null lines should contain no
  information, which is provably false except in very trivial cases. It is
  possible that some of the difficulty arises from the fact that my
  spacetimes can be foliated into Cauchy surfaces in such a way that the set
  of these surfaces (for a particular foliation, e.g. the set of all time
  slices in my space+time formulation) is a subshift of finite type. 't
  Hooft~(\cite{hooft2}) has described a type of inherently two-dimensional
  (rather than 1+1-dimensional) system which I believe typically cannot be
  foliated in this way. (The set of \emph{all} Cauchy surfaces in 't Hooft's
  systems seems to be a subshift of finite type; whether this is true of my
  systems as well is not clear.) This may be important because it allows,
  for example, a particle to be emitted, bounce off something and return to
  the emitting body in such a manner that the body has evolved considerably
  in the interim, while the particle itself during that period consists
  simply of two graphical edges. This is a relativistic sort of thing, and
  it seems impossible in my systems due to the foliation. My guess is that
  the set of two-dimensional combinatorial spacetimes is much bigger than
  the one I have studied; there may also be transformations I have missed,
  having to do with causal structure, which would make all of these systems
  transparently relativistic. However, at this point I have no answers.}

\newpage\vspace*{40pt}\section[Invariants and particles]{\LARGE Invariants and
particles}\bigskip
\label{invar}

By far the most useful method I have found for analyzing the behavior of
combinatorial spacetime laws of evolution is to investigate their constants of
motion. I will call these constants of motion \emph{invariants}. Typically the
presence of an invariant implies the existence of a local flow, which in turn
(due to the combinatorial nature of the system) may often be interpreted in
terms of particles.

Again I will focus on the 1+1-dimensional case. Let \X\ be a set of
one-dimensional spaces, let $K$ be the set of colors in \X, let $\X'$ be the
set of finite spaces in \X, and let \Seg{\X} be the set of (finite) segments
that occur in elements of \X. Let $\#(S,x)$ denote the number of occurrences of
a segment $S$ in a space $x$. Let $G$ be an abelian group, and consider a
function $f:\Seg{\X} \mapsto G$ where $f(S)\neq 0$ for only finitely many $S\in
\Seg{\X}$. The function $f$ induces a function $F:\X' \mapsto G$ as follows: if
$x \in \X'$ then $F(x)=\sum_{S\in\Seg{\X}}\#(S,x)f(S)$. Since $f$ is nonzero
for only finitely many $S$, only finite sums are involved; hence $F$ is well
defined. If $T:\X\mapsto\X$ is a local invertible map and $F(T(x))=F(x)$ for
all $x\in\X'$, I will say that $F$ is a $G$-invariant of $T$.

The set of functions $\F_G(T)$ that induce $G$-invariants of $T$ form a group
under addition. The zero of this group is the function $z_G^{\X}$ which assigns
zero to every segment in \Seg{\X}. The invariant induced by this function is a
function $Z_G^{\X}$ which assigns zero to every space in $\X'$. The function
$Z_G^{\X}$ will be called the \emph{zero G-invariant} of \X. It is a
$G$-invariant of \emph{every} local invertible map $T:\X\mapsto\X$. Though
$Z_G^{\X}$ itself is trivial, the structure of the set $\zf_G(\X)$ of
functions that induce $Z_G^{\X}$ typically is not trivial, and this is what
makes the zero $G$-invariant of \X\ important.

For example, if $\X=\x{2}$ and $f:\Seg{\X}\mapsto\Z$ sends \seg{aa} to 1,
\seg{ab} to 1, \seg{a} to $-1$ and all other segments to zero, then $f$ induces
the zero \Z-invariant of \X. This is true because every segment \seg{a} in any
space $x \in \x{2}$ is the leftmost cell in a two-cell segment whose rightmost
cell is colored $a$ or $b$. Similarly, if $g:\Seg{\X}\mapsto\Z$ sends \seg{aa}
to 1, \seg{ba} to 1, \seg{a} to $-1$ and all other segments to zero, then $g$
induces the zero \Z-invariant of \X.

The set $\zf_G(\X)$ is a group under addition. (Thus, in the example described
above, the function $f-g$ also induces the zero \Z-invariant on \x{2}. This is
the function that sends \seg{ab} to 1, \seg{ba} to $-1$, and all other segments
to zero.) For any $T$, $\zf_G(\X)$ is a subgroup of $\F_G(T)$. If $f \in
\F_G(T)$ and $f$ induces the $G$-invariant $F$, then so does $f+g$ for any $g
\in \zf_G(\X)$. The set $\I_G(T)$ of $G$-invariants of $T$ is itself a group:
it is simply the quotient group $\F_G(T)/\zf_G(\X)$.

If $f$ is in $\F_G(T)$, let $w$ equal the maximum width of the (finitely many)
segments $S$ in \Seg{\X} such that $f(S)\neq 0$. (I will call $w$ the
\emph{width} of $f$.) Suppose there is a segment $S_0 \in \Seg{\X}$ whose width
is less than $w$ such that $f(S_0)=g\neq 0$. Consider the function
$h:\Seg{\X}\mapsto G$ which maps $[S_0c]$ to $g$ for each $c\in K$ that is
allowed to follow $S_0$, maps $S_0$ to $-g$, and maps all other segments to
zero. Then $h$ is in $\zf_G(\X)$, so $f$ and $f+h$ induce the same
$G$-invariant $F$ of $T$. I will say that $f+h$ is obtained from $f$ by
``extending $S_0$ to the right.'' By repeating this process of extending
segments to the right, one can construct a function $f'$ which also induces $F$
such that every segment $S$ with $f'(S)\neq 0$ has width $w$.

Let $F$ be a $G$-invariant of $T$, and let $w$ be the minimum width of all
functions that induce $F$. I will call $F$ a \emph{width-w} $G$-invariant of
$T$. By the above argument, there exists an $f$ inducing $F$ which is zero on
all segments that do not have width $w$. Thus, in order to find all
$G$-invariants of $T$ it suffices to examine, for each $w>0$, the set of
functions $f:\Seg{\X}\mapsto G$ such that $f(S)=0$ if the width of $S$ is not
$w$.

Given $\X=\x{A}$, it is easy to list some functions $f\in\zf_{\Z}(\X)$ such
that $f(S)=0$ if $|S|\neq w$. Let $U$ be any allowed segment in \X\ of width
$w-1$. Let $g_U$ be the function such that $g_U(U)=-1$, $g(\seg{Uc})=1$ for
each $c$ that may follow $U$, and $g_U(S)=0$ for all other segments $S$. Let
$h_U$ be the function such that $h_U(U)=-1$, $h(\seg{dU})=1$ for all $d$ that
may precede $U$, and $h_U(S)=0$ for all other $S$. Then $g_U$ and $h_U$ are in
$\zf_{\Z}(\X)$. It follows that$f_U\equiv g_U-h_U$ is in $\zf_{\Z}(\X)$; and
$f_U(S)=0$ whenever $|S|\neq w$.

In fact, the above functions $f_U$ generate the space of all $f\in\zf_{\Z}(\X)$
having the property that $f(S)=0$ if $|S|\neq w$. To see this, consider a
finite collection of width-$w$ segments. Suppose that we wish to assemble them
to form a space, or to form several spaces, in such a way that no segments are
left over. Then the number of segments of the form \seg{d_jU} must equal the
number of segments of the form \seg{Uc_i}, so that these may be glued to one
another to form segments of the form \seg{d_jUc_i}. This is exactly the
condition specified by the relation $f_U$. Furthermore, if these conditions are
met then one may glue the segments together to form spaces without any segment
being left over. So the segments cannot satisfy any other independent linear
relations.

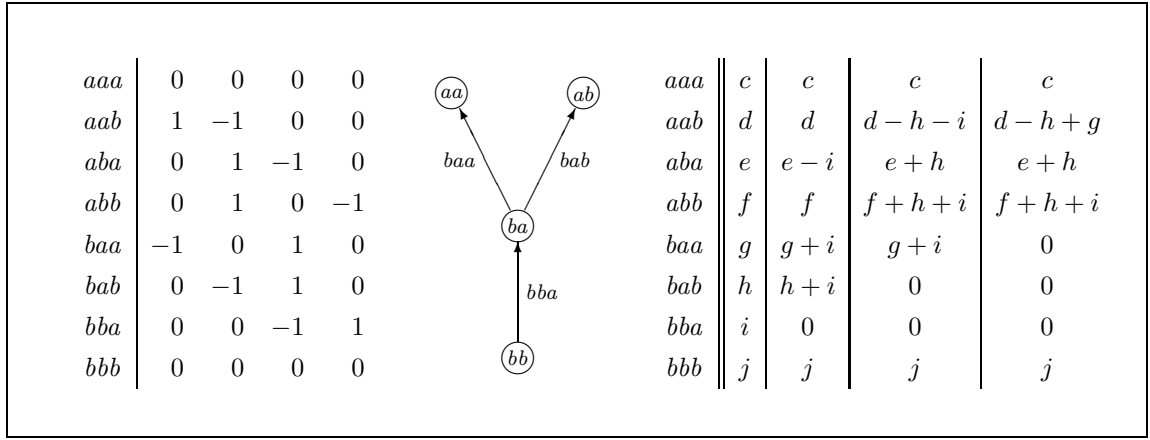
\begin{figure}
\fbox{\begin{picture}(425,158)\footnotesize
\put(20,20){$\begin{array}[b]{c|rrrr}
\mathit{aaa}& 0& 0& 0& 0\\
\mathit{aab}& 1&-1& 0& 0\\
\mathit{aba}& 0& 1&-1& 0\\
\mathit{abb}& 0& 1& 0&-1\\
\mathit{baa}&-1& 0& 1& 0\\
\mathit{bab}& 0&-1& 1& 0\\
\mathit{bba}& 0& 0&-1& 1\\
\mathit{bbb}& 0& 0& 0& 0
\end{array}$}
\put(190,27){\scriptsize
\put(0,0){\makebox(0,0){$\mathit{bb}$}\put(0,0){\circle{12}}}
\put(0,50){\makebox(0,0){$\mathit{ba}$}\put(0,0){\circle{12}}}
\put(-25,100){\makebox(0,0){$\mathit{aa}$}\put(0,0){\circle{12}}}
\put(25,100){\makebox(0,0){$\mathit{ab}$}\put(0,0){\circle{12}}}
\put(0,6){\vector(0,1){38}}\put(3,25){\makebox(0,0)[l]{$\mathit{bba}$}}
\put(-2.68,55.77){\vector(-1,2){19.23}}
\put(2.68,55.77){\vector(1,2){19.23}}
\put(-15.5,75){\makebox(0,0)[r]{$\mathit{baa}$}}
\put(15.5,75){\makebox(0,0)[l]{$\mathit{bab}$}}
}
\put(240,20){$\begin{array}[b]{c||c|c|c|c}
\mathit{aaa}&c&c  &c    &c\\
\mathit{aab}&d&d  &d-h-i&d-h+g\\
\mathit{aba}&e&e-i&e+h  &e+h\\
\mathit{abb}&f&f  &f+h+i&f+h+i\\
\mathit{baa}&g&g+i&g+i  &0\\
\mathit{bab}&h&h+i&0    &0\\
\mathit{bba}&i&0  &0    &0\\
\mathit{bbb}&j&j  &j    &j
\end{array}$}
\end{picture}}\linespread{1}\caption{Zero $w=3$ invariants for \x{2}. At left
the functions $f_{\seg{aa}}$, $f_{\seg{ab}}$, $f_{\seg{ba}}$ and $f_{\seg{bb}}$
are displayed. These generate the space of zero $w=3$ invariants. Note that the
sum of these functions is zero; they are dependent. Their span has dimension
one less than the number of functions. At center is the spanning tree of \G. At
right \G\ and the functions $f_U$ are used to reduce an arbitrary function $f$
to canonical form.}
\label{zeroinvars}
\end{figure}

Consider the directed graph \G\ whose vertices are associated with the
width-\mbox{$(w-1)$} segments and whose edges are associated with the width-$w$
segments; an edge associated with $S$ points from the vertex associated with
$U$ to the vertex associated with $V$ if and only if $U$ is embedded in $S$ at
the beginning of $S$ and $V$ is embedded in $S$ at the end of $S$. Suppose that
\G\ is strongly connected. Then we may construct a directed spanning tree of
this graph, with the edges in the tree pointing away from the root. Let
$f:\Seg{\X}\mapsto G$ have the property that $f(S)=0$ if $|S|\neq w$, and let
$F$ be the function induced by $f$. We may use \G\ to put $f$ in a canonical
form, as follows. Choose a vertex that is distance 1 from the root. This vertex
is associated with a width-$(w-1)$ segment $U$; the edge pointing towards it is
associated with a segment \seg{dU}. Let $f'=f+f_Uf(\seg{dU})$, and then let
$f=f'$. Now $f(\seg{dU})=0$. Now repeat the process, each time choosing a new
vertex that is not the root but is as close to the root as possible. One may
easily verify that, once $f(\seg{dU})$ is set to zero during this process, its
value is left unchanged thereafter. The end result is that $f(S)=0$ for each
$S$ associated with an edge in \G. An example is illustrated in
Figure~\ref{zeroinvars}.

Suppose $f$ is in $\F_G(T)$ and we wish to know if it is in $\zf_G(\X)$. We can
find out in the following way. First, by extending to the left and right,
express $f$ so that $f(S)=0$ whenever $|S|\neq w$ for some $w$. Second,
construct \G and a spanning tree of \G\ (as above), construct the set of
generators of the zero width-$w$ invariants, and use these objects to reduce
$f$ to canonical form. Then $f\in\zf_G(\X)$ if and only if $f(S)=0$ for all
$S$.

Let $T$ be given by \ang{L,R,G^L,N,A} where each segment $L_i$ has width
$d$. Recall that we can associated with the $j$th length-1 segment embedded in
$R_i$ a set of $k$ forcing strings $R_{ijk}$ using an algorithm described in
Section~\ref{revalg}. If $w\geq d$ then we can associate with each length-$w$
segment $S$ a set of forcing strings $S_i$ as follows. Suppose
$S=[L_{i_1}L_{i_2}\ldots L_{i_k}]$, where here the segments are glued together
with $(d-1)$-cell overlap. Then the forcing strings associated with $S$ are
obtained by concatenating the forcing strings associated with each length-1
segment embedded in $[R_{i_1}R_{i_2}\ldots R_{i_k}]$ (where here the segments
are glued together with no overlap). The concatenation is done in the obvious
way: the strings for the leftmost cell are overlapped in all possible ways with
the strings of the next cell to the right so that the two marked cells appear
next to one another in the proper order and so long as overlapping cells match
one another; then the resulting segments are overlapped with those associated
with the next cell to the right, and so on.

For example, suppose we are considering the law whose forcing strings are
described in Figure~\ref{force}. If we are given the segment
$S=\seg{baa}=[L_4L_1]$, then the image $[R_4R_1]$ is the segment
$[R'_5R'_1]$. To find the associated forcing strings, we attempt to concatenate
the nine forcing strings associated with $R'_5$ with the three forcing strings
associated with $R'_1$. Start with the strings associated with $R'_5$. The
strings are to be concatenated by overlapping them so that the underlined cell
in the string associated with $R'_5$ is directly to the left of the underlined
cell in the string associated with $R'_1$. The first string associated with
$R'_5$ is \seg{a\underline{a}b}. Since $b$ directly follows the underlined cell
in this string, it cannot be concatenated with any of the three strings
associated with $R'_1$, whose underlined cells are $c$ in all cases. On the
other hand, \seg{a\underline{a}cb} can be concatenated with
\seg{\underline{c}b} to produce \seg{a\underline{ac}b}; \seg{a\underline{a}cc}
can be concatenated with \seg{\underline{c}cb} and \seg{\underline{c}cc} to
produce \seg{a\underline{ac}cb} and \seg{a\underline{ac}cc}, and so on. The
result is that there are nine forcing strings associated with $S$.

Let $f$ be a function that induces a $G$-invariant $F$ of $T$. Suppose that
$f(S)=0$ if $|S|\neq w$. For each width-$w$ segment $S\in\Seg{\X}$, let
$f'_S(U)=1$ for each forcing string $U$ associated with $S$, and let
$f'_S(U)=0$ otherwise. Let $f'$ be the sum of the functions $f'_S$. Let $F$ be
the function on $\X'$ induced by $f$, and let $F'$ be the function on $\X'$
induced by $f'$. Then it is easy to verify that $F(x)=F'(T(x))$ for each
$x\in\X$. The reason for this is basically that for each width-$w$ segment $S$
in $x$ there is associated exactly one forcing string $U$ in $T(x)$. (Actually
the same $U$ in $T(x)$ may be associated with several width-$w$ segments in
$x$. But if so, these width-$w$ segments must be adjacent to one another, and
the presence of any one of them must force the rest of them to be present. In
that case the sum of $f$ over those width-$w$ segments is exactly the value of
$f'(U)$.)

The important fact to notice here is the following. If $F$ is a $G$-invariant
of $T$, then it must be the case that $F=F'$. Hence the function $f'-f$ must be
in $\zf_G(\X)$.

We may now characterize all width-$w$ invariants of $T$ as follows. For each
allowed width-$w$ segment $S$ assign a commutative generator $h_S$ such that
$f(S)=h_S$, and let $f(S)=0$ for all other segments. Compute the forcing
strings, and compute $f'-f$. Extend to the left and right until $f'-f$ is
nonzero only on segments of width $y$ for some fixed $y$. Then put $f'-f$ in
canonical form. Since $f'-f\in\zf_G(\X)$, all of the entries of the canonical
form of $f'-f$ must be zero. This gives a set of relations on the $h_S$'s. The
resulting set of generators and relations describes an abelian group $H$. Let
$G$ be an abelian group and $h:H\mapsto G$ be any homomorphism. Let
$g:\Seg{\X}\mapsto G$ be given by $g(S)=h(f(S))$ for each segment $S$. Then $g$
induces a width-$w$ $G$-invariant of $T$, and all width-$w$ invariants of $T$
may be obtained in this way.

Though this procedure is straightforward, it does tend to produce many
redundant relations. I suspect that another procedure can be found which
does not do this.

Invariants of the sort described here were first introduced in the context of
cellular automata by Pomeau~\cite{pomeau}, and more recently were discussed in
a series of papers by Takesue~\cite{takesue1,takesue2,takesue3} and in a paper
by Hattori and Takesue~\cite{hattori/takesue}. The latter paper includes a
procedure for computing all width-$w$ \R-invariants for one-dimensional
cellular automaton rules defined on spaces \x{k}. The above procedure is
somewhat different from theirs: it is specially adapted to invertible rules,
and it is valid for general 1+1-dimensional combinatorial spacetimes. (I have
an algorithm to compute the width-$w$ invariants for any rule given by a
nonoverlap rule table, even if it is not invertible; however, I will not
describe this here.)

Figure~\ref{invarfig}
\begin{figure}
\fbox{\begin{picture}(425,290)\footnotesize
\put(0,0){\makebox(425,290){$\begin{array}{c|c||c|c|c|ccccc|c}
\mathit{aa}&a&w+r    &(w-k      ,k+r    )&1&00&00&00&00&00&(1,0)\\
\mathit{ab}&d&w+y    &(w-z-k    ,k+y+z  )&3&01&00&10&01&01&(1,-1)\\
\mathit{ac}&a&w+r+y+u&(w-k      ,k+y+u+r)&2&00&00&00&00&01&(1,0)\\
\mathit{ad}&d&w+r-z  &(w-z-k    ,k+r    )&2&00&00&10&00&00&(1,0)\\
\mathit{ba}&a&w+r-y  &(w-y-k    ,k+r    )&1&00&10&00&00&00&(1,0)\\
\mathit{bb}&d&w      &(w-y-z-k  ,k+y+z  )&3&01&10&10&01&01&(1,-1)\\
\mathit{bc}&a&w+r+u  &(w-y-k    ,k+y+u+r)&2&00&10&00&00&01&(1,0)\\
\mathit{bd}&d&w+r-z-y&(w-y-z-k  ,k+r    )&2&00&10&10&00&00&(1,0)\\
\mathit{ca}&c&w+r+z+s&(w-k      ,k+s+r+z)&1&00&00&00&01&00&(1,0)\\
\mathit{cb}&b&w-u    &(w-u-r-k  ,k+r    )&2&10&00&10&10&00&(0,0)\\
\mathit{cc}&b&w      &(w-u-r-k  ,k+u+r  )&3&10&01&10&10&01&(0,0)\\
\mathit{cd}&c&w+r+s  &(w-k      ,k+s+r  )&2&00&00&01&01&00&(1,0)\\
\mathit{da}&c&w+z    &(w-s-r-k  ,k+s+r+z)&2&10&10&00&01&10&(0,0)\\
\mathit{db}&b&w-u-s  &(w-u-s-r-k,k+r    )&1&10&00&00&00&00&(0,0)\\
\mathit{dc}&b&w-s    &(w-u-s-r-k,k+u+r  )&2&10&01&00&00&01&(0,0)\\
\mathit{dd}&c&w      &(w-s-r-k  ,k+s+r  )&3&10&10&01&01&10&(0,0)\\
\end{array}$}}
\end{picture}}\linespread{1}\caption{Invariants and particles for a reversible
cellular automaton on \x{4}. The two columns at left give the automaton rule
table. The next column lists the functions that induce $w=2$ invariants. The
column entries are the group elements assigned by the function to the segment
in the leftmost column. The space of functions is spanned by the six functions
obtained by setting any five of the six variables to zero. The next column
describes the local flow for these functions. Here there are seven
variables. The leftmost entry in an ordered pair is the quantity sent to the
left; the rightmost entry is the quantity sent to the right. The next column is
the period invariant. The next five columns are the five solitons. Here
``$ij$'' means that $i$ is sent to the left and $j$ is sent to the right
($i,j\in\{0,1\}$). The last column is a local flow for the rotation invariant.}
\label{invarfig}
\end{figure}
shows an example of the results of a width-2 invariants computation for a
cellular automaton rule $T$. The rule is given in the first two columns; the
invariants are given in the third column. The six variables $w$, $r$, $y$, $z$,
$s$ and $u$ are free variables. Suppose that we are interested in
\Z-invariants. Then these invariants are spanned by those that are obtained by
setting one of the six variables to 1 and the rest to 0. Let us call the
associated functions for these six invariants $f_w$, $f_r$, $f_y$, $f_z$,
$f_s$ and $f_u$.

Consider, for instance, the function $f_w$ obtained by setting $w=1$ and
$r=y=z=s=u=0$. Then $f_w(S)=1$ for each width-2 segment $S$. So the induced
invariant $F_w(x)$ is the number of width-2 segments in $x$, which is just
the width of $x$. Hence the width is an invariant of $T$. (Of course this
must be true, since $T$ is a cellular automaton.)

Now set $w=3$, $s=u=1$, $z=-1$, $r=-2$ and $y=0$. The resulting function $f$
is given in the fifth column of the figure. It turns out that in this case
$F(x)$ is the period of the orbit of $x$ under $T$. In other words,
$T^{F(x)}(x)=x$. Note that the average value of $f(S)$ is 2. Thus the period
of the orbit of $x$ under $T$ is approximately two times the width of $x$.
That is: for random $x$ of a fixed width $w$, the periods cluster around
$2w$.

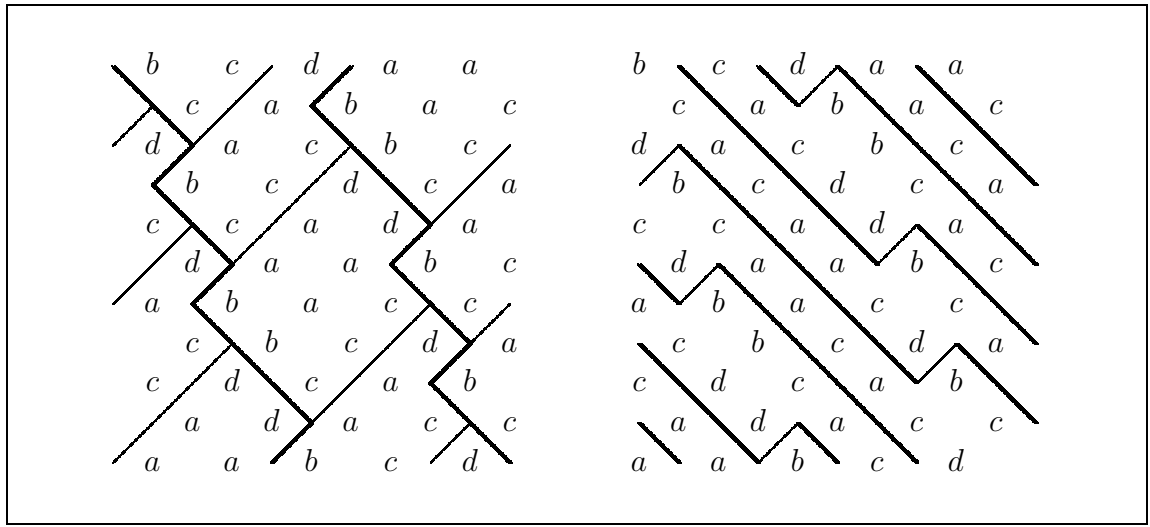
\begin{figure}
\fbox{\begin{picture}(425,190)
\put(52,20){
\put(0,-3){
\put(0,0){\makebox(0,0)[b]{$a$}}
\put(30,0){\makebox(0,0)[b]{$a$}}
\put(60,0){\makebox(0,0)[b]{$b$}}
\put(90,0){\makebox(0,0)[b]{$c$}}
\put(120,0){\makebox(0,0)[b]{$d$}}

\put(15,15){\makebox(0,0)[b]{$a$}}
\put(45,15){\makebox(0,0)[b]{$d$}}
\put(75,15){\makebox(0,0)[b]{$a$}}
\put(105,15){\makebox(0,0)[b]{$c$}}
\put(135,15){\makebox(0,0)[b]{$c$}}

\put(0,30){\makebox(0,0)[b]{$c$}}
\put(30,30){\makebox(0,0)[b]{$d$}}
\put(60,30){\makebox(0,0)[b]{$c$}}
\put(90,30){\makebox(0,0)[b]{$a$}}
\put(120,30){\makebox(0,0)[b]{$b$}}

\put(15,45){\makebox(0,0)[b]{$c$}}
\put(45,45){\makebox(0,0)[b]{$b$}}
\put(75,45){\makebox(0,0)[b]{$c$}}
\put(105,45){\makebox(0,0)[b]{$d$}}
\put(135,45){\makebox(0,0)[b]{$a$}}

\put(0,60){\makebox(0,0)[b]{$a$}}
\put(30,60){\makebox(0,0)[b]{$b$}}
\put(60,60){\makebox(0,0)[b]{$a$}}
\put(90,60){\makebox(0,0)[b]{$c$}}
\put(120,60){\makebox(0,0)[b]{$c$}}

\put(15,75){\makebox(0,0)[b]{$d$}}
\put(45,75){\makebox(0,0)[b]{$a$}}
\put(75,75){\makebox(0,0)[b]{$a$}}
\put(105,75){\makebox(0,0)[b]{$b$}}
\put(135,75){\makebox(0,0)[b]{$c$}}

\put(0,90){\makebox(0,0)[b]{$c$}}
\put(30,90){\makebox(0,0)[b]{$c$}}
\put(60,90){\makebox(0,0)[b]{$a$}}
\put(90,90){\makebox(0,0)[b]{$d$}}
\put(120,90){\makebox(0,0)[b]{$a$}}

\put(15,105){\makebox(0,0)[b]{$b$}}
\put(45,105){\makebox(0,0)[b]{$c$}}
\put(75,105){\makebox(0,0)[b]{$d$}}
\put(105,105){\makebox(0,0)[b]{$c$}}
\put(135,105){\makebox(0,0)[b]{$a$}}

\put(0,120){\makebox(0,0)[b]{$d$}}
\put(30,120){\makebox(0,0)[b]{$a$}}
\put(60,120){\makebox(0,0)[b]{$c$}}
\put(90,120){\makebox(0,0)[b]{$b$}}
\put(120,120){\makebox(0,0)[b]{$c$}}

\put(15,135){\makebox(0,0)[b]{$c$}}
\put(45,135){\makebox(0,0)[b]{$a$}}
\put(75,135){\makebox(0,0)[b]{$b$}}
\put(105,135){\makebox(0,0)[b]{$a$}}
\put(135,135){\makebox(0,0)[b]{$c$}}

\put(0,150){\makebox(0,0)[b]{$b$}}
\put(30,150){\makebox(0,0)[b]{$c$}}
\put(60,150){\makebox(0,0)[b]{$d$}}
\put(90,150){\makebox(0,0)[b]{$a$}}
\put(120,150){\makebox(0,0)[b]{$a$}}
}
\multiput(-.5,0)(1,0){2}{
\qbezier(45,0)(52.5,7.5)(60,15)
\qbezier(105,30)(112.5,37.5)(120,45)
\qbezier(90,75)(97.5,82.5)(105,90)
\qbezier(60,135)(67.5,142.5)(75,150)
\qbezier(15,60)(22.5,67.5)(30,75)
\qbezier(0,105)(7.5,112.5)(15,120)
\qbezier(135,0)(120,15)(105,30)
\qbezier(60,15)(37.5,37.5)(15,60)
\qbezier(120,45)(105,60)(90,75)
\qbezier(30,75)(15,90)(0,105)
\qbezier(105,90)(82.5,112.5)(60,135)
\qbezier(15,120)(0,135)(-15,150)}

\qbezier(60,15)(82.5,37.5)(105,60)
\qbezier(105,90)(120,105)(135,120)
\qbezier(-15,120)(-7.5,127.5)(0,135)
\qbezier(105,0)(112.5,7.5)(120,15)
\qbezier(120,45)(127.5,52.5)(135,60)
\qbezier(-15,60)(0,75)(15,90)
\qbezier(15,120)(30,135)(45,150)
\qbezier(-15,0)(7.5,22.5)(30,45)
\qbezier(30,75)(52.5,97.5)(75,120)

}
\put(236,20){
\put(0,-3){
\put(0,0){\makebox(0,0)[b]{$a$}}
\put(30,0){\makebox(0,0)[b]{$a$}}
\put(60,0){\makebox(0,0)[b]{$b$}}
\put(90,0){\makebox(0,0)[b]{$c$}}
\put(120,0){\makebox(0,0)[b]{$d$}}

\put(15,15){\makebox(0,0)[b]{$a$}}
\put(45,15){\makebox(0,0)[b]{$d$}}
\put(75,15){\makebox(0,0)[b]{$a$}}
\put(105,15){\makebox(0,0)[b]{$c$}}
\put(135,15){\makebox(0,0)[b]{$c$}}

\put(0,30){\makebox(0,0)[b]{$c$}}
\put(30,30){\makebox(0,0)[b]{$d$}}
\put(60,30){\makebox(0,0)[b]{$c$}}
\put(90,30){\makebox(0,0)[b]{$a$}}
\put(120,30){\makebox(0,0)[b]{$b$}}

\put(15,45){\makebox(0,0)[b]{$c$}}
\put(45,45){\makebox(0,0)[b]{$b$}}
\put(75,45){\makebox(0,0)[b]{$c$}}
\put(105,45){\makebox(0,0)[b]{$d$}}
\put(135,45){\makebox(0,0)[b]{$a$}}

\put(0,60){\makebox(0,0)[b]{$a$}}
\put(30,60){\makebox(0,0)[b]{$b$}}
\put(60,60){\makebox(0,0)[b]{$a$}}
\put(90,60){\makebox(0,0)[b]{$c$}}
\put(120,60){\makebox(0,0)[b]{$c$}}

\put(15,75){\makebox(0,0)[b]{$d$}}
\put(45,75){\makebox(0,0)[b]{$a$}}
\put(75,75){\makebox(0,0)[b]{$a$}}
\put(105,75){\makebox(0,0)[b]{$b$}}
\put(135,75){\makebox(0,0)[b]{$c$}}

\put(0,90){\makebox(0,0)[b]{$c$}}
\put(30,90){\makebox(0,0)[b]{$c$}}
\put(60,90){\makebox(0,0)[b]{$a$}}
\put(90,90){\makebox(0,0)[b]{$d$}}
\put(120,90){\makebox(0,0)[b]{$a$}}

\put(15,105){\makebox(0,0)[b]{$b$}}
\put(45,105){\makebox(0,0)[b]{$c$}}
\put(75,105){\makebox(0,0)[b]{$d$}}
\put(105,105){\makebox(0,0)[b]{$c$}}
\put(135,105){\makebox(0,0)[b]{$a$}}

\put(0,120){\makebox(0,0)[b]{$d$}}
\put(30,120){\makebox(0,0)[b]{$a$}}
\put(60,120){\makebox(0,0)[b]{$c$}}
\put(90,120){\makebox(0,0)[b]{$b$}}
\put(120,120){\makebox(0,0)[b]{$c$}}

\put(15,135){\makebox(0,0)[b]{$c$}}
\put(45,135){\makebox(0,0)[b]{$a$}}
\put(75,135){\makebox(0,0)[b]{$b$}}
\put(105,135){\makebox(0,0)[b]{$a$}}
\put(135,135){\makebox(0,0)[b]{$c$}}

\put(0,150){\makebox(0,0)[b]{$b$}}
\put(30,150){\makebox(0,0)[b]{$c$}}
\put(60,150){\makebox(0,0)[b]{$d$}}
\put(90,150){\makebox(0,0)[b]{$a$}}
\put(120,150){\makebox(0,0)[b]{$a$}}
}
\multiput(-.5,0)(1,0){2}{\qbezier(15,0)(7.5,7.5)(0,15)
\qbezier(150,15)(135,30)(120,45)
\qbezier(45,0)(22.5,22.5)(0,45)
\qbezier(75,0)(67.5,7.5)(60,15)
\qbezier(150,45)(127.5,67.5)(105,90)
\qbezier(105,30)(60,75)(15,120)
\qbezier(90,75)(52.5,112.5)(15,150)
\qbezier(60,135)(52.5,142.5)(45,150)
\qbezier(105,0)(67.5,37.5)(30,75)
\qbezier(15,60)(7.5,67.5)(0,75)
\qbezier(150,75)(112.5,112.5)(75,150)
\qbezier(150,105)(127.5,127.5)(105,150)}

\qbezier(45,0)(52.5,7.5)(60,15)
\qbezier(105,30)(112.5,37.5)(120,45)
\qbezier(15,60)(22.5,67.5)(30,75)
\qbezier(90,75)(97.5,82.5)(105,90)
\qbezier(0,105)(7.5,112.5)(15,120)
\qbezier(60,135)(67.5,142.5)(75,150)

}
\end{picture}}\linespread{1}\caption{Particles associated with the rule of
Figure~\ref{invarfig}. A complete period is given for the evolution of the
width-5 space \cir{aabcd} (note that the horizontal rows of the figure wrap
around). Time moves up. In the left-hand figure the first and fourth soliton
types given in Figure~\ref{invarfig} are shown. There are two solitons of each
type. The first type of soliton is given by the darker lines; they tend to move
to the left. When they collide with the second type of soliton (which tends to
move to the right), the two solitons move along the same path for two time
steps before continuing on their separate ways. The right-hand figure shows
particles associated with the rotation invariant. One can view them either as
particles and antiparticles which are created and annihilated in pairs, or as
particles which sometimes travel briefly backwards in time and whose net
velocity is faster than ``light.''}
\label{law11}
\end{figure}

Now consider $f_r$. The induced invariant $F_r$ turns out to give the
amount that $x$ is \emph{rotated} in a single orbit under $T$. That is: if
one displays the cellular automaton on a lattice as shown in
Figure~\ref{law11}, so that for each length-2 segment $L_i$ at time $t$ the
associated length-1 segment $R_i$ at time $t+1$ is situated between the
lattice points associated with $L_i$ at time $t$, then after one orbit the
original space $x$ has been rotated $F_r(x)$ cells to the right.

In the example pictured in Figure~\ref{law11},
the original space $x$ contains the segments \seg{aa}, \seg{ab}, \seg{bc},
\seg{cd} and \seg{da}. Thus the invariant associated with this space has the
value $(w+r)+(w+y)+(w+r+u)+(w+r+s)+(w+z)=5w+3r+u+y+s+z$. Setting $w=1$ and the
rest to zero gives the width (5); setting $r=1$ and the rest to zero gives the
rotation (3); and the period is $5(3)+3(-2)+1+0+1+(-1)=10$. The evolution of
$x$ shown in the figure verifies that these numbers are in fact the width,
rotation and period.

Note also that the width-2 zero invariants are spanned by the standard zero
invariants, which are induced by $f_y-f_z$, $f_u-f_y$, $f_s-f_u$ and
$f_z-f_s$. (The space of the zero invariants has dimension 3; the sum of any
three of these equals the fourth. Also it follows from this that $f_y$, $f_z$,
$f_s$ and $f_u$ all induce the same invariant of $T$.)

The reason for these various aspects of $T$ becomes evident when one examines
the \emph{local flow} associated with the width-2 \Z-invariants of $T$.

Let $f$ be a function that induces a $G$-invariant. A local flow for this
function is a local rule about how a group element associated with some segment
$S$ in $x$ is transmitted to the group elements associated with segments in
$T(x)$. Such a rule is described as follows. For a finite set of segments $S$
that includes (but is not necessarily limited to) those $S$ with the property
that $f(S)\neq 0$, one chooses a finite set of neighborhoods \seg{USV}. If this
neighborhood is embedded in $x$, then some segment $W$ is forced to be embedded
in $T(x)$. The transmission rule assigns a group element $g_U$ to each segment
$U$ embedded in $W$ so that the sum of these group elements is $f(S)$. (Thus
the group element assigned to $S$ locally spreads out as the space evolves.)
Now let $x$ be any space, and let $S$ be any segment in $T(x)$. If the
transmission rule has the property that the sum of the group elements
transmitted to $S$ is always equal to $f(S)$, then the rule is a local flow of
$f$.

A simple type of local flow exists for any $f$ if the rule $T$ is a cellular
automaton rule with the property that the left-hand segments in its rule table
have width 2. Suppose for simplicity that the adjacency matrix associated with
the rule table is strongly connected. Let $f$ be a function such that $f(S)=0$
whenever $|S|\neq w$. Let $S$ be any width-$w$ segment located in $x$. Then
there is a width-$(w-1)$ segment $U$ with the property that \seg{cUd} is
present in $T(x)$ for some $c$ and $d$ that depend on $S$'s immediate neighbors
in $x$. Consider the type of path-designating rule which transmits group
elements $S_l$ to \seg{cU} and $S_r$ to \seg{Ud}, where $S_l$ and $S_r$ depend
only on $S$. It turns out that there is always a one-parameter group of rules
of this sort which are local flows for $f$.

To prove this, begin by choosing any width-$w$ segment $S$. Let $S_r=k$ where
$k$ is an unknown group element. Suppose $S=\seg{dU}$. Consider
$\seg{Sc}=\seg{dUc}$ for each color $c$.  If this is present in $x$, then $T$
forces a width-$w$ segment $V$ to be present in $T(x)$. The sum of group
elements transmitted to $V$ must equal $f(V)$. So $\seg{dU}_r+\seg{Uc}_l=f(V)$,
so $\seg{Uc}_l=f(V)-k$. Also, the sum of group elements transmitted from
\seg{Uc} must equal $f(\seg{Uc})$. So
$\seg{Uc}_r=f(\seg{Uc})-f(V)-k$. Proceeding in a similar way by extending $S$
further to the right, one eventually computes a value of $U_l$ and $U_r$ for
each width-$w$ segment $U$. The values depend only on $f$ and on $k$. Using the
fact that $f$ induces a $G$-invariant of $T$, one may easily check that any
transmission rule derived in this way is consistent: given any space $x$ the
transmitted values that arrive at or leave a width-$w$ segment $U$ are
guaranteed to sum to $f(U)$.

The width-2 local flow of this sort for our example rule is given in column 4
of Figure~\ref{invarfig}. The ordered pair in each row is $(U_l,U_r)$ where $U$
is the segment in the left-hand column of the table. Clearly
$U_l+U_r=f(U)$. The presence of the new variable $k$ corresponds to the fact
that one can add any fixed group element to each of the left-hand entries of
the ordered pairs provided that one subtracts the same group element from each
of the right-hand entries.

To see how this flow works, consider for instance \seg{aab} at time $t$. This
forces \seg{ad} to be present at time $t+1$. The transmission rule says that
from \seg{aa} at time $t$ the element $k+r$ is transmitted to the right,
arriving at \seg{ad} at time $t+1$; and from \seg{ab} at time $t$ the element
$w-z-k$ is transmitted to the left, arriving at \seg{ad} at time $t+1$. The sum
of these two elements arriving at \seg{ad} is $w-z+r$, which is exactly the
group element associated with \seg{ad}.

I have been able to prove that the width-$w$ invariants of any rule given by a
nonoverlap rule table where the left-hand segments have width 2 are always
associated with a local flow (even when the rule is not a cellular
automaton). A result of Hattori and Takesue~\cite{hattori/takesue} implies that
there is guaranteed to be a local flow associated with the width-$w$ invariants
of any cellular automaton. I am convinced that there is a simple proof that
this also holds for any 1+1-dimensional oriented combinatorial spacetime, but
have not found this proof.

Certain local flows may easily be interpreted in terms of particles. For
example, suppose that we find an integer solution to the local flow in which
each ordered pair is of the form $(0,1)$, $(1,0)$ or $(0,0)$. The value 1
may be thought of in this case as a particle travelling through the lattice.
Each such particle moves either to the left or to the right; it sometimes
may change direction, but it never vanishes. So it is a soliton.

For example, in Figure~\ref{invarfig} five integer flows are given that
correspond to five solitons. These are the only five width-2 solutions of
this type. The first three solitons tend to move to the left, but
occasionally move to the right. The last two solitons tend to move to the
right, but occasionally move to the left. The number of left-moving
particles and the number of right-moving particles are invariants, induced
respectively by $f_w-f_r$ and $f_w-f_r+f_s$.

Examine the behavior of solitons of types 1 and 4 in Figure~\ref{law11}. Given
a list of which solitons are present at time $0$ and which direction they are
headed in at that time, one may easily compute the paths of the solitons in
accordance with a simple set of propagation and interaction rules. Moreover,
given the pattern of solitons on the lattice one may compute the vertex colors
as follows. Each vertex may be bordered by solitons on any of four possible
sides: top left, top right, bottom left and bottom right. A vertex is bordered
at bottom right by a type-4 soliton and not a type-1 soliton if and only if its
color is $c$. A vertex whose color isn't $c$ is bordered at top right by a
type-1 soliton if and only if its color is $d$. A vertex whose color isn't $c$
or $d$ is bordered at bottom left by a type-1 soliton if and only if its color
is $b$. And all other vertices are colored $a$. Thus these solitons are really
the whole story for this system.

Note that the system is also completely described by choosing other pairs of
solitons, where one is a left-moving soliton and one is a right-moving
soliton. Not all such pairs work, however.

Also note that the period of the system is equal to the number of solitons plus
the width, and the rotation to the \emph{left} (which is the width invariant
minus the rotation invariant) is equal to the number of left-moving
solitons. In the course of one period, each left-moving soliton passes through
each right-moving soliton exactly once, and at each passing-through there is a
slight delay in its travel to the left. By taking this into account, one can
easily prove that the so-called period invariant is in fact the period, and
that the so-called rotation invariant is in fact the rotation.

Solitons abound in these systems, but there are also other kinds of
particles. For example, consider a local flow where each ordered pair is of the
form (1,0), (0,-1), (1,-1) and (0,0). Here the 1's can be considered as
particles and the $-1$'s as antiparticles. The particles move to the left; the
antiparticles move to the right. Particle-antiparticle pairs may be
produced. And it may happen that a particle will collide with an antiparticle
and they will annihilate one another.

An interesting example of this type of particle is a certain local flow
associated with the rotation invariant of our example law; this is given in the
last column of Figure~\ref{invarfig}. An evolution of these particles is
pictured in Figure~\ref{law11}. The particle travels to the left. Occasionally
a particle-antiparticle pair is produced. But this can only happen when there
is also a particle situated one cell to the right of the pair
production. (Since the particles and antiparticles have width 2, an adjacent
particle and antiparticle overlap one another by one cell. Here the rightmost
cell of an antiparticle is, according to the rules, always the leftmost cell of
a particle.) Thus the antiparticle can only travel rightwards for one time step
before it hits the particle. The pair annihilate one another, while the
left-moving particle continues on its way.

One may think of the situation as described above. Or, following Feynman, one
may think of the new left-moving particle as really being the same particle as
the left-moving particle that was just annihilated. In the new description the
original particle simply jogged briefly backward in time and then moved forward
again. In this case Feynman's description seems particularly appropriate. Since
causal effects propagating at maximum velocity travel along the diagonals for
this rule, we may think of a particle moving along a diagonal as travelling at
the speed of light. If we think of the particle as sometimes briefly moving
backwards in time, then this type of particle (when looked at over large time
intervals) moves faster than light.

I will conclude this chapter by analyzing the behavior of the 1+1-dimensional
system described in Chapter~\ref{examples}. Recall that empirically I found
that orbits in this system were always finite (though typically quite
large). Here I will prove that the orbits are finite by analyzing the system in
terms of invariants and particles. While in the case analyzed above one can
understand the behavior of the system in terms of a finite number of particle
types, here it seems that an infinite sequence of particle types is required.

My first step was to compute the width-3 invariants using the methods described
earlier. The result, after simplification, is that the function $f$ with the
property that $f(\seg{ab})=f(\seg{acc})=f(\seg{acb})=1$ (and $f(S)=0$ otherwise)
induces an invariant. Because of the structure of $\zf_G(\x{3})$, it follows
that the function $g$ such that $g(\seg{ba})=g(\seg{cca})=g(\seg{bca})=1$ (and
$g(S)=0$ otherwise) induces the same invariant.

Since I do not have a general method for determining how invariants are
propagated locally, I tried to do this by hand. To do this, one evolves a
segment forward in time to see what all possible results are, given boundary
conditions. The table of forcing strings in Figure~\ref{force} is useful for
this purpose. In this case the solution is simple. The segment \seg{ab} forces
\seg{cca}, no matter what surrounds it. Similarly, \seg{acc} always evolves to
\seg{bca}, and \seg{acb} always evolves to \seg{ba}. Moreover, it is simple to
check that this process is invertible: \seg{cca}, when evolved backwards,
always produces \seg{ab}, and so on. Also, each of the segments \seg{ba},
\seg{cca} and \seg{bca} evolves to something involving \seg{ab}, \seg{acc} and
\seg{acb} (for example, the nine forcing strings for $R'_5$, which are the
strings associated with \seg{ba}, each end in either \seg{ab}, \seg{acc} or
\seg{acb}), and this process is also invertible.

Thus we may view this invariant locally as being represented by a single type
of soliton which may be represented in six different ways. Let
$A=\{\seg{ab},\seg{acc},\seg{acb}\}$, and let
$B=\{\seg{ba},\seg{cca},\seg{bca}\}$. A soliton is present wherever one finds
one of the segments in $A$ or in $B$. Whenever the soliton is represented by a
segment in $A$, it evolves to one in $B$, and vice versa.

The solitons break a space $x$ up into two types of sections, which I call the
\emph{ac} sections and the \emph{bc} sections.

An \emph{ac} section is a sequence of $a$'s and $c$'s which does not contain
two consecutive $c$'s. Either the entire space is a section of this sort, or
else the section is bounded on the left and right by solitons. A soliton that
bounds it on the left takes the form \seg{ba}, \seg{bca} or \seg{cca}; a
soliton that bounds it on the right takes the form \seg{ab}, \seg{acb} or
\seg{acc}. If the section is bounded, then it begins and ends with $a$, and
these $a$'s are also part of the bounding solitons.

A \emph{bc} section is a sequence of $b$'s and $c$'s. Either the entire space
is a section of this sort, or else the section is bounded on the left and right
by solitons. A soliton that bounds it on the left takes the form \seg{ab},
\seg{acb} or \seg{acc}; a soliton that bounds it on the right take the form
\seg{ba}, \seg{bca} or \seg{cca}. If the section is bounded, it must either
have length greater than one or it must consist of a single $b$. A bounded
section includes the $b$'s and $c$'s that are part of its bounding solitons.

Hence every space consists either of a single \emph{ac} or \emph{bc} section,
or an alternating sequence of sections separated by solitons. In the latter case
the sequence is like this: a soliton represented by a segment in $A$; then a
\emph{bc} section; then a soliton represented by a segment in $B$; then an
\emph{ac} section; then a soliton represented by a segment in $A$, and so
on. There is always an even number of solitons, half represented by segments
in $A$ and half represented by segments in $B$.

Since a segment in $A$ evolves to one in $B$ and vice versa, it follows that an
\emph{ac} section evolves to a \emph{bc} section and vice versa.

In another example of always-periodic evolution that I examined, it turned out
that there was a soliton present, and that this soliton placed a restriction
on the size of the space. That is: given the nature of the soliton, it was not
possible for two consecutive solitons to be arbitrarily far apart. Hence the
number of possible spaces containing $n$ solitons was bounded for any $n$;
since this number was conserved, evolutions were forced to be periodic. This is
not the case here, however. Two consecutive solitons of the sort that we have
looked at so far may be separated from one another by an arbitrarily large
distance, since an \emph{ac} section and a \emph{bc} section may be as large as
you like. My hope was that the periodicity of evolution of this system could be
explained by another particle that lived in the \emph{ac} and \emph{bc}
sections and put restrictions on their widths.

This did not quite turn out to be the case. Instead, I found that the bounded
sections are themselves solitons. It turns out that this system contains an
infinite sequence of types of solitons; each bounded section belongs to one of
these types. I shall call these soliton types $P_n$, $0 \leq n<\infty$ (and I
will call the soliton type we discussed earlier the Q soliton to distinguish it
from these). Each soliton type $P_n$ has only a finite number of states (since
it corresponds to a width-$w$ invariant for some $w$), and any space $x$
containing a $Q$-type soliton consists entirely of solitons (these alternate
between $P$-type solitons and $Q$-type solitons). Since the number of solitons
of each type is conserved, the width of the space in this case is bounded and
evolution is forced to be periodic. In the case where $x$ does not contain any
$Q$-type solitons and consists of a single section, it is easy to verify that
its evolution is also periodic; hence it is periodic in every case.

In what follows I will focus attention on the \emph{interior} of a bounded
\emph{bc} section. This refers to that portion of a \emph{bc} section that
remains when the $b$ and $c$ cells belonging to its two bounding $Q$ solitons
are stripped away. The definition only makes sense if the \emph{bc} section is
large enough so that the bounding $Q$ solitons do not overlap one another;
I will at first assume that this is the case, and treat exceptions later.
 
\begin{figure}
\fbox{\begin{picture}(425,322)\linespread{1}\footnotesize\put(0,12){\makebox(425,322)[bl]{\hfill\begin{tabular}{rcccccl}
$ab$&$\overbrace{b\cdots b}^{n_1}$&$\overbrace{c\cdots c}^{n_2}$&$\overbrace{b
\cdots b}^{n_3}$&$\cdots$&$\overbrace{c\cdots c}^{n_k}$&$ba$\\
$acb$&&&&&&$bca$\\
$acc$&&&&&&$cca$\\
$ba$&$\overbrace{a\cdots a}^{n_1}$&$\overbrace{ca\cdots ca}^{n_2}$&$\overbrace{a
\cdots a}^{n_3}$&$\cdots$&$\overbrace{ca\cdots ca}^{n_k}$&$ab$\\
$bca$&&&&&&$acb$\\
$cca$&&&&&&$acc$\\
&&&&&&$acab$\\
&&&&&&$acacb$\\
&&&&&&$acacc$\\
&&&&&&$cacab$\\
&&&&&&$cacacb$\\
&&&&&&$cacacc$\\
$a$&$\overbrace{c\cdots c}^{n_1}$&$\overbrace{b\cdots b}^{n_2}$&$\overbrace{c
\cdots c}^{n_3}$&$\cdots$&$\overbrace{b\cdots b}^{n_k}$&$cba$\\
&&&&&&$cbca$\\
&&&&&&$ccca$\\
&&&&&&$cbba$\\
&&&&&&$cbbca$\\
&&&&&&$cbcca$\\
&&&&&&$bbba$\\
&&&&&&$bbbca$\\
&&&&&&$bbcca$\\
\end{tabular}\hfill\begin{tabular}{ccc}
$A$&\hspace{18pt}&$B_1$\\ \cline{1-1} \cline{3-3}
$bc$&&$c$\\
$bcb$&&$bb$\\
$bbb$&&$bc$\\
$bcc$&&$cb$\\
$bccb$&&$cc$\\
$bcbb$&&$bbb$\\
$ccc$&&$bbc$\\
$cccb$&&$bcb$\\
$ccbb$&&$bcc$\\
&&$cbb$\\
$A'$&&$ccb$\\ \cline{1-1}
$cb$&&$ccc$\\
$bcb$&&$bbbb$\\
$bbb$&&$bbcb$\\
$ccb$&&$bbcc$\\
$bccb$&&$bcbb$\\
$bbcb$&&$bccb$\\
$ccc$&&$bccc$\\
$bccc$&&$ccbb$\\
$bbcc$&&$cccb$\\
&&$bbcbb$\\
&&$bbccb$\\
&&$bccbb$\\
&&$bcccb$\\
\end{tabular}\hfill
}}\end{picture}} \linespread{1}\caption[An infinite family of solitons]{An
infinite family of solitons. On the left, the evolution of the interior of a
\emph{bc} section through two time steps is shown. The multiple strings listed
to the right and left of the interior at each state represent the various
boundary conditions being considered and how they evolve. The result is that
the colors are permuted, $c$, $cb$ or $bb$ is added on the right, and a $Q$
particle is removed on the left (since now there is an $a$ immediately to the
left of what used to be in the interior). If we evolve through two more time
steps, the net result is that the colors of the interior are no longer
permuted, but a string from column $A$ has been added to the right and a string
from column $A'$ has been removed from the left. Note that it does not matter
whether the interior begins or ends with $b$'s or $c$'s. The strings in column
$B_1$ are the interiors of the $P_1$ soliton type; the $P_n$ soliton types
($n>1$) are generated using columns $A$ and~$B_1$.}
\label{step4}
\end{figure}
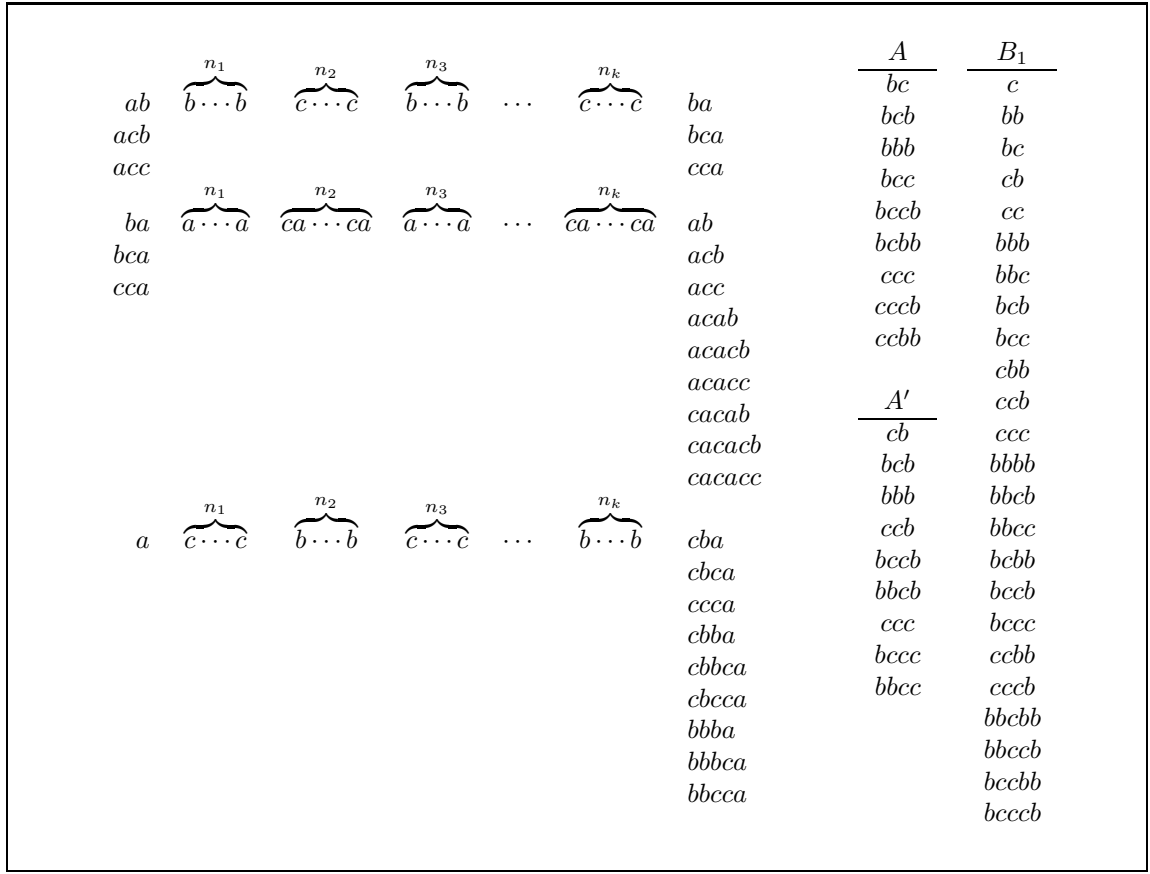

Consider Figure~\ref{step4},
which describes the evolution of an arbitrary nonempty interior of a bounded
\emph{bc} section through four time steps. Here it is assumed that we are given
a fixed nonempty interior $I$ of a \emph{bc} section bounded by two $Q$
solitons. It is not assumed that we know anything about what the representation
of those $Q$ solitons is, or about what lies beyond them. All possible
evolutions are computed. It turns out that, if $I$ does not consist of a single
cell $b$, then the result is the interior of a bounded \emph{bc} section which
may be obtained from $I$ in the following way: first append one of a set $A$ of
nine strings of cells to the right-hand end of $I$; then delete one of a set
$A'$ of nine strings of cells from the left-hand end of the resulting string.

The set $A$ has this property: given a sufficiently long bounded string of
$b$'s and $c$'s, the right-hand end of that string will be identical to exactly
one element of $A$. The strings in $A'$ are the left-right reflections of those
in $A$. Hence the set $A'$ has this property: given a sufficiently long bounded
string of $b$'s and $c$'s, the left-hand end of that string will be identical
to exactly one element of $A'$.

In the same figure another set $B_1$ of 24 strings of cells is displayed. These
are the interiors of bounded \emph{bc} sections that correspond to $P_1$
solitons. The set $B_1$ has this property: if $s \in B_1$ and you add one of
the strings in $A$ to its right-hand end and then delete a string in $A'$ from
its left-hand end, you get back an element of $B_1$.

For example: \emph{bbc} is in $B_1$. The string \emph{bccb} is in $A$; if we
add this to the right-hand end of \emph{bbc} we get \emph{bbcbccb}. If we now
wish to delete an element of $A'$ from the left-hand end, there is only one way
to do this: we must delete \emph{bbcb}; the result is \emph{ccb}, which is also
in $B_1$.

We now inductively define sets of strings $B_n$ for each $n>1$. The set $B_n$
is obtained by appending an element of $A$ to the right-hand end of an element
of $B_{n-1}$ in all possible ways. Since there are 9 elements in $A$ and 24 in
$B_1$, there are 9 ways to append an element of $A$ to the right-hand end of
each element of $B_1$, and hence $24 \times 9 = 216$ elements in $B_2$; in
general there are $24 \times 9^{n-1}$ elements in $B_n$.

The key fact is that each set $B_n$ has the same property as did the set $B_1$.

\begin{theorem} For each $n>0$, if we append an element of $A$ to the
right-hand end of any string in $B_n$ then there is always a unique way to
delete an element of $A'$ from the left-hand end of the resulting string, and
the result is again an element of $B_n$.
\end{theorem}
\textsc{Proof.} This is true of $B_1$. Assume that it is true of $B_{n-1}$. Let
$s$ be an element of $B_n$. Then $s=\mathit{uv}$ where $u \in B_{n-1}$ and $v
\in A$. By the inductive hypothesis we may delete an element of $A'$ from the
left-hand end of \emph{uv} in a unique way and obtain some string $w \in
B_{n-1}$. Now suppose we add an element $z$ of $A$ to the right-hand end of
$s$, obtaining \emph{sz}. This equals \emph{uvz}, so there is a unique way to
delete an element of $A'$ from the left-hand end of this string. The result is
\emph{wz}, which is obtained by adding an element of $A$ to the right-hand end
of an element of $B_{n-1}$, and hence is\linebreak in $B_n$.\proofend\bigskip

Consider now a \emph{bc} section whose interior is in $B_n$. Since $b$ is not
in $B_n$, the above theorem implies that evolving this section for four time
steps leads to a \emph{bc} section whose interior is in $B_n$. It turns out
that evolving the original section for two time steps also leads to a \emph{bc}
section whose interior is in $B_n$. Thus the soliton type $P_n$, $n>0$, is
represented by the \emph{bc} sections containing elements of $B_n$ as
interiors, and by the \emph{ac} sections which may be obtained from these
\emph{bc} sections by evolving them for one time step.

Now consider any interior $I$ of a \emph{bc} section. We wish to know if the
interior is in $B_n$ for some $n$. If $I$ is long enough, one may obtain a
shorter string from it by deleting an element of $A$ from its right-hand end
(there is always one way to do this). If enough such deletions are performed,
one ends up with one of the following strings: $c$, $cc$, $bb$, $cbb$, $cb$,
$ccb$, $b$, or the empty string. In the first six cases, these strings are in
$B_1$ (and hence $I$ is in $B_n$ for some $n$). In the latter two cases, adding
any element of $A$ to the right-hand end of the remainder string produces an
element of $B_1$. Hence $I$ is in $B_n$ for some $n$ unless no element of $A$
may be deleted from its right-hand end and its interior (if it has a defined
interior) is $b$ or the empty string. There are 24 \emph{bc} sections which
fall into this category (they have interior $b$, or empty interior, or else
they are made of two overlapping $Q$ solitons); these constitute the soliton
$P_0$.

It is natural to wonder whether the set of $P_n$ solitons ($n\in\N$) might be
generated in some fashion by a single invariant. Perhaps this is possible with
some new type of invariant, but it is not possible with the type of invariant
considered here. The reason is illustrated in the following theorem.

\begin{theorem} There does not exist a width-$m$ invariant $I$ (for some $m$)
such that the value of $I$ associated with a $P_n$ soliton is $n$.
\end{theorem}
\textsc{Proof.} A $P_n$ soliton may be represented by a \emph{bc} segment whose
interior is a string of $c$'s having length $3n-2$, $3n-1$ or $3n$. Let $3n-2$
be larger than $m$. If one adds a $c$ to the interior of a \emph{bc} segment
consisting of $3n-2$ $c$'s, the same soliton $P_n$ is still represented. Hence
the value of $I$ should not change; it follows that the value of $I$ on a
string of $m$ $c$'s must be zero. However, if one adds a $c$ to the interior of
a \emph{bc} segment consisting of $3n$ $c$'s, the soliton represented by that
segment changes from $P_n$ to $P_{n+1}$. Hence the value of $I$ should increase
by 1; it follows that the value of $I$ on a string of $m$ $c$'s must be 1. This
is a contradiction.\proofend\bigskip

While I did manage to use invariants and particles to solve the problem of the
periodicity of this system, my computer would not have been so lucky. My
current algorithms to find invariants can only do so for a specified
width. This width has to be pretty small; otherwise there are too many
equations. In addition, the algorithm cannot find infinite sequences of
invariants, as I have done here. It would not surprise me, however, if there
are better algorithms for finding invariants of a specified width, and
algorithms for finding infinite sequences of invariants. Perhaps there even
exists an algorithm for finding all invariants of a 1+1-dimensional system in
finite time; this remains to be seen. At any rate, I suspect there is much room
for progress in this area.

\cleardoublepage\vspace*{40pt}\section[Higher dimensions]{\LARGE Higher
dimensions}\bigskip
\label{higherD}

The end result of Chapter~\ref{orient} was a simple, attractive model which (if
the conjecture holds) describes all 1+1-dimensional combinatorial
spacetimes. The main characteristics of this model are:
\begin{itemize}\linespread{1}\normalsize
\item A set of combinatorial objects, each of which describes the set of
allowed spaces;
\item A set of elementary transformations of these objects which correspond to
local, invertible maps on the spaces and which generate all such maps.
\end{itemize}
It is natural to hope that a similar model can be found in higher
dimensions. This chapter describes my preliminary attempts to find such a
model.

\subsection{Space sets defined by lists of allowed neighborhoods}

In Chapter~\ref{examples} I suggested that our $n$-dimensional spaces be given
by vertex-colored $n$-graphs, and that sets of $n$-dimensional spaces be
given by choosing a nonnegative integer $r$, listing all allowed vertex
neighborhoods of radius $r$, and specifying that the allowed spaces are exactly
those which have the property that the radius-$r$ neighborhood of each vertex
in the space is in this list. An equivalent formulation (and one which is more
in line with our procedure in the one-dimensional case) is expressed in terms
of neighborhoods, rather than vertex neighborhoods. In this case a set of
spaces is defined by choosing a positive integer $d$, listing all allowed
neighborhoods of diameter $d$, and specifying that the allowed spaces are
exactly those which have the property that each diameter-$d$ neighborhood in
the space is in this list.

This sounds reasonable enough, but it contains an ambiguity. What does it mean,
exactly, to specify a neighborhood? A seemingly natural choice is to define a
neighborhood in a graph \G\ to be a particular kind of subgraph of \G. More
precisely, we might say that a neighborhood of diameter $d$ in a graph \G\ is
defined to be a connected subgraph $S$ of \G\ that is maximal in the property
that the maximum distance between vertices in $S$ is equal to $d$. (By
``maximal in the property'' I mean that there does not exist a subgraph $S'$ of
$G$ such that $S$ is contained properly in $S$ and $S'$ also has this
property.)

However, this choice of definition does not precisely match our procedure in the
one-dimensional case. There our neighborhoods were not subgraphs; they were
\emph{segments}. A segment is similar to a subgraph; however, it does not
contain a certain sort of information that a subgraph does contain. For
example, consider the space \cir{ab}. If we travel from the $a$-colored vertex
to the right, we pass through the $b$-colored vertex and arrive back at the
\emph{same} $a$-colored vertex from which we began the trip. Segments ignore
this notion of sameness; they are concerned only with the sequence of colors
encountered as we travel to the right. The segment \seg{aba} is contained in
\cir{ab}, but there is no subgraph of \cir{ab} containing three vertices. And
certainly there is no segment that looks anything like the subgraph \cir{ab} of
\cir{ab}. In fact, a subgraph and a segment do not mean the same thing even if
their appearances are similar. If $S$ is a subgraph of \G\ then every vertex and
edge in $S$ is assumed to be a distinct vertex or edge of \G; this is not
assumed for segments.

We may in fact view segments as finite connected subgraphs, not of the graph
\G, but of the \emph{universal covering space} of \G. This is the infinite
directed 2-regular graph that looks like \G\ locally but which contains no
closed loops. (The universal covering space of \cir{ab} is the infinite
directed 2-regular graph in which $a$'s and $b$'s alternate; it may be
represented as $(\dots\mathit{ababab}\dots)$, where here the parentheses need
not be taken to mean that somehow the graph closes up at infinity.)

Note that our choice to define sets of one-dimensional spaces in terms of
allowed segments, rather than allowed subgraphs, seems not to be dictated by
any of the original set of defining principles for combinatorial
spacetimes. And there is nothing to prevent us from defining space sets in
terms of subgraphs. For example, if we define a set of spaces by specifying
that the allowed width-2 subgraphs are \seg{aa}, \seg{ab}, \seg{ba} and
\seg{bb} (where here I mean \seg{ab} to be the subgraph consisting of an
$a$-colored vertex connected by a directed edge to a $b$-colored vertex), we
get the same spaces that are in \x{2} \emph{except} that no length-1 spaces are
included. We might also include the subgraph \cir{cd} in our list; this would
mean that the single space \cir{cd} is also in our space set. It is natural in
this case to extend our definition: instead of only specifying allowed
subgraphs of diameter $d$, we may also specify allowed \emph{closed} subgraphs
of diameter less than $d$. Hence we may also specify that \cir{e} is in our set
of spaces.

The space sets that we have just defined are not definable using segments. One
may easily verify that, by using the subgraph method of defining space sets,
one may define all of the space sets that can be defined using segments. Thus
the subgraph method is the more general one. However, we did obtain an elegant
formulation of 1+1-dimensional combinatorial spacetimes using segments. This
suggests that segments may be preferable; these other space sets obtained by
using subgraphs may not be worth considering.

Now let us examine the analogous question in higher dimensions. Again, if our
neighborhoods are connected subgraphs then a neighborhood is allowed to be an
entire space. Even if a connected subgraph is not an entire space, its topology
can be complex. For example, in two dimensions a connected subgraph which is
not a closed space might have any number of handles or crosscaps. Also, the
boundary of a connected subgraph might not be connected; for instance, in two
dimensions the subgraph might be a cylinder. In addition, a connected subgraph
is not required to be oriented; thus in two dimensions an allowed neighborhood
would be the M\"{o}bius strip. In summary, if one uses this definition then a
neighborhood may have the topology of any pseudomanifold with or without
boundary.

The use of segments in the one-dimensional case suggests that this definition
of neighborhood might not be the most useful one. The idea instead is that a
neighborhood should be defined to be a subgraph of the universal covering space
of some allowed space. But what should we mean in the higher-dimensional
context by a covering space? Two possibilities suggest themselves. On the one
hand, we might mean the universal covering space of the graph that describes
the pseudomanifold. On the other hand, we might mean the universal covering
space of the pseudomanifold itself.

This question has already been discussed in Chapter~\ref{examples}, from a
slightly different point of view. If we define the set of allowed neighborhoods
to be the set of finite connected subgraphs of universal covering spaces of
allowed spaces, then these neighborhoods are finite colored trees. If the
dimension of our set of spaces is $n$, then the edges of the trees have $n+1$
colors and no vertex meets two edges of the same color. We define a set of
spaces by listing the allowed neighborhoods of a certain size; the allowed
spaces are exactly those whose universal covering spaces contain only those
neighborhoods.

This setup is well defined, and perhaps it is a useful one. However, the space
sets obtained in this manner have the characteristic that if $x$ is an allowed
space then so is any space $y$ whose graph is a covering graph of the graph of
$x$. What this means is that in our space sets there are infinitely many types
of allowed faces. In Chapter~\ref{examples} I gave some of the reasons why I
believe that this is not a desirable property. An additional reason is given in
the next section.

Consider, then, our second option, which is that we think about covering spaces
of pseudomanifolds rather than covering spaces of the associated graphs. In the
one-dimensional case we obtain a universal covering space by making infinitely
many copies of the one-dimensional faces of our original space and piecing them
together so that the adjacency relations of the faces in the original space are
maintained, but so that no closed loop is formed. We might try a similar
procedure to form the covering space of an $n$-dimensional pseudomanifold,
using $n$-dimensional faces rather than one-dimensional faces. However, this
procedure fails whenever the original space is not flat. For example, consider
the 2-sphere represented by the upper left-hand graph in
Figure~\ref{twographs}. There is only one space that can be obtained by
assembling copies of the three faces in this 2-sphere (namely, the 2-sphere
itself). The reason for this is that in our setup faces are not simply
topological in nature; they are geometrical. In this case, they contain
curvature. If you attach them, there is only one way to do so, and they close
up automatically into a sphere.

So it seems that both of these attempts to generalize the notion of segment to
higher dimensions run into difficulties.

Of course, there are other possibilities to consider. One natural idea is to
require that the neighborhoods in our list of allowed neighborhoods are
homeomorphic to $n$-dimensional disks. This procedure is analogous to that used
to define manifolds in differential geometry. I am somewhat reluctant to accept
this idea because it makes life complicated. In order to insure that our
proposed neighborhoods are of the desired type, it is necessary to compute
their homology groups. In addition, this will not be sufficient in the
four-dimensional case unless in the meantime someone proves the Poincar\'{e}
conjecture. I would prefer a simple combinatorial criterion for whether a
neighborhood is allowed or not. However, there is no denying the importance of
manifolds to mathematics and to physics, so this path is probably worth
pursuing.

So far I have concentrated my research on two other approaches to this
problem. The starting point for each of these approaches is the formulation of
the 1+1-dimensional case in terms of matrix descriptions. A matrix description
may be thought of as a list of one-dimensional faces and zero-dimensional
hinges. The hinges are branch points. The natural generalization of this
procedure is to define $n$-dimensional space sets using faces and hinges. This
idea is explored in Section~\ref{hinge}. A second approach is to generalize the
notion of bialgebra. Earlier we found that the scalars in a commutative
cocommutative bialgebra describe sets of one-dimensional spaces, and that the
axioms of a bialgebra correspond to equivalence maps between these space
sets. One might hope to generalize this setup by finding a sequence of
algebraic systems in which the scalars correspond to sets of $n$-dimensional
spaces and the axioms again correspond to equivalence maps between space
sets. This approach is discussed in Section~\ref{highdimbialgebras}.

\subsection[The spacetime picture of 1+1-dimensional dynamics]{The spacetime
picture of 1+1-dimensional dynamics}

Let $A_i$ be a matrix of nonnegative integers for each $i\in\{0,\dots,n\}$,
with $A=A_0=A_n$ and $\x{A}=\X$. Let $T_i:A_{i-1}\mapsto A_i$ be an elementary
map for each $i\in\{1,\dots,n\}$. Let $T=T_nT_{n-1}\dots T_1$ be a law of
evolution on $A$. Our goal is to describe a set \Y\ of two-dimensional spaces
in which there is a local invertible map between \Y\ and the 1+1-dimensional
evolutions of \X\ under $T$. This map should have two properties. Firstly, we
should be able to foliate each $Y$ into one-dimensional spaces $Y_t$ in such a
way that, for some fixed $x\in\X$, $Y_t$ is locally equivalent to $T^t(x)$ for
each $t\in\Z$. Secondly, if neighborhoods of $T^t(x)$ and $T^{t+1}(x)$ are
causally linked, the corresponding neighborhoods in $Y_t$ and $Y_{t+1}$ should
be near one another in $Y$.

\begin{figure}
\fbox{\begin{picture}(425,340)\footnotesize\put(42.5,150){
\multiput(0,0)(60,0){6}{\put(10,0){\circle{5}}\put(30,0){\circle*{5}}
        \put(10,40){\circle*{5}}\put(30,40){\circle{5}}
        \multiput(12.5,0)(0,40){2}{\line(1,0){15}}
        \put(0,10){\circle*{5}}\put(40,10){\circle{5}}
        \multiput(0,12.5)(40,0){2}{\line(0,1){15}}\put(0,30){\circle{5}}
        \put(40,30){\circle*{5}}
        \multiput(30,0)(-30,30){2}{\qbezier(1.77,1.77)(5,5)(8.23,8.23)
                \qbezier(6.41,3.59)(5,5)(3.59,6.41)}
        \multiput(10,0)(30,30){2}{\qbezier(-1.77,1.77)(-5,5)(-8.23,8.23)
                \qbezier(-6.41,3.59)(-5,5)(-3.59,6.41)}}
\multiput(42.5,10)(60,0){5}{\multiput(0,0)(0,20){2}{
        \put(0,0){\line(1,0){15}}\multiput(6.5,-2)(2,0){2}{\line(0,1){4}}}}
\multiput(0,60)(300,0){2}{\multiput(0,0)(0,60){2}{
        \put(10,0){\circle{5}}\put(30,0){\circle*{5}}
        \put(10,40){\circle*{5}}\put(30,40){\circle{5}}
        \multiput(12.5,0)(0,40){2}{\line(1,0){15}}
        \put(0,10){\circle*{5}}\put(40,10){\circle{5}}
        \multiput(0,12.5)(40,0){2}{\line(0,1){15}}\put(0,30){\circle{5}}
        \put(40,30){\circle*{5}}
        \multiput(30,0)(-30,30){2}{\qbezier(1.77,1.77)(5,5)(8.23,8.23)
                \qbezier(6.41,3.59)(5,5)(3.59,6.41)}
        \multiput(10,0)(30,30){2}{\qbezier(-1.77,1.77)(-5,5)(-8.23,8.23)
                \qbezier(-6.41,3.59)(-5,5)(-3.59,6.41)}}}
\multiput(300,-120)(0,60){2}{
        \put(10,0){\circle{5}}\put(30,0){\circle*{5}}
        \put(10,40){\circle*{5}}\put(30,40){\circle{5}}
        \multiput(12.5,0)(0,40){2}{\line(1,0){15}}
        \put(0,10){\circle*{5}}\put(40,10){\circle{5}}
        \multiput(0,12.5)(40,0){2}{\line(0,1){15}}\put(0,30){\circle{5}}
        \put(40,30){\circle*{5}}
        \multiput(30,0)(-30,30){2}{\qbezier(1.77,1.77)(5,5)(8.23,8.23)
                \qbezier(6.41,3.59)(5,5)(3.59,6.41)}
        \multiput(10,0)(30,30){2}{\qbezier(-1.77,1.77)(-5,5)(-8.23,8.23)
                \qbezier(-6.41,3.59)(-5,5)(-3.59,6.41)}}
\put(240,-60){
        \put(10,0){\circle{5}}\put(30,0){\circle*{5}}
        \put(10,40){\circle*{5}}\put(30,40){\circle{5}}
        \multiput(12.5,0)(0,40){2}{\line(1,0){15}}
        \put(0,10){\circle*{5}}\put(40,10){\circle{5}}
        \multiput(0,12.5)(40,0){2}{\line(0,1){15}}\put(0,30){\circle{5}}
        \put(40,30){\circle*{5}}
        \multiput(30,0)(-30,30){2}{\qbezier(1.77,1.77)(5,5)(8.23,8.23)
                \qbezier(6.41,3.59)(5,5)(3.59,6.41)}
        \multiput(10,0)(30,30){2}{\qbezier(-1.77,1.77)(-5,5)(-8.23,8.23)
                \qbezier(-6.41,3.59)(-5,5)(-3.59,6.41)}}
\multiput(60,120)(60,0){2}{
        \put(10,0){\circle{5}}\put(30,0){\circle*{5}}
        \put(10,40){\circle*{5}}\put(30,40){\circle{5}}
        \multiput(12.5,0)(0,40){2}{\line(1,0){15}}
        \put(0,10){\circle*{5}}\put(40,10){\circle{5}}
        \multiput(0,12.5)(40,0){2}{\line(0,1){15}}\put(0,30){\circle{5}}
        \put(40,30){\circle*{5}}
        \multiput(30,0)(-30,30){2}{\qbezier(1.77,1.77)(5,5)(8.23,8.23)
                \qbezier(6.41,3.59)(5,5)(3.59,6.41)}
        \multiput(10,0)(30,30){2}{\qbezier(-1.77,1.77)(-5,5)(-8.23,8.23)
                \qbezier(-6.41,3.59)(-5,5)(-3.59,6.41)}}
\put(60,60){
        \multiput(10,0)(60,0){2}{\put(0,0){\circle{5}}\put(20,0){\circle*{5}}
                \put(2.5,0){\line(1,0){15}}}
        \multiput(130,0)(60,0){2}{\put(0,0){\circle{5}}\put(20,0){\circle*{5}}
                \put(2.5,0){\line(1,0){15}}}
        \multiput(10,40)(60,0){2}{\put(0,0){\circle*{5}}\put(20,0){\circle{5}}
                \put(2.5,0){\line(1,0){15}}}
        \multiput(30,0)(0,40){2}{\put(2.5,0){\line(1,0){35}}
                \put(20,-2){\line(0,1){4}}}
        \multiput(0,0)(120,0){2}{\put(0,10){\circle*{5}}\put(100,10){\circle{5}}
                \put(0,30){\circle{5}}\put(100,30){\circle*{5}}
                \put(0,12.5){\line(0,1){15}}\put(100,12.5){\line(0,1){15}}}
        \multiput(0,0)(120,0){2}{
                \multiput(90,0)(-90,30){2}{\qbezier(1.77,1.77)(5,5)(8.23,8.23)
                        \qbezier(6.41,3.59)(5,5)(3.59,6.41)}}
        \multiput(0,0)(120,0){2}{
                \multiput(10,0)(90,30){2}{\qbezier(-1.77,1.77)(-5,5)(-8.23,8.23)
                        \qbezier(-6.41,3.59)(-5,5)(-3.59,6.41)}}
        \put(152.5,0){\line(1,0){35}}\put(170,-2){\line(0,1){4}}
        \put(130,40){\circle*{5}}\put(210,40){\circle{5}}
        \put(132.5,40){\line(1,0){75}}}
\multiput(0,-120)(180,240){2}{
        \put(10,0){\circle{5}}\put(90,0){\circle*{5}}
        \put(10,40){\circle*{5}}\put(90,40){\circle{5}}
        \put(0,10){\circle*{5}}\put(100,10){\circle{5}}
        \put(0,30){\circle{5}}\put(100,30){\circle*{5}}
        \multiput(12.5,0)(0,40){2}{\line(1,0){75}}
        \multiput(0,12.5)(100,0){2}{\line(0,1){15}}
        \multiput(90,0)(-90,30){2}{\qbezier(1.77,1.77)(5,5)(8.23,8.23)
                \qbezier(6.41,3.59)(5,5)(3.59,6.41)}
        \multiput(10,0)(90,30){2}{\qbezier(-1.77,1.77)(-5,5)(-8.23,8.23)
                \qbezier(-6.41,3.59)(-5,5)(-3.59,6.41)}}
\multiput(0,-60)(120,0){2}{
        \put(10,0){\circle{5}}\put(90,0){\circle*{5}}
        \put(10,40){\circle*{5}}\put(90,40){\circle{5}}
        \put(0,10){\circle*{5}}\put(100,10){\circle{5}}
        \put(0,30){\circle{5}}\put(100,30){\circle*{5}}
        \put(30,40){\circle{5}}\put(70,40){\circle*{5}}
        \put(12.5,0){\line(1,0){75}}
        \multiput(12.5,40)(60,0){2}{\line(1,0){15}}
        \put(32.5,40){\line(1,0){35}}\put(50,38){\line(0,1){4}}
        \multiput(0,12.5)(100,0){2}{\line(0,1){15}}
        \multiput(90,0)(-90,30){2}{\qbezier(1.77,1.77)(5,5)(8.23,8.23)
                \qbezier(6.41,3.59)(5,5)(3.59,6.41)}
        \multiput(10,0)(90,30){2}{\qbezier(-1.77,1.77)(-5,5)(-8.23,8.23)
                \qbezier(-6.41,3.59)(-5,5)(-3.59,6.41)}}
\put(120,-120){
        \put(10,0){\circle{5}}\put(150,0){\circle*{5}}
        \put(10,40){\circle*{5}}\put(150,40){\circle{5}}
        \put(0,10){\circle*{5}}\put(160,10){\circle{5}}
        \put(0,30){\circle{5}}\put(160,30){\circle*{5}}
        \put(90,40){\circle{5}}\put(130,40){\circle*{5}}
        \put(12.5,0){\line(1,0){135}}
        \put(12.5,40){\line(1,0){75}}\put(132.5,40){\line(1,0){15}}
        \put(92.5,40){\line(1,0){35}}\put(110,38){\line(0,1){4}}
        \multiput(0,12.5)(160,0){2}{\line(0,1){15}}
        \multiput(150,0)(-150,30){2}{\qbezier(1.77,1.77)(5,5)(8.23,8.23)
                \qbezier(6.41,3.59)(5,5)(3.59,6.41)}
        \multiput(10,0)(150,30){2}{\qbezier(-1.77,1.77)(-5,5)(-8.23,8.23)
                \qbezier(-6.41,3.59)(-5,5)(-3.59,6.41)}}
\multiput(0,0)(0,60){2}{\multiput(10,-20)(60,0){6}{\multiput(0,0)(20,0){2}{
        \put(0,2.5){\line(0,1){15}}\multiput(-2,9)(0,2){2}{\line(1,0){4}}}}}
\multiput(0,60)(0,60){2}{\multiput(0,0)(120,0){3}{\multiput(40,10)(0,20){2}{
        \put(2.5,0){\line(1,0){15}}\multiput(9,-2)(2,0){2}{\line(0,1){4}}}}}
\multiput(100,-110)(180,0){2}{\multiput(0,0)(0,60){2}{\multiput(0,0)(0,20){2}{
        \put(2.5,0){\line(1,0){15}}\multiput(9,-2)(2,0){2}{\line(0,1){4}}}}}
\multiput(100,130)(0,20){2}{\multiput(0,0)(120,-180){2}{
        \put(2.5,0){\line(1,0){15}}\multiput(9,-2)(2,0){2}{\line(0,1){4}}}}
\multiput(10,-80)(80,0){2}{\multiput(0,0)(120,0){2}{
        \put(0,2.5){\line(0,1){15}}\multiput(-2,9)(0,2){2}{\line(1,0){4}}}}
\multiput(250,-80)(20,0){2}{\multiput(0,0)(60,0){2}{
        \put(0,2.5){\line(0,1){15}}\multiput(-2,9)(0,2){2}{\line(1,0){4}}}}
\multiput(10,100)(20,0){2}{\multiput(0,0)(60,0){3}{
        \put(0,2.5){\line(0,1){15}}\multiput(-2,9)(0,2){2}{\line(1,0){4}}}}
\multiput(190,100)(80,0){2}{
        \put(0,2.5){\line(0,1){15}}\multiput(-2,9)(0,2){2}{\line(1,0){4}}}
\multiput(310,100)(20,0){2}{
        \put(0,2.5){\line(0,1){15}}\multiput(-2,9)(0,2){2}{\line(1,0){4}}}
\put(50,-100){\makebox(0,0){$a$}}
\put(200,-100){\makebox(0,0){$b$}}
\put(320,-100){\makebox(0,0){$c$}}
\put(50,-40){\makebox(0,0){$d$}}
\put(170,-40){\makebox(0,0){$e$}}
\put(260,-40){\makebox(0,0){$f$}}
\put(320,-40){\makebox(0,0){$g$}}
\put(20,20){\makebox(0,0){$h$}}
\put(80,20){\makebox(0,0){$i$}}
\put(140,20){\makebox(0,0){$j$}}
\put(200,20){\makebox(0,0){$k$}}
\put(260,20){\makebox(0,0){$l$}}
\put(320,20){\makebox(0,0){$m$}}
\put(20,80){\makebox(0,0){$n$}}
\put(110,80){\makebox(0,0){$o$}}
\put(230,80){\makebox(0,0){$p$}}
\put(320,80){\makebox(0,0){$q$}}
\put(20,140){\makebox(0,0){$r$}}
\put(80,140){\makebox(0,0){$s$}}
\put(140,140){\makebox(0,0){$t$}}
\put(230,140){\makebox(0,0){$u$}}
\put(320,140){\makebox(0,0){$v$}}

}\end{picture}}\linespread{1}\caption{A method for embedding a sequence of
elementary maps of one-dimensional space sets in a set of two-dimensional
spaces. Each cell in a one-dimensional space is associated with a 01-face. The
color of the cell is written in the associated face. Each horizontal row of
01-faces is associated with a space, with the orientation of the space going
from left to right. Time moves up. The first elementary map sends \seg{a} to
\seg{d}, \seg{b} to \seg{ef}, and \seg{c} to \seg{g}; the second map sends
\seg{d} to \seg{hi}, \seg{e} to \seg{jk}, \seg{f} to \seg{l}, and \seg{g} to
\seg{m}; and so on. Different colors are used for each space set in the
sequence. If one is given elementary maps $T_i$ such that $T=T_n\dots T_1$ is a
map from a space set to itself, one may use this procedure to define a
two-dimensional space set corresponding to the evolutions of spaces in \X\
under $T$ by writing down a set of allowed faces of each type and a set of
adjacency rules for those faces. The 01-faces and 12-faces should be thought of
as having no symmetries. Each 01-face is identified by its associated color and
orientation with respect to space and time; the 12-faces are identified by the
configuration of their neighboring 01-faces. The 02-faces may all be taken to
be identical and symmetrical.}
\label{embed}
\end{figure}
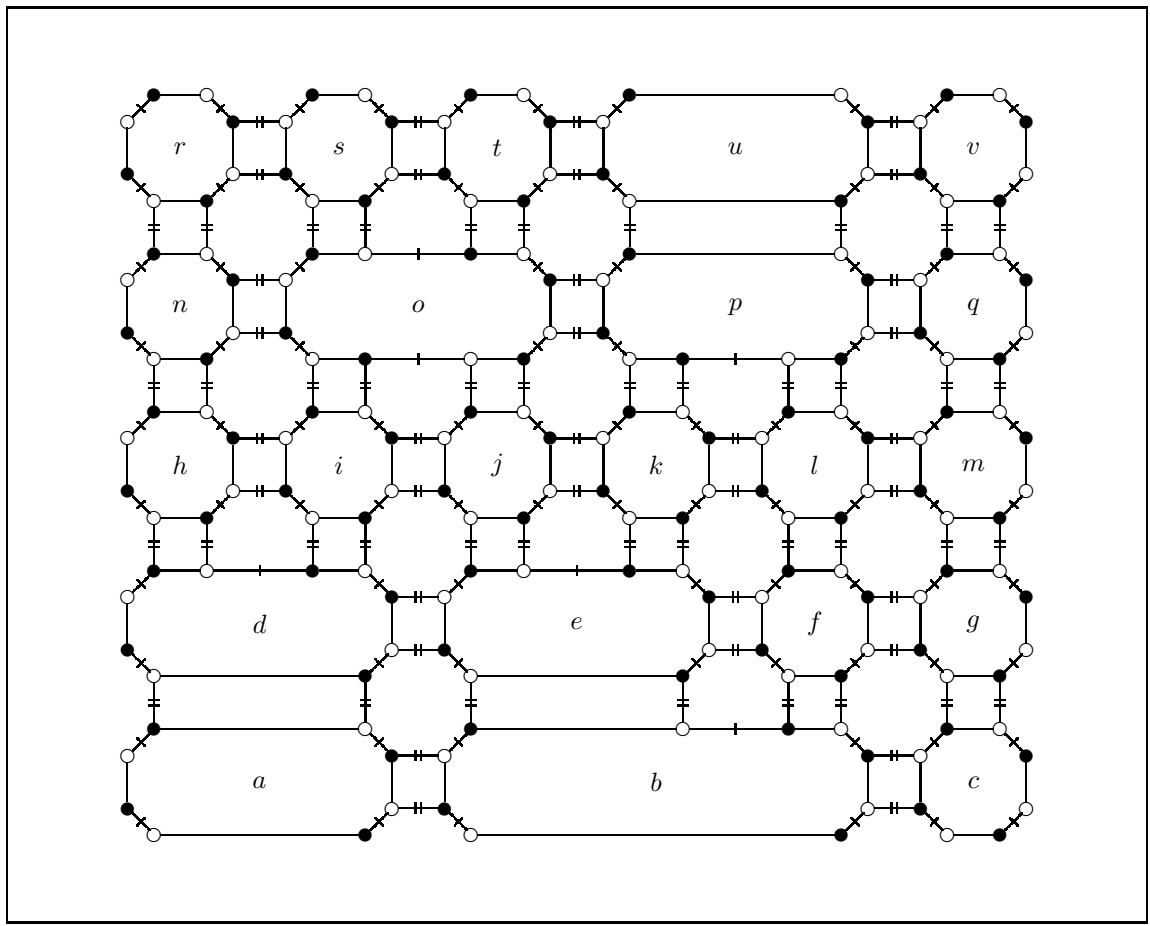

One method for doing this is illustrated in Figure~\ref{embed}.
Each space set $A_i$ corresponds to a horizontal row of 01-faces in the
figure. Each map $T_i:A_{i-1}\mapsto A_i$ corresponds to a row of 12-faces
between the row of 01-faces corresponding to $A_{i-1}$ and the row of 01-faces
corresponding to $A_i$. The entire set of two-dimensional spaces can be
completely specified by writing down the set of allowed 01-faces, 02-faces and
12-faces and the set of adjacency rules for these faces. There is only one kind
of 02-face; it has four sides and can be adjacent to any 01-face or
12-face. There are $n$ distinct rows of 01-faces and of 12-faces (corresponding
to the $n$ space sets $A_1$ through $A_n$ and the $n$ maps $T_1$ through
$T_n$); faces in distinct rows are distinct from one another. In the row of
01-faces associated with $A_i$ there is exactly one type of face for each color
in $A_i$. The face has either 8, 10 or 12 sides, depending on whether the
associated color is mapped into two colors by the maps $T_i$ and $T_{i+1}$. For
each possible arrangement of neighboring 01-faces in $A_{i-1}$ and $A_i$ there
is a distinct 12-face in the row of 12-faces corresponding to $T_i$. The
12-face has six sides if an expansion or contraction is taking place at that
point, and eight sides otherwise.

Given these rules for constructing \Y, it follows that if a row of 01-faces in
a spacetime $Y\in\Y$ corresponds to a space $x\in\X$, then if one moves up $n$
rows of 01-faces in $Y$ one arrives at a row of 01-faces corresponding to
$T(x)$.

Note that each space set $A_i$ is oriented: there is a distinction between left
and right. Also time is oriented: there is a distinction between up and
down. Hence each face should be thought of as having two kinds of arrows
painted on it: an arrow pointing to the left which indicates the direction of
spatial orientation, and an arrow pointing up which indicates the direction of
time. It follows that the faces are completely asymmetrical; one cannot
nontrivially rotate or reflect a face and obtain the same face. (Since these
spacetimes are oriented manifolds, the two arrows are actually more than enough
to achieve this purpose: the direction of one of the arrows may be inferred
from the direction of the other arrow and the orientation of the face.)

Note also that the set of faces and adjacency rules have the following
property: each 01-face can meet several different 12-faces at a 1-edge, but
each 12-face can meet only one 01-face at a 1-edge. Thus the branching in this
description of a set of spaces essentially takes place only at the
01-faces. (Here I am ignoring the 02-faces, which do not play any role apart
from the fact that they each have four sides.)

The figure shows how a neighborhood of a 1+1-dimensional evolution can be
mapped onto a two-dimensional spacetime. However, my goal was to find a set of
two-dimensional spacetimes that was in one-to-one correspondence with a set of
1+1-dimensional evolutions. To make this correspondence work, we need a
slightly more sophisticated notion of ``1+1-dimensional evolution'' than has
been mentioned up to~now.

For example: suppose that $x$ is a space and that the orbit of $x$ under $T$
has size $m$. We usually think of $m$ as the minimum positive integer such that
$T^m(x)=x$. But here we need to think of there being a family of
1+1-dimensional evolutions associated with this orbit. For if we examine all
spacetimes in our set which contain $x$ as a space, we will find one which has
$nm$ rows of 01-faces for each $n\in\N$.

Secondly, suppose that $x$ is equal to \cir{S^k} for some segment $S$ and some
positive integer $k$. If one evolves $x$ using $T$, then after $nm$ time steps
one arrives again at $x$. But now if one wishes to identify this $x$ with the
original $x$ to form a closed spacetime, there are $k$ different ways of making
the identification. Each of the resulting objects must be considered to be a
distinct $1+1$-dimensional evolution.

\begin{figure}
\fbox{\begin{picture}(425,180)\footnotesize\put(0,-10){
\put(50,80){\put(10,0){\circle{5}}\put(30,0){\circle*{5}}
\put(10,40){\circle*{5}}\put(30,40){\circle{5}}
\multiput(12.5,0)(0,40){2}{\line(1,0){15}}
\put(0,10){\circle*{5}}\put(40,10){\circle{5}}
\multiput(0,12.5)(40,0){2}{\line(0,1){15}}\put(0,30){\circle{5}}
\put(40,30){\circle*{5}}
\multiput(30,0)(-30,30){2}{\qbezier(1.77,1.77)(5,5)(8.23,8.23)
\qbezier(6.41,3.59)(5,5)(3.59,6.41)}
\multiput(10,0)(30,30){2}{\qbezier(-1.77,1.77)(-5,5)(-8.23,8.23)
\qbezier(-6.41,3.59)(-5,5)(-3.59,6.41)} \put(20,20){\makebox(0,0){$a$}}
\put(10,-22){\makebox(0,0){$w$}} \put(30,-22){\makebox(0,0){$x$}}
\put(10,62){\makebox(0,0){$w$}} \put(30,62){\makebox(0,0){$x$}}
\put(-22,10){\makebox(0,0){$y$}} \put(-22,30){\makebox(0,0){$z$}}
\put(62,10){\makebox(0,0){$y$}} \put(62,30){\makebox(0,0){$z$}}
\multiput(-2.5,10)(0,20){2}{\multiput(0,0)(60,0){2}{ \put(0,0){\line(-1,0){15}}
\multiput(-6.5,-2)(-2,0){2}{\line(0,1){4}}}}
\multiput(10,-2.5)(20,0){2}{\multiput(0,0)(0,60){2}{ \put(0,0){\line(0,-1){15}}
\multiput(-2,-6.5)(0,-2){2}{\line(1,0){4}}}}} \put(162.5,0){
\multiput(0,50)(0,60){2}{\put(10,0){\circle{5}}\put(30,0){\circle*{5}}
\put(10,40){\circle*{5}}\put(30,40){\circle{5}}
\multiput(12.5,0)(0,40){2}{\line(1,0){15}}
\put(0,10){\circle*{5}}\put(40,10){\circle{5}}
\multiput(0,12.5)(40,0){2}{\line(0,1){15}}\put(0,30){\circle{5}}
\put(40,30){\circle*{5}} \put(20,20){\makebox(0,0){$a$}}
\multiput(30,0)(-30,30){2}{\qbezier(1.77,1.77)(5,5)(8.23,8.23)
\qbezier(6.41,3.59)(5,5)(3.59,6.41)}
\multiput(10,0)(30,30){2}{\qbezier(-1.77,1.77)(-5,5)(-8.23,8.23)
\qbezier(-6.41,3.59)(-5,5)(-3.59,6.41)}
\multiput(-2.5,10)(0,20){2}{\multiput(0,0)(60,0){2}{ \put(0,0){\line(-1,0){15}}
\multiput(-6.5,-2)(-2,0){2}{\line(0,1){4}}}}}
\put(0,50){\put(10,-22){\makebox(0,0){$w$}} \put(30,-22){\makebox(0,0){$x$}}
\put(10,122){\makebox(0,0){$w$}} \put(30,122){\makebox(0,0){$x$}}
\put(-22,10){\makebox(0,0){$y$}} \put(-22,30){\makebox(0,0){$z$}}
\put(62,10){\makebox(0,0){$y$}} \put(62,30){\makebox(0,0){$z$}}
\put(-22,70){\makebox(0,0){$u$}} \put(-22,90){\makebox(0,0){$v$}}
\put(62,70){\makebox(0,0){$u$}} \put(62,90){\makebox(0,0){$v$}}}
\multiput(10,47.5)(20,0){2}{\multiput(0,0)(0,60){3}{
        \put(0,0){\line(0,-1){15}}\multiput(-2,-6.5)(0,-2){2}{\line(1,0){4}}}}}
\put(275,0){\multiput(0,80)(60,0){2}{
        \put(10,0){\circle{5}}\put(30,0){\circle*{5}}
        \put(10,40){\circle*{5}}\put(30,40){\circle{5}}
        \multiput(12.5,0)(0,40){2}{\line(1,0){15}}
        \put(0,10){\circle*{5}}\put(40,10){\circle{5}}
        \multiput(0,12.5)(40,0){2}{\line(0,1){15}}\put(0,30){\circle{5}}
        \put(40,30){\circle*{5}}
        \put(20,20){\makebox(0,0){$a$}}
        \multiput(30,0)(-30,30){2}{\qbezier(1.77,1.77)(5,5)(8.23,8.23)
                \qbezier(6.41,3.59)(5,5)(3.59,6.41)}
        \multiput(10,0)(30,30){2}{\qbezier(-1.77,1.77)(-5,5)(-8.23,8.23)
                \qbezier(-6.41,3.59)(-5,5)(-3.59,6.41)}
\multiput(10,-2.5)(20,0){2}{\multiput(0,0)(0,60){2}{
        \put(0,0){\line(0,-1){15}}\multiput(-2,-6.5)(0,-2){2}{\line(1,0){4}}}}}
\put(0,80){\put(10,-22){\makebox(0,0){$w$}}\put(30,-22){\makebox(0,0){$x$}}
        \put(10,62){\makebox(0,0){$q$}}\put(30,62){\makebox(0,0){$r$}}
        \put(70,-22){\makebox(0,0){$y$}}\put(90,-22){\makebox(0,0){$z$}}
        \put(70,62){\makebox(0,0){$s$}}\put(90,62){\makebox(0,0){$t$}}
        \put(-22,10){\makebox(0,0){$u$}}\put(-22,30){\makebox(0,0){$v$}}
        \put(122,10){\makebox(0,0){$u$}}\put(122,30){\makebox(0,0){$v$}}}
\multiput(-2.5,90)(0,20){2}{\multiput(0,0)(60,0){3}{
        \put(0,0){\line(-1,0){15}}
        \multiput(-6.5,-2)(-2,0){2}{\line(0,1){4}}}}}

}\end{picture}}\linespread{1}\caption{Related 1+1-dimensional evolutions as
represented in two dimensions. In the figure certain pairs of edges are
labelled by the same label; these edges are to be identified to form a closed
graph. The leftmost graph represents the evolution of a space \cir{a} under a
map $T$ which sends this space to itself. If $T$ is applied twice, this also
maps \cir{a} to itself; this evolution is shown in the second graph. If one
starts instead with a space \cir{aa}, $T$ again maps this space to itself;
however, it can do this in two ways, which correspond to two different
evolutions. The right-hand graph represents one of these evolutions if one
identifies $w$ with $q$, $x$ with $r$, $y$ with $s$ and $z$ with $t$; it
represents the other evolution if one identifies $w$ with $s$, $x$ with $t$,
$y$ with $q$ and $z$ with $r$. Note that the center and right-hand graphs are
covering graphs of the left-hand graph in which polygon types and gluing
relations are preserved.}
\label{evolutions}
\end{figure}
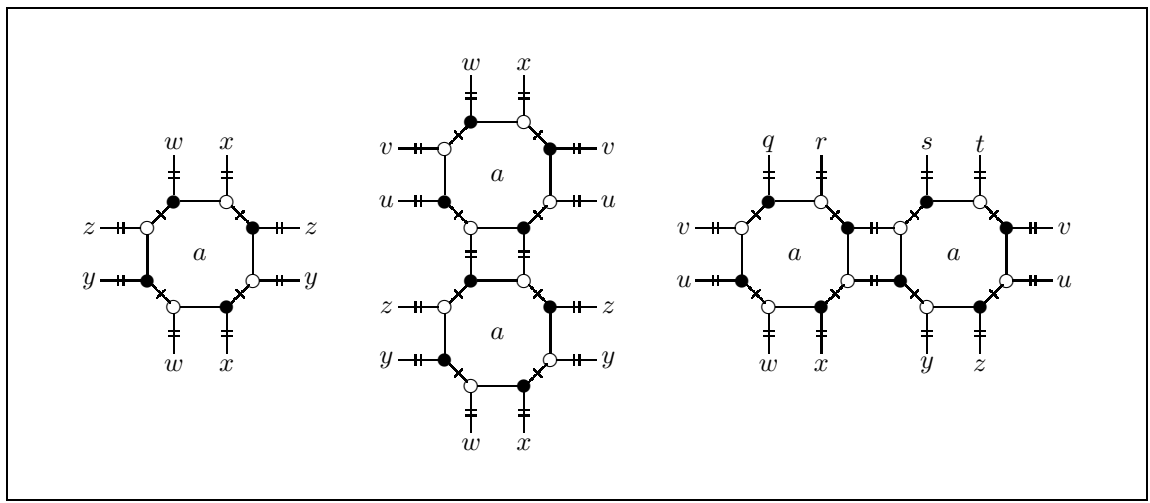

These matters are illustrated in Figure~\ref{evolutions}.
It is easy to see that if we define our set of 1+1-dimensional evolutions in
this way, then there is indeed an invertible map between these evolutions and
the set of two-dimensional spacetimes described in Figure~\ref{embed}.

There are, of course, many other ways in which 1+1-dimensional evolutions may be
described as sets of two-dimensional spacetimes. One would hope that, given an
as-yet-to-be-developed sophisticated notion of equivalence among sets of
two-dimensional spacetimes, it would follow that each of these distinct ways
produce sets of spacetimes that are equivalent.

The above scheme is a useful model to keep in mind when we examine
ways to define sets of two-dimensional spaces in Sections~\ref{hinge}
and~\ref{highdimbialgebras}. Note that in each resulting set of spacetimes
there are finitely many allowed types of faces. This is further evidence that
spacetimes having this property form an important class of objects.

\subsection{Two-dimensional spacetime dynamics}

Each of the spacetimes obtainable by the method of embedding described in the
above section can, of course, be foliated into spaces in such a way that each
space determines the remainder of the spacetime in a local manner. I will say
that any set of spacetimes having this property \emph{contains dynamics}. The
question briefly addressed here is: are there other two-dimensional spacetimes,
not equivalent to these, which contain dynamics?

For example, each closed two-dimensional graph corresponding to one of our
1+1-dimensional evolutions has the topology of a torus. But it is not
immediately obvious that this is a necessary property of two-dimensional
dynamics-containing spacetimes.

Consider, for instance, a sphere. If one foliates this sphere with horizontal
circles and views each circle as a Cauchy surface, then there seemingly is a
problem. The north and south poles are degenerate circles; more importantly, if
time moves up then the south pole has no predecessor and the north pole has no
successor. Hence this foliation is not compatible with our space+time view of a
law of evolution as an invertible map on a set of spaces. However, one may
still imagine in this case that the state of each space determines the
remainder of the spacetime. The set of possible poles would in this case be a
set of disks (rather than circles). Certain disks might evolve to circles;
other circles might evolve to disks.

In this case one could obtain both spherical and toroidal spacetimes. Note,
though, that there could be only finitely many spherical spacetimes in such a
set of spacetimes. This is because a spherical spacetime would be completely
determined by a Cauchy disk that it contained. Each such disk would be an
allowed neighborhood, and the number of allowed neighborhoods is finite.

Hence it is not clear that allowing for spherical spacetimes of this sort adds
anything significant to the picture. I have not yet considered other sorts of
foliations or other spacetime topologies.

If the above reasoning is correct and extends to the rest of the
two-dimensional picture and also to higher dimensions, then it would seem that
time must extend both forwards and backwards from an ``initial'' singularity in
a set of combinatorial spacetimes (except perhaps in the case of finitely many
elements of the spacetime set). This does not fit the usual picture of the big
bang in general relativity. Of course, in general relativity the initial
singularity truly is singular; we don't know what really goes on there.

If this reasoning does hold up in the two-dimensional case, one should not take
it to indicate that our 1+1-dimensional evolutions essentially tell the story
of two-dimensional spacetime dynamics. It is possible to imagine a foliation of
a set of two-dimensional spacetimes into Cauchy surfaces, each of which is a
circle, which has the property that the set of allowed circles is \emph{not} a
subshift of finite type. It will be important to investigate this possibility
further.

\subsection{Space sets defined by systems of faces and hinges}
\label{hinge}

In our matrix descriptions, one-dimensional spaces are described by specifying
a set of allowed edges and vertices and a set of adjacency rules for these
objects. The adjacency rules are described by uniquely specifying the vertex
neighbors of each edge. In a one-dimensional space a vertex has exactly one
edge coming in and one edge going out; here, though, a vertex may have many
edges coming in and many going out. Hence the vertices are branch points, or
``hinges,'' in the space set description.

This suggests that we might try to describe $n$-dimensional space sets by
specifying faces and hinges. As in the one-dimensional case the faces are
one-dimensional and the hinges are zero-dimensional, it is natural to try to
generalize this to the $n$-dimensional case by letting the faces be
$n$-dimensional and the hinges be $(n-1)$-dimensional. In this section I will
describe several such generalizations along these lines.

If $n=2$, then the faces are polygons and the hinges are edges. Forget for a
moment the idea of representing two-dimensional spaces by 2-graphs. Let a set
of spaces be made up of a set of polygons and edges and a set of rules which
specify uniquely how the edges are to be glued to the sides of a
polygon. Suppose there are $k$ edges, represented by the numbers 1 through
$k$. An edge has two ends; label them $a$ and $b$. Hence $5b$ refers to the
$b$th end of edge 5. We may specify our set of spaces by writing down a list of
all allowed polygons and rules for how they attach to the edges. Each polygon
and its associated rules can be specified by a circular list of the edges
adjacent to the polygon. The edges are listed in order as they are encountered
when traversing the boundary of the polygon in a particular direction. It is
necessary to specify which end of the edge is encountered first as we traverse
the boundary. For example, $(3a,5a,3b,8b)$ indicates a polygon with four sides
which hits edge 3 twice. It is the same as $(5a,3b,8b,3a)$, and also as
$(3b,8a,3a,5b)$ (in the second case we are traversing the boundary of the
polygon in the opposite direction; hence the order of encountering the ends of
the edges is reversed). The same circular list may be listed as often as we
like; this simply means that there are a number of differently colored polygons
having the same size and adjacency rules.

If we restrict ourselves to oriented spaces, then we can specify that the order
of edges in our circular lists corresponds to the direction of the orientation
of the associated polygons. If two polygons are to be glued together at edge
$m$, this must be done in such a way that their orientations are
compatible. This means that, when travelling in the direction of the
orientations of these two polygons, one polygon must meet $ma$ first and the
other must meet $mb$ first.

\begin{figure}
\fbox{\begin{picture}(425,180)\scriptsize
\put(30,120){
\put(0,0){\vector(1,0){40}}\put(40,0){\vector(0,1){40}}
\put(40,40){\vector(-1,0){40}}\put(0,40){\vector(0,-1){40}}
\multiput(20,-5)(0,50){2}{\makebox(0,0){1}}
\multiput(-5,20)(50,0){2}{\makebox(0,0){2}}}
\put(10,60){
\put(0,0){\vector(1,0){40}}\put(40,0){\vector(1,0){40}}
\put(40,40){\vector(-1,-1){40}}\put(80,0){\vector(-1,1){40}}
\put(0,-10){\vector(1,0){40}}\put(40,-10){\vector(1,0){40}}
\put(40,-50){\vector(-1,1){40}}\put(80,-10){\vector(-1,-1){40}}
\put(20,-5){\makebox(0,0){1}}\put(60,-5){\makebox(0,0){2}}
\put(16,24){\makebox(0,0){2}}\put(64,24){\makebox(0,0){1}}
\put(16,-34){\makebox(0,0){2}}\put(64,-34){\makebox(0,0){1}}}
\put(127.15,45){\multiput(0,0)(50,50){2}{
\put(0,0){\vector(1,0){40}}\put(40,0){\vector(0,1){40}}
\put(40,40){\vector(-1,0){40}}\put(0,40){\vector(0,-1){40}}
\multiput(20,-5)(0,50){2}{\makebox(0,0){1}}
\multiput(-5,20)(50,0){2}{\makebox(0,0){2}}}
\multiput(50,0)(-50,50){2}{
\put(40,0){\vector(-1,0){40}}\put(40,40){\vector(0,-1){40}}
\put(0,40){\vector(1,0){40}}\put(0,0){\vector(0,1){40}}}
\multiput(20,95)(50,-100){2}{\makebox(0,0){1}}
\multiput(-5,70)(100,-50){2}{\makebox(0,0){2}}}
\put(334.65,90){
\multiput(5,12.07)(-38.28,-92.42){2}{
\put(0,0){\vector(0,1){40}}\put(0,40){\vector(1,1){28.28}}
\put(28.28,68.28){\vector(0,-1){40}}\put(28.28,28.28){\vector(-1,-1){28.28}}}
\multiput(-5,12.07)(38.28,-92.42){2}{
\put(0,0){\vector(0,1){40}}\put(0,40){\vector(-1,1){28.28}}
\put(-28.28,68.28){\vector(0,-1){40}}\put(-28.28,28.28){\vector(1,-1){28.28}}}
\multiput(12.07,5)(-92.42,-38.28){2}{
\put(0,0){\vector(1,0){40}}\put(40,0){\vector(1,1){28.28}}
\put(68.28,28.28){\vector(-1,0){40}}\put(28.28,28.28){\vector(-1,-1){28.28}}}
\multiput(12.07,-5)(-92.42,38.28){2}{
\put(0,0){\vector(1,0){40}}\put(40,0){\vector(1,-1){28.28}}
\put(68.28,-28.28){\vector(-1,0){40}}\put(28.28,-28.28){\vector(-1,1){28.28}}}
\multiput(-32.07,0)(64.14,0){2}{\makebox(0,0){1}}
\multiput(0,-32.07)(0,64.14){2}{\makebox(0,0){1}}
\multiput(-22.67,-22.67)(45.34,45.34){2}{\makebox(0,0){2}}
\multiput(22.67,-22.67)(-45.34,45.34){2}{\makebox(0,0){2}}
\multiput(15.14,70.21)(-30.28,0){2}{\makebox(0,0){2}}
\multiput(70.21,15.14)(0,-30.28){2}{\makebox(0,0){2}}
\multiput(38.38,60.35)(-76.76,0){2}{\makebox(0,0){1}}
\multiput(60.35,38.38)(0,-76.76){2}{\makebox(0,0){1}}
\multiput(15.14,-70.21)(-30.28,0){2}{\makebox(0,0){2}}
\multiput(-70.21,15.14)(0,-30.28){2}{\makebox(0,0){2}}
\multiput(38.38,-60.35)(-76.76,0){2}{\makebox(0,0){1}}
\multiput(-60.35,38.38)(0,-76.76){2}{\makebox(0,0){1}}
}

\end{picture}}\linespread{1}\caption{A two-dimensional space set with no
vertex restrictions. The space set is described by the single polygon shown at
upper left; it may also be described by the single circular list
$(1a,2a,1a,2a)$. The tail of an arrow represents the $a$ end of an edge; the
head represents the $b$ end. No orientability is assumed. To assemble spaces
one makes copies of the polygon and then glues edges together in pairs so that
edge labels and arrows match. Many spaces can be created in this way. For
example, two polygons can be glued together to make a sphere; in this case the
vertices resemble the one pictured at bottom left. The polygons can tile the
plane by letting every vertex have degree four, as shown at center. In fact,
for any $n$ there exists a space in which a vertex has degree $2n$. A vertex
with degree eight is shown at right.}
\label{badpolygons}
\end{figure}
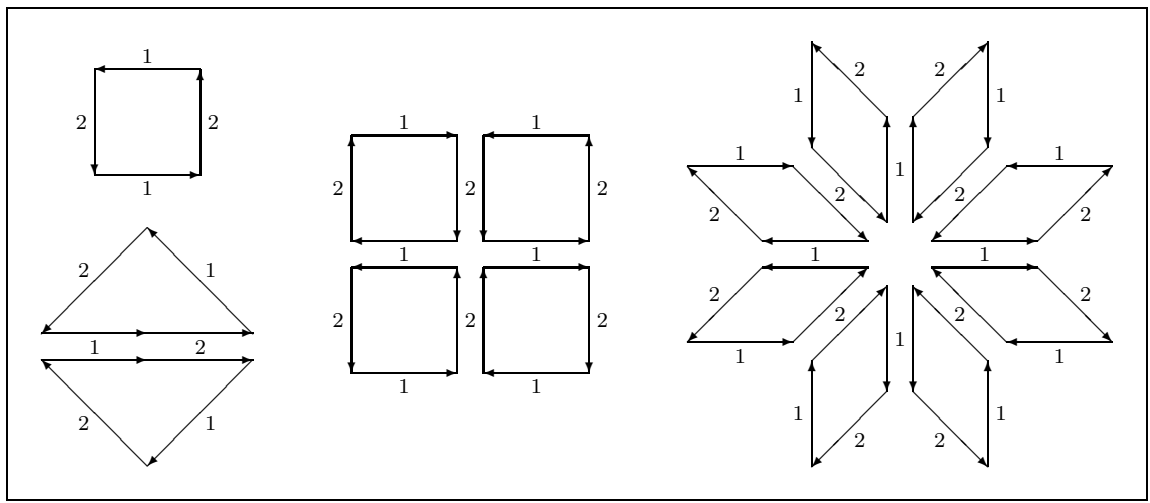

This scheme is illustrated in Figure~\ref{badpolygons}.
It is simple, but it has one feature which I consider to be undesirable. The
undesirable feature is that, while the size of faces is controlled in this
setup, the size of vertices is not. That is: given a set of polygons and
adjacency rules, one can assemble a space containing vertices having
arbitrarily large degree. This is undesirable since it means that a vertex has
infinitely many allowed neighborhoods, hence violating the local simplicity
requirement. It is also undesirable from the point of view of Poincar\'{e}
duality: faces are dual to vertices, so if it is important to allow only
finitely many faces one might suppose that it must be equally important to
allow only finitely many vertices. Finally, note that in our scheme for
constructing a spacetime description of 1+1-dimensional dynamics the size of
vertices is restricted.

I claim, then, that in order to define space sets of the desired type, one must
have control over the size of faces \emph{and} of vertices. This applies in $n$
dimensions.

Another scheme which suggests itself is derived from $n$-graphs. In the
two-dimensional case we have three types of polygons and three types of
edges. We may carry out the same procedure as above with these polygons: the
edges of type $i$ may be numbered by integers from 1 to $k_i$; then circular
lists may be provided for polygons of each type with the lowest-numbered edge type
given first (e.g., the list $(3a,5a,3b,8b)_{01}$ represents a 01-polygon which
hits the third 0-edge, then the fifth 1-edge, then the third 0-edge again, then
the eighth 1-edge). A set of circular lists of this sort constitutes a
description of a set of spaces.

If we restrict ourselves to sets of oriented spaces, this simplifies
somewhat. Again the order of our circular lists corresponds to the direction of
orientation of the polygons. In 2-graphs, the polygons that hit a given edge
may be divided into two groups such that any gluing at the edge attaches a
polygon from one group to a polygon from the other group. For example, at a
0-edge the two groups are the 01-polygons and the 02-polygons. We may specify
that each 01-polygon always hits the $a$ end of 0-edges first as one travels
around in the direction of orientation. It follows that each 02-polygon must
hit the $b$ ends of 0-edges first as one travels around in the direction of
orientation (so that the orientations match when gluing a 01-polygon to an
02-polygon at a 0-edge). The same goes for 1-edges and 2-edges. The result is
that the $a$'s and $b$'s in our circular lists are implicit, and can be
omitted.

\begin{figure}
\fbox{\setlength{\unitlength}{.9pt}\begin{picture}(472.22,207.84)\scriptsize

\put(59.30,38.97){\multiput(0,0)(0,64.95){3}{
\put(-30,0){\circle{5}}\put(-15,25.98){\circle*{5}}\put(15,25.98){\circle{5}}
\put(30,0){\circle*{5}}\put(15,-25.98){\circle{5}}\put(-15,-25.98){\circle*{5}}
\qbezier(-28.75,2.17)(-22.5,12.99)(-16.25,23.81)\put(-12.5,25.98){\line(1,0){25}}\qbezier(16.25,23.81)(22.5,12.99)(28.75,2.17)\qbezier(28.75,-2.17)(22.5,-12.99)(16.25,-23.81)\put(12.5,-25.98){\line(-1,0){25}}\qbezier(-16.25,-23.81)(-22.5,-12.99)(-28.75,-2.17)}
\put(0,0){
        \put(0,-27.98){\line(0,1){4}}
        \qbezier(-24.23,13.99)(-22.5,12.99)(-20.77,11.99)
        \qbezier(24.23,13.99)(22.5,12.99)(20.77,11.99)
        \put(0,-17.98){\makebox(0,0){1}}
        \put(-22.5,12.99){\put(6.92,-4){\makebox(0,0){3}}}
        \put(22.5,12.99){\put(-6.92,-4){\makebox(0,0){2}}}
\put(0,64.95){
        \multiput(-1,23.98)(2,0){2}{\line(0,1){4}}
        \multiput(22,-13.855)(1,1.73){2}{\qbezier(1.73,-1)(0,0)(-1.73,1)}
        \multiput(-22,-13.855)(-1,1.73){2}{\qbezier(-1.73,-1)(0,0)(1.73,1)}
\put(0,64.95){
        \multiput(-1,-23.98)(2,0){2}{\line(0,-1){4}}
        \qbezier(24.23,-13.99)(22.5,-12.99)(20.77,-11.99)
        \qbezier(-24.23,-13.99)(-22.5,-12.99)(-20.77,-11.99)
        \put(0,17.98){\makebox(0,0){1}}
        \put(-22.5,-12.99){\put(6.92,4){\makebox(0,0){2}}}
        \put(22.5,-12.99){\put(-6.92,4){\makebox(0,0){3}}}
        \put(0,27.98){\line(0,-1){4}}
        \multiput(22,13.855)(1,-1.73){2}{\qbezier(1.73,1)(0,0)(-1.73,-1)}
        \multiput(-22,13.855)(-1,-1.73){2}{\qbezier(-1.73,1)(0,0)(1.73,-1)}}}}}

\put(277.92,38.97){\multiput(0,0)(45,77.94){2}{\multiput(0,0)(90,0){2}{
\put(-30,0){\circle{5}}\put(-15,25.98){\circle*{5}}\put(15,25.98){\circle{5}}
\put(30,0){\circle*{5}}\put(15,-25.98){\circle{5}}\put(-15,-25.98){\circle*{5}}
\qbezier(-28.75,2.17)(-22.5,12.99)(-16.25,23.81)\put(-12.5,25.98){\line(1,0){25}}\qbezier(16.25,23.81)(22.5,12.99)(28.75,2.17)\qbezier(28.75,-2.17)(22.5,-12.99)(16.25,-23.81)\put(12.5,-25.98){\line(-1,0){25}}\qbezier(-16.25,-23.81)(-22.5,-12.99)(-28.75,-2.17)
\put(0,0){
        \put(0,-27.98){\line(0,1){4}}
        \qbezier(-24.23,13.99)(-22.5,12.99)(-20.77,11.99)
        \qbezier(24.23,13.99)(22.5,12.99)(20.77,11.99)
        \put(0,-17.98){\makebox(0,0){1}}
        \put(-22.5,12.99){\put(6.92,-4){\makebox(0,0){3}}}
        \put(22.5,12.99){\put(-6.92,-4){\makebox(0,0){2}}}}}}
\multiput(0,51.96)(45,77.94){2}{\multiput(0,0)(90,0){2}{
\put(-30,0){\circle{5}}\put(-15,25.98){\circle*{5}}\put(15,25.98){\circle{5}}
\put(30,0){\circle*{5}}\put(15,-25.98){\circle{5}}\put(-15,-25.98){\circle*{5}}
\qbezier(-28.75,2.17)(-22.5,12.99)(-16.25,23.81)\put(-12.5,25.98){\line(1,0){25}}\qbezier(16.25,23.81)(22.5,12.99)(28.75,2.17)\qbezier(28.75,-2.17)(22.5,-12.99)(16.25,-23.81)\put(12.5,-25.98){\line(-1,0){25}}\qbezier(-16.25,-23.81)(-22.5,-12.99)(-28.75,-2.17)
\put(0,0){
        \multiput(-1,23.98)(2,0){2}{\line(0,1){4}}
        \multiput(22,-13.855)(1,1.73){2}{\qbezier(1.73,-1)(0,0)(-1.73,1)}
        \multiput(-22,-13.855)(-1,1.73){2}{\qbezier(-1.73,-1)(0,0)(1.73,1)}}}}
\multiput(45,25.98)(-45,77.94){2}{\multiput(0,0)(90,0){2}{
\put(-30,0){\circle{5}}\put(-15,25.98){\circle*{5}}\put(15,25.98){\circle{5}}
\put(30,0){\circle*{5}}\put(15,-25.98){\circle{5}}\put(-15,-25.98){\circle*{5}}
\qbezier(-28.75,2.17)(-22.5,12.99)(-16.25,23.81)\put(-12.5,25.98){\line(1,0){25}}\qbezier(16.25,23.81)(22.5,12.99)(28.75,2.17)\qbezier(28.75,-2.17)(22.5,-12.99)(16.25,-23.81)\put(12.5,-25.98){\line(-1,0){25}}\qbezier(-16.25,-23.81)(-22.5,-12.99)(-28.75,-2.17)
\put(0,0){
        \multiput(-1,-23.98)(2,0){2}{\line(0,-1){4}}
        \qbezier(24.23,-13.99)(22.5,-12.99)(20.77,-11.99)
        \qbezier(-24.23,-13.99)(-22.5,-12.99)(-20.77,-11.99)
        \put(0,17.98){\makebox(0,0){1}}
        \put(-22.5,-12.99){\put(6.92,4){\makebox(0,0){2}}}
        \put(22.5,-12.99){\put(-6.92,4){\makebox(0,0){3}}}
        \put(0,27.98){\line(0,-1){4}}
        \multiput(22,13.855)(1,-1.73){2}{\qbezier(1.73,1)(0,0)(-1.73,-1)}
        \multiput(-22,13.855)(-1,-1.73){2}{\qbezier(-1.73,1)(0,0)(1.73,-1)}}}}}

\put(138.61,43.92){\put(0,0){\circle{5}}\put(60,0){\circle*{5}}\put(0,60){\circle*{5}}
\put(60,60){\circle{5}}\put(0,120){\circle{5}}\put(60,120){\circle*{5}}
\multiput(2.5,0)(0,60){3}{\line(1,0){55}}
\multiput(0,0)(60,0){2}{\put(0,2.5){\line(0,1){55}}\multiput(-2,29)(0,2){2}{\line(1,0){4}}}
\multiput(0,60)(60,0){2}{\put(0,2.5){\line(0,1){55}}\put(-2,30){\line(1,0){4}}}
\put(1.12,2.24){\line(1,2){57.76}}\put(58.88,2.24){\line(-1,2){57.76}}
\put(30,-8){\makebox(0,0){1}}
\put(-8,90){\makebox(0,0){3}}
\put(68,90){\makebox(0,0){2}}
\put(30,-2){\line(0,1){4}}
\multiput(29,118)(2,0){2}{\line(0,1){4}}}

\end{picture}}\linespread{1}\caption{Defining space sets on $n$-graphs by
listing the allowed $n$-faces and their adjacency relations. Here an oriented
two-dimensional space set is described by the three polygonal faces at
left. The hinges are edges. There are three 1-edge hinges (numbered 1 to 3),
one 0-edge hinge and one 2-edge hinge. The space set may be described by the
circular lists $(1,1,1,2,1,3)_{01},$ $(1,1,1,1,1,1)_{02}$ and
$(1,1,2,1,3,1)_{12}$. The graph at center is an example of a space in this set
which has three faces, one of each type. Here $F-V/2=0$, so this is a
torus. The polygons may be used to tile the plane as shown at right.}
\label{threepolygons}
\end{figure}
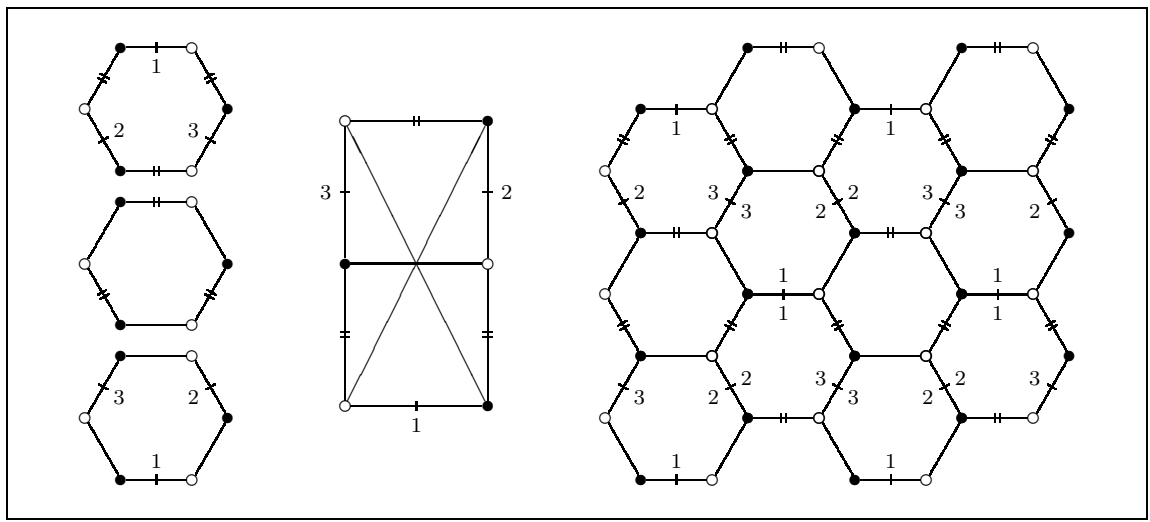

This scheme is illustrated in Figure~\ref{threepolygons}.
Note that in this case we do not have the undesirable feature of the previous
scheme, since each vertex has degree 3.

I have thus far omitted one important detail from this discussion, which has to
do with symmetry. Suppose in the case of oriented 2-graphs that we have a
four-sided 01-polygon given by $(1,1,1,1)_{01}$. Should we consider this
polygon to possess a nontrivial symmetry, or not? This is up to us to
choose. On the one hand, we may suppose that the only relevant properties of
polygons are their sizes, orientations and adjacency relations. Then the above
polygon has a nontrivial symmetry: you may rotate it by 180 degrees and obtain
the same polygon. On the other hand, we may suppose that each polygon is marked
in some way: e.g., imagine that the polygons are white except for a black dot
near one of the edges. In the oriented case this insures that the polygon has
no nontrivial symmetries. A third alternative is to suppose that some polygons
are marked and some are not.

Consider two space sets \X\ and \Y\ corresponding to the same set of circular
lists of allowed polygons. Let the properties of polygons in \X\ be determined
by sizes, orientations and adjacency relations only. Let the polygons in \Y\
each contain a black dot near one of the edges, as described above. Then
there may be polygons in \X\ with nontrivial symmetries, but there are no such
polygons in \Y. Given $x\in\X$ we may obtain a corresponding space $y\in\Y$ by
replacing each polygon $p$ in $x$ with a corresponding polygon $p'$, where $p'$
now has a dot near one of its edges. If $p$ contains nontrivial symmetries,
then there is more than one way to place the dot in $p'$. Hence if $x$ contains
a polygon with nontrivial symmetries then it corresponds to more than one space
in \Y. It follows that \X\ and \Y\ are typically quite different.

For example, the 02-face in the space set pictured in
Figure~\ref{threepolygons} may be considered to be symmetrical under 120-degree
rotations, or it may be considered to be asymmetrical. In the former case there
is only one torus and one plane tiling of the sort shown in the figure. In the
latter case there are three tori and infinitely many plane tilings.

In my research I have run into difficulties when trying to allow
symmetries. The difficulties arise when trying to write down local equivalence
transformations. But I am new at this, and it is quite possible that these
difficulties are imaginary or easily surmountable. Symmetries seem to be quite
important in physics, and it would be nice to know how to deal with them
here. However, as a result of these difficulties I have focused most of my
efforts on the case where all polygons are taken to have no nontrivial
symmetries. Perhaps this is the natural case to treat first. This was my
procedure in one dimension, where oriented space sets have no symmetries; it
turned out there that adding symmetries to the space set (in the unoriented
case) required an extra element to be added to the bialgebra. Thus in a sense
symmetries made the problem more complex; perhaps this is true in general.

From now on, then, we will assume that all polygons have no nontrivial
symmetries. Note that the sets of two-dimensional spaces corresponding to
oriented 1+1-dimensional dynamics have this property.

Consider again our formulation of a two-dimensional space set as being
determined by a list of polygons and edges of all three types, in which the
edges are branch points in the sense that many polygonal sides may be adjacent
to the same edge, but only one edge may be adjacent to a given polygonal
side. This may be easily generalized to $n$ dimensions. In that case, one
provides a list of $n$-faces and $(n-1)$-hinges. Each $(n-1)$-facet of an
$n$-face is mapped by a graph isomorphism to an $(n-1)$-hinge.

When $n=2$ there are always two possible isomorphisms; hence our $a$ and $b$
notation. In general the situation is more complex. However, it is simplified
if we choose to assume that each hinge contains no nontrivial symmetries. Then
we may place a special mark on one of the vertices of each hinge. The
isomorphism is then determined by indicating the vertex of the $(n-1)$-facet
that is to be mapped to the marked vertex of the hinge.

Again the formulation is not complete until we decide whether or not to treat our
$n$-faces as possessing nontrivial symmetries.

I will now add two wrinkles to this scheme. Both wrinkles are present in the
two-dimensional space sets corresponding to 1+1-dimensional dynamics that we
described earlier.

First consider the two-dimensional case. The first wrinkle is to specify that
each 02-face has only four sides and no other significant properties. Then a
space set is described by giving circular lists for the 01-faces and 12-faces.
The only hinges in this case are located at the 1-edges. One is allowed to
assemble a space from the 01-faces and 12-faces in any manner such that the
adjacency rules at 1-edges are followed and such that each resulting 02-face
has four sides. (One may consider there to be only one type of 02-face,
symmetrical under a 180-degree rotation; alternatively, one may consider that
there is a unique asymmetrical 02-face for each way that a 02-face can be
surrounded by the 01-faces and 12-faces. It does not matter; in either case the
only role played by 02-faces is that they have four sides.)

This restriction may seem arbitrary at first glance. However, it is based on
the following general construction.

In what follows I will use the notation ``$k$-face'' to mean a $k$-dimensional
face in the usual way. The notation ``$j$-component'' will mean a component of
the subgraph obtained from the original $n$-graph by deleting all
$j$-colored edges. The notation ``$jk$-component'' will mean a component of the
subgraph obtained from the original $n$-graph by deleting all $j$-colored
and $k$-colored edges. The notation ``$ij$-polygon'' will mean a component of
the subgraph obtained from the original $n$-graph by deleting all edges
whose colors are not $i$ or $j$. Hence if $n=2$ an $ij$-polygon is the
same as what we previously called an $ij$-face (though these could also
now be referred to as $k$-components). The notation ``$j$-edge'' will continue to
mean a $j$-colored edge.

Now instead of considering the set of all $n$-graphs, consider the set of
$n$-graphs in which each $ij$-polygon has four sides whenever $|i-j|>1$. I
will call these the \emph{Poincar\'{e} n-graphs}. They may be viewed as
pseudomanifolds in the following way.

If $n=1$ a closed graph represents a circle or a line as before, only now we
view the 0-edges as 1-faces (``edges'') and the 1-edges as 0-faces
(``vertices''). If $n=2$ then the 2-components (i.e., the 01-polygons) are
taken to be the two-dimensional faces whose one-dimensional facets are glued
together in pairs to form the two-dimensional manifold. The one-dimensional
facets of a 2-component are its 0-edges. The 2-edges indicate the isomorphism
between these facets. This is why each 02-polygon has four sides: two of them
are the 0-edges to be glued together, and the gluing is accomplished by
shrinking the 2-edges to points and performing the resulting identification of
the 0-edges. It follows that each 02-polygon represents a 1-face (``edge'') of
the manifold: each edge is in two 2-faces, and this structure is indicated
explicitly by the 02-polygon. Similarly, each 12-polygon represents a 0-face
(``vertex'') of the manifold: each vertex is surrounded by a circle of 2-faces,
and again this structure is indicated explicitly by the 12-polygon. Said
another way, the $k$-faces of a Poincar\'{e} 2-graph are given by its
$k$-components.

The construction generalizes inductively to arbitrary $n$. Each $n$-component
of an $n$-graph is an $(n-1)$-graph, which therefore represents an
$(n-1)$-dimensional pseudo\-manifold. The $n$-faces of the desired
$n$-dimensional pseudomanifold are taken to be the cones over these
$(n-1)$-dimensional pseudomanifolds. The $(n-1)$-dimensional facets of these
$n$-faces are represented by $(n-1,n)$-components. These are $(n-2)$-graphs,
which therefore represent $(n-2)$-dimensional pseudomanifolds; the facets are
the cones over these pseudomanifolds. The requirement that each $jn$-polygon
has four sides for $j\leq n-2$ implies that the $n$-edges join together two
facets whose graphs are isomorphic. The isomorphism between facets is given by
shrinking the $n$-edges to points and identifying the vertices and edges of the
subgraphs in the indicated manner. When the $n$-faces are glued together at
$(n-1)$-facets via these isomorphisms, the result is an $n$-dimensional
pseudomanifold. It turns out that it is natural to consider the $k$-components
to be the $k$-faces of this pseudomanifold.

Note that we may also view a Poincar\'{e} $n$-graph as an $n$-graph in
the usual way. That is: there are two ways to construct pseudomanifolds from
these graphs. The resulting pseudomanifolds are homeomorphic. From the
$n$-graph point of view the $k$-faces of the Poincar\'{e} graph are now
blown up into $n$-faces; each such $n$-face may be seen to represent
topologically the $n$-dimensional neighborhood of the corresponding $k$-face.

I believe that it is possible to generalize the definition of Poincar\'{e}
$n$-graphs slightly by allowing the $jn$-polygons ($j\leq n-2$) to have
either two or four sides. What this means in practice is that an
$(n-1)$-dimensional facet may be glued to itself. I think that these
self-gluings may be precisely defined in a consistent manner, though I have not
worked out the details in higher dimensions and will not use this
generalization in what follows.

\begin{figure}
\fbox{\begin{picture}(425,180)\scriptsize
\multiput(48,20)(189,0){2}{\multiput(0,0)(0,100){2}{
\qbezier(0,20)(30,10)(60,0)\qbezier(60,0)(90,0)(120,0)\qbezier(120,0)(130,10)(140,20)
\qbezier(140,20)(110,30)(80,40)\qbezier(80,40)(50,40)(20,40)\qbezier(20,40)(10,30)(0,20)}
\put(0,20){\line(0,1){100}}\put(60,0){\line(0,1){100}}\put(120,0){\line(0,1){100}}
\put(140,20){\line(0,1){100}}\put(80,40){\line(0,1){100}}\put(20,40){\line(0,1){100}}}
\put(48,20){\multiput(0,0)(0,100){2}{
\put(90,-2){\line(0,1){4}}\qbezier(109.36,28.08)(110,30)(110.64,31.92)\qbezier(11.42,28.58)(10,30)(8.58,31,42)}
\put(0,50){\put(0,20){\multiput(-2,-2)(0,2){3}{\line(1,0){4}}}
\put(60,0){\multiput(-2,-2)(0,2){3}{\line(1,0){4}}}
\put(120,0){\multiput(-2,-2)(0,2){3}{\line(1,0){4}}}
\put(140,20){\multiput(-2,-2)(0,2){3}{\line(1,0){4}}}
\put(80,40){\multiput(-2,-2)(0,2){3}{\line(1,0){4}}}
\put(20,40){\multiput(-2,-2)(0,2){3}{\line(1,0){4}}}}}
\put(237,20){\multiput(0,0)(0,100){2}{
\multiput(89,-2)(2,0){2}{\line(0,1){4}}\qbezier(110.31,27.76)(110.95,29.68)(111.59,31.60)\qbezier(108.41,28.40)(109.05,30.32)(109.69,32.24)\qbezier(12.13,29.29)(10.71,30.71)(9.29,32.13)\qbezier(10.71,27.87)(9.29,29.29)(7.87,30.71)
\qbezier(27.44,8.72)(28.08,10.64)(28.72,12.56)\qbezier(29.36,8.08)(30,10)(30.64,11.92)\qbezier(31.28,7.44)(31.92,9.36)(32.56,11.28)\qbezier(127.16,10)(128.58,8.58)(130,7.16)\qbezier(128.58,11.42)(130,10)(131.42,8.58)\qbezier(130,12.84)(131.42,11.42)(132.84,10)\multiput(48,38)(2,0){3}{\line(0,1){4}}}}

\end{picture}}\linespread{1}\caption{A 2-face and 1-face from a
three-dimensional Poincar\'{e} graph. A 2-face is a two-dimensional boundary
between two 3-faces. It is bordered not only by the 3-faces, but also by the
1-faces and 0-faces which constitute the polygonal border of the region. This
can be seen explicitly in the left-hand figure. The 3-faces intersect the
2-face at the top and bottom 01-polygons. These polygons in this case have six
sides, which means that they represent triangles (the 0-edges represent sides
of the triangle; the 1-edges represent vertices of the triangle). Hence the
2-face meets three 1-faces (at 03-polygons) and three 0-faces (at
13-polygons). Now for the dual case. A 1-face is an edge. It is bounded by two
vertices at either end, as well as by the 3-faces which meet at that edge, and
by an equal number of two-dimensional boundaries between those 3-faces. This
can be seen explicitly in the right-hand figure. The two 0-faces meet the
1-face at the top and bottom 23-polygons. In this case there are three 3-faces
meeting the 1-face at 02-polygons, and three 2-faces meeting the 1-face at
03-polygons. The structure of these faces, and the duality relation between
them, is thus made very readily apparent by this graphical representation.}
\label{poincarethreed}
\end{figure}
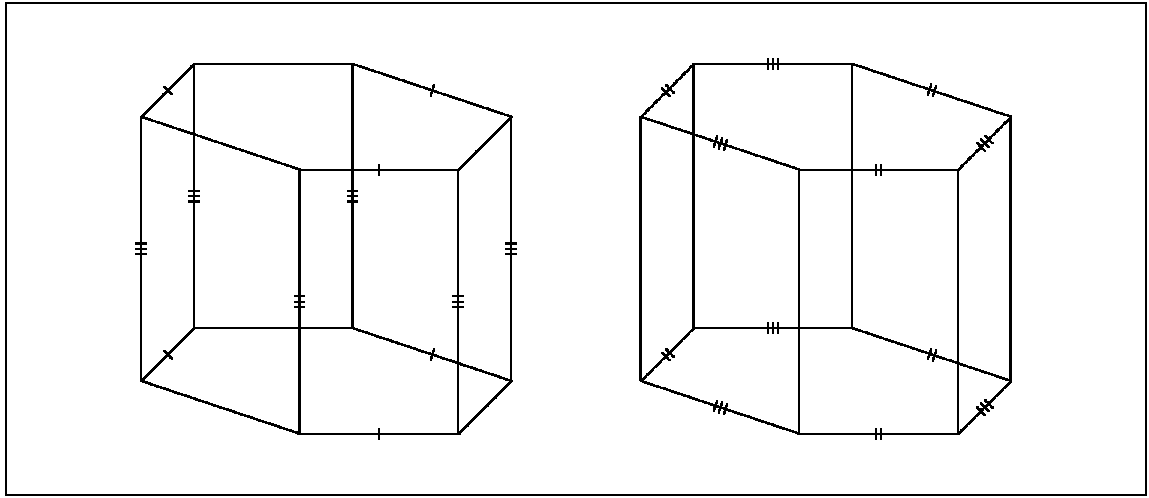

Though an extra complication is involved in the definition of Poincar\'{e}
$n$-graphs (as compared to $n$-graphs), the former have the property
that they behave very nicely with regards to Poincar\'{e} duality (hence their
name). This is because the Poincar\'{e} dual of a Poincar\'{e} $n$-graph is
given simply by permuting the colors of the edges in the graph (the permutation
sends $i$ to $n-i$ for each $i$). Graphically speaking, a $k$-face of a
Poincar\'{e} $n$-graph looks like an $n-k$ face of that graph; hence the
representation of duality in these graphs is quite explicit. For an example,
see Figure~\ref{poincarethreed}.

My first wrinkle, then, in $n$ dimensions, is to use Poincar\'{e}
$n$-graphs rather than $n$-graphs. With $n$-graphs we defined space
sets by describing all allowed $n$-faces and their connectivity
properties. There were $n+1$ different types of $n$-faces (for these were
represented by $i$-components, and there are $n+1$ possible values of
$i$). Here we will define space sets by describing all allowed $n$-faces
\emph{and} all allowed 0-faces. There is only one type of $n$-face (an
$n$-component) and one type of 0-face (a 0-component); hence there are fewer
types of objects to describe. An $n$-face and a 0-face meet at
$0n$-components. Hence these components may be used as the hinges when
describing connectivity. (Note that in two dimensions these components are just
the 1-edges.)

The second wrinkle is to modify the Poincar\'{e}-graph picture by disallowing a
certain type of branching. We do this by imposing the requirement that each of
the hinge-components described in the above paragraph may be adjacent to any
number of $n$-faces, but may only be adjacent to one 0-face. Thus the 0-faces
that border a given $n$-face are uniquely determined, while the $n$-faces that
border a given 0-face are not. Since each hinge is now identified with the
unique $0n$-component in a 0-face that attaches to the hinge, we may think of
the hinge as being part of the 0-face.

Note that our representation of oriented one-dimensional space sets by a matrix
description $A$ has this property. The $n$-faces are the edges of
\g{A}; the 0-faces are the vertices of \g{A}. Using Poincar\'{e} graphs we
would represent this by letting the edges of \g{A} be 1-edges and the vertices
of \g{A} be 0-edges. In this case the branch points ($0n$-components) are
vertices of the Poincar\'{e} graph. The adjacency rules specify that any number
of 1-faces may be adjacent to a 0-face at a given vertex of the graph, but that
only one 0-face is adjacent to a given 1-face at a vertex.

Our representation of 1+1-dimensional dynamics by two-dimensional space sets
also took exactly this form, except that the roles of 0-faces and $n$-faces
were reversed. It follows from Poincar\'{e} duality that such role reversal
does not matter; the choice of roles for 0-faces and for $n$-faces is only a
matter of convention.

In this section I have described three methods of describing $n$-dimensional
space sets using faces and hinges (apart from the method in which there was no
control over vertex size). It is possible that all three methods will turn out
to be equivalent from the point of view of combinatorial dynamics. I have
presented all of them since at present I do not know whether they are
equivalent or not, and since even if they are equivalent it may turn out that
one is easier to use than another. In my research I am currently focusing on
the third of these methods, since less information is required to specify a
space set using this method and it is therefore easier to implement on a
computer.

\begin{figure}
\fbox{\setlength{\unitlength}{.95pt}\begin{picture}(447.37,180)\scriptsize
\put(41.21,130){
\put(0,0){\circle*{5}}\put(30,30){\circle*{5}}\put(0,30){\circle{5}}
\put(30,0){\circle{5}}\multiput(2.5,0)(0,30){2}{\put(0,0){\line(1,0){25}}\put(12.5,-2){\line(0,1){4}}}
\multiput(0,2.5)(30,0){2}{\put(0,0){\line(0,1){25}}\multiput(-2,11.5)(0,2){2}{\line(1,0){4}}}\put(15,-7){\makebox(0,0){1}}\put(15,37){\makebox(0,0){2}}}
\put(56.21,56.21){
\put(-15,36.21){\circle{5}}\put(15,36.21){\circle*{5}}
\put(36.21,15){\circle{5}}\put(36.21,-15){\circle*{5}}
\put(15,-36.21){\circle{5}}\put(-15,-36.21){\circle*{5}}
\put(-36.21,-15){\circle{5}}\put(-36.21,15){\circle*{5}}
\multiput(-12.5,36.21)(0,-72.42){2}{\put(0,0){\line(1,0){25}}\put(12.5,-2){\line(0,1){4}}}
\multiput(36.21,-12.5)(-72.42,0){2}{\put(0,0){\line(0,1){25}}\put(-2,12.5){\line(1,0){4}}}
\put(-34.44,16.77){\line(1,1){17.68}}\put(-34.44,-16.77){\line(1,-1){17.68}}
\put(34.44,-16.77){\line(-1,-1){17.68}}\put(34.44,16.77){\line(-1,1){17.68}}
\multiput(0,43.21)(0,-86.42){2}{\makebox(0,0){1}}
\multiput(43.21,0)(-86.42,0){2}{\makebox(0,0){2}}
\put(0,43.21){\circle{8.5}}}
\put(361.16,90){\put(73.21,0){\circle{8.5}}}
\put(191.16,90){\put(-73.21,0){\circle{8.5}}}
\multiput(361.16,90)(-170,0){2}{
\put(-15,36.21){\circle{5}}\put(15,36.21){\circle*{5}}
\put(36.21,15){\circle{5}}\put(36.21,-15){\circle*{5}}
\put(15,-36.21){\circle{5}}\put(-15,-36.21){\circle*{5}}
\put(-36.21,-15){\circle{5}}\put(-36.21,15){\circle*{5}}
\multiput(-12.5,36.21)(0,-72.42){2}{\put(0,0){\line(1,0){25}}\put(12.5,-2){\line(0,1){4}}}
\multiput(36.21,-12.5)(-72.42,0){2}{\put(0,0){\line(0,1){25}}\put(-2,12.5){\line(1,0){4}}}
\qbezier(-34.44,16.77)(-25.61,25.61)(-16.77,34.44)
\qbezier(-34.44,-16.77)(-25.61,-25.61)(-16.77,-34.44)
\qbezier(34.44,-16.77)(25.61,-25.61)(16.77,-34.44)
\qbezier(34.44,16.77)(25.61,25.61)(16.77,34.44)
\multiput(0,29.21)(0,-58.42){2}{\makebox(0,0){1}}
\put(0,29.21){\circle{8.5}}
\multiput(29.21,0)(-58.42,0){2}{\makebox(0,0){2}}
\multiput(0,73.21)(0,-146.42){2}{\makebox(0,0){2}}
\multiput(73.21,0)(-146.42,0){2}{\makebox(0,0){1}}
\multiput(-15,38.71)(30,0){2}{\multiput(0,0)(0,-102.42){2}{\put(0,0){\line(0,1){25}}\multiput(-2,11.5)(0,2){2}{\line(1,0){4}}}}
\multiput(38.71,-15)(0,30){2}{\multiput(0,0)(-102.42,0){2}{\put(0,0){\line(1,0){25}}\multiput(11.5,-2)(2,0){2}{\line(0,1){4}}}}
\multiput(-66.21,-12.5)(132.42,0){2}{\put(0,0){\line(0,1){25}}\put(-2,12.5){\line(1,0){4}}}
\multiput(-12.5,-66.21)(0,132.42){2}{\put(0,0){\line(1,0){25}}\put(12.5,-2){\line(0,1){4}}}
\put(-15,66.21){\circle*{5}}\put(15,66.21){\circle{5}}
\put(66.21,15){\circle*{5}}\put(66.21,-15){\circle{5}}
\put(15,-66.21){\circle*{5}}\put(-15,-66.21){\circle{5}}
\put(-66.21,-15){\circle*{5}}\put(-66.21,15){\circle{5}}
\qbezier(-64.44,16.77)(-40.605,40.605)(-16.77,64.44)
\qbezier(64.44,16.77)(40.605,40.605)(16.77,64.44)
\qbezier(64.44,-16.77)(40.605,-40.605)(16.77,-64.44)
\qbezier(-64.44,-16.77)(-40.605,-40.605)(-16.77,-64.44)}

\end{picture}}\linespread{1}\caption{An oriented two-dimensional space set
defined by the third method. The space set is described by the 0-face and
2-face shown at left (or by the set of circular lists containing the single
element $((1,1),(1,2),(1,1),(1,2))$). An edge number on the 2-face is circled
in order to make the face asymmetrical. One may verify that there is only one
2-graph such that each 12-polygon and 02-polygon has four sides and each
01-polygon has eight sides. It follows that there are two spaces in the set;
these are shown.}
\label{poincarespaceset}
\end{figure}
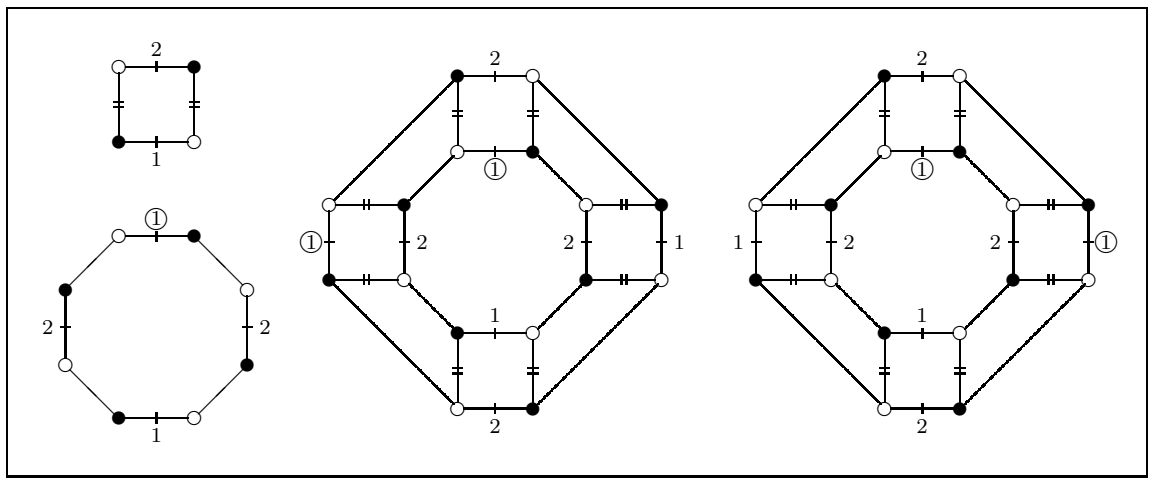

For example, in the oriented two-dimensional case a space set of this sort may
be described by a set of circular lists of ordered pairs. Each list corresponds
to a 2-face; each ordered pair corresponds to a 1-edge in the 2-face. The edge
information is listed in the order of orientation. The first element of the
ordered pair indicates a 0-face; the second indicates an edge of that
0-face. For example, if a space set is defined by the single circular list
$((1,1),(1,2),(1,1),(1,2))$, this means that there is one 2-face with four
sides, and that there is one 0-face with two sides (since the first elements of
the ordered pairs are all 1, and since the second elements of ordered pairs of
the form $(1,x)$ are 1 and 2). This space set is illustrated in
Figure~\ref{poincarespaceset}.

Let us use this third method, then, to describe space sets. The next step is to
determine simple transformations of space set descriptions which generate local
equivalences between space sets. In the one-dimensional case we found that one
set of moves that seemed to be sufficient for our purposes consisted of the
splitting, doubling and reducing moves (and their inverses). In higher
dimensions there are natural analogues of these moves. I will call them by the
same names, except that the doubling moves will now be called the
\emph{topological equivalence} moves.

In the splitting moves, we select a 0-face $V\!$. Let $\{H_i\,|\,1\leq i\leq
m\}$ denote the set of hinges contained in $V\!$. We select one of these hinges
(call it $H_j$). Let $S$ be the set of all $0n$-components in $n$-faces which
are attached to $H_j$. We break $S$ into disjoint subsets $S_1$ and $S_2$ such
that $S_1\cup S_2=S$. Now we remove our original 0-face $V$ and replace it with
two copies of itself, $V_1$ and $V_2$. In so doing, we have removed each hinge
$H_i$ in $V$ and replaced it with $H_{i1}$ in $V_1$ and with $H_{i2}$ in
$V_2$. We attach each $0n$-component in $S_1$ to $H_{j1}$ and each component in
$S_2$ to $H_{j2}$. Next we replace each $n$-face $F_k$ with $2^{r_k}$
$n$-faces, where $r_k$ is the number of $0n$-components of $F_k$ that were
adjacent to some $H_i$ for $i\neq j$. These faces are obtained by replacing
each such $H_i$ by $H_{i1}$ or $H_{i2}$ in all possible ways.

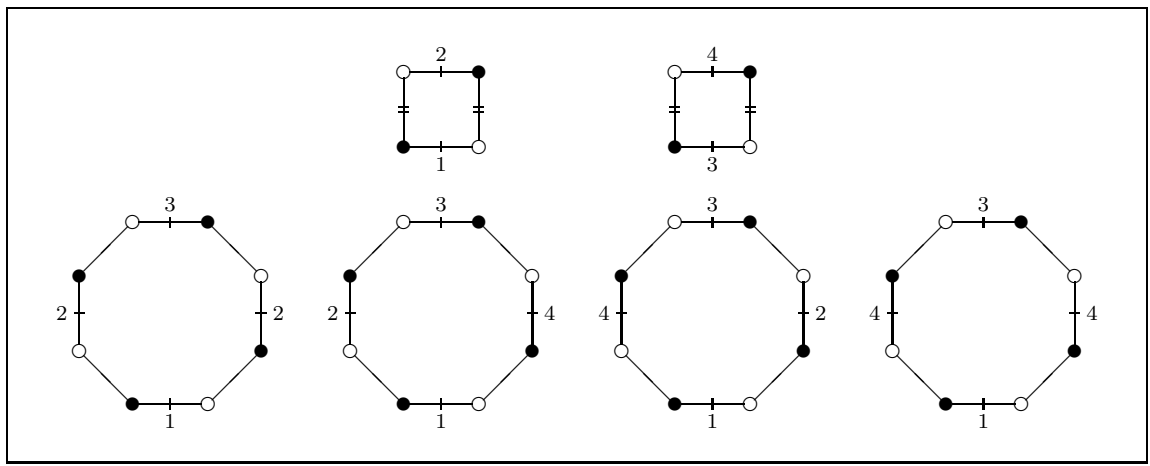
\begin{figure}
\fbox{\setlength{\unitlength}{.95pt}\begin{picture}(447.37,174.42)\scriptsize
\multiput(154.71,122.42)(107.96,0){2}{
\put(0,0){\circle*{5}}\put(30,30){\circle*{5}}\put(0,30){\circle{5}}
\put(30,0){\circle{5}}\multiput(2.5,0)(0,30){2}{\put(0,0){\line(1,0){25}}\put(12.5,-2){\line(0,1){4}}}
\multiput(0,2.5)(30,0){2}{\put(0,0){\line(0,1){25}}\multiput(-2,11.5)(0,2){2}{\line(1,0){4}}}}
\put(154.71,122.42){\put(15,-7){\makebox(0,0){1}}\put(15,37){\makebox(0,0){2}}
\put(107.96,0){\put(15,-7){\makebox(0,0){3}}\put(15,37){\makebox(0,0){4}}}}
\multiput(61.75,56.21)(107.96,0){4}{
\put(-15,36.21){\circle{5}}\put(15,36.21){\circle*{5}}
\put(36.21,15){\circle{5}}\put(36.21,-15){\circle*{5}}
\put(15,-36.21){\circle{5}}\put(-15,-36.21){\circle*{5}}
\put(-36.21,-15){\circle{5}}\put(-36.21,15){\circle*{5}}
\multiput(-12.5,36.21)(0,-72.42){2}{\put(0,0){\line(1,0){25}}\put(12.5,-2){\line(0,1){4}}}
\multiput(36.21,-12.5)(-72.42,0){2}{\put(0,0){\line(0,1){25}}\put(-2,12.5){\line(1,0){4}}}
\put(-34.44,16.77){\line(1,1){17.68}}\put(-34.44,-16.77){\line(1,-1){17.68}}
\put(34.44,-16.77){\line(-1,-1){17.68}}\put(34.44,16.77){\line(-1,1){17.68}}}
\put(61.75,56.21){\put(0,-43.21){\makebox(0,0){1}}
\put(0,43.21){\makebox(0,0){3}}
\put(-43.21,0){\makebox(0,0){2}}
\put(43.21,0){\makebox(0,0){2}}
\put(107.96,0){\put(0,-43.21){\makebox(0,0){1}}
\put(0,43.21){\makebox(0,0){3}}
\put(-43.21,0){\makebox(0,0){2}}
\put(43.21,0){\makebox(0,0){4}}
\put(107.96,0){\put(0,-43.21){\makebox(0,0){1}}
\put(0,43.21){\makebox(0,0){3}}
\put(-43.21,0){\makebox(0,0){4}}
\put(43.21,0){\makebox(0,0){2}}
\put(107.96,0){\put(0,-43.21){\makebox(0,0){1}}
\put(0,43.21){\makebox(0,0){3}}
\put(-43.21,0){\makebox(0,0){4}}
\put(43.21,0){\makebox(0,0){4}}}}}}

\end{picture}}\linespread{1}\caption{Splitting the space set pictured in
Figure~\ref{poincarespaceset}. The edge labelled 1 in that figure occurs twice
in the 2-face. The goal is to create two different edges, one which hits the
2-face in the first location and one which hits the 2-face in the second
location. To do this we must replace the original 0-face with two 0-faces. The
edges of 0-faces are all distinct, so now we have four edges. Edge 1 in the
original figure has now become edges 1 and 3. The two edges labelled 1 in the
2-face are relabelled 1 and 3. But also edge 2 has now become edges 2 and 4. So
it is necessary to make four copies of the 2-face so that the edges that were
formerly labelled 2 now are labelled 2 or 4 in every possible way.}
\label{twodsplit}
\end{figure}

The procedure is illustrated in Figure~\ref{twodsplit}.
Note that our splitting procedure in one dimension is a special case of this
more general definition of splitting. The move is invertible; the inverse of
this move is also an allowed move.

The topological equivalence moves on space set descriptions are analogues of
the moves that generate topological equivalence on spaces (e.g., in
$n$-graphs or in Poincar\'{e} graphs). In fact, the moves are exactly the
same in both cases except that if we have a space set description then we
cannot apply the move unless there is no branching in a sufficiently large
region of the description. Whenever there is no branching in a region of a
description, then in that region the description looks like a portion of a
single space. We may then apply a topological equivalence move to that
region. The result is that the move in question is applied to every space that
contains the region.

An example of this is the doubling move in the one-dimensional case. Suppose
there is a vertex in a one-dimensional space set description such that only one
edge comes into the vertex and only one edge goes out. Then there is no
branching in this region of the description; the edge-vertex-edge sequence
looks like a piece of a single space. This is exactly when we may apply the
undoubling move: we remove the vertex and combine the two edges into a single
edge. The result is that the same move is applied to each space that contains
those two edges and vertex. The move is valid since if any space contains any
part of the edge-vertex-edge sequence, then it contains the rest of that
sequence. It is easy to see that the doubling and undoubling moves generate
topological equivalence on 1-graphs.

It would be useful, then, to possess a finite set of local moves that generate
topological equivalence on $n$-graphs or on Poincar\'{e} $n$-graphs for
any $n$. At the moment I have only solved this problem in two
dimensions. Recall from Chapter~\ref{examples} that the $k$-dipole moves were
proven by Ferri and Gagliardi to generate topological equivalence in the case
where one is restricted to manifolds. I would prefer not to impose this
restriction, for reasons mentioned earlier: differentiating manifolds from
pseudomanifolds is a complex business involving computation of homology groups
at best, and a solution of the Poincar\'{e} conjecture at worst. So it seems
desirable to find a set of moves which are suitable to the general case of
pseudomanifolds.

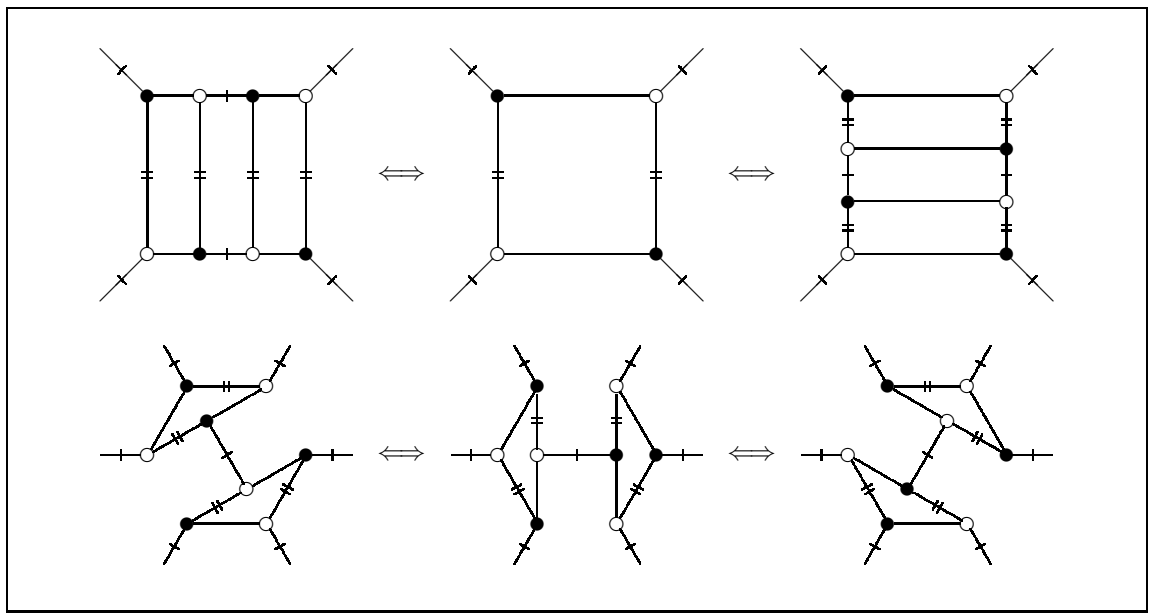
\begin{figure}
\fbox{\begin{picture}(425,222)\footnotesize
\multiput(50,132)(132.5,0){3}{\put(0,0){\circle{5}}\put(0,60){\circle*{5}}
\put(60,0){\circle*{5}}\put(60,60){\circle{5}}
\put(-1.77,-1.77){\put(0,0){\line(-1,-1){16}}\put(-8,-8){\qbezier(1.41,-1.41)(0,0)(-1.41,1.41)}}
\put(-1.77,61.77){\put(0,0){\line(-1,1){16}}\put(-8,8){\qbezier(1.41,1.41)(0,0)(-1.41,-1.41)}}
\put(61.77,-1.77){\put(0,0){\line(1,-1){16}}\put(8,-8){\qbezier(1.41,1.41)(0,0)(-1.41,-1.41)}}
\put(61.77,61.77){\put(0,0){\line(1,1){16}}\put(8,8){\qbezier(1.41,-1.41)(0,0)(-1.41,1.41)}}}
\put(50,132){\put(20,0){\circle*{5}}\put(40,0){\circle{5}}\put(20,60){\circle{5}}\put(40,60){\circle*{5}}\multiput(0,0)(0,60){2}{\multiput(2.5,0)(20,0){3}{\line(1,0){15}}}\multiput(30,-2)(0,60){2}{\line(0,1){4}}
\multiput(0,0)(20,0){4}{\put(0,2.5){\line(0,1){55}}\multiput(-2,29)(0,2){2}{\line(1,0){4}}}
\put(132.5,0){\multiput(0,0)(0,60){2}{\put(2.5,0){\line(1,0){55}}}
\multiput(0,0)(60,0){2}{\put(0,2.5){\line(0,1){55}}\multiput(-2,29)(0,2){2}{\line(1,0){4}}}
\put(132.5,0){\multiput(0,0)(0,20){4}{\put(2.5,0){\line(1,0){55}}}
\put(0,20){\circle*{5}}\put(0,40){\circle{5}}\put(60,20){\circle{5}}\put(60,40){\circle*{5}}\multiput(0,0)(60,0){2}{\multiput(0,2.5)(0,20){3}{\line(0,1){15}}}\multiput(-2,30)(60,0){2}{\line(1,0){4}}\multiput(0,0)(60,0){2}{\multiput(-2,9)(0,40){2}{\multiput(0,0)(0,2){2}{\line(1,0){4}}}}}}}
\multiput(146.25,162)(132.5,0){2}{\makebox(0,0){$\Longleftrightarrow$}}

\multiput(146.25,56)(132.5,0){2}{\makebox(0,0){$\Longleftrightarrow$}}
\put(0,56){\multiput(80,0)(132.5,0){3}{
\put(-30,0){\circle{5}}\put(-15,25.98){\circle*{5}}\put(15,25.98){\circle{5}}
\put(30,0){\circle*{5}}\put(15,-25.98){\circle{5}}\put(-15,-25.98){\circle*{5}}
\put(-32.5,0){\line(-1,0){15}}\put(32.5,0){\line(1,0){15}}\multiput(-40,-2)(80,0){2}{\line(0,1){4}}\qbezier(-16.25,28.15)(-20,34.65)(-23.75,41.15)\qbezier(16.25,28.15)(20,34.65)(23.75,41.15)\qbezier(-16.25,-28.15)(-20,-34.65)(-23.75,-41.15)\qbezier(16.25,-28.15)(20,-34.65)(23.75,-41.15)
\put(-20,34.65){\qbezier(-1.73,-1)(0,0)(1.73,1)}
\put(20,34.65){\qbezier(1.73,-1)(0,0)(-1.73,1)}
\put(-20,-34.65){\qbezier(-1.73,1)(0,0)(1.73,-1)}
\put(20,-34.65){\qbezier(-1.73,-1)(0,0)(1.73,1)}}
\put(80,0){\put(-7.5,12.99){\circle*{5}}\put(7.5,-12.99){\circle{5}}
\multiput(0,0)(15,-25.98){2}{\qbezier(-27.83,1.25)(-18.75,6.495)(-9.67,11.74)\qbezier(-5.33,14.24)(3.75,19.485)(12.83,24.73)\put(-18.75,6.495){\multiput(-.865,-.5)(1.73,1){2}{\qbezier(1,-1.73)(0,0)(-1,1.73)}}}\qbezier(-6.25,10.82)(0,0)(6.25,-10.82)\qbezier(-1.73,-1)(0,0)(1.73,1)
\qbezier(-28.75,2.17)(-22.5,12.99)(-16.25,23.81)\put(-12.5,25.98){\line(1,0){25}}\qbezier(28.75,-2.17)(22.5,-12.99)(16.25,-23.81)\put(12.5,-25.98){\line(-1,0){25}}
        \multiput(-1,27.98)(2,0){2}{\line(0,-1){4}}
        \multiput(22,-13.855)(1,1.73){2}{\qbezier(1.73,-1)(0,0)(-1.73,1)}
\put(132.5,0){\put(-15,0){\circle{5}}\put(15,0){\circle*{5}}
\multiput(-15,-23.48)(30,0){2}{\multiput(0,0)(0,25.48){2}{\line(0,1){20,98}}}\put(-12.5,0){\line(1,0){25}}\put(0,-2){\line(0,1){4}}
\qbezier(-28.75,2.17)(-22.5,12.99)(-16.25,23.81)\qbezier(16.25,23.81)(22.5,12.99)(28.75,2.17)\qbezier(28.75,-2.17)(22.5,-12.99)(16.25,-23.81)\qbezier(-16.25,-23.81)(-22.5,-12.99)(-28.75,-2.17)
        \multiput(-17,11.99)(0,2){2}{\multiput(0,0)(30,0){2}{\line(1,0){4}}}
        \multiput(22,-13.855)(1,1.73){2}{\qbezier(1.73,-1)(0,0)(-1.73,1)}
        \multiput(-22,-13.855)(-1,1.73){2}{\qbezier(-1.73,-1)(0,0)(1.73,1)}
\put(132.5,0){\put(7.5,12.99){\circle{5}}\put(-7.5,-12.99){\circle*{5}}
\put(-12.5,25.98){\line(1,0){25}}\qbezier(16.25,23.81)(22.5,12.99)(28.75,2.17)\put(12.5,-25.98){\line(-1,0){25}}\qbezier(-16.25,-23.81)(-22.5,-12.99)(-28.75,-2.17)
\multiput(0,0)(-15,-25.98){2}{\qbezier(27.83,1.25)(18.75,6.495)(9.67,11.74)\qbezier(5.33,14.24)(-3.75,19.485)(-12.83,24.73)\put(18.75,6.495){\multiput(.865,-.5)(-1.73,1){2}{\qbezier(-1,-1.73)(0,0)(1,1.73)}}}\qbezier(6.25,10.82)(0,0)(-6.25,-10.82)\qbezier(1.73,-1)(0,0)(-1.73,1)
        \multiput(-1,27.98)(2,0){2}{\line(0,-1){4}}
        \multiput(-22,-13.855)(-1,1.73){2}{\qbezier(-1.73,-1)(0,0)(1.73,1)}}}}}

\end{picture}}\linespread{1}\caption{Topological equivalence moves on
Poincar\'{e} 2-graphs. The two moves at top involving the central configuration
and any single move from the bottom (and their inverses) are sufficient to
generate topological equivalence.}
\label{twodtopol}
\end{figure}

Meanwhile, we may examine the two-dimensional case. The $k$-dipole moves
clearly are sufficient for 2-graphs. However, we are considering Poincar\'{e}
graphs; the $k$-dipole moves do not preserve the property of being a
Poincar\'{e} graph. Nevertheless it is not difficult to find a set of moves
which generate topological equivalence of Poincar\'{e} graphs. Such a set is
shown in Figure~\ref{twodtopol}.
(I shall omit the proof of the sufficiency of these moves; it is not difficult.)

Consider the move from the upper row of this figure which goes from the central
configuration to the right-hand configuration. In order to apply this move to a
space set description, we need to find a region of the description that looks
like the central configuration. This may be done as follows. Choose a 0-edge on
a 2-face. Draw in the four-sided 02-polygon containing this edge. The 1-edges
on either side of this 0-edge in the 2-face are adjacent to hinges in
0-faces. Draw these 0-faces in. Now the 02-polygon is bordered on three
sides. It remains to see how one may glue a 2-face to the fourth side. If there
is only one way to do this, then locally the space set looks like the central
configuration, and the move may be applied. The procedure is similar for the
other moves.

Now for the reducing moves. In one dimension this move deletes a 0-face (and
all adjacent 1-faces) if it cannot be ``completed'': that is, if it cannot be
surrounded by 1-faces. In $n$~dimensions one wishes also to be allowed to
remove a 0-face (and all adjacent \mbox{$n$-faces}) if it cannot be surrounded
by $n$-faces. One such situation occurs when there is a hinge in the 0-face
which is not attached to any $0n$-components of $n$-faces; the removal of the
0-face is certainly allowed in this case. Unfortunately, this move is not
sufficient for reducing purposes if $n$ is greater than one. The problem is
that, while each hinge of the 0-face may be attached to at least one
$0n$-component of an $n$-face, there may be no set of $0n$-components of
$n$-faces which can be attached to each of the hinges of the 0-face
simultaneously without causing some sort of incompatibility. A similar problem
may occur while trying to complete an $n$-face. These situations do not arise
in one dimension.

\begin{figure}
\fbox{\setlength{\unitlength}{.95pt}\begin{picture}(447.37,172.42)\scriptsize

\multiput(123.72,86.21)(199.93,0){2}{
\put(-15,36.21){\circle{5}}\put(15,36.21){\circle*{5}}
\put(36.21,15){\circle{5}}\put(36.21,-15){\circle*{5}}
\put(15,-36.21){\circle{5}}\put(-15,-36.21){\circle*{5}}
\put(-36.21,-15){\circle{5}}\put(-36.21,15){\circle*{5}}
\multiput(-12.5,36.21)(0,-72.42){2}{\put(0,0){\line(1,0){25}}\put(12.5,-2){\line(0,1){4}}}
\multiput(36.21,-12.5)(-72.42,0){2}{\put(0,0){\line(0,1){25}}\put(-2,12.5){\line(1,0){4}}}
\qbezier(-34.44,16.77)(-25.61,25.61)(-16.77,34.44)
\qbezier(-34.44,-16.77)(-25.61,-25.61)(-16.77,-34.44)
\qbezier(34.44,-16.77)(25.61,-25.61)(16.77,-34.44)
\qbezier(34.44,16.77)(25.61,25.61)(16.77,34.44)
\multiput(-15,38.71)(30,0){2}{\multiput(0,0)(0,-102.42){2}{\put(0,0){\line(0,1){25}}\multiput(-2,11.5)(0,2){2}{\line(1,0){4}}}}
\multiput(38.71,-15)(0,30){2}{\multiput(0,0)(-102.42,0){2}{\put(0,0){\line(1,0){25}}\multiput(11.5,-2)(2,0){2}{\line(0,1){4}}}}
\multiput(-66.21,-12.5)(132.42,0){2}{\put(0,0){\line(0,1){25}}\put(-2,12.5){\line(1,0){4}}}
\multiput(-12.5,-66.21)(0,132.42){2}{\put(0,0){\line(1,0){25}}\put(12.5,-2){\line(0,1){4}}}
\put(-15,66.21){\circle*{5}}\put(15,66.21){\circle{5}}
\put(66.21,15){\circle*{5}}\put(66.21,-15){\circle{5}}
\put(15,-66.21){\circle*{5}}\put(-15,-66.21){\circle{5}}
\put(-66.21,-15){\circle*{5}}\put(-66.21,15){\circle{5}}
\qbezier(-64.44,16.77)(-40.605,40.605)(-16.77,64.44)
\qbezier(64.44,16.77)(40.605,40.605)(16.77,64.44)
\qbezier(64.44,-16.77)(40.605,-40.605)(16.77,-64.44)
\qbezier(-64.44,-16.77)(-40.605,-40.605)(-16.77,-64.44)}
\put(123.72,86.21){
\put(0,29.21){\makebox(0,0){3}}
\put(0,-29.21){\makebox(0,0){1}}
\put(29.21,0){\makebox(0,0){2}}
\put(-29.21,0){\makebox(0,0){2}}
\put(0,73.21){\makebox(0,0){4}}
\put(0,-73.21){\makebox(0,0){2}}
\put(73.21,0){\makebox(0,0){1}}
\put(-73.21,0){\makebox(0,0){1}}
\put(199.93,0){\put(0,29.21){\makebox(0,0){3}}
\put(0,-29.21){\makebox(0,0){1}}
\put(29.21,0){\makebox(0,0){4}}
\put(-29.21,0){\makebox(0,0){4}}
\put(0,73.21){\makebox(0,0){4}}
\put(0,-73.21){\makebox(0,0){2}}
\put(73.21,0){\makebox(0,0){3}}
\put(-73.21,0){\makebox(0,0){3}}}}

\end{picture}}\linespread{1}\caption{Reducing the space set in
Figure~\ref{twodsplit} by completing 2-faces. The first and last 2-faces in
that figure appear here as the central 2-face in the left and right diagrams,
respectively. Then 0-faces are added in every possible way (there is only one
way to do this compatible with the requirement that all 02-polygons have four
sides). The result in each case is that another complete 2-face is forced to be
part of the diagram. But these additional 2-faces are not part of the space
set. Hence the first and last 2-faces in Figure~\ref{twodsplit} may be deleted
without changing the space set.}
\label{reduce}
\end{figure}
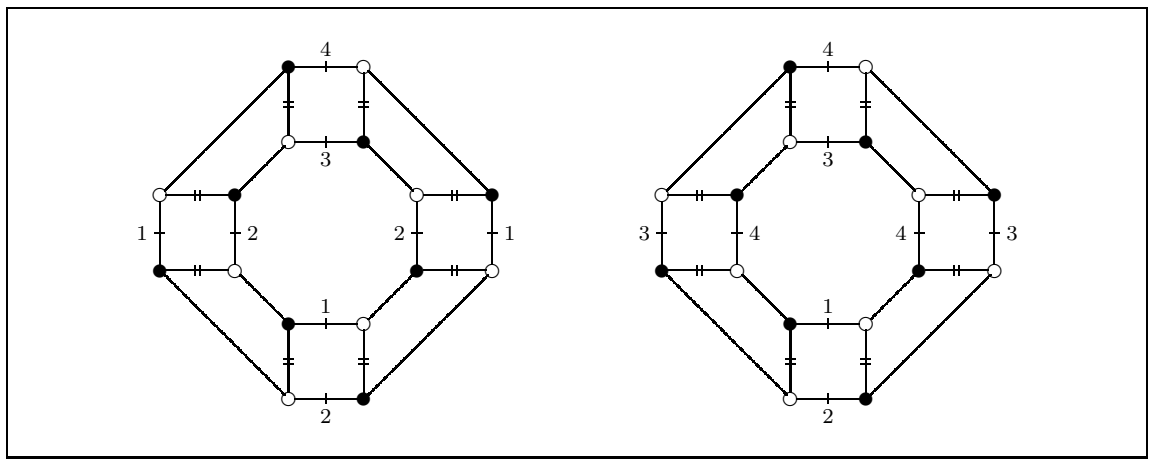

For example, in two dimensions when we surround a 2-face by 0-faces it is
implied that 2-faces with certain properties need to exist. If they do not,
then the original 2-face cannot be surrounded by 0-faces in that way. If there
is no way to surround the 2-face with 0-faces, then the 2-face cannot be part
of any space in the space set; hence it may be removed. This situation is
illustrated in Figure~\ref{reduce}.

A simpler reducing move in two dimensions involves the completion of an edge,
rather than a vertex. Recall that an edge is a 02-polygon, which is required
to have four sides. Given a 0-edge on a 2-face, one must be able to surround
the four-sided 02-polygon that is adjacent to that 0-edge by 0-faces and
2-faces. If this cannot be done, then the 2-face may not be contained in a
space, so it may be removed. I do not know whether this edge-reducing move (in
concert with splitting moves and topological-equivalence moves) is sufficient
to generate all other sorts of reducing moves in two dimensions.

In summary, the situation is rather complex. In the one-dimensional setting we
were able to generate all transformations using a single elementary operation:
namely, removing a 0-face. One might hope that this same sort of procedure
would be possible in $n$ dimensions as well, and that it would make life
simpler. However, in two dimensions I have found that removing a 0-face is a
pretty complicated operation.

Currently I am in the process of writing a computer program which generates
two-dimensional space set descriptions and transforms them using various
moves. From this I hope to gain some intuitions concerning whether the moves I
am considering do everything I want them to do, and what a minimal set of
generators for these moves might be.

One test for whether a set of moves is sufficient is that all space set
descriptions which describe the empty space set ought to be transformable into
one another using elementary moves. If they are not, I will conclude that my
set of moves needs to be expanded.

\begin{figure}
\fbox{\setlength{\unitlength}{.95pt}\begin{picture}(447.37,507.26)\scriptsize

\multiput(100.73,457.26)(107.96,0){3}{
\put(0,0){\circle*{5}}\put(30,30){\circle*{5}}\put(0,30){\circle{5}}
\put(30,0){\circle{5}}\multiput(2.5,0)(0,30){2}{\put(0,0){\line(1,0){25}}\put(12.5,-2){\line(0,1){4}}}
\multiput(0,2.5)(30,0){2}{\put(0,0){\line(0,1){25}}\multiput(-2,11.5)(0,2){2}{\line(1,0){4}}}}
\put(100.73,457.26){\put(15,-7){\makebox(0,0){1}}\put(15,37){\makebox(0,0){2}}
\put(107.96,0){\put(15,-7){\makebox(0,0){3}}\put(15,37){\makebox(0,0){4}}
\put(107.96,0){\put(15,-7){\makebox(0,0){5}}\put(15,37){\makebox(0,0){6}}}}}
\multiput(61.75,391.05)(107.96,0){4}{
\put(-15,36.21){\circle{5}}\put(15,36.21){\circle*{5}}
\put(36.21,15){\circle{5}}\put(36.21,-15){\circle*{5}}
\put(15,-36.21){\circle{5}}\put(-15,-36.21){\circle*{5}}
\put(-36.21,-15){\circle{5}}\put(-36.21,15){\circle*{5}}
\multiput(-12.5,36.21)(0,-72.42){2}{\put(0,0){\line(1,0){25}}\put(12.5,-2){\line(0,1){4}}}
\multiput(36.21,-12.5)(-72.42,0){2}{\put(0,0){\line(0,1){25}}\put(-2,12.5){\line(1,0){4}}}
\put(-34.44,16.77){\line(1,1){17.68}}\put(-34.44,-16.77){\line(1,-1){17.68}}
\put(34.44,-16.77){\line(-1,-1){17.68}}\put(34.44,16.77){\line(-1,1){17.68}}}
\put(61.75,391.05){\put(0,-43.21){\makebox(0,0){1}}
\put(0,43.21){\makebox(0,0){3}}
\put(-43.21,0){\makebox(0,0){2}}
\put(43.21,0){\makebox(0,0){4}}
\put(107.96,0){\put(0,-43.21){\makebox(0,0){1}}
\put(0,43.21){\makebox(0,0){3}}
\put(-43.21,0){\makebox(0,0){6}}
\put(43.21,0){\makebox(0,0){4}}
\put(107.96,0){\put(0,-43.21){\makebox(0,0){5}}
\put(0,43.21){\makebox(0,0){3}}
\put(-43.21,0){\makebox(0,0){4}}
\put(43.21,0){\makebox(0,0){2}}
\put(107.96,0){\put(0,-43.21){\makebox(0,0){5}}
\put(0,43.21){\makebox(0,0){3}}
\put(-43.21,0){\makebox(0,0){4}}
\put(43.21,0){\makebox(0,0){6}}}}}}

\multiput(46.75,284.84)(107.96,0){4}{
\put(0,0){\circle*{5}}\put(30,30){\circle*{5}}\put(0,30){\circle{5}}
\put(30,0){\circle{5}}\multiput(2.5,0)(0,30){2}{\put(0,0){\line(1,0){25}}\put(12.5,-2){\line(0,1){4}}}
\multiput(0,2.5)(30,0){2}{\put(0,0){\line(0,1){25}}\multiput(-2,11.5)(0,2){2}{\line(1,0){4}}}}
\put(46.75,284.84){\put(15,-7){\makebox(0,0){1}}\put(15,37){\makebox(0,0){2}}
\put(107.96,0){\put(15,-7){\makebox(0,0){3}}\put(15,37){\makebox(0,0){4}}
\put(107.96,0){\put(15,-7){\makebox(0,0){5}}\put(15,37){\makebox(0,0){6}}
\put(107.96,0){\put(15,-7){\makebox(0,0){7}}\put(15,37){\makebox(0,0){8}}}}}}
\multiput(61.75,218.63)(107.96,0){4}{
\put(-15,36.21){\circle{5}}\put(15,36.21){\circle*{5}}
\put(36.21,15){\circle{5}}\put(36.21,-15){\circle*{5}}
\put(15,-36.21){\circle{5}}\put(-15,-36.21){\circle*{5}}
\put(-36.21,-15){\circle{5}}\put(-36.21,15){\circle*{5}}
\multiput(-12.5,36.21)(0,-72.42){2}{\put(0,0){\line(1,0){25}}\put(12.5,-2){\line(0,1){4}}}
\multiput(36.21,-12.5)(-72.42,0){2}{\put(0,0){\line(0,1){25}}\put(-2,12.5){\line(1,0){4}}}
\put(-34.44,16.77){\line(1,1){17.68}}\put(-34.44,-16.77){\line(1,-1){17.68}}
\put(34.44,-16.77){\line(-1,-1){17.68}}\put(34.44,16.77){\line(-1,1){17.68}}}
\put(61.75,218.63){\put(0,-43.21){\makebox(0,0){1}}
\put(0,43.21){\makebox(0,0){3}}
\put(-43.21,0){\makebox(0,0){2}}
\put(43.21,0){\makebox(0,0){4}}
\put(107.96,0){\put(0,-43.21){\makebox(0,0){1}}
\put(0,43.21){\makebox(0,0){3}}
\put(-43.21,0){\makebox(0,0){2}}
\put(43.21,0){\makebox(0,0){8}}
\put(107.96,0){\put(0,-43.21){\makebox(0,0){5}}
\put(0,43.21){\makebox(0,0){7}}
\put(-43.21,0){\makebox(0,0){4}}
\put(43.21,0){\makebox(0,0){6}}
\put(107.96,0){\put(0,-43.21){\makebox(0,0){5}}
\put(0,43.21){\makebox(0,0){7}}
\put(-43.21,0){\makebox(0,0){8}}
\put(43.21,0){\makebox(0,0){6}}}}}}

\multiput(123.72,86.21)(199.93,0){2}{
\put(-15,36.21){\circle{5}}\put(15,36.21){\circle*{5}}
\put(36.21,15){\circle{5}}\put(36.21,-15){\circle*{5}}
\put(15,-36.21){\circle{5}}\put(-15,-36.21){\circle*{5}}
\put(-36.21,-15){\circle{5}}\put(-36.21,15){\circle*{5}}
\multiput(-12.5,36.21)(0,-72.42){2}{\put(0,0){\line(1,0){25}}\put(12.5,-2){\line(0,1){4}}}
\multiput(36.21,-12.5)(-72.42,0){2}{\put(0,0){\line(0,1){25}}\put(-2,12.5){\line(1,0){4}}}
\qbezier(-34.44,16.77)(-25.61,25.61)(-16.77,34.44)
\qbezier(-34.44,-16.77)(-25.61,-25.61)(-16.77,-34.44)
\qbezier(34.44,-16.77)(25.61,-25.61)(16.77,-34.44)
\qbezier(34.44,16.77)(25.61,25.61)(16.77,34.44)
\multiput(-15,38.71)(30,0){2}{\multiput(0,0)(0,-102.42){2}{\put(0,0){\line(0,1){25}}\multiput(-2,11.5)(0,2){2}{\line(1,0){4}}}}
\multiput(38.71,-15)(0,30){2}{\multiput(0,0)(-102.42,0){2}{\put(0,0){\line(1,0){25}}\multiput(11.5,-2)(2,0){2}{\line(0,1){4}}}}
\multiput(-66.21,-12.5)(132.42,0){2}{\put(0,0){\line(0,1){25}}\put(-2,12.5){\line(1,0){4}}}
\multiput(-12.5,-66.21)(0,132.42){2}{\put(0,0){\line(1,0){25}}\put(12.5,-2){\line(0,1){4}}}
\put(-15,66.21){\circle*{5}}\put(15,66.21){\circle{5}}
\put(66.21,15){\circle*{5}}\put(66.21,-15){\circle{5}}
\put(15,-66.21){\circle*{5}}\put(-15,-66.21){\circle{5}}
\put(-66.21,-15){\circle*{5}}\put(-66.21,15){\circle{5}}
\qbezier(-64.44,16.77)(-40.605,40.605)(-16.77,64.44)
\qbezier(64.44,16.77)(40.605,40.605)(16.77,64.44)
\qbezier(64.44,-16.77)(40.605,-40.605)(16.77,-64.44)
\qbezier(-64.44,-16.77)(-40.605,-40.605)(-16.77,-64.44)}
\put(123.72,86.21){
\put(0,29.21){\makebox(0,0){3}}
\put(0,-29.21){\makebox(0,0){1}}
\put(29.21,0){\makebox(0,0){4}}
\put(-29.21,0){\makebox(0,0){2}}
\put(0,73.21){\makebox(0,0){4}}
\put(0,-73.21){\makebox(0,0){2}}
\put(73.21,0){\makebox(0,0){3}}
\put(-73.21,0){\makebox(0,0){1}}
\put(199.93,0){\put(0,29.21){\makebox(0,0){7}}
\put(0,-29.21){\makebox(0,0){5}}
\put(29.21,0){\makebox(0,0){6}}
\put(-29.21,0){\makebox(0,0){8}}
\put(0,73.21){\makebox(0,0){8}}
\put(0,-73.21){\makebox(0,0){6}}
\put(73.21,0){\makebox(0,0){5}}
\put(-73.21,0){\makebox(0,0){7}}}}

\end{picture}}\linespread{1}\caption{Separating a finite space set. The space
set in Figure~\ref{poincarespaceset} was split in Figure~\ref{twodsplit} and
reduced in Figure~\ref{reduce}. Edge 1 occurs twice in the resulting space set;
this set is now split to produce the description shown in the first two rows of
this figure. The middle two 2-faces may then be removed by edge-reducing. Edge
3 occurs twice in the resulting space set; this set is now split to produce the
description shown in the next two rows. Again the middle two 2-faces may be
removed by edge-reducing. The two spaces in the resulting space set are
disjoint: each 0-face and 2-face is used in only one of the spaces.}
\label{finitespaceset}
\end{figure}
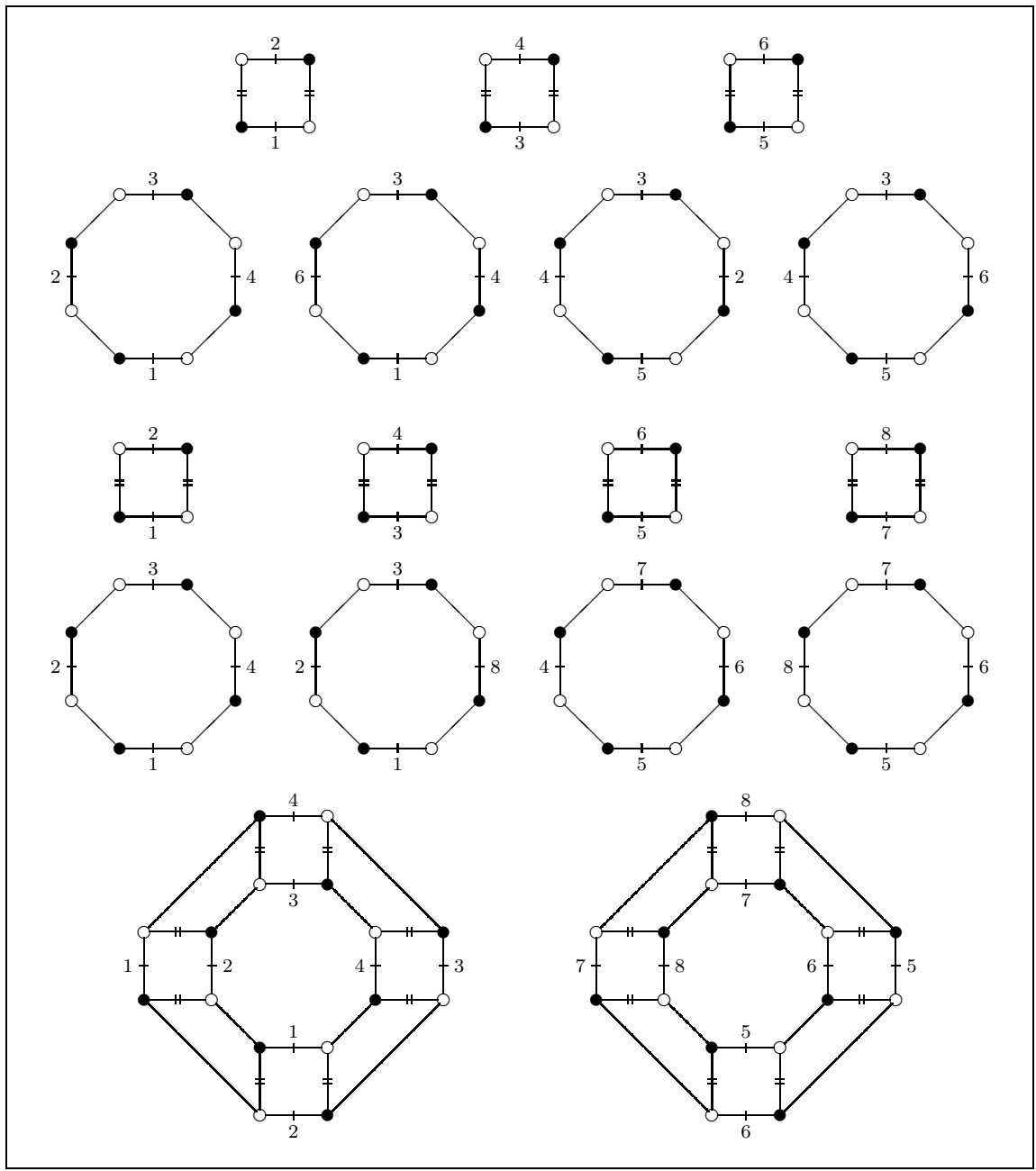

More generally, I suspect (but am by no means certain) that all space set
descriptions which describe a finite set of spaces should be transformable into
one another if and only if the sets contain equal number of spaces having the
same topology. To see why I suspect this, examine Figure~\ref{finitespaceset}.
Here a space set containing two spaces is transformed so that the descriptions
of these spaces are disjoint from one another. (This possibility of ``freezing
out'' certain finite spaces so that their descriptions are disjoint from the
descriptions of other spaces was not present in the one-dimensional case.) If
such disjointness can always be achieved for finite space sets, then the
topological equivalence moves imply that the above conjecture about space set
equivalence does in fact hold.

\subsection{Multidimensional bialgebras}
\label{highdimbialgebras}

My other approach to defining higher-dimensional space sets is tensorial in
nature. It is an attempt to generalize the bialgebra approach that proved
successful in the oriented one-dimensional case.

Consider an $n$-graph which represents an $n$-dimensional
pseudomanifold. In the oriented case there are two kinds of vertices (colored
white and black), and each edge connects one kind of vertex to the
other. Thus the edges are directed (we may adopt the convention that edges
point from black vertices to white vertices). Let each vertex be represented by
a tensor. The $n+1$ different colors of edges refer to $n+1$ different
vector spaces. So we need two types of tensors: $V^{a_0\ldots a_n}$ represents
a black vertex (since edges point away from black vertices) and $V_{a_0\ldots
a_n}$ represents a white vertex (since edges point towards white vertices). The
position of an index in one of these tensors is important, since each position
is associated with a different vector space. Hence an upper index in the $i$th
position may only be traced with a lower index in the $i$th position. The
scalars in these $V$'s correspond exactly to the set of $n$-graphs.

For example, if $n=2$ then $V_{abc}V^{abc}$ is a sphere; it is the upper
left-hand space in Figure~\ref{twographs}. Similarly
$V_{afb}V_{cdh}V_{ile}V_{kgj}V^{ade}V^{cgb}V^{ifj}V^{klh}$ is a torus; it is
the bottom right-hand space in that figure.

Our aim, though, is to represent \emph{sets} of spaces tensorially, not just
single spaces. Thus it is desirable to introduce some sort of branching into
this picture. One natural idea is the following: instead of allowing a tensor
to possess only one index per vector space, allow it to possess many such
indices. From the graphical point of view, we are allowing many edges of each
color to hit a vertex. Call this an $n$-multigraph. Each vertex and edge in
the multigraph is given a distinct label. We may now potentially construct more
than one space from such a multigraph: select vertices in the multigraph, and
at each selected vertex select one edge of each color in such a way that each
edge is selected either at both of its adjacent vertices or at neither. The
resulting subgraph is an $n$-graph, and thus represents an
$n$-pseudomanifold.

But clearly this is not what we are after. Space sets of this sort can only
contain finitely many spaces. In a matrix description a given edge or vertex in
the graph \g{A} can be present many times in a single space. We want the same
to be true here.

So, a natural next step is to suppose that the spaces associated with an
$n$-multigraph are those labelled $n$-graphs having the properties that
the labelled vertices and edges in the graph are copies of those in the
multigraph, and that a vertex and edge are adjacent in the graph only if their
counterparts are adjacent in the multigraph. But here we run into the same
problem found earlier: there is no limitation on the size of $n$-faces if this
method is used. If a four-sided 2-face is present in the multigraph, it is
possible that a graph in the space set might contain an associated $4k$-sided
2-face for any positive $k$.

This problem can be eliminated if one supposes that the space set associated
with a multigraph is made up of those spaces that can be constructed from the
$n$-faces present in the multigraph. An $n$-face in an $n$-multigraph is a
$k$-component contained in the multigraph $(0\leq k\leq n)$. That is: it is a
connected subgraph containing no $k$-edges and having the property that every
vertex is adjacent to exactly one $i$-edge for each $i\neq k$.

Note that adjacency relations between faces are given automatically by the fact
that each vertex and edge of the multigraph has a unique label. If an
$n$-graph (with labelled vertices and edges) can be written down in such a
way that each $n$-face in the graph is also an $n$-face in the multigraph, then
any two adjacent $n$-faces in the graph must also be adjacent in the
multigraph. For example, in the one-dimensional case a $k$-component consists
of two vertices connected either by a 0-edge or a 1-edge. Two such components
can be adjacent only if their common vertex has the same label (and of course
the rules of 1-graphs must be obeyed; hence for instance a 0-component cannot
be next to a 0-component).

\begin{figure}
\fbox{\begin{picture}(425,205)\scriptsize
\put(62.5,115){\begin{picture}(80,70)
\multiput(0,0)(50,0){2}{
        \qbezier(15,10)(5,25)(5,35)\qbezier(5,35)(5,45)(15,60)
        \qbezier(15,10)(25,20)(25,35)\qbezier(25,35)(25,45)(15,60)}
\put(0,0){\put(15,5){\circle{10}}\put(15,65){\circle*{10}}
        \qbezier(15,10)(-15,10)(-15,35)\qbezier(-15,35)(-15,60)(15,60)}
\put(50,0){\put(15,5){\circle*{10}}\put(15,65){\circle{10}}
        \qbezier(15,10)(45,10)(45,35)\qbezier(45,35)(45,60)(15,60)}
\multiput(0,0)(0,60){2}{\qbezier(20,5)(40,5)(60,5)\put(39,3){\line(0,1){4}}\put(41,3){\line(0,1){4}}}
\put(23,35){\line(1,0){4}}\put(53,35){\line(1,0){4}}
\put(-20,35){\makebox(0,0){$a$}}
\put(0,35){\makebox(0,0){$b$}}
\put(19,35){\makebox(0,0){$c$}}
\put(62,35){\makebox(0,0){$d$}}
\put(80,35){\makebox(0,0){$e$}}
\put(100,35){\makebox(0,0){$f$}}
\put(40,72){\makebox(0,0){$g$}}
\put(40,-2){\makebox(0,0){$h$}}
\end{picture}}

\put(210,115){
\multiput(0,0)(50,0){2}{\put(15,5){\circle{10}}\put(15,65){\circle*{10}}
        \qbezier(15,10)(5,25)(5,35)\qbezier(5,35)(5,45)(15,60)
        \qbezier(15,10)(25,20)(25,35)\qbezier(25,35)(25,45)(15,60)}
\multiput(100,0)(50,0){2}{\put(15,5){\circle*{10}}\put(15,65){\circle{10}}
        \qbezier(15,10)(5,25)(5,35)\qbezier(5,35)(5,45)(15,60)
        \qbezier(15,10)(25,20)(25,35)\qbezier(25,35)(25,45)(15,60)}
\put(0,35){\makebox(0,0){$a$}}
\put(31,35){\makebox(0,0){$c$}}
\put(23,35){\line(1,0){4}}
\put(50,0){\put(0,35){\makebox(0,0){$b$}}
\put(31,35){\makebox(0,0){$c$}}
\put(23,35){\line(1,0){4}}
\put(50,0){\put(-1,35){\makebox(0,0){$d$}}
\put(30,35){\makebox(0,0){$e$}}
\put(3,35){\line(1,0){4}}
\put(50,0){\put(-1,35){\makebox(0,0){$d$}}
\put(30,35){\makebox(0,0){$f$}}
\put(3,35){\line(1,0){4}}}}}}

\put(37.5,25){\put(0,0){\circle{10}}\put(40,0){\circle*{10}}
        \put(0,60){\circle*{10}}\put(40,60){\circle{10}}
        \put(-6,30){\makebox(0,0){$c$}}\put(47,30){\makebox(0,0){$d$}}
        \put(20,-7){\makebox(0,0){$h$}}\put(20,67){\makebox(0,0){$g$}}
        \multiput(5,0)(0,60){2}{\put(0,0){\line(1,0){30}}\multiput(14,-2)(2,0){2}{\line(0,1){4}}}
        \multiput(-2,30)(40,0){2}{\line(1,0){4}}
        \multiput(0,5)(40,0){2}{\line(0,1){50}}}
\multiput(136.5,25)(72,0){4}{\put(0,0){\circle{10}}\put(40,0){\circle*{10}}
        \put(0,60){\circle*{10}}\put(40,60){\circle{10}}
        \put(20,-7){\makebox(0,0){$h$}}\put(20,67){\makebox(0,0){$g$}}
        \multiput(5,0)(0,60){2}{\put(0,0){\line(1,0){30}}\multiput(14,-2)(2,0){2}{\line(0,1){4}}}
        \multiput(0,5)(40,0){2}{\line(0,1){50}}}
\put(136.5,25){\put(-5,30){\makebox(0,0){$a$}}\put(45,30){\makebox(0,0){$e$}}
\put(72,0){\put(-5,30){\makebox(0,0){$a$}}\put(45,30){\makebox(0,0){$f$}}
\put(72,0){\put(-5,30){\makebox(0,0){$b$}}\put(45,30){\makebox(0,0){$e$}}
\put(72,0){\put(-5,30){\makebox(0,0){$b$}}\put(45,30){\makebox(0,0){$f$}}}}}}
\end{picture}}\linespread{1}\caption{A two-dimensional space set represented
by a multigraph. The multigraph is displayed at upper left. The remainder of
the figure shows the nine 2-faces contained in the multigraph. The resulting
space set contains four spaces.}
\label{multigraph}
\end{figure}
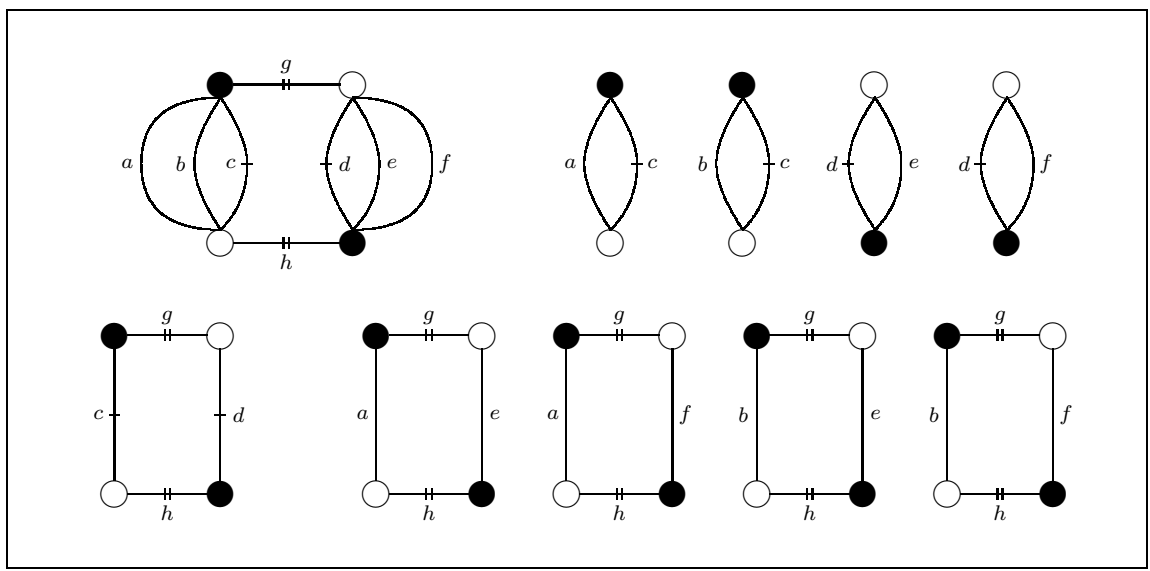

A two-dimensional example is given in Figure~\ref{multigraph}.
Note that the sort of branching in multigraphs seems to have a different
character than the sort of branching that occurs in the various face/hinge
formulations. It may still be the case that the multigraph formulation is
equivalent to one or more of the face/hinge formulations, but such an
equivalence would seem to be nontrivial.

Tensorially, we may attempt to express a multigraph by writing down one tensor
for each type of multigraph vertex. This means that there must be infinitely
many tensor types: for any sequence of positive integers $k_i$, $0\leq i\leq
n$, we need a tensor with $k_i$ indices associated with vector space $i$ for
each $i$. It would be nicer if, as in the case of standard bialgebras, one
could express the desired algebraic system using finitely many tensor types.

A natural way to attempt to do this is to introduce symmetric branching tensors
$T(i)_{a_1\ldots a_k}^c$ and $T(i)^{a_1\ldots a_k}_c$ for each vector space $i$
and for each nonnegative integer $k$. Now we may use the original $V$'s and
these $T$'s in combination to indicate branching. If a white vertex is adjacent
to $k_i$ edges of type $i$ for each $i$, then we will write this vertex as
$V_{x_0\ldots x_n}T(0)^{x_0}_{a_{01}\ldots a_{0k_0}}\cdots
T(n)^{x_n}_{a_{n1}\ldots a_{nk_n}}$. Black vertices are treated similarly. For
example, the tensor
$V^{wcg}V_{xch}V^{ydh}V_{zdg}T_w^{ab}T^{ab}_xT_y^{ef}T^z_{ef}$ represents the
multigraph in Figure~\ref{multigraph}. Note that it is not necessary to
indicate the vector space associated with each $T$ here, since it is implicit:
if a $T$ is traced with the $i$th index of a $V$ tensor, then it must be
associated with the $i$th vector space.

The situation is similar to that of a commutative cocommutative bialgebra. Each
$T(i)^c_{a_1\ldots a_k}$ can be thought of as a commutative multiplication
operator. Then if $k>1$ we may write $T(i)$ in terms of tensors of the form
$T(i)^x_{ab}$ satisfying the commutativity and associativity axioms. If we reverse
upper and lower indices we get tensors of the form $T(i)_x^{ab}$ satisfying the
cocommutativity and coassociativity axioms. If $k=1$ then we have tensors of
the form $T(i)^c_a$. These indicate that there is a single edge going from $a$
to $c$; hence $T(i)^c_a=\delta(i)^c_a$ where $\delta(i)$ refers to the
Kronecker delta on vector space $i$. If $k=0$ then we have tensors of the form
$T(i)^c$. These play the role of the unit in a bialgebra. Reversing upper
and lower indices gives the counit $T(i)_c$. Both of these tensors indicate
that a pathway comes to a dead end. Hence it is reasonable to suppose that
$T(i)_{ab}^cT(i)_c=T(i)_aT(i)_b$, $T(i)_{ab}^cT(i)^a=\delta(i)_b^c$, and
$T(i)_aT(i)^a=1$.

In short, for each vector space $i$ we have abstract tensors $T(i)_{ab}^c$,
$T(i)^{ab}_c$, $T(i)^c$, $T(i)_c$ and $\delta(i)_a^b$ satisfying all of the
axioms of a commutative cocommutative bialgebra \emph{except} for the main
bialgebra axiom $T(i)_{ab}^cT(i)_c^{de}=T_a^{wx}T_b^{yz}T_{wy}^dT_{xz}^e$
relating multiplication and comultiplication. It is tempting to think that this
axiom should be included also. But this turns out to be mistaken, as we shall
see.

The next step is to introduce additional axioms which generate local
equivalence transformations of these space sets. We already have axioms which
describe relationships among the $T$'s. We still need axioms that describe
relationships among the $V$'s, and axioms that relate the $T$'s to the $V$'s.

The former sort of axioms are the topological equivalence moves. Just as in the
previous section, the idea is to perform these moves at a place where no
branching is involved. Thus these are moves on the $V$'s alone. The moves need
to be tensorial in nature; that is, they must be local on the tensor
diagram. (The moves should be of the form ``the product of these tensors traced
with one another in this way is equal to the product of those tensors traced
with one another in that way.'') The dipole moves do not satisfy this
requirement, since they involve checking the properties of components of
subgraphs. Hence new moves are required. At present I know some promising moves
in $n$ dimensions, but have only proven their sufficiency in the
two-dimensional case.

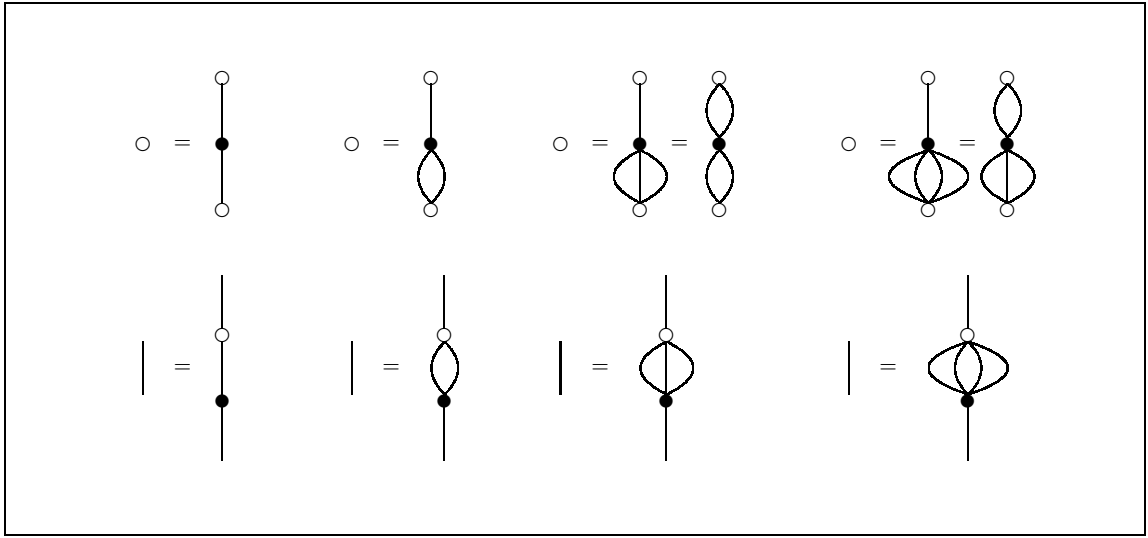
\begin{figure}[t!]
\fbox{\begin{picture}(425,195)\scriptsize

\put(49,145){
\put(0,0){\circle{5}}
\put(15,0){\makebox(0,0){$=$}}
\put(30,0){\put(0,0){\circle*{5}}\put(0,2.5){\line(0,1){20}}
\put(0,-2.5){\line(0,-1){20}}\put(0,25){\circle{5}}\put(0,-25){\circle{5}}}
}
\put(49,60){
\put(0,-10){\line(0,1){20}}
\put(15,0){\makebox(0,0){$=$}}
\put(30,0){\put(0,-12.5){\circle*{5}}\put(0,12.5){\circle{5}}
\put(0,-10){\line(0,1){20}}\put(0,-15){\line(0,-1){20}}\put(0,15){\line(0,1){20}}}
}
\put(128,145){
\put(0,0){\circle{5}}
\put(15,0){\makebox(0,0){$=$}}
\put(30,0){\put(0,0){\circle*{5}}\put(0,2.5){\line(0,1){20}}
\qbezier(0,-2.5)(-5,-7.5)(-5,-12.5)\qbezier(-5,-12.5)(-5,-17.5)(0,-22.5)
\qbezier(0,-2.5)(5,-7.5)(5,-12.5)\qbezier(5,-12.5)(5,-17.5)(0,-22.5)
\put(0,25){\circle{5}}\put(0,-25){\circle{5}}}
}
\put(128,60){
\put(0,-10){\line(0,1){20}}
\put(15,0){\makebox(0,0){$=$}}
\put(35,0){\put(0,-12.5){\circle*{5}}\put(0,12.5){\circle{5}}
\qbezier(0,-10)(-5,-5)(-5,0)\qbezier(-5,0)(-5,5)(0,10)
\qbezier(0,-10)(5,-5)(5,0)\qbezier(5,0)(5,5)(0,10)
\put(0,-15){\line(0,-1){20}}\put(0,15){\line(0,1){20}}}
}
\put(207,145){
\put(0,0){\circle{5}}
\put(15,0){\makebox(0,0){$=$}}
\put(30,0){\put(0,0){\circle*{5}}\put(0,2.5){\line(0,1){20}}
\qbezier(0,-2.5)(-10,-7.5)(-10,-12.5)\qbezier(-10,-12.5)(-10,-17.5)(0,-22.5)
\qbezier(0,-2.5)(10,-7.5)(10,-12.5)\qbezier(10,-12.5)(10,-17.5)(0,-22.5)
\put(0,-2.5){\line(0,-1){20}}
\put(0,25){\circle{5}}\put(0,-25){\circle{5}}}
\put(45,0){\makebox(0,0){$=$}}
\put(60,0){\put(0,0){\circle*{5}}
\qbezier(0,2.5)(5,7.5)(5,12.5)\qbezier(5,12.5)(5,17.5)(0,22.5)
\qbezier(0,2.5)(-5,7.5)(-5,12.5)\qbezier(-5,12.5)(-5,17.5)(0,22.5)
\qbezier(0,-2.5)(-5,-7.5)(-5,-12.5)\qbezier(-5,-12.5)(-5,-17.5)(0,-22.5)
\qbezier(0,-2.5)(5,-7.5)(5,-12.5)\qbezier(5,-12.5)(5,-17.5)(0,-22.5)
\put(0,25){\circle{5}}\put(0,-25){\circle{5}}}
}
\put(207,60){
\put(0,-10){\line(0,1){20}}
\put(15,0){\makebox(0,0){$=$}}
\put(40,0){\put(0,-12.5){\circle*{5}}\put(0,12.5){\circle{5}}
\qbezier(0,-10)(-10,-5)(-10,0)\qbezier(-10,0)(-10,5)(0,10)
\qbezier(0,-10)(10,-5)(10,0)\qbezier(10,0)(10,5)(0,10)
\put(0,-10){\line(0,1){20}}
\put(0,-15){\line(0,-1){20}}\put(0,15){\line(0,1){20}}}
}
\put(316,145){
\put(0,0){\circle{5}}
\put(15,0){\makebox(0,0){$=$}}
\put(30,0){\put(0,0){\circle*{5}}\put(0,2.5){\line(0,1){20}}
\qbezier(0,-2.5)(-5,-7.5)(-5,-12.5)\qbezier(-5,-12.5)(-5,-17.5)(0,-22.5)
\qbezier(0,-2.5)(5,-7.5)(5,-12.5)\qbezier(5,-12.5)(5,-17.5)(0,-22.5)
\qbezier(0,-2.5)(-15,-7.5)(-15,-12.5)\qbezier(-15,-12.5)(-15,-17.5)(0,-22.5)
\qbezier(0,-2.5)(15,-7.5)(15,-12.5)\qbezier(15,-12.5)(15,-17.5)(0,-22.5)
\put(0,25){\circle{5}}\put(0,-25){\circle{5}}}
\put(45,0){\makebox(0,0){$=$}}
\put(60,0){\put(0,0){\circle*{5}}
\qbezier(0,2.5)(5,7.5)(5,12.5)\qbezier(5,12.5)(5,17.5)(0,22.5)
\qbezier(0,2.5)(-5,7.5)(-5,12.5)\qbezier(-5,12.5)(-5,17.5)(0,22.5)
\qbezier(0,-2.5)(-10,-7.5)(-10,-12.5)\qbezier(-10,-12.5)(-10,-17.5)(0,-22.5)
\qbezier(0,-2.5)(10,-7.5)(10,-12.5)\qbezier(10,-12.5)(10,-17.5)(0,-22.5)
\put(0,-2.5){\line(0,-1){20}}
\put(0,25){\circle{5}}\put(0,-25){\circle{5}}}
}
\put(316,60){
\put(0,-10){\line(0,1){20}}
\put(15,0){\makebox(0,0){$=$}}
\put(45,0){\put(0,-12.5){\circle*{5}}\put(0,12.5){\circle{5}}
\qbezier(0,-10)(-15,-5)(-15,0)\qbezier(-15,0)(-15,5)(0,10)
\qbezier(0,-10)(15,-5)(15,0)\qbezier(15,0)(15,5)(0,10)
\qbezier(0,-10)(-5,-5)(-5,0)\qbezier(-5,0)(-5,5)(0,10)
\qbezier(0,-10)(5,-5)(5,0)\qbezier(5,0)(5,5)(0,10)
\put(0,-15){\line(0,-1){20}}\put(0,15){\line(0,1){20}}}
}
\end{picture}}\linespread{1}\caption{The first group of topological equivalence
moves on $n$-graphs. Here the moves for $n=1$ through 4 are shown (in column 1
through 4 of the figure). In the first row, color the edges in any way you like
such that the rules of $n$-graphs are maintained. This determines the free
edges of each diagram (I did not draw in the free edges). Each diagram contains
one free edge of each color. Each move exchanges two of these diagrams with the
free edges of the same color identified. A similar set of moves is obtained by
reversing vertex colors. In the second row, color the edges of the right-hand
diagrams in any way consistent with $n$-graph rules. Then the two free edges
have the same color $i$. The single edge in the left-hand diagrams denotes
$\delta(i)_a^b$. The bottom-row move generates the top-row move (in the same
column) that exchanges the first two diagrams.}
\label{topmovesone}
\end{figure}

The moves that I have found fall into two groups. Those in the first group
(see Figure~\ref{topmovesone})
are the following: \[V_{a_0\ldots a_rb_{r+1}\ldots b_n}V^{c_0\ldots
c_rb_{r+1}\ldots b_n}V_{c_0\ldots c_rd_{r+1}\ldots d_n}=V_{a_0\ldots
a_rd_{r+1}\ldots d_n}\] for $0\leq r<n$. These moves are special cases of
dipole moves in which the checking of components may be done locally. Note that
the $r=0$ move is implied by the simpler move
\[V_{ab_1\ldots b_n}V^{cb_1\ldots b_n}=\delta_a^c,\] so it is better to
substitute this move for the $r=0$ move.

Suppose we are about to apply one of these moves to an $n$-graph. Define
the face vector $f=(f_0\ldots f_n)$ where $f_k$ is the number of $k$-faces in
the pseudomanifold before applying a move. Let $f'$ be the face vector of the
pseudomanifold after the move is applied. Due to the symmetry of the situation
there are $\lfloor (n+1)/2\rfloor$ different types of moves. For each type of
move there is an associated constant vector $g_k$, independent of the
particular $n$-graph being operated on, such that $f'_k=f_k+g_k$. It turns
out that the $\lfloor (n+1)/2\rfloor$ vectors $g_k$ are independent. This means
that these moves cannot be sufficient to generate topological equivalence
unless there are $n+1-\lfloor (n+1)/2\rfloor=\lfloor n/2\rfloor+1$ independent
linear relations on face vectors that are topological invariants of
$n$-graphs. The Euler characteristic is one such relation; the relation
$(n+1)V=2E$ ($V$ is the number of vertices; $E$ is the number of edges) is
another. It is possible that $n$-graphs are such that other linear
invariants of this sort exist, but I doubt it. My guess is that if one imposes
restrictions on the types of allowed pseudomanifold, more such relations might
show up. For example, the Dehn-Sommerville relations (see~\cite{grunbaum}) are
linear relations among faces of simplicial complexes that hold when the
complexes are required to be Eulerian manifolds. Something like this might hold
also for $n$-graphs. But, without such a restriction, more
face-number-changing moves are probably needed.

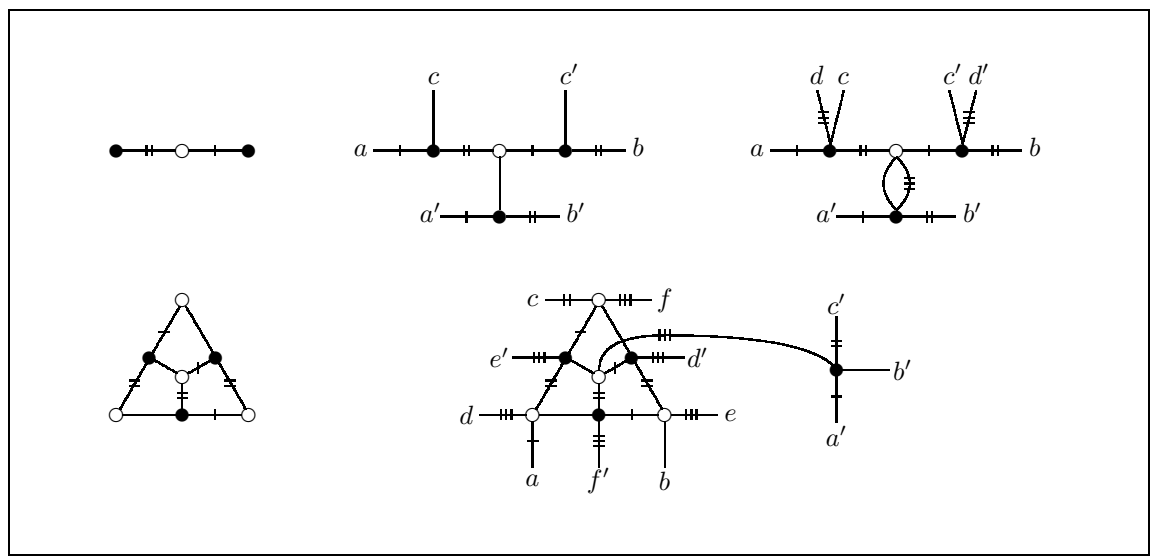
\begin{figure}\fbox{\begin{picture}(425,200)\footnotesize\settowidth{\wod}{$'$}

\put(62.5,150){
\put(-25,0){\circle*{5}}\put(0,0){\circle{5}}\put(25,0){\circle*{5}}
\put(-2.5,0){\line(-1,0){20}}\put(2.5,0){\line(1,0){20}}
\put(12.5,-2){\line(0,1){4}}\put(-11.5,-2){\line(0,1){4}}\put(-13.5,-2){\line(0,1){4}}
}
\put(182.5,150){
\put(-25,27.5){\makebox(0,0){\raisebox{0ex}[1ex][0ex]{$c$}}}
\put(25,27.5){\makebox(0,0){\raisebox{0ex}[1ex][0ex]{\hspace*{\wod}$c'$}}}
\put(-52.5,0){\makebox(0,0){\raisebox{0ex}[1ex][0ex]{$a$}}}
\put(52.5,0){\makebox(0,0){\raisebox{0ex}[1ex][0ex]{$b$}}}
\put(-27.5,-25){\makebox(0,0){\raisebox{0ex}[1ex][0ex]{\hspace*{\wod}$a'$}}}
\put(27.5,-25){\makebox(0,0){\raisebox{0ex}[1ex][0ex]{\hspace*{\wod}$b'$}}}
\put(-25,0){\circle*{5}}\put(0,0){\circle{5}}\put(25,0){\circle*{5}}
\put(0,-25){\circle*{5}}\put(0,-2.5){\line(0,-1){20}}\put(2.5,0){\line(1,0){20}}
\put(-2.5,0){\line(-1,0){20}}\put(27.5,0){\line(1,0){20}}\put(-27.5,0){\line(-1,0){20}}
\put(-25,2.5){\line(0,1){20}}\put(25,2.5){\line(0,1){20}}
\put(-2.5,-25){\line(-1,0){20}}\put(2.5,-25){\line(1,0){20}}
\put(12.5,-2){\line(0,1){4}}\put(-11.5,-2){\line(0,1){4}}\put(-13.5,-2){\line(0,1){4}}
\put(-12.5,-27){\line(0,1){4}}\put(11.5,-27){\line(0,1){4}}\put(13.5,-27){\line(0,1){4}}
\put(-37.5,-2){\line(0,1){4}}\put(36.5,-2){\line(0,1){4}}\put(38.5,-2){\line(0,2){4}}
}
\put(332.5,150){
\put(-52.5,0){\makebox(0,0){\raisebox{0ex}[1ex][0ex]{$a$}}}
\put(52.5,0){\makebox(0,0){\raisebox{0ex}[1ex][0ex]{$b$}}}
\put(-27.5,-25){\makebox(0,0){\raisebox{0ex}[1ex][0ex]{\hspace*{\wod}$a'$}}}
\put(27.5,-25){\makebox(0,0){\raisebox{0ex}[1ex][0ex]{\hspace*{\wod}$b'$}}}
\put(-25,0){\circle*{5}}\put(0,0){\circle{5}}\put(25,0){\circle*{5}}
\put(0,-25){\circle*{5}}
\put(0,-2.5){\qbezier(0,0)(-5,-5)(-5,-10)\qbezier(-5,-10)(-5,-15)(0,-20)
\qbezier(0,0)(5,-5)(5,-10)\qbezier(5,-10)(5,-15)(0,-20)}
\multiput(3,-10.5)(0,-2){3}{\line(1,0){4}}
\multiput(-29.5,10.5)(0,2){3}{\line(1,0){4}}
\multiput(29.5,10.5)(0,2){3}{\line(-1,0){4}}
\put(2.5,0){\line(1,0){20}}\put(-2.5,0){\line(-1,0){20}}
\put(27.5,0){\line(1,0){20}}\put(-27.5,0){\line(-1,0){20}}
\put(-25,2.5){\qbezier(0,0)(0,0)(-5,20)\qbezier(0,0)(0,0)(5,20)}
\put(25,2.5){\qbezier(0,0)(0,0)(-5,20)\qbezier(0,0)(0,0)(5,20)}
\put(-30,27.5){\makebox(0,0){\raisebox{0ex}[1ex][0ex]{$d$}}}
\put(-20,27.5){\makebox(0,0){\raisebox{0ex}[1ex][0ex]{$c$}}}
\put(20,27.5){\makebox(0,0){\raisebox{0ex}[1ex][0ex]{\hspace*{\wod}$c'$}}}
\put(30,27.5){\makebox(0,0){\raisebox{0ex}[1ex][0ex]{\hspace*{\wod}$d'$}}}
\put(-2.5,-25){\line(-1,0){20}}\put(2.5,-25){\line(1,0){20}}
\put(12.5,-2){\line(0,1){4}}\put(-11.5,-2){\line(0,1){4}}\put(-13.5,-2){\line(0,1){4}}
\put(-12.5,-27){\line(0,1){4}}\put(11.5,-27){\line(0,1){4}}\put(13.5,-27){\line(0,1){4}}
\put(-37.5,-2){\line(0,1){4}}\put(36.5,-2){\line(0,1){4}}\put(38.5,-2){\line(0,2){4}}
}
\put(62.5,50){
\put(0,0){\circle*{5}}\put(25,0){\circle{5}}\put(-25,0){\circle{5}}
\put(12.5,-2){\line(0,1){4}}\multiput(-2,6.215)(0,2){2}{\line(1,0){4}}
\multiput(-20.125,10.905)(0,2){2}{\line(1,0){4}}
\multiput(20.125,10.905)(0,2){2}{\line(-1,0){4}}
\put(-8.875,31.245){\line(1,0){4}}\put(6,15.8){\line(0,1){4}}
\put(0,14.43){\circle{5}}\put(0,43.3){\circle{5}}\put(12.5,21.65){\circle*{5}}
\put(-12.5,21.65){\circle*{5}}\put(-22.5,0){\line(1,0){45}}\put(0,2.5){\line(0,1){9.43}}
\qbezier(-23.75,2.16)(-12.5,21.65)(-1.25,40.84)\qbezier(-12.5,21.65)(-12.5,21.65)(-2.16,15.68)
\qbezier(23.75,2.16)(12.5,21.65)(1.25,40.84)\qbezier(12.5,21.65)(12.5,21.65)(2.16,15.68)
}
\put(220,50){
\put(0,0){\circle*{5}}\put(25,0){\circle{5}}\put(-25,0){\circle{5}}
\put(12.5,-2){\line(0,1){4}}\multiput(-2,6.215)(0,2){2}{\line(1,0){4}}
\multiput(-20.125,10.905)(0,2){2}{\line(1,0){4}}
\multiput(20.125,10.905)(0,2){2}{\line(-1,0){4}}
\put(-8.875,31.245){\line(1,0){4}}\put(6,15.8){\line(0,1){4}}
\put(0,14.43){\circle{5}}\put(0,43.3){\circle{5}}\put(12.5,21.65){\circle*{5}}
\put(-12.5,21.65){\circle*{5}}\put(-22.5,0){\line(1,0){45}}\put(0,2.5){\line(0,1){9.43}}
\qbezier(-23.75,2.16)(-12.5,21.65)(-1.25,40.84)\qbezier(-12.5,21.65)(-12.5,21.65)(-2.16,15.68)
\qbezier(23.75,2.16)(12.5,21.65)(1.25,40.84)\qbezier(12.5,21.65)(12.5,21.65)(2.16,15.68)
\qbezier(0,16.93)(0,30)(25,30)\qbezier(25,30)(80,30)(90,16.93)
\put(90,16.93){\circle*{5}}\multiput(23,28)(2,0){3}{\line(0,1){4}}
\put(90,16.93){\line(1,0){20}}\put(90,16.93){\line(0,1){20}}
\put(90,16.93){\line(0,-1){20}}\put(88,6.93){\line(1,0){4}}
\multiput(88,25.93)(0,2){2}{\line(1,0){4}}\put(27.5,0){\line(1,0){17.5}}
\put(25,-2.5){\line(0,-1){17.5}}\put(-27.5,0){\line(-1,0){17.5}}
\put(-25,-2.5){\line(0,-1){17.5}}\put(0,0){\line(0,-1){20}}
\put(-2.5,43.3){\line(-1,0){17.5}}\put(2.5,43.3){\line(1,0){17.5}}
\put(12.5,21.65){\line(1,0){20}}\put(-12.5,21.65){\line(-1,0){20}}
\put(-27,-10){\line(1,0){4}}\multiput(-2,-12)(0,2){3}{\line(1,0){4}}
\multiput(-37,-2)(2,0){3}{\line(0,1){4}}
\multiput(-13,41.3)(2,0){2}{\line(0,1){4}}
\multiput(-24.5,19.65)(2,0){3}{\line(0,1){4}}
\multiput(37,-2)(-2,0){3}{\line(0,1){4}}
\multiput(12,41.3)(-2,0){3}{\line(0,1){4}}
\multiput(24.5,19.65)(-2,0){3}{\line(0,1){4}}
\put(-25,-25){\makebox(0,0){$a$}}
\put(25,-25){\makebox(0,0){$b$}}
\put(-25,43.3){\makebox(0,0){$c$}}
\put(115,16.93){\makebox(0,0){$b'$}}
\put(90,-8.07){\makebox(0,0){$a'$}}
\put(90,41.93){\makebox(0,0){$c'$}}
\put(-50,0){\makebox(0,0){$d$}}
\put(50,0){\makebox(0,0){$e$}}
\put(25,43.3){\makebox(0,0){$f$}}
\put(37.5,21.65){\makebox(0,0){$d'$}}
\put(-37.5,21.65){\makebox(0,0){$e'$}}
\put(0,-25){\makebox(0,0){$f'$}}
}

\end{picture}}\linespread{1}\caption{Topological equivalence moves from group
two. Pictured at left in row one is a 1-ball and in row two is a 2-ball
(free vertices are not drawn). To the right in row 1 are two moves using the
1-ball (for $n=2$ and $n=3$). To the right in row 2 is a move using the 2-ball
(for $n=3$).}
\label{topmovestwo}
\end{figure}

The moves in the second group (see Figure~\ref{topmovestwo}) do not affect face
vectors. It is difficult to write down a general formula for these moves;
instead I will describe an inductive method of generating them.

To begin, consider the following construction in $n$ dimensions. We start by
drawing an $n$-simplex in the usual way. Now construct a graphical diagram
associated with this simplex. Place vertices at the center of each $k$-face of
the simplex, and color them white if $k$ is even and black if $k$ is odd. If a
$k$-face is contained in a $(k+1)$-face, draw an edge of the graph connecting
the vertex at the center of the $k$-face to the vertex at the center of the
$(k+1)$-face. For each white vertex at the corners of the simplex (i.e., the
vertices at the centers of the 0-faces) draw a ``free'' edge emerging from the
vertex and pointing away from the simplex.

Now each vertex in the diagram has degree $n+1$. It remains to color the
edges. Arbitrarily assign each of the colors 0 through $n$ to the free
edges. Next consider any other edge $e$. It joins the vertex at the center of a
$k$-face to the vertex at the center of a $(k+1)$-face that contains
it. Consider the free edges at the corners of the $(k+1)$-face. Exactly one of
these edges is not at a corner of the $k$-face. Give edge $e$ the same color as
the color of that free edge. The result is that every vertex in the diagram
meets exactly one edge of each color. Topologically, the diagram represents an
$n$-ball.

Call any such $k$-dimensional diagram a ``$k$-ball.'' For $0<k<n$ we construct
$n$-dimensional moves from these $k$-balls as follows. Let
$N=\{0,\ldots,n\}$. Choose a $k$-ball whose edge colors lie in some set
$K\subset N$ (hence $|K|=k+1$). Create a new vertex outside the $k$-ball. For
each of the $n-k$ colors in $N\!\setminus\!K$, attach an edge of that color
connecting the central vertex of the $k$-ball (i.e., the vertex that was placed
at the center of the $k$-face of the $k$-simplex) to the new vertex, and attach
a free edge of that color to each of the other vertices in the $k$-ball. For
each color in $K$, attach a free edge of that color to the new vertex. Now we
pair up the free edges and label them. The pairs of edges are labelled $a$ and
$a'$, $b$ and $b'$, and so on; each pair is labelled with a distinct
symbol. The pairing is as follows. If the color of a free edge attached to a
vertex on the $k$-ball is in $K$, it is paired with the free edge with the same
color that is attached to the new vertex. Otherwise, each free edge is paired
with the free edge of the same color that is attached to the opposite vertex of
the $k$-ball. (A vertex of the graph is the center of a face of the
simplex. This face is determined by $m$ corner vertices of the simplex. The
vertex of the graph that is at the center of the face determined by the
remaining corner vertices of the simplex is the opposite vertex.) The resulting
diagram is the left-hand side of the move. The right-hand side is a copy of the
same diagram, with primed and unprimed indices reversed, and with the colors of
the vertices reversed if $k$ is even.

Examples are given in Figure~\ref{topmovestwo}.
It is by no means self-evident that there exists a finite set of local moves
that generate topological equivalence in $n$ dimensions, let alone that these
groups of moves are sufficient for this purpose. In fact, I do not even know
whether or not the moves in group 2 preserve the topology when $n>2$. (They
seemed such a nice generalization of the $n=2$ group-2 move that I could not
resist writing them down.) What I do know is that these moves are indeed
sufficient in two dimensions, in which case there is one type of move in each
group. The moves in this case are similar to those for two-dimensional
Poincar\'{e} graphs (see Figure~\ref{twodtopol}).

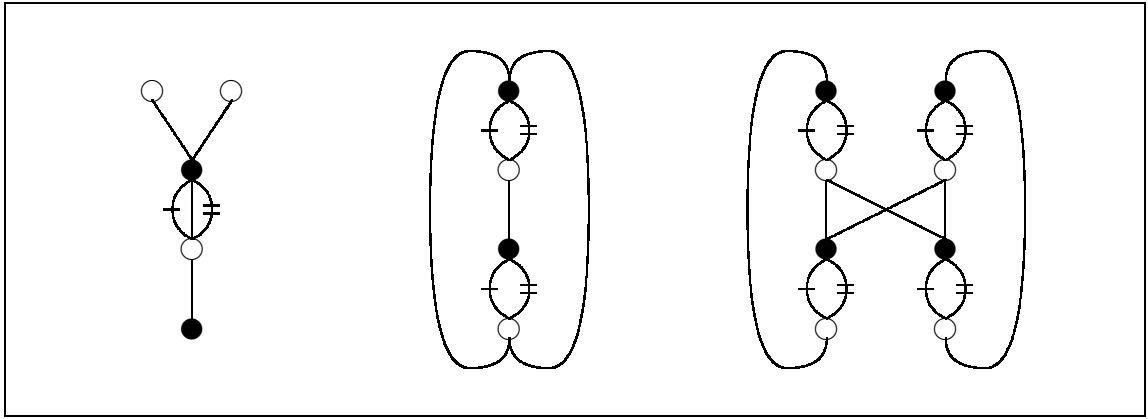
\begin{figure}
\fbox{\setlength{\unitlength}{1.5pt}\begin{picture}(283.33,100)\scriptsize
\put(35,20){
\put(10,0){\circle*{5}}\put(10,20){\circle{5}}\put(10,40){\circle*{5}}
\multiput(0,60)(20,0){2}{\circle{5}}
\put(10,0){\line(0,1){17.5}}\put(10,40){\line(0,-1){17.5}}
\qbezier(10,42.5)(15,50)(20,57.5)\qbezier(10,42.5)(5,50)(0,57.5)
\qbezier(10,37.5)(5,35)(5,30)\qbezier(5,30)(5,25)(10,22.5)
\qbezier(10,37.5)(15,35)(15,30)\qbezier(15,30)(15,25)(10,22.5)
\put(3,30){\line(1,0){4}}\multiput(13,29)(0,2){2}{\line(1,0){4}}
}
\put(125,50){
\put(0,-30){\put(0,0){\circle{5}}\put(0,20){\circle*{5}}
        \qbezier(0,2.5)(5,5)(5,10)\qbezier(5,10)(5,15)(0,17.5)
        \qbezier(0,2.5)(-5,5)(-5,10)\qbezier(-5,10)(-5,15)(0,17.5)
        \put(-7,10){\line(1,0){4}}\multiput(3,9)(0,2){2}{\line(1,0){4}}}
\put(0,10){\put(0,0){\circle{5}}\put(0,20){\circle*{5}}
        \qbezier(0,2.5)(5,5)(5,10)\qbezier(5,10)(5,15)(0,17.5)
        \qbezier(0,2.5)(-5,5)(-5,10)\qbezier(-5,10)(-5,15)(0,17.5)
        \put(-7,10){\line(1,0){4}}\multiput(3,9)(0,2){2}{\line(1,0){4}}}
\put(0,-7.5){\line(0,1){15}}
\qbezier(0,32.5)(0,40)(-10,40)\qbezier(-10,40)(-20,40)(-20,0)
\qbezier(-20,0)(-20,-40)(-10,-40)\qbezier(-10,-40)(0,-40)(0,-32.5)
\qbezier(0,32.5)(0,40)(10,40)\qbezier(10,40)(20,40)(20,0)
\qbezier(20,0)(20,-40)(10,-40)\qbezier(10,-40)(0,-40)(0,-32.5)
}
\put(205,50){
\multiput(0,-30)(0,40){2}{\multiput(0,0)(30,0){2}{\put(0,0){\circle{5}}\put(0,20){\circle*{5}}
        \qbezier(0,2.5)(5,5)(5,10)\qbezier(5,10)(5,15)(0,17.5)
        \qbezier(0,2.5)(-5,5)(-5,10)\qbezier(-5,10)(-5,15)(0,17.5)
        \put(-7,10){\line(1,0){4}}\multiput(3,9)(0,2){2}{\line(1,0){4}}}}
\multiput(0,-7.5)(30,0){2}{\line(0,1){15}}
\qbezier(0,-7.5)(15,0)(30,7.5)\qbezier(30,-7.5)(15,0)(0,7.5)
\qbezier(0,32.5)(0,40)(-10,40)\qbezier(-10,40)(-20,40)(-20,0)
\qbezier(-20,0)(-20,-40)(-10,-40)\qbezier(-10,-40)(0,-40)(0,-32.5)
\put(30,0){\qbezier(0,32.5)(0,40)(10,40)\qbezier(10,40)(20,40)(20,0)
\qbezier(20,0)(20,-40)(10,-40)\qbezier(10,-40)(0,-40)(0,-32.5)}
}
\end{picture}}\linespread{1}\caption{An interpretation of $T$'s in terms of
multigraphs. The leftmost figure is a way to picture a tensor $T(0)_a^{bc}$
that is consistent with our foregoing discussion. This gives a means for
associating a multigraph with the tensors $T(0)_{ab}^cT(0)_c^{ab}$ (center) and
$T(0)_a^{wx}T(0)_b^{yz}T(0)_{wy}^aT(0)_{xz}^b$ (right). The center multigraph
contains two spaces (one containing the leftmost 0-edge and one containing the
rightmost 0-edge). The right multigraph contains three spaces (one containing
the leftmost 0-edge and not the rightmost 0-edge, one containing the rightmost
0-edge and not the leftmost 0-edge, and one containing both the leftmost and
rightmost 0-edges). Thus these multigraphs are not equivalent.}
\label{tscalars}
\end{figure}

A question which I have avoided thus far is this: what is the meaning of a
scalar that involves only $T$'s and no $V$'s? A clue is provided by the bottom
row of moves in the figure. They suggest a way of interpreting $T$'s which
gives a means to directly translate any scalar in $T$'s and $V$'s (whether
$V$'s are present or not) into a multigraph. This is shown in
Figure~\ref{tscalars}.
Also shown there are the multigraphs associated with the
scalars $T_{ab}^cT_c^{ab}$ and $T_a^{wx}T_b^{yz}T_{wy}^aT_{xz}^b$. The second
scalar is obtained from the first by applying the main bialgebra
axiom. However, the first multigraph describes a set of two spaces, and the
second graph describes a set of three spaces. So this axiom is not correct in
the present context.

Finally we need axioms relating the $V$'s and $T$'s. An obvious axiom of this
sort arises from the fact that if a vertex doesn't connect to anything along a
certain vector space, then it cannot be present in any pseudomanifold in the
space set. Hence it may be removed, and its connections to other tensors may be
replaced by units or counits. Thus $V_{a_0\ldots
a_n}T(0)^{a_0}=T(1)_{a_1}\cdots T(n)_{a_n}$. (For each axiom there is a
corresponding symmetry class of axioms generated by orientation reversal [i.e.,
switching upper indices with lower indices] and by permuting vector spaces. I
will only list one element of each symmetry class.)

This relates vertices to units; now it would be nice to relate vertices to the
multiplication and comultiplication tensors. A simple axiom of this sort is the
following analogue of the splitting axioms described previously. If the paths
emerging from a vertex along vector space $i$ branches into two parts, we may
replace this vertex with two vertices, one whose path along $i$ goes along one
branch, and one whose path along $i$ goes along the other branch. The paths
along vector spaces other than $i$ which led to the original vertex must now
themselves branch and go to each of the two new vertices. Thus we have
$V_{a_0\ldots a_n}T^{a_0}_{xy}=V_{x b_1\ldots b_n}V_{y c_1\ldots
c_n}T^{b_1c_1}_{a_1}\cdots T^{b_nc_n}_{a_n}$.

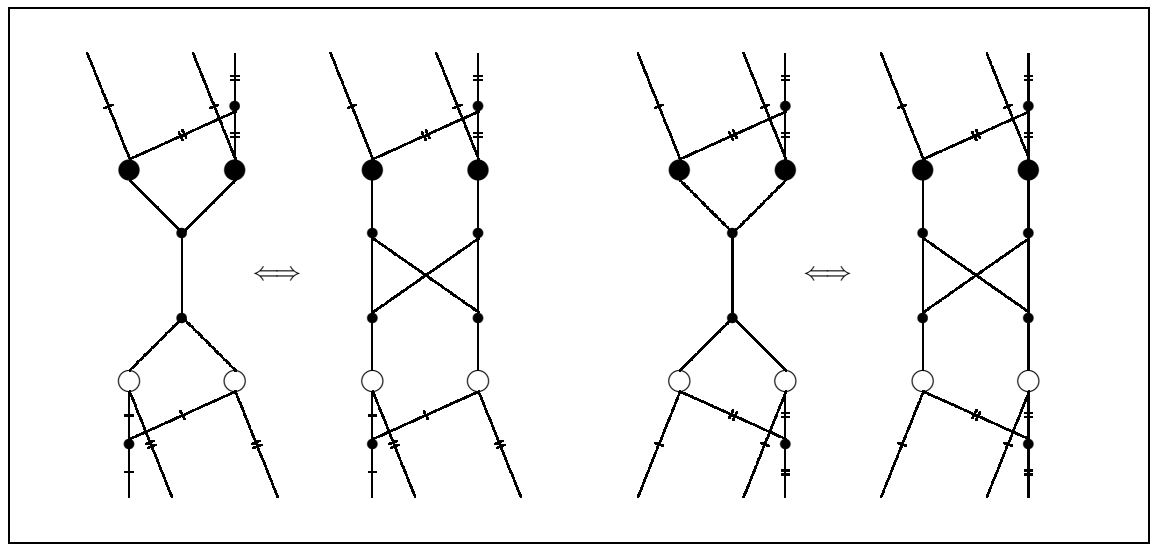
\begin{figure}
\fbox{\setlength{\unitlength}{.8pt}\begin{picture}(531.25,245)\footnotesize
\put(63,0){\put(15,122.5){
\multiput(-25,-50)(50,0){2}{\circle{10}}
\multiput(-25,50)(50,0){2}{\circle*{10}}
\multiput(0,-20)(0,40){2}{\circle*{5}}
\qbezier(-25,-45)(-12.5,-32.5)(0,-20)\qbezier(25,-45)(12.5,-32.5)(0,-20)
\qbezier(-25,45)(-12.5,32.5)(0,20)\qbezier(25,45)(12.5,32.5)(0,20)
\put(0,-20){\line(0,1){40}}
\put(-25,-55){\line(0,-1){50}}\put(-25,-80){\circle*{5}}
\put(-27,-66.25){\line(1,0){4}}\put(-27,-93.25){\line(1,0){4}}
\qbezier(25,-55)(0,-66.25)(-25,-77.5)\put(0,-66.25){\qbezier(-.89,1.78)(0,0)(.89,-1.78)}
\qbezier(-25,-55)(-15,-80)(-5,-105)\qbezier(25,-55)(35,-80)(45,-105)
\multiput(-15,-80)(50,0){2}{\multiput(-.37,.93)(.74,-1.86){2}{\qbezier(1.86,.74)(0,0)(-1.86,-.74)}}
\put(25,55){\line(0,1){50}}\put(25,80){\circle*{5}}
\multiput(27,65.25)(0,2){2}{\line(-1,0){4}}\multiput(27,92.25)(0,2){2}{\line(-1,0){4}}
\qbezier(-25,55)(0,66.25)(25,77.5)\put(0,66.25){\multiput(-.89,-.445)(1.78,.89){2}{\qbezier(.89,-1.78)(0,0)(-.89,1.78)}}
\qbezier(25,55)(15,80)(5,105)\qbezier(-25,55)(-35,80)(-45,105)
\multiput(15,80)(-50,0){2}{\qbezier(1.86,.74)(0,0)(-1.86,-.74)}
}
\put(60,122.5){\makebox(0,0){$\Longleftrightarrow$}}
\put(130,122.5){
\multiput(-25,-50)(50,0){2}{\circle{10}}
\multiput(-25,50)(50,0){2}{\circle*{10}}
\multiput(-25,-45)(50,0){2}{\line(0,1){90}}
\multiput(-25,-20)(50,0){2}{\multiput(0,0)(0,40){2}{\circle*{5}}}
\qbezier(-25,-17.5)(0,0)(25,17.5)\qbezier(25,-17.5)(0,0)(-25,17.5)
\put(-25,-55){\line(0,-1){50}}\put(-25,-80){\circle*{5}}
\put(-27,-66.25){\line(1,0){4}}\put(-27,-93.25){\line(1,0){4}}
\qbezier(25,-55)(0,-66.25)(-25,-77.5)\put(0,-66.25){\qbezier(-.89,1.78)(0,0)(.89,-1.78)}
\qbezier(-25,-55)(-15,-80)(-5,-105)\qbezier(25,-55)(35,-80)(45,-105)
\multiput(-15,-80)(50,0){2}{\multiput(-.37,.93)(.74,-1.86){2}{\qbezier(1.86,.74)(0,0)(-1.86,-.74)}}
\put(25,55){\line(0,1){50}}\put(25,80){\circle*{5}}
\multiput(27,65.25)(0,2){2}{\line(-1,0){4}}\multiput(27,92.25)(0,2){2}{\line(-1,0){4}}
\qbezier(-25,55)(0,66.25)(25,77.5)\put(0,66.25){\multiput(-.89,-.445)(1.78,.89){2}{\qbezier(.89,-1.78)(0,0)(-.89,1.78)}}
\qbezier(25,55)(15,80)(5,105)\qbezier(-25,55)(-35,80)(-45,105)
\multiput(15,80)(-50,0){2}{\qbezier(1.86,.74)(0,0)(-1.86,-.74)}
}}

\put(323.25,0){\put(15,122.5){
\multiput(-25,-50)(50,0){2}{\circle{10}}
\multiput(-25,50)(50,0){2}{\circle*{10}}
\multiput(0,-20)(0,40){2}{\circle*{5}}
\qbezier(-25,-45)(-12.5,-32.5)(0,-20)\qbezier(25,-45)(12.5,-32.5)(0,-20)
\qbezier(-25,45)(-12.5,32.5)(0,20)\qbezier(25,45)(12.5,32.5)(0,20)
\put(0,-20){\line(0,1){40}}
\put(25,-55){\line(0,-1){50}}\put(25,-80){\circle*{5}}
\multiput(27,-65.25)(0,-2){2}{\line(-1,0){4}}\multiput(27,-92.25)(0,-2){2}{\line(-1,0){4}}
\qbezier(-25,-55)(0,-66.25)(25,-77.5)\put(0,-66.25){\multiput(-.89,.445)(1.78,-.89){2}{\qbezier(.89,1.78)(0,0)(-.89,-1.78)}}
\qbezier(25,-55)(15,-80)(5,-105)\qbezier(-25,-55)(-35,-80)(-45,-105)
\multiput(15,-80)(-50,0){2}{\qbezier(1.86,-.74)(0,0)(-1.86,.74)}
\put(25,55){\line(0,1){50}}\put(25,80){\circle*{5}}
\multiput(27,65.25)(0,2){2}{\line(-1,0){4}}\multiput(27,92.25)(0,2){2}{\line(-1,0){4}}
\qbezier(-25,55)(0,66.25)(25,77.5)\put(0,66.25){\multiput(-.89,-.445)(1.78,.89){2}{\qbezier(.89,-1.78)(0,0)(-.89,1.78)}}
\qbezier(25,55)(15,80)(5,105)\qbezier(-25,55)(-35,80)(-45,105)
\multiput(15,80)(-50,0){2}{\qbezier(1.86,.74)(0,0)(-1.86,-.74)}
}
\put(60,122.5){\makebox(0,0){$\Longleftrightarrow$}}
\put(130,122.5){
\multiput(-25,-50)(50,0){2}{\circle{10}}
\multiput(-25,50)(50,0){2}{\circle*{10}}
\multiput(-25,-45)(50,0){2}{\line(0,1){90}}
\multiput(-25,-20)(50,0){2}{\multiput(0,0)(0,40){2}{\circle*{5}}}
\qbezier(-25,-17.5)(0,0)(25,17.5)\qbezier(25,-17.5)(0,0)(-25,17.5)
\put(25,-55){\line(0,-1){50}}\put(25,-80){\circle*{5}}
\multiput(27,-65.25)(0,-2){2}{\line(-1,0){4}}\multiput(27,-92.25)(0,-2){2}{\line(-1,0){4}}
\qbezier(-25,-55)(0,-66.25)(25,-77.5)\put(0,-66.25){\multiput(-.89,.445)(1.78,-.89){2}{\qbezier(.89,1.78)(0,0)(-.89,-1.78)}}
\qbezier(25,-55)(15,-80)(5,-105)\qbezier(-25,-55)(-35,-80)(-45,-105)
\multiput(15,-80)(-50,0){2}{\qbezier(1.86,-.74)(0,0)(-1.86,.74)}
\put(25,55){\line(0,1){50}}\put(25,80){\circle*{5}}
\multiput(27,65.25)(0,2){2}{\line(-1,0){4}}\multiput(27,92.25)(0,2){2}{\line(-1,0){4}}
\qbezier(-25,55)(0,66.25)(25,77.5)\put(0,66.25){\multiput(-.89,-.445)(1.78,.89){2}{\qbezier(.89,-1.78)(0,0)(-.89,1.78)}}
\qbezier(25,55)(15,80)(5,105)\qbezier(-25,55)(-35,80)(-45,105)
\multiput(15,80)(-50,0){2}{\qbezier(1.86,.74)(0,0)(-1.86,-.74)}
}}

\end{picture}}\linespread{1}\caption{Two two-dimensional moves suggested by the
main bialgebra move. The left one works; the right one doesn't. Here large
vertices (colored black and white) represent the $V$'s; small vertices
represent $T$'s. In the left-hand diagram of the first move, two 01-polygons
meet two 02-polygons at a 0-edge. In the right-hand diagram, each 01-polygon
and 02-polygons now corresponds to two polygons, and each 01-polygon can meet
exactly one 02-polygon. Thus in both cases there are four ways for the polygons
to meet at this juncture. In the second move, on the left four 01-polygons meet
one 02-polygon. On the right, it is not clear anymore how many 01-polygons
there are; there now might be new polygons that pass through this piece of the
multigraph twice. This is the source of the problem with the second move.}
\label{twodbialgmove}
\end{figure}

I suspect that these axioms are not sufficient. Presently I am puzzling over
the situation in Figure~\ref{twodbialgmove}. Here there are two versions of
moves in two dimensions which resemble the main bialgebra axiom. One of the
moves seems to be correct, and the other doesn't, as explained in the figure. I
do not know how to obtain the correct-seeming move using the moves mentioned so
far. So there is more work to be done~here.

Note that in one dimension the topological equivalence moves are
$V_{ab}V^{ac}=\delta_b^c$ and $V_{ba}V^{ca}=\delta_b^c$. We may add to this,
for example, the move
\[V_{ab}V^{ac}T^b_{de}T_c^{fg}=V_{rd}V_{se}V^{tf}V^{ug}T^r_{wx}T^s_{yz}T_t^{wy}T_u^{xz}.\]
It follows that all bialgebra moves hold on the $T$'s, that each scalar may be
expressed\linebreak\clearpage
\noindent entirely in terms of $T$'s, and that from the combinatorial spacetime
point of view this algebraic formulation is equivalent to the bialgebra
formulation.

To summarize, in $n$ dimensions our algebraic structure contains $n+1$ vector
spaces, the tensors $V$, and multiplication, comultiplication, unit and counit
tensors on each vector space $i$. The latter satisfy all axioms of a
commutative cocommutative bialgebra except for the main bialgebra axiom. There
is an axiom relating the $V$'s and the units and counits, and other axioms
(perhaps including generalizations of the main bialgebra axiom) relating the
$V$'s and the multiplication and comultiplication tensors. Finally there are
topological equivalence axioms on the $V$'s.

The construction of these algebraic systems is a work in progress. My hope is
that the finished systems will have the property that scalars will describe
sets of oriented $n$-dimensional spaces and that the axioms will generate all
local, invertible, topology-preserving transformations on these space sets. Of
course, this may not turn out to be possible. If it does turn out to be
possible, then it seems that these systems might be considered to be natural
$n$-dimensional generalizations of commutative cocommutative bialgebras. If we
then drop the commutativity and cocommutativity axioms from the $T$'s, perhaps
the resulting systems might be considered to be natural $n$-dimensional
generalizations of bialgebras.

\cleardoublepage\vspace*{40pt}\section[Conclusion]{\LARGE Conclusion}\bigskip

Clearly no unified field theory has emerged from these pages. In fact, no
direct contact has been made with any part of mainstream physics whatsoever.

However, worse things could have happened. The research program could have come
to a dead end. There might have been no obtainable general results of any
significant sort.

Quite the opposite has occurred. Repeated advances have been made. New and
elegant structure has emerged. Points of contact (often unexpected ones) have
been established with various more established areas of mathematics (including
Hopf algebras, which seemingly have some relevance to physics). No roadblocks
have appeared to prevent further progress.

As I see it, the evidence indicates that the point of view which I have tried
to put forth here (indeed, it is this point of view which to my mind
constitutes the main substance of this dissertation) is a mathematically
fruitful one. It remains to determine whether it will be a physically fruitful
one as well.

I suggest, then, that the project is worth continuing. Any and all interested
readers are encouraged to join in.

\cleardoublepage
 
\end{document}